\documentclass[10pt]{article}
\usepackage{graphicx}
\usepackage{amsmath}
\usepackage{amssymb}
\usepackage[figuresright]{rotating}
\usepackage{caption2}
\begin{document}
\begin{center}
{\large {\bf \sc{  Review of the QCD sum rules for exotic states
  }}} \\[2mm]
Zhi-Gang  Wang \footnote{E-mail: zgwang@aliyun.com.  }   \\
 Department of Physics, North China Electric Power University, Baoding 071003, P. R. China
\end{center}

\begin{abstract}
We review the exotic states, such as the $X$, $Y$, $Z$, $T$ and $P$ states, and present their possible assignments based on the QCD sum rules. We present many predictions which can be confronted to the experimental data in the future to diagnose the exotic states. Furthermore, we also mention other theoretical methods.
\end{abstract}

\tableofcontents

\section{Introduction}
In 1964, Gell-Mann suggested  that  multiquark states beyond the  minimal valence quark constituents $q\bar{q}$ and $qqq$  might  exist \cite{Gell-Mann-1964},
a quantitative model for the tetraquark states with the quark constituents $qq\bar{q}\bar{q}$ was developed by Jaffe using the MIT bag model in 1977 \cite{Jaffe-1977,Jaffe-1977-2}. Later, the five-quark baryons with the quark constituents $qqqq\bar{q}$ were developed \cite{Strottman-1979}, while the name pentaquark  was introduced by Lipkin \cite{Lipkin-1987}.
Also in 1964, Dyson and Xuang studied the dibaryon or six-quark states based on   the $SU(6)$ symmetry \cite{Dyson-1964}, for more literatures on this subject, one can consult Refs.\cite{Dibaryon-review-2017,Dibaryon-review-2000}.
The QCD  allows the existence of multiquark states and hybrid states which contain not
only quarks but also gluonic degrees of freedom \cite{Jaffe-1976-Glueball,Morgan-1987-Glueball-S,Close-1983-qqg}.

Before observation of the $X(3872)$ by the Belle collaboration in 2003 \cite{X3872-2003}, the most promising and most hot subject is the nature of the light mesons below $1\,\rm{GeV}$, are they traditional ${}^3P_0$ states, tetraquark states or molecular states? \cite{Jaffe-1977,Jaffe-1977-2,Scalar-1GeV-Weinstein-PRL-1982, Scalar-1GeV-KK-Weinstein-PRD-1990,Scalar-1GeV-KK-Achasov-PRD-1997,
 Scalar-1GeV-KK-Pennington-PRD-2002}. In fact, the observation of the $X(3872)$ stimulates more motivations and curiosities in exploring the nature of the light scalar mesons \cite{Maiani-2004-PRL-scalar,Jaffe-2003-diquark,
Scalar-1GeV-Nielsen-PLB-2005-QCDSR,
 Scalar-1GeV-WangZG-EPJC-2005-QCDSR,
Scalar-1GeV-WangZG-JPG-2005-QCDSR,Scalar-1GeV-Lee-EPJA-2006-QCDSR,
Scalar-1GeV-Lee-PLB-2006-QCDSR,Scalar-1GeV-Groote-PRD-2014-QCDSR,
Scalar-1GeV-WangZG-EPJC-2016-QCDSR,Scalar-1GeV-Azizi-PLB-2018-QCDSR,
Scalar-1GeV-Azizi-PLB-2019-QCDSR,Scalar-1GeV-HJLee-PRD-2019-QCDSR,Weinberg-2013-PRL},  for more references, see the reviews
 \cite{Review-Amsler-2004-tetra,Review-Close-2002-tetra-mole,
 Review-Klempt-2007-tetra-hybrid}. However, it is an  un-resoled problem until now. There have been several excellent  reviews of the exotic states with emphasis on different aspects  \cite{Review-XYZ-Hosaka-PTEP-2016,Review-XYZ-ChenHX-PRT-2016,Review-XYZ-Lebed-PPNP-2017,
Review-XYZ-Ali-PPNP-2017,Review-XYZ-Esposito-PRT-2017,Review-XYZ-GuoFK-RMP-2018,
Review-XYZ-Olsen-RMP-2018,Review-XYZ-Karliner-ARNPS-2018,Review-XYZ-Brambilla-PRT-2019,Review-XYZ-LiuYR-PPNP-2019,Review-XYZ-ChenHX-RPP-2023,
Review-XYZ-Wangbo-PRT-2023,Review-XYZ-GengLS-PRT-2025,Review-QCDSR-Nielsen-PRT-2010,
Review-QCDSR-Nielsen-JPG-2019}.

 Firstly, let us see the experimental data on the exotic states pedagogically and dogmatically  in the order $X$, $Y$, $Z$, $T$ and $P$ sequentially, and sort out of the exotic states according to the masses from low to high roughly. In fact, there are relations among those $X$, $Y$ and $Z$ states in one way or the other, it is impossible to sort out them distinctly, the most related ones are grouped together into  one sub-section. Furthermore,  we would like to emphasize the assignment based on the QCD sum rules at the end of every sub-section if there exists a room, as the predicted spectroscopy based on the QCD sum rules cannot accommodate all the exotic states. Other possible assignments would be presented in Sects.{\bf\ref{33-4-quark}}, {\bf\ref{11-4-quark}}, {\bf\ref{333-11-5-quark}} and {\bf \ref{Singly-Q-tetraquark}}.

\subsection{$X(3872)$}
In 2003, the  Belle collaboration   observed   a narrow charmonium-like state $X(3872)$ with the mass $3872.0\pm0.6\pm0.5\,\rm{MeV}$ near the $D\bar{D}^*$ threshold in the $\pi^+ \pi^- J/\psi$ mass spectrum in the exclusive  processes  $B^\pm \to K^\pm \pi^+ \pi^- J/\psi$  \cite{X3872-2003}.
 The evidences for the decay modes $X(3872) \to \gamma J/\psi, \, \gamma \psi^{\prime}$ observed  by the Belle and BaBar collaborations imply the positive charge conjugation $C=+$ \cite{X3872-Jpsi-gamma-Belle-2005,X3872-Jpsi-gamma-BaBar-2006,
 X3872-Jpsi-gamma-BaBar-2009}.  Angular correlations between final state particles $\pi^+ \pi^- J/\psi$ analyzed by the CDF, Belle and LHCb collaborations favor the $J^{PC}=1^{++}$ assignment \cite{X3872-JPC-CDF-2007,X3872-JPC-Belle-2011,X3872-JPC-LHCb-2013,X3872-JPC-LHCb-2015}.
  It is a possible candidate for the tetraquark state \cite{Maiani-2005-PRD-tetra-X3872,Maiani-1405-Tetra-model-2,Lebed-dynamical-PRL-2014,
  X3872-tetra-Narison-PRD-2007,
  X3872-tetra-Ebert-PLB-2006,X3872-tetra-WangZG-HuangT-PRD-2014,WZG-HC-PRD-2020,
  WZG-X3872-decay-PRD-2024}, (not) molecular state (\cite{X3872-Not-mole-LiuX-EPJC-2008})
  \cite{X3872-mole-Tornqvist-PLB-2004,X3872-mole-Swanson-PLB-2004,
X3872-mole-Swanson-PLB-2004-decay,X3872-mole-Mehen-PRD-2007,
X3872-mole-Grinstein-PRL-2009,X3872-mole-Close-PLB-2004,
X3872-mole-Oset-PRD-2010,X3872-mole-Voloshin-PLB-2004,
X3872-mole-FKGuo-PRD-2013,X3872-mole-FKGuo-PLB-2013,
X3872-mole-Wong-PRC-2004,X3872-mole-Petrov-PLB-2006,X3872-mole-Nieves-PRD-2012,
X3872-mole-Nefediev-PRD-2007,X3872-mole-Kalashnikova-PRD-2005,
X3872-mole-Braaten-PRD-2010,X3872-mole-Lyubovitskij-PRD-2008,
X3872-mole-WangZG-HT-EPJC-2014,WZG-tetra-mole-IJMPA-2021,
WZG-tetra-mole-AAPPS-2022,X3872-mole-Mutuk-EPJC-2018},
(not) traditional charmonium $\chi_{c1}(2{\rm P})$ (\cite{X3872-Not-charmonium-Eichten-2006-PRD}) \cite{ X3872-charmonium-Godfrey-2004-PRD,
X3872-charmonium-Suzuki-2005-PRD,X3872-charmonium-KTChao-2009-PRD}, threshold cusp \cite{X3872-cusps-Bugg-2004-PLB}, etc. However, none of those available assignment  has won an overall consensus, its nature is still under  heated debates, the very narrow width and exotic branching fractions make it a hot potato, the door remains open for another binding
mechanism.  The QCD sum rules allow both the color  $\bar{\mathbf{3}}\mathbf{3}$-type and $\mathbf{1}\mathbf{1}$-type tetraquark assignments, see Table \ref{Identifications-Table-cqcq-positive} in Sect.{\bf\ref{Tetra-Positive}}  and Table \ref{Assignments-mole-tetra} in Sect.{\bf \ref{11-tetra-states}}.

\subsection{$X(3915)$, $X(3940)$, $Y(3940)$, $Z(3930)$, $X(4160)$,  $X(3860)$, $X(3960)$}
In 2005, the  Belle collaboration observed the $Y(3940)$ in the $B$-decays
$B\to K Y(3940)\to K \omega J/\psi$ \cite{Y3940-Belle-PRL-2005}, later, the BaBar
collaboration confirmed the $Y(3940)$ in the $\omega J/\psi$ mass spectrum with a mass about $3.915\,\rm{GeV}$ in the decays $B\to K \omega J/\psi$ \cite{
Y3940-BaBar-PRL-2008,Y3940-BaBar-PRD-2010}.

In 2007, the Belle collaboration observed the $X(3940)$ in the process
$e^+ e^- \to J/\psi X(3940)$ with the subprocess $X(3940)\to D^* \bar{D}$ \cite{X3940-Belle-PRL-2007}. In 2008, the Belle collaboration confirmed the $X(3940)$ in the same process, and observed the $X(4160)$ in the subprocess
$X(4160)\to D^* \bar{D}^*$ \cite{X3940-Belle-PRL-2008}.

In 2010, the $X(3915)$ was observed in the process $\gamma\gamma \to J/\psi\omega$ by the Belle  collaboration  \cite{X3915-Belle-PRL-2010}, then the BaBar collaboration determined its quantum numbers $J^P=0^+$ \cite{X3915-BaBar-PRL-2012}, it is a good candidate for the conventional charmonium $\chi_{c0}(2{\rm P})$.
The $Y(3940)$ and $X(3915)$ could be the same particle, and is denoted as the $\chi_{c0}(3915)$ with the assignment $J^{PC}=0^{++}$ in {\it The Review of Particle Physics} \cite{PDG-2024}.

In 2006, the $Z(3930)$ was observed in the  process $\gamma\gamma \to J/\psi\omega$ by  the Belle  collaboration \cite{Z3930-Belle-PRL-2006}, then confirmed by the BaBar collaboration in the same process \cite{Z3930-BaBar-PRD-2010}, the two collaborations both determined its quantum numbers to be $J^{PC}=2^{++}$, and it is widely assigned as the $\chi_{c2}(3930)$ \cite{PDG-2024}.

In 2017, the Belle collaboration  performed a full amplitude analysis of the process $e^+ e^- \to J/\psi D \bar{D}$, and observed  a new charmonium-like state $X^*(3860)$  which decays to the $D \bar{D}$ pair, the measured mass and width are
   $3862^{+26}_{-32}{}^{+40}_{-13}\,\rm{MeV}$  and $201^{+154}_{-67}{}^{+88}_{-82}\,\rm{MeV}$, respectively  \cite{Belle-X3860-PRD-2017}. The $J^{PC} =0^{++}$   hypothesis is favored over the $2^{++}$   hypothesis at the level of $2.5\sigma$, and the Belle collaboration assigned the $X^*(3860)$ as an alternative  $\chi_{c0}(\rm 2P)$ state   \cite{Belle-X3860-PRD-2017}.

 In 2020, the LHCb collaboration performed  an amplitude analysis of the $B^+\to D^+D^-K^+$ decays and  observed
that it is necessary to include the $\chi_{c0}(3930)$ and $\chi_{c2}(3930)$ with the $J^{PC}=0^{++}$ and $2^{++}$ respectively in the $D^+D^-$ channel, and to include the $X_0(2900)$ and $X_1(2900)$ with the $J^{P}=0^{+}$ and $1^{-}$ respectively in the $D^-K^+$ channel \cite{LHCb-X3930-PRL-2020,LHCb-X3930-PRD-2020}. The measured Breit-Wigner masses and widths are
\begin{eqnarray}
\chi_{c0}(3930) &:& M = 3923.8 \pm 1.5 \pm 0.4\, {\rm MeV} ,\,\,\, \Gamma = 17.4 \pm 5.1 \pm 0.8\,{\rm  MeV} \, ,\nonumber\\
\chi_{c2}(3930) &:& M = 3926.8 \pm 2.4 \pm 0.8 \, {\rm MeV} , \,\,\, \Gamma = 34.2 \pm 6.6 \pm 1.1\, {\rm  MeV} \, ,
\end{eqnarray}
and the $\chi_{c0}(3915)$ and $\chi_{c0}(3930)$ can be  identified as the $\chi_{c0}(2\rm{P})$.

In 2023,  the LHCb collaboration announced the observation of the $X(3960)$ in the $D_s^+D_s^-$  mass spectrum in the $B^{+} \to D^{+}_s D^{-}_s K^{+}$ decays, and  the assignment $J^{PC}=0^{++}$ is  favored \cite{LHCb-X3960-PRL-2023},  the measured Breit-Wigner mass and width are $ 3956 \pm 5\pm 10 $ MeV and $ 43\pm 13\pm 8 $ MeV, respectively.

The thresholds of the $D_s^+D_s^-$ and $D^+D^-$ are $3938\,\rm{MeV}$ and $3739\,\rm{MeV}$, respectively, which favors assigning  the $X(3960)$ and $\chi_{c0}(3930)$ as the same particle.  However, the ratio of the branching fractions \cite{LHCb-X3960-PRL-2023},
\begin{eqnarray}
\frac{\Gamma(X\to D^+D^-)}{\Gamma(X\to D_s^+D_s^-)}&=&0.29\pm0.09\pm0.10\pm0.08\, ,
\end{eqnarray}
 implies the exotic nature of this state, as it is harder to excite an $s\bar{s}$ pair from the vacuum compared with the $u\bar{u}$
 or $d\bar{d}$ pair, and the traditional  charmonium states predominantly decay into the $D\bar{D}$ and $D^*\bar{D}^*$ states rather than into the $D_s\bar{D}_s$ and $D_s^*\bar{D}^*_s$ states. In addition, there is no room for the $X(3860)$, so,  at least one of the $X(3860)$, $X(3915)$, $\chi_{c0}(3930)$ and $X(3960)$ should be  exotic state \cite{No-Chi0-GuoFK-PRD-2012,No-Chi0-GuoFK-SB-2023,WangZG-X3860-EPJA-2017,
 X3915-X3930-ZYZhou-PRL-2015}.

For example, the updated  nonrelativistic potential model (NR) and  Godfrey-Isgur relativized potential model (GI) indicate that the 2P charmonium states have the masses (NR;\,GI)\cite{Godfrey-PRD-2005-charmonium},
\begin{flalign}
 & \chi_{c2}(2{\rm P}) : 3972 \mbox{ MeV} \, , \, 3979 \mbox{ MeV} \, , \nonumber \\
 & \chi_{c1}(2{\rm P}) : 3925 \mbox{ MeV}\, , \, 3953 \mbox{ MeV} \, , \nonumber \\
 & \chi_{c0}(2{\rm P}) : 3852 \mbox{ MeV} \, ,\, 3916 \mbox{ MeV} \, ,
\end{flalign}
even the assignments of the $\chi_{c0}(2{\rm P})$ and $\chi_{c2}(2{\rm P})$ are not comfortable enough.

We can identify the $X(3915)$, $Y(3940)$ and $\chi_{c0}(3930)$ as the same particle tentatively, and assign it as the color $\bar{\mathbf{3}}\mathbf{3}$-type scalar  tetraquark state based on the QCD sum rules, see Table \ref{Identifications-Table-cqcq-positive} in Sect.{\bf\ref{Tetra-Positive}}, furthermore, there exists a room for the $X(3860)$. On the other hand,  we can assign the $X(3960)$  as the color $\bar{\mathbf{3}}\mathbf{3}$-type or $\mathbf{1}\mathbf{1}$-type tetraquark state, see Table \ref{Identifications-Table-cscs-positive} in Sect.{\bf\ref{Tetra-Positive}}  and Table \ref{Assignments-mole-tetra} in Sect.{\bf \ref{11-tetra-states}}.

\subsection{$\eta_c(3945)$, $h_c(4000)$, $\chi_{c1}(4010)$, $h_c(4300)$}
In 2024, the LHCb collaboration explored the decays   $B^{+}\to {D^{\ast+}D^{-}K^{+}}$ and ${D^{\ast-}D^{+}K^{+}}$, and observed four charmonium(-like) states
  $\eta_c(3945)$, $h_c(4000)$, $\chi_{c1}(4010)$ and $h_c(4300)$ with the quantum numbers $J^{PC}=0^{-+}$,  $1^{+-}$, $1^{++}$ and $1^{+-}$ respectively  in the $D^{\ast\pm}D^{\mp}$  mass spectrum \cite{LHCb-hc4000-PRL-2024}.
The measured Breit-Wigner masses and widths are
\begin{flalign}
 & \eta_c(3945) : M = 3945\,_{-17}^{+28}{}\,_{-28}^{+37} \mbox{ MeV}\, , \, \Gamma = 130\,_{-49}^{+92}{}\,_{-70}^{+101} \mbox{ MeV} \, , \nonumber\\
 & h_c(4000) : M = 4000\,_{-14}^{+17}{}\,_{-22}^{+29} \mbox{ MeV}\, , \, \Gamma = 184\,_{-45}^{+71}{}\,_{-61}^{+97} \mbox{ MeV}\, ,\nonumber \\
 & \chi_{c1}(4010) : M = 4012.5\,_{-3.9}^{+3.6}{}\,_{-3.7}^{+4.1} \mbox{ MeV} \, ,\, \Gamma = 62.7\,_{-6.4}^{+7.0}{}\,_{-6.6}^{+6.4}\mbox{ MeV} \, , \nonumber\\
 & h_c(4300) : M = 4307.3\,_{-6.6}^{+6.4}{}\,_{-4.1}^{+3.3} \mbox{ MeV} \, ,\, \Gamma = 58\,_{-16}^{+28}{}\,_{-25}^{+28} \mbox{ MeV} \, .
\end{flalign}

The updated  nonrelativistic potential model  (NR) and  Godfrey-Isgur relativized potential model (GI) indicate that the 3S/2P/3P charmonium states have the masses (NR;\,GI) \cite{Godfrey-PRD-2005-charmonium},
\begin{flalign}
& \eta_c(3{\rm S})~~ : 4043 \mbox{ MeV} \, , \, 4064 \mbox{ MeV} \, , \nonumber \\
& \chi_{c1}(2{\rm P}) : 3925 \mbox{ MeV}\, , \, 3953 \mbox{ MeV} \, , \nonumber \\
& \chi_{c1}(3{\rm P}) : 4271 \mbox{ MeV}\, , \, 4317 \mbox{ MeV} \, , \nonumber \\
& h_c(2{\rm P})~ : 3934 \mbox{ MeV} \, ,\, 3956 \mbox{ MeV} \, , \nonumber \\
  & h_c(3{\rm P})~ : 4279 \mbox{ MeV} \, ,\, 4318 \mbox{ MeV} \, ,
\end{flalign}
the assignments $\eta_c(3945)$, $h_c(4000)$ and $\chi_{c1}(4010)$ in Ref.\cite{LHCb-hc4000-PRL-2024} are rather marginal.

Based on the predictions of the QCD sum rules, we can assign the $h_c(4000)$ and $\chi_{c1}(4010)$ as the color  $\bar{\mathbf{3}}\mathbf{3}$-type  tetraquark states tentatively based on the QCD sum rules, see Table \ref{Identifications-Table-cqcq-positive} in Sect.{\bf\ref{Tetra-Positive}}.

\subsection{$X(4140)$, $X(4274)$, $X(4500/4475)$, $X(4700/4710)$, $X(4685/4650)$, $X(4630/4800)$}
In 2009, the CDF collaboration observed the $X(4140)$ in the $J/\psi\phi$ mass spectrum in the $B^+\rightarrow J/\psi\,\phi K^+$ decays with a significance larger than $3.8 \sigma$  \cite{X4140-CDF-PRL-0903}.
 In 2011, the CDF collaboration confirmed the
$Y(4140)$ in the $B^\pm\rightarrow J/\psi\,\phi K^\pm$ decays  with
 a significance greater  than $5\sigma$, and  observed an evidence for the  $X(4274)$ with an approximate significance of $3.1\sigma$
\cite{X4140-CDF-MPLA-1101}.
In 2013, the CMS collaboration also  confirmed the $X(4140)$  in the $B^\pm \to J/\psi \phi K^\pm$ decays \cite{X4140-CMS-PLB-1309}.

In 2016, the LHCb collaboration performed the first full amplitude analysis of the $B^+\to J/\psi \phi K^+$   decays,
confirmed the $X(4140)$ and $X(4274)$ in the $J/\psi \phi$ mass spectrum,  and determined the    spin-parity to be $J^{P} =1^{+}$ \cite{X4140-X4500-LHCb-PRL-16061,X4140-X4500-LHCb-PRD-16062}. Moreover, the LHCb collaboration observed  two new particles  $X(4500)$ and $X(4700)$ in the $J/\psi \phi$ mass spectrum,  and determined the  spin-parity to be $J^{P} =0^{+}$ \cite{X4140-X4500-LHCb-PRL-16061,X4140-X4500-LHCb-PRD-16062}. The measured  masses and widths are
\begin{flalign}
 & X(4140) : M = 4146.5 \pm 4.5 ^{+4.6}_{-2.8} \mbox{ MeV}
\, , \, \Gamma = 83 \pm 21 ^{+21}_{-14} \mbox{ MeV} \, , \nonumber \\
 & X(4274) : M = 4273.3 \pm 8.3 ^{+17.2}_{-3.6} \mbox{ MeV}
\, , \, \Gamma = 56 \pm 11 ^{+8}_{-11} \mbox{ MeV} \, , \nonumber \\
 & X(4500) : M = 4506 \pm 11 ^{+12}_{-15} \mbox{ MeV} \, ,
\, \Gamma = 92 \pm 21 ^{+21}_{-20} \mbox{ MeV} \, , \nonumber \\
 & X(4700) : M = 4704 \pm 10 ^{+14}_{-24} \mbox{ MeV} \, ,
\, \Gamma = 120 \pm 31 ^{+42}_{-33} \mbox{ MeV} \, .
\end{flalign}
If they are tetraquark states,  their quark constituents must be $cs\bar{c}\bar{s}$. The S-wave   $J/\psi\phi$ systems  have the  $J^{PC}=0^{++}$, $1^{++}$, $2^{++}$, while the P-wave $ J/\psi\phi$ systems have the  $J^{PC}=0^{-+}$, $1^{-+}$, $2^{-+}$, $3^{-+}$.
The LHCb's data rule out the $0^{++}$ or $2^{++}$ $D_s^{*+}D_s^{*-}$ molecule assignments.

In 2021, the LHCb collaboration performed an improved full amplitude analysis of the exclusive process $B^+\to J/\psi \phi K^+$,  observed  the $X(4685)$  ($X(4630)$) in the $J/\psi \phi$  mass spectrum with the $J^P=1^+$ ($1^-$) and Breit-Wigner masses and widths,
\begin{flalign}
 & X(4685) : M = 4684\pm 7_{-16}^{+13} \mbox{ MeV}\, , \, \Gamma = 126\pm15 _{-41}^{+37} \mbox{ MeV} \, ,\nonumber \\
 & X(4630) : M = 4626\pm 16_{-110}^{+18} \mbox{ MeV} \, ,\, \Gamma = 174\pm27 _{-73}^{+134} \mbox{ MeV} \, ,
\end{flalign}
furthermore, they observed the $Z_{cs}(4000)$ and $Z_{cs}(4220)$ with the $J^P=1^+$ in the $J/\psi K^+$ mass spectrum and  confirmed the four old particles \cite{LHCb-Zcs4000-PRL-2021}.

In 2024, the LHCb collaboration performed the first full amplitude analysis of the decays $B^+ \to \psi(2S) K^+ \pi^+ \pi^-$, and they developed   an amplitude model with 53 components  comprising 11 hidden-charm exotic states, for example, the $X(4475)$,  $X(4650)$, $X(4710)$  and $X(4800)$  in the $\psi(\rm 2S)\rho^0(770)$ mass spectrum with the $J^{P}=0^+$, $1^+$, $0^+$ and $1^-$, respectively, while the  $X(4800)$   is just  an  effective description of  generic partial wave-function  \cite{LHCb-JHEP-11-tetra-2407}.

The $X(4475)$, $X(4650)$, $X(4710)$ and $X(4800)$ have the isospin $(I,I_3)=(1,0)$, while the $X(4500)$, $X(4685)$, $X(4700)$ and $X(4630)$ have the isospin $(I,I_3)=(0,0)$ according to the final states $\psi(2S)\rho^0(770)$
and $J/\psi\phi$.
A possible  explanation is that those states are genuinely different states, if the $X(4475)$ state is
the $c\bar{c}(u\bar{u} -d\bar{d})$ isospin partner of the $X(4500)$ interpreted as the $c\bar{c}s\bar{s}$ state, we  would generally
expect a larger mass difference of $M_{X(4500)}-M_{X(4475)}\approx 200\,\rm{MeV}$ rather than several $\rm{MeV}$. The un-normal light-flavor $SU(3)$ breaking effects make them difficult to assign in the scenario of tetraquark states.

Based on the predictions of the QCD sum rules, we  can assign the $X(4140)$, $X(4274)$, $X(4500)$, $X(4685)$ and $X(4700)$  as the color $\bar{\mathbf{3}}\mathbf{3}$-type  tetraquark states with the positive parity  tentatively, see Table \ref{Identifications-Table-cscs-positive} in Sect.{\bf\ref{Tetra-Positive}}, and assign the $X(4630)$
as the color $\bar{\mathbf{3}}\mathbf{3}$-type  tetraquark state with the negative parity tentatively,  see Table \ref{Assignments-Table-Y-cscs} in Sect.{\bf \ref{Tetraqurk-Negative}}.

\subsection{$X(4350)$}
In 2010,  the Belle collaboration measured the process $\gamma
\gamma \to \phi J/\psi$ for the $\phi J/\psi$  mass
distributions and observed a narrow peak of
$8.8^{+4.2}_{-3.2}$  events with a significance of $3.2\,\sigma$  \cite{Belle-X4350-PRL-2010}. The mass and width are $(4350.6^{+4.6}_{-5.1}\pm 0.7)\,\rm{MeV}$ and $(13.3^{+17.9}_{-9.1}\pm 4.1)\,\rm{MeV}$ respectively. However, the $X(4350)$ is not confirmed by other experiments.

\subsection{$X_0(2900)$, $X_1(2900)$, $T_{c\bar{s}}(2900)$}
In 2020, the LHCb collaboration reported a narrow peak in the $D^- K^+$
invariant mass spectrum in the decays $B^\pm\to D^+ D^- K^\pm$ \cite{LHCb-X3930-PRL-2020,LHCb-X3930-PRD-2020}.  The peak could
be  reasonably  parameterized in terms of two Breit-Wigner resonances:
\begin{eqnarray}
X_0(2900) &:& J^P=0^+,~M=2866\pm 7 \pm 2~{\rm MeV},~~\Gamma=\phantom{1}57\pm 12 \pm 4~{\rm MeV}~,  \nonumber\\
X_1(2900) &:& J^P=1^-,~M=2904\pm 5 \pm 1~{\rm MeV},~~\Gamma=110 \pm 11 \pm 4~{\rm MeV}~.
\end{eqnarray}
They are the first exotic hadrons with fully open flavor,  the valence quarks  are $ud\bar{c}\bar{s}$ \cite{LHCb-X3930-PRL-2020,LHCb-X3930-PRD-2020}. The narrow peak can be assigned as the color $\bar{\mathbf{3}}\mathbf{3}$-type tetraquark state with the $J^P=0^+$ \cite{X2900-tetra-Karliner-PRD-2020,X2900-tetra-WangZG-IJMPA-2020,
X2900-tetra-ZhangJR-PRD-2021,X2900-tetra-mole-ChenHX-CPL-2020},  its radial/orbital excitation \cite{X2900-tetra-2S-HeXG-EPJC-2020},  non-tetraquark state \cite{X2900-No-tetra-LuQF-PRD-2020},
  $D^*\bar{K}^*$ molecular state \cite{X2900-tetra-mole-ChenHX-CPL-2020,X2900-DvKv-LiuMZ-PRD-2020,X2900-mole-WangQ-CPC-2021,
 X2900-mole-Oset-PLB-2020,X2900-mole-Azizi-JPG-2021,X2900-mole-Narison-NPA-2021},  triangle singularity \cite{X2900-Triangle-LiuXH-PRD-2020}, etc.

In 2023, the LHCb collaboration observed the tetraquark candidates $T_{c\bar{s}}^{0/++}(2900)$ with the spin-parity  $J^P=0^+$ in the processes $B^+ \to D^- D_s^+ \pi^+$ and $B^0 \to \bar{D}^0 D_s^+ \pi^-$ with the  significance larger than $9\sigma$ \cite{Tcs2900-LHCb-PRL-2023,Tcs2900-LHCb-PRD-2023}. The measured Breit-Wigner masses and widths  are
\begin{eqnarray}
T_{c\bar{s}}^{0}(2900)&:& M=2.892\pm0.014\pm0.015\,\rm{GeV}\, , \,\Gamma=0.119\pm0.026\pm0.013\,\rm{GeV}\, , \nonumber\\
T_{c\bar{s}}^{++}(2900)&:& M=2.921\pm0.017\pm0.020\,\rm{GeV}\, , \, \Gamma=0.137\pm0.032\pm0.017\,\rm{GeV}\, ,
\end{eqnarray}
respectively, and they  belong to a new type of
open-charm tetraquark states with the $c$ and $\bar{s}$ quarks.
 The $\bar{X}_0(2900)$, $T^{0}_{c\bar{s}}(2900)$ and $T^{++}_{c\bar{s}}(2900)$   can be accommodated in the light flavor $SU(3)_F$ symmetry sextet.

We can assign the $X_0(2900)$, $T^{0}_{c\bar{s}}(2900)$ and $T^{++}_{c\bar{s}}(2900)$ as the color $\bar{\mathbf{3}}\mathbf{3}$-type  tetraquark states with the $J^P=0^+$ tentatively based on the QCD sum rules, see Sect.{\bf\ref{Singly-Q-tetraquark}}.

\subsection{$X(5568)$}
In 2016,  the  D0 collaboration  observed   a narrow structure $X(5568)$ in the decays  $X(5568) \to B_s^0 \pi^{\pm}$ with significance  of $5.1\sigma$ \cite{X5568-D0-PRL-2016}.   The mass and natural width  are  $
5567.8 \pm 2.9 { }^{+0.9}_{-1.9}\,\rm{MeV} $   and $ 21.9 \pm 6.4 {}^{+5.0}_{-2.5}\,\rm{MeV}$,  respectively.
The $B_s^0 \pi^{\pm}$ systems consist of two quarks and two antiquarks of four different flavors, just like the $T_{c\bar{s}}(2900)$ with the  $J^P=0^+$ observed 7 years later in the   $D_s^+\pi^+/D_s^+\pi^-$ mass spectrum by the LHCb collaboration \cite{Tcs2900-LHCb-PRL-2023,Tcs2900-LHCb-PRD-2023}.
The D0 collaboration fitted the $B_s^0 \pi^{\pm}$ systems with the S-wave Breit-Wigner parameters,  the favored assignments  are $J^P =0^+$, but
the assignments  $J^P = 1^+$ cannot be excluded according to  decays  $X(5568) \to B_s^*\pi^+ \to B_s^0 \pi^+ \gamma$, where the low-energy photon is not
detected. It can be assigned as a $us\bar{b}\bar{d}$ tetraquark state
with the $J^{PC}=0^{++}$
\cite{X5568-WangZG-CTP-2016,X5568-ZhuSL-PRL-2016,X5568-Azizi-PRD-2016,
X5568-WangW-CPC-2016,X5568-Nielsen-PRD-2016}, however, the $X(5568)$ is not confirmed by the LHCb, CMS, ATLAS and CDF collaborations
\cite{X5568-No-LHCb-PRL-2016,X5568-No-CMS-PRL-2018,
  X5568-No-ATLAS-PRL-2018,X5568-No-CDF-PRL-2018}.

\subsection{$X(6600)$, $X(6900)$, $X(7300)$}
In 2020, the LHCb collaboration reported evidences of two fully-charm  tetraquark  candidates  in the $J/\psi J/\psi$  mass spectrum \cite{LHCb-cccc-SB-2020}. They observed  a broad structure  above the $J/\psi J/\psi$ threshold ranging from 6.2 to 6.8 GeV and a narrow structure at about 6.9 GeV  with the significance of larger  than $5\,\sigma$. In addition, they also observed  some vague structures around 7.2 GeV.

In 2023, the ATLAS collaboration  observed statistically significant excesses in the $J/\psi J/\psi$ channel, which are consistent with a narrow resonance at about $6.9\,\rm{GeV}$ and a broader structure at much lower mass. And they also observed a statistically significant excess at about $7.0\,\rm{GeV}$ in the $J/\psi \psi^\prime$ channel \cite{ATLAS-cccc-PRL-2023}.

In 2024, the CMS collaboration observed three resonant structures in the $J/\psi J/\psi$ mass spectrum with the masses $6638^{+43}_{-38}{}^{+16}_{-31}\,\rm{MeV}$,
 $6847^{+44}_{-28}{}^{+48}_{-20}\,\rm{MeV}$ and $7134^{+48}_{-25} {}^{+41}_{-15}\, \rm{ MeV}$, respectively \cite{CMS-cccc-PRL-2024}. While in the no-interference model, the measured Breit-Wigner masses and widths are \cite{CMS-cccc-PRL-2024},
\begin{flalign}
 & X(6600) : M = 6552\pm10\pm12 \mbox{ MeV}\, , \, \Gamma = 124^{+32}_{-26} \pm 33 \mbox{ MeV} \, ,\nonumber \\
 & X(6900) : M = 6927\pm9\pm4 \mbox{ MeV} \, ,\, \Gamma = 122^{+24}_{-21} \pm 18 \mbox{ MeV} \, , \nonumber \\
  & X(7300) : M = 7287^{+20}_{-18}\pm 5 \mbox{ MeV} \, ,\, \Gamma =95^{+59}_{-40} \pm 19 \mbox{ MeV} \, .
\end{flalign}
The two-meson pairs  $J/\psi J/\psi$, $J/\psi \psi^\prime$, $\psi^\prime \psi^\prime$, $h_c h_c$, $\chi_{c0}\chi_{c0}$, $\chi_{c1}\chi_{c1}$ and $\chi_{c2}\chi_{c2}$ lie at $6194\,\rm{MeV}$, $6783\,\rm{MeV}$, $7372\,\rm{MeV}$, $7051\,\rm{MeV}$, $6829\,\rm{MeV}$, $7021\,\rm{MeV}$ and $7112\,\rm{MeV}$, respectively \cite{PDG-2024}, it is difficult to assign the $X(6600)$, $X(6900)$ and $X(7300)$ as the charmonium-charmonium molecular states without introducing coupled channel effects.

We can assign the $X(6600)$, $X(6900)$ and $X(7300)$ as the color  $\bar{\mathbf{3}}\mathbf{3}$-type  tetraquark states tentatively based on the QCD sum rules, see Table \ref{cccc-mass-Regge-update} in Sect.{\bf\ref{QQQQ-tetraquark}}.

\subsection{$Y(4260/4230)$, $Y(4360/4320)$, $Y(4390)$, $Y(4500)$, $Y(4660)$, $Y(4710/4750/4790)$}
In 2005, the BaBar collaboration  studied the initial-state radiation process  $e^+ e^- \to \gamma_{ISR} \pi^+\pi^- J/\psi$ and observed the $Y(4260)$   in the $\pi^+\pi^- J/\psi$ mass spectrum, the measured mass and width are  $\left(4259 \pm 8 {}^{+2}_{-6}\right) \,\rm{MeV}$ and  $\left(88 \pm 23 {}^{+6}_{-4} \right)\, \rm{MeV}$, respectively \cite{BaBar-Y4260-PRL-0506}. Subsequently the $Y(4260)$ was confirmed by the Belle and CLEO collaborations \cite{Belle-Y4260-PRL-0707,CLEO-Y4260-PRD-0606},
the Belle collaboration also observed an evidence for a very broad structure $Y(4008)$ in the $\pi^+\pi^- J/\psi$ mass spectrum.

In 2007, the  Belle collaboration  studied the initial-state radiation process $e^+e^- \to \gamma_{ISR}\pi^+ \pi^- \psi^{\prime}$, and  observed the $Y(4360)$ and $Y(4660)$ in the $\pi^+ \pi^- \psi^{\prime}$  mass spectrum   at $(4361\pm 9\pm 9)\, \rm{MeV}$ with a width of  $(74\pm 15\pm 10)\,\rm{ MeV}$ and   $(4664\pm 11\pm 5)\,\rm{ MeV}$ with a width of  $(48\pm 15\pm 3) \,\rm{MeV}$, respectively \cite{Belle-Y4660-PRL-0707,Belle-Y4660-PRD-1410}, then the $Y(4660)$ was confirmed by the BaBar collaboration \cite{BaBar-Y4360-Y4660-PRD-2014}.

In 2008, the Belle collaboration studied  the initial-state radiation process $e^+e^- \to \gamma_{ISR} \Lambda_c^+ \Lambda_c^-$, observed a clear peak $Y(4630)$   in the $\Lambda_c^+ \Lambda_c^-$  mass spectrum just above the $\Lambda_c^+ \Lambda_c^-$ threshold, and determined  the mass and width to be   $\left(4634^{+8}_{-7}{}^{+5}_{-8}\right)\,\rm{MeV}$ and $\left(92^{+40}_{-24}{}^{+10}_{-21}\right)\,\rm{MeV}$, respectively \cite{Belle-Y4630-PRL-0807}. Thereafter, the $Y(4660)$ and $Y(4630)$ are taken as  the same particle according to the uncertainties of the masses and widths, for example, in {\it The Review of Particle Physics} \cite{PDG-2024}.

In 2014, the BESIII collaboration searched for the production of $e^+e^-\to \omega\chi_{cJ}$ with $J=0,1,2$, and observed a resonance in the $\omega\chi_{c0}$ cross section,  the measured mass and width  are $4230\pm 8\pm 6\, \rm{ MeV}$    and $ 38\pm 12\pm 2\,\rm{MeV}$, respectively \cite{BES-2014-4230-PRL}.

In 2016, the BESIII collaboration measured the cross sections of the process $e^+ e^- \to \pi^+\pi^- h_c$, and observed two structures, the $Y(4220)$ has a mass of $4218.4\pm4.0\pm0.9\,\rm{MeV}$ and a width of $66.0\pm9.0\pm0.4\,\rm{MeV}$ respectively, and the $Y(4390)$ has a mass  of $4391.6\pm6.3\pm1.0\,\rm{MeV}$ and a width of $139.5\pm16.1\pm0.6\,\rm{MeV}$ respectively \cite{BES-Y4390-PRL-2017}.

Also in 2016, the BESIII collaboration precisely measured the cross section of the process $e^+ e^- \to  \pi^+\pi^- J/\psi$  and observed  two resonant structures, which agree with the $Y(4260)$ and $Y(4360)$, respectively. The first resonance has a mass of $4222.0\pm3.1\pm 1.4\,  \rm{MeV}$ and a width of $44.1 \pm 4.3\pm 2.0 \,\rm{MeV}$, while the second one has a mass of $4320.0\pm 10.4 \pm 7.0\, \rm{MeV}$   and a width of $101.4^{+25.3}_{-19.7}\pm 10.2\,\rm{MeV}$ \cite{BES-Y4220-Y4320-PRL-2017}.

In 2022, the BESIII collaboration observed two resonant structures in the $K^+K^-J/\psi$ mass spectrum, one is the $Y(4230)$ and the other is the $Y(4500)$, which was  observed for the first time with the Breit-Wigner  mass and width $4484.7\pm 13.3\pm 24.1\,\rm{MeV}$ and $111.1\pm 30.1\pm 15.2\, \rm{MeV}$, respectively \cite{BESIII-Y4500-KK-CPC-2022}.

In 2023, the BESIII collaboration observed three enhancements in the  $ D^{*-}D^{*0}\pi^+$  mass spectrum in the Born cross sections of the process $e^{+}e^{-}\to D^{*0}D^{*-}\pi^{+}$, the first and third resonances are the $Y(4230)$ and $Y(4660)$, respectively, while the second resonance has the Breit-Wigner  mass and width  $4469.1\pm26.2\pm3.6\,\rm{MeV}$ and  $246.3\pm 36.7\pm 9.4\,\rm{MeV}$, respectively, and is roughly compatible with the $Y(4500)$ \cite{BESIII-Y4500-DvDvpi-PRL-2023}.

Also in 2023, the BESIII collaboration observed three resonance structures in the  $ D_{s}^{\ast+}D_{s}^{\ast-}$ mass spectrum, the two significant structures are consistent  with the $\psi(4160)$ and $\psi(4415)$, respectively,  while the third structure is new, and has the Breit-Wigner  mass and width
 $4793.3\pm7.5\,\rm{MeV}$ and $27.1\pm7.0\,\rm{MeV}$, respectively, therefore is named as $Y(4790)$ \cite{BESIII-Y4790-DvsDvs-PRL-2023}.

Also in 2023,  the BESIII collaboration observed a new resonance $Y(4710)$ in the $K^+K^-J/\psi$ mass spectrum with a significance over $5\sigma$, the measured  Breit-Wigner  mass and width are  $ 4708_{-15}^{+17}\pm21\, \rm{MeV}$  and $ 126_{-23}^{+27}\pm30\, \rm{MeV}$, respectively \cite{BESIII-Y4710-KK-PRL-2023}.

In 2024, the BESIII collaboration measured the Born cross sections for  the processes $e^+e^-\to\omega\chi_{c1}$ and $\omega\chi_{c2}$, and  observed the well established $\psi(4415)$ in the $\omega\chi_{c2}$ mass spectrum \cite{BESIII-Y4544-omegachi-2024}. In addition, they   observed a new resonance in the $\omega\chi_{c1}$ mass spectrum, and measured  the mass and width as   $4544.2\pm18.7\pm1.7\, \rm{MeV}$ and $116.1\pm33.5\pm1.7\, \rm{MeV}$, respectively, which are also roughly compatible with the $Y(4500)$.

Also in 2024,  the BESIII collaboration studied the processes  $e^+e^-\to\omega X(3872)$ and $\gamma X(3872)$, and observed that the relatively large cross section for the $e^+e^-\to\omega X(3872)$ process is mainly due to the  enhancement about  4.75 GeV, which maybe  indicate a potential structure in the $e^+e^-\to\omega X(3872)$  cross section  \cite{BESIII-Y4750-omegaX-PRD-2024}. If the enhancement is confirmed in the future by enough experimental data, there maybe exist another $Y$ state, the $Y(4750)$.

We should bear in mind, in 2023, the BESIII collaboration studied  the process $e^+e^- \to  \Lambda_c^+ \Lambda_c^-$ at twelve center-of-mass energies from 4.6119 to 4.9509 GeV, determined the Born cross sections and effective form-factors  with unprecedented precision, and obtained flat cross sections about 4.63 GeV, which does not  indicate the resonant structure $Y(4630)$ \cite{BESIII-No-Y4630-PRL-2023}.

The charmonium-like candidates $Y(4260/4230)$, $Y(4360/4320)$, $Y(4390)$, $Y(4500)$, $Y(4660)$ and $Y(4710/4750/4790)$ with the $J^{PC}=1^{--}$ overwhelm the accommodating capacity of the traditional $c\bar{c}$ model, some of them should be multiquark states.

Based on the predictions of the QCD sum rules, we can assign the $Y(4260/4230)$, $Y(4360/4320)$,  $Y(4390)$ and $Y(4750)$ as the color
$\bar{\mathbf{3}}\mathbf{3}$-type tetraquark states with an explicit P-wave between the diquark and antidiquark tentatively, see Table \ref{Assignment-Y} in Sect.{\bf\ref{Tetraquark-P-wave-explicit}}, and assign the $Y(4360)$, $Y(4390)$, $Y(4500)$, $Y(4660)$, $Y(4710)$ and $Y(4790)$ as the color $\bar{\mathbf{3}}\mathbf{3}$-type tetraquark states with an implicit P-wave in the diquark or antidiquark tentatively, see Tables \ref{Assignments-Table-Y-cqcq}-\ref{Assignments-Table-Y-cscs} in Sect.{\bf \ref{Tetraqurk-Negative}}.

\subsection{$Z_c(3900/3885)$, $Z_c(4020/4025)$,  $Z_{cs}(3985/4000)$, $Z_{cs}(4123)$, $Z_{cs}(4220)$}
In 2013, the BESIII collaboration studied  the process  $e^+e^- \to \pi^+\pi^-J/\psi$ at a center-of-mass energy of $4.260\,\rm{GeV}$, and observed a structure $Z_c^\pm(3900)$ in the $\pi^\pm J/\psi$ mass spectrum with a mass of $(3899.0\pm 3.6\pm 4.9)\,\rm{ MeV}$
and a width of $(46\pm 10\pm 20) \,\rm{MeV}$ \cite{BES-Z3900-PRL-2013}, at the same time, the Belle collaboration studied  the process
$e^+e^- \to \gamma_{ISR}\, \pi^+\pi^-J/\psi$  using initial-state radiation, and observed  a structure $Z_c^\pm(3900)$ in the $\pi^\pm J/\psi$ mass spectrum with a mass of $(3894.5\pm 6.6\pm 4.5)\,\rm{ MeV}$
and a width of $(63\pm 24\pm 26) \,\rm{MeV}$ \cite{Belle-Z3900-PRL-2013}.
Then this structure was confirmed by the  CLEO collaboration \cite{CLEO-Z3900-PLB-2013}.

Also in 2013, the BESIII collaboration  studied the process $e^+e^- \to (D^{*} \bar{D}^{*})^{\pm} \pi^\mp$ at $\sqrt{s}=4.26\,\rm{ GeV}$, and observed
 a structure $Z^{\pm}_c(4025)$ near the $(D^{*} \bar{D}^{*})^{\pm}$ threshold in the $\pi^\mp$ recoil mass spectrum \cite{BES1308-Z4025-PRL-2014}.
 The measured mass and width  are $(4026.3\pm2.6\pm3.7)\,\rm{MeV}$  and $(24.8\pm5.6\pm7.7)\,\rm{MeV}$, respectively \cite{BES1308-Z4025-PRL-2014}.
Slightly later, the  BESIII collaboration studied the process $e^+e^- \to \pi^+\pi^- h_c$ at $\sqrt{s}$    from $3.90\,\rm{ GeV}$ to $4.42\,\rm{GeV}$, and observed a distinct structure  $Z_c(4020)$   in the $\pi^\pm h_c$ mass spectrum,  the measured mass and width  are $(4022.9\pm 0.8\pm 2.7)\,\rm{MeV}$   and $(7.9\pm 2.7\pm 2.6)\,\rm{MeV}$, respectively \cite{BES1309-Z4020-PRL-2013}.

In 2014, the BESIII collaboration studied  the process $e^+e^- \to \pi D \bar{D}^*$ at $\sqrt{s}=4.26\,\rm{ GeV}$, and observed a distinct charged structure $Z_c(3885)$  in the $(D \bar{D}^*)^{\pm}$
 mass spectrum  \cite{BES-Z3885-PRL-2014}. The measured mass and width are $(3883.9 \pm 1.5 \pm 4.2)\,\rm{ MeV}$ and  $(24.8 \pm 3.3 \pm 11.0)\,\rm{ MeV}$,
respectively, and   the angular distribution of the $\pi Z_c(3885)$ system favors the  assignment $J^P=1^+$ \cite{BES-Z3885-PRL-2014}.

We tentatively  identify the $Z_c(3900)$
 and $Z_c(3885)$ as the same particle according to  the uncertainties of the masses and widths \cite{X3872-tetra-WangZG-HuangT-PRD-2014}. In 2017, the BESIII collaboration established the spin-parity of the $Z_c(3900)$
to be $J^P = 1^+$ \cite{BES-Zc3900-JP-PRL-2017}.

In 2021, the BESIII collaboration observed an excess  near the $D_s^- D^{*0}$ and $D^{*-}_s D^0$ thresholds in the $K^{+}$ recoil-mass spectrum with the significance of  5.3 $\sigma$ in the processes $e^+e^-\to K^+ (D_s^- D^{*0} + D^{*-}_s D^0)$ \cite{BES-Zcs3985-PRL-2021}.
 The Breit-Wigner  mass and width of the new structure $Z_{cs}(3985)$ were measured  as  $3985.2^{+2.1}_{-2.0}\pm1.7\,\rm{MeV}$   and $13.8^{+8.1}_{-5.2}\pm4.9\,\rm{MeV}$, respectively.

The $Z_c(3885)$ and $Z_{cs}(3985)$ have similar production modes,
 \begin{eqnarray}
 e^{+}e^{-} &\to& Z_c^-(3885) \, \pi^+ \to (D\bar{D}^{*})^-\,\pi^+\, ,\nonumber\\
 e^+e^- &\to& Z_{cs}^-(3985)\, K^+\to  (D_s^- D^{*0} + D^{*-}_s D^0)\,K^+\, ,
 \end{eqnarray}
and they should be cousins and have similar properties.

Also in 2021, the LHCb collaboration reported  two new  exotic states with the valence quarks  $c\bar{c}u\bar{s}$  in the $J/\psi  K^+$  mass spectrum  in the decays   $B^+ \to J/\psi \phi K^+$ \cite{LHCb-Zcs4000-PRL-2021}.  The most significant state $Z^+_{cs}(4000)$ has a mass of $4003 \pm 6 {}^{+4}_{-14}\,\rm{MeV}$, a width of $131 \pm 15 \pm 26\,\rm{MeV}$, and the spin-parity
$J^P =1^+$, while the broader state $Z^+_{cs}(4220)$ has a mass of $4216 \pm 24{}^{ +43}_{-30}\,\rm{MeV}$, a width of $233 \pm 52 {}^{+97}_{-73}\,\rm{MeV}$, and the spin-parity $J^P=1^+$ or $1^-$ (with a $2\sigma$  difference in favor of the first hypothesis) \cite{LHCb-Zcs4000-PRL-2021}. Considering the large difference between the widths, the $Z_{cs}(3985)$ and $Z_{cs}(4000)$ are unlikely to be the same particle.

In 2023, the BESIII collaboration reported an excess of the $Z^-_{cs}(4123)\to D_{s}^{*-}D^{*0}$ candidate at  a mass of $(4123.5\pm0.7\pm4.7) \, \rm{MeV}$ with a significance of $2.1\sigma$ in the process $e^{+} e^{-}\rightarrow K^{+}D_{s}^{*-}D^{* 0}+c.c.$ \cite{BES-Zcs4123-CPC-2023}. The $Z^-_{cs}(4123)$ is consistent with the tetraquark state with the valence quarks $c\bar{c}s\bar{u}$, spin-parity-charge-conjugation  $J^{PC}=1^{+-}$, a mass $4.11\pm0.08\,\rm{GeV}$ and a width $22.71 \pm 1.65\, \rm{MeV}$ predicted in previous work based on the QCD sum rules \cite{WangZG-Zcs4123-tetra-CPC-2022}.

The charmonium-like states $Z_c(3900/3885)$, $Z_c(4020/4025)$,  $Z_{cs}(3985/4000)$, $Z_{cs}(4123)$ and $Z_{cs}(4220)$ have non-zero electric charge, and are excellent candidates for the tetraquark (molecular) states \cite{WangZG-Zcs4123-tetra-CPC-2022,WangZG-Zcs3985-mass-tetra-CPC-2021}.

Based on the predictions of the QCD sum rules, we can assign the $Z_c(3900/3885)$, $Z_c(4020)$, $Z_{cs}(3985/4000)$ and $Z_{cs}(4123)$ as the
color $\bar{\mathbf{3}}\mathbf{3}$-type tetraquark states, see Table  \ref{Identifications-Table-cqcq-positive} and Table \ref{Assignments-Zcs-mass} in Sect.{\bf\ref{Tetra-Positive}}, or   $\mathbf{1}\mathbf{1}$-type tetraquark states, see Table \ref{Assignments-mole-tetra} in Sect.{\bf \ref{11-tetra-states}}.

\subsection{$Z_1(4050)$, $Z_1(4250)$, $Z_c(4100)$}
In 2008, the Belle collaboration  reported the first observation of
two resonance-like structures $Z^+_1(4050)$ and $Z^+_2(4250)$ exceeding  $5\, \sigma$ in the $\pi^+\chi_{c1}$  mass spectrum  near $4.1 \,\rm{GeV}$ in the exclusive decays $\bar{B}^0\to K^- \pi^+ \chi_{c1}$  \cite{Belle-Z1-Z2-PRD-2008}.  The Breit-Wigner  masses and widths are
$M_1=4051\pm14^{+20}_{-41} \,\rm{MeV}$,
$\Gamma_1=82^{+21}_{-17}$$^{+47}_{-22}\, \rm{MeV}$,
$M_2=4248^{+44}_{-29}$$^{+180}_{-35}\,\rm{MeV}$ and
$\Gamma_2=177^{+54}_{-39}$$^{+316}_{-61}\,\rm{MeV}$, respectively.
 However, the BaBar collaboration observed no evidence for the $Z_1^+(4050)$ and $Z^+_2(4250)$ states in the $\pi^+\chi_{c1}$ mass spectrum in the exclusive decays   $\bar{B}^0 \to \pi^+\chi_{c1} K^- $ and $B^+ \to \pi^+\chi_{c1} K^0_S $ \cite{BaBar-Z1-Z2-PRD-2012}.

In 2018, the LHCb collaboration observed an evidence for the $\eta_c \pi^-$ resonant structure  $Z_c(4100)$ with the significance larger than $3\,\sigma$ in a Dalitz plot analysis of the  $B^0 \to \eta_c  K^+\pi^- $ decays, the measured mass and width are $4096 \pm 20^{+18}_{-22}\,\rm{MeV}$ and $ 152 \pm 58^{+60}_{-35}\,\rm{MeV}$ respectively \cite{LHCb-Z4100-EPJC-1809}. The assignments $J^P =0^+$ and $1^-$  are both consistent with the experimental data. However, the $Z_c(4100)$ is not confirmed by other experiments until now.

\subsection{$Z_c(4430)$, $Z_c(4600)$}
In 2007, the  Belle collaboration   observed a distinct peak   in the   $\pi^{\pm} \psi^\prime$  mass spectrum  in the decays $B\to K \pi^{\pm} \psi^{\prime}$,
 the mass and width are $ \left(4433\pm4\pm2\right)\,\rm{ MeV}$ and $ \left(45^{+18}_{-13} {}^{+30}_{-13}\right)\,\rm {MeV}$, respectively \cite{Belle-Zc4430-PRL-2007}.
In 2009, the  Belle collaboration observed a signal for the decay $Z(4430)^+ \to \pi^+ \psi^\prime$ from a Dalitz plot analysis of the decays  $B \to K \pi^+ \psi^\prime$ \cite{Belle-Zc4430-PRD-2009}.
In 2013, the Belle collaboration performed a full amplitude analysis of the  decays $B^0 \to \psi^{\prime} K^+ \pi^-$
 to reach the favored assignments  $J^P=1^+$  \cite{Belle-Zc4430-PRD-2013}.

In 2014, the LHCb collaboration analyzed the $B^0\to\psi'\pi^-K^+$ decays  by performing a four-dimensional fit of the amplitude, and  provided the first independent confirmation of the  $Z^-(4430)$ resonance
and established its spin-parity  $J^P=1^+$. The measured Breit-Wigner mass and width are $\left(4475\pm7\,{_{-25}^{+15}}\right)\,\rm {MeV}$ and $\left(172\pm13\,{_{-34}^{+37}}\right)\,\rm {MeV}$, respectively \cite{LHCb-Zc4430-PRL-2014}, which excludes the possibility of  assigning the $Z_c(4430)$ as the $D^*D_1$ molecular state with the spin-parity $J^P=0^-$ \cite{Nielsen-Z4430-PLB-2008}, although it lies near the $D^*D_1$ threshold.

The Okubo-Zweig-Iizuka supper-allowed decays
\begin{eqnarray}
Z_c(3900)&\to&J/\psi\pi\, , \nonumber \\
Z_c^\prime(4430)&\to&\psi^\prime\pi\,
\end{eqnarray}
are expected to take place easily, and the energy gaps  have the relation $M_{Z_c^\prime}-M_{Z_c}=m_{\psi^\prime}-m_{J/\psi}$,  the $Z_c(4430)$ can be assigned as the first radial excitation of the $Z_c(3900)$  \cite{Maiani-1405-Tetra-model-2,
Nielsen-MPLA-2014-Zc,WangZG-Zc4430-tetra-CTP-2015}, which was proposed before the $J^P$ of the $Z_c(3900)$ were determined by the BESIII collaboration \cite{BES-Zc3900-JP-PRL-2017}.

In 2019, the LHCb collaboration performed an angular analysis of the weak decays $B^0\to J/\psi K^+\pi^-$, examined   the $m(J/\psi \pi^-)$ versus the $m(K^+\pi^-)$ plane, and observed two possible resonant structures in the vicinity of  the energies $m(J/\psi \pi^-)=4200 \,\rm{MeV}$ and $4600\,\rm{MeV}$, respectively \cite{LHCb-Zc4200-Zc4600-PRL-2019}, the structure $Z_c(4600)$ has not been confirmed by other experiments  yet.
According to the mass gaps $M_{Z_c(4600)}-M_{Z_c(4020)}\approx M_{Z_c(4430)}-M_{Z_c(3900)}$, we can tentatively assign the $Z_c(4600)$ as the first radial excitation of the $Z_c(4020)$ \cite{WangZG-Zc4600-tetra-CPC-2020,ChenHX-Z4600-PRD-2019}.

Based on the predictions of the QCD sum rules, we can assign the $Z_c(4430)$ and $Z_c(4600)$ as the first radial excitations of the color
$\bar{\mathbf{3}}\mathbf{3}$-type tetraquark states, see Table  \ref{Identifications-Table-cqcq-positive} in Sect.{\bf\ref{Tetra-Positive}} and Table \ref{Excitation-mass-Negatvie} in Sect.{\bf\ref{Tetraquark excitations}}.

\subsection{$Z_c(4200)$, $Z_{\bar{c}\bar{s}}(4600)$, $Z_{\bar{c}\bar{s}}(4900)$, $Z_{\bar{c}\bar{s}}(5200)$}
In 2014, the Belle collaboration analyzed the decays $\bar{B}^0\to K^- \pi^+ J/\psi$ and observed a resonance  $Z_c(4200)$ in the $J/\psi \pi^+$ mass spectrum   with a statistical significance  more than $6.2\,\sigma$, the measured    mass and width are
$ 4196^{+31}_{-29}{}^{+17}_{-13} \,\rm{MeV}$
and  $ 370^{+70}_{-70}{}^{+70}_{-132}\,\rm{MeV}$, respectively, the preferred assignment  is $J^P = 1^+$ \cite{Zc4200-Belle-PRD-2014}.

In 2019, the LHCb collaboration performed an angular analysis of the  decays $B^0\to J/\psi K^+\pi^-$, examined the $m(J/\psi \pi^-)$ versus the $m(K^+\pi^-)$ plane, and observed two  structures in the vicinity of  the energies $m(J/\psi \pi^-)=4200 \,\rm{MeV}$ and $4600\,\rm{MeV}$, respectively \cite{LHCb-Zc4200-Zc4600-PRL-2019}.

In 2024, the LHCb collaboration performed the first full amplitude analysis of the decays $B^+ \to \psi(2S) K^+ \pi^+ \pi^-$, and they developed   an amplitude model with 53 components  comprising 11 hidden-charm exotic states, for example,
the $Z_c(4200)$ and $Z_c(4430)$ in the $\psi(2S) \pi^+$ mass spectrum with the $J^P=1^+$; the $Z_{\bar{c}\bar{s}}(4600)$ and $Z_{\bar{c}\bar{s}}(4900)$ in the $\psi(2S)K^*(892)$ mass spectrum with the $J^P=1^+$, which might be the radial excitations of the $Z_{\bar{c}\bar{s}}(4000)$ in the scenario of tetraquark states with the valence quarks $c\bar{c}d\bar{s}$;
the $Z_{\bar{c}\bar{s}}(4000)$, $Z_c(4055)$ and $Z_{\bar{c}\bar{s}}(5200)$  are  effective descriptions of  generic partial wave-functions with the $J^P=1^+$, $1^-$ and $1^-$, respectively
  \cite{LHCb-JHEP-11-tetra-2407}. The spin-parity of the $Z_c(4200)$ is
     determined to be $1^+$ for the first time with a significance exceeding $5\sigma$.

We group the $Z_c(4200)$ with the $Z_{\bar{c}\bar{s}}(4600)$, $Z_{\bar{c}\bar{s}}(4900)$ and $Z_{\bar{c}\bar{s}}(5200)$ together into one-subsection as its assignment is  still an open problem, and we would like to revisit this subject to discuss the possible assignment based on the QCD sum rules in Sect.{\bf\ref{HH-tetraquark-33}}.

\subsection{$Z_b(10610)$, $Z_b(10650)$, $Y(10750)$}
In 2011, the Belle collaboration reported the first observation of the $Z_b(10610)$ and $Z_b(10650)$ in the $\pi^{\pm}\Upsilon({\rm 1,2,3S})$  and $\pi^{\pm} h_b({\rm 1,2P})$  mass spectra associated  with a single charged pion in the  $\Upsilon({\rm 5S})$ decays,  the quantum numbers  $I^G(J^P)=1^+(1^+)$ are favored \cite{Belle-Zb10610-1105}.
Subsequently, the Belle collaboration updated the measured parameters
$ M_{Z_b(10610)}=(10607.2\pm2.0)\,\rm{ MeV}$, $M_{Z_b(10650)}=(10652.2\pm1.5)\,\rm{MeV}$, $\Gamma_{Z_b(10610)}=(18.4\pm2.4) \,\rm{MeV}$ and
$\Gamma_{Z_b(10650)}=(11.5\pm2.2)\,\rm{ MeV}$, respectively  \cite{Belle-Zb10610-PRL-1110}.
In 2013, the Belle collaboration observed the $\Upsilon(5{\rm S}) \to \Upsilon ({\rm 1,2,3S}) \pi^0 \pi^0$ decays for the first time, and obtained  the neutral $Z_b^0(10610)$ in a Dalitz
analysis of the decays to the final states $\Upsilon(2,3{\rm S}) \pi^0$ \cite{Belle-Zb10610-neutral-PRD-1308}.

In 2019, the Belle collaboration observed  a resonance structure $Y(10750)$  in the  $e^+e^-\to\Upsilon(nS)\pi^+\pi^-$ ($n=1,\,2,\,3$) cross sections \cite{Belle-Y10750-JHEP-2019}. The Breit-Wigner  mass and width are
$10752.7\pm5.9\,{}^{+0.7}_{-1.1}\,\rm{MeV}$ and $35.5^{+17.6}_{-11.3}\,{}^{+3.9}_{-3.3}\,\rm{MeV}$, respectively. The $Y(10750)$ is observed in the  $\Upsilon(n{\rm S})\pi^+\pi^-
$  mass spectrum with $n=1,\,2,\,3$, its quantum numbers are  $J^{PC}=1^{--}$.

The Belle II collaboration confirmed the $Y(10750)$  in the processes $e^+e^-\to \omega\chi_{b1}(1\rm{P})$, $\omega\chi_{b2}(1\rm{P})$ \cite{BelleII-Y10750-PRL-2023}, $\pi^{+}\pi^{-}\Upsilon(1\rm{S})$, $\pi^{+}\pi^{-}\Upsilon(2\rm{S})$ \cite{BelleII-Y10750-JHEP-2024}, and observed no evidence in the processes
$e^+e^-\to \omega\eta_b(1{\rm S})$, $ \omega\chi_{b0}(1{\rm P})$ \cite{BelleII-Y10750-PRD-2024}, $\pi^{+}\pi^{-}\Upsilon(3\rm{S})$ \cite{BelleII-Y10750-JHEP-2024}.

Based on the predictions of the QCD sum rules, we can assign the $Z_b(10610)$ and $Z_b(10650)$ as the color
$\bar{\mathbf{3}}\mathbf{3}$-type or $\mathbf{1}\mathbf{1}$-type tetraquark states tentatively, see  Sect.{\bf\ref{Tetra-Positive}} and  Sect.{\bf\ref{11-tetra-states}}, and assign the $Y(10750)$ as the color  $\bar{\mathbf{3}}\mathbf{3}$-type tetraquark state with an explicit P-wave between the diquark and antidiquark tentatively, see Sect.{\bf \ref{Tetraquark-P-wave-explicit}}.

\subsection{$T_{cc}(3875)$}
In 2021, the LHCb collaboration  formally announced  observation of the exotic state $T_{cc}^+(3875)$   just below  the $D^0D^{*+}$ threshold \cite{LHCb-Tcc-NatureP-2022,LHCb-Tcc-NatureC-2022}.  The Breit-Wigner mass and width are $\delta M_{BW} = -273\pm 61\pm 5^{+11}_{-14}~\text{KeV}$ below the $D^0D^{*+}$ threshold and $\Gamma_{BW} = 410\pm 165\pm 43^{+18}_{-38}~\text{KeV}$  \cite{LHCb-Tcc-NatureP-2022,LHCb-Tcc-NatureC-2022}. The exotic state $T^+_{cc}(3875)$ is consistent with  the ground  isoscalar  tetraquark state with  the valence quarks  $cc\bar{u}\bar{d}$   and spin-parity  $J^{P}=1^+$, and  exploring the $DD$ mass spectrum  disfavors
 interpreting  the  $T_{cc}^+(3875)$ as an isovector state.
 The observation of the  $T_{cc}^+(3875)$ is a great breakthrough  beyond the $\Xi_{cc}^{++}$ for hadron physics, and it is the first doubly-charmed tetraquark candidate  with the typical quark configuration $cc\bar{u}\bar{d}$.

Based on the predictions of the QCD sum rules, we can assign the $T_{cc}(3875)$  as the color
$\bar{\mathbf{3}}\mathbf{3}$-type or $\mathbf{1}\mathbf{1}$-type tetraquark state tentatively based on the QCD sum rules, see  Sect.{\bf\ref{Doubly H tetraquark}} and  Sect.{\bf\ref{11-doubly-tetra-states}}.

\subsection{$P_c(4312)$, $P_c(4380)$, $P_c(4440)$, $P_c(4457)$, $P_c(4337)$, $P_{cs}(4338)$, $P_{cs}(4459)$}
In 2015,  the  LHCb collaboration  observed  two exotic structures $P_c(4380)$ and $P_c(4450)$ in the $J/\psi p$ mass spectrum in the $\Lambda_b^0\to J/\psi K^- p$ decays \cite{LHCb-Pc4380}.  The  $P_c(4380)$ has a mass of $4380\pm 8\pm 29\,\rm{MeV}$  and a width of $205\pm 18\pm 86\,\rm{MeV}$, while the $P_c(4450)$ has a mass of $4449.8\pm 1.7\pm 2.5\,\rm{MeV}$ and a width of $39\pm 5\pm 19\,\rm{MeV}$. The preferred spin-parity assignments of the $P_c(4380)$ and $P_c(4450)$ are  $J^P={\frac{3}{2}}^-$ and ${\frac{5}{2}}^+$, respectively  \cite{LHCb-Pc4380}.

In 2019, the LHCb collaboration studied the $\Lambda_b^0\to J/\psi K^- p$ decays with a data sample, which is an order of magnitude larger than that previously analyzed, and observed a  narrow pentaquark candidate $P_c(4312)$ in the $J/\psi p$ mass spectrum. Furthermore,
 the LHCb collaboration confirmed the  pentaquark structure $P_c(4450)$, and observed that it consists  of two narrow overlapping peaks $P_c(4440)$ and $P_c(4457)$ \cite{LHCb-Pc4312}.    The measured  masses and widths are
\begin{flalign}
 &P_c(4312) : M = 4311.9\pm0.7^{+6.8}_{-0.6} \mbox{ MeV}\, , \, \Gamma = 9.8\pm2.7^{+ 3.7}_{- 4.5} \mbox{ MeV} \, , \nonumber \\
 & P_c(4440) : M = 4440.3\pm1.3^{+4.1}_{-4.7} \mbox{ MeV}\, , \, \Gamma = 20.6\pm4.9_{-10.1}^{+ 8.7} \mbox{ MeV} \, , \nonumber \\
 &P_c(4457) : M = 4457.3\pm0.6^{+4.1}_{-1.7} \mbox{ MeV} \, ,\, \Gamma = 6.4\pm2.0_{- 1.9}^{+ 5.7} \mbox{ MeV} \,   .
\end{flalign}

In 2021, the LHCb collaboration reported an evidence of a hidden-charm pentaquark candidate $P_{cs}(4459)$ with the strangeness $S=-1$ in the $J/\psi \Lambda$ mass spectrum with  a significance of  $3.1\sigma$ in the $\Xi_b^- \to J/\psi K^- \Lambda$ decays  \cite{LHCb-Pcs4459},
the Breit-Wigner mass and width are
\begin{flalign}
 &P_{cs}(4459) : M = 4458.8 \pm 2.9 {}^{+4.7}_{-1.1} \mbox{ MeV}\, , \, \Gamma = 17.3 \pm 6.5 {}^{+8.0}_{-5.7} \mbox{ MeV} \, ,
\end{flalign}
and the spin-parity  have not been determined yet up to now.

In 2022,  the LHCb collaboration observed an evidence for a structure $P_c(4337)$ in the $J/\psi p$ and $J/\psi \bar{p}$ systems in the $B_s^0 \to J/\psi p \bar{p}$ decays with  a significance about $3.1-3.7\sigma$ depending on the $J^P$ hypothesis \cite{LHCb-Pc4337}, the Breit-Wigner mass and width are
\begin{flalign}
 &P_{c}(4337) : M =4337^{+7}_{-4} {}^{+2}_{-2}\, \mbox{ MeV}\, , \, \Gamma = 29^{+26}_{-12} {}^{+14}_{-14}\, \mbox{ MeV} \, .
\end{flalign}
Its existence  is still need confirmation and its spin-parity are not measured yet.

 In 2023, the LHCb collaboration observed an evidence for a new structure $P_{cs}(4338)$ in the $J/\psi \Lambda$ mass distribution in the $B^- \to J/\psi \Lambda \bar{p}$ decays \cite{LHCb-Pcs4338}, the measured  Breit-Wigner mass and width are
 \begin{flalign}
 &P_{cs}(4338) : M =4338.2\pm0.7\pm0.4\, \mbox{ MeV}\, , \, \Gamma = 7.0\pm1.2\pm1.3\, \mbox{ MeV} \, ,
\end{flalign}
  and the favored  spin-parity is $J^P={\frac{1}{2}}^-$.

The $P_{cs}(4338)$ and $P_{cs}(4459)$ are observed in the $J/\psi \Lambda$  mass spectrum,  they have the isospin $I=0$, as the strong decays conserve isospin.
 The  $P_c(4312)$, $P_c(4380)$, $P_c(4440)$, $P_c(4457)$, $P_{cs}(4459)$ and $P_{cs}(4338)$ lie slightly below or above the thresholds of the charmed  meson-baryon pairs $\bar{D}\Sigma_c$, $\bar{D}\Sigma_c^*$, $\bar{D}^*\Sigma_c$, $\bar{D}^*\Sigma_c$, $\bar{D}\Xi_c^\prime$ ($\bar{D}\Xi_c^*$, $\bar{D}^*\Xi_c$, $\bar{D}^*\Xi_c^\prime$) and $\bar{D}\Xi_c$, respectively.  It is difficult to identify  the $P_c(4337)$ as the  molecular state without resorting to the help of  large coupled-channel effects  due to lacking nearby  meson-baryon thresholds. Or the $P_c(4312)$ and $P_c(4337)$ are the same particle, such a possibility cannot be excluded at the present time.

Based on the predictions of the QCD sum rules, we can assign the $P_c(4312)$,  $P_c(4337)$, $P_c(4380)$, $P_c(4440)$, $P_c(4457)$ and $P_{cs}(4459)$ as the color $\bar{\mathbf 3}\bar{\mathbf 3} \bar{\mathbf 3}$-type  pentaquark states tentatively, see Table \ref{mass-penta-Pc4312} and Table
\ref{mass-1508-et al-Pcs}
 in Sect.{\bf\ref{333-penta-Sect}}, and assign the $P_c(4312)$, $P_c(4380)$, $P_c(4440)$, $P_c(4457)$,  $P_{cs}(4338)$ and $P_{cs}(4459)$ as the color $\mathbf 1\mathbf 1 $-type pentaquark states, see Table \ref{mass-residue-penta-mole} in Sect.{\bf\ref{11-penta-Sect}}.

\subsection{$d^*(2380)$}
In 2014, the scientists in the WASA-at-COSY collaboration and SAID
data analysis center  performed
exclusive and kinematically complete high-statistics measurements of the
polarized $\vec{n}p$ scattering through the quasifree process $\vec{d}p \to np + p_{\rm spectator}$ in the energy region of the
narrow resonance-like structure $d^*$ with the $I\,(J^P) = 0\,(3^+)$,
and confirmed their (WASA-at-COSY collaboration) early observation of the $d^*(2380)$  in the
double-pionic fusion channels,
they produced a resonance pole in the ${}^3D_3-{}^3G_3$
coupled partial waves at $2380\pm10 - i40\pm5$ MeV \cite{WASA-at-COSY-PRL-2014,WASA-at-COSY-PRC-2014}, --- in accordance with the
$\Delta\Delta$ dibaryon resonance \cite{d2380-HuanF-PRC-2015,d2380-SHLee-PRD-2015,d2380-HXChen-PRC-2015,
d2380-Gal-PLB-2017,d2380-HuangF-review-2022}. And we will revisit this subject at the end of Sect.{\bf \ref{11-doubly-tetra-states}}.

\section{Theoretical foundations}
In this section, we would like to review the typical theoretical methods and related possible assignments concisely, then focus on the QCD sum rules in the subsequent sub-sections, see Sects.{\bf \ref{Tetra-QCDSR}}, {\bf\ref{reliable?}} and {\bf\ref{Energy-scale-dependence}}.

\subsection{Typical theoretical methods and possible assignments}
There have been tremendous progresses on the hadron spectrum containing two heavy quarks experimentally since the observation of the $X(3872)$. It is surprising that many resonant structures lie around thresholds of a pair of heavy hadrons.
A natural  conjecture is that  they are possible deuteron-like two-particle bound states bound via attractive interactions induced by one-pion exchange or one-boson exchange \cite{X3872-Not-mole-LiuX-EPJC-2008,X3872-mole-Swanson-PLB-2004,
X3872-mole-Mehen-PRD-2007,
X3872-mole-FKGuo-PRD-2013,X3872-mole-Nieves-PRD-2012,
Tornqvist-ZPC-1994,Karliner-cc-mole-2015-PRL,
One-pion-X3872-Close-PRD-2008,One-pion-Nieves-PRD-2013,One-boson-Oka-PRD-2012,
One-pion-GuoFK-PRD-2015,One-pion-Hosaka-PRD-2017,One-pion-Close-PRD-2010}, it is only a possibility.
In the heavy quark limit, the $Q_{\mathbf 3}\bar{q}_{\bar{\mathbf 3}}$ mesons and $Q_{\mathbf 3} [qq]_{\bar{\mathbf 3}}$ baryons have the antiquark-diquark symmetry, $\bar{q}_{\bar{\mathbf 3}}\leftrightarrow [qq]_{\bar{\mathbf 3}}$, therefore the $Q\bar{q}\, q^\prime \bar{Q}$  and $Q[qq]\, q^\prime \bar{Q}$ systems could be analyzed  in the same theoretical scheme, in this sub-section, we would like to focus on the tetraquark systems.
Someone maybe wonder: are they  threshold cusps, triangle singularities  or genuine resonances? As there always exist threshold cusps at the S-wave thresholds or triangle singularities near the thresholds.
Firstly, let us see the outcomes  based on the (non)relativistic effective field theory.

\subsubsection{Threshold cusps, Triangle singularities or genuine Resonances}
Not all peaks in the invariant mass distributions are genuine resonances, they often arise  due to the nearby kinematical singularities of the transition amplitudes in the complex energy plane. Those singularities (or Landau singularities) occur when the intermediate particles  are on the mass-shell.
The simplest case is the cusp at the normal two-body threshold, there always exists a cusp at the S-wave threshold of two particles coupling to the final states,
while a more complicated case is the so-called triangle singularity. They
maybe produce observable effects if the involved interactions are strong enough,
sometimes, even  mimic the behavior of a resonance. It is important to distinguish  kinematic singularities from genuine resonances. We would like to give an example concerning the exotic states $Y(4260)$ and $Z_c(3900)$ to illustrate their possible assignments in the scenarios  of threshold cusps, triangle singularities and genuine resonances.

The threshold cusp is determined by masses of the involved particles,
 how strong the cusp  depends on detailed dynamics and the cusp could be rather dramatic if there is a nearby pole, thus it plays an important role in studying the exotic states \cite{GuoFK-cusp-triangle-PPNP-2020}. For example, the $X(3872)$, $Z_c(3900)$, $Z_c(4020)$, $Z_b(10610)$ and $Z_b(10650)$ lie near the $D^0\bar{D}^{*0}$, $D\bar D^*$, $D^*\bar D^*$, $B\bar B^*$ and $B^*\bar B^*$ thresholds, respectively, their quantum numbers are the same as the corresponding S-wave meson pairs although the $J^{PC}$ of the $Z_c(4020)$ have not been fully determined yet \cite{PDG-2024}.

The $X(3872)$ was assigned to be a threshold cusp by Bugg \cite{X3872-cusps-Bugg-2004-PLB}, subsequently, he realized  that the very narrow line shapes  in the $J/\psi\rho$ and $D^0\bar{D}^{*0}$ channels  could not be fitted with only a threshold cusp, and a resonance or virtual state pole was necessary \cite{Bugg-X3872-no-cusp-JPG-2008}.

\begin{figure}
 \centering
 \includegraphics[totalheight=4cm,width=6cm]{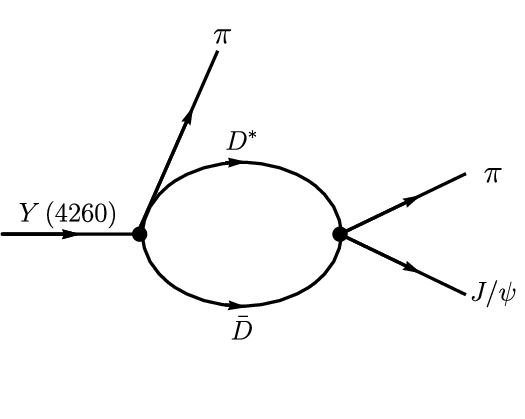}
 \caption{ A Feynman diagram for the  decay $Y(4260)\to J/\psi \pi^+ \pi^-$ maybe  lead to  threshold cusp.  }\label{Y4260-Swan-cusp-Feynman}
\end{figure}

In a modified threshold cusp model \cite{Swanson-Zc-Zb-cusp-PRD-2015,Swanson-Zc3900-cusp-IJMPE-2016}, see the Feynman diagram shown in Fig.\ref{Y4260-Swan-cusp-Feynman} as an example, both the inelastic ($J/\psi\pi,h_c\pi$) and elastic ($D\bar D^*$, $D^*\bar D^*$) decay modes were considered for the $Z_c(3900)$ and $Z_c(4020)$, analogous discussions are applied to the $Z_b(10610)$ and $Z_b(10650)$.
A Gaussian form-factor  was chosen for all the vertices including the tree-level ones. Then the experimental data for the $J/\psi\pi$ and $D\bar D^*$ mass spectra for the $Z_c(3900)$ and the $D^*\bar D^*$ and $h_c\pi$ mass spectra for the $Z_c(4020)$ could be fitted very good, exotic resonances are
not required to account for the experimental data. However, the fitting quality depends crucially on the cutoff parameter in the Gaussian form-factor.

The triangle singularity is determined by the masses of the intermediate particles plus the invariant masses of the external ones, therefore the triangle singularities are  sensitive to the kinematic variables, the peak position and peak shape change according to the variations of the external energies.
More precisely, the triangle singularities are determined by the scalar triangle loop integral, which does not depend on the orbital angular momentum for each vertex,  however, sharp triangle singularity peaks are constrained to  the S-wave internal particles, as momentum  power factor  weakens the singular behavior in other  cases \cite{GuoFK-cusp-triangle-PPNP-2020}.

\begin{figure}
 \centering
 \includegraphics[totalheight=4cm,width=6cm]{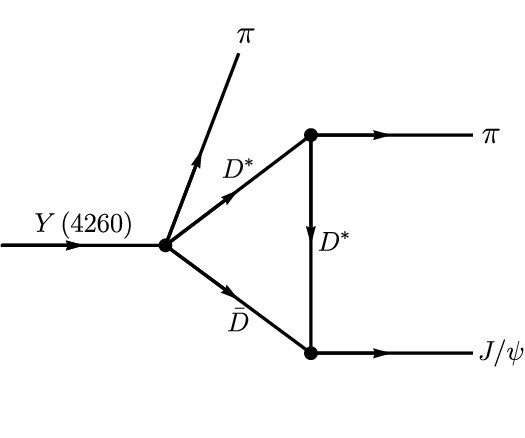}
 \caption{ A Feynman diagram for the  decay $Y(4260)\to J/\psi \pi^+ \pi^-$ in the ISPE mechanism.  }\label{Y4260-LiuX-TS-Feynman}
\end{figure}

In the initial single-pion emission (ISPE) mechanism, the triangle diagrams contribute to  threshold cusps, see Fig.\ref{Y4260-LiuX-TS-Feynman} for a typical Feynman diagram. This mechanism was suggested firstly to study the exotic structures $Z_b(10610)$ and $Z_b(10650)$,
 Chen and Liu introduced a dipole form-factor  to accompany  the exchanged $B$-meson propagator and took account of the $B\bar{B}$, $B\bar{B}^*$, $B^*\bar{B}$ and $B^*\bar{B}^*$ triangle loop diagrams, and produced sharp cusps right around the $Z_b(10610)$ and $Z_b(10650)$ structures in the $\Upsilon(1 {\rm S},2{\rm S},3 {\rm S})\pi$ and $h_b(1 {\rm P},2{\rm P})\pi$ mass spectra, but  observed no cusp at the $B\bar B$ threshold \cite{LiuX-Zb-cusp-PRD-2011}. Similarly, Chen, Liu and Matsuki took account of the $D\bar{D}$, $D\bar{D}^*$, $D^*\bar{D}$ and $D^*\bar{D}^*$ triangle loop diagrams and the intermediate $f_0(600)$ and $f_0(980)$ to study the decays $Y(4260)\to J/\psi \pi\pi$, and observed two peaks, the $Z_c(3900)$ and its reflection \cite{LiuX-Zc-cusp-PRD-2013}. And they studied   other processes with possible triangle singularities \cite{LiuX-Zcs-cusp-PRL-2013}.

\begin{figure}
 \centering
 \includegraphics[totalheight=4cm,width=6cm]{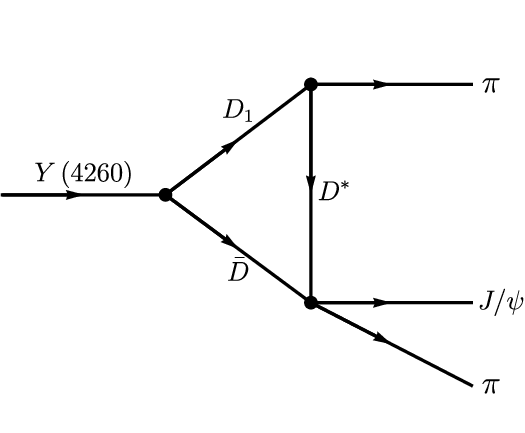}
 \caption{ The Feynman diagram for the  decays $Y(4260)\to J/\psi \pi^+ \pi^-$ in the molecule scenario with re-scattering mechanism.  }\label{Y4260-GFK-TS-Feynman}
\end{figure}

In Ref.\cite{GuoFK-Threshold-Str-PRL-2021}, Dong, Guo and Zou  show that the threshold cusp appears as a peak only for channels with attractive interaction, and the cusp's width is inversely proportional to the reduced mass  for the relevant threshold. There should be threshold structures at any threshold of a  $q\bar{Q}$ and $Q\bar{q}$ ($Qqq$) pair, which have attractive interaction at threshold, in the invariant mass distribution of a $Q\bar{Q}$ state and a $q\bar{q}$ ($qqq$) state coupling to the $q\bar{Q}$ and $Q\bar{q}$ ($Qqq$) pair, and the structure becomes more pronounced if there is a near-threshold pole.

In Ref.\cite{LiuXH-tri-Y4260-Zc3900-PRD-2013}, Liu and Li suppose the $Y(4260)$ as a $D_1\bar{D}$ molecular state to study its decays, see Fig.\ref{Y4260-GFK-TS-Feynman} as an example, and observe that under  special kinematic configurations, the triangle singularity  maybe occur in the re-scattering amplitude, which can change the threshold behavior significantly. Obvious threshold enhancements or narrow cusp structures appear quite
naturally without introducing a genuine resonance, but cannot exclude  existence of a genuine resonance, such a mechanism  also works for the pentaquark structures  \cite{LiuXH-tri-Pc-PLB-2016,GuoFK-tri-Pc-EPJA-2016}.

If the $Y(4260)$ have a large $D_1\bar{D}$  molecular component, the decays $Y(4260)\to J/\psi\pi^+\pi^-$  can occur through  the re-scattering process \cite{ZhaoQ-Y4260-Zc3900-Tri-pole-PRL-2013,ZhaoQ-Y4260-Tri-pole-PLB-2013,
GuoFK-Y4260-mole-PRD-2014}, see Fig.\ref{Y4260-GFK-TS-Feynman}. The singularity regions provide an ideal environment for  forming  bound states or resonances.
Although the  $Y(4260)$ lies slightly below the $D_1(2420)\bar{D}$ threshold, the triangle singularity in the  ${J/\psi\pi^\pm}$ invariant mass distribution of the re-scattering amplitude is still near the physical boundary and can influence the $J/\psi\pi^\pm$ invariant mass distribution around the $D^*\bar{D}$ threshold significantly \cite{ZhaoQ-Y4260-Zc3900-Tri-pole-PRL-2013}. Despite the importance of the triangle diagram contribution, it is insisted  that a $Z_c(3900)$ resonance was still needed in order to fit to the narrow peak observed in experiments \cite{LiuXH-tri-Y4260-Zc3900-PRD-2013,
ZhaoQ-Y4260-Zc3900-Tri-pole-PRL-2013,GuoFK-Zc-Zcs-mole-PRD-2022,
GuoFK-Zc3900-mole-PRD-2024}. The diagrams similar  to Fig.\ref{Y4260-GFK-TS-Feynman}  also play an important role in the hidden-bottom sector \cite{Zb-produ-GuoFK-PRD-2019,Zb-produ-GuoFK-PRD-2017}.

The resonance pole can be incorporated by constructing a unitarized coupled-channel scattering $T$-matrix by fitting to the experimental  data,  the best fit still demands  the $T$-matrix to have a resonant or virtual pole near the $D\bar D^*$ threshold, which can be interpreted as the $Z_c(3900)$ \cite{GuoFK-Y4260-resonace-virtual-PLB-2014}.
The molecule assignment provides a natural explanation for the resonance-like structure $Z_c(3900)$ in the $Y(4260)$ decays \cite{GuoFK-Y4260-mole-PRD-2014,GuoFK-Y4260-resonace-virtual-PLB-2014},  the kinematical  threshold cusp cannot produce
a narrow peak in the invariant mass distribution in the elastic channel in contrast with a genuine S-matrix pole \cite{GuoFK-Y4260-cusp-PRD-2015}.

In a similar scenario, Szczepaniak suggests that the $Z_c(3900)$ peak could be attributed to the $D_0^*(2300)\bar{D}^*D$ loop instead of the $D_1(2420)\bar{D}D^*$ loop, which is in the physical region by neglecting  the width of the $D_0^*(2300)$ \cite{Triangle-Szczepaniak-PLB-2015}. The  triangle singularities can produce enhancement potentially in the amplitude  consistent with the experimental data qualitatively.

In Ref.\cite{GuoFK-Zc3900-pole-SCPMA-2024}, Chen,  Du and  Guo perform a unified description of the $\pi^+\pi^-$ and $J/\psi\pi^\pm$  mass distributions  for the $e^+e^- \to J/\psi \pi^+\pi^-$ and the $D^0 D^{\ast-}$ mass distribution for the $e^+e^-\to D^0 D^{\ast-} \pi^+$ at $\sqrt{s}=$  4.23 and 4.26~GeV.
They take  account of the open-charm meson loops containing triangle singularities, the $J/\psi\pi$-$D\bar D^*$ coupled-channel interaction respecting unitarity,
and the strong $\pi\pi$-$K\bar K$ final-state interaction using dispersion relations, which  lead to a precise determination of the  pole mass and width $(3880.7 \pm 1.7\pm 22.4)$ MeV and $(35.9 \pm 1.4\pm 15.3)$ MeV, respectively, and indicate the molecular and non-molecular components are of similar importance for the structure $Z_c(3900)$.

Precisely measuring the near threshold structures  plays  an important role in diagnosing the heavy-hadron interactions, therefore understanding the puzzling hidden-charm and hidden-bottom structures. Furthermore, it is important to search for the resonant structures in processes free of  triangle singularities, such as the photo-production and pion-induced production  processes in the $e^+e^-$ and $pp$ collisions \cite{PhotonProd-LiBA-PLB-2005,PhotonProd-LiuXH-PRD-2008,PionPro-HeJ-PLB-2019}.
For a recent review on the production of the exotic hadrons in the $pp$ and nuclear collisions, see Ref.\cite{Product-exotic-GuoFK}.

\subsubsection{Dynamical Generated Resonances and Molecular States}
\begin{figure}
 \centering
 \includegraphics[totalheight=4cm,width=9cm]{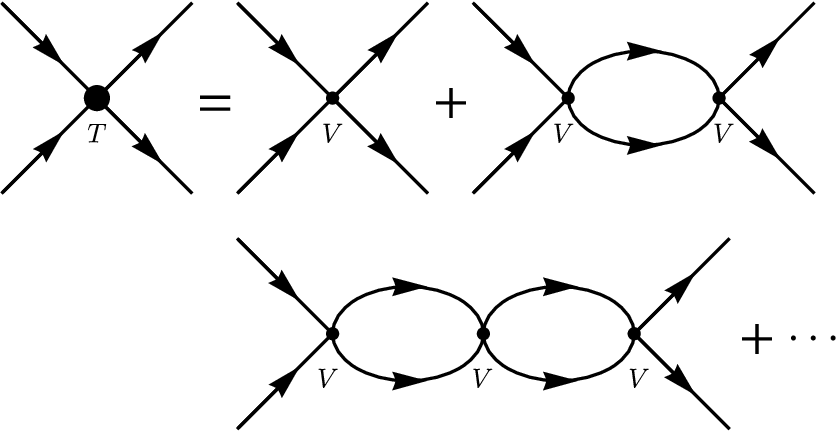}
 \caption{ Summing the two-meson loops through the Bethe-Salpeter equation.  }\label{BSE-sum}
\end{figure}

If we take the traditional heavy mesons as the elementary degrees of freedom,
then we construct the heavy meson effective Lagrangian according to the chiral symmetry, hidden-local symmetry and heavy quark symmetry \cite{HMEFT-Gatto-PRT-1996,HMEFT-WangZG-EPJP-2014,HMEFT-Colangelo-PRD-2018}. It is easy to obtain the    two-meson scattering amplitudes $V$. Then we have three choices:

{\bf Firstly},
we unitarize the amplitudes by taking account of the intermediate two-meson loops with the coupled channel effects through the Bethe-Salpeter or Lippmann-Schwinger equation with on-shell factorization \cite{BSE-Oset-PPNP-2000,BSE-Hosaka-dynamically-PRC-2012},
\begin{eqnarray}
T &=&V+VGV+VGVGV+\cdots\, ,\nonumber\\
&=& [1 - V \, G]^{-1}\, V \, ,
\end{eqnarray}
where the $G$ is the loop function, see Fig.\ref{BSE-sum} for a diagrammatical representation. Then we explore the analytical properties of the full amplitudes $T$, and try to find the poles in the complex Riemann sheets, such as the bound states, virtual states and resonances. Such discussions are applied to the baryon-meson systems directly.

Bound states appear as poles on the physical sheet, and
 only appear on the real $s$-axis below the lowest threshold by causality.
 Virtual states also appear on the real $s$-axis, however, on the unphysical Riemann sheet.  Resonances appear as poles on an unphysical Riemann sheet close
to the physical one with non-zero imaginary part, and they appear in conjugate pairs. For example, the loosely bound states $X(3872)$, $Y(3940)$, $Z_c(3900)$, $Z_b(10610)$, $Z_b(10650)$ \cite{X3872-mole-Oset-PRD-2010,X3872-mole-Petrov-PLB-2006,BSE-X3872-Oset-EPJA-2007,
BSE-Y3940-Oset-PRD-2009,BSE-Zc3900-Nielsen-PRD-2014,BSE-Zc-Zcs-GuoFK-PRD-2022,
BSE-Oset-Zb-PRD-2015,
BSE-FKGuo-Zb-PRD-2013,BSE-FKGuo-Zb-EPJA-2011,BSE-FKGuo-Zb-NPA-2005},  the hidden-charm pentaquark resonances  \cite{BSE-WuJJ-PRC-2011,BSE-CWXiao-PRD-2013,BSE-Oset-penta-mole-PRD-2015,
BSE-LiangWH-EPJA-2016,BSE-CWXiao-PRD-2019}. We usually apply Weinberg's  compositeness condition to estimate the hadronic molecule components \cite{Composite-Hyodo-PRL-2013,Composite-GuoZH-PRD-2016,Composite-Hyodo-PRC-2016,
Composite-KangXW-PRD-2016,Composite-GuoFK-PRD-2022,
Composite-X3872-GuoFK-JHEP-2024,Composite-X3872-GuoFK-2025}.

{\bf Secondly}, we take the scattering amplitudes $V$ as interaction kernels, solve the quasi-potential Bethe-Salpeter or Lippmann-Schwinger equation with the coupled channel effects directly, then explore the analytical properties of the full amplitudes \cite{Quasi-BSE-HeJ-Zc3900-PRD-2015,Quasi-BSE-HeJ-Pc-PLB-2016,
Quasi-BSE-HeJ-Pc-EPJC-2019,
Quasi-LSE-Ortega-Zc-EPJC-2019,BSE-tetra-mole-LiXQ-JHEP-2012,
BSE-tetra-mole-LiXQ-PRD-2020,
BSE-tetra-mole-GuoXH-PRD-2022},  or obtain the bound energies directly to estimate the bound states \cite{X3872-mole-Nieves-PRD-2012,LSE-momentum-Nieves-PRD-2011,LSE-momentum-Nieves-PRD-2018,
LSE-momentum-GengLS-PRD-2019,
LSE-momentum-GengLS-PRL-2019,LSE-momentum-GengLS-PRD-2021}.

{\bf Thirdly}, we reduce the scattering amplitudes to interaction potentials in the momentum space in terms of the Breit approximation and introduce monopole form-factors associated with the exchanged particles. Generally, we should introduce  form-factors  in each
interaction vertex, which reflects the off-shell effect of the
exchanged meson and the structure effect, because the components of
the molecular states  and exchanged mesons are not point
particles.  Then we perform the Fourier
transformation to obtain the potential in the coordinate space, finally we solve the Schrodinger equation directly to obtain the  binding energy \cite{X3872-Not-mole-LiuX-EPJC-2008,OBE-X3872-mole-ZhuSL-EPJC-2009,OBE-Y4260-mole-DingGJ-PRD-2009,
OBE-X3872-mole-Faessler-PRD-2009,OBE-Zb-mole-ZhuSL-PRD-2011,
OBE-penta-mole-ZhuSL-CPC-2012,OBE-tetra-mole-Hosaka-PRD-2012,
OBE-tetra-mole-ZhuSL-PRD-2013,OBE-tetra-mole-WangBo-PRD-2019,
OBE-penta-mole-WangBo-PRD-2019,OBE-penta-mole-WangBo-JHEP-2019,OBE-penta-mole-GengLS-PRD-2021}.

\subsubsection{$Q\bar{Q}$ states with coupled channel effects}
In the famous GI model, the charmonium $\chi_{c1}(2^3{\rm P}_1)$ has the mass about $3953\,\rm{MeV}$ \cite{X3872-charmonium-Godfrey-2004-PRD}, which is about $100\, \rm{MeV}$ above the $X(3872)$ lying  near the $D^0 \bar{D}^{*0}$ threshold, the strong coupling to the nearby threshold maybe lead to some $DD^*$ molecular configuration. Furthermore, pure charmonium assignment cannot
 interpret the  high $\gamma \psi^\prime$ decay rate \cite{X3872-KangXW-psi2-PRD-2024}.

We can extend the constituent quark models to include the meson-meson Fock components, and write the physical charmonium (bottomonium) states in terms of $| \Psi \rangle$,
\begin{equation}
 | \Psi \rangle = \sum_\alpha c_\alpha | \psi_\alpha \rangle
 + \sum_\beta \chi_\beta(P) |\phi_{1} \phi_{2} \beta \rangle
\end{equation}
where the $|\psi_\alpha\rangle$ are the $Q\bar{Q}$ eigenstates,
the $\phi_{i}$ are $Q\bar{q}$ or $q\bar{Q}$ eigenstates,
$|\phi_{1} \phi_{2} \beta \rangle$ are the two-meson state with $\beta$ quantum
numbers, and the $\chi_\beta(P)$ is the relative wave
function between the two mesons. Then we solve the Schrodinger equation directly \cite{QbarQ-MM-Braaten-PRD-2004,QbarQ-MM-X3940-Ortega-PRD-2010,
QbarQ-MM-X3872-Rupp-EPJC-2013,QbarQ-MM-Ortega-PLB-2018,QbarQ-MM-ZhouZY-PRD-2017,
QbarQ-MM-ZhouZY-PRD-2018,
QbarQ-MM-X3872-Ferretti-PLB-2019,QbarQ-MM-X3872-Nefediev-PU-2019,
QbarQ-MM-Hosaka-JPG-2020,QbarQ-MM-ZhongXH-PRD-2024}.

\subsubsection{Hybrid and Tetraquark states in Born-Oppenheimer approximation}
Due to the large ratio of the mass of a nucleus to that of the electron,
the electrons respond almost instantaneously to the motion of the nuclei.
The energy of the electrons combined with the repulsive
Coulomb energy of the nuclei defines a Born-Oppenheimer  potential.
Accordingly,  due to the large ratio
of the heavy-quark mass $m_Q$ to the energy scale
$\Lambda_{\rm QCD}$ associated with the gluon field,
the gluons respond almost instantaneously to the motion of the heavy quarks $Q$ and $\bar Q$ \cite{Born-Braaten-PRL-2013,Born-Braaten-PRL-2014,Born-Braaten-PRD-2014,
Born-Brambilla-PRD-2015,
Born-Soto-PRD-2017,Born-Brambilla-PRD-2019}.

In the static limit, the $Q$ and $ \bar Q$ serve as two color-sources separated by a distance $r$,  the ground-state flavor-singlet Born-Oppenheimer potential
$V_{\Sigma_g^+}(r)$ is defined by the minimal energy of the gluonic
configurations, whose small and large $r$ limiting behaviors  are qualitatively compatible with the simple phenomenological Cornell potential, thus the $V_{\Sigma_g^+}(r)$ describes the traditional heavy quarkonium states.
The excited Born-Oppenheimer potentials $V_\Gamma(r)$ are defined as the minimal energies of the excited configurations for the gluon and light-quark fields with the quantum numbers $\Gamma$ \cite{Born-Braaten-PRL-2013,Born-Braaten-PRL-2014,Born-Braaten-PRD-2014}.

The  hybrid states $\bar{Q}GQ$ are energy levels  of a heavy quark  pair $Q \bar Q$ in the excited flavor-singlet Born-Oppenheimer potentials, the hybrid potentials. Juge, Kuti and Morningstar calculated many  flavor-singlet potentials
using the quenched lattice QCD \cite{Born-Juge-PRL-1999}.
There have been some works on the lowest lying  hybrid potentials
using lattice QCD with two flavors of dynamical Wilson fermions \cite{Born-Bali-PRD-2000,Born-Bali-PRD-2005}.
At large $r$, the hybrid potential $V_{\Gamma}(r)$ is a flux-tube extending between the $Q$ and $\bar Q$.
At small $r$, the hybrid potential approaches the repulsive
color-Coulomb potential between a $Q$ and $\bar Q$ in a color-octet state.
In the limit $r \to 0$, the $Q$ and $\bar Q$ sources  reduce to a
single local color-octet $Q \bar Q$ source.
The energy levels of the flavor-singlet gluon and light-quark field configurations
bound to a static color-octet source are called
 static hybrid mesons. The most effective pictorial representation of the hybrid states is the flux-tube model. Lattice QCD simulations show that two static quarks $Q$ and $\bar{Q}$ at large distances are
confined by approximately cylindrical regions of the color fields \cite{Born-Juge-PRL-1999,Born-Bali-PRD-2000,Born-Bali-PRD-2005}.

The $\bar{Q}Q\bar{q}q$ tetraquark states are energy levels in the Born-Oppenheimer potentials with  nonsinglet (excited singlet) flavor quantum numbers, the tetraquark potentials,
which are  distinguished  by the quantum numbers,
$u\bar{u}\pm d\bar{d}$, $u\bar{d}$, $d\bar{u}$, $s \bar q$, $s \bar s$, etc.
At large $r$, the minimal-energy configuration consists of
two static mesons localized near the $Q$ and $\bar Q$ sources.
At small $r$, the minimal-energy configuration is the
flavor-singlet $\Sigma_g^+$ potential accompanied by one or two pions (two or three pions), depending on the quantum numbers $\Gamma$. There have been works on the energies of static adjoint mesons using the quenched lattice QCD \cite{Born-Foster-PRD-1999}.
The static adjoint mesons are energy levels of the light-quark and gluon fields with nonsinglet flavor quantum numbers
bound to a static color-octet source.

The heavy quark motion is  restored by solving the Schrodinger equation in each of those potentials, and many  $X$, $Y$ and $Z$ mesons could  be assigned  as the bound states with  the Born-Oppenheimer potentials \cite{Born-Braaten-PRL-2013,Born-Braaten-PRL-2014,Born-Braaten-PRD-2014,
Born-Brambilla-PRD-2015,
Born-Soto-PRD-2017,Born-Brambilla-PRD-2019}.

\subsubsection{Tetraquarks in diquark models}\label{33-66-diquark-model}
If we take the quarks in color triplet $\mathbf{3}$ as the basic constituents, then we could construct the hadrons  according to the $SU(3)$ symmetry.
For the
 traditional  mesons,
\begin{eqnarray}
 q_{\mathbf{3}} \otimes \bar{q}_{\mathbf{\bar 3}} \to [\bar{q}q]_{\mathbf{1}} \, . \nonumber
\end{eqnarray}
For the traditional baryons,
\begin{eqnarray}
q_{\mathbf{3}} \otimes q_{\mathbf{3}}  \otimes q_{\mathbf{3}}  \to [q q q]_{\mathbf{1}} \, . \nonumber
\end{eqnarray}
For the tetraquark molecular states,
\begin{eqnarray}
 q_{\mathbf{3}} \otimes \bar{q}_{\mathbf{\bar 3}} \otimes q_{\mathbf{3}} \otimes \bar{q}_{\mathbf{\bar 3}} \to [\bar{q}q]_{\mathbf{1}}[\bar{q}q]_{\mathbf{1}}\oplus [\bar{q}q]_{\mathbf{8}}[\bar{q}q]_{\mathbf{8}}\to[\bar{q}q\bar{q}q]_{\mathbf{1}}\, \oplus [\bar{q}q\bar{q}q]_{\mathbf{1}}\, ,\nonumber
\end{eqnarray}
and we usually call the color $\mathbf{1}\mathbf{1}$ type structures as the molecular states.
For the  tetraquark states,
\begin{eqnarray}
  q_{\mathbf{3}} \otimes  q_{\mathbf{3}}\otimes \bar{q}_{\mathbf{\bar 3}} \otimes \bar{q}_{\mathbf{\bar 3}}
\to [q q]_{\mathbf{\bar 3}} [\bar q \bar q]_{\mathbf{3}} \oplus [q q]_{\mathbf{6}} [\bar q \bar q]_{\mathbf{\bar 6}} \to[qq\bar{q}\bar{q}]_{\mathbf{1}} \oplus [qq\bar{q}\bar{q}]_{\mathbf{1}}\, ,\nonumber
\end{eqnarray}
and we usually call the color $\bar{\mathbf{3}}\mathbf{3}$ type structures as the tetraquark states.
For the  pentaquark molecular states,
\begin{eqnarray}
 q_{\mathbf{3}}\otimes q_{\mathbf{3}} \otimes q_{\mathbf{3}} \otimes q_{\mathbf{3}} \otimes \bar{q}_{\mathbf{\bar 3}} \to [q q q]_\mathbf{1} [\bar q q]_{\mathbf{1}} \to [q q q q\bar{q}]_\mathbf{1}\, .\nonumber
\end{eqnarray}
For the  pentaquark states,
\begin{eqnarray}
  q_{\mathbf{3}} \otimes q_{\mathbf{3}} \otimes q_{\mathbf{3}} \otimes q_{\mathbf{3}} \otimes \bar{q}_{\mathbf{\bar 3}}
\to [q q]_{\mathbf{\bar 3}} [q q]_{\mathbf{\bar 3}} [\bar q]_{\mathbf{\bar 3}}
\to [qqqq\bar{q}]_{\mathbf{1}}  \, .\nonumber
\end{eqnarray}

If we take the viewpoint of the quantum field theory, the scattering amplitude for one-gluon exchange  is proportional to,
\begin{eqnarray}\label{T-to-33-66}
\left(\frac{\lambda^a}{2}\right)_{ij}\left(\frac{\lambda^a}{2}\right)_{kl}
&=&-\frac{N_c+1}{4N_c}\left(\delta_{ij}\delta_{kl}-\delta_{il}\delta_{kj} \right)+\frac{N_c-1}{4N_c}\left( \delta_{ij}\delta_{kl}+\delta_{il}\delta_{kj}\right) \, ,
\end{eqnarray}
where
the $\lambda^a$ is the  Gell-Mann matrix,  the $i$, $j$, $k$, $m$ and $l$ are color indexes, the $N_c$ is the color number, and $N_c=3$ in the real world.  The negative sign in front of the antisymmetric  antitriplet $\bar{\mathbf{3}}$ indicates the interaction
is attractive, which favors  formation of
the diquarks in  color antitriplet,  while the positive sign in front of the symmetric sextet $\mathbf{6}$ indicates
 the interaction  is repulsive,  which  disfavors  formation of
the diquarks in  color sextet.

In this sub-sub-section, we would like to focus on
the tetraquark states, as the extension to the pentaquark states is straightforward.  Now we define the color factor,
\begin{eqnarray}\label{lambda-lambda}
\widehat{C}_i \cdot \widehat{C}_j &=& \frac{\lambda_i^a}{2} \cdot \frac{\lambda_j^a}{2} \, ,
\end{eqnarray}
where the subscripts $i$ and $j$ denote the quarks,  $\langle\widehat{C}_i \cdot \widehat{C}_j \rangle=-\frac{2}{3}$ and  $\frac{1}{3}$ for the $\bar{\mathbf{3}}$ and $\mathbf{6}$ diquark $[qq]$, respectively,  and $\langle\widehat{C}_i \cdot \widehat{C}_j \rangle=-\frac{4}{3}$ and  $\frac{1}{6}$
for the  $\mathbf{1}$ and $\mathbf{8}$ quark-antiquark $\bar{q}q$, respectively.
If we define $\widehat{C}_{12}\cdot \widehat{C}_{34}=(\widehat{C}_1+\widehat{C}_2 )\cdot (\widehat{C}_3+\widehat{C}_4 )$, then $\langle \widehat{C}_{12}\cdot \widehat{C}_{34}\rangle=-\frac{4}{3}$ and $-\frac{10}{3}$ for the $\bar{\mathbf{3}}\mathbf{3}$ and $\mathbf{6}\bar{\mathbf{6}}$ type tetraquark states. It is feasible to take both the $\bar{\mathbf{3}}\mathbf{3}$ and $\mathbf{6}\bar{\mathbf{6}}$ diquark configurations to explore the tetraquark states, while the preferred or usually chosen  configuration is of the $\bar{\mathbf{3}}\mathbf{3}$ type.

The color-spin Hamiltonian can be written as \cite{Jaffe-diquark-PRT-2005},
\begin{eqnarray} \label{color-spin}
H=-\sum_{i\neq j} \kappa_{ij} \; S_i\cdot S_j\;\frac{\lambda^a_i}{2}\cdot \frac{\lambda^a_j}{2}\, ,
\end{eqnarray}
  the  color factor $\frac{\lambda^a_i}{2}\cdot \frac{\lambda^a_j}{2}$  can be absorbed into the chromomagnetic couplings $\kappa_{ij}$ after taking matrix elements between the  $\bar{\mathbf{3}}\mathbf{3}$ type tetraquark states.

In 2004, Maiani et al introduced the simple spin-spin Hamiltonian,
\begin{eqnarray} \label{Type-I-model}
H&=&2m_{[cq]}+2(\kappa_{cq})_{\bf \bar 3}\left(S_c \cdot S_q+S_{\bar c}
\cdot S_{\bar q}\right) + 2\kappa_{q{\bar q}}\left(S_q \cdot S_{\bar q}\right) + 2\kappa_{c{\bar q}}\left(S_c \cdot S_{\bar q}+S_{\bar c} \cdot S_q\right)
\nonumber \\
&&+ 2\kappa_{c{\bar c}}\left(S_c \cdot S_{\bar c}\right)\, ,
\end{eqnarray}
 to study the
hidden-charm tetraquark states in the diquark model, where the $m_{cq}$ is the charmed diquark mass
\cite{Maiani-2005-PRD-tetra-X3872}. They took the $X(3872)$ with the $J^{PC}=1^{++}$  as the basic input and predicted a mass spectrum for the $\bar{\mathbf{3}}\mathbf{3}$ type hidden-charm tetraquark states with the $J^{PC}=0^{++}$, $1^{+-}$ and $2^{++}$. Maiani et al assigned the $J^{PC}=1^{+-}$ charged resonance in the $Y(4260) \to \pi^+\pi^- J/\psi$ decays as the $X(3882)$ or $Z(3882)$ according to the BESIII and Belle data, however, there is no evidence for the lower resonance $X(3754)$ or $Z(3754)$  \cite{Maiani-4260-Z3900-tetra-PRD-2013}.
In fact, in the Type-I diquark model, see the Hamiltonian in Eq.\eqref{Type-I-model} \cite{Maiani-2005-PRD-tetra-X3872}, the predicted masses $3754\, \rm{MeV}$ and $3882\,\rm{MeV}$ for the $J^{PC}=1^{+-}$ states are smaller than that of the tetraquark candidates $Z_c(3900)$ and $Z_c(4020)$ observed later, respectively \cite{BES-Z3900-PRL-2013,Belle-Z3900-PRL-2013,
BES1308-Z4025-PRL-2014,BES1309-Z4020-PRL-2013,BES-Z3885-PRL-2014}.

In 2005, Maiani et al assigned  the $Y(4260)$ to be the first orbital excitation of the $[cs][\bar{c}\bar{s}]$ state by including the spin-orbit interaction, and obtained a crucial prediction  that the $Y(4260)$ should decay predominantly in the $D_s \bar{D}_s$ channel \cite{Maiani-4260-tetra-PRD-2005}. The decay model $Y(4260)\to D_s\bar{D}_s$ has not been observed yet up to now. In 2009, Drenska,  Faccini and Polosa studied  the $[cs][\bar{c}\bar{s}]$ tetraquark states with the $J^{PC}=0^{++}$, $0^{-+}$, $0^{--}$, $1^{++}$, $1^{+-}$, $1^{-+}$ and $1^{--}$   by computing the mass spectrum and decay modes \cite{Polosa-cscs-PRD-2009}.

In 2014, Maiani et al restricted the dominant spin-spin interactions to
the ones within each diquark, and simplify the effective Hamiltonian,
\begin{eqnarray} \label{diquark-type-II}
H&=&2m_{[cq]}+2\kappa_{cq}\left(S_c \cdot S_q+S_{\bar c}
\cdot S_{\bar q}\right)\, ,
\end{eqnarray}
which could describe the hierarchy of the masses of the  $X(3872)$, $Z_c(3900)$, $Z_c(4020)$ very well in the scenario of tetraquark states \cite{Maiani-1405-Tetra-model-2},
furthermore, they introduced a spin-orbit interaction to interpret the $Y$ states,
\begin{eqnarray}\label{Maiani-model-2-L1-H}
H&=&M_{00}+B_c\frac{L^2}{2}-2a L\cdot S+2\kappa_{cq}\left(S_c \cdot S_q+S_{\bar c}\cdot S_{\bar q}\right)\, ,
\end{eqnarray}
where the $M_{00}$, $B_c$ and $a$ are parameters to be fitted experimentally.
Then they sorted the tetraquark states in terms of $|S_{qc}, S_{\bar{q}\bar{c}}; S, L; J\rangle$, where the $L$ is the angular momentum between the diquark and antidiquark, $\vec{S}_{qc}=\vec{S}_{q}+\vec{S}_{c}$, $\vec{S}_{\bar{q}\bar{c}}=\vec{S}_{\bar{q}}+\vec{S}_{\bar{c}}$, $\vec{S}=\vec{S}_{qc}+ \vec{S}_{\bar{q}\bar{c}}$, $\vec{J}=\vec{S}+ \vec{L}$, and assigned the $Y(4008)$, $Y(4260)$, $Y(4290/4220)$ and $Y(4630)$         to be the  tetraquark states $|0, 0; 0, 1; 1\rangle$,
$\frac{1}{\sqrt{2}}\left(|1, 0; 1, 1; 1\rangle+|0, 1; 1, 1; 1\rangle\right)$,       $|1, 1; 0, 1; 1\rangle$  and
$|1, 1; 2, 1; 1\rangle$, respectively.
The effective  Hamiltonian, see Eq.\eqref{diquark-type-II}, is referred to as the Type-II diquark model. Then the  mass spectrum of the $[cs][\bar{c}\bar{s}]$ tetraquark states was explored \cite{Lebed-cscs-PRD-2016}, and applied to study the LHCb's $J/\psi\phi$ resonances \cite{Maiani-Jpsi-phi-PRD-2016}.

In 2017,  Maiani,  Polosa and Riquer  introduced a  hypothesis that the diquarks and antidiquarks in tetraquarks are separated by a potential barrier to answer the  long standing questions challenging the diquark-antidiquark model of exotic resonances \cite{Maiani-barrier-PLB-2018}.

In 2018, Ali et al  analyzed the P-wave hidden-charm  tetraquark states in the diquark model using an effective Hamiltonian incorporating the dominant spin-spin, spin-orbit and tensor interactions,
\begin{eqnarray}\label{Ali-model-tensor-H}
H&=&2m_{\mathcal D}+ \frac{B_{\mathcal D}}{2}  L^2
+2a_Y  L\cdot S  +b_Y \left(3{S}_1\cdot {\vec{n}} {S}_2\cdot {\vec{n}}-{S}_1\cdot{ S_2}\right)+  2\kappa_{cq} \left( S_{q}\cdot S_{c}
 +  S_{\bar q}\cdot S_{\bar c} \right) \, ,
\end{eqnarray}
where $\vec{n}=\frac{\vec{r}}{r}$, the $ S_{1}$ and $ S_{2}$ are the spins of the $\bar{\mathbf{3}}$ diquark $\mathcal D$ ($[qc]$) and $\mathbf{3}$  antidiquark $\bar{\mathcal D}$ ($[\bar{q}\bar{c}]$), respectively, the $m_{\mathcal D}$, $B_Y$, $a_Y$ and $b_Y$ are parameters to be fitted experimentally \cite{Ali-Maiani-Y-EPJC-2018}.  And their updated analysis indicate  that it is favorable to assign the $Y(4220)$, $Y(4330)$,  $Y(4390)$, $Y(4660)$  as the  tetraquark states
$|0, 0; 0, 1; 1\rangle$,
$\frac{1}{\sqrt{2}}\left(|1, 0; 1, 1; 1\rangle+|0, 1; 1, 1; 1\rangle\right)$,        $|1, 1; 0, 1; 1\rangle$ and
$|1, 1; 2, 1; 1\rangle$, respectively.

In 2021, Maiani, Polosa and   Riquer  suggested that the $Z_{cs} (3985)$ and $Z_{cs}(4003)$ are two different particles, and there exist two $SU(3)_f$ nonets with  the $J^{PC}=1^{++}$ and $1^{+-}$, respectively, thus they could assign the  $X(3872)$, $Z_c(3900)$, $Z_{cs}(3985)$, $Z_{cs}(4003)$ and $X(4140)$ consistently \cite{Maiani-Zcs-SB-2021}.

Again, let us turn to the chromomagnetic interaction model, see Eq.\eqref{color-spin}, and choose the $\bar{\mathbf{3}}\mathbf{3}$ plus $\mathbf{6}\bar{\mathbf{6}}$  configurations and $\mathbf{1}\mathbf{1}$ plus $\mathbf{8}\mathbf{8}$ configurations as two independent representations (or basis) respectively  to explore the mass spectrum of the exotic states and their decay channels, and have obtained  many successful descriptions  \cite{Chromomagnetic-Hogaasen-PRD-2006,Chromomagnetic-Hogaasen-EPJC-2007,
Chromomagnetic-ZhuSL-PRD-2014,Chromomagnetic-WuJ-PRD-2018,
Chromomagnetic-GuoT-PRD-2022}.

In the dynamical diquark picture,  Brodsky,  Hwang and Lebed
 assume that the
${\mathcal D}\bar{\mathcal D}$ pair forms promptly at the production point, and
rapidly separates due to the kinematics of the production process,  as the diquark and antidiquark are colored objects, they cannot separate asymptotically far apart; they create a  color flux tube or  string between them.  If
sufficient energy is available, the string would break to create  an additional
$q\bar q$ pair, and rearrange into a baryon-antibaryon pair, for example, the $\Lambda_c {\bar \Lambda}_c$ pair.
The overlap of the wave-functions between the quark and antiquark is suppressed greatly, due to the large spatial separation between the diquark and antidiquark pair, therefore, the  transition rate is
suppressed and leads to small exotic widths \cite{Lebed-dynamical-PRL-2014}. The exotic mass spectrum is calculated in this picture \cite{Lebed-dynamical-PRD-2017,Lebed-dynamical-JHEP-2019,Lebed-dynamical-JHEP-2020,
Lebed-dynamical-PRD-2020,Lebed-dynamical-PRD-2020-2,Lebed-dynamical-PRD-2021,
Lebed-dynamical-PRD-2024}.

In the relativized quark model, the Hamiltonian can be written as,
\begin{eqnarray}
H=\sum_{i=1}^{4}(p_{i}^{2}+m_{i}^{2})^{1/2}+\sum_{i<j}V_{ij}^{\mathrm{conf}}+\sum_{i<j}V_{ij}^{\mathrm{oge}}
\end{eqnarray}
where the $V_{ij}^{\mathrm{conf}}$ is the linear confining potential,
the $V_{ij}^{\mathrm{oge}}$ is the one-gluon
exchange potential including a Coulomb and a hyperfine term. Then the $\bar{\mathbf{3}}\mathbf{3}$ type configurations or both the $\bar{\mathbf{3}}\mathbf{3}$ and $\mathbf{6}\bar{\mathbf{6}}$ type configurations are taken into account  to solve the Schrodinger equations to obtain the mass spectrum \cite{Tetra-33-RQM-LuQF-PRD-2016,Tetra-33-RQM-Ferretti-PRD-2018,Tetra-33-RQM-Roberts-EPJC-2020,Tetra-33-RQM-Ferretti-JHEP-2020,
Tetra-33-66-RQM-LuQF-PRD-2020,After-cccc-33-LuQF-EPJC-2020,
Tetra-33-66-RQM-YuGL-EPJC-2023,
Tetra-33-RQM-DongWC-PRD-2023,
Tetra-33-66-RQM-YuGL-EPJC-2024,Tetra-33-RQM-DongWC-arXiv-2023}.

In the quasipotential approach, Ebert et al take the $\bar{\mathbf{3}}\mathbf{3}$-type configurations  to study the hidden-charm (hidden-bottom, charm-bottom or fully-heavy) tetraquark mass spectrum by solving the Schrodinger type equations, where  an effective one-gluon exchange potential plus a linear confining potential are adopted  \cite{X3872-tetra-Ebert-PLB-2006,Tetra-33-Ebert-PRD-2007,Tetra-33-Ebert-EPJC-2008,
Tetra-33-QQQQ-Faustov-PRD-2020}.

In the  constituent quark model, all possible quark configurations  satisfying  the Pauli principle are explored by solving the Schrodinger equation with the potential kernel containing  the confinement plus one-gluon-exchange plus (or not plus) one-meson-exchange interactions \cite{CQM-Tetra-Vijande-EPJA-2004,CQM-Tetra-Vijande-PRD-2006,CQM-Tetra-Vijande-PRD-2007,
CQM-Tetra-ZhangZY-PRC-2009,CQM-Tetra-ZhuSL-PRD-2019,
CQM-Tetra-Penta-YangG-Symmetry-2020,
CQM-Tetra-SHLee-PRD-2021,CQM-Tetra-ZhaoZ-PRD-2021}, while in the color flux-tube model, a multi-body interacting confinement potential  instead of a two-body interacting  confinement potential is chosen \cite{FluxTube-Tetra-PingJL-PRD-2018,FluxTube-Tetra-ChenH-EPJA-2020}.

\subsubsection{Tetraquark states with Lattice QCD}
Lattice QCD provides rather  accurate and reliable calculations for the hadrons which lie well below strong-decay threshold and do not decay strongly, the physical  information is commonly extracted from the discrete energy spectrum.
The physical system with  specified  quantum numbers is created from the vacuum $|0\rangle$ using an operator ${\cal O}_j^\dagger$  at time $t\!=\!0$, then this system propagates for a time $t$ before being annihilated by an operator ${\cal O}_i$.   The spectral decomposition is performed to express  the correlators $C_{ij}(t)$ in terms of the energies  $E_n$  and overlaps $Z_j^n$ of the eigenstates $|n\rangle$,
\begin{eqnarray}
C_{ij}(t)&=& \langle 0|{\cal O}_i (t) {\cal O}_j^\dagger (0)|0 \rangle=\sum_{n}Z_i^nZ_j^{n*}~e^{-E_n t}~,
\end{eqnarray}
$ Z_i^n\equiv \langle 0|{\cal O}_i|n\rangle$.
 The correlators $C_{ij}(t)$ are calculated on the lattice  and  their time-dependence is used to extract the $E_n$ and $Z_n^i$  \cite{Latt-En-Michael-NPB-1985,Latt-En-Blossier-JHEP-2009}. The lattice QCD has been applied extensively to study the exotic states \cite{Latt-Zc3900-mole-No-Prelovsek-PRD-2015,Latt-X3872-Y4140-Prelovsek-PRD-2015,
Latt-QQ-tetra-Francis-PRL-2017,Latt-QQ-tetra-Cheung-JHEP-2017,
Latt-QQ-tetra-Junnarkar-PRD-2019,Latt-QQ-tetra-Leskovec-PRD-2019,
Latt-cb-tetra-Francis-PRD-2019,Latt-Review-QQ-tetra-Francis-2025}.

In the energy region near or above the strong decay thresholds,  the masses of the bound states and resonances are inferred  from the finite-volume scattering matrix of  one-channel elastic or multiple-channel inelastic scattering.
Various approaches with varying degrees of mathematical rigour have been used in the simulations \cite{Latt-method-Bicudo-PR-2023}.
The simplest example is a one-channel elastic scattering with the partial wave $l$, where   the scattering matrix $S(p)$ satisfying unitarity $SS^\dagger=1$ is parameterized in terms of the phase shift $\delta_l(p)$,
\begin{eqnarray}
S(p)&=&e^{2i\delta_l(p)}=1+2iT(p)\, , \nonumber\\
T(p)&=&\frac{1}{\cot(\delta_l(p))-i}\, ,
\end{eqnarray}
 the phase shift $\delta_l(p)$ for the S-wave scattering is extracted using the well-established and rigorous Luscher's relation \cite{Latt-Luscher-NPB-1991, Latt-Luscher-NPB-1991-2,Latt-Luscher-Review-2018}, which applies for the elastic scattering below and above threshold. The phase shifts $\delta(p)$   provide copious information  about the masses of resonances and bound states.
  In the vicinity of a  hadronic resonance with a mass $m_R$ and a width $\Gamma$, the  cross section  has a Breit-Wigner-type shape with the value  $\delta(s=m_R^2)=\tfrac{\pi}{2}$,
\begin{equation}
T(p)=\frac{-\sqrt{s}~ \Gamma(p)}{s-m_R^2+i \sqrt{s}\, \Gamma(p)}=\frac{1}{\cot\delta(p)-i}\, .
\end{equation}
Below and above threshold, the $p\cot \delta(p)$ can be expanded by the effective range approximation,
\begin{equation}
p\cot \delta(p)=\frac{1}{a}+\frac{1}{2}r p^2\, ,
\end{equation}
where the $a$ is the scattering length and the $r$ is the effective range.
On the other hand, the bound state ($B$) is realized when the scattering amplitude $T(p)$ has a pole at the value  $p_B= i|p_B|$,
\begin{equation}
T=  \frac{1}{\cot(\delta(p_B))-i}=\infty\, ,
\end{equation}
the  location of this shallow bound state can be obtained by parameterizing $\delta(p)$ near the threshold and finding the $p_B$ which satisfies $\cot(\delta(p_B))=i$.
Most of the exotic candidates are  above several two-hadron thresholds, and have more than one decay channels, which  requires determining the scattering matrix for the coupled-channel nonelastic scattering matrix elements \cite{Latt-X3872-mole-Prelovsek-PRL-2013,
Latt-Zc3900-mole-ChenY-PRD-2014,Latt-Tc-mole-Padmanath-PRL-2022,
Latt-Tc-mole-ChenY-PLB-2022,Latt-Tc-mole-DuML-PRL-2023}.

The HALQCD collaboration use  the HALQCD approach to extract  the coupled-channel scattering matrix \cite{HALQCD-Aoki-PTEP-2012,HALQCD-Aoki-PRL-2023}, the HALQCD approach  is based on the lattice determination of the potential $V(r)$ between different channels, then employs the Nambu-Bethe-Salpeter equation to extract the masses of the bound states.

\subsection{Multiquark states with the QCD sum rules}\label{Tetra-QCDSR}
The QCD sum rules were introduced by  Shifman,
Vainstein and Zakharov in 1979 to study the conventional mesons \cite{SVZ-NPB-1979-1,SVZ-NPB-1979-2},  then they were extended to study the conventional  baryons  by Ioffe \cite{Ioffe-NPB-1981}. The QCD sum rules are  analytic and fully relativistic, and  approach the bound state problem in QCD from short distances and move to longer distances
step by step by including the non-perturbative effects so as to  extract information on the hadronic properties.

In the past years, the QCD sum rules  have been applied widely  to
study the hadronic properties, such as the masses of the quarks;  masses and decay constants of the light and heavy mesons and baryons;
 form-factors of the mesons and baryons;
 valence quark distributions and spin
structure functions of the nucleons;
structure functions of the photon, pseudoscalar, vector and axialvector  mesons;
  hadronic matrix elements  for the $K^0-\bar{K}^0$, $B_d-\bar{B}_d$,
$B_s-\bar{B}_s$ mixing;  strong coupling constants and magnetic moments of the mesons and baryons;   parameters of the effective field theories;
spectroscopy and properties of the exotic states;
hadrons in the nuclear matter; properties of hadronic matter at
high temperature and  density, etc  \cite{Review-QCDSR-Nielsen-PRT-2010,Review-QCDSR-Nielsen-JPG-2019,
Nucleon-WF-Chernyak-NPB-1984,Nucleon-WF-Chernyak-PRT-1984,
Reinders85,LCQCDSR-Balitsky-NPB-1989,LCQCDSR-Braun-ZPC-1989,
Shifman-1992,
Colangelo-review,Narison-Book-2007,Condensate-Gubler-PPNP-2019,Leinweber-AP-2997}.
Especially since 2007, the QCD sum rules have been  applied extensively to study the $X$, $Y$, $Z$, $T$ and $P$ states, which are the typical multiquark candidates \cite{Review-QCDSR-Nielsen-PRT-2010,Review-QCDSR-Nielsen-JPG-2019,
X3872-tetra-Narison-PRD-2007,X3872-tetra-WangZG-HuangT-PRD-2014,
X3872-mole-WangZG-HT-EPJC-2014}.  For the early works on the exotic states, we  can consult Refs.\cite{Early-QCDSR-exotic-Balitsky-PLB-1982,
Early-QCDSR-exotic-Govaerts-PLB-1983,
Early-QCDSR-exotic-Govaerts-NPB-1984,Early-QCDSR-exotic-Narison-PLB-1984,
Early-QCDSR-exotic-Govaerts-NPB-1985,Early-QCDSR-exotic-Larin-SJNP-1986,
Early-QCDSR-exotic-Balitsky-ZPC-1986,
Early-QCDSR-exotic-Braun-PLB-1986,Early-QCDSR-exotic-Narison-ZPC-1987,
Early-QCDSR-exotic-Govaerts-NPB-1987,
Early-QCDSR-exotic-Braun-SJNP-1989,Early-QCDSR-exotic-Kisslinger-PRD-1995}.

In this sub-section, we would like to illustrate the general procedure of the QCD sum rules for the masses of the conventional hadrons and multiquark states concisely.

At the beginning  point, let us write down the  general  two-point vacuum correlation functions,
\begin{eqnarray}\label{CF-first}
\Pi(p)&=& i\int d^4 x\, e^{ip\cdot x}
 \langle 0| T \left\{J(x)J^\dagger(0)\right\}|0\rangle\, ,
\end{eqnarray}
where the $J(x)$ are  the local currents consist of quark-gluon fields with specified  quantum numbers, and the $T$ denotes the time-ordering operation.
For the conventional mesons and baryons, the currents $J(x)$ have been explored extensively \cite{Reinders85}, for the multiquark states, the currents $J(x)$ can be constructed straightforwardly.

At the large squared momentum region  $P^2= -p^2 \gg \Lambda_{QCD}^2$, the integral in  Eq.\eqref{CF-first} is dominated by small spatial distances and time intervals,
\begin{eqnarray}
 t \sim  |\vec{x}| \sim \frac{1}{\sqrt{P^2}} \ll r\, ,
\end{eqnarray}
to avoid fast exponential  oscillating,
where the hadron size $r\sim \frac{1}{\Lambda_{QCD}}$. If we set $\Lambda_{QCD}=(200-300)\,\rm{MeV}$, then the hadron size $r\sim 0.7-1.0 \, \rm{fm}$. Therefore, at the condition of large hadron size, say $r \gg  1\,\rm{fm}$, the local currents $J(x)$ are questionable to interpolate the  corresponding hadrons.

Now let us take it for granted that the exotic states have the size $r \leq 1\,\rm{fm}$, just like the conventional mesons and baryons, could be interpolated by the local currents $J(x)$ tacitly. For example, the charge radii    of the $\pi^\pm$, $K^\pm$ and $p$ are
$\sqrt{\langle r^2\rangle}=0.659 \pm0.004\, \rm{fm}$, $0.560\pm0.031 \,\rm{fm}$ and $0.8409 \pm 0.0004\,\rm{fm}$ respectively from the Particle Data Group \cite{PDG-2024}. We extend the QCD sum rules on the conventional mesons and baryons  to study the multiquark states directly with a simple replacement of the interpolating currents, and would like to come back to this subject again in Sect.{\bf \ref{reliable?}}.

A Lorentz invariant vacuum average can be expressed as
\begin{eqnarray}
\langle 0| T \left\{J(x)J^\dagger(0)\right\}|0\rangle&=& \int d\tau \, \exp\left(i\tau x^2\right) f(\tau)\, ,
\end{eqnarray}
 where the $f(\tau)$ is a function.
Then
\begin{eqnarray}
 \Pi(p^2)&=&i\int d \tau \int d^4 x  \exp\left(i\tau x^2\right) \exp\left(iP^2/4\tau\right)f(\tau)\, .
\end{eqnarray}
The dominant contributions to the $\Pi(p^2)$ come
from the region,
\begin{eqnarray}
x^2\sim \frac{1}{\tau} \sim \frac{1}{P^2}\, .
\end{eqnarray}
In the limit $P^2 \to \infty$, we reach the  light-cone $x^2\sim 0$, which
 is a necessary but not yet sufficient condition
for the short-distance dominance, we have to constrain $|\vec{x}| \sim \frac{1}{\sqrt{P^2}}$.

Now we focus on  the quark-gluon degrees of freedom and calculate the correlation functions  using Wilson's operator product expansion   to separate the physics of short and long distances,
\begin{equation}
\Pi(p^2)= \sum_n C_n(p^2,\mu)\langle{\mathcal{O}}_n(\mu)\rangle \, ,
\end{equation}
where the $C_n(p^2,\mu)$ are the Wilson's coefficients  encoding  short-distance  contributions,  the $\langle{\mathcal{O}}_n(\mu)\rangle$ are  vacuum  expectations   of the local operators with  dimension $n$. The short-distance contributions at  $p^2>\mu^2$ are encoded  in the coefficients
$C_n(p^2,\mu)$, the long-distance contributions at $p^2<\mu^2$ are absorbed into the vacuum condensates  $\langle{\mathcal{O}}_n(\mu)\rangle$ \cite{Colangelo-review}.
If $\mu\gg \Lambda_{QCD}$, the Wilson coefficients $C_n(p^2,\mu)$ depend only on short-distance dynamics,   the vacuum condensates $\langle{\mathcal{O}}_n(\mu)\rangle$ embody the long-distance effects.
 The lowest condensate  is vacuum expectation of the unit operator $\langle{\mathcal{O}}_0(\mu)\rangle$ associated with the perturbative contributions.
The vacuum condensates with dimensions $n=3$, $4$, $5$, $6$,  $\cdots$ are quark condensate $\langle \bar{q}q\rangle$, gluon condensate $\langle \frac{\alpha_sGG}{\pi}\rangle$, mixed condensate $\langle \bar{q}g_s \sigma Gq\rangle$, four-quark condensate $\langle \bar{q}q\rangle^2$, $\cdots$, which parameterize  the non-perturbative effects or soft gluons and quarks. We can consult Refs.\cite{Reinders85,Colangelo-review} for the basic techniques  in performing the operator product expansion.

\begin{figure}
 \centering
  \includegraphics[totalheight=6cm,width=8cm]{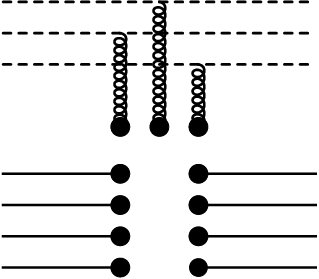}
 \caption{ The  counting role for the truncation of the operator product expansion, where the solid lines and dashed lines represent  the light quarks and heavy quarks, respectively. }\label{One-One-Lowest-High}
\end{figure}

If there exist $m$ heavy quark lines and $n$ light-quark lines in the correlation functions $\Pi(p^2)$,   each heavy quark line emits a gluon and each light quark line contributes a quark-antiquark pair, we obtain a quark-gluon operator,
\begin{eqnarray}\label{quark-gluon-operator}
\underbrace{G_{\mu\nu} \cdots G_{\alpha\beta}} \underbrace{\bar{q}q\, \bar{q}q \cdots \bar{q}q\,\bar{q}q}\, ,
\end{eqnarray}
which is of dimension $2m+3n$, we should perform the operator product expansion up to the  vacuum condensates  of dimension $2m+3n$ at least \cite{X3872-tetra-WangZG-HuangT-PRD-2014,WZG-HC-PRD-2020,
X3872-mole-WangZG-HT-EPJC-2014,WZG-tetra-mole-IJMPA-2021,WZG-tetra-mole-AAPPS-2022}. For example, $m=3$ and $n=4$, we should calculate the vacuum condensates of dimension 18, see Fig.\ref{One-One-Lowest-High}. When  $m=n=1$, we obtain the conventional $Q\bar{q}$ mesons, it is obvious that we should calculate the mixed condensate $\langle \bar{q}g_s \sigma G q\rangle$.

Then we obtain the K\"allen-Lehmann representation through dispersion relation  at the quark-gluon degrees of freedom,
\begin{eqnarray}
\Pi_{QCD}(p^2) &=& \frac{1}{\pi}\int_{\Delta^2}^{\infty} ds\frac{{\rm Im}\Pi_{QCD}(s)}{s - p^2}   \, ,
\end{eqnarray}
where the $\Delta^2$ denotes the thresholds.

At the hadron degrees of freedom, we insert  a complete set of intermediate hadronic states $|n\rangle$ with the same quantum numbers as the currents $J(x)$ into the correlation functions $\Pi(p^2)$, and take account of the current-hadron couplings to obtain the analytical expressions, again we obtain the K\"allen-Lehmann representation through dispersion relation,
\begin{eqnarray}
\Pi_{H}(p^2) &=& \frac{1}{\pi}\int_{\Delta^2}^{\infty} ds\frac{{\rm Im}\Pi_{H}(s)}{s - p^2}   \, ,
\end{eqnarray}
where
\begin{eqnarray}
 \rho_H(s)=\frac{{\rm Im}\Pi_{H}(s)}{\pi} =\sum_n |\langle0|J(0)|n\rangle|^2 \delta \left(s-M_n^2\right)\, ,
 \end{eqnarray}
 the subscript $H$ denotes the hadron side.

According to the Quark-Hadron duality, we introduce the continuum threshold parameters $s_0$, and match the QCD side with hadron side of the correlation functions $\Pi(p^2)$,
\begin{eqnarray}\label{duality}
 \frac{1}{\pi}\int_{\Delta^2}^{s_0} ds\frac{{\rm Im}\Pi_{H}(s)}{s - p^2}&=& \frac{1}{\pi}\int_{\Delta^2}^{s_0} ds\frac{{\rm Im}\Pi_{QCD}(s)}{s - p^2}    \, .
\end{eqnarray}
An important point is the choice of the continuum threshold $s_0$, which  is a physical parameter that
should be determined from the hadronic spectrum.
Then we perform the Borel transformation,
\begin{eqnarray}
\Pi(T^2)=\mathcal{B}[\Pi(P^2)]\equiv
\lim\limits_{\tiny \begin{matrix} P^2, n
\rightarrow \infty \\ P^2/n = T^2 \end{matrix}}
\frac{(P^2)^{n+1}}{(n)!}\Big(-\frac{d}{dP^2}\Big)^n\, \Pi(P^2)\,,
\end{eqnarray}
with $P^2=-p^2$ to obtain the QCD sum rules,
\begin{eqnarray}
 \frac{1}{\pi}\int_{\Delta^2}^{s_0} ds\,{\rm Im}\Pi_{H}(s)\exp\left(-\frac{s}{T^2} \right)&=& \frac{1}{\pi}\int_{\Delta^2}^{s_0} ds\,{\rm Im}\Pi_{QCD}(s)\exp\left(-\frac{s}{T^2} \right)   \, ,
\end{eqnarray}
where the $T^2$ is  the Borel parameter. Some typical and useful examples of the Borel transformation are given in the Appendix. If only the ground state is taken, then
\begin{eqnarray}
 \frac{{\rm Im}\Pi_{H}(s)}{\pi}&=&\lambda_H^2\left(s-M_H^2\right)\, ,
 \end{eqnarray}
  we obtain the QCD sum rules,
 \begin{eqnarray}
 \lambda_H^2\exp\left(-\frac{M_H^2}{T^2} \right)&=& \frac{1}{\pi}\int_{\Delta^2}^{s_0} ds\,{\rm Im}\Pi_{QCD}(s)\exp\left(-\frac{s}{T^2} \right)   \, ,
\end{eqnarray}
where the $M_H$ is the mass of the ground state of the conventional hadron or multiquark state, the $\lambda_H$ is the pole residue.

It is obvious  that the Borel transformation wipes out any eventual subtraction terms in the correlation functions  and suppresses
the continuum contributions exponentially, therefore,  it improves the convergent behavior  of the dispersion integral. Furthermore, it suppresses
the higher-dimensional  operators in the operator product expansion factorially, which contain inverse powers of the $P^2$, see Eq.\eqref{Some-Borel}, thus justifies truncation of the operator product expansion and favors  a good convergent behavior.

Finally, we eliminate the pole residue $\lambda_H$ to obtain the QCD sum rules for the ground state mass,
\begin{eqnarray}
 M_H^2&=& \frac{-\frac{d}{d\tau}\int_{\Delta^2}^{s_0} ds\,{\rm Im}\Pi_{QCD}(s)\exp\left(- \tau s  \right) }{\int_{\Delta^2}^{s_0} ds\,{\rm Im}\Pi_{QCD}(s)\exp\left(-\tau s  \right) } \mid_{\tau=\frac{1}{T^2}} \, .
\end{eqnarray}
In the QCD sum rules, we choose some phenomenological inputs which
 limit  the accuracy of this method to be around $10\%-20\%$
\cite{Leinweber-AP-2997}.

\subsection{Are QCD sum rules reliable to study  multiquark states} \label{reliable?}
Any color singlet four-quark and five-quark currents $J(x)$ can be written as
$J(x)=J_A^i(x)J_B^i(x)$, where the $J_A^i(x)$ and  $J_B^i(x)$ are color singlet clusters with $i=1$, $2$, $3$, $\cdots$, for example,
\begin{eqnarray}\label{Fierz-Zc}
J_{\mu}(x)&=&\frac{\varepsilon^{ijk}\varepsilon^{imn}}{\sqrt{2}}
\Big\{u^{T}_j(x)C\gamma_5 c_k(x)\bar{d}_m(x)\gamma_\mu C \bar{c}^{T}_n(x)  -u^{T}_j(x)C\gamma_\mu c_k(x)\bar{d}_m(x)\gamma_5 C \bar{c}^{T}_n(x)  \Big\} \, , \nonumber\\
 &=&\frac{1}{2\sqrt{2}}\Big\{\,i\bar{c}i\gamma_5 c\,\bar{d}\gamma_\mu u-i\bar{c} \gamma_\mu c\,\bar{d}i\gamma_5 u+\bar{c} u\,\bar{d}\gamma_\mu\gamma_5 c
-\bar{c} \gamma_\mu \gamma_5u\,\bar{d}c  \nonumber\\
&&  - i\bar{c}\gamma^\nu\gamma_5c\, \bar{d}\sigma_{\mu\nu}u+i\bar{c}\sigma_{\mu\nu}c\, \bar{d}\gamma^\nu\gamma_5u
- i \bar{c}\sigma_{\mu\nu}\gamma_5u\,\bar{d}\gamma^\nu c+i\bar{c}\gamma^\nu u\, \bar{d}\sigma_{\mu\nu}\gamma_5c   \,\Big\} \, ,
\end{eqnarray}
for the four-quark current \cite{X3872-tetra-WangZG-HuangT-PRD-2014,Two-particle-Zc3900-WangZG-IJMPA-2020},
and
 \begin{eqnarray}\label{Fierz-Pc}
 J(x)&=&\varepsilon^{ila} \varepsilon^{ijk}\varepsilon^{lmn}  u^T_j(x) C\gamma_5 d_k(x)\,u^T_m(x) C\gamma_5 c_n(x)\,  C\bar{c}^{T}_{a}(x) \, , \nonumber\\
&=&-\frac{1}{4}\mathcal{S}_{ud}\gamma_5c\,\bar{c}u+\frac{1}{4}\mathcal{S}_{ud}\gamma^\lambda\gamma_5c\,\bar{c}\gamma_{\lambda}u
+\frac{1}{8}\mathcal{S}_{ud}\sigma^{\lambda\tau}\gamma_5c\,\bar{c}\sigma_{\lambda\tau}u+\frac{1}{4}\mathcal{S}_{ud}\gamma^{\lambda}c\,\bar{c}\gamma_{\lambda}\gamma_5u
+\frac{i}{4}\mathcal{S}_{ud}c\,\bar{c}i\gamma_5 u \nonumber\\
&&+\frac{1}{4}\mathcal{S}_{ud}\gamma_5u\,\bar{c}c-\frac{1}{4}\mathcal{S}_{ud}\gamma^\lambda\gamma_5u\,\bar{c}\gamma_{\lambda}c
-\frac{1}{8}\mathcal{S}_{ud}\sigma^{\lambda\tau}\gamma_5u\,\bar{c}\sigma_{\lambda\tau}c-\frac{1}{4}\mathcal{S}_{ud}\gamma^{\lambda}u\,\bar{c}\gamma_{\lambda}\gamma_5c
-\frac{i}{4}\mathcal{S}_{ud}u\,\bar{c}i\gamma_5 c\, , \nonumber \\
\end{eqnarray}
for the five-quark current
 \cite{WZG-HC-penta-IJMPA-2020}, where the components $\mathcal{S}_{ud}\Gamma c=\varepsilon^{ijk}u^{T}_iC\gamma_5d_j \Gamma c_k$.

According to the pioneer works \cite{Two-hadron-Kondo-PLB-2005,Two-hadron-LeeSH-PLB-2005,
 Two-hadron-LeeHJ-PRD-2008}, in the coordinate space, we write the two-hadron-reducible contributions as
 \begin{eqnarray}
 \Pi_{RE}(x)&=&\langle 0|T\left\{ J_A^i(x)J_A^{j\dagger}(0) \right\}|0\rangle\langle 0|T\left\{ J_B^i(x)J_B^{j\dagger}(0) \right\}|0\rangle \, ,
 \end{eqnarray}
 and the two-hadron-irreducible contributions as
 \begin{eqnarray}
 \Pi_{IR}(x)&=&\langle 0|T\left\{ J(x)J^{\dagger}(0) \right\}|0\rangle-\langle 0|T\left\{ J_A^i(x)J_A^{j\dagger}(0) \right\}|0\rangle\langle 0|T\left\{ J_B^i(x)J_B^{j\dagger}(0) \right\}|0\rangle \, .
 \end{eqnarray}
 At the phenomenological side of the correlation functions, we can write
 the two-hadron-reducible contributions as
 \begin{eqnarray}
\Pi_{RE}(p^2)=|\lambda_{AB}|^2 \frac{i}{(2\pi)^4}\int d^4q \frac{i}{q^2-M_{A}^2}\frac{i}{(p-q)^2-M_{B}^2}\, ,
\end{eqnarray}
 where the couplings
 \begin{eqnarray}
 \langle 0|J^i_A(0)|A(q)\rangle\langle 0|J^i_B(0)|B(p-q)\rangle &=&\lambda_{AB} \, ,
 \end{eqnarray}
and the $\lambda_{AB} $ could be estimated phenomenologically  \cite{Two-hadron-Kondo-PLB-2005,Two-hadron-LeeSH-PLB-2005,
 Two-hadron-LeeHJ-PRD-2008,Two-hadron-ChenHX-PRD-2010}.

\begin{figure}
 \centering
  \includegraphics[totalheight=4cm,width=5cm]{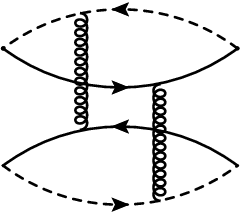}
 \caption{ The nonfactorizable  Feynman diagrams of the order $\mathcal{O}(\alpha_s^2)$ for the color singlet-singlet  type currents, other diagrams obtained by interchanging of the heavy quark lines (dashed lines) and light quark lines (solid lines) are implied.}\label{meson-afs2}
\end{figure}

In the correlation functions for the color singlet-singlet type currents \cite{Lucha-Two-PRD-2019,Lucha-Two-PRD-2019-2},   Lucha, Melikhov and Sazdjian assert that  the Feynman diagrams can be divided into  factorizable  and nonfactorizable diagrams in the color space,
 the contributions  at the order $\mathcal{O}(\alpha_s^k)$ with $k\leq1$, which are factorizable in the color space, are exactly  canceled out    by the meson-meson scattering states at the hadron side,
the nonfactorizable diagrams, if have a Landau singularity, begin to make contributions  to the tetraquark (molecular) states,  the tetraquark (molecular) states begin to receive contributions at the order $\mathcal{O}(\alpha_s^2)$, see Fig.\ref{meson-afs2}.

In Ref.\cite{WZG-Landau-PRD-2020}, we examine the assertion of Lucha, Melikhov and Sazdjian in details  and use two examples for the currents  $J_\mu(x)$ and $J_{\mu\nu}(x)$ to illustrate that the Landau equation is of no use in the QCD sum rule for the tetraquark molecular states, where
\begin{eqnarray}\label{two-example-current}
J_\mu(x)&=&\frac{1}{\sqrt{2}}\Big[\bar{u}(x)i\gamma_5 c(x)\bar{c}(x)\gamma_\mu d(x)+\bar{u}(x)\gamma_\mu c(x)\bar{c}(x)i\gamma_5 d(x) \Big]  \, , \nonumber \\
J_{\mu\nu}(x)&=&\frac{1}{\sqrt{2}}\Big[\bar{s}(x)\gamma_\mu c(x) \bar{c}(x)\gamma_\nu\gamma_5  s(x)-\bar{s}(x)\gamma_\nu\gamma_5 c(x) \bar{c}(x)\gamma_\mu s(x)\Big]\, .
\end{eqnarray}

{\bf Firstly}, we cannot assert that the factorizable  Feynman diagrams in color space are exactly canceled out by the meson-meson scattering states, because the meson-meson scattering state and tetraquark molecular state both have four valence quarks, which can be divided into  two color-neutral clusters. We cannot distinguish which Feynman diagrams contribute to the  meson-meson scattering state or tetraquark molecular state based on the two color-neutral clusters \cite{WZG-Landau-PRD-2020}.

{\bf Secondly}, the quarks and gluons are confined objects, they cannot be put on the mass-shell, it is questionable  to assert  that the Landau equation is applicable  for the quark-gluon bound states \cite{Landau}.

If we insist on applying  the Landau equation to study the Feynman diagrams, we should choose the pole masses rather than the $\overline{MS}$ masses to warrant  mass poles.
As the tetraquark (molecular) states begin to receive contributions at the order $\mathcal{O}(\alpha_s^2)$ \cite{Lucha-Two-PRD-2019,Lucha-Two-PRD-2019-2},
it is reasonable to take the pole masses $\hat{m}_Q$ as,
\begin{eqnarray}
\hat{m}_Q&=&m_Q(m_Q)\left[1+\frac{4}{3}\frac{\alpha_s(m_Q)}{\pi}+f\left(\frac{\alpha_s(m_Q)}{\pi}\right)^2+g\left(\frac{\alpha_s(m_Q)}{\pi}\right)^3\right]\, ,
\end{eqnarray}
see Refs.\cite{PDG-2018,Three-loop-mass} for the explicit expressions of the $f$  and $g$. If the Landau equation is applicable  for the tetraquark (molecular) states, it is certainly applicable for
the traditional  charmonium and bottomonium states.
In the case of the $c$-quark ($b$-quark), the pole mass $\hat{m}_c=1.67\pm0.07\,\rm{GeV}$ ($\hat{m}_b=4.78\pm0.06\,\rm{GeV}$) from the Particle Data Group \cite{PDG-2018}, the Landau singularity appears at the $s$-channel  $\sqrt{s}=\sqrt{p^2}=2\hat{m}_c=3.34\pm0.14\,{\rm{GeV}}>M_{\eta_c}/M_{J/\psi}$ ($2\hat{m}_b=9.56\pm0.12\,{\rm{GeV}}>M_{\eta_b}/M_{\Upsilon}$).  It is  unreliable  that the  masses of the charmonium (bottomonium) states lie below the threshold $2\hat{m}_c$ ($2\hat{m}_b$)  for the $\eta_c$ and $J/\psi$ ($\eta_b$ and $\Upsilon$) \cite{WZG-Landau-PRD-2020}.

{\bf Thirdly}, the nonfactorizable Feynman diagrams which have the Landau singularities begin to appear at the order $\mathcal{O}(\alpha_s^0/\alpha_s^1)$ rather than at the order $\mathcal{O}(\alpha_s^2)$, and make contributions  to the tetraquark molecular states, if the assertion (only  nonfactorizable Feynman diagrams which have Landau singularities make contributions to the tetraquark molecular states) of Lucha, Melikhov and Sazdjian is right.

\begin{figure}
 \centering
  \includegraphics[totalheight=4cm,width=5cm]{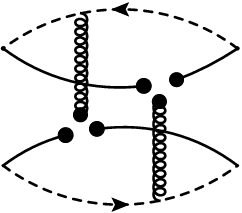}
 \caption{ The nonfactorizable   Feynman diagrams contribute to the vacuum condensates
$\langle \bar{q}g_s\sigma G q \rangle^2$ for the color singlet-singlet  type currents, where the solid lines and dashed lines denote the light quarks and heavy quarks, respectively. }\label{meson-qqg-qqg}
\end{figure}

The nonfactorizable contributions appear at the  order $\mathcal{O}(\alpha_s)$ due to the   operators  $\bar{q}g_sGq\bar{q}g_sGq$, which  come from the Feynman diagrams shown in Fig.\ref{meson-qqg-qqg}.
If we insist on choosing the pole mass and applying the landau equation to study the diagrams, we obtain a sub-leading Landau singularity at the $s$-channel $s=p^2=4\hat{m}_c^2$.
From the operators  $\bar{q}g_sGq\bar{q}g_sGq$, we  obtain the vacuum condensate $\langle\bar{q}g_s\sigma Gq\rangle^2$,  where the $g_s^2$ is absorbed into the vacuum condensate. The  nonfactorizable Feynman diagrams appear at the order  $\mathcal{O}(\alpha_s^0)$
or $\mathcal{O}(\alpha_s^1)$, not at the order $\mathcal{O}(\alpha_s^2)$ asserted in Refs.\cite{Lucha-Two-PRD-2019,Lucha-Two-PRD-2019-2}.

\begin{figure}
 \centering
  \includegraphics[totalheight=4cm,width=10cm]{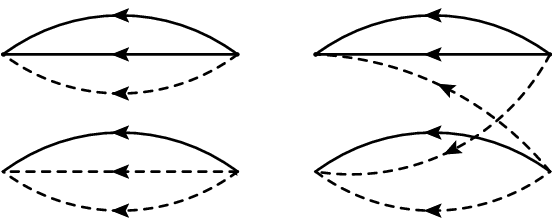}
   \vglue+9mm
   \includegraphics[totalheight=4cm,width=10cm]{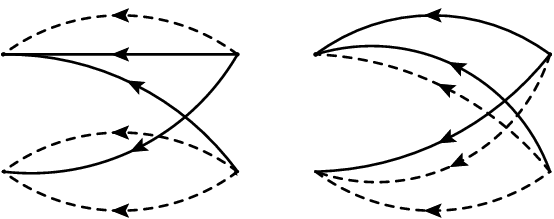}
 \caption{ The lowest order Feynman diagrams  for the triply-heavy dibaryon states, where the solid lines and dashed lines represent  the light quarks and heavy quarks, respectively. }\label{dibaryon-Lowest-diagram}
\end{figure}

\begin{figure}
 \centering
  \includegraphics[totalheight=4cm,width=10cm]{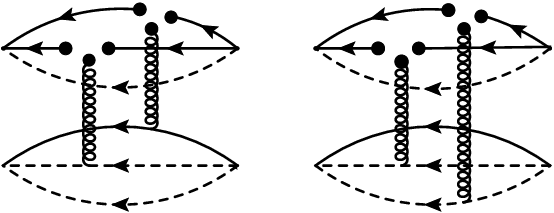}
    \caption{ The  connected Feynman diagrams originate from the first diagram in Fig.\ref{dibaryon-Lowest-diagram}, other
diagrams obtained by interchanging of the heavy quark lines (dashed lines) or light quark lines (solid lines) are implied. }\label{dibaryon-qqg-qqg}
\end{figure}

In fact, for the triply-heavy dibaryon-type currents $\eta(x)$ and $\eta_\mu(x)$, \begin{eqnarray}
 \eta(x)&=&  J^T_c(x) C\gamma_5 J_{cc}(x) \, , \nonumber\\
 \eta_\mu(x)&=&  J^T_c(x) C\gamma_\mu J_{cc}(x) \, , \nonumber\\
 J_{c}(x)&=& \varepsilon^{ijk} q^T_i(x) C\gamma_\alpha q_{j}(x)\gamma^\alpha\gamma_5c_k(x) \, , \nonumber\\
  J_{cc}(x)&=& \varepsilon^{ijk} c^T_i(x) C\gamma_\alpha c_{j}(x)\gamma^\alpha\gamma_5q_k(x) \, ,
\end{eqnarray}
even in the lowest order Feynman diagrams, there are both connected and  disconnected contributions  in the color space \cite{WangZG-dibaryon-PRD-2020}, see Fig.\ref{dibaryon-Lowest-diagram}. From the first diagram in  Fig.\ref{dibaryon-Lowest-diagram}, we can obtain both connected  and disconnected  Feynman diagrams,  the connected contributions appear due to the   quark-gluon operators  $\bar{q}g_sGq\bar{q}g_sGq$  \cite{WangZG-dibaryon-PRD-2020}, which are of the order $\mathcal{O}(\alpha_s^1)$ and come from the Feynman diagrams shown in Fig.\ref{dibaryon-qqg-qqg}.

{\bf Fourthly}, the Landau equation serves  as a kinematical equation in the momentum space, and does not depend  on the factorizable and nonfactorizable properties of the Feynman diagrams in the color space.

\begin{figure}
 \centering
  \includegraphics[totalheight=6cm,width=8cm]{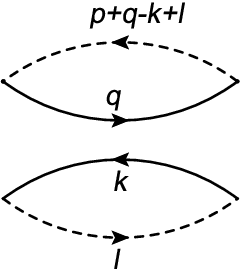}
 \caption{ The  Feynman diagrams  for the lowest order  contributions, where the solid lines and dashed lines represent  the light quarks and heavy quarks, respectively. }\label{Lowest-diagram}
\end{figure}

In the leading order, the factorizable Feynman diagrams shown in Fig.\ref{Lowest-diagram} can be divided into  two color-neutral clusters, however, in the momentum space, they are nonfactorizable diagrams, the basic integrals are of the form,
\begin{eqnarray}\label{Basic-Integral}
\int d^4q d^4k d^4l \frac{1}{\left(p+q-k+l\right)^2-m_c^2}\frac{1}{q^2-m_q^2}\frac{1}{k^2-m_q^2}\frac{1}{l^2-m_c^2}\, .
\end{eqnarray}
If we choose the pole masses, there exists a  Landau singularity  at $s=p^2=(\hat{m}_u+\hat{m}_d+\hat{m}_c+\hat{m}_c)^2$, which is just a signal of a four-quark intermediate state. We cannot assert that it is a signal of a two-meson scattering state or a tetraquark molecular state, because the meson-meson scattering state and  tetraquark molecular state both have four valence quarks,  $q$, $\bar{q}$, $c$ and $\bar{c}$, which form  two color-neutral clusters.

{\bf Fifthly},  only formal QCD sum rules for the tetraquark (molecular) states are obtained based on the assertion of Lucha, Melikhov and Sazdjian
in Refs.\cite{Lucha-Two-PRD-2019,Lucha-Two-PRD-2019-2}, no feasible QCD sum rules  are obtained up to now.

{\bf Sixthly},  we carry out the operator product expansion in  the deep Euclidean space, $-p^2 \to \infty$,  then obtain the physical spectral densities at the quark-gluon level  through dispersion relation \cite{SVZ-NPB-1979-1,SVZ-NPB-1979-2},
 \begin{eqnarray}
 \rho_{QCD}(s)&=&\frac{1}{\pi}{\rm Im}\,\Pi(s+i\epsilon)\mid_{\epsilon\to 0}\, ,
 \end{eqnarray}
 where the $\Pi(s)$ denotes the correlation functions. The Landau singularities require that the squared momentum $p^2=(\hat{m}_u+\hat{m}_d+\hat{m}_c+\hat{m}_c)^2$ in the Feynman diagrams, see Fig.\ref{Lowest-diagram} and Eq.\eqref{Basic-Integral}, it is questionable  to perform the operator product expansion.

{\bf Seventhly},  we choose the local four-quark or five-quark currents, while the traditional mesons and baryons are spatial extended objects and have mean spatial sizes $\sqrt{\langle r^2\rangle} \neq 0$, for example, $\sqrt{\langle r^2\rangle_{E,\Sigma_{c}^{++}}}=0.48\,\rm{fm}$,
$\sqrt{\langle r^2\rangle_{M,\Sigma_{c}^{++}}}=0.83\,\rm{fm}$, $\sqrt{\langle r^2\rangle_{M,\Sigma_{c}^{0}}}=0.81\,\rm{fm}$ from the lattice QCD,
where the subscripts $E$ and $M$ stand for the electric  and magnetic radii, respectively \cite{Can-Sigmac-2015},
$\sqrt{\langle r^2\rangle_{M,\Sigma_{c}^{++}}}=0.77\,\rm{fm}$, $\sqrt{\langle r^2\rangle_{M,\Sigma_{c}^{0}}}=0.52\,\rm{fm}$, $\sqrt{\langle r^2\rangle_{M,\Sigma_{c}^{+}}}=0.81\,\rm{fm}$, $\sqrt{\langle r^2\rangle_{M,\Xi_{c}^{\prime+}}}=0.55\,\rm{fm}$, $\sqrt{\langle r^2\rangle_{M,\Xi_{c}^{\prime0}}}=0.79\,\rm{fm}$ from the self-consistent $SU(3)$
chiral quark-soliton model \cite{Kim-Radii-soliton},  $\sqrt{\langle r^2\rangle_{D^+}}=0.43\,\rm{fm}$, $\sqrt{\langle r^2\rangle_{D^0}}=0.55\,\rm{fm}$ from the light-front quark model \cite{Hwang-Radii-LF},
$\sqrt{\langle r^2\rangle_{J/\psi}}=0.41\,\rm{fm}$, $\sqrt{\langle r^2\rangle_{\chi_{c2}}}=0.71\,\rm{fm}$ from the screened potential  model \cite{X3872-charmonium-KTChao-2009-PRD}. Local currents couple potentially to the compact objects having the average spatial sizes as that of the typical heavy mesons and baryons, not to the two-particle scattering states with average  spatial size $\sqrt{\langle r^2\rangle}\geq 1.0\,\rm{fm}$, which is too large to be interpolated  by the local currents \cite{WZG-local-current,WangZG-Regge-Ms-CPC-2021}.

Now we take a short digression to give {\bf a short notice}. In the QCD sum rules, as we choose the local currents, the four-quark and five-quark states are all compact objects, they are $\bar{\mathbf{3}}\mathbf{3}$-type, $\mathbf{6}\bar{\mathbf{6}}$-type, $\mathbf{1}\mathbf{1}$-type or $\mathbf{8}\mathbf{8}$-type tetraquark states, and  $\bar{\mathbf{3}}\bar{\mathbf{3}}\bar{\mathbf{3}} $-type or $\mathbf{1}\mathbf{1}$-type pentaquark states, although we usually call the $\mathbf{1}\mathbf{1}$-type states as the molecular states.

Now, let us write down  the correlation functions $\Pi_{\mu\nu}(p)$ and $\Pi_{\mu\nu\alpha\beta}(p)$ for the two currents shown in Eq.\eqref{two-example-current},
\begin{eqnarray}
\Pi_{\mu\nu}(p)&=&i\int d^4x e^{ip \cdot x} \langle0|T\Big\{J_\mu(x)J_\nu^{\dagger}(0)\Big\}|0\rangle \, , \\
\Pi_{\mu\nu\alpha\beta}(p)&=&i\int d^4x e^{ip \cdot x} \langle0|T\left\{J_{\mu\nu}(x)J_{\alpha\beta}^{\dagger}(0)\right\}|0\rangle \, .
\end{eqnarray}
If we assume the four-quark currents couple potentially both to the two-meson scattering states and molecular states, then  we can express  the $J_\mu(x)$ and $J_{\mu\nu}(x)$ in terms of the heavy meson fields,
\begin{eqnarray}
J_\mu(x)&=&\frac{1}{\sqrt{2}}\frac{f_D m_D^2}{m_c}f_{D^*}m_{D^*}\left[D^0(x)D^{*-}_\mu(x)+D^{*0}_\mu(x)D^{-}(x)\right]\nonumber\\
&&+\frac{1}{\sqrt{2}}\frac{f_D m_D^2}{m_c}f_{D_0}\left[D^0(x)i\partial_{\mu}D_{0}^-(x)+i\partial_{\mu}D^{0}_{0}(x)D^{-}(x)\right]+\lambda_{Z}Z_{c,\mu}(x)+\cdots\, ,
\end{eqnarray}

\begin{eqnarray}
J_{\mu\nu}(x)&=&\frac{1}{\sqrt{2}}f_{D_s^*}m_{D_s^*}f_{D_{s1}}m_{D_{s1}}\left[D_{s,\mu}^{*+}(x)D^{-}_{s1,\nu}(x) -D_{s1,\nu}^{+}(x)D^{*-}_{s,\mu}(x)\right]\nonumber\\
&&-\frac{1}{\sqrt{2}}f_{D_s^*}m_{D_s^*}f_{D_{s}}\left[D_{s,\mu}^{*+}(x)\partial_{\nu}D_s^{-}(x) -\partial_{\nu}D_{s}^{+}(x)D^{*-}_{s,\mu}(x)\right]\nonumber\\
&&+\frac{1}{\sqrt{2}}f_{D_{s0}}f_{D_{s1}}m_{D_{s1}}\left[i\partial_{\mu} D_{s0}^{+}(x)D^{-}_{s1,\nu}(x) -D_{s1,\nu}^{+}(x)i\partial_{\mu}D^{-}_{s0}(x)\right]\nonumber\\
&&-\frac{1}{\sqrt{2}}f_{D_{s0}}f_{D_{s}}\left[i\partial_{\mu} D_{s0}^{+}(x)\partial_{\nu}D^{-}_{s}(x) -\partial_{\nu}D_{s}^{+}(x)i\partial_{\mu}D^{-}_{s0}(x)\right]\nonumber\\
&&-\tilde{\lambda}_{X^-}\varepsilon_{\mu\nu\alpha\beta}i\partial^{\alpha}X_c^{-\beta}(x)
-\tilde{\lambda}_{X^+}\left[i\partial_{\mu}X^{+}_{c,\nu}(x)-i\partial_{\nu}X^{+}_{c,\mu}(x)\right]+\cdots\, ,
\end{eqnarray}
where we have taken the standard definitions for the decay constants of the traditional mesons and  pole residues of the tetraquark molecular states,
\begin{eqnarray}\label{JA-MM-Z}
 \langle 0|J_{\mu}(0)|Z_c(p)\rangle &=& \lambda_{Z} \,  \varepsilon_{\mu}(p)\, ,
\end{eqnarray}
\begin{eqnarray}\label{JT-MM-Z}
  \langle 0|J_{\mu\nu}(0)|X_c^-(p)\rangle &=& \tilde{\lambda}_{X^-} \, \varepsilon_{\mu\nu\alpha\beta} \, \varepsilon^{\alpha}(p)p^{\beta}\, , \nonumber\\
 \langle 0|J_{\mu\nu}(0)|X_c^+(p)\rangle &=&\tilde{\lambda}_{X^+} \left[\varepsilon_{\mu}(p)p_{\nu}-\varepsilon_{\nu}(p)p_{\mu} \right]\, ,
\end{eqnarray}
$\lambda_{X^\pm}=\tilde{\lambda}_{X^\pm}M_{X^\pm}$,  the $Z_c$, $X^-$ and $X^+$ have the $J^{PC}=1^{+-}$, $1^{-+}$ and $1^{++}$, respectively, the  $\varepsilon_\mu(p)$ are the polarization vectors, we introduce the superscripts $\mp$ to denote the parity.

It is straightforward to obtain the hadronic representation,
\begin{eqnarray}
\Pi_{\mu\nu}(p)&=&\Pi(p^2)\left(-g_{\mu\nu} +\frac{p_\mu p_\nu}{p^2}\right) +\cdots\, \, , \\
\Pi_{\mu\nu\alpha\beta}(p)&=&\Pi_{-}(p^2)\left(g_{\mu\alpha}g_{\nu\beta} -g_{\mu\beta}g_{\nu\alpha} -g_{\mu\alpha}\frac{p_{\nu}p_{\beta}}{p^2}-g_{\nu\beta}\frac{p_{\mu}p_{\alpha}}{p^2}+g_{\mu\beta}\frac{p_{\nu}p_{\alpha}}{p^2}+g_{\nu\alpha}\frac{p_{\mu}p_{\beta}}{p^2}\right) \nonumber\\
&&+\Pi_{+}(p^2)\left( -g_{\mu\alpha}\frac{p_{\nu}p_{\beta}}{p^2}-g_{\nu\beta}\frac{p_{\mu}p_{\alpha}}{p^2}+g_{\mu\beta}\frac{p_{\nu}p_{\alpha}}{p^2}+g_{\nu\alpha}\frac{p_{\mu}p_{\beta}}{p^2}\right) \, ,
\end{eqnarray}
where
\begin{eqnarray}
\Pi(p^2)&=&\frac{\lambda_Z^2}{M_Z^2-p^2}+\Pi_{TW}(p^2)+\cdots\, , \nonumber\\
\Pi_{-}(p^2)&=&P_{-}^{\mu\nu\alpha\beta}\Pi_{\mu\nu\alpha\beta}(p) =\frac{\lambda_{X^-}^2}{M_{X^-}^2-p^2}+\Pi^-_{TW}(p^2)+\cdots\, , \nonumber\\
\Pi_{+}(p^2)&=&P_{+}^{\mu\nu\alpha\beta}\Pi_{\mu\nu\alpha\beta}(p) =\frac{\lambda_{X^+}^2}{M_{X^+}^2-p^2}+\cdots\, ,
\end{eqnarray}
we project out the components $\Pi_{-}(p^2)$ and $\Pi_{+}(p^2)$ by introducing the operators $P_{-}^{\mu\nu\alpha\beta}$ and $P_{+}^{\mu\nu\alpha\beta}$ respectively,
\begin{eqnarray}
P_{-}^{\mu\nu\alpha\beta}&=&\frac{1}{6}\left( g^{\mu\alpha}-\frac{p^\mu p^\alpha}{p^2}\right)\left( g^{\nu\beta}-\frac{p^\nu p^\beta}{p^2}\right)\, , \nonumber\\
P_{+}^{\mu\nu\alpha\beta}&=&\frac{1}{6}\left( g^{\mu\alpha}-\frac{p^\mu p^\alpha}{p^2}\right)\left( g^{\nu\beta}-\frac{p^\nu p^\beta}{p^2}\right)-\frac{1}{6}g^{\mu\alpha}g^{\nu\beta}\, ,
\end{eqnarray}
\begin{eqnarray}\label{Pi-TW}
\Pi_{TW}(p^2)&=&\frac{\lambda_{DD^*}^2}{16\pi^2}\int_{\Delta_1^2}^{s_0}ds \frac{1}{s-p^2}\frac{\sqrt{\lambda(s,m_{D}^2,m_{D^*}^2)}}{s}\left[1+\frac{\lambda(s,m_{D}^2,m_{D^*}^2)}{12sm_{D^*}^2} \right]\nonumber\\
&&+\frac{\lambda_{DD_0}^2}{16\pi^2}\int_{\Delta_2^2}^{s_0}ds \frac{1}{s-p^2}\frac{\sqrt{\lambda(s,m_{D}^2,m_{D_0}^2)}}{s}\frac{\lambda(s,m_{D}^2,m_{D_0}^2)}{12s}
+\cdots\, ,
\end{eqnarray}
\begin{eqnarray}\label{Pi-TW-N}
\Pi^-_{TW}(p^2)&=&\frac{\lambda_{D^*_sD_{s1}}^2}{16\pi^2}\int_{\Delta_3^2}^{s_0}ds \frac{1}{s-p^2}\frac{\sqrt{\lambda(s,m_{D_s^*}^2,m_{D_{s1}}^2)}}{s}\left[1+\frac{\lambda(s,m_{D^*_s}^2,m_{D_{s1}}^2)}{12sm_{D^*_s}^2}  \frac{\lambda(s,m_{D^*_s}^2,m_{D_{s1}}^2)}{12sm_{D_{s1}}^2}\right] \nonumber\\
&&+\frac{\lambda_{D_s^*D_s}^2}{16\pi^2}\int_{\Delta_4^2}^{s_0}ds \frac{1}{s-p^2}\frac{\sqrt{\lambda(s,m_{D_s^*}^2,m_{D_s}^2)}}{s}\frac{\lambda(s,m_{D_s^*}^2,m_{D_s}^2)}{12s}\nonumber\\
&&+\frac{\lambda_{D_{s0}D_{s1}}^2}{16\pi^2}\int_{\Delta_5^2}^{s_0}ds \frac{1}{s-p^2}\frac{\sqrt{\lambda(s,m_{D_{s0}}^2,m_{D_{s1}}^2)}}{s}\frac{\lambda(s,m_{D_{s0}}^2,m_{D_{s1}}^2)}{12s}
+\cdots\, ,
\end{eqnarray}
where $\lambda_{DD^*}^2=\frac{f_{D}^2m_{D}^4f_{D^*}^2m_{D^*}^2}{m_c^2}$, $\Delta_1^2=(m_{D}+m_{D^*})^2$,
$\lambda_{DD_0}^2=\frac{f_{D}^2m_{D}^4f_{D_0}^2}{m_c^2}$, $\Delta_2^2=(m_{D}+m_{D_0})^2$,
$\lambda_{D_s^*D_{s1}}^2=f_{D^*_s}^2m_{D^*_s}^2f_{D_{s1}}^2m_{D_{s1}}^2$, $\Delta_3^2=(m_{D^*_s}+m_{D_{s1}})^2$,
$\lambda_{D_s^*D_{s}}^2=f_{D^*_s}^2m_{D^*_s}^2f_{D_{s}}^2$, $\Delta_4^2=(m_{D_s^*}+m_{D_s})^2$,
$\lambda_{D_{s0}D_{s1}}^2=f_{D_{s0}}^2f_{D_{s1}}^2m_{D_{s1}}^2$,
$\Delta_5^2=(m_{D_{s0}}+m_{D_{s1}})^2$
and $\lambda(a,b,c)=a^2+b^2+c^2-2ab-2bc-2ca$.

The traditional hidden-flavor mesons have the normal quantum numbers, $J^{PC}=0^{-+}$, $0^{++}$, $1^{--}$, $1^{+-}$, $1^{++}$, $2^{--}$, $2^{-+}$, $2^{++}$, $\cdots$.
The component $\Pi_{-}(p^2)$ receives  contributions with the exotic quantum numbers $J^{PC}=1^{-+}$,  the component $\Pi_{+}(p^2)$  receives contributions with the normal quantum numbers $J^{PC}=1^{++}$. We  choose the component $\Pi_{-}(p^2)$ with the exotic quantum numbers $J^{PC}=1^{-+}$ and discard the  component $\Pi_{+}(p^2)$ with the normal quantum numbers $J^{PC}=1^{++}$. Thereafter, we will neglect  the superscript $-$ in the $X_c^-$ for simplicity.

According to the assertion of Lucha, Melikhov and Sazdjian \cite{Lucha-Two-PRD-2019,Lucha-Two-PRD-2019-2}, all contributions of the order $\mathcal{O}(\alpha_s^k)$ with $k\leq 1$ are exactly canceled out by the two-meson scattering states, we  set
\begin{eqnarray}
\Pi(p^2)&=&\Pi_{TW}(p^2)+\cdots\, , \nonumber\\
\Pi_{-}(p^2)&=&\Pi^-_{TW}(p^2)+\cdots\, ,
\end{eqnarray}
at the hadron side \cite{WZG-Landau-PRD-2020}.
 Then let us   take the
quark-hadron duality below the continuum threshold $s_0$ and perform Borel transformation   with respect to
the variable $P^2=-p^2$ to obtain  the  QCD sum rules:
\begin{eqnarray}\label{TW-A-QCDSR}
\Pi_{TW}(T^2)&=&\frac{\lambda_{DD^*}^2}{16\pi^2}\int_{\Delta_1^2}^{s_0}ds \frac{\sqrt{\lambda(s,m_{D}^2,m_{D^*}^2)}}{s}\left[1+\frac{\lambda(s,m_{D}^2,m_{D^*}^2)}{12sm_{D^*}^2} \right]\exp\left(-\frac{s}{T^2}\right)\nonumber\\
&&+\frac{\lambda_{DD_0}^2}{16\pi^2}\int_{\Delta_2^2}^{s_0}ds \frac{\sqrt{\lambda(s,m_{D}^2,m_{D_0}^2)}}{s}\frac{\lambda(s,m_{D}^2,m_{D_0}^2)}{12s}\exp\left(-\frac{s}{T^2}\right)\nonumber\\
&=&\kappa\int_{4m_c^2}^{s_0}ds\,\rho_{Z,QCD}(s)\exp\left(-\frac{s}{T^2}\right)\, ,
\end{eqnarray}
\begin{eqnarray}\label{TW-Negative-V-QCDSR}
\Pi^-_{TW}(T^2)&=&\frac{\lambda_{D^*_sD_{s1}}^2}{16\pi^2}\int_{\Delta_3^2}^{s_0}ds \frac{\sqrt{\lambda(s,m_{D_s^*}^2,m_{D_{s1}}^2)}}{s}\left[1+\frac{\lambda(s,m_{D^*_s}^2,m_{D_{s1}}^2)}{12sm_{D^*_s}^2}  \frac{\lambda(s,m_{D^*_s}^2,m_{D_{s1}}^2)}{12sm_{D_{s1}}^2}\right]\exp\left(-\frac{s}{T^2}\right) \nonumber\\
&&+\frac{\lambda_{D_s^*D_s}^2}{16\pi^2}\int_{\Delta_4^2}^{s_0}ds \frac{\sqrt{\lambda(s,m_{D_s^*}^2,m_{D_s}^2)}}{s}\frac{\lambda(s,m_{D_s^*}^2,m_{D_s}^2)}{12s}\exp\left(-\frac{s}{T^2}\right)\nonumber\\
&&+\frac{\lambda_{D_{s0}D_{s1}}^2}{16\pi^2}\int_{\Delta_5^2}^{s_0}ds \frac{\sqrt{\lambda(s,m_{D_{s0}}^2,m_{D_{s1}}^2)}}{s}\frac{\lambda(s,m_{D_{s0}}^2,m_{D_{s1}}^2)}{12s}\exp\left(-\frac{s}{T^2}\right)\nonumber\\
&=&\kappa\int_{4m_c^2}^{s_0}ds\,\rho_{X,QCD}(s)\exp\left(-\frac{s}{T^2}\right)\, ,
\end{eqnarray}
 the   explicit expressions of the QCD spectral densities $\rho_{Z,QCD}(s)$ and $\rho_{X,QCD}(s)$ are given in Ref.\cite{WZG-Landau-PRD-2020}. We introduce the parameter $\kappa$ to measure the deviations from $1$, if $\kappa\approx1$, we could  get the conclusion tentatively that
the two-meson scattering states can  saturate the QCD sum rules.
Then we  differentiate   Eqs.\eqref{TW-A-QCDSR}-\eqref{TW-Negative-V-QCDSR} with respect to  $\frac{1}{T^2}$,   and obtain two additional  QCD sum rules,
 \begin{eqnarray}\label{TW-A-QCDSR-Dr}
-\frac{d\Pi_{TW}(T^2)}{d(1/T^2)}&=&-\kappa\frac{d}{d(1/T^2)}\int_{4m_c^2}^{s_0}ds\,\rho_{Z,QCD}(s)\exp\left(-\frac{s}{T^2}\right)\, ,
\end{eqnarray}
\begin{eqnarray}\label{TW-Negative-V-QCDSR-Dr}
-\frac{d\Pi^-_{TW}(T^2)}{d(1/T^2)}&=&-\kappa\frac{d}{d(1/T^2)}\int_{4m_c^2}^{s_0}ds\,\rho_{X,QCD}(s)\exp\left(-\frac{s}{T^2}\right)\, .
\end{eqnarray}
Thereafter, we will denote the QCD sum rules in Eqs.\eqref{TW-A-QCDSR-Dr}-\eqref{TW-Negative-V-QCDSR-Dr} as the QCDSR I, and
the QCD sum rules in Eqs.\eqref{TW-A-QCDSR}-\eqref{TW-Negative-V-QCDSR} as the QCDSR II.

On the other hand, if the two-meson scattering states cannot saturate the QCD sum rules, we have to introduce the tetraquark molecular states to saturate the QCD sum rules,
\begin{eqnarray}\label{TetraQ-A-QCDSR}
\lambda_{Z/X}^2\exp\left(-\frac{M_{Z/X}^2}{T^2}\right)&=&\int_{4m_c^2}^{s_0}ds\,\rho_{Z/Z,QCD}(s)\exp\left(-\frac{s}{T^2}\right)\, ,
\end{eqnarray}
then  we differentiate   Eq.\eqref{TetraQ-A-QCDSR} with respect to  $\frac{1}{T^2}$,   and obtain two QCD sum rules for the masses of the  tetraquark molecular states,
\begin{eqnarray}\label{TetraQ-A-QCDSR-Dr}
M_{Z/X}^2&=&\frac{-\frac{d}{d(1/T^2)}\int_{4m_c^2}^{s_0}ds\,\rho_{Z/X,QCD}(s)\exp\left(-\frac{s}{T^2}\right)}
{\int_{4m_c^2}^{s_0}ds\,\rho_{Z/X,QCD}(s)\exp\left(-\frac{s}{T^2}\right)}\, .
\end{eqnarray}

\begin{figure}
\centering
\includegraphics[totalheight=6cm,width=7cm]{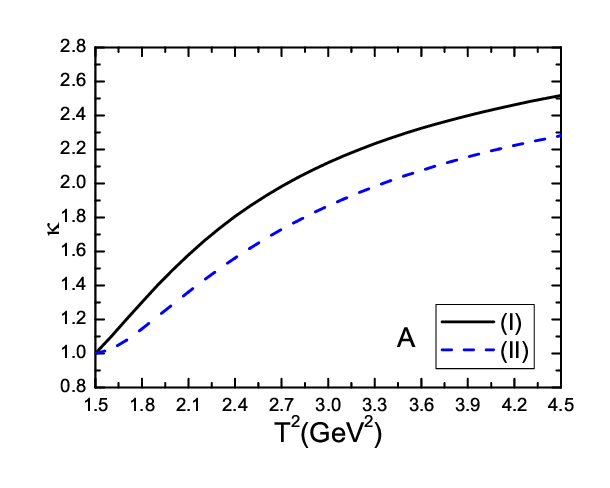}
\includegraphics[totalheight=6cm,width=7cm]{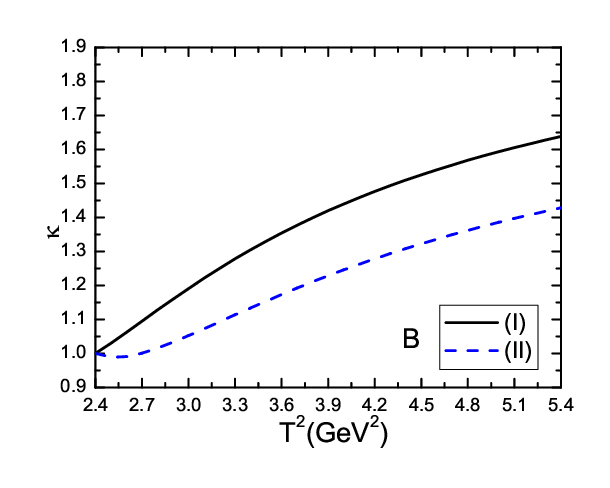}
  \caption{ The $\kappa$  with variations of the  Borel parameters $T^2$, where the $A$ and $B$ correspond to the $\bar{u}c\bar{c}d$ and  $\bar{s}c\bar{c}s$ meson-meson scattering states, respectively,  the (I) and (II) correspond to  QCDSR I and II, respectively, the  $\kappa$ values are normalized to be $1$ for the Borel parameters $T^2=1.5\,\rm{GeV}^2$ and $2.4\,\rm{GeV}^2$, respectively. }\label{kappa-Borel}
\end{figure}

In Fig.\ref{kappa-Borel}, we  plot the values of the $\kappa$  with variations of the  Borel parameters $T^2$ with the continuum threshold parameters
   $\sqrt{s_0}=4.40\,\rm{GeV}$  and  $5.15\,\rm{GeV}$ for the $\bar{u}c\bar{c}d$ and  $\bar{s}c\bar{c}s$ two-meson scattering states, respectively.
From Fig.\ref{kappa-Borel}, we can see explicitly  that the values of the $\kappa$ increase monotonically and
 quickly  with the increase of the Borel parameters $T^2$, no platform appears, which indicates  that the QCD sum rules in Eqs.\eqref{TW-A-QCDSR}-\eqref{TW-Negative-V-QCDSR} obtained according to the assertion of  Lucha, Melikhov and Sazdjian are unreasonable. Reasonable QCD sum rules lead to platforms flat enough  or not flat enough, rather than  no evidence of platforms, the two-meson scattering states cannot saturate the QCD sum rules.

\begin{figure}
\centering
\includegraphics[totalheight=6cm,width=7cm]{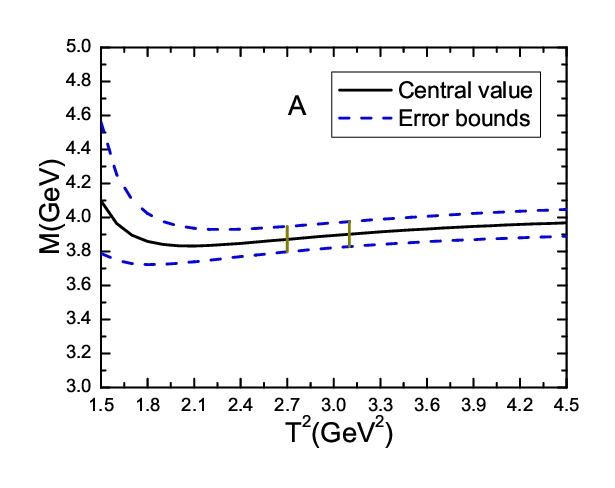}
\includegraphics[totalheight=6cm,width=7cm]{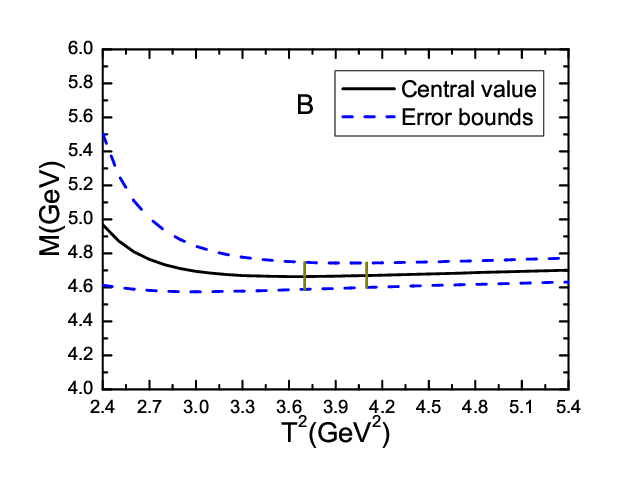}
  \caption{ The masses  with variations of the  Borel parameters $T^2$, where the $A$ and $B$ correspond to the $\bar{u}c\bar{c}d$ and  $\bar{s}c\bar{c}s$ tetraquark molecular states, respectively, the regions between the two vertical lines are the Borel windows. }\label{mass-Borel}
\end{figure}

We saturate the hadron side of the QCD sum rules with the tetraquark molecular states alone, and study the QCD sum rues shown in Eqs.\eqref{TetraQ-A-QCDSR}-\eqref{TetraQ-A-QCDSR-Dr}.
In Fig.\ref{mass-Borel}, we  plot tetraquark molecule masses  with variations of the  Borel parameters $T^2$. From the Fig.\ref{mass-Borel}, we observe that  there appear Borel platforms in the Borel windows indeed, the relevant results are shown explicitly in Table \ref{Borel-pole-mass}, we adopt the energy scale formula
$\mu=\sqrt{M_{X/Z}^2-4\times(1.84\,\rm{GeV})^2}$  to choose the best energy scales of the QCD spectral densities. The tetraquark molecular states alone can satisfy the QCD sum rules \cite{WZG-Landau-PRD-2020}. We obtain the prediction $M_Z=3.89\pm0.09\,\rm{GeV}$ and $M_{X}=4.67\pm0.08\,\rm{GeV}$, which are consistent with the $Z_c(3900)$ observed by the BESIII and Belle collaborations \cite{BES-Z3900-PRL-2013,Belle-Z3900-PRL-2013} and
the $X(4630)$ {\bf observed one-year latter} by the LHCb collaboration \cite{LHCb-Zcs4000-PRL-2021}. If we have taken account of the light-flavor $SU(3)$ breaking effects of the energy scale formula, the fit between the theoretical calculation and experimental measurement would be better for the $X(4630)$.

The local currents do suppress but do not forbid the couplings between the four-quark currents and two-meson scattering states, as the overlaps of the wave-functions are very small \cite{WZG-local-current}, furthermore, the quantum field theory  does not forbid the couplings between the four-quark currents and two-meson scattering states if they have the same quantum numbers. We  study the contributions of the intermediate   meson-meson  scattering states  $D
\bar{D}^\ast$, $J/\psi \pi$, $J/\psi \rho$, etc besides the tetraquark molecular state $Z_c$ to the correlation function $\Pi_{\mu\nu}(p)$ as an example,
\begin{eqnarray}\label{Self-Energy}
\Pi_{\mu\nu}(p) &=&-\frac{\widehat{\lambda}_{Z}^{2}}{ p^2-\widehat{M}_{Z}^2-\Sigma_{DD^*}(p^2)-\Sigma_{J/\psi\pi}(p^2)-\Sigma_{J/\psi\rho}(p^2)+\cdots}\widetilde{g}_{\mu\nu}(p)+\cdots \, ,
\end{eqnarray}
where
$\widetilde{g}_{\mu\nu}(p)=-g_{\mu\nu}+\frac{p_{\mu}p_{\nu}}{p^2}$. We choose the bare quantities $\widehat{\lambda}_{Z}$ and $\widehat{M}_{Z}$  to absorb the divergences in the self-energies $\Sigma_{D\bar{D}^*}(p^2)$, $\Sigma_{J/\psi \pi}(p^2)$, $\Sigma_{J/\psi \rho}(p^2)$, etc. The renormalized energies  satisfy  the relation $p^2-M_{Z}^2-\overline{\Sigma}_{DD^*}(p^2)-\overline{\Sigma}_{J/\psi\pi}(p^2)-\overline{\Sigma}_{J/\psi\rho}(p^2)+\cdots=0$, where the overlines above the
self-energies denote that the divergent terms have been subtracted. As the tetraquark molecular state $Z_c$ is unstable, the relation should be modified,
$p^2-M_{Z}^2-{\rm Re}\overline{\Sigma}_{DD^*}(p^2)-{\rm Re}\overline{\Sigma}_{J/\psi\pi}(p^2)-{\rm Re}\overline{\Sigma}_{J/\psi\rho}(p^2)+\cdots=0$, and
$-{\rm Im}\overline{\Sigma}_{DD^*}(p^2)-{\rm Im}\overline{\Sigma}_{J/\psi\pi}(p^2)-{\rm Im}\overline{\Sigma}_{J/\psi\rho}(p^2)+\cdots=\sqrt{p^2}\Gamma(p^2)$.
The renormalized self-energies  contribute  a finite imaginary part to modify the dispersion relation,
\begin{eqnarray}\label{Modify-width}
\Pi_{\mu\nu}(p) &=&-\frac{\lambda_{Z}^{2}}{ p^2-M_{Z}^2+i\sqrt{p^2}\Gamma(p^2)}\widetilde{g}_{\mu\nu}(p)+\cdots \, .
 \end{eqnarray}
If we assign the $Z_c(3900)$ to be the $D\bar{D}^*+D^*\bar{D}$  tetraquark molecular state with the $J^{PC}=1^{+-}$ \cite{X3872-mole-WangZG-HT-EPJC-2014,WZG-tetra-mole-IJMPA-2021,WZG-tetra-mole-AAPPS-2022},  the physical  width $\Gamma_{Z_c(3900)}(M_Z^2)=(28.2 \pm 2.6)\, \rm{MeV}$ from the Particle Data Group \cite{PDG-2018}.

We  take  account of the finite width effect by the  simple replacement of the hadronic spectral density,
\begin{eqnarray}
\lambda^2_{Z}\delta \left(s-M^2_{Z} \right) &\to& \lambda^2_{Z}\frac{1}{\pi}\frac{M_{Z}\Gamma_{Z}(s)}{(s-M_{Z}^2)^2+M_{Z}^2\Gamma_{Z}^2(s)}\, ,
\end{eqnarray}
where
\begin{eqnarray}
\Gamma_{Z}(s)&=&\Gamma_{Z} \frac{M_{Z}}{\sqrt{s}}\sqrt{\frac{s-(M_{D}+M_{D^*})^2}{M^2_{Z}-(M_{D}+M_{D^*})^2}} \, .
\end{eqnarray}
Then the hadron  sides of  the QCD sum rules in Eq.\eqref{TetraQ-A-QCDSR} and Eq.\eqref{TetraQ-A-QCDSR-Dr} undergo the following changes,
\begin{eqnarray}
\lambda^2_{Z}\exp \left(-\frac{M^2_{Z}}{T^2} \right) &\to& \lambda^2_{Z}\int_{(m_{D}+m_{D^*})^2}^{s_0}ds\frac{1}{\pi}\frac{M_{Z}\Gamma_{Z}(s)}{(s-M_{Z}^2)^2+M_{Z}^2\Gamma_{Z}^2(s)}\exp \left(-\frac{s}{T^2} \right)\, , \nonumber\\
&=&(0.78\sim0.79)\,\lambda^2_{Z}\exp \left(-\frac{M^2_{Z}}{T^2} \right)\, , \\
\lambda^2_{Z}M^2_{Z}\exp \left(-\frac{M^2_{Z}}{T^2} \right) &\to& \lambda^2_{Z}\int_{(m_{D}+m_{D^*})^2}^{s_0}ds\,s\,\frac{1}{\pi}\frac{M_{Z}\Gamma_{Z}(s)}{(s-M_{Z}^2)^2+M_{Z}^2\Gamma_{Z}^2(s)}\exp \left(-\frac{s}{T^2} \right)\, , \nonumber\\
&=&(0.80\sim0.81)\,\lambda^2_{Z}M^2_{Z}\exp \left(-\frac{M^2_{Z}}{T^2} \right)\, ,
\end{eqnarray}
for  the value $\sqrt{s_0}=4.40\,\rm{GeV}$.
We can absorb the numerical factors  $0.78\sim0.79$ and $0.80\sim0.81$ into the pole residue with the simple replacement $\lambda_{Z}\to 0.89\lambda_Z$ safely, the intermediate   meson-loops cannot  affect  the mass $M_{Z}$ significantly, but affect the pole residue remarkably, which are consistent with the fact that we obtain the masses of the tetraquark molecular states from a fraction, see Eq.\eqref{TetraQ-A-QCDSR-Dr}.

\begin{table}
\begin{center}
\begin{tabular}{|c|c|c|c|c|c|c|c|}\hline\hline
$J^{PC}$                     &$T^2 (\rm{GeV}^2)$ &$\sqrt{s_0} (\rm{GeV})$&$\mu(\rm GeV)$    &pole         & $M(\rm{GeV})$           & $\lambda(10^{-2}\rm{GeV}^5)$ \\ \hline

$1^{+-}\,(\bar{u}c\bar{c}d)$ &$2.7-3.1$          &$4.40\pm0.10$          &$1.3$             &$(40-63)\%$  & $3.89\pm0.09$           & $1.72\pm0.30$ \\ \hline

$1^{-+}\,(\bar{s}c\bar{c}s)$ &$3.7-4.1$          &$5.15\pm0.10$          &$2.9$             &$(42-60)\%$  & $4.67\pm0.08$           & $6.87\pm0.84$ \\ \hline
 \hline
\end{tabular}
\end{center}
\caption{ The Borel windows, continuum threshold parameters, energy scales of the QCD spectral densities, pole contributions, masses and pole residues of the $\bar{u}c\bar{c}d$ and  $\bar{s}c\bar{c}s$ tetraquark molecular states \cite{WZG-Landau-PRD-2020}. }\label{Borel-pole-mass}
\end{table}

We obtain the conclusion confidently that it is reliable to study the multiquark states with the QCD sum rules, the contaminations from the two-particle scattering states play a tiny role \cite{WZG-Landau-PRD-2020}.

\subsection{Energy scale dependence of the QCD sum rules} \label{Energy-scale-dependence}
In calculating the Feynman diagrams, we usually   adopt  the dimensional regularization to regularize the divergences, and resort to wave-function, quark-mass and current renormalizations to absorb the ultraviolet divergences, and resort to the vacuum condensate redefinitions to absorb  the infrared divergences. Thus, the correlation functions $\Pi(p^2)$ are free of divergences. And we expect to calculate the $\Pi(p^2)$ at any energy scale $\mu$ at which perturbative calculations are feasible, and the physical quantities are independent on the specified energy scale $\mu$. Roughly speaking, the correlation functions $\Pi(p^2)$ are independent on the energy scale approximately,
\begin{eqnarray}
\frac{d}{d\mu}\Pi(p^2)&=&0\, ,
\end{eqnarray}
at least the bare $\Pi(p^2)$ are independent on the energy scale.

We write down the correlation functions $\Pi(p^2)$ for the hidden-charm (or hidden-bottom) four-quark currents, the most commonly chosen currents in studying the $X$, $Y$ and $Z$ states,  in the K\"allen-Lehmann representation,
\begin{eqnarray}\label{CT-spectral-mu}
\Pi(p^2)&=&\int_{4m^2_Q(\mu)}^{s_0} ds \frac{\rho_{QCD}(s,\mu)}{s-p^2}+\int_{s_0}^\infty ds \frac{\rho_{QCD}(s,\mu)}{s-p^2} \, .
\end{eqnarray}
In fact, there are subtraction terms neglected  at the right side of Eq.\eqref{CT-spectral-mu}, which could be deleted after performing the Borel transformation.
The  $\Pi(p^2)$ should  be independent on the energy scale we adopt to perform the operator product expansion,
but which does not mean
\begin{eqnarray}
\frac{d}{d\mu}\int_{4m^2_Q(\mu)}^{s_0} ds \frac{\rho_{QCD}(s,\mu)}{s-p^2}\rightarrow 0 \, ,
\end{eqnarray}
 due to the two features inherited from the QCD sum rules:\\
$\bullet$ Perturbative corrections are neglected, even in the QCD sum rules for the traditional mesons, we cannot take  account of the radiative corrections up to arbitrary  orders, for example, we only have calculated  the radiative corrections up to the order $\mathcal{O}(\alpha_s^2)$ for the pseudoscalar $D/B$ mesons up to now \cite{WZG-D-meson-EPJC-2015,Tree-loop-Chetyrkin-PLB-2001,Tree-loop-Chetyrkin-EPJC-2001}.  The higher dimensional vacuum condensates are factorized into lower dimensional ones based on the vacuum saturation,
for example,
\begin{eqnarray}\label{Four-quark-fact}
\langle \bar{\psi}\Gamma \psi  \bar{\psi}\Gamma \psi \rangle &=&\frac{1}{144}\langle \bar{\psi} \psi\rangle^2
\left[ {\rm Tr}(\Gamma){\rm Tr}(\Gamma)-{\rm Tr}(\Gamma\Gamma)\right]\, ,
\end{eqnarray}
where $\psi=u$, $d$ or $s$,
${\rm Tr}={\rm Tr}_{D}{\rm Tr}_C$, the subscripts denote the Dirac spinor  and color spaces, respectively,
 therefore  the energy scale dependence of the higher dimensional vacuum condensates is modified.\\
$\bullet$ Truncations $s_0$ which are physical quantities determined by the experimental data set in, the correlations  between the thresholds $4m^2_Q(\mu)$ and continuum thresholds $s_0$ are unknown. Quark-hadron duality is just an assumption. \\

After performing the Borel transformation, we obtain the integrals
 \begin{eqnarray}
 \int_{4m_Q^2(\mu)}^{s_0} ds \rho_{QCD}(s,\mu)\exp\left(-\frac{s}{T^2} \right)\, ,
 \end{eqnarray}
which are sensitive to the $Q$-quark mass $m_Q(\mu)$, in other words,   the energy scale $\mu$.  Variations of the energy scale $\mu$ can lead to changes of integral ranges $4m_Q^2(\mu)-s_0$ of the variable
$ds$ besides the QCD spectral densities $\rho_{QCD}(s,\mu)$, therefore changes of the Borel windows and predicted hadron masses and pole residues. The strong fine-structure $\alpha_s=\frac{g_s^2}{4\pi}$ appears  even in the tree-level,
\begin{eqnarray}
\langle \bar{q}\gamma_\mu t^a q D_\eta g_sG^a_{\lambda\tau}\rangle
&=&\frac{g_{\eta\lambda}g_{\mu\tau}-g_{\eta\tau}g_{\mu\lambda}}{27} 4\pi\alpha_s \langle \bar{q}q\rangle^2\, ,
\end{eqnarray}
where $t^a=\frac{\lambda^a}{2}$. Thus we have to deal with the energy scale dependence of the QCD sum rules.

Let us take a short digression and perform some phenomenological analysis.
We can describe the heavy four-quark systems  $Q\bar{Q}q\bar{q}$ by a double-well potential with two light quarks $q\bar{q}$ lying in the two wells, respectively.
   In the heavy quark limit,  the $Q$-quark serves as a static well potential and  attracts with the light quark $q$  to form a heavy diquark $\mathcal{D}^i_{qQ}$ in  color antitriplet,
\begin{eqnarray}
q+Q &\to & \mathcal{D}^i_{qQ} \, ,
\end{eqnarray}
or attracts with the light antiquark $\bar{q}$ to form a meson-like color-singlet cluster (meson-like
color-octet cluster),
\begin{eqnarray}
\bar{q}+Q &\to & \bar{q}Q\,\, (\bar{q}\lambda^{a}Q) \, ,
\end{eqnarray}
 the $\bar{Q}$-quark serves  as another static well potential and attracts with the light antiquark $\bar{q}$  to form a heavy antidiquark $\mathcal{D}^i_{\bar{q}\bar{Q}}$ in  color triplet,
\begin{eqnarray}
\bar{q}+\bar{Q} &\to & \mathcal{D}^i_{\bar{q}\bar{Q}} \, ,
\end{eqnarray}
or attracts with the light quark $q$ to form a heavy meson-like color-singlet cluster (meson-like
color-octet-cluster),
\begin{eqnarray}
q+\bar{Q} &\to & \bar{Q}q\,\, (\bar{Q}\lambda^{a}q) \, .
\end{eqnarray}
 Then
\begin{eqnarray}
 \mathcal{D}^i_{qQ}+\mathcal{D}^i_{\bar{q}\bar{Q}} &\to &  {\rm \bar{\mathbf{3}}\mathbf{3}-type \,\,\, tetraquark \,\,\, states}\, , \nonumber\\
 \bar{q}Q+\bar{Q}q &\to & {\rm \mathbf{1}\mathbf{1}-type \,\,\, tetraquark \,\,\, states}\, , \nonumber\\
  \bar{q}\lambda^aQ+\bar{Q}\lambda^a q &\to &{\rm \mathbf{8}\mathbf{8}-type \,\,\, tetraquark \,\,\, states}\, ,
\end{eqnarray}
the two heavy quarks $Q$ and $\bar{Q}$ stabilize the four-quark systems $q\bar{q}Q\bar{Q}$, just as in the case
of the $(\mu^-e^+)(\mu^+ e^-)$ molecule in QED \cite{Lebed-dynamical-PRL-2014}.

We can also describe the hidden-charm (or hidden-bottom) five-quark systems  $qq_1q_2Q\bar{Q}$
by a double-well potential. In the heavy quark limit, the $Q$-quark ($\bar{Q}$-quark) serves as a static well potential, the diquark $\mathcal{D}_{q_1q_2}^j$ and quark $q$  lie in the two wells,  respectively,
\begin{eqnarray}
q_1+q_2 +\bar{Q}&\to & \mathcal{D}_{q_1q_2}^j +\bar{Q}^k \to  \mathcal{T}^i_{q_1q_2\bar{Q}} \, , \nonumber\\
q+Q &\to & \mathcal{D}^i_{qQ} \, ,
\end{eqnarray}
or
\begin{eqnarray}
q_1+q_2 +Q&\to & \mathcal{D}_{q_1q_2}^j +Q^j \to  q_1q_2Q \, , \nonumber\\
q+\bar{Q} &\to & q\bar{Q} \, ,
\end{eqnarray}
 where the $\mathcal{T}^i_{q_1q_2\bar{Q}}$ denotes the heavy
 triquark in the color triplet. Then
\begin{eqnarray}
 \mathcal{D}^i_{qQ}+\mathcal{T}^i_{q_1q_2\bar{Q}} &\to &  {\rm \bar{\mathbf{3}}\bar{\mathbf{3}}\bar{\mathbf{3}}-type \,\,\, pentaquark \,\,\, states}\, , \nonumber\\
 q_1q_2Q+\bar{Q}q &\to & {\rm \mathbf{1}\mathbf{1}-type \,\,\, pentaquark \,\,\, states}\, .
\end{eqnarray}

Now we can obtain the conclusion tentatively that  the heavy tetraquark states  are characterized by the effective heavy quark masses ${\mathbb{M}}_Q$ (or constituent quark masses) and
the virtuality $V=\sqrt{M^2_{X/Y/Z}-(2{\mathbb{M}}_Q)^2}$ (or bound energy not as robust) \cite{X3872-mole-WangZG-HT-EPJC-2014,WangZG-X4140-Zb10650-mole-EPJC-2014,WangZG-Penta-EPJC-2016-70,
 WangZG-formula-Vect-tetra-EPJC-2014}.

In summary, the QCD sum rules have three typical energy scales $\mu^2$, $T^2$, $V^2$.  It is natural to set the energy  scales  as,
 \begin{eqnarray}
 \mu^2&=&V^2={\mathcal{O}}(T^2)\, ,
 \end{eqnarray}
and we obtain the energy scale formula \cite{WangZG-formula-Vect-tetra-EPJC-2014},
\begin{eqnarray}\label{formula}
\mu&=&\sqrt{M^2_{X/Y/Z/P}-(2{\mathbb{M}}_Q)^2}\, ,
\end{eqnarray}
which works very well for the tetraquark states and pentaquark states. It can improve the convergence of the operator product expansion remarkably  and enhance the pole contributions  remarkably.

We usually set the small $u$ and $d$ quark masses to be zero, and take account of the light-flavor $SU(3)$
breaking effects by introducing an effective $s$-quark mass ${\mathbb{M}}_s$, thus we reach the modified energy scale formula,
\begin{eqnarray} \label{modify-formula}
\mu&=&\sqrt{M^2_{X/Y/Z}-(2{\mathbb{M}}_Q)^2}-\kappa\,{\mathbb{M}}_s\, ,
\end{eqnarray}
to choose the suitable  energy scales of the QCD spectral densities
\cite{WZG-HC-PRD-2020,WZG-tetra-mole-IJMPA-2021,WZG-tetra-mole-AAPPS-2022,WangZG-Zcs3985-mass-tetra-CPC-2021,WZG-HC-ss-NPB-2024}, where the $\kappa=0$, $1$ and $2$ denote the numbers of the $s$-quarks,  the ${\mathbb{M}}_Q$ and ${\mathbb{M}}_s$  have universal values to be commonly used elsewhere.

\begin{figure}
\centering
\includegraphics[totalheight=10cm,width=14cm]{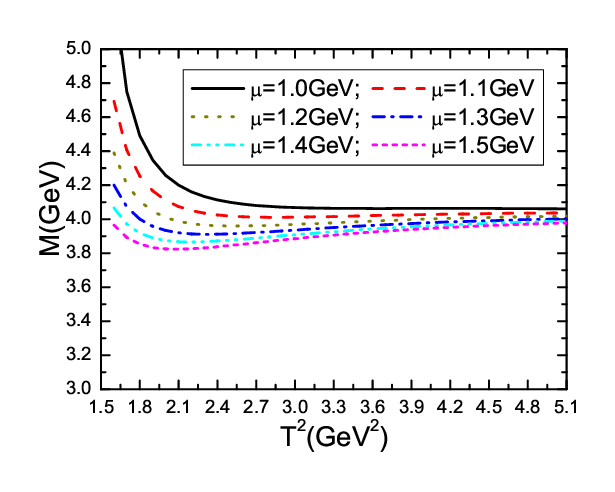}
  \caption{ The mass  with variation of the  Borel parameter $T^2$ and energy scale $\mu$ for the $Z_c(3900)$. }\label{Zc3900-mu}
\end{figure}

We can rewrite the energy scale formula in Eq.\eqref{formula} in the following form,
\begin{eqnarray}\label{formula-Regge}
M^2_{X/Y/Z/P}&=&\mu^2+{\rm Constants}\, ,
\end{eqnarray}
where the Constants have the values $4{\mathbb{M}}_Q^2$ \cite{WangZG-Regge-CTP-2021}.  As we cannot obtain energy scale independent QCD sum rules, we conjecture that the predicted multiquark masses and the pertinent energy scales of the QCD spectral densities have a
 Regge-trajectory-like relation, see  Eq.\eqref{formula-Regge}, where the Constants are free parameters and fitted by the QCD sum rules. Direct calculations have proven that the Constants  have universal values and work well.
 We take account of the light-flavor $SU(3)$ breaking effects, and write down the modified energy scale formula \cite{WangZG-Regge-Ms-CPC-2021},
 \begin{eqnarray}\label{Modify-formula-Regge}
M^2_{X/Y/Z/P}&=&(\mu+\kappa {\mathbb{M}}_s) ^2+{\rm Constants}\, .
\end{eqnarray}

In Ref.\cite{X3872-tetra-WangZG-HuangT-PRD-2014}, we take the $X(3872)$ and $Z_c(3900)$ as the hidden-charm tetraquark with the $J^{PC}=1^{++}$ and $1^{+-}$, respectively, and explore the energy scale dependence of the QCD sum rules for the exotic states  for the first time.
In Fig.\ref{Zc3900-mu},  we plot the mass of the $Z_c(3900)$  with variations of the  Borel parameter $T^2$ and energy scale $\mu$ for the continuum
threshold parameter $\sqrt{s_0}=4.4\,\rm{GeV}$. From the figure, we can see clearly  that the mass decreases monotonously
with increase of the energy scale. The energy scale $\mu=1.4\,\rm{GeV}$ is the lowest energy scale to reproduce the experimental data \cite{BES-Z3900-PRL-2013,Belle-Z3900-PRL-2013}.

There are three schemes to choose the input parameters at the QCD side of the QCD sum rules: \\
{\bf Scheme I}. We take the energy scale formula and its modifications, see Eqs.\eqref{formula}-\eqref{modify-formula}, to choose the energy scales of the QCD spectral densities in a consistent way. \\
{\bf Scheme II}. We take the $\overline{MS}$
(modified-minimal-subtraction) masses for the heavy quarks $m_Q(m_Q)$, and take the light quark masses and vacuum condensates at the energy scale $\mu=1\,\rm{GeV}$.\\
{\bf Scheme III}. We take all the input parameters at the energy scale $\mu=1\,\rm{GeV}$.

The {\bf Scheme II} is adopted  in most QCD sum rules \cite{Review-QCDSR-Nielsen-PRT-2010,
Review-QCDSR-Nielsen-JPG-2019,X3872-tetra-Narison-PRD-2007,Nielsen-Z4430-PLB-2008,Nielsen-mole-width-PRD-2008,Nielsen-Zcs-mole-Korean-2009,
Nielsen-Zc4430-PLB-2009,Nielsen-Z4050-mole-NPA-2009,Y4660-Nielsen-NPA-2009,
Nielsen-Y4140-mole-PLB-2009,Nielsen-cc-mole-mix-X3872-PRD-2009,JRZhang-mole-PRD-2009,
JRZhang-Y4140-mole-JPG-2010,JRZhang-X4350-mole-CTP-2010,WChen-JPC-0-tetra-PRD-2010,
Nielsen-X3872-mix-PRD-2010,Nielsen-X3872-tetra-mole-PRD-2011,Nielsen-Y-mole-PRD-2011,
Nielsen-Tcc-PLB-2011,Nielsen-Y4140-PLB-2011,JRZhang-Zb10610-mole-PLB-2011,WChen-Vect-axial-tetra-PRD-2011,
YLLiu-Zb10610-PRD-2012,
Nielsen-Zc3900-decay-PRD-2013,WChen-X3872-mix-PRD-2013,JRZhang-Zc3900-mole-PRD-2013,CFQiao-HC-tetra-EPJC-2014,
CFQiao-Zc4025-tetra-EPJC-2014,WChen-Zc-Zb-mole-PRD-2015,Narison-tetra-mole-IJMPA-2016,CFQiao-mole-like-EPJC-2016,
Narison-mole-IJMPA-2018,CFQiao-QQ-tetra-PRD-2020,Narison-QQ-mole-tetra-PRD-2020,CFQiao-Zcs-tetra-NPB-2021,WChen-Zcs-tetra-NPB-2021,CFQiao-QQQQ-EPJC-2021,
JRZhang-cccc-PRD-2021,Narison-mole-PRD-2021} (\cite{Azizi-Zc3900-decay-PRD-2016,Azizi-Zc3900-mag-PRD-2017,Azizi-Zc3900-Zc4430-PRD-2017,
Azizi-Z4100-tetra-EPJC-2019,Azizi-Tccss-tetra-NPB-2019,Azizi-Review-2020,Ozdem-Zcs3985-EPJP-2021,
Ozdem-Zc-PRD-2021,Azizi-Zcs3985-tetra-EPJC-2021,Azizi-Tcc-mole-JHEP-2022,
Azizi-Tcc-tetra-NPB-2022,Mutuk-X3960-mole-EPJC-2022,Mutuk-X2900-mole-JPG-2021}), where the $\overline{MS}$ masses  $m_Q(m_Q)$ are usually smaller than (or equal to) the values from the Particle Data Group (with much smaller pole contributions) \cite{PDG-2024}.

The {\bf Scheme III} was adopted  in early works of Wang and his collaborators in 2009-2011 \cite{WZG-Vect-HQ-JPG-2009,WZG-Scalar-HQ-PRD-2009,WZG-Y4140-EPJC-2009,
WZG-Y4140-EPJC-2009-2,WZG-pi-chi-EPJC-2009,WZG-pi-chi-EPJC-2009-2,
WZG-Scalar-HQ-EPJC-2010,WZG-pseud-Y4660-EPJC-2010,WZG-axial-HQ-EPJC-2010,
WZG-X4350-PLB-2010,WZG-Y4660-CTP-2010,WZG-Y4274-IJMPA-2011}, where many elegant four-quark currents were constructed originally.

In Ref.\cite{WangZG-Scheme-1-2-3-IJMPA-2019}, we study the pentaquark molecular states in the three schemes in details and examine their advantages and shortcomings,
in {\bf Scheme I} and {\bf III},  we truncate the operator product expansion up to the vacuum condensates of $D=13$, while in {\bf Scheme II}, we truncate the operator product expansion up to the vacuum condensates of $D=10$, which is commonly adopted in this case.

We write down  the  correlation functions $\Pi(p)$, $\Pi_{\mu\nu}(p)$ and $\Pi_{\mu\nu\alpha\beta}(p)$ firstly,
\begin{eqnarray}
\Pi(p)&=&i\int d^4x e^{ip \cdot x} \langle0|T\left\{J(x)\bar{J}(0)\right\}|0\rangle \, , \nonumber \\
\Pi_{\mu\nu}(p)&=&i\int d^4x e^{ip \cdot x} \langle0|T\left\{J_{\mu}(x)\bar{J}_{\nu}(0)\right\}|0\rangle \, ,\nonumber \\
\Pi_{\mu\nu\alpha\beta}(p)&=&i\int d^4x e^{ip \cdot x} \langle0|T\left\{J_{\mu\nu}(x)\bar{J}_{\alpha\beta}(0)\right\}|0\rangle \, ,
\end{eqnarray}
where the currents $J(x)=J^{\bar{D}\Sigma_c}(x)$, $J_\mu(x)=J^{\bar{D}\Sigma_c^*}_{\mu}(x)$, $ J^{\bar{D}^*\Sigma_c}_{\mu}(x)$, $J_{\mu\nu}(x)=J^{\bar{D}^*\Sigma_c^*}_{\mu\nu}(x)$,
\begin{eqnarray}
 J^{\bar{D}\Sigma_c}(x)&=& \bar{c}(x)i\gamma_5 u(x)\, \varepsilon^{ijk}  u^T_i(x) C\gamma_\alpha d_j(x)\, \gamma^\alpha\gamma_5 c_{k}(x) \, ,\nonumber \\
 J^{\bar{D}\Sigma_c^*}_{\mu}(x)&=& \bar{c}(x)i\gamma_5 u(x)\, \varepsilon^{ijk}  u^T_i(x) C\gamma_\mu d_j(x)\, c_{k}(x) \, ,\nonumber \\
 J^{\bar{D}^*\Sigma_c}_{\mu}(x)&=& \bar{c}(x)\gamma_\mu u(x)\, \varepsilon^{ijk}  u^T_i(x) C\gamma_\alpha d_j(x)\, \gamma^\alpha\gamma_5 c_{k}(x) \, ,\nonumber \\
 J^{\bar{D}^*\Sigma_c^*}_{\mu\nu}(x)&=& \bar{c}(x)\gamma_\mu u(x)\, \varepsilon^{ijk}  u^T_i(x) C\gamma_\nu d_j(x)\,   c_{k}(x) +(\mu\leftrightarrow\nu)\, .
\end{eqnarray}

We separate the contributions of the molecular states with the $J^P={\frac{1}{2}}^\pm$, ${\frac{3}{2}}^\pm$ and
${\frac{5}{2}}^\pm$ unambiguously,
then we introduce the  weight functions $\sqrt{s}\exp\left(-\frac{s}{T^2}\right)$ and $\exp\left(-\frac{s}{T^2}\right)$ to obtain the QCD sum rules at the  hadron side,
\begin{eqnarray}
\int_{4m_c^2}^{s_0}ds \left[\sqrt{s}\rho^1_{j,H}(s)\pm\rho^0_{j,H}(s)\right]\exp\left( -\frac{s}{T^2}\right)
&=&2M_{\mp}{\lambda^{\mp}_{j}}^2\exp\left( -\frac{M_{\mp}^2}{T^2}\right) \, ,
\end{eqnarray}
where the $s_0$ are the continuum threshold parameters and the $T^2$ are the Borel parameters, the $\rho^1_{j,H}(s)$ and $\rho^0_{j,H}(s)$ with the $j=\frac{1}{2}$, $\frac{3}{2}$ and $\frac{5}{2}$ are hadronic spectral densities, the $\lambda_j^\pm$ are the pole residues.

We perform the operator product expansion to obtain  the analytical QCD spectral densities $\rho^1_{j,QCD}(s)$ and $\rho^0_{j,QCD}(s)$ through dispersion relation,  then we take the
quark-hadron duality below the continuum thresholds  $s_0$ and introduce the weight functions $\sqrt{s}\exp\left(-\frac{s}{T^2}\right)$ and $\exp\left(-\frac{s}{T^2}\right)$ to obtain  the  QCD sum rules:
\begin{eqnarray}\label{QCDN}
2M_{-}{\lambda^{-}_{j}}^2\exp\left( -\frac{M_{-}^2}{T^2}\right)
&=& \int_{4m_c^2}^{s_0}ds \left[\sqrt{s}\rho^1_{j,QCD}(s)+\rho^0_{j,QCD}(s)\right]\exp\left( -\frac{s}{T^2}\right)\, .
\end{eqnarray}
For the technical details, one can consult Ref.\cite{WangZG-Scheme-1-2-3-IJMPA-2019} or Sect.{\bf \ref{333-penta-Sect}}.

We differentiate  Eq.\eqref{QCDN} with respect to  $\tau=\frac{1}{T^2}$, then eliminate the
 pole residues to obtain the  molecule   masses,
 \begin{eqnarray}\label{QCDSR-M}
 M^2_{-} &=& \frac{-\frac{d}{d \tau}\int_{4m_c^2}^{s_0}ds \,\left[\sqrt{s}\,\rho^1_{QCD}(s)+\,\rho^0_{QCD}(s)\right]\exp\left(- \tau s\right)}{\int_{4m_c^2}^{s_0}ds \left[\sqrt{s}\,\rho_{QCD}^1(s)+\,\rho^0_{QCD}(s)\right]\exp\left( -\tau s\right)}\, ,
 \end{eqnarray}
where $\rho_{QCD}^1(s)=\rho_{j,QCD}^1(s)$ and $\rho^0_{QCD}(s)=\rho^0_{j,QCD}(s)$.

\begin{table}
\begin{center}
\begin{tabular}{|c|c|c|c|c|c|c|c|}\hline\hline
                                     &$J^P$                &$D$      &$\mu(\rm GeV)$    &$T^2 (\rm{GeV}^2)$  &$s_0 (\rm{GeV}^2)$   &pole     \\ \hline
$\bar{D}^0\,\Sigma_c^{+}(2455)$      &${\frac{1}{2}}^-$    &13       &2.2               &$3.1-3.5$           &$25.0\pm1.0$         &$(41-62)\%$  \\
                                     &                     &10       &1.0               &$2.7-3.1$           &$24.5\pm1.0$         &$(38-63)\%$     \\
                                     &                     &13       &1.0               &$3.4-4.2$           &$21.5\pm1.0$         &$(7-24)\%$    \\ \hline

$\bar{D}^0\,\Sigma_c^{*+}(2520)$     &${\frac{3}{2}}^-$    &13       &2.4               &$3.3-3.7$           &$25.5\pm1.0$         &$(39-59)\%$  \\
                                     &                     &10       &1.0               &$2.8-3.2$           &$25.5\pm1.0$         &$(40-64)\%$     \\
                                     &                     &13       &1.0               &$3.5-4.3$           &$22.0\pm1.0$         &$(7-23)\%$    \\ \hline

$\bar{D}^{*0}\,\Sigma_c^+(2455)$     &${\frac{3}{2}}^-$    &13       &2.5               &$3.3-3.7$           &$26.5\pm1.0$         &$(40-60)\%$  \\
                                     &                     &10       &1.0               &$2.8-3.2$           &$26.5\pm1.0$         &$(40-65)\%$     \\
                                     &                     &13       &1.0               &$3.7-4.6$           &$23.0\pm1.0$         &$(5-19)\%$    \\  \hline

$\bar{D}^{*0}\,\Sigma_c^{*+}(2520)$  &${\frac{5}{2}}^-$    &13       &2.6               &$3.4-3.8$           &$27.0\pm1.0$         &$(39-59)\%$  \\
                                     &                     &10       &1.0               &$3.0-3.4$           &$27.5\pm1.0$         &$(39-62)\%$     \\
                                     &                     &13       &1.0               &$3.5-4.4$           &$23.5\pm1.0$         &$(7-25)\%$    \\  \hline  \hline
\end{tabular}
\end{center}
\caption{ The truncations of the operator product expansion $D$, energy scales $\mu$, Borel parameters $T^2$, continuum threshold parameters $s_0$ and
 pole contributions for the hidden-charm pentaquark molecular states, where the $\mu=1\,\rm{GeV}$ denote the energy scale of the vacuum condensates, and the data are listed in {\bf Scheme I, II, III} sequentially \cite{WangZG-Scheme-1-2-3-IJMPA-2019}.}\label{Borel-Scheme-1-2-3}
\end{table}

We show the  Borel windows $T^2$, continuum threshold parameters $s_0$, energy scales of the QCD spectral densities and pole contributions of the ground states  explicitly in Table \ref{Borel-Scheme-1-2-3}. In the {\bf Scheme III}, the pole contributions are less than $25\%$, which are too small, and the {\bf Scheme III} could be abandoned. On the other hand,
the convergent behaviors of the operator product expansion have relation  {\bf Scheme I}\,$>$\,{\bf Scheme II}\,$>$\,{\bf Scheme III}.

\begin{figure}
 \centering
  \includegraphics[totalheight=5cm,width=7cm]{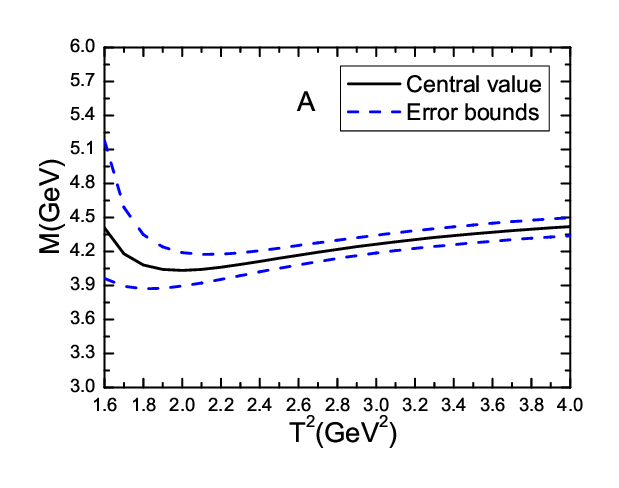}
 \includegraphics[totalheight=5cm,width=7cm]{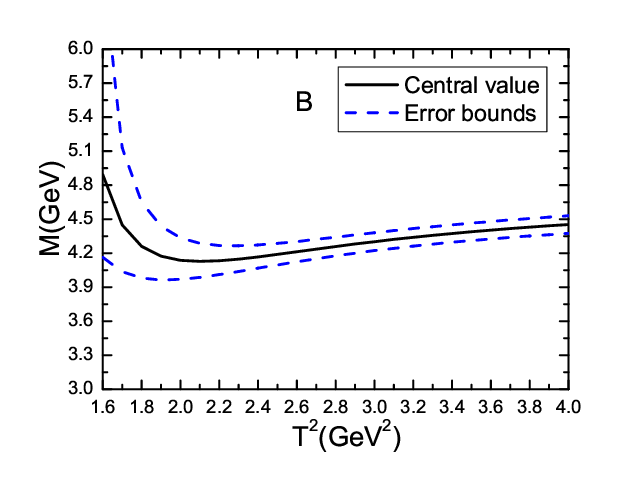}
 \includegraphics[totalheight=5cm,width=7cm]{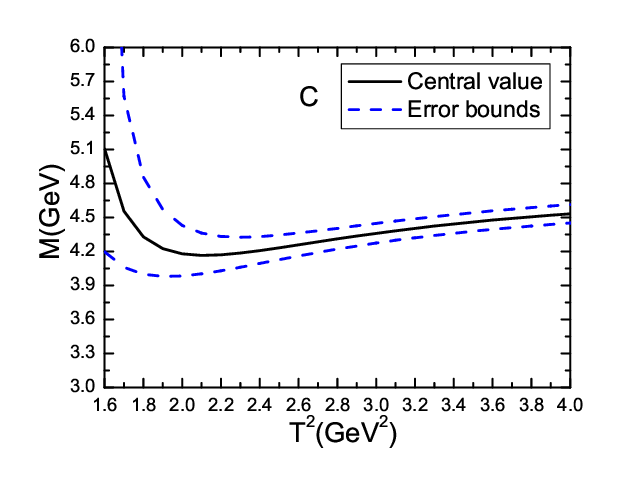}
 \includegraphics[totalheight=5cm,width=7cm]{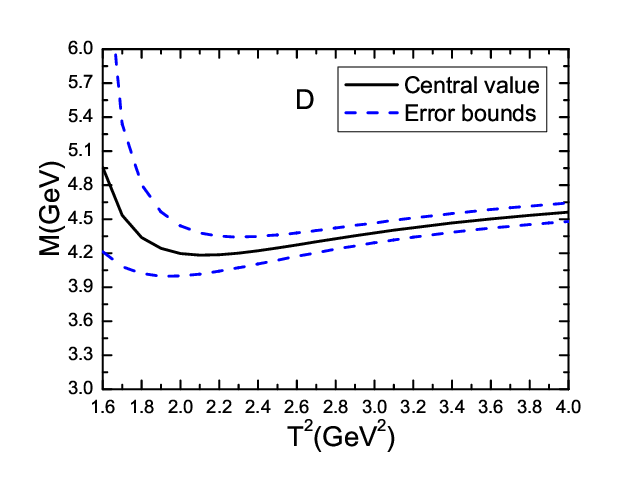}
  \caption{ The masses of the  molecular states  with variations of the Borel parameter $T^2$  in the {\bf Scheme I},  where  the $A$, $B$, $C$ and $D$  denote the  molecular  states  $\bar{D}\Sigma_c$, $\bar{D}\Sigma_c^*$, $\bar{D}^{*}\Sigma_c$ and $ \bar{D}^{*}\Sigma_c^*$, respectively.   }\label{massDSigma}
\end{figure}

\begin{figure}
 \centering
\includegraphics[totalheight=5cm,width=7cm]{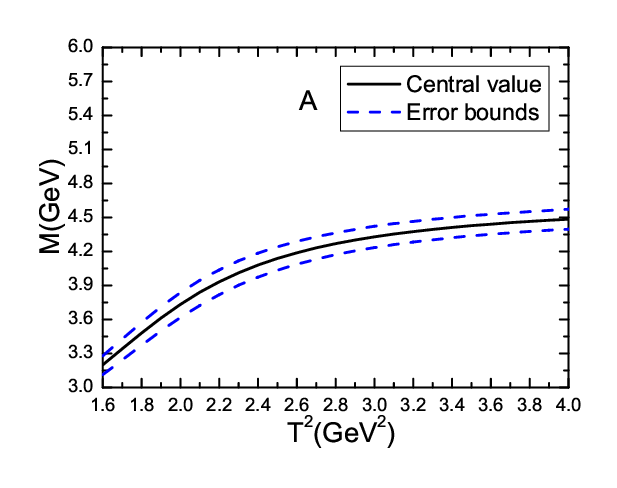}
\includegraphics[totalheight=5cm,width=7cm]{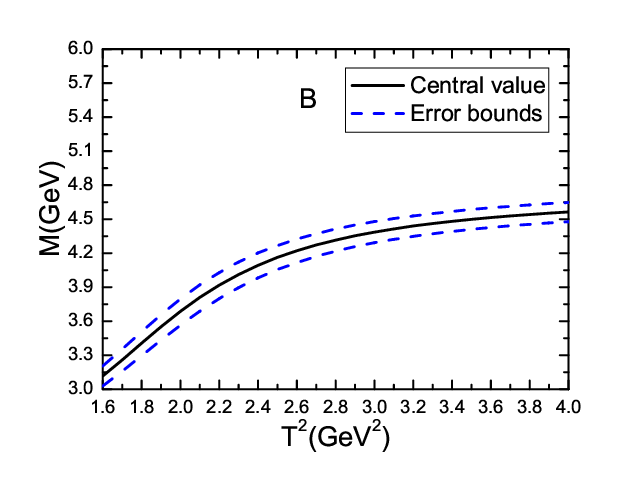}
\includegraphics[totalheight=5cm,width=7cm]{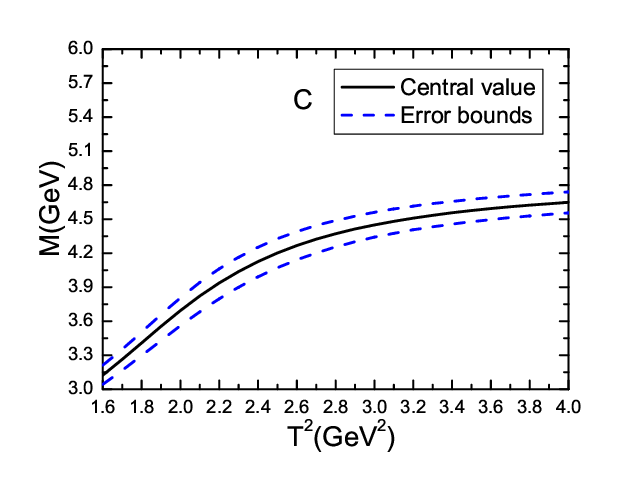}
\includegraphics[totalheight=5cm,width=7cm]{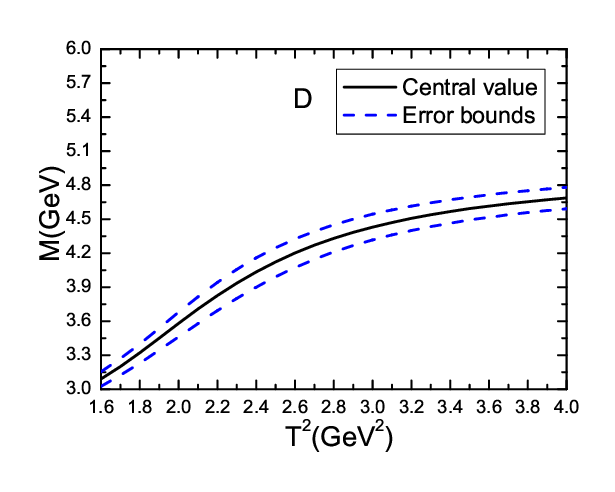}
\caption{ The masses of the molecular states  with variations of the Borel parameter $T^2$  in the {\bf Scheme II},  where  the $A$, $B$, $C$ and $D$  denote the  molecular  states  $\bar{D}\Sigma_c$, $\bar{D}\Sigma_c^*$, $\bar{D}^{*}\Sigma_c$ and $ \bar{D}^{*}\Sigma_c^*$, respectively.   }\label{massDSigmaNiel}
\end{figure}

\begin{table}
\begin{center}
\begin{tabular}{|c|c|c|c|c|c|c|c|}\hline\hline
                                  &$J^P$                &$D$      &$\mu(\rm GeV)$ &$M (\rm{GeV})$          &$\lambda (10^{-3}\rm{GeV}^6)$ & Thresholds (MeV)     \\  \hline
$\bar{D}^0\,\Sigma_c^+(2455)$     &${\frac{1}{2}}^-$    &13       &2.2            &$4.32^{+0.11}_{-0.11}$  &$1.95^{+0.37}_{-0.33}  $      &4318       \\
                                  &                     &10       &1.0            &$4.30^{+0.14}_{-0.17}$  &$1.00^{+0.27}_{-0.24} $       &            \\
                                  &                     &13       &1.0            &$4.30^{+0.09}_{-0.10}$  &$0.77^{+0.19}_{-0.15} $       &           \\ \hline

$\bar{D}^0\,\Sigma_c^+(2520)$     &${\frac{3}{2}}^-$    &13       &2.4            &$4.39^{+0.10}_{-0.11}$  &$1.23^{+0.21}_{-0.20}  $      &4382       \\
                                  &                     &10       &1.0            &$4.39^{+0.14}_{-0.17}$  &$0.64^{+0.16}_{-0.15} $       &            \\
                                  &                     &13       &1.0            &$4.38^{+0.10}_{-0.09}$  &$0.47^{+0.11}_{-0.09} $       &           \\ \hline

$\bar{D}^{*0}\,\Sigma_c^+(2455)$  &${\frac{3}{2}}^-$    &13       &2.5            &$4.46^{+0.11}_{-0.12}$  &$2.31^{+0.41}_{-0.38} $       &4460           \\
                                  &                     &10       &1.0            &$4.45^{+0.16}_{-0.20}$  &$1.15^{+0.31}_{-0.29} $       &         \\
                                  &                     &13       &1.0            &$4.48^{+0.11}_{-0.10}$  &$0.88^{+0.19}_{-0.18} $       &        \\ \hline

$\bar{D}^{*0}\,\Sigma_c^+(2520)$  &${\frac{5}{2}}^-$    &13       &2.6            &$4.50^{+0.12}_{-0.12}$  &$1.74^{+0.31}_{-0.28} $       &4524    \\
                                  &                     &10       &1.0            &$4.51^{+0.16}_{-0.19}$  &$0.93^{+0.25}_{-0.22} $       &        \\
                                  &                     &13       &1.0            &$4.51^{+0.11}_{-0.10}$  &$0.66^{+0.15}_{-0.12} $       &       \\ \hline  \hline
\end{tabular}
\end{center}
\caption{ The  masses and pole residues of the hidden-charm pentaquark molecular states, where the data are listed in {\bf Scheme I, II, III} sequentially \cite{WangZG-Scheme-1-2-3-IJMPA-2019}.}\label{mass-residue-scheme-1-2-3}
\end{table}

At last, we take  account of  all uncertainties  of the input   parameters,
and obtain  the masses and pole residues of  the  molecular states, which are shown explicitly in Table \ref{mass-residue-scheme-1-2-3} and Figs.\ref{massDSigma}-\ref{massDSigmaNiel}.

In Figs.\ref{massDSigma}-\ref{massDSigmaNiel}, we plot the masses  at much larger ranges of the Borel parameters than the Borel windows. The predicted masses  in the  {\bf Scheme I} decrease monotonously and quickly  with
 increase of the Borel parameters at the region $T^2\leq 2.0\,\rm{GeV}^2$, then reach small platforms and increase slowly with increase of the Borel parameters. The predicted masses  in the  {\bf Scheme II} increase monotonously and quickly  with increase of the Borel parameters at the region $T^2<2.6\,\rm{GeV}^2$, then  increase slowly and  steadily with increase of the Borel parameters.
  It is obvious that the flatness of the platforms   have relation  {\bf Scheme I}\,$>$\,{\bf Scheme II}.

 Both the predictions  in the {\bf Scheme I} and {\bf II} support assigning the $P_c(4312)$ as the $\bar{D}\Sigma_c$  molecular state with the $J^P={\frac{1}{2}}^-$, assigning the $P_c(4380)$ as the $\bar{D}\Sigma_c^*$  molecular state with the $J^P={\frac{3}{2}}^-$,  assigning the $P_c(4440/4457)$ as the $\bar{D}^{*}\Sigma_c$  molecular state with the $J^P={\frac{3}{2}}^-$ or the $\bar{D}^{*}\Sigma_c^*$  molecular state with the $J^P={\frac{5}{2}}^-$. However, if we take account of the vacuum condensates of the dimensions of
 $D=11$ and $13$ in the {\bf Scheme II}, we would obtain a mass about $200\,\rm{MeV}$ larger than  the corresponding one given in Table \ref{mass-residue-scheme-1-2-3}, so we prefer the {\bf Scheme I} \cite{WangZG-Scheme-1-2-3-IJMPA-2019}.

\section{$\bar{\mathbf 3}3$ type  tetraquark states}\label{33-4-quark}

\subsection{Hidden-heavy tetraquark states}\label{HH-tetraquark-33}
The scattering amplitude for one-gluon exchange  is proportional to,
\begin{eqnarray}
\left(\frac{\lambda^a}{2}\right)_{ij}\left(\frac{\lambda^a}{2}\right)_{kl}&=&-\frac{1}{3}
\left(\delta_{ij}\delta_{kl}-\delta_{il}\delta_{kj} \right)
+\frac{1}{6}\left(\delta_{ij}\delta_{kl}+\delta_{il}\delta_{kj} \right) \, ,
\end{eqnarray}
the negative (positive) sign in front of the antisymmetric  antitriplet $\bar{\mathbf{3}}$ (symmetric sextet $\mathbf{6}$) indicates the interaction
is attractive (repulsive), which favors (disfavors)  formation of
the diquarks in  color $\bar{\mathbf{3}}$ ($\mathbf{6}$).
We usually construct the $\bar{\mathbf{3}}\mathbf{3}$ type color-singlet four-quark currents $J_{\bar{\mathbf{3}}\mathbf{3}}$ to interpolate the tetraquark states,
\begin{eqnarray}
J_{\bar{\mathbf{3}}\mathbf{3}}&=&\varepsilon^{kij}\varepsilon^{kmn} q^T_i C\Gamma q^\prime_j \,\bar{q}_m \Gamma^\prime C \bar{q}^{\prime T}_n \, ,
\end{eqnarray}
where the $q$ and $q^\prime$ are the quarks, the $\Gamma$ and $\Gamma^\prime$
 are the Dirac $\gamma$-matrixes.
 The color factor has the relation,
 \begin{eqnarray}
 \varepsilon^{kij}\varepsilon^{kmn}&=& \delta_{im}\delta_{jn}-\delta_{in}\delta_{jm}\, .
 \end{eqnarray}
With the simple replacement,
\begin{eqnarray}
\delta_{im}\delta_{jn}-\delta_{in}\delta_{jm} &\to& \delta_{im}\delta_{jn}+\delta_{in}\delta_{jm}\, ,
\end{eqnarray}
we obtain  the corresponding $\mathbf{6}\bar{\mathbf{6}}$ type currents $J_{\mathbf{6}\bar{\mathbf{6}}}$,
\begin{eqnarray}
J_{\mathbf{6}\bar{\mathbf{6}}}&=&q^T_i C\Gamma q^\prime_j \,\bar{q}_i \Gamma^\prime C \bar{q}^{\prime T}_j+q^T_i C\Gamma q^\prime_j \,\bar{q}_j \Gamma^\prime C \bar{q}^{\prime T}_i\, .
\end{eqnarray}
If the $J_{\mathbf{6}\bar{\mathbf{6}}}$ type currents satisfy the Fermi-Dirac statistics, the quantum field theory does not forbid their existence. In fact, in the potential quark models, we usually take both the
$\bar{\mathbf{3}}\mathbf{3}$ and $\mathbf{6}\bar{\mathbf{6}}$ diquark configurations, see Sect.{\bf \ref{33-66-diquark-model}}. The $\mathbf{6}\bar{\mathbf{6}}$ type currents are also widely used in literatures \cite{Review-XYZ-ChenHX-PRT-2016,Review-XYZ-LiuYR-PPNP-2019,
X2900-tetra-mole-ChenHX-CPL-2020,
X5568-ZhuSL-PRL-2016,ChenHX-Z4600-PRD-2019,WChen-JPC-0-tetra-PRD-2010,
ChenHX-Light-Scalar-PRD-2007,ChenHX-Light-66-PRD-2008,
ChenHX-Light-66-PRD-2009,ChenW-Light-66-PRD-2017,
ChenW-Tetra-0-2-PRD-2017,cccc-ChenW-PLB-2017}.

Or we construct the $\mathbf{8}\mathbf{8}$ type currents directly to interpolate the exotic states \cite{CFQiao-QQ-tetra-PRD-2020,CFQiao-QQQQ-EPJC-2021,WangZG-Y2175-NPA-2007,WangZG-Y4274-88-EPJC-2017,
WangZG-Zc4200-tetra-IJMPA-2015,88-cccc-TangL-EPJC-2024}, although the one-gluon exchange induced interaction is repulsive in this channel, see Sect.{\bf{\ref{33-66-diquark-model}}}. In Ref.\cite{WangZG-Zc4200-tetra-IJMPA-2015},  we take  the $Z_c(4200)$ as the   $\mathbf{8}\mathbf{8}$  type axial-vector molecule-like state, and construct the  $\mathbf{8}\mathbf{8}$  type axial-vector
current to study its mass and width with the QCD sum rules,  the numerical results  support assigning  the $Z_c(4200)$  as  the $\mathbf{8}\mathbf{8}$   type  molecule-like  state with the $J^{PC}=1^{+-}$. Furthermore,  we discuss the possible assignments of the $Z_c(3900)$, $Z_c(4200)$ and $Z(4430)$ as the $\bar{\mathbf 3}{\mathbf 3}$ type tetraquark states with the $J^{PC}=1^{+-}$.

\subsubsection{Tetraquark states with positive parity}\label{Tetra-Positive}
The one-gluon-exchange  induced attractive  interactions  favor  formation of
the diquarks in the color antitriplet, flavor
antitriplet  and spin singlet  \cite{Jaffe-diquark-PRT-2005},
 while the most favored configurations are the scalar  and axialvector  diquark states \cite{WangZG-Heavy-diquark-EPJC-2011,TangL-Heavy-diquark-CPC-2012,ZhangAL-Heavy-diquark-PRD-2013,WangZG-Light-diquark-CTP-2013,Light-Diquark-Dosch-ZPC-1989,
Light-Diquark-Jamin-PLB-1990,Light-Diquark-ZhangAL-PRD-2007}.
  The QCD sum rules indicate that the  heavy-light scalar and axialvector  diquark states have almost  degenerate masses \cite{WangZG-Heavy-diquark-EPJC-2011,TangL-Heavy-diquark-CPC-2012,ZhangAL-Heavy-diquark-PRD-2013},  while  the masses of the
 light  axialvector  diquark states lie about  $(150-200)\,\rm{MeV}$ above that of the  light scalar diquark states \cite{WangZG-Light-diquark-CTP-2013,Light-Diquark-Dosch-ZPC-1989,
Light-Diquark-Jamin-PLB-1990,Light-Diquark-ZhangAL-PRD-2007}, if they have the same valence quarks.

The diquarks $\varepsilon^{ijk}q^{T}_j C\Gamma q^{\prime}_k$ in the $\bar{\mathbf{3}}$ have  five  structures  in the Dirac spinor space, where $C\Gamma=C\gamma_5$, $C$, $C\gamma_\mu \gamma_5$,  $C\gamma_\mu $ and $C\sigma_{\mu\nu}$ for the scalar, pseudoscalar, vector, axialvector  and  tensor diquarks, respectively.
In the  non-relativistic quark model, a P-wave changes the parity by contributing a factor $(-)^L=-$ with  the angular momentum $L=1$.
The $C\gamma_5$ and $C\gamma_\mu$ diquark states have the spin-parity $J^P=0^+$ and $1^+$, respectively, the corresponding  $C$ and $C\gamma_\mu\gamma_5$ diquark states have the spin-parity $J^P=0^-$ and $1^-$, respectively, the effects of the  P-waves   are embodied in the underlined  $\gamma_5$ in the  $C\gamma_5 \underline{\gamma_5} $ and $C\gamma_\mu \underline{\gamma_5} $.  The tensor diquark states have both the $J^P=1^+$ and $1^-$ components, we project out the $1^+$ and $1^-$ components explicitly, and introduce the symbols $\widetilde{A}$ and $\widetilde{V}$ to represent them respectively. We would like to give an example on the heavy-light tensor diquarks to
illustrate how to perform the projection.

Under parity transformation  $\widehat{P}$, the tensor diquarks have the  properties,
\begin{eqnarray}
\widehat{P}\varepsilon^{ijk}q^{T}_j(x)C\sigma_{\mu\nu}\gamma_5Q_k(x)\widehat{P}^{-1}
&=&\varepsilon^{ijk}q^{T}_j(\tilde{x})C\sigma^{\mu\nu}\gamma_5Q_k(\tilde{x}) \, , \nonumber\\
\widehat{P}\varepsilon^{ijk}q^{T}_j(x)C\sigma_{\mu\nu} Q_k(x)\widehat{P}^{-1}
&=&-\varepsilon^{ijk}q^{T}_j(\tilde{x})C\sigma^{\mu\nu} Q_k(\tilde{x}) \, ,
\end{eqnarray}
where the four vectors $x^\mu=(t,\vec{x})$ and $\tilde{x}^\mu=(t,-\vec{x})$.
We introduce the four vector $t^\mu=(1,\vec{0})$ and project out the $1^+$ and $1^-$ components explicitly \cite{WangZG-tensor-diquark-CTP-2019},
\begin{eqnarray}
\widehat{P}\varepsilon^{ijk}q^{T}_j(x)C\sigma^t_{\mu\nu}\gamma_5Q_k(x)\widehat{P}^{-1}&=&
+\varepsilon^{ijk}q^{T}_j(\tilde{x})C\sigma^t_{\mu\nu}\gamma_5Q_k(\tilde{x}) \, , \nonumber\\
\widehat{P}\varepsilon^{ijk}q^{T}_j(x)C\sigma^v_{\mu\nu}\gamma_5Q_k(x)\widehat{P}^{-1}&=&
-\varepsilon^{ijk}q^{T}_j(\tilde{x})C\sigma^v_{\mu\nu}\gamma_5Q_k(\tilde{x}) \, ,\nonumber\\
\widehat{P}\varepsilon^{ijk}q^{T}_j(x)C\sigma^v_{\mu\nu} Q_k(x)\widehat{P}^{-1}&=&+\varepsilon^{ijk}q^{T}_j(\tilde{x})C\sigma^v_{\mu\nu} Q_k(\tilde{x}) \, , \nonumber\\
\widehat{P}\varepsilon^{ijk}q^{T}_j(x)C\sigma^t_{\mu\nu} Q_k(x)\widehat{P}^{-1}&=&-\varepsilon^{ijk}q^{T}_j(\tilde{x})C\sigma^t_{\mu\nu} Q_k(\tilde{x}) \, ,
\end{eqnarray}
where $\sigma^t_{\mu\nu} =\frac{i}{2}\Big[\gamma^t_\mu, \gamma^t_\nu \Big]$,
$\sigma^v_{\mu\nu} =\frac{i}{2}\Big[\gamma^v_\mu, \gamma^t_\nu \Big]$,
$\gamma^v_\mu =  \gamma \cdot t t_\mu$,
$\gamma^t_\mu=\gamma_\mu-\gamma \cdot t t_\mu$.
We can also introduce the  P-wave explicitly in the  $C\gamma_5$ and $C\gamma_\mu$ diquarks and obtain  the vector diquarks  $\varepsilon^{ijk}q^{T}_j C\gamma_5 \stackrel{\leftrightarrow}{\partial}_\mu q^{\prime}_k$ and  tensor diquarks  $\varepsilon^{ijk}q^{T}_j C\gamma_\mu \stackrel{\leftrightarrow}{\partial}_\nu q^{\prime}_k$, where the derivative  $\stackrel{\leftrightarrow}{\partial}_\mu=\stackrel{\rightarrow}{\partial}_\mu-\stackrel{\leftarrow}{\partial}_\mu$ embodies  the P-wave effects.

We can also adopt  the covariant derivative  with the simple replacement $\partial_\mu \to D_\mu=\partial_\mu-ig_s G_\mu$, then the four-quark currents are gauge covariant, however, gluonic components are introduced, we have to deal with both valence quarks and gluons.

Now let us construct the $\bar{\mathbf{3}}\mathbf{3}$-type four-quark currents to interpolate the hidden-charm tetraquark states with the $J^{PC}=0^{++}$, $1^{+-}$, $1^{++}$ and $2^{++}$,
\begin{eqnarray}\label{Current-SS}
J_{SS}(x)&=&\varepsilon^{ijk}\varepsilon^{imn}u^{T}_j(x)C\gamma_5 c_k(x)  \bar{d}_m(x)\gamma_5 C \bar{c}^{T}_n(x) \, ,\nonumber \\
J_{AA}(x)&=&\varepsilon^{ijk}\varepsilon^{imn}u^{T}_j(x)C\gamma_\mu c_k(x)  \bar{d}_m(x)\gamma^\mu C \bar{c}^{T}_n(x) \, ,\nonumber \\
J_{\tilde{A}\tilde{A}}(x)&=&\varepsilon^{ijk}\varepsilon^{imn}u^{T}_j(x)C\sigma^v_{\mu\nu} c_k(x)  \bar{d}_m(x)\sigma_v^{\mu\nu} C \bar{c}^{T}_n(x) \, ,\nonumber \\
J_{VV}(x)&=&\varepsilon^{ijk}\varepsilon^{imn}u^{T}_j(x)C\gamma_\mu\gamma_5 c_k(x)  \bar{d}_m(x)\gamma_5\gamma^\mu C \bar{c}^{T}_n(x) \, ,\nonumber \\
J_{\tilde{V}\tilde{V}}(x)&=&\varepsilon^{ijk}\varepsilon^{imn}u^{T}_j(x)C\sigma^t_{\mu\nu} c_k(x)  \bar{d}_m(x)\sigma_t^{\mu\nu} C \bar{c}^{T}_n(x) \, ,\nonumber \\
J_{PP}(x)&=&\varepsilon^{ijk}\varepsilon^{imn}u^{T}_j(x)Cc_k(x)  \bar{d}_m(x) C \bar{c}^{T}_n(x) \, ,
\end{eqnarray}

\begin{eqnarray}\label{Current-SA-negative}
J^{SA}_{-,\mu}(x)&=&\frac{\varepsilon^{ijk}\varepsilon^{imn}}{\sqrt{2}}\Big[u^{T}_j(x)C\gamma_5c_k(x) \bar{d}_m(x)\gamma_\mu C \bar{c}^{T}_n(x)-u^{T}_j(x)C\gamma_\mu c_k(x)\bar{d}_m(x)\gamma_5C \bar{c}^{T}_n(x) \Big] \, ,\nonumber\\
J^{AA}_{-,\mu\nu}(x)&=&\frac{\varepsilon^{ijk}\varepsilon^{imn}}{\sqrt{2}}\Big[u^{T}_j(x) C\gamma_\mu c_k(x) \bar{d}_m(x) \gamma_\nu C \bar{c}^{T}_n(x)  -u^{T}_j(x) C\gamma_\nu c_k(x) \bar{d}_m(x) \gamma_\mu C \bar{c}^{T}_n(x) \Big] \, , \nonumber\\
J^{S\widetilde{A}}_{-,\mu\nu}(x)&=&\frac{\varepsilon^{ijk}\varepsilon^{imn}}{\sqrt{2}}\Big[u^{T}_j(x)C\gamma_5 c_k(x)  \bar{d}_m(x)\sigma_{\mu\nu} C \bar{c}^{T}_n(x)- u^{T}_j(x)C\sigma_{\mu\nu} c_k(x)  \bar{d}_m(x)\gamma_5 C \bar{c}^{T}_n(x) \Big] \, , \nonumber\\
J_{-,\mu}^{\widetilde{A}A}(x)&=&\frac{\varepsilon^{ijk}\varepsilon^{imn}}{\sqrt{2}}\Big[u^{T}_j(x)C\sigma_{\mu\nu}\gamma_5 c_k(x)\bar{d}_m(x)\gamma^\nu C \bar{c}^{T}_n(x)-u^{T}_j(x)C\gamma^\nu c_k(x)\bar{d}_m(x)\gamma_5\sigma_{\mu\nu} C \bar{c}^{T}_n(x) \Big] \, , \nonumber\\
J_{-,\mu}^{\widetilde{V}V}(x)&=&\frac{\varepsilon^{ijk}\varepsilon^{imn}}{\sqrt{2}}\left[u^{T}_j(x)C\sigma_{\mu\nu} c^k(x)\bar{d}_m(x)\gamma_5\gamma^\nu C \bar{c}^{T}_n(x)+u^{T}_j(x)C\gamma^\nu \gamma_5c_k(x)\bar{d}_m(x) \sigma_{\mu\nu} C \bar{c}^{T}_n(x) \right] \, , \nonumber\\
J^{VV}_{-,\mu\nu}(x)&=&\frac{\varepsilon^{ijk}\varepsilon^{imn}}{\sqrt{2}}\Big[u^{T}_j(x) C\gamma_\mu \gamma_5c_k(x) \bar{d}_m(x) \gamma_5\gamma_\nu C \bar{c}^{T}_n(x)  -u^{T}_j(x) C\gamma_\nu\gamma_5 c_k(x) \bar{d}_m(x) \gamma_5\gamma_\mu C \bar{c}^{T}_n(x) \Big] \, , \nonumber\\
J^{PV}_{-,\mu}(x)&=&\frac{\varepsilon^{ijk}\varepsilon^{imn}}{\sqrt{2}}\Big[u^{T}_j(x)Cc_k(x) \bar{d}_m(x)\gamma_5\gamma_\mu C \bar{c}^{T}_n(x)+u^{T}_j(x)C\gamma_\mu \gamma_5c_k(x)\bar{d}_m(x)C \bar{c}^{T}_n(x) \Big] \, ,
\end{eqnarray}

\begin{eqnarray}\label{Current-SA-positive}
J^{SA}_{+,\mu}(x)&=&\frac{\varepsilon^{ijk}\varepsilon^{imn}}{\sqrt{2}}\Big[u^{T}_j(x)C\gamma_5c_k(x) \bar{d}_m(x)\gamma_\mu C \bar{c}^{T}_n(x)+u^{T}_j(x)C\gamma_\mu c_k(x)\bar{d}_m(x)\gamma_5C \bar{c}^{T}_n(x) \Big] \, ,\nonumber\\
J^{S\widetilde{A}}_{+,\mu\nu}(x)&=&\frac{\varepsilon^{ijk}\varepsilon^{imn}}{\sqrt{2}}\Big[u^{T}_j(x)C\gamma_5 c_k(x)  \bar{d}_m(x)\sigma_{\mu\nu} C \bar{c}^{T}_n(x)+ u^{T}_j(x)C\sigma_{\mu\nu} c_k(x)  \bar{d}_m(x)\gamma_5 C \bar{c}^{T}_n(x) \Big] \, , \nonumber\\
J_{+,\mu}^{\widetilde{V}V}(x)&=&\frac{\varepsilon^{ijk}\varepsilon^{imn}}{\sqrt{2}}\left[u^{T}_j(x)C\sigma_{\mu\nu} c_k(x)\bar{d}_m(x)\gamma_5\gamma^\nu C \bar{c}^{T}_n(x)-u^{T}_j(x)C\gamma^\nu \gamma_5c_k(x)\bar{d}_m(x) \sigma_{\mu\nu} C \bar{c}^{T}_n(x) \right] \, , \nonumber\\
J_{+,\mu}^{\widetilde{A}A}(x)&=&\frac{\varepsilon^{ijk}\varepsilon^{imn}}{\sqrt{2}}\left[u^{T}_j(x)C\sigma_{\mu\nu}\gamma_5 c_k(x)\bar{d}_m(x)\gamma^\nu C \bar{c}^{T}_n(x)+u^{T}_j(x)C\gamma^\nu c_k(x)\bar{d}_m(x)\gamma_5\sigma_{\mu\nu} C \bar{c}^{T}_n(x) \right] \, , \nonumber\\
J^{PV}_{+,\mu}(x)&=&\frac{\varepsilon^{ijk}\varepsilon^{imn}}{\sqrt{2}}\Big[u^{T}_j(x)Cc_k(x) \bar{d}_m(x)\gamma_5\gamma_\mu C \bar{c}^{T}_n(x)-u^{T}_j(x)C\gamma_\mu \gamma_5c_k(x)\bar{d}_m(x)C \bar{c}^{T}_n(x) \Big] \, ,\nonumber\\
J^{AA}_{+,\mu\nu}(x)&=&\frac{\varepsilon^{ijk}\varepsilon^{imn}}{\sqrt{2}}\Big[u^{T}_j(x) C\gamma_\mu c_k(x) \bar{d}_m(x) \gamma_\nu C \bar{c}^{T}_n(x)  +u^{T}_j(x) C\gamma_\nu c_k(x) \bar{d}_m(x) \gamma_\mu C \bar{c}^{T}_n(x) \Big] \, , \nonumber\\
J^{VV}_{+,\mu\nu}(x)&=&\frac{\varepsilon^{ijk}\varepsilon^{imn}}{\sqrt{2}}\Big[u^{T}_j(x) C\gamma_\mu \gamma_5c_k(x) \bar{d}_m(x) \gamma_5\gamma_\nu C \bar{c}^{T}_n(x)  +u^{T}_j(x) C\gamma_\nu\gamma_5 c_k(x) \bar{d}_m(x) \gamma_5\gamma_\mu C \bar{c}^{T}_n(x) \Big] \, , \nonumber\\
\end{eqnarray}
where the subscripts $\pm$ denote the positive and negative charge conjugation, respectively, the superscripts or subscripts $P$, $S$, $A$($\widetilde{A}$) and $V$($\widetilde{V}$) denote the pseudoscalar, scalar, axialvector and vector diquark and antidiquark operators, respectively \cite{WZG-HC-PRD-2020}.
With the simple replacements,
\begin{eqnarray}
u &\to& q\, ,\nonumber\\
\bar{d} &\to& \bar{s}\, ,
\end{eqnarray}
where $q=u$, $d$, we obtain the corresponding currents for the $c\bar{c}q\bar{s}$ states \cite{WangZG-Zcs3985-mass-tetra-CPC-2021}.
Again, with the simple replacements,
\begin{eqnarray}
u &\to& s\, ,\nonumber\\
\bar{d} &\to& \bar{s}\, ,
\end{eqnarray}
we obtain the corresponding currents for the $c\bar{c}s\bar{s}$ states \cite{WZG-HC-ss-NPB-2024,WZG-Y4140-Y4274-X4350-IJMPA-2015,WangZG-X4140-decay-EPJC-2019}.

We introduce the symbols,
\begin{eqnarray}
J(x)&=&J_{SS}(x)\, , \,\,J_{AA}(x)\, , \,\,J_{\widetilde{A}\widetilde{A}}(x)\, , \, \,J_{VV}(x)\, ,
\,\, J_{\widetilde{V}\widetilde{V}}(x)\, , \,\, J_{PP}(x)\, , \nonumber\\
 J_\mu(x)&=&J^{SA}_{-,\mu}(x)\, , \, \,J_{-,\mu}^{\widetilde{A}A}(x)\, , \,\, J_{-,\mu}^{\widetilde{V}V}(x)\, , \,\, J^{PV}_{-,\mu}(x)\, , \,\, J^{SA}_{+,\mu}(x)\, , \,\, J_{+,\mu}^{\widetilde{V}V}(x)\, , \,\, J_{+,\mu}^{\widetilde{A}A}(x)\, , \,\, J^{PV}_{+,\mu}(x)\, , \nonumber\\
J_{\mu\nu}(x)&=&J^{AA}_{-,\mu\nu}(x)\, , \,\,J^{S\widetilde{A}}_{-,\mu\nu}(x)\, , \,\,J^{VV}_{-,\mu\nu}(x)\, , \, \,J^{S\widetilde{A}}_{+,\mu\nu}(x)\, , \, \, J^{AA}_{+,\mu\nu}(x)\, , \, \,J^{VV}_{+,\mu\nu}(x)\, ,
 \end{eqnarray}
 for simplicity.

Under parity transformation $\widehat{P}$, the currents have the  properties,
\begin{eqnarray}
\widehat{P} J(x)\widehat{P}^{-1}&=&+J(\tilde{x}) \, , \nonumber\\
\widehat{P} J_\mu(x)\widehat{P}^{-1}&=&-J^\mu(\tilde{x}) \, , \nonumber\\
\widehat{P} J^{S\widetilde{A}}_{\mu\nu}(x)\widehat{P}^{-1}&=&-J_{S\widetilde{A}}^{\mu\nu}(\tilde{x}) \, , \nonumber\\
\widehat{P} J^{AA/VV}_{\mu\nu}(x)\widehat{P}^{-1}&=&+J_{AA/VV}^{\mu\nu}(\tilde{x}) \, ,
\end{eqnarray}
where  $x^\mu=(t,\vec{x})$ and $\tilde{x}^\mu=(t,-\vec{x})$, and we have neglected other superscripts and subscripts.

The currents $J(x)$, $J_\mu(x)$ and $J_{\mu\nu}(x)$ have the  symbolic quark constituent  $\bar{c}c\bar{d}u$ with the isospin $I=1$ and $I_3=1$,
 other currents in the isospin multiplets can be constructed analogously,  for example, we write down the isospin singlet current for  the $J_{SS}(x)$ directly,
\begin{eqnarray}
J^{I=0}_{SS}(x)&=&\frac{\varepsilon^{ijk}\varepsilon^{imn}}{\sqrt{2}}\Big[u^{T}_j(x)C\gamma_5 c_k(x)  \bar{u}_m(x)\gamma_5 C \bar{c}^{T}_n(x)+d^{T}_j(x)C\gamma_5 c_k(x)  \bar{d}_m(x)\gamma_5 C \bar{c}^{T}_n(x)\Big] \, . \nonumber \\
\end{eqnarray}
In the isospin limit, the currents with the  symbolic quark constituents $\bar{c}c\bar{d}u$, $\bar{c}c\bar{u}d$, $\bar{c}c\frac{\bar{u}u-\bar{d}d}{\sqrt{2}}$, $\bar{c}c\frac{\bar{u}u+\bar{d}d}{\sqrt{2}}$ couple potentially  to the hidden-charm
tetraquark states with degenerated  masses, the currents with the isospins $I=1$ and $0$ lead to the same QCD sum rules.  Thereafter, we will denote the $Z_c$ states as the isospin triplet, and the $X$ states as the isospin singlet,
\begin{eqnarray}
Z_c&:&\bar{c}c\bar{d}u\, ,\, \bar{c}c\bar{u}d\, ,\, \bar{c}c\frac{\bar{u}u-\bar{d}d}{\sqrt{2}}\, , \nonumber\\
X&:& \bar{c}c\frac{\bar{u}u+\bar{d}d}{\sqrt{2}}\, .
\end{eqnarray}
Accordingly, the $Z_{cs}$ states with the symbolic quark constituents $\bar{c}c\bar{s}q$ and isospin $I=\frac{1}{2}$  have degenerated  masses.

The currents with the symbolic quark constituents  $\bar{c}c\frac{\bar{u}u-\bar{d}d}{\sqrt{2}}$ and $\bar{c}c\frac{\bar{u}u+\bar{d}d}{\sqrt{2}}$ have definite
charge conjugation. We would like to assume  that the $\bar{c}c\bar{d}u$ type tetraquark states have the same charge conjugation as their  charge-neutral cousins.

Under charge conjugation transformation  $\widehat{C}$, the currents $J(x)$, $J_\mu(x)$ and $J_{\mu\nu}(x)$ have the properties,
\begin{eqnarray}
\widehat{C}J(x)\widehat{C}^{-1}&=&+ J(x)\mid_{u\leftrightarrow d} \, , \nonumber\\
\widehat{C}J_{\pm,\mu}(x)\widehat{C}^{-1}&=&\pm J_{\pm,\mu}(x)\mid_{u\leftrightarrow d}  \, , \nonumber\\
\widehat{C}J_{\pm,\mu\nu}(x)\widehat{C}^{-1}&=&\pm J_{\pm,\mu\nu}(x)\mid_{u\leftrightarrow d}  \, ,
\end{eqnarray}
where we have neglected other superscripts and subscripts.

Now we write down  the correlation functions $\Pi(p)$, $\Pi_{\mu\nu}(p)$ and $\Pi_{\mu\nu\alpha\beta}(p)$,
\begin{eqnarray}\label{CF-Pi}
\Pi(p)&=&i\int d^4x e^{ip \cdot x} \langle0|T\Big\{J(x)J^{\dagger}(0)\Big\}|0\rangle \, ,\nonumber\\
\Pi_{\mu\nu}(p)&=&i\int d^4x e^{ip \cdot x} \langle0|T\Big\{J_\mu(x)J_{\nu}^{\dagger}(0)\Big\}|0\rangle \, ,\nonumber\\
\Pi_{\mu\nu\alpha\beta}(p)&=&i\int d^4x e^{ip \cdot x} \langle0|T\Big\{J_{\mu\nu}(x)J_{\alpha\beta}^{\dagger}(0)\Big\}|0\rangle \, .
\end{eqnarray}

At the hadron side, we  insert  a complete set of intermediate hadronic states with
the same quantum numbers as the currents $J(x)$, $J_\mu(x)$ and $J_{\mu\nu}(x)$ into the
correlation functions $\Pi(p)$, $\Pi_{\mu\nu}(p)$ and $\Pi_{\mu\nu\alpha\beta}(p)$   to obtain the hadronic representation
\cite{SVZ-NPB-1979-1,SVZ-NPB-1979-2}, and isolate the ground state hidden-charm tetraquark contributions,
\begin{eqnarray}
\Pi(p)&=&\frac{\lambda_{Z^+}^2}{M_{Z^+}^2-p^2} +\cdots =\Pi_{+}(p^2) \, ,\nonumber\\
\Pi_{\mu\nu}(p)&=&\frac{\lambda_{Z^+}^2}{M_{Z^+}^2-p^2}\left( -g_{\mu\nu}+\frac{p_{\mu}p_{\nu}}{p^2}\right) +\cdots \nonumber \\
&=&\Pi_{+}(p^2)\left( -g_{\mu\nu}+\frac{p_{\mu}p_{\nu}}{p^2}\right)+\cdots \, ,\nonumber\\
\Pi^{AA,-}_{\mu\nu\alpha\beta}(p)&=&\frac{\tilde{\lambda}_{ Z^+}^2}{M_{Z^+}^2-p^2}\left(p^2g_{\mu\alpha}g_{\nu\beta} -p^2g_{\mu\beta}g_{\nu\alpha} -g_{\mu\alpha}p_{\nu}p_{\beta}-g_{\nu\beta}p_{\mu}p_{\alpha}+g_{\mu\beta}p_{\nu}p_{\alpha}+g_{\nu\alpha}p_{\mu}p_{\beta}\right) \nonumber\\
&&+\frac{\tilde{\lambda}_{ Z^-}^2}{M_{Z^-}^2-p^2}\left( -g_{\mu\alpha}p_{\nu}p_{\beta}-g_{\nu\beta}p_{\mu}p_{\alpha}+g_{\mu\beta}p_{\nu}p_{\alpha}+g_{\nu\alpha}p_{\mu}p_{\beta}\right) +\cdots  \nonumber\\
&=&\widetilde{\Pi}_{+}(p^2)\left(p^2g_{\mu\alpha}g_{\nu\beta} -p^2g_{\mu\beta}g_{\nu\alpha} -g_{\mu\alpha}p_{\nu}p_{\beta}-g_{\nu\beta}p_{\mu}p_{\alpha}+g_{\mu\beta}p_{\nu}p_{\alpha}+g_{\nu\alpha}p_{\mu}p_{\beta}\right) \nonumber\\
&&+\widetilde{\Pi}_{-}(p^2)\left( -g_{\mu\alpha}p_{\nu}p_{\beta}-g_{\nu\beta}p_{\mu}p_{\alpha}+g_{\mu\beta}p_{\nu}p_{\alpha}+g_{\nu\alpha}p_{\mu}p_{\beta}\right) \, ,\nonumber
\end{eqnarray}
\begin{eqnarray}
\Pi^{S\widetilde{A},\pm}_{\mu\nu\alpha\beta}(p)&=&\frac{\tilde{\lambda}_{ Z^-}^2}{M_{Z^-}^2-p^2}\left(p^2g_{\mu\alpha}g_{\nu\beta} -p^2g_{\mu\beta}g_{\nu\alpha} -g_{\mu\alpha}p_{\nu}p_{\beta}-g_{\nu\beta}p_{\mu}p_{\alpha}+g_{\mu\beta}p_{\nu}p_{\alpha}+g_{\nu\alpha}p_{\mu}p_{\beta}\right) \nonumber\\
&&+\frac{\tilde{\lambda}_{ Z^+}^2}{M_{Z^+}^2-p^2}\left( -g_{\mu\alpha}p_{\nu}p_{\beta}-g_{\nu\beta}p_{\mu}p_{\alpha}+g_{\mu\beta}p_{\nu}p_{\alpha}+g_{\nu\alpha}p_{\mu}p_{\beta}\right) +\cdots  \nonumber\\
&=&\widetilde{\Pi}_{-}(p^2)\left(p^2g_{\mu\alpha}g_{\nu\beta} -p^2g_{\mu\beta}g_{\nu\alpha} -g_{\mu\alpha}p_{\nu}p_{\beta}-g_{\nu\beta}p_{\mu}p_{\alpha}+g_{\mu\beta}p_{\nu}p_{\alpha}+g_{\nu\alpha}p_{\mu}p_{\beta}\right) \nonumber\\
&&+\widetilde{\Pi}_{+}(p^2)\left( -g_{\mu\alpha}p_{\nu}p_{\beta}-g_{\nu\beta}p_{\mu}p_{\alpha}+g_{\mu\beta}p_{\nu}p_{\alpha}+g_{\nu\alpha}p_{\mu}p_{\beta}\right) \, , \nonumber\\
\Pi_{\mu\nu\alpha\beta}^{AA,+}(p)&=&\frac{\lambda_{ Z^+}^2}{M_{Z^+}^2-p^2}\left( \frac{\widetilde{g}_{\mu\alpha}\widetilde{g}_{\nu\beta}+\widetilde{g}_{\mu\beta}\widetilde{g}_{\nu\alpha}}{2}-\frac{\widetilde{g}_{\mu\nu}\widetilde{g}_{\alpha\beta}}{3}\right) +\cdots \, \, , \nonumber \\
&=&\Pi_{+}(p^2)\left( \frac{\widetilde{g}_{\mu\alpha}\widetilde{g}_{\nu\beta}+\widetilde{g}_{\mu\beta}\widetilde{g}_{\nu\alpha}}{2}-\frac{\widetilde{g}_{\mu\nu}\widetilde{g}_{\alpha\beta}}{3}\right) +\cdots\, ,
\end{eqnarray}
where $\widetilde{g}_{\mu\nu}=g_{\mu\nu}-\frac{p_{\mu}p_{\nu}}{p^2}$.  We add the superscripts $\pm$ in the $\Pi^{AA,-}_{\mu\nu\alpha\beta}(p)$, $\Pi^{S\widetilde{A},\pm}_{\mu\nu\alpha\beta}(p)$ and $\Pi_{\mu\nu\alpha\beta}^{AA,+}(p)$ to denote the positive and negative charge conjugation, respectively,
 and add the superscripts (subscripts) $\pm$ in the $Z_c^{\pm}$ (the $\Pi_{\pm}(p^2)$ and $\widetilde{\Pi}_{\pm}(p^2)$ components) to denote the positive  and negative parity, respectively. And the $\Pi^{VV,\pm}_{\mu\nu\alpha\beta}(p)$ and $\Pi^{AA,\pm}_{\mu\nu\alpha\beta}(p)$ have the same tensor structures.  The  pole residues $\lambda_{Z^\pm}$ are defined by
\begin{eqnarray}\label{define-pole-residue}
 \langle 0|J(0)|Z_c^+(p)\rangle &=&\lambda_{Z^+}\, , \nonumber\\
 \langle 0|J_\mu(0)|Z_c^+(p)\rangle &=&\lambda_{Z^+}\varepsilon_\mu\, , \nonumber\\
  \langle 0|J_{\pm,\mu\nu}^{S\widetilde{A}}(0)|Z_c^-(p)\rangle &=& \tilde{\lambda}_{Z^-} \, \varepsilon_{\mu\nu\alpha\beta} \, \varepsilon^{\alpha}p^{\beta}\, , \nonumber\\
 \langle 0|J_{\pm,\mu\nu}^{S\widetilde{A}}(0)|Z_c^+(p)\rangle &=&\tilde{\lambda}_{Z^+} \left(\varepsilon_{\mu}p_{\nu}-\varepsilon_{\nu}p_{\mu} \right)\, , \nonumber\\
  \langle 0|J_{-,\mu\nu}^{AA/VV}(0)|Z_c^+(p)\rangle &=& \tilde{\lambda}_{Z^+} \, \varepsilon_{\mu\nu\alpha\beta} \, \varepsilon^{\alpha}p^{\beta}\, , \nonumber\\
 \langle 0|J_{-,\mu\nu}^{AA/VV}(0)|Z_c^-(p)\rangle &=&\tilde{\lambda}_{Z^-} \left(\varepsilon_{\mu}p_{\nu}-\varepsilon_{\nu}p_{\mu} \right)\, , \nonumber\\
  \langle 0|J_{+,\mu\nu}^{AA/VV}(0)|Z_c^+(p)\rangle &=& \lambda_{Z^+}\, \varepsilon_{\mu\nu} \, ,
\end{eqnarray}
where $\lambda_{Z^\pm}=\tilde{\lambda}_{Z^\pm}M_{Z^\pm}$,
the  $\varepsilon_{\mu/\alpha}$ and $\varepsilon_{\mu\nu}$ are the polarization vectors.
We choose the components $\Pi_{+}(p^2)$ and $p^2\widetilde{\Pi}_{+}(p^2)$ to study the   hidden-charm tetraquark states.

In Table \ref{Current-Table}, we present the quark constituents and corresponding currents explicitly.
\begin{table}
\begin{center}
\begin{tabular}{|c|c|c|c|c|c|c|c|c|}\hline\hline
 $Z_c$                                                                            & $J^{PC}$  & Currents              \\ \hline

$[uc]_{S}[\overline{dc}]_{S}$                                                     & $0^{++}$  & $J_{SS}(x)$              \\

$[uc]_{A}[\overline{dc}]_{A}$                                                     & $0^{++}$  & $J_{AA}(x)$               \\

$[uc]_{\tilde{A}}[\overline{dc}]_{\tilde{A}}$                                     & $0^{++}$  & $J_{\widetilde{A}\widetilde{A}}(x)$             \\

$[uc]_{V}[\overline{dc}]_{V}$                                                     & $0^{++}$  & $J_{VV}(x)$               \\

$[uc]_{\tilde{V}}[\overline{dc}]_{\tilde{V}}$                                     & $0^{++}$  & $J_{\widetilde{V}\widetilde{V}}(x)$           \\

$[uc]_{P}[\overline{dc}]_{P}$                                                     & $0^{++}$  & $J_{PP}(x)$              \\ \hline

$[uc]_S[\overline{dc}]_{A}-[uc]_{A}[\overline{dc}]_S$                             & $1^{+-}$  & $J^{SA}_{-,\mu}(x)$         \\

$[uc]_{A}[\overline{dc}]_{A}$                                                     & $1^{+-}$  & $J^{AA}_{-,\mu\nu}(x)$        \\

$[uc]_S[\overline{dc}]_{\widetilde{A}}-[uc]_{\widetilde{A}}[\overline{dc}]_S$     & $1^{+-}$  & $J^{S\widetilde{A}}_{-,\mu\nu}(x)$     \\

$[uc]_{\widetilde{A}}[\overline{dc}]_{A}-[uc]_{A}[\overline{dc}]_{\widetilde{A}}$ & $1^{+-}$  & $J_{-,\mu}^{\widetilde{A}A}(x)$   \\

$[uc]_{\widetilde{V}}[\overline{dc}]_{V}+[uc]_{V}[\overline{dc}]_{\widetilde{V}}$ & $1^{+-}$  & $J_{-,\mu}^{\widetilde{V}V}(x)$      \\

$[uc]_{V}[\overline{dc}]_{V}$                                                     & $1^{+-}$  & $J^{VV}_{-,\mu\nu}(x)$        \\

$[uc]_P[\overline{dc}]_{V}+[uc]_{V}[\overline{dc}]_P$                             & $1^{+-}$  & $J^{PV}_{-,\mu}(x)$         \\
\hline

$[uc]_S[\overline{dc}]_{A}+[uc]_{A}[\overline{dc}]_S$                             & $1^{++}$  & $J^{SA}_{+,\mu}(x)$        \\

$[uc]_S[\overline{dc}]_{\widetilde{A}}+[uc]_{\widetilde{A}}[\overline{dc}]_S$     & $1^{++}$  & $J^{S\widetilde{A}}_{+,\mu\nu}(x)$     \\

$[uc]_{\widetilde{V}}[\overline{dc}]_{V}-[uc]_{V}[\overline{dc}]_{\widetilde{V}}$ & $1^{++}$  & $J_{+,\mu}^{\widetilde{V}V}(x)$      \\

$[uc]_{\widetilde{A}}[\overline{dc}]_{A}+[uc]_{A}[\overline{dc}]_{\widetilde{A}}$ & $1^{++}$  & $J_{+,\mu}^{\widetilde{A}A}(x)$       \\
$[uc]_P[\overline{dc}]_{V}-[uc]_{V}[\overline{dc}]_P$                             & $1^{++}$  & $J^{PV}_{+,\mu}(x)$         \\ \hline

$[uc]_{A}[\overline{dc}]_{A}$                                                     & $2^{++}$  & $J^{AA}_{+,\mu\nu}(x)$       \\

$[uc]_{V}[\overline{dc}]_{V}$                                                     & $2^{++}$  & $J^{VV}_{+,\mu\nu}(x)$        \\
\hline\hline
\end{tabular}
\end{center}
\caption{ The quark constituents and corresponding currents  for the hidden-charm tetraquark states \cite{WZG-HC-PRD-2020}. }\label{Current-Table}
\end{table}

At the QCD side, we carry out the operator product expansion for the correlation functions $\Pi(p)$, $\Pi_{\mu\nu}(p)$ and $\Pi_{\mu\nu\alpha\beta}(p)$.
For example, we contract the quark fields in the $\Pi_{\mu\nu}(p)$ with the Wick's theorem, and obtain the results,
\begin{eqnarray}\label{Wick-CF}
\Pi_{\mu\nu}(p)&=&-\frac{i\varepsilon^{ijk}\varepsilon^{imn}\varepsilon^{i^{\prime}j^{\prime}k^{\prime}}\varepsilon^{i^{\prime}m^{\prime}n^{\prime}}}{2}\int d^4x e^{ip \cdot x}   \nonumber\\
&&\left\{{\rm Tr}\left[ \gamma_5S_c^{kk^{\prime}}(x)\gamma_5 CU^{T}_{jj^{\prime}}(x)C\right] {\rm Tr}\left[ \gamma_\nu S_c^{n^{\prime}n}(-x)\gamma_\mu C D^{T}_{m^{\prime}m}(-x)C\right] \right. \nonumber\\
&&+{\rm Tr}\left[ \gamma_\mu S_c^{kk^{\prime}}(x)\gamma_\nu CU^{T}_{jj^{\prime}}(x)C\right] {\rm Tr}\left[ \gamma_5 S_c^{n^{\prime}n}(-x)\gamma_5 C D^{T}_{m^{\prime}m}(-x)C\right] \nonumber\\
&&\mp{\rm Tr}\left[ \gamma_\mu S_c^{kk^{\prime}}(x)\gamma_5 CU^{T}_{jj^{\prime}}(x)C\right] {\rm Tr}\left[ \gamma_\nu S_c^{n^{\prime}n}(-x)\gamma_5 C D^{T}_{m^{\prime}m}(-x)C\right] \nonumber\\
 &&\left.\mp{\rm Tr}\left[ \gamma_5 S_c^{kk^{\prime}}(x)\gamma_\nu CU^{T}_{jj^{\prime}}(x)C\right] {\rm Tr}\left[ \gamma_5 S_c^{n^{\prime}n}(-x)\gamma_\mu C D^{T}_{m^{\prime}m}(-x)C\right] \right\} \, ,
\end{eqnarray}
where the full quark propagators, $S^{ij}(x)=U^{ij}(x)=D^{ij}(x)$,
\begin{eqnarray}\label{S-progator}
S^{ij}(x)&=& \frac{i\delta_{ij}\!\not\!{x}}{ 2\pi^2x^4}
-\frac{\delta_{ij}m_q}{4\pi^2x^2}-\frac{\delta_{ij}\langle
\bar{q}q\rangle}{12} +\frac{i\delta_{ij}\!\not\!{x}m_q
\langle\bar{q}q\rangle}{48}-\frac{\delta_{ij}x^2\langle \bar{q}g_s\sigma Gq\rangle}{192}+\frac{i\delta_{ij}x^2\!\not\!{x} m_q\langle \bar{q}g_s\sigma
 Gq\rangle }{1152}\nonumber\\
&& -\frac{ig_s G^{a}_{\alpha\beta}t^a_{ij}(\!\not\!{x}
\sigma^{\alpha\beta}+\sigma^{\alpha\beta} \!\not\!{x})}{32\pi^2x^2} -\frac{i\delta_{ij}x^2\!\not\!{x}g_s^2\langle \bar{q} q\rangle^2}{7776} -\frac{\delta_{ij}x^4\langle \bar{q}q \rangle\langle g_s^2 GG\rangle}{27648}-\frac{1}{8}\langle\bar{q}_j\sigma^{\mu\nu}q_i \rangle \sigma_{\mu\nu} \nonumber\\
&&   -\frac{1}{4}\langle\bar{q}_j\gamma^{\mu}q_i\rangle \gamma_{\mu }+\cdots \, ,
\end{eqnarray}
and $S_c^{ij}(x)=S_Q^{ij}(x)$,
\begin{eqnarray}\label{Q-progator}
S_Q^{ij}(x)&=&\frac{i}{(2\pi)^4}\int d^4k e^{-ik \cdot x} \left\{
\frac{\delta_{ij}}{\!\not\!{k}-m_Q}
-\frac{g_sG^n_{\alpha\beta}t^n_{ij}}{4}\frac{\sigma^{\alpha\beta}(\!\not\!{k}+m_Q)+(\!\not\!{k}+m_Q)
\sigma^{\alpha\beta}}{(k^2-m_Q^2)^2}\right.\nonumber\\
&&\left. +\frac{g_s D_\alpha G^n_{\beta\lambda}t^n_{ij}(f^{\lambda\beta\alpha}+f^{\lambda\alpha\beta}) }{3(k^2-m_Q^2)^4}-\frac{g_s^2 (t^at^b)_{ij} G^a_{\alpha\beta}G^b_{\mu\nu}(f^{\alpha\beta\mu\nu}+f^{\alpha\mu\beta\nu}+f^{\alpha\mu\nu\beta}) }{4(k^2-m_Q^2)^5}+\cdots\right\} \, ,\nonumber\\
f^{\lambda\alpha\beta}&=&(\!\not\!{k}+m_Q)\gamma^\lambda(\!\not\!{k}+m_Q)\gamma^\alpha(\!\not\!{k}+m_Q)\gamma^\beta(\!\not\!{k}+m_Q)\, ,\nonumber\\
f^{\alpha\beta\mu\nu}&=&(\!\not\!{k}+m_Q)\gamma^\alpha(\!\not\!{k}+m_Q)\gamma^\beta(\!\not\!{k}+m_Q)\gamma^\mu(\!\not\!{k}+m_Q)\gamma^\nu(\!\not\!{k}+m_Q)\, ,
\end{eqnarray}
with $Q=c$, $D_\alpha=\partial_\alpha-ig_sG^n_\alpha t^n$ \cite{X3872-tetra-WangZG-HuangT-PRD-2014,Reinders85,Pascual-1984}, and the $\mp$ correspond to  the currents $J^{SA}_{+,\mu}(x)$ and $J^{SA}_{-,\mu}(x)$, respectively.
In Eq.\eqref{S-progator}, we retain the terms $\langle\bar{q}_j\sigma_{\mu\nu}q_i \rangle$ and $\langle\bar{q}_j\gamma_{\mu}q_i\rangle$ come from the Fierz transformation  of the $\langle q_i \bar{q}_j\rangle$ to  absorb the gluons  emitted from the other quark lines to form $\langle\bar{q}_j g_s G^a_{\alpha\beta} t^a_{mn}\sigma_{\mu\nu} q_i \rangle$ and $\langle\bar{q}_j\gamma_{\mu}q_i D_\nu g_s G^a_{\alpha\beta}t^a_{mn}\rangle$  to extract the mixed condensate $\langle\bar{q}g_s\sigma G q\rangle$ and four-quark condensate $g_s^2\langle\bar{q}q\rangle^2$, respectively \cite{X3872-tetra-WangZG-HuangT-PRD-2014}, where $q=u$, $d$ or $s$.
The condensate $g_s^2\langle \bar{q}q\rangle^2$ comes from the
$\langle \bar{q}\gamma_\mu t^a q g_s D_\eta G^a_{\lambda\tau}\rangle$, $\langle\bar{q}_jD^{\dagger}_{\mu}D^{\dagger}_{\nu}D^{\dagger}_{\alpha}q_i\rangle$  and
$\langle\bar{q}_jD_{\mu}D_{\nu}D_{\alpha}q_i\rangle$  rather than comes from the radiative  $\mathcal{O}(\alpha_s)$ corrections to the $\langle \bar{q}q\rangle^2$.

In fact, the method of adopting the full propagators in Eq.\eqref{Wick-CF} implies vacuum saturation implicitly, the factorization of the higher dimensional vacuum condensates, for example,
 see Eq.\eqref{Four-quark-fact}, is already performed. We  calculate the higher  dimensional vacuum condensates using the formula $t^a_{ij}t^a_{mn}=-\frac{1}{6}\delta_{ij}\delta_{mn}+\frac{1}{2}\delta_{jm}\delta_{in}$ rigorously.

Let us see Eq.\eqref{Wick-CF} again,  there are two $Q$-quark propagators and two $q$-quark propagators, if each $Q$-quark line emits a gluon and each $q$-quark line contributes  a quark-antiquark  pair, we obtain an  operator $G_{\mu\nu}G_{\alpha\beta}\bar{u}u\bar{d}d$ (or $G_{\mu\nu}G_{\alpha\beta}\bar{q}q\bar{s}s$ or $G_{\mu\nu}G_{\alpha\beta}\bar{s}s\bar{s}s$), which is of dimension 10, see the Feynman diagram  in Fig.\ref{meson-qqg-qqg} for example.  We should take account of the vacuum condensates at least up to dimension $10$.  The higher dimensional vacuum condensates are  associated with the $\frac{1}{T^2}$, $\frac{1}{T^4}$ or $\frac{1}{T^6}$, which manifest themselves at small Borel parameter $T^2$ and play an important role in determining the Borel windows, where they play a minor important role. Therefore,
 we take account of the vacuum condensates $\langle\bar{q}q\rangle$, $\langle\frac{\alpha_{s}GG}{\pi}\rangle$, $\langle\bar{q}g_{s}\sigma Gq\rangle$, $\langle\bar{q}q\rangle^2$, $g_s^2\langle\bar{q}q\rangle^2$,
$\langle\bar{q}q\rangle \langle\frac{\alpha_{s}GG}{\pi}\rangle$,  $\langle\bar{q}q\rangle  \langle\bar{q}g_{s}\sigma Gq\rangle$,
$\langle\bar{q}g_{s}\sigma Gq\rangle^2$ and $\langle\bar{q}q\rangle^2 \langle\frac{\alpha_{s}GG}{\pi}\rangle$, which are vacuum expectations of the quark-gluon operators of the order $\mathcal{O}(\alpha_s^k)$ with $k\leq1$.
We  truncate the operator product expansion in such a way consistently and rigorously. The condensates $\langle g_s^3 GGG\rangle$, $\langle \frac{\alpha_s GG}{\pi}\rangle^2$,
 $\langle \frac{\alpha_s GG}{\pi}\rangle\langle \bar{q} g_s \sigma Gq\rangle$ have the dimensions 6, 8, 9 respectively,  but they are  vacuum expectations
of the quark-gluon operators of the order    $\mathcal{O}( \alpha_s^{3/2})$, $\mathcal{O}(\alpha_s^2)$, $\mathcal{O}( \alpha_s^{3/2})$ respectively, and discarded. Furthermore, direct calculations indicate such contributions are tiny indeed \cite{WangXW-LamLam-EPJA-2021}.

We accomplish the integrals in Eqs.\eqref{Wick-CF}-\eqref{Q-progator} sequentially by simply setting $d^4xd^4kd^4q\to d^Dxd^Dkd^Dq$ with $D=4-2\epsilon$ to regularize the divergences, such a simple scheme misses  many subtraction terms, which make no contribution  in obtaining the imaginary parts through $p^2\to p^2+i\epsilon$.
Finally, we obtain the QCD spectral densities $\rho_{QCD}(s)$ through dispersion relation. Now we take a short digression to give an example,
\begin{eqnarray}
{\rm Int}(p^2)&=&\int^{+\infty}_{-\infty} d^Dxd^Dkd^Dq \frac{\exp\left[i (p+q+k)\cdot x \right]}{(q^2-m_Q^2)^\lambda(k^2-m_Q^2)^\tau x^{2n} } \nonumber\\
&=&-i\frac{2^{4-2n}\pi^2}{\Gamma(n)}\int^{+\infty}_{-\infty} d^Dkd^Dq \frac{\Gamma(\alpha)}{(q^2-m_Q^2)^\lambda(k^2-m_Q^2)^\tau (p+q+k)^{2\alpha} } \nonumber\\
&=&-i\frac{2^{4-2n}\pi^2}{\Gamma(n)} \frac{i\pi^2}{\Gamma(\tau)}\int_{0}^{1}dx  \int^{+\infty}_{-\infty} d^Dq \frac{x^{\alpha-1}(1-x)^{\tau-1} \Gamma(\alpha+\tau-\frac{D}{2})}{(q^2-m_Q^2)^\lambda (x(1-x)(p+q)^2-(1-x)m_Q^2)^{\alpha+\tau-\frac{D}{2}} } \nonumber\\
&=&\frac{i\,2^{4-2n}\pi^6}{\Gamma(\lambda)\Gamma(\tau)\Gamma(n)} \int_{0}^{1}dx \frac{x^{\alpha-1}(1-x)^{\tau-1}}{[x(1-x)]^{\alpha+\tau-\frac{D}{2}}} \int^{1}_{0} dy \frac{y^{\alpha+\tau-\frac{D}{2}-1}(1-y)^{\lambda-1} }{ [y(1-y)]^{\alpha+\lambda+\tau-D} } \frac{\Gamma(\alpha+\lambda+\tau-D)}{(p^2-\tilde{m}_Q^2)^{\alpha+\lambda+\tau-D}} \, ,  \nonumber\\
\end{eqnarray}
with $\alpha=\frac{D}{2}-n$. We set $D=4-2\epsilon$ and $z=x(1-y)$, then we obtain,
\begin{eqnarray}
{\rm Int}(p^2)&=&\frac{i\,2^{4-2n}\pi^6}{\Gamma(\lambda)\Gamma(\tau)\Gamma(n)} \int^{y_f}_{y_i} dy  y^{1-\lambda}  \int_{z_i}^{1-y} dz z^{1-\tau}(1-y-z)^{n-1} \frac{\Gamma(\lambda+\tau-n-2+\epsilon)}{(p^2-\tilde{m}_Q^2)^{
\lambda+\tau-n-2+\epsilon}} \, . \nonumber\\
\end{eqnarray}
In the case $\lambda=\tau=1$ and $n=2$,  we obtain,
\begin{eqnarray}
{\rm Int}(p^2)&=&i\,\pi^6 \int^{y_f}_{y_i} dy \, \int_{z_i}^{1-y}\, dz \,(1-y-z) \frac{\Gamma(\epsilon-2)}{(p^2-\tilde{m}_Q^2)^{\epsilon-2}} \, ,
\end{eqnarray}
and
\begin{eqnarray}
\frac{1}{\pi}{\rm Im\, Int}(s+i\epsilon)&=&i\,\frac{\pi^6}{2} \int^{y_f}_{y_i} dy \, \int_{z_i}^{1-y}\, dz \,(1-y-z) \,(s-\tilde{m}_Q^2)^{2} \, ,
\end{eqnarray}
where $y_{f}=\frac{1+\sqrt{1-4m_Q^2/s}}{2}$,
$y_{i}=\frac{1-\sqrt{1-4m_Q^2/s}}{2}$, $z_{i}=\frac{ym_Q^2}{y s -m_Q^2}$ and
$\tilde{m}_Q^2=\frac{m_Q^2}{y}+\frac{m_Q^2}{z}$.

Then we match  the hadron side with the QCD  side of the components $\Pi_{+}(p^2)$ and $p^2\widetilde{\Pi}_{+}(p^2)$ of the correlation functions $\Pi(p)$, $\Pi_{\mu\nu}(p)$ and $\Pi_{\mu\nu\alpha\beta}(p)$ below the continuum thresholds   $s_0$ and perform Borel transformation   with respect to
 $P^2=-p^2$ to obtain  the  QCD sum rules:
\begin{eqnarray}\label{QCDSR}
\lambda^2_{Z^+}\, \exp\left(-\frac{M^2_{Z^+}}{T^2}\right)= \int_{4m_c^2}^{s_0} ds\, \rho_{QCD}(s) \, \exp\left(-\frac{s}{T^2}\right) \, .
\end{eqnarray}

We derive Eq.\eqref{QCDSR} with respect to  $\tau=\frac{1}{T^2}$,  and obtain the QCD sum rules for
 the masses of the hidden-charm tetraquark states $Z_c$ or $X$,
 \begin{eqnarray}
 M^2_{Z^+}&=& -\frac{\int_{4m_c^2}^{s_0} ds\frac{d}{d \tau}\rho_{QCD}(s)\exp\left(-\tau s \right)}{\int_{4m_c^2}^{s_0} ds \rho_{QCD}(s)\exp\left(-\tau s\right)}\, .
\end{eqnarray}

Now let us begin to perform numerical analysis, and write down the energy-scale dependence of  the input parameters,
\begin{eqnarray}
\langle\bar{q}q \rangle(\mu)&=&\langle\bar{q}q \rangle({\rm 1GeV})\left[\frac{\alpha_{s}({\rm 1GeV})}{\alpha_{s}(\mu)}\right]^{\frac{12}{33-2n_f}}\, , \nonumber\\
\langle\bar{s}s \rangle(\mu)&=&\langle\bar{s}s \rangle({\rm 1GeV})\left[\frac{\alpha_{s}({\rm 1GeV})}{\alpha_{s}(\mu)}\right]^{\frac{12}{33-2n_f}}\, , \nonumber\\
\langle\bar{q}g_s \sigma Gq \rangle(\mu)&=&\langle\bar{q}g_s \sigma Gq \rangle({\rm 1GeV})\left[\frac{\alpha_{s}({\rm 1GeV})}{\alpha_{s}(\mu)}\right]^{\frac{2}{33-2n_f}}\, , \nonumber\\
 \langle\bar{s}g_s \sigma Gs \rangle(\mu)&=&\langle\bar{s}g_s \sigma Gs \rangle({\rm 1GeV})\left[\frac{\alpha_{s}({\rm 1GeV})}{\alpha_{s}(\mu)}\right]^{\frac{2}{33-2n_f}}\, , \nonumber\\
 m_c(\mu)&=&m_c(m_c)\left[\frac{\alpha_{s}(\mu)}{\alpha_{s}(m_c)}\right]^{\frac{12}{33-2n_f}} \, ,\nonumber\\
m_s(\mu)&=&m_s({\rm 2GeV} )\left[\frac{\alpha_{s}(\mu)}{\alpha_{s}({\rm 2GeV})}\right]^{\frac{12}{33-2n_f}}\, ,\nonumber\\
\alpha_s(\mu)&=&\frac{1}{b_0t}\left[1-\frac{b_1}{b_0^2}\frac{\log t}{t} +\frac{b_1^2(\log^2{t}-\log{t}-1)+b_0b_2}{b_0^4t^2}\right]\, ,
\end{eqnarray}
 where $t=\log \frac{\mu^2}{\Lambda_{QCD}^2}$, $b_0=\frac{33-2n_f}{12\pi}$, $b_1=\frac{153-19n_f}{24\pi^2}$, $b_2=\frac{2857-\frac{5033}{9}n_f+\frac{325}{27}n_f^2}{128\pi^3}$,  $\Lambda_{QCD}=210\,\rm{MeV}$, $292\,\rm{MeV}$  and  $332\,\rm{MeV}$ for the flavors  $n_f=5$, $4$ and $3$, respectively  \cite{PDG-2012,Narison-mix}. Because  the $c$-quark is concerned, we choose the flavor number to be $n_f=4$.

At the beginning  points, we adopt  the commonly-used values  $\langle
\bar{q}q \rangle=-(0.24\pm 0.01\, \rm{GeV})^3$, $\langle
\bar{s} s \rangle=(0.8 \pm 0.1)\langle \bar{q}q \rangle$,  $\langle
\bar{q}g_s\sigma G q \rangle=m_0^2\langle \bar{q}q \rangle$, $\langle
\bar{s}g_s\sigma G s \rangle=m_0^2\langle \bar{s}s \rangle$,
$m_0^2=(0.8 \pm 0.1)\,\rm{GeV}^2$,  $\langle \frac{\alpha_s
GG}{\pi}\rangle=(0.012\pm0.004)\,\rm{GeV}^4 $    at the typical energy scale  $\mu=1\, \rm{GeV}$
\cite{SVZ-NPB-1979-1,SVZ-NPB-1979-2,Reinders85,Colangelo-review}, and adopt  the $\overline{MS}$
(modified-minimal-subtraction) masses   $m_{c}(m_c)=(1.275\pm0.025)\,\rm{GeV}$ and $m_{s}({\rm 2 GeV})=(0.095\pm0.005)\,\rm{GeV}$ from the Particle Data Group \cite{PDG-2012}.

We apply the modified energy scale formula $\mu=\sqrt{M^2_{X/Y/Z}-(2{\mathbb{M}}_c)^2}-\kappa\,{\mathbb{M}}_s$ to choose the suitable  energy scales of the QCD spectral densities
\cite{WZG-HC-PRD-2020,WangZG-Zcs3985-mass-tetra-CPC-2021,WZG-HC-ss-NPB-2024}, where the ${\mathbb{M}}_c$ and ${\mathbb{M}}_s$ are the effective $c$ and $s$-quark masses respectively, and have universal values to be commonly used elsewhere. We adopt  the updated value  ${\mathbb{M}}_c=1.82\,\rm{GeV}$ \cite{WangZG-Y-tetra-EPJC-1601}, and take  the collective  light-flavor $SU(3)$-breaking effects into account  by
introducing  an  effective $s$-quark mass ${\mathbb{M}}_s=0.20\,\rm{GeV}$ ($0.12\,\rm{GeV}$) for the S-wave (P-wave) tetraquark  states  \cite{WZG-tetra-mole-IJMPA-2021,WZG-tetra-mole-AAPPS-2022,
WangZG-Zcs4123-tetra-CPC-2022,WangZG-Zcs3985-mass-tetra-CPC-2021,
WangZG-Regge-Ms-CPC-2021,
WZG-HC-ss-NPB-2024,WZG-cc-mole-EPJA-2022,
WangZG-Pcs4459-penta-IJMPA-2021,WZG-HC-Pseudo-NPB-2022}.

The continuum threshold parameters are not completely free parameters, and cannot be determined by the QCD sum rules themselves completely. We often consult the experimental data in choosing the continuum threshold parameters.
The  $Z_c(4430)$ can be assigned to be the first radial excitation of the $Z_c(3900)$ according to the
analogous decays,
\begin{eqnarray}
Z_c^\pm(3900)&\to&J/\psi\pi^\pm\, , \nonumber \\
Z_c^\pm(4430)&\to&\psi^\prime\pi^\pm\, ,
\end{eqnarray}
and the  analogous mass gaps  $M_{Z_c(4430)}-M_{Z_c(3900)}=591\,\rm{MeV}$ and $M_{\psi^\prime}-M_{J/\psi}=589\,\rm{MeV}$ from the Particle Data Group \cite{Maiani-1405-Tetra-model-2,Nielsen-MPLA-2014-Zc,WangZG-Zc4430-tetra-CTP-2015,PDG-2018}.

We tentatively choose the continuum threshold parameters as $\sqrt{s_0}=M_{X/Z}+0.60\,\rm{GeV}$ and vary the continuum threshold parameters $s_0$ and Borel parameters $T^2$ to satisfy
the following four   criteria:\\
$\bullet$ Pole dominance at the hadron  side;\\
$\bullet$ Convergence of the operator product expansion;\\
$\bullet$ Appearance of the Borel platforms;\\
$\bullet$ Fulfillment of the  modified energy scale formula,\\
  via trial  and error.

Thereafter, such criteria are adopted  for all the hidden-charm (hidden-bottom) tetraquark (molecular) states, hidden-charm (hidden-bottom) pentaquark (molecular) states, doubly-charm (doubly-bottom) tetraquark (molecular) states, doubly-charm (doubly-bottom) pentaquark (molecular) states, etc.

The pole dominance at the hadron  side and convergence of the operator product expansion at the QCD side are two basic  criteria, we should satisfy them to obtain reliable QCD sum rules.
In the QCD sum rules for the hidden-charm tetraquark and pentaquark states, the largest power of the energy variable  $s$ in the QCD spectral densities $\rho_{QCD}(s)\sim s^4$ and $s^5$, respectively, which make the integrals,
\begin{eqnarray}
\int_{4m_c^2}^{s_0} ds s^{4}\exp\left( -\frac{s}{T^2} \right)\, ,\,\,\,\int_{4m_c^2}^{s_0} ds s^{5}\exp\left( -\frac{s}{T^2} \right)\, ,
\end{eqnarray}
converge more slowly compared to the traditional hadrons, {\bf it is very difficult to satisfy the two  basic  criteria at the same time}.
We define the pole contributions (PC) by,
\begin{eqnarray}
{\rm{PC}}&=&\frac{\int_{4m_{c}^{2}}^{s_{0}}ds\,\rho_{QCD}\left(s\right)\exp\left(-\frac{s}{T^{2}}\right)} {\int_{4m_{c}^{2}}^{\infty}ds\,\rho_{QCD}\left(s\right)\exp\left(-\frac{s}{T^{2}}\right)}\, ,
\end{eqnarray}
 while we define the contributions of the vacuum condensates $D(n)$ of dimension $n$ by,
\begin{eqnarray}\label{Dn-strong}
D(n)&=&\frac{\int_{4m_{c}^{2}}^{s_{0}}ds\,\rho_{QCD,n}(s)\exp\left(-\frac{s}{T^{2}}\right)}
{\int_{4m_{c}^{2}}^{s_{0}}ds\,\rho_{QCD}\left(s\right)\exp\left(-\frac{s}{T^{2}}\right)}\, ,
\end{eqnarray}
or
\begin{eqnarray}\label{Dn-weak}
D(n)&=&\frac{\int_{4m_{c}^{2}}^{\infty}ds\,\rho_{QCD,n}(s)\exp\left(-\frac{s}{T^{2}}\right)}
{\int_{4m_{c}^{2}}^{\infty}ds\,\rho_{QCD}\left(s\right)\exp\left(-\frac{s}{T^{2}}\right)}\, ,
\end{eqnarray}
sometimes we would like to use the notation $D_n$ in stead of $D(n)$.
Compared to the criterion in Eq.\eqref{Dn-weak}, the criterion in Eq.\eqref{Dn-strong} is strong and leads to
larger Borel parameter $T^2$.
If we only study the ground state contributions, the Eq.\eqref{Dn-strong} is preferred.

After trial and error, we obtain the Borel windows, continuum threshold parameters, energy scales of the QCD spectral densities,  pole contributions, and contributions of the vacuum condensates of dimension $10$, which are shown explicitly in Tables \ref{BorelP-cqcq-positive}-\ref{BorelP-cscs-positive}.

\begin{table}
\begin{center}
\begin{tabular}{|c|c|c|c|c|c|c|c|c|}\hline\hline
 $Z_c$($X_c$)                                         & $J^{PC}$ & $T^2 (\rm{GeV}^2)$ & $\sqrt{s_0}(\rm GeV) $      &$\mu(\rm{GeV})$   &pole         &$|D(10)|$ \\ \hline

$[uc]_{S}[\overline{dc}]_{S}$                         & $0^{++}$ & $2.7-3.1$          & $4.40\pm0.10$               &$1.3$             &$(40-63)\%$  &$<1\%$   \\

$[uc]_{A}[\overline{dc}]_{A}$                         & $0^{++}$ & $2.8-3.2$          & $4.52\pm0.10$               &$1.5$             &$(40-63)\%$  &$\leq1\%$    \\

$[uc]_{\tilde{A}}[\overline{dc}]_{\tilde{A}}$         & $0^{++}$ & $3.1-3.5$          & $4.55\pm0.10$               &$1.6$             &$(42-62)\%$  &$<1\%$    \\

$[uc]_{V}[\overline{dc}]_{V}$                         & $0^{++}$ & $3.7-4.1$          & $5.22\pm0.10$               &$2.9$             &$(41-60)\%$  &$\ll1\%$    \\

$[uc]_{\tilde{V}}[\overline{dc}]_{\tilde{V}}$         & $0^{++}$ & $4.9-5.7$          & $5.90\pm0.10$               &$3.9$             &$(41-61)\%$  &$\ll 1\%$   \\

$[uc]_{P}[\overline{dc}]_{P}$                         & $0^{++}$ & $5.2-6.0$          & $6.03\pm0.10$               &$4.1$             &$(40-60)\%$  &$\ll1\%$    \\ \hline

$[uc]_S[\overline{dc}]_{A}-[uc]_{A}[\overline{dc}]_S$ & $1^{+-}$ & $2.7-3.1$          & $4.40\pm0.10$               &$1.4$             &$(40-63)\%$  &$<1\%$    \\

$[uc]_{A}[\overline{dc}]_{A}$                         & $1^{+-}$ & $3.3-3.7$          & $4.60\pm0.10$               &$1.7$             &$(40-59)\%$  &$\ll 1\%$  \\

$[uc]_S[\overline{dc}]_{\widetilde{A}}-[uc]_{\widetilde{A}}[\overline{dc}]_S$     & $1^{+-}$ & $3.3-3.7$     & $4.60\pm0.10$     &$1.7$      &$(40-59)\%$ &$\ll 1\%$  \\

$[uc]_{\widetilde{A}}[\overline{dc}]_{A}-[uc]_{A}[\overline{dc}]_{\widetilde{A}}$ & $1^{+-}$ & $3.2-3.6$     & $4.60\pm0.10$     &$1.7$      &$(41-61)\%$ &$\ll 1\%$ \\

$[uc]_{\widetilde{V}}[\overline{dc}]_{V}+[uc]_{V}[\overline{dc}]_{\widetilde{V}}$ & $1^{+-}$ & $3.7-4.1$     & $5.25\pm0.10$     &$2.9$      &$(41-60)\%$ &$\ll 1\%$ \\

$[uc]_{V}[\overline{dc}]_{V}$                         & $1^{+-}$ & $5.1-5.9$          & $6.00\pm0.10$               &$4.1$             &$(41-60)\%$  &$\ll 1\%$  \\

$[uc]_P[\overline{dc}]_{V}+[uc]_{V}[\overline{dc}]_P$ & $1^{+-}$ & $5.1-5.9$          & $6.00\pm0.10$               &$4.1$             &$(41-60)\%$  &$\ll1\%$    \\
\hline

$[uc]_S[\overline{dc}]_{A}+[uc]_{A}[\overline{dc}]_S$ & $1^{++}$ & $2.7-3.1$          & $4.40\pm0.10$               &$1.4$             &$(40-62)\%$  &$\ll 1\%$  \\

$[uc]_S[\overline{dc}]_{\widetilde{A}}+[uc]_{\widetilde{A}}[\overline{dc}]_S$     & $1^{++}$ & $3.3-3.7$     & $4.60\pm0.10$     &$1.7$      &$(40-59)\%$ &$\ll 1\%$  \\

$[uc]_{\widetilde{V}}[\overline{dc}]_{V}-[uc]_{V}[\overline{dc}]_{\widetilde{V}}$ & $1^{++}$ & $2.8-3.2$     & $4.62\pm0.10$     &$1.8$      &$(40-63)\%$ &$<2\%$ \\

$[uc]_{\widetilde{A}}[\overline{dc}]_{A}+[uc]_{A}[\overline{dc}]_{\widetilde{A}}$ & $1^{++}$ & $4.6-5.3$     & $5.73\pm0.10$     &$3.7$      &$(40-60)\%$ &$\ll 1\%$ \\

$[uc]_P[\overline{dc}]_{V}-[uc]_{V}[\overline{dc}]_P$ & $1^{++}$ & $5.1-5.9$          & $6.00\pm0.10$               &$4.1$             &$(40-60)\%$  &$\ll1\%$    \\
\hline

$[uc]_{A}[\overline{dc}]_{A}$                         & $2^{++}$ & $3.3-3.7$          & $4.65\pm0.10$               &$1.8$             &$(40-60)\%$       &$<1\%$ \\

$[uc]_{V}[\overline{dc}]_{V}$                         & $2^{++}$ & $5.0-5.8$          & $5.95\pm0.10$               &$4.0$             &$(40-60)\%$       &$\ll1\%$ \\
\hline\hline
\end{tabular}
\end{center}
\caption{ The Borel windows, continuum threshold parameters, energy scales of the QCD spectral densities,  pole contributions, and contributions of the vacuum condensates of dimension $10$  for the ground state $c\bar{c}u\bar{d}$ tetraquark states \cite{WZG-HC-PRD-2020}. }\label{BorelP-cqcq-positive}
\end{table}

\begin{table}
\begin{center}
\begin{tabular}{|c|c|c|c|c|c|c|c|c|}\hline\hline
 $X_c$              &$J^{PC}$ & $T^2 (\rm{GeV}^2)$ & $\sqrt{s_0}(\rm GeV) $      &$\mu(\rm{GeV})$   &pole         &$|D(10)|$ \\ \hline

$[sc]_{S}[\overline{sc}]_{S}$  &$0^{++}$  &$3.1-3.5$  &$4.65\pm0.10$               &$1.4$   &$(40-61)\%$  &$\ll 1\%$   \\

$[sc]_{A}[\overline{sc}]_{A}$  &$0^{++}$  &$3.1-3.5$  &$4.70\pm0.10$               &$1.5$   &$(39-60)\%$  &$\ll1\%$    \\

$[sc]_{\tilde{A}}[\overline{sc}]_{\tilde{A}}$  &$0^{++}$  &$3.4-3.9$  &$4.75\pm0.10$    &$1.6$    &$(39-61)\%$    &$\ll1\%$    \\

$[sc]_{V}[\overline{sc}]_{V}$   &$0^{++}$  &$4.0-4.5$  &$5.40\pm0.10$               &$2.8$     &$(41-60)\%$  &$\ll1\%$    \\

$[sc]_{\tilde{V}}[\overline{sc}]_{\tilde{V}}$  &$0^{++}$  &$5.2-6.1$
&$6.05\pm0.10$   &$3.7$   &$(40-61)\%$  &$\ll 1\%$   \\

$[sc]_{P}[\overline{sc}]_{P}$  &$0^{++}$  &$5.2-6.1$   &$6.10\pm0.10$               &$3.8$     &$(40-61)\%$  &$\ll1\%$    \\ \hline

$[sc]_{S}[\overline{sc}]_{S}{}^*$  &$0^{++}$  &$2.8-3.2$  &$4.50\pm0.10$               &$1.3$   &$(39-61)\%$  &$\ll 1\%$   \\

$[sc]_{A}[\overline{sc}]_{A}{}^*$  &$0^{++}$  &$2.7-3.1$  &$4.55\pm0.10$               &$1.3$   &$(39-62)\%$  &$\leq1\%$    \\

$[sc]_{\tilde{A}}[\overline{sc}]_{\tilde{A}}{}^*$  &$0^{++}$  &$3.0-3.5$  &$4.60\pm0.10$    &$1.4$    &$(39-62)\%$    &$\ll1\%$    \\

$[sc]_{V}[\overline{sc}]_{V}{}^*$   &$0^{++}$  &$3.6-4.1$  &$5.25\pm0.10$               &$2.6$     &$(40-62)\%$  &$\ll1\%$    \\

$[sc]_{\tilde{V}}[\overline{sc}]_{\tilde{V}}{}^*$  &$0^{++}$  &$4.7-5.4$
&$5.86\pm0.10$   &$3.5$   &$(41-60)\%$  &$\ll 1\%$   \\

$[sc]_{P}[\overline{sc}]_{P}{}^*$  &$0^{++}$  &$4.8-5.6$   &$5.95\pm0.10$               &$3.7$     &$(40-61)\%$  &$\ll1\%$    \\ \hline

$[sc]_S[\overline{sc}]_{A}-[sc]_{A}[\overline{sc}]_S$  &$1^{+-}$  &$3.2-3.7$          &$4.70\pm0.10$   &$1.5$  &$(39-61)\%$  &$\ll1\%$    \\

$[sc]_{A}[\overline{sc}]_{A}$     &$1^{+-}$  &$3.4-3.8$   &$4.76\pm0.10$               &$1.6$   &$(41-60)\%$    &$\ll 1\%$  \\

$[sc]_S[\overline{sc}]_{\widetilde{A}}-[sc]_{\widetilde{A}}[\overline{sc}]_S$
&$1^{+-}$ & $3.4-3.9$    &$4.76\pm0.10$   &$1.6$  &$(39-61)\%$  &$\ll 1\%$  \\

$[sc]_{\widetilde{A}}[\overline{sc}]_{A}-[sc]_{A}[\overline{sc}]_{\widetilde{A}}$
&$1^{+-}$  &$3.4-3.8$    &$4.76\pm0.10$    &$1.6$   &$(40-60)\%$ &$\ll 1\%$ \\

$[sc]_{\widetilde{V}}[\overline{sc}]_{V}+[sc]_{V}[\overline{sc}]_{\widetilde{V}}$
&$1^{+-}$  &$4.0-4.5$   &$5.40\pm0.10$  &$2.8$   &$(40-60)\%$  &$\ll 1\%$ \\

$[sc]_{V}[\overline{sc}]_{V}$  &$1^{+-}$   &$5.4-6.3$   &$6.16\pm0.10$               &$3.8$  &$(41-61)\%$  &$\ll 1\%$  \\

$[sc]_P[\overline{sc}]_{V}+[sc]_{V}[\overline{sc}]_P$   &$1^{+-}$  &$4.6-5.3$          &$5.70\pm0.10$   &$3.2$   &$(41-61)\%$  &$\ll1\%$   \\ \hline

$[sc]_S[\overline{sc}]_{A}+[sc]_{A}[\overline{sc}]_S$  &$1^{++}$   &$3.2-3.6$          &$4.70\pm0.10$    &$1.5$  &$(41-61)\%$  &$\ll 1\%$  \\

$[sc]_S[\overline{sc}]_{\widetilde{A}}+[sc]_{\widetilde{A}}[\overline{sc}]_S$
&$1^{++}$ &$3.4-3.9$    &$4.76\pm0.10$   &$1.6$   &$(39-60)\%$ &$\ll 1\%$  \\

$[sc]_{\widetilde{V}}[\overline{sc}]_{V}-[sc]_{V}[\overline{sc}]_{\widetilde{V}}$
&$1^{++}$ & $3.2-3.6$   &$4.86\pm0.10$   &$1.9$   &$(40-61)\%$ &$\ll1\%$ \\

$[sc]_{\widetilde{A}}[\overline{sc}]_{A}+[sc]_{A}[\overline{sc}]_{\widetilde{A}}$
&$1^{++}$ &$4.9-5.8$    &$5.93\pm0.10$    &$3.5$   &$(39-61)\%$ &$\ll 1\%$ \\

$[sc]_P[\overline{sc}]_{V}-[sc]_{V}[\overline{sc}]_P$  &$1^{++}$  &$4.6-5.4$          &$5.70\pm0.10$   &$3.2$   &$(39-61)\%$  &$\ll1\%$   \\ \hline

$[sc]_{A}[\overline{sc}]_{A}$    &$2^{++}$  &$3.5-4.0$   &$4.82\pm0.10$               &$1.8$   &$(40-60)\%$     &$\ll1\%$ \\

$[sc]_{V}[\overline{sc}]_{V}$    &$2^{++}$  &$5.1-6.0$   &$6.05\pm0.10$               &$3.7$     &$(40-61)\%$   &$\ll1\%$ \\
\hline\hline
\end{tabular}
\end{center}
\caption{ The Borel windows, continuum threshold parameters, energy scales of the QCD spectral densities,  pole contributions, and contributions of the vacuum condensates of dimension $10$  for the ground state $cs\bar{c}\bar{s}$ tetraquark states \cite{WZG-HC-ss-NPB-2024}. }\label{BorelP-cscs-positive}
\end{table}

At the hadron  side, the pole contributions are about $(40-60)\%$, the pole dominance  is well satisfied.
It is more easy to obtain the pole contributions $(40-60)\%$ with the help of the modified energy scale formula.
In Table \ref{BorelP-cscs-positive}, we  provide  two sets of data for the scalar tetraquark states, one set data is based on the continuum threshold parameters about  $\sqrt{s_0}=M_X+0.60\pm 0.10\,\rm{GeV}$ with the central values $\sqrt{s_0}<M_Y+0.60\,\rm{GeV}$ (this  criterion is also adopted for the tetraquark states with the $J^{PC}=1^{+-}$, $1^{++}$ and $2^{++}$), the other set data is based on  the  continuum threshold parameters about $\sqrt{s_0}=M_X+0.55\pm 0.10\,\rm{GeV}$ with the central values $\sqrt{s_0}<M_Y+0.55\,\rm{GeV}$, which will be characterized by the additional symbol $*$ in Tables \ref{BorelP-cscs-positive}, \ref{mass-Table-cscs-positive} and \ref{Identifications-Table-cscs-positive}.

At the QCD side, the contributions of the vacuum condensates of dimension $10$ are $|D(10)|\leq 1 \%$ or $\ll 1\%$ except for $|D(10)|< 2 \%$ for the $[uc]_{\widetilde{V}}[\overline{dc}]_{V}-[uc]_{V}[\overline{dc}]_{\widetilde{V}}$ tetraquark state with the $J^{PC}=1^{++}$, the convergent behavior of the operator product  expansion is very good.

\begin{figure}
\centering
\includegraphics[totalheight=10cm,width=14cm]{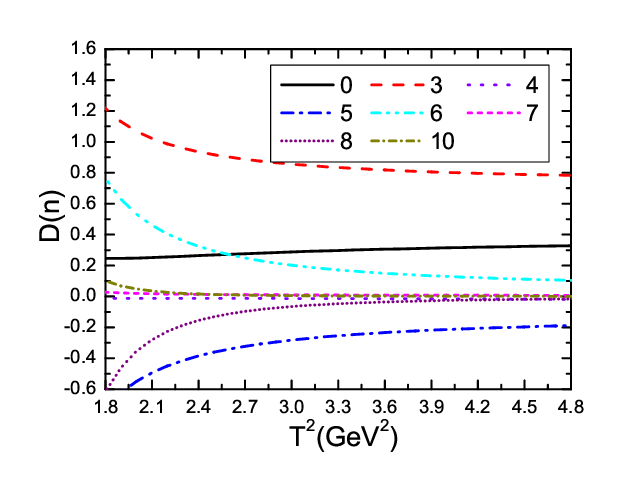}
  \caption{ The $D(n)$  with variation of the  Borel parameter $T^2$  for the current $J^{SA}_{-,\mu}(x)$. }\label{Zc3900-fr}
\end{figure}

In Fig.\ref{Zc3900-fr}, we plot the contributions of the vacuum condensates $D(n)$ with variation of the Borel parameter $T^2$ for the current $J^{SA}_{-,\mu}(x)$. From the figure, we can see clearly that the main contributions come from the $D(0)$, $D(3)$, $D(5)$, $D(6)$ and $D(8)$, while the largest contribution comes from the $D(3)$. Therefore, if we would like to calculate the radiative $\mathcal{O}(\alpha_s)$ corrections, we should calculate the radiative $\mathcal{O}(\alpha_s)$ corrections to the $D(3)$, not just to the $D(0)$. At the present time, only the radiative $\mathcal{O}(\alpha_s)$ corrections to the $D(0)$ are partially calculated \cite{X2900-mole-Narison-NPA-2021,
Narison-tetra-mole-IJMPA-2016,Narison-QQ-mole-tetra-PRD-2020,
Narison-mole-PRD-2021,afs-tetra-Narison-IJMPA-2016,afs-tetra-Narison-PRD-2022,
afs-tetra-KTChao-JHEP-2024,afs-tetra-KTChao-JHEP-2022}.

\begin{figure}
 \centering
 \includegraphics[totalheight=6cm,width=7cm]{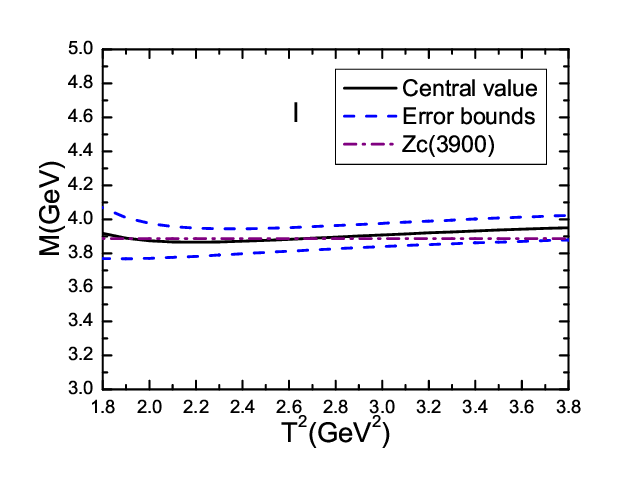}
 \includegraphics[totalheight=6cm,width=7cm]{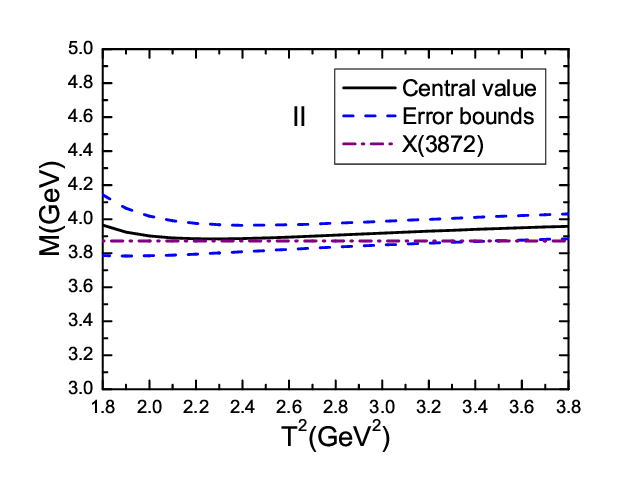}
 \caption{ The masses of the  $[uc]_S[\overline{dc}]_{A}-[uc]_{A}[\overline{dc}]_S $ (I) and $[uc]_S[\overline{dc}]_{A}+[uc]_{A}[\overline{dc}]_S$ (II) axialvector tetraquark states   with variations  of the Borel parameters $T^2$.  }\label{mass-1-fig}
\end{figure}

We take account of all the uncertainties of the input parameters and obtain the masses and pole residues of the scalar, axialvector and tensor hidden-charm  tetraquark states, which are shown explicitly in Tables \ref{mass-Table-cqcq-positive}-\ref{mass-Table-cscs-positive}. From  Tables \ref{BorelP-cqcq-positive}--\ref{mass-Table-cscs-positive}, we could see clearly  that the modified  energy scale formula $\mu=\sqrt{M^2_{X/Y/Z}-(2{\mathbb{M}}_c)^2}-\kappa \,{\mathbb{M}}_s $ is well satisfied.
 In  Fig.\ref{mass-1-fig}, we plot the masses of the  $[uc]_S[\overline{dc}]_{A}-[uc]_{A}[\overline{dc}]_S$ and $[uc]_S[\overline{dc}]_{A}+[uc]_{A}[\overline{dc}]_S$ tetraquark states having the spin-parity $J^P=1^+$ with variations of the Borel parameters at much larger ranges than the Borel widows as an example, there appear platforms in the Borel windows indeed.

\begin{table}
\begin{center}
\begin{tabular}{|c|c|c|c|c|c|c|c|c|}\hline\hline
$Z_c$($X_c$)                                                            & $J^{PC}$  & $M_Z (\rm{GeV})$   & $\lambda_Z (\rm{GeV}^5) $             \\ \hline

$[uc]_{S}[\overline{dc}]_{S}$                                           & $0^{++}$  & $3.88\pm0.09$      & $(2.07\pm0.35)\times 10^{-2}$           \\

$[uc]_{A}[\overline{dc}]_{A}$                                           & $0^{++}$  & $3.95\pm0.09$      & $(4.49\pm0.77)\times 10^{-2}$           \\

$[uc]_{\tilde{A}}[\overline{dc}]_{\tilde{A}}$                           & $0^{++}$  & $3.98\pm0.08$      & $(4.30\pm0.63)\times 10^{-2}$           \\

$[uc]_{V}[\overline{dc}]_{V}$                                           & $0^{++}$  & $4.65\pm0.09$      & $(1.35\pm0.22)\times 10^{-1}$           \\

$[uc]_{\tilde{V}}[\overline{dc}]_{\tilde{V}}$                           & $0^{++}$  & $5.35\pm0.09$      & $(4.87\pm0.51)\times 10^{-1}$             \\

$[uc]_{P}[\overline{dc}]_{P}$                                           & $0^{++}$  & $5.49\pm0.09$      & $(2.11\pm0.21)\times 10^{-1}$       \\ \hline

$[uc]_S[\overline{dc}]_{A}-[uc]_{A}[\overline{dc}]_S$                   & $1^{+-}$  & $3.90\pm0.08$      & $(2.09\pm0.33)\times 10^{-2}$        \\

$[uc]_{A}[\overline{dc}]_{A}$                                           & $1^{+-}$  & $4.02\pm0.09$      & $(3.00\pm0.45)\times 10^{-2}$           \\

$[uc]_S[\overline{dc}]_{\widetilde{A}}-[uc]_{\widetilde{A}}[\overline{dc}]_S$     & $1^{+-}$   & $4.01\pm0.09$    & $(3.02\pm0.45)\times 10^{-2}$    \\

$[uc]_{\widetilde{A}}[\overline{dc}]_{A}-[uc]_{A}[\overline{dc}]_{\widetilde{A}}$ & $1^{+-}$   & $4.02\pm0.09$    & $(6.09\pm0.90)\times 10^{-2}$    \\

$[uc]_{\widetilde{V}}[\overline{dc}]_{V}+[uc]_{V}[\overline{dc}]_{\widetilde{V}}$ & $1^{+-}$   & $4.66\pm0.10$    & $(1.18\pm0.21)\times 10^{-1}$    \\

$[uc]_{V}[\overline{dc}]_{V}$                                           & $1^{+-}$  & $5.46\pm0.09$      & $(1.72\pm0.17)\times 10^{-1}$           \\

$[uc]_P[\overline{dc}]_{V}+[uc]_{V}[\overline{dc}]_P$                   & $1^{+-}$  & $5.45\pm0.09$      & $(1.87\pm0.19)\times 10^{-1}$        \\
\hline

$[uc]_S[\overline{dc}]_{A}+[uc]_{A}[\overline{dc}]_S$                   & $1^{++}$  & $3.91\pm0.08$      & $(2.10\pm0.34)\times 10^{-2}$        \\

$[uc]_S[\overline{dc}]_{\widetilde{A}}+[uc]_{\widetilde{A}}[\overline{dc}]_S$     & $1^{++}$   & $4.02\pm0.09$    & $(3.01\pm0.45)\times 10^{-2}$    \\

$[uc]_{\widetilde{V}}[\overline{dc}]_{V}-[uc]_{V}[\overline{dc}]_{\widetilde{V}}$ & $1^{++}$   & $4.08\pm0.09$    & $(3.67\pm0.67)\times 10^{-2}$    \\

$[uc]_{\widetilde{A}}[\overline{dc}]_{A}+[uc]_{A}[\overline{dc}]_{\widetilde{A}}$ & $1^{++}$   & $5.19\pm0.09$    & $(2.12\pm0.24)\times 10^{-1}$    \\

$[uc]_P[\overline{dc}]_{V}-[uc]_{V}[\overline{dc}]_P$                   & $1^{++}$  & $5.46\pm0.09$      & $(1.89\pm0.19)\times 10^{-1}$       \\
\hline

$[uc]_{A}[\overline{dc}]_{A}$                                           & $2^{++}$  & $4.08\pm0.09$      & $(4.67\pm0.68)\times 10^{-2}$      \\

$[uc]_{V}[\overline{dc}]_{V}$                                           & $2^{++}$  & $5.40\pm0.09$      & $(2.32\pm0.25)\times 10^{-1}$      \\
\hline\hline
\end{tabular}
\end{center}
\caption{ The masses and pole residues of the ground state hidden-charm tetraquark states \cite{WZG-HC-PRD-2020}. }\label{mass-Table-cqcq-positive}
\end{table}

\begin{table}
\begin{center}
\begin{tabular}{|c|c|c|c|c|c|c|c|c|}\hline\hline
$X_c$    &$J^{PC}$  &$M_X (\rm{GeV})$   &$\lambda_X (\rm{GeV}^5) $   \\ \hline

$[sc]_{S}[\overline{sc}]_{S}$  &$0^{++}$  &$4.08\pm0.09$  &$(3.18\pm0.52)\times 10^{-2}$  \\

$[sc]_{A}[\overline{sc}]_{A}$  &$0^{++}$  &$4.13\pm0.09$  &$(6.02\pm1.02)\times 10^{-2}$  \\

$[sc]_{\tilde{A}}[\overline{sc}]_{\tilde{A}}$  &$0^{++}$  &$4.16\pm0.09$  &$(5.83\pm0.87)\times 10^{-2}$  \\

$[sc]_{V}[\overline{sc}]_{V}$   &$0^{++}$  &$4.82\pm0.09$ &$(1.70\pm0.25)\times 10^{-1}$   \\

$[sc]_{\tilde{V}}[\overline{sc}]_{\tilde{V}}$ &$0^{++}$   &$5.46\pm0.10$
&$(5.37\pm0.58)\times 10^{-1}$  \\

$[sc]_{P}[\overline{sc}]_{P}$  &$0^{++}$  &$5.54\pm0.10$  &$(2.13\pm0.22)\times 10^{-1}$    \\ \hline

$[sc]_{S}[\overline{sc}]_{S}{}^*$  &$0^{++}$  &$3.99\pm0.09$  &$(2.41\pm0.42)\times 10^{-2}$  \\

$[sc]_{A}[\overline{sc}]_{A}{}^*$  &$0^{++}$  &$4.04\pm0.09$  &$(4.24\pm0.76)\times 10^{-2}$  \\

$[sc]_{\tilde{A}}[\overline{sc}]_{\tilde{A}}{}^*$  &$0^{++}$  &$4.08\pm0.08$  &$(4.39\pm0.69)\times 10^{-2}$  \\

$[sc]_{V}[\overline{sc}]_{V}{}^*$   &$0^{++}$  &$4.70\pm0.09$ &$(1.31\pm0.23)\times 10^{-1}$   \\

$[sc]_{\tilde{V}}[\overline{sc}]_{\tilde{V}}{}^*$ &$0^{++}$   &$5.37\pm0.11$
&$(4.41\pm0.47)\times 10^{-1}$  \\

$[sc]_{P}[\overline{sc}]_{P}{}^*$  &$0^{++}$  &$5.47\pm0.11$  &$(1.86\pm0.19)\times 10^{-1}$       \\  \hline

$[sc]_S[\overline{sc}]_{A}-[sc]_{A}[\overline{sc}]_S$   &$1^{+-}$  &$4.11\pm0.10$      &$(2.91\pm0.48)\times 10^{-2}$        \\

$[sc]_{A}[\overline{sc}]_{A}$     &$1^{+-}$  &$4.17\pm0.08$      &$(3.65\pm0.55)\times 10^{-2}$           \\

$[sc]_S[\overline{sc}]_{\widetilde{A}}-[sc]_{\widetilde{A}}[\overline{sc}]_S$
&$1^{+-}$   &$4.17\pm0.09$    &$(3.71\pm0.57)\times 10^{-2}$    \\

$[sc]_{\widetilde{A}}[\overline{sc}]_{A}-[sc]_{A}[\overline{sc}]_{\widetilde{A}}$
&$1^{+-}$   &$4.18\pm0.09$    & $(7.50\pm1.12)\times 10^{-2}$    \\

$[sc]_{\widetilde{V}}[\overline{sc}]_{V}+[sc]_{V}[\overline{sc}]_{\widetilde{V}}$
&$1^{+-}$   &$4.82\pm0.09$    &$(1.45\pm0.23)\times 10^{-1}$    \\

$[sc]_{V}[\overline{sc}]_{V}$    &$1^{+-}$  &$5.57\pm0.11$  &$(1.89\pm0.19)\times 10^{-1}$    \\

$[sc]_P[\overline{sc}]_{V}+[sc]_{V}[\overline{sc}]_P$  &$1^{+-}$  &$5.13\pm0.10$      &$(1.33\pm0.16)\times 10^{-1}$  \\ \hline

$[sc]_S[\overline{sc}]_{A}+[sc]_{A}[\overline{sc}]_S$   &$1^{++}$  &$4.11\pm0.09$      &$(2.88\pm0.46)\times 10^{-2}$     \\

$[sc]_S[\overline{sc}]_{\widetilde{A}}+[sc]_{\widetilde{A}}[\overline{sc}]_S$
 &$1^{++}$   &$4.17\pm0.09$    &$(3.67\pm0.57)\times 10^{-2}$    \\

$[sc]_{\widetilde{V}}[\overline{sc}]_{V}-[sc]_{V}[\overline{sc}]_{\widetilde{V}}$
&$1^{++}$   &$4.29\pm0.09$    &$(5.49\pm0.92)\times 10^{-2}$    \\

$[sc]_{\widetilde{A}}[\overline{sc}]_{A}+[sc]_{A}[\overline{sc}]_{\widetilde{A}}$
&$1^{++}$   &$5.34\pm0.10$    &$(2.52\pm0.30)\times 10^{-1}$    \\

$[sc]_P[\overline{sc}]_{V}-[sc]_{V}[\overline{sc}]_P$  &$1^{++}$  &$5.12\pm0.10$      &$(1.33\pm0.17)\times 10^{-1}$    \\ \hline

$[sc]_{A}[\overline{sc}]_{A}$   &$2^{++}$  &$4.24\pm0.09$   &$(6.03\pm0.88)\times 10^{-2}$      \\

$[sc]_{V}[\overline{sc}]_{V}$    &$2^{++}$  &$5.49\pm0.11$   &$(2.43\pm0.26)\times 10^{-1}$      \\
\hline\hline
\end{tabular}
\end{center}
\caption{ The masses and pole residues of the ground state hidden-charm-hidden-strange  tetraquark states \cite{WZG-HC-ss-NPB-2024}. }\label{mass-Table-cscs-positive}
\end{table}

\begin{table}
\begin{center}
\begin{tabular}{|c|c|c|c|c|c|c|c|c|}\hline\hline
$Z_c$($X_c$)                                                            & $J^{PC}$  & $M_{X/Z}(\rm{GeV})$   & Assignments        &$Z_c^\prime$ ($X_c^\prime$)      \\ \hline

$[uc]_{S}[\overline{dc}]_{S}$                                           & $0^{++}$  & $3.88\pm0.09$      & ?\,$X(3860)$       &       \\

$[uc]_{A}[\overline{dc}]_{A}$                                           & $0^{++}$  & $3.95\pm0.09$      & ?\,$X(3915)$       & \\

$[uc]_{\tilde{A}}[\overline{dc}]_{\tilde{A}}$                           & $0^{++}$  & $3.98\pm0.08$      &                    & \\

$[uc]_{V}[\overline{dc}]_{V}$                                           & $0^{++}$  & $4.65\pm0.09$      &                    & \\

$[uc]_{\tilde{V}}[\overline{dc}]_{\tilde{V}}$                           & $0^{++}$  & $5.35\pm0.09$      &                    &  \\

$[uc]_{P}[\overline{dc}]_{P}$                                           & $0^{++}$  & $5.49\pm0.09$      &                    &  \\ \hline

$[uc]_S[\overline{dc}]_{A}-[uc]_{A}[\overline{dc}]_S$                   & $1^{+-}$  & $3.90\pm0.08$      & ?\,$Z_c(3900)$      &?\,$Z_c(4430)$    \\

$[uc]_{A}[\overline{dc}]_{A}$                                           & $1^{+-}$  & $4.02\pm0.09$      & ?\,$Z_c(4020/4055)$ &?\,$Z_c(4600)$        \\

$[uc]_S[\overline{dc}]_{\widetilde{A}}-[uc]_{\widetilde{A}}[\overline{dc}]_S$     & $1^{+-}$   & $4.01\pm0.09$    & ?\,$Z_c(4020/4055)$ &?\,$Z_c(4600)$     \\

$[uc]_{\widetilde{A}}[\overline{dc}]_{A}-[uc]_{A}[\overline{dc}]_{\widetilde{A}}$ & $1^{+-}$   & $4.02\pm0.09$    & ?\,$Z_c(4020/4055)$ &?\,$Z_c(4600)$    \\

$[uc]_{\widetilde{V}}[\overline{dc}]_{V}+[uc]_{V}[\overline{dc}]_{\widetilde{V}}$ & $1^{+-}$   & $4.66\pm0.10$    & ?\,$Z_c(4600)$      &    \\

$[uc]_{V}[\overline{dc}]_{V}$                                           & $1^{+-}$  & $5.46\pm0.09$      &                    &  \\

$[uc]_P[\overline{dc}]_{V}+[uc]_{V}[\overline{dc}]_P$                   & $1^{+-}$  & $5.45\pm0.09$      &                    &  \\
\hline

$[uc]_S[\overline{dc}]_{A}+[uc]_{A}[\overline{dc}]_S$                   & $1^{++}$  & $3.91\pm0.08$      & ?\,$X(3872)$       &   \\

$[uc]_S[\overline{dc}]_{\widetilde{A}}+[uc]_{\widetilde{A}}[\overline{dc}]_S$     & $1^{++}$   & $4.02\pm0.09$    &?\,$Z_c(4050)$ &   \\

$[uc]_{\widetilde{V}}[\overline{dc}]_{V}-[uc]_{V}[\overline{dc}]_{\widetilde{V}}$ & $1^{++}$   & $4.08\pm0.09$    &?\,$Z_c(4050)$ &    \\

$[uc]_{\widetilde{A}}[\overline{dc}]_{A}+[uc]_{A}[\overline{dc}]_{\widetilde{A}}$ & $1^{++}$   & $5.19\pm0.09$    &               & \\

$[uc]_P[\overline{dc}]_{V}-[uc]_{V}[\overline{dc}]_P$                   & $1^{++}$  & $5.46\pm0.09$      &                    &  \\
\hline

$[uc]_{A}[\overline{dc}]_{A}$                                           & $2^{++}$  & $4.08\pm0.09$      &?\,$Z_c(4050)$      & \\

$[uc]_{V}[\overline{dc}]_{V}$                                           & $2^{++}$ & $5.40\pm0.09$      &                    & \\ \hline \hline

$[uc]_{A}[\overline{dc}]_{A}$                                           & $1^{+-}$  & $4.02\pm0.09$      & ?\,$h_c(4000)$ &        \\

$[uc]_S[\overline{dc}]_{\widetilde{A}}-[uc]_{\widetilde{A}}[\overline{dc}]_S$     & $1^{+-}$   & $4.01\pm0.09$    & ?\,$h_c(4000)$ &      \\

$[uc]_{\widetilde{A}}[\overline{dc}]_{A}-[uc]_{A}[\overline{dc}]_{\widetilde{A}}$ & $1^{+-}$   & $4.02\pm0.09$    & ?\,$h_c(4000)$ &    \\

$[uc]_S[\overline{dc}]_{\widetilde{A}}+[uc]_{\widetilde{A}}[\overline{dc}]_S$     & $1^{++}$   & $4.02\pm0.09$    &?\,$\chi_{c1}(4010)$ &   \\

\hline\hline
\end{tabular}
\end{center}
\caption{ The possible  assignments  of the  hidden-charm tetraquark states, the isospin limit is implied \cite{WZG-HC-PRD-2020,WangZG-Zc4600-tetra-CPC-2020,WZG-HC-ss-NPB-2024}. }\label{Identifications-Table-cqcq-positive}
\end{table}

\begin{table}
\begin{center}
\begin{tabular}{|c|c|c|c|c|c|c|c|c|}\hline\hline
$X_c$      &$J^{PC}$  & $M_X(\rm{GeV})$   & Assignments    &$X_c^\prime$ \\ \hline

$[sc]_{S}[\overline{sc}]_{S}$  &$0^{++}$  &$4.08\pm0.09$   &     & \\

$[sc]_{A}[\overline{sc}]_{A}$  &$0^{++}$  &$4.13\pm0.09$   &       &?\,$X(4700)$ \\

$[sc]_{\tilde{A}}[\overline{sc}]_{\tilde{A}}$  &$0^{++}$  &$4.16\pm0.09$  &                     &?\,$X(4700)$ \\

$[sc]_{V}[\overline{sc}]_{V}$  &$0^{++}$  &$4.82\pm0.09$  &    & \\

$[sc]_{\tilde{V}}[\overline{sc}]_{\tilde{V}}$  &$0^{++}$  &$5.46\pm0.10$  &                   &  \\

$[sc]_{P}[\overline{sc}]_{P}$  &$0^{++}$  &$5.54\pm0.10$     &  &  \\ \hline

$[sc]_{S}[\overline{sc}]_{S}{}^*$  &$0^{++}$  &$3.99\pm0.09$   &?\,$X(3960)$      &?\,$X(4500)$   \\

$[sc]_{A}[\overline{sc}]_{A}{}^*$  &$0^{++}$  &$4.04\pm0.09$   &     & \\

$[sc]_{\tilde{A}}[\overline{sc}]_{\tilde{A}}{}^*$  &$0^{++}$  &$4.08\pm0.08$  &                    & \\

$[sc]_{V}[\overline{sc}]_{V}{}^*$  &$0^{++}$  &$4.70\pm0.09$  &?\,$X(4700)$  & \\

$[sc]_{\tilde{V}}[\overline{sc}]_{\tilde{V}}{}^*$  &$0^{++}$  &$5.37\pm0.11$  &                   &  \\

$[sc]_{P}[\overline{sc}]_{P}{}^*$  &$0^{++}$  &$5.47\pm0.11$     &  &  \\ \hline

$[sc]_S[\overline{sc}]_{A}-[sc]_{A}[\overline{sc}]_S$  &$1^{+-}$  &$4.11\pm0.10$      &       &   \\

$[sc]_{A}[\overline{sc}]_{A}$     &$1^{+-}$  &$4.17\pm0.08$    & &   \\

$[sc]_S[\overline{sc}]_{\widetilde{A}}-[sc]_{\widetilde{A}}[\overline{sc}]_S$
 &$1^{+-}$   &$4.17\pm0.09$   &  &    \\

$[sc]_{\widetilde{A}}[\overline{sc}]_{A}-[sc]_{A}[\overline{sc}]_{\widetilde{A}}$
&$1^{+-}$   &$4.18\pm0.09$    &  &   \\

$[sc]_{\widetilde{V}}[\overline{sc}]_{V}+[sc]_{V}[\overline{sc}]_{\widetilde{V}}$
&$1^{+-}$   &$4.82\pm0.09$    &        &    \\

$[sc]_{V}[\overline{sc}]_{V}$   &$1^{+-}$  &$5.57\pm0.11$   &   &  \\

$[sc]_P[\overline{sc}]_{V}+[sc]_{V}[\overline{sc}]_P$   &$1^{+-}$  &$5.13\pm0.10$        &   &  \\ \hline

$[sc]_S[\overline{sc}]_{A}+[sc]_{A}[\overline{sc}]_S$   &$1^{++}$  &$4.11\pm0.09$     &?\,$X(4140)$       &?\,$X(4685)$   \\

$[sc]_S[\overline{sc}]_{\widetilde{A}}+[sc]_{\widetilde{A}}[\overline{sc}]_S$
 &$1^{++}$   &$4.17\pm0.09$    &?\,$X(4140)$  &?\,$X(4685)$   \\

$[sc]_{\widetilde{V}}[\overline{sc}]_{V}-[sc]_{V}[\overline{sc}]_{\widetilde{V}}$
&$1^{++}$   &$4.29\pm0.09$    &?\,$X(4274)$ &    \\

$[sc]_{\widetilde{A}}[\overline{sc}]_{A}+[sc]_{A}[\overline{sc}]_{\widetilde{A}}$
&$1^{++}$   &$5.34\pm0.10$    &      & \\

$[sc]_P[\overline{sc}]_{V}-[sc]_{V}[\overline{sc}]_P$  &$1^{++}$  &$5.12\pm0.10$      &   &  \\ \hline

$[sc]_{A}[\overline{sc}]_{A}$   &$2^{++}$  &$4.24\pm0.09$     &  & \\

$[sc]_{V}[\overline{sc}]_{V}$   &$2^{++}$  &$5.49\pm0.11$      &   & \\
\hline\hline
\end{tabular}
\end{center}
\caption{ The possible  assignments of the hidden-charm-hidden-strange tetraquark states \cite{WZG-HC-ss-NPB-2024}. }\label{Identifications-Table-cscs-positive}
\end{table}

In Tables \ref{Identifications-Table-cqcq-positive}-\ref{Identifications-Table-cscs-positive}, we present the possible assignments of the ground state hidden-charm tetraquark states, and revisit the assignments based on the tetraquark scenario, there are rooms for the $X(3860)$, $X(3872)$, $X(3915)$,  $X(3960)$, $X(4140)$, $X(4274)$, $X(4500)$, $X(4685)$, $X(4700)$, $Z_c(3900)$, $Z_c(4020)$, $Z_c(4050)$, $Z_c(4055)$, $Z_c(4430)$ and $Z_c(4600)$.

It is not difficult to reproduce the mass of the $X(3872)$ in the scenario of tetraquark state \cite{X3872-tetra-Narison-PRD-2007,X3872-tetra-WangZG-HuangT-PRD-2014,WZG-HC-PRD-2020}, it is a great challenge to reproduce the tiny width $1.19\pm 0.21 \, \rm{MeV}$ from  the Particle Data Group consistently \cite{PDG-2024}.
In Ref.\cite{WZG-X3872-decay-PRD-2024}, we study the strong decays $X(3872)\to J/\psi \pi^+\pi^-$, $J/\psi\omega$,  $\chi_{c1}\pi^0$, $D^{*0}\bar{D}^0$ and $D^{0}\bar{D}^0\pi^0$ via the QCD sum rules according to rigorous quark-hadron duality, and obtain the total decay width about $1\,\rm{MeV}$, it is  the first time  to reproduce the tiny  width of the $X(3872)$ via the QCD sum rules.
The thresholds of the $J/\psi\rho$ and $J/\psi\omega$ are $3872.16\,\rm{MeV}$ and $3879.56\,\rm{MeV}$ respectively, which are larger than the value $3871.64\,\rm{MeV}$ of the mass of the $X(3872)$ from the Particle Data Group \cite{PDG-2024} and also lead to the possibility that it may be a $J/\psi\rho$ or $J/\psi \omega$ molecular state \cite{X3872-mole-Mutuk-EPJC-2018}, the decays $X(3872) \to J/\psi \pi^+\pi^-$ and $J/\psi \pi^+\pi^-\pi^-$ take place through the virtual $\rho$ and $\omega$ mesons, respectively, we introduce form-factors to parameterize the off-shell-ness, however, the arbitrariness in choosing the form-factors would impair the predictive ability.

As far as the $Z_c(3900)$ is concerned, it is not difficult to reproduce its mass
and width via the QCD sum rules \cite{X3872-tetra-WangZG-HuangT-PRD-2014,WZG-HC-PRD-2020,WangZhang-Solid}. It is the benchmark for the  $\bar{\mathbf 3}\mathbf{3}$ type hidden-charm tetraquark states in the simple diquark models  \cite{Maiani-1405-Tetra-model-2,Born-Braaten-PRL-2013,Maiani-4260-Z3900-tetra-PRD-2013,
Maiani-4260-tetra-PRD-2005,Zc3900-tetra-Esposito-PLB-2015}.

In Table \ref{Identifications-Table-cqcq-positive}, there are enough rooms to accommodate the $h_c(4000)$ and $\chi_{c1}(4010)$ in the scenario of tetraquark states, as the central values of the predicted  tetraquark masses happen to coincide with the experimental data from the LHCb collaboration \cite{LHCb-hc4000-PRL-2024}. We should bear in mind that the predictions were made {\bf before} the experimental observation, therefore the calculations are robust enough \cite{WZG-HC-PRD-2020}.  While in the traditional charmonium scenario, the $h_c^\prime$ and $\chi_{c1}^\prime$ have the masses $3956\,\rm{MeV}$ and $3953\,\rm{MeV}$, respectively \cite{Godfrey-PRD-2005-charmonium}, there exist  discrepancies    about $50-60\,\rm{MeV}$.

The predictions $M_X=4.11\pm0.09\,\rm{GeV}$ and $4.17\pm0.09\,\rm{GeV}$ for the 1S  $[sc]_S[\overline{sc}]_{A}+[sc]_{A}[\overline{sc}]_S$  and
$[sc]_S[\overline{sc}]_{\widetilde{A}}+[sc]_{\widetilde{A}}[\overline{sc}]_S$ tetraquark  states with the $J^{PC}=1^{++}$ respectively support assigning  the  $X(4140)$ and $X(4685)$   as the 1S and 2S $[sc]_S[\overline{sc}]_{A}+[sc]_{A}[\overline{sc}]_S$  or
$[sc]_S[\overline{sc}]_{\widetilde{A}}+[sc]_{\widetilde{A}}[\overline{sc}]_S$ tetraquark states respectively. In Refs.\cite{WangZG-X4140-decay-EPJC-2019,2S-WZG-X4630-X4685-AHEP-2021}, we obtain the prediction $M_X=4.14 \pm 0.10\,\rm{GeV}$ for the 1S $[sc]_{\widetilde{V}}[\overline{sc}]_{V}-[sc]_{V}[\overline{sc}]_{\widetilde{V}}$
tetraquark state with the  $J^{PC}=1^{++}$ in the old scheme, which supports  assigning  the $X(4140)$ as the $[sc]_{\widetilde{V}}[\overline{sc}]_{V}-[sc]_{V}[\overline{sc}]_{\widetilde{V}}$ tetraquark state.

The prediction $M_X=4.29\pm0.09\,\rm{GeV}$ for the  1S $[sc]_{\widetilde{V}}[\overline{sc}]_{V}-[sc]_{V}[\overline{sc}]_{\widetilde{V}}$
tetraquark state supports assigning the $X(4274)$ as the $[sc]_{\widetilde{V}}[\overline{sc}]_{V}-[sc]_{V}[\overline{sc}]_{\widetilde{V}}$
tetraquark state with the  $J^{PC}=1^{++}$.  The calculations in Refs.\cite{WangZG-X4140-decay-EPJC-2019,2S-WZG-X4630-X4685-AHEP-2021} are updated, as where the light-flavor $SU(3)$ mass-breaking effects in the energy scale formula are  not taken into account.

 In Ref.\cite{WangZG-X4274-decay-APPB-2020},  the predictions  support assigning  the $X(4274)$ as the   $[sc]_A[\bar{s}\bar{c}]_V-[sc]_V[\bar{s}\bar{c}]_A$   tetraquark state with a relative P-wave between the diquark and antidiquark pairs, such an assignment   does not suffer from shortcomings in sense of treating scheme. The $X(4274)$ maybe have two important  Fock components.

The $X(4140)$ and $X(4274)$ are also play an important role in establishing the hidden-charm tetraquark states, especially the $X(4140)$. There have been several possible assignments of the $X(4140)$, such  as  the tetraquark state \cite{Lebed-cscs-PRD-2016,
Maiani-Jpsi-phi-PRD-2016,
Tetra-33-RQM-LuQF-PRD-2016,Tetra-33-RQM-Ferretti-PRD-2018,WZG-HC-ss-NPB-2024,
WangZG-X4140-decay-EPJC-2019,
2S-WZG-X4630-X4685-AHEP-2021,
X4140-tetraquark-Stancu,X4140-tetra-RLZhu-PRD-2016,
X4140-tetra-WuJ-PRD-2016,X4140-Y4700-tetra-HXChen-EPJC-2017}, hybrid state \cite{WZG-Y4140-EPJC-2009,WZG-Y4140-EPJC-2009-2,X4140-hybrid-Mahajan} or  rescattering effect \cite{X4140-rescat-LiuXH-PLB-2017}, etc.

In Table \ref{Identifications-Table-cscs-positive}, we assign the $X(3960)$ and $X(4500)$  as the 1S and 2S $[cs]_S[\bar{c}\bar{s}]_S$ tetraquark states with the $J^{PC}=0^{++}$, the $X(4700)$  as the 1S $[cs]_V[\bar{c}\bar{s}]_V$ tetraquark state
with the $J^{PC}=0^{++}$; or assign the $X(4700)$ as the 2S $[cs]_A[\bar{c}\bar{s}]_A$ or $[cs]_{\tilde{A}}[\bar{c}\bar{s}]_{\tilde{A}}$ tetraquark state with the $J^{PC}=0^{++}$. In Ref.\cite{X4775-Zcs4600-WangZG-2409}, we assign the $X(4500)$ as the $[sc]_{\widehat{V}}[\overline{sc}]_{\widehat{V}}$ tetraquark state with the $J^{PC}=0^{++}$, where the $\widehat{V}$ denotes the diquark with an explicit P-wave.
 In Ref.\cite{X4140-Y4700-tetra-HXChen-EPJC-2017}, Chen et al assign the $X(4500)$ and $X(4700)$  as the $D$-wave tetraquark states with
the quark content $cs\bar{c}\bar{s}$ and $J^P = 0^+$: the $X(4500)$
consists of one $D$-wave  diquark and one $S$-wave antidiquark, with the antisymmetric color structure $\bar{\mathbf{3}} \mathbf{3}$; the $X(4700)$
consists of similar diquarks but with the symmetric color structure
$\mathbf{6}\bar{\mathbf{6}}$.

However, there is no room for the $X(4350)$, we should introduce mixing effect  to interpret its nature \cite{WZG-X4350-PLB-2010}.

In fact, the predictions of the QCD sum rules have arbitrariness, and depend on the interpolating currents, truncations of the operator product expansion, convergent behavior,  pole contributions, input parameters, etc. Therefore, the predictions maybe quite different,  we have to perform systematic investigations with uniform criterion to outcome the shortcomings.

We suggest to study the two-body  strong decays to diagnose  those hidden-charm tetraquark states \cite{WZG-HC-PRD-2020,WZG-HC-ss-NPB-2024}, for example,
 \begin{eqnarray}
Z_c^\pm(1^{+-}) &\to&\pi^{\pm}J/\psi\,  ,  \,\pi^{\pm}\psi^\prime\,  ,  \, \pi^{\pm} h_c \, ,  \, \rho^{\pm} \eta_c  \, , \, (D\bar{D}^{*})^\pm\, ,\, (D^{*}\bar{D})^\pm\, ,\, (D^{*}\bar{D}^{*})^\pm\, , \nonumber\\
Z_c^\pm(0^{++}) &\to&\pi^{\pm}\eta_c\,  ,  \, \pi^{\pm} \chi_{c1} \, ,  \, \rho^{\pm} J/\psi \, , \, \rho^{\pm} \psi^\prime \, , \,(D\bar{D})^\pm\, ,\, (D^{*}\bar{D}^{*})^\pm\, , \nonumber\\
Z_c^\pm(1^{++}) &\to& \pi^{\pm} \chi_{c1} \, ,  \, \rho^{\pm} J/\psi \, , \, \rho^{\pm} \psi^\prime \, , \,(D\bar{D}^{*})^\pm\, ,\, (D^{*}\bar{D})^\pm\, ,\, (D^{*}\bar{D}^{*})^\pm\, , \nonumber\\
Z_c^\pm(2^{++}) &\to&\pi^{\pm}\eta_c \,  ,  \, \pi^{\pm} \chi_{c1}  \, ,  \, \rho^{\pm} J/\psi \, , \, \rho^{\pm} \psi^\prime \, , \,(D\bar{D})^\pm\, ,\, (D^{*}\bar{D}^{*})^\pm\, ,
\end{eqnarray}
 \begin{eqnarray}\label{Zc0-aixal-decay}
Z_c^0(1^{+-}) &\to&\pi^{0}J/\psi\,  ,  \,\pi^{0}\psi^\prime\,  ,  \, \pi^{0} h_c \, ,  \, \rho^{0} \eta_c  \, , \, (D\bar{D}^{*})^0\, ,\, (D^{*}\bar{D})^0\, ,\, (D^{*}\bar{D}^{*})^0\, , \nonumber\\
Z_c^0(0^{++}) &\to&\pi^{0}\eta_c\,  ,  \, \pi^{0} \chi_{c1} \, ,  \, \rho^{0} J/\psi \, , \, \rho^{0} \psi^\prime \, , \,(D\bar{D})^0\, ,\, (D^{*}\bar{D}^{*})^0\, , \nonumber\\
Z_c^0(1^{++}) &\to& \pi^{0} \chi_{c1} \, ,  \, \rho^{0} J/\psi \, , \, \rho^{0} \psi^\prime \, , \,(D\bar{D}^{*})^0\, ,\, (D^{*}\bar{D})^0\, ,\, (D^{*}\bar{D}^{*})^0\, , \nonumber\\
Z_c^0(2^{++}) &\to&\pi^{0}\eta_c \,  ,  \, \pi^{0} \chi_{c1}  \, ,  \, \rho^{0} J/\psi \, , \, \rho^{0} \psi^\prime \, , \,(D\bar{D})^0\, ,\, (D^{*}\bar{D}^{*})^0\, ,
\end{eqnarray}
\begin{eqnarray}
X(1^{+-}) &\to&\eta J/\psi\,  ,  \,\eta\psi^\prime\,  ,  \,\eta h_c \, ,  \, \omega \eta_c  \, , \, (D\bar{D}^{*})^0\, ,\, (D^{*}\bar{D})^0\, ,\, (D^{*}\bar{D}^{*})^0\, , \nonumber\\
X(0^{++}) &\to&\eta\eta_c\,  ,  \, \eta \chi_{c1} \, ,  \, \omega J/\psi \, , \, \omega\psi^\prime \, , \,(D\bar{D})^0\, ,\, (D^{*}\bar{D}^{*})^0\, , \nonumber\\
X(1^{++}) &\to& \eta \chi_{c1} \, ,  \, \omega J/\psi \, , \, \omega \psi^\prime \, , \,(D\bar{D}^{*})^0\, ,\, (D^{*}\bar{D})^0\, ,\, (D^{*}\bar{D}^{*})^0\, , \nonumber\\
X(2^{++}) &\to&\eta\eta_c \,  ,  \, \eta \chi_{c1}  \, ,  \, \omega J/\psi \, , \, \omega \psi^\prime \, , \,(D\bar{D})^0\, ,\, (D^{*}\bar{D}^{*})^0\, ,
\end{eqnarray}
 \begin{eqnarray}
X(1^{+-}) &\to&\eta J/\psi\,  ,  \,\eta\psi^\prime\,  ,  \,\eta h_c \, ,  \, \phi \eta_c  \, , \, D_s\bar{D}_s^{*}\, ,\, D_s^{*}\bar{D}_s\, ,\, D_s^{*}\bar{D}^{*}_s\, , \nonumber\\
X(0^{++}) &\to&\eta\eta_c\,  ,  \, \eta \chi_{c1} \, ,  \, \phi J/\psi \, , \, \phi\psi^\prime \, , \,D_s\bar{D}_s\, ,\, D^{*}_s\bar{D}^{*}_s\, , \nonumber\\
X(1^{++}) &\to& \eta \chi_{c1} \, ,  \, \phi J/\psi \, , \, \phi \psi^\prime \, , \,D_s\bar{D}^{*}_s\, ,\, D^{*}_s\bar{D}_s\, ,\, D^{*}_s\bar{D}^{*}_s\, , \nonumber\\
X(2^{++}) &\to&\eta\eta_c \,  ,  \, \eta \chi_{c1}  \, ,  \, \phi J/\psi \, , \, \phi \psi^\prime \, , \,D_s\bar{D}_s\, ,\, D_s^{*}\bar{D}^{*}_s\, .
\end{eqnarray}
The LHCb collaboration observed the $h_c(4000)$ and $\chi_{c1}(4010)$ in the $D^{\ast\pm}D^{\mp}$  mass spectrum \cite{LHCb-hc4000-PRL-2024}, see the two-body strong decays of the $Z_c^0$ shown in Eq.\eqref{Zc0-aixal-decay}.

As the $c\bar{c}q\bar{s}$ tetraquark states are concerned, we present the predictions based on the direct calculations plus light-flavor $SU(3)$-breaking effects in Table \ref{Assignments-Zcs-mass} \cite{WangZG-Zcs4123-tetra-CPC-2022,WangZG-Zcs3985-mass-tetra-CPC-2021}. In Ref.\cite{WangZG-Zcs4123-tetra-CPC-2022},  we tentatively assign the $Z_c(4020/4025)$ as the $A\bar{A}$-type hidden-charm tetraquark state with the $J^{PC}=1^{+-}$ according to the analogous properties of the $Z_c(3900/3885)$ and $Z_{cs}(3985/4000)$,  and study the $A\bar{A}$-type tetraquark states without strange, with strange and with hidden-strange via the QCD sum rules in a consistent way. Then we study  the hadronic  coupling constants  of the tetraquark states without strange and with strange via the QCD sum rules based on rigorous quark-hadron duality, and obtain the total decay widths,
\begin{eqnarray}
\Gamma_{Z_{cs}} &=& 22.71\pm1.65\, ({\rm or}\, \pm 6.60)\,\rm{MeV}\, ,\nonumber\\
\Gamma_{Z_{c}} &=&29.57\pm2.30\, ({\rm or}\, \pm 9.20)\,\,\rm{MeV}\, ,
\end{eqnarray}
and suggest to search for the $Z_{cs}$ state in the mass spectrum of the $h_c K$, $J/\psi K$, $\eta_c K^*$, $D^*\bar{D}_s^*$, $D_s^*\bar{D}^*$. Slightly later, the BESIII collaboration observed an evidence for the $Z_{cs}(4123)$ \cite{BES-Zcs4123-CPC-2023}.

\begin{table}
\begin{center}
\begin{tabular}{|c|c|c|c|c|c|c|c|c|}\hline\hline
$Z_c$($X_c$)                                                            & $J^{PC}$  & $M_Z (\rm{GeV})$   & Assignments          \\ \hline

$[uc]_{S}[\overline{sc}]_{S}$                                           & $0^{++}$  & $3.97\pm0.09$      &                           \\

$[uc]_{A}[\overline{sc}]_{A}$                                           & $0^{++}$  & $4.04\pm0.09$      &                     \\

$[uc]_{\tilde{A}}[\overline{sc}]_{\tilde{A}}$                           & $0^{++}$  & $4.07\pm0.08$      &                     \\

$[uc]_{V}[\overline{sc}]_{V}$                                           & $0^{++}$  & $4.74\pm0.09$      &                     \\

$[uc]_{\tilde{V}}[\overline{sc}]_{\tilde{V}}$                           & $0^{++}$  & $5.44\pm0.09$      &                      \\

$[uc]_{P}[\overline{sc}]_{P}$                                           & $0^{++}$  & $5.58\pm0.09$      &                      \\ \hline

$[uc]_S[\overline{sc}]_{A}-[uc]_{A}[\overline{sc}]_S$                   & $1^{+-}$  & $3.99\pm0.09$      & ?\,$Z_{cs}(3985)$        \\

$[uc]_{A}[\overline{sc}]_{A}$                                           & $1^{+-}$  & $4.11\pm0.09$      & ?\,$Z_{cs}(4123)$         \\

$[uc]_S[\overline{sc}]_{\widetilde{A}}-[uc]_{\widetilde{A}}[\overline{sc}]_S$     & $1^{+-}$   & $4.10\pm0.09$    & ?\,$Z_{cs}(4123)$     \\

$[uc]_{\widetilde{A}}[\overline{sc}]_{A}-[uc]_{A}[\overline{sc}]_{\widetilde{A}}$ & $1^{+-}$   & $4.11\pm0.09$    & ?\,$Z_{cs}(4123)$     \\

$[uc]_{\widetilde{V}}[\overline{sc}]_{V}+[uc]_{V}[\overline{sc}]_{\widetilde{V}}$ & $1^{+-}$   & $4.75\pm0.10$    &      \\

$[uc]_{V}[\overline{sc}]_{V}$                                           & $1^{+-}$  & $5.55\pm0.09$      &                      \\

$[uc]_P[\overline{sc}]_{V}+[uc]_{V}[\overline{sc}]_P$                   & $1^{+-}$  & $5.54\pm0.09$      &                     \\
\hline

$[uc]_S[\overline{sc}]_{A}+[uc]_{A}[\overline{sc}]_S$                   & $1^{++}$  & $3.99\pm0.09$      &?\,$Z_{cs}(3985)$         \\

$[uc]_S[\overline{sc}]_{\widetilde{A}}+[uc]_{\widetilde{A}}[\overline{sc}]_S$     & $1^{++}$   & $4.11\pm0.09$    &?\,$Z_{cs}(4123)$     \\

$[uc]_{\widetilde{V}}[\overline{sc}]_{V}-[uc]_{V}[\overline{sc}]_{\widetilde{V}}$ & $1^{++}$   & $4.17\pm0.09$    &     \\

$[uc]_{\widetilde{A}}[\overline{sc}]_{A}+[uc]_{A}[\overline{sc}]_{\widetilde{A}}$ & $1^{++}$   & $5.28\pm0.09$    &                \\

$[uc]_P[\overline{sc}]_{V}-[uc]_{V}[\overline{sc}]_P$                   & $1^{++}$  & $5.55\pm0.09$      &                      \\
\hline

$[uc]_{A}[\overline{sc}]_{A}$                                           & $2^{++}$  & $4.17\pm0.09$      & \\

$[uc]_{V}[\overline{sc}]_{V}$                                           & $2^{++}$  & $5.49\pm0.09$      &                  \\
\hline\hline
\end{tabular}
\end{center}
\caption{ The possible assignments of the ground state hidden-charm tetraquark states with strangeness \cite{WangZG-Zcs4123-tetra-CPC-2022,WangZG-Zcs3985-mass-tetra-CPC-2021}. }\label{Assignments-Zcs-mass}
\end{table}

With the simple replacement,
\begin{eqnarray}
c \to b\, ,
\end{eqnarray}
we obtain the corresponding QCD sum rules for the hidden-bottom tetraquark states.
And we would like to present the results from the QCD sum rules in Ref.\cite{WZG-HB-tetra-EPJC-2019} directly, see Tables \ref{Hidden-bottom-Borel}-\ref{Hidden-bottom-mass-residue}.

 \begin{table}
\begin{center}
\begin{tabular}{|c|c|c|c|c|c|c|c|c|}\hline\hline
 $Z_b$                                                & $J^{PC}$ & $T^2 (\rm{GeV}^2)$ & $\sqrt{s_0}(\rm GeV) $      &$\mu(\rm{GeV})$   &pole         &$|D(10)|$ \\ \hline

$[ub]_{S}[\overline{db}]_{S}$                         & $0^{++}$ & $7.0-8.0$          & $11.16\pm0.10$              &$2.40$            &$(44-66)\%$  &$\leq3\%$   \\

$[ub]_{A}[\overline{db}]_{A}$                         & $0^{++}$ & $6.4-7.4$          & $11.14\pm0.10$              &$2.30$            &$(44-68)\%$  &$\leq11\%$    \\

$[ub]_{\tilde{A}}[\overline{db}]_{\tilde{A}}$         & $0^{++}$ & $7.2-8.2$          & $11.17\pm0.10$              &$2.40$            &$(45-66)\%$  &$\leq4\%$    \\

$[ub]_{\tilde{V}}[\overline{db}]_{\tilde{V}}$         & $0^{++}$ & $11.4-12.8$        & $12.22\pm0.10$              &$5.40$            &$(44-61)\%$  &$\ll 1\%$   \\

$[ub]_S[\overline{db}]_{A}-[ub]_{A}[\overline{db}]_S$ & $1^{+-}$ & $7.0-8.0$          & $11.16\pm0.10$              &$2.40$            &$(44-66)\%$  &$<4\%$    \\

$[ub]_{A}[\overline{db}]_{A}$                         & $1^{+-}$ & $7.1-8.1$          & $11.17\pm0.10$              &$2.40$            &$(44-65)\%$  &$\leq4\%$  \\

$[ub]_{\widetilde{A}}[\overline{db}]_{A}-[ub]_{A}[\overline{db}]_{\widetilde{A}}$ & $1^{+-}$ & $6.9-7.9$     & $11.17\pm0.10$     &$2.40$      &$(44-66)\%$ &$\leq7\%$ \\

$[ub]_S[\overline{db}]_{\widetilde{A}}-[ub]_{\widetilde{A}}[\overline{db}]_S$     & $1^{+-}$ & $7.1-8.1$     & $11.17\pm0.10$     &$2.40$      &$(44-66)\%$ &$\leq4\%$ \\

$[ub]_S[\overline{db}]_{A}+[ub]_{A}[\overline{db}]_S$ & $1^{++}$ & $7.1-8.1$          & $11.18\pm0.10$              &$2.45$            &$(44-65)\%$  &$\leq3\%$   \\

$[ub]_{\widetilde{V}}[\overline{db}]_{V}-[ub]_{V}[\overline{db}]_{\widetilde{V}}$ & $1^{++}$ & $6.8-7.8$     & $11.19\pm0.10$     &$2.50$      &$(44-66)\%$ &$\leq4\%$ \\

$[ub]_{\widetilde{A}}[\overline{db}]_{A}+[ub]_{A}[\overline{db}]_{\widetilde{A}}$ & $1^{++}$ & $9.7-11.1$    & $11.99\pm0.10$     &$4.90$      &$(44-63)\%$ &$\ll 1\%$ \\

$[ub]_{A}[\overline{db}]_{A}$                         & $2^{++}$ & $7.2-8.2$          & $11.19\pm0.10$              &$2.50$            &$(44-65)\%$         &$<4\%$ \\ \hline\hline
\end{tabular}
\end{center}
\caption{ The Borel parameters, continuum threshold parameters, energy scales of the QCD spectral densities,  pole contributions, and contributions of the vacuum condensates of dimension $10$  for the ground state hidden-bottom tetraquark states \cite{WZG-HB-tetra-EPJC-2019}. }\label{Hidden-bottom-Borel}
\end{table}

\begin{table}
\begin{center}
\begin{tabular}{|c|c|c|c|c|c|c|c|c|}\hline\hline
 $Z_b$                                                                  & $J^{PC}$  & $M_Z (\rm{GeV})$   & $\lambda_Z (\rm{GeV}^5) $             \\ \hline

$[ub]_{S}[\overline{db}]_{S}$                                           & $0^{++}$  & $10.61\pm0.09$     & $(1.10\pm0.17)\times 10^{-1}$           \\

$[ub]_{A}[\overline{db}]_{A}$                                           & $0^{++}$  & $10.60\pm0.09$     & $(1.61\pm0.25)\times 10^{-1}$           \\

$[ub]_{\tilde{A}}[\overline{db}]_{\tilde{A}}$                           & $0^{++}$  & $10.61\pm0.09$     & $(1.81\pm0.27)\times 10^{-1}$           \\

$[ub]_{\tilde{V}}[\overline{db}]_{\tilde{V}}$                           & $0^{++}$  & $11.66\pm0.12$     & $3.03\pm0.31    $                      \\

$[ub]_S[\overline{db}]_{A}-[ub]_{A}[\overline{db}]_S$                   & $1^{+-}$  & $10.61\pm0.09$     & $(1.08\pm0.16)\times 10^{-1}$        \\

$[ub]_{A}[\overline{db}]_{A}$                                           & $1^{+-}$  & $10.62\pm0.09$     & $(1.07\pm0.16)\times 10^{-1}$           \\

$[ub]_{\widetilde{A}}[\overline{db}]_{A}-[ub]_{A}[\overline{db}]_{\widetilde{A}}$ & $1^{+-}$   & $10.62\pm0.09$    & $(2.12\pm0.31)\times 10^{-1}$    \\

$[ub]_S[\overline{db}]_{\widetilde{A}}-[ub]_{\widetilde{A}}[\overline{db}]_S$     & $1^{+-}$   & $10.62\pm0.09$    & $(1.08\pm0.16)\times 10^{-1}$    \\

$[ub]_S[\overline{db}]_{A}+[ub]_{A}[\overline{db}]_S$                   & $1^{++}$  & $10.63\pm0.09$     & $(1.17\pm0.17)\times 10^{-1}$        \\

$[ub]_{\widetilde{V}}[\overline{db}]_{V}-[ub]_{V}[\overline{db}]_{\widetilde{V}}$ & $1^{++}$   & $10.63\pm0.09$    & $(1.22\pm0.20)\times 10^{-1}$    \\

$[ub]_{\widetilde{A}}[\overline{db}]_{A}+[ub]_{A}[\overline{db}]_{\widetilde{A}}$ & $1^{++}$   & $11.45\pm0.14$    & $(8.52\pm1.02)\times 10^{-1}$    \\

$[ub]_{A}[\overline{db}]_{A}$                                           & $2^{++}$  & $10.65\pm0.09$     & $(1.72\pm0.24)\times 10^{-1}$      \\ \hline\hline
\end{tabular}
\end{center}
\caption{ The masses and pole residues of the ground state hidden-bottom tetraquark states \cite{WZG-HB-tetra-EPJC-2019}. }\label{Hidden-bottom-mass-residue}
\end{table}

In calculations, we use the energy scale formula $\mu=\sqrt{M^2_{X/Y/Z}-(2{\mathbb{M}}_b)^2}$ with the effective $b$-quark mass $\mathbb{M}_b=5.17\,\rm{GeV}$ to determine the ideal energy scales of the QCD spectral densities \cite{WangZG-Zb10610-tetra-NPA-2014}, and  choose the continuum threshold parameters  $\sqrt{s_0}=Z_b+0.55\pm0.10\,\rm{GeV}$ as a constraint to extract the masses and pole residues from the QCD sum rules.
The predicted masses $10.61\pm0.09\,\rm{GeV}$ and $10.62\pm0.09\,\rm{GeV}$ for the $1^{+-}$ tetraquark states support assigning the $Z_b(10610)$ and $Z_b(10650)$ to be the  hidden-bottom tetraquark states with the $J^{PC}=1^{+-}$, more theoretical and experimental works are still needed to assign the $Z_b(10610)$ and $Z_b(10650)$ unambiguously according to the partial decay widths.

In the diquark-model, the $Z_b^{\pm}(10610)$ and $Z_b^{\pm}(10650)$ are also assigned as the $S\bar{A}-A\bar{S}$-type and $A\bar{A}$-type hidden-bottom  tetraquark states, respectively
\cite{Zb-Ali-WangW-PRD-2012,Zb-Ali-PRD-2015}. And we could extend this subsection to study the $B_c$-like tetraquark states \cite{WChen-Zcs-tetra-NPB-2021,WangZG-Bc-tetra-EPL-2019,
Azizi-Bc-tetra-PRD-2017,Ozdem-Bc-tetra-PLB-2023}.

\subsubsection{Tetraquark states with the first radial excitations}\label{Tetraquark excitations}
In this sub-section, we would like to study the first radial excitations of the hidden-charm tetraquark states and make possible assignments of the exotic states.
In Ref.\cite{Baxi-Gui}, M. S. Maior de Sousa and R. Rodrigues da Silva suggest a theoretical scheme to study the double-pole QCD sum rules, and study the quarkonia $\rho( \rm 1S,2S)$, $\psi(\rm 1S,2S)$, $\Upsilon(\rm 1S,2S)$ and $\psi_t(\rm 1S,2S)$, the predicted hadron masses are not good enough, they adopt the experimental  masses except for the $\psi_t(\rm 1S,2S)$ to study the decay constants. In Ref.\cite{WangZG-Zc4430-tetra-CTP-2015}, we extend this scheme to study the $Z_c(3900)$ and $Z_c(4430)$  as the ground state and its first radial excitation, respectively, and observed that it is impossible to reproduce the experimental masses at the same energy scale, just as in the case of the $\rho(\rm 1S,2S)$, $\psi(\rm 1S,2S)$ and $\Upsilon(\rm 1S,2S)$ \cite{Baxi-Gui}, we should resort to the energy scale formula, see Eq.\eqref{formula}, to choose the optimal energy scales independently.

 We  adopt  the correlation functions $\Pi_{\mu\nu}(p)$ and $\Pi_{\mu\nu\alpha\beta}(p)$ \cite{WangZG-Zc4600-tetra-CPC-2020}, see Eq.\eqref{CF-Pi},
and set  $J_\mu(x)=J^1_\mu(x)$, $J^2_\mu(x)$, $J^3_\mu(x)$, and
$J^1_\mu(x)=J^{SA}_{-,\mu}(x)$, $J^2_\mu(x)=J_{-,\mu}^{\widetilde{A}A}(x)$, $ J^3_\mu(x)=J_{-,\mu}^{\widetilde{V}V}(x)$ and $J_{\mu\nu}(x)=J^{AA}_{-,\mu\nu}(x)$, see Eq.\eqref{Current-SA-negative}.

At the hadron side, if we only take account of the ground state hidden-charm tetraquark states, we obtain the  QCD sum rules:
\begin{eqnarray}\label{QCDSR-I1}
\lambda^2_{Z}\, \exp\left(-\frac{M^2_{Z}}{T^2}\right)= \int_{4m_c^2}^{s_0} ds\, \rho_{QCD}(s) \, \exp\left(-\frac{s}{T^2}\right) \, ,
\end{eqnarray}
\begin{eqnarray}\label{QCDSR-I2}
M_{Z}^2&=&- \frac{\int_{4m_c^2}^{s_0} ds \, \frac{d}{d \tau }\,\rho_{QCD}(s)e^{-\tau s}}{\int_{4m_c^2}^{s_0} ds \rho_{QCD}(s)e^{-\tau s}}\mid_{\tau=\frac{1}{T^2}}\, .
\end{eqnarray}
Thereafter, we will refer the QCD sum rules in Eq.\eqref{QCDSR-I1} and Eq.\eqref{QCDSR-I2} as QCDSR I.

If we take into account the contributions of the first radially excited tetraquark states $Z_c^\prime$ in the hadronic representation, we can obtain the QCD sum rules,
\begin{eqnarray}\label{QCDSR-II}
\lambda^2_{Z}\, \exp\left(-\frac{M^2_{Z}}{T^2}\right)+\lambda^2_{Z^\prime}\, \exp\left(-\frac{M^2_{Z^\prime}}{T^2}\right)&=& \int_{4m_c^2}^{s_0^\prime} ds\, \rho_{QCD}(s) \, \exp\left(-\frac{s}{T^2}\right) \, ,
\end{eqnarray}
where the $s_0^\prime$ is continuum threshold parameter, then we   introduce the notations $\tau=\frac{1}{T^2}$, $D^n=\left( -\frac{d}{d\tau}\right)^n$, and  resort to the subscripts $1$ and $2$ to represent  the ground  state  $Z_c$ and first radially excited  state $Z_c^\prime$ respectively for simplicity.
 We rewrite the   QCD sum rules as
\begin{eqnarray}\label{QCDSR-II-re-N}
\lambda_1^2\exp\left(-\tau M_1^2 \right)+\lambda_2^2\exp\left(-\tau M_2^2 \right)&=&\Pi_{QCD}(\tau) \, ,
\end{eqnarray}
here we introduce the subscript $QCD$ to represent the QCD representation.
We derive  the QCD sum rules in Eq.\eqref{QCDSR-II-re-N} in regard  to $\tau$ to obtain
\begin{eqnarray}\label{QCDSR-II-Dr-N}
\lambda_1^2M_1^2\exp\left(-\tau M_1^2 \right)+\lambda_2^2M_2^2\exp\left(-\tau M_2^2 \right)&=&D\Pi_{QCD}(\tau) \, .
\end{eqnarray}
From Eqs.\eqref{QCDSR-II-re-N}-\eqref{QCDSR-II-Dr-N}, we obtain the QCD sum rules,
\begin{eqnarray}\label{QCDSR-II-Residue-N}
\lambda_i^2\exp\left(-\tau M_i^2 \right)&=&\frac{\left(D-M_j^2\right)\Pi_{QCD}(\tau)}{M_i^2-M_j^2} \, ,
\end{eqnarray}
where the indexes $i \neq j$.
Let us derive   the QCD sum rules in Eq.\eqref{QCDSR-II-Residue-N} in regard  to $\tau$ to obtain
\begin{eqnarray}
M_i^2&=&\frac{\left(D^2-M_j^2D\right)\Pi_{QCD}(\tau)}{\left(D-M_j^2\right)\Pi_{QCD}(\tau)} \, , \nonumber\\
M_i^4&=&\frac{\left(D^3-M_j^2D^2\right)\Pi_{QCD}(\tau)}{\left(D-M_j^2\right)\Pi_{QCD}(\tau)}\, .
\end{eqnarray}
 The squared masses $M_i^2$ satisfy the  equation,
\begin{eqnarray}
M_i^4-b M_i^2+c&=&0\, ,
\end{eqnarray}
where
\begin{eqnarray}
b&=&\frac{D^3\otimes D^0-D^2\otimes D}{D^2\otimes D^0-D\otimes D}\, , \nonumber\\
c&=&\frac{D^3\otimes D-D^2\otimes D^2}{D^2\otimes D^0-D\otimes D}\, , \nonumber\\
D^j \otimes D^k&=&D^j\Pi_{QCD}(\tau) \,  D^k\Pi_{QCD}(\tau)\, ,
\end{eqnarray}
the indexes $i=1,2$ and $j,k=0,1,2,3$.
Finally we solve above equation analytically to obtain two solutions \cite{WangZG-Zc4430-tetra-CTP-2015,Baxi-Gui},
\begin{eqnarray}\label{QCDSR-II-M1-N}
M_1^2&=&\frac{b-\sqrt{b^2-4c} }{2} \, ,
\end{eqnarray}
\begin{eqnarray}\label{QCDSR-II-M2-N}
M_2^2&=&\frac{b+\sqrt{b^2-4c} }{2} \, .
\end{eqnarray}
Thereafter, we will denote  the QCD sum rules in Eq.\eqref{QCDSR-II} and Eqs.\eqref{QCDSR-II-M1-N}-\eqref{QCDSR-II-M2-N} as QCDSR II.
In calculations, we observe that if we specify the energy scales of the  spectral densities in the QCD representation,   only one solution satisfies the energy scale formula $\mu=\sqrt{M^2_{X/Y/Z}-(2{\mathbb{M}}_c)^2}$ in the QCDSR II,  we have to abandon the other solution, i.e. the mass $M_1$ ($M_Z$). It is the unique feature  of our works \cite{WangZG-Zc4430-tetra-CTP-2015,WangZG-Zc4600-tetra-CPC-2020}, which is in contrast to Ref.\cite{Baxi-Gui}.

The Okubo-Zweig-Iizuka supper-allowed decays,
\begin{eqnarray}\label{1S-2S-3S-gaps}
Z_c&\to&J/\psi\pi\, , \nonumber \\
Z_c^\prime&\to&\psi^\prime\pi\, \nonumber \\
Z_c^{\prime\prime}&\to&\psi^{\prime\prime}\pi\, ,
\end{eqnarray}
are expected to take place easily. The energy gaps maybe have the relations $M_{Z^\prime}-M_{Z}=m_{\psi^\prime}-m_{J/\psi}$ and $M_{Z^{\prime\prime}}-M_{Z^\prime}=m_{\psi^{\prime\prime}}-m_{\psi^\prime}$.
The charmonium masses are $m_{J/\psi}=3.0969\,\rm{GeV}$,   $m_{\psi^\prime}=3.686097\,\rm{GeV}$ and $m_{\psi^{\prime\prime}}=4.039\,\rm{GeV}$ from the Particle Data Group \cite{PDG-2018}, $m_{\psi^\prime}-m_{J/\psi}=0.59\,\rm{GeV}$, $m_{\psi^{\prime\prime}}-m_{J/\psi}=0.94\,\rm{GeV}$, we can  choose the continuum threshold parameters to be  $\sqrt{s_0}=M_{Z}+0.59\,\rm{GeV}$ and $\sqrt{s_0^\prime}=M_{Z}+0.95\,\rm{GeV}$ tentatively and vary the continuum threshold parameters and Borel parameters to satisfy
the  four   criteria in Sect {\bf \ref{Tetra-Positive}}.

 After  trial and error,  we  obtain the continuum threshold parameters,  Borel windows,  best energy scales, and  contributions of the ground states for the QCDSR I, see Table \ref{Borel-QCDSR-I-netative}.
  In Table \ref{Borel-QCDSR-I-netative}, we write the continuum threshold parameters as $s_0=21.0\pm 1.0\,\rm{GeV}^2$  rather than as $s_0=(4.58\pm 0.11\,\rm{GeV})^2$ for the $[uc]_{\tilde{A}}[\bar{d}\bar{c}]_A-[uc]_A[\bar{d}\bar{c}]_{\tilde{A}}$ and $[uc]_A[\bar{d}\bar{c}]_A $ tetraquark states to remain  the same form as in Ref.\cite{WangZG-Y-tetra-EPJC-1601}.
Again we obtain the  parameters for the QCDSR II using trial  and error, see Table \ref{Borel-QCDSR-II-netative}. We take the energy scale formula $\mu=\sqrt{M^2_{X/Y/Z}-(2{\mathbb{M}}_c)^2}$  to obtain  the ideal energy scales of the spectral densities  \cite{WangZG-Zc4430-tetra-CTP-2015,WangZG-Zc4600-tetra-CPC-2020}.

From Tables \ref{Borel-QCDSR-I-netative}-\ref{Borel-QCDSR-II-netative}, we can see clearly that the contributions of the single-pole  terms  are about $(40-60)\%$ for the QCDSR I, the contributions of the two-pole  terms  are about $(70-80)\%$ for the QCDSR II, which satisfy the pole dominance very well.
In the QCDSR II, the contributions of the ground states are about $(30-45)\%$, which are much less than the  ground state  contributions in the QCDSR I,  we prefer  the QCDSR I for the ground states.

We take account of all the uncertainties of the parameters, and obtain the masses and pole residues, which are shown in Tables \ref{Ground-mass-Negatvie}-\ref{Excitation-mass-Negatvie}. From those Tables, we can see that the ground state tetraquark masses from the QCDSR I and the radially excited  tetraquark  masses from the QCDSR II satisfy the energy scale formula $\mu=\sqrt{M^2_{X/Y/Z}-(2{\mathbb{M}}_c)^2}$, where the updated value of the effective
$c$-quark mass ${\mathbb{M}}_c=1.82\,\rm{GeV}$ is adopted  \cite{WangZG-Y-tetra-EPJC-1601}. In Table \ref{Excitation-mass-Negatvie}, we also present the central values of the ground state  masses and pole residues extracted from the QCDSR II at the same energy scales. If we examine Table \ref{Excitation-mass-Negatvie}, we would observe   the ground state masses cannot satisfy the energy scale formula, and should be discarded.

\begin{figure}
\centering
\includegraphics[totalheight=6cm,width=7cm]{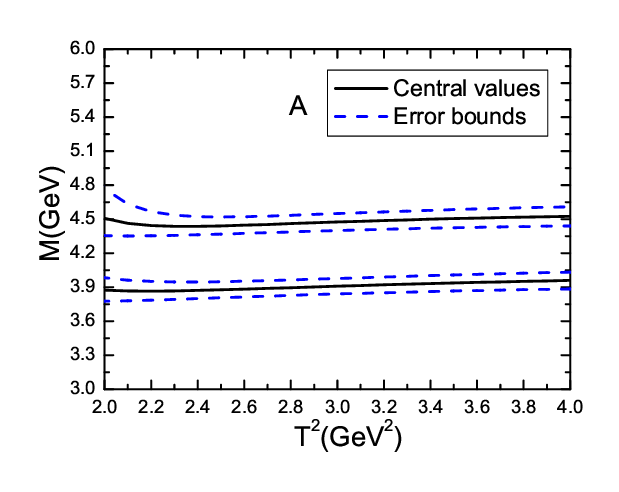}
\includegraphics[totalheight=6cm,width=7cm]{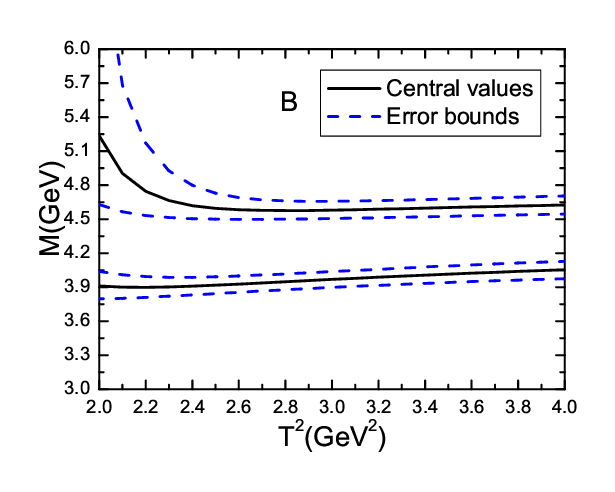}
  \caption{ The masses  with variations of the  Borel parameters $T^2$ for  the axialvector hidden-charm tetraquark states, the $A$ and $B$ represent  the $[uc]_S[\bar{d}\bar{c}]_A-[uc]_A[\bar{d}\bar{c}]_S$ and $[uc]_{\tilde{A}}[\bar{d}\bar{c}]_A-[uc]_A[\bar{d}\bar{c}]_{\tilde{A}}$ tetraquark states, respectively.  }\label{Zc-mass-1S2S}
\end{figure}

For example, in Fig.\ref{Zc-mass-1S2S}, we plot the ground state  masses from the QCDSR I and the first radially excited tetraquark  masses from the QCDSR II with variations of the Borel parameters for  the  $[uc]_S[\bar{d}\bar{c}]_A-[uc]_A[\bar{d}\bar{c}]_S$ and $[uc]_{\tilde{A}}[\bar{d}\bar{c}]_A-[uc]_A[\bar{d}\bar{c}]_{\tilde{A}}$ states. From the figure, we observe  that there indeed appear very flat platforms in the Borel windows.

We present the possible assignments explicitly in Table \ref{Identifications-Table-cqcq-positive}.
 The predicted mass $M_{Z}=4.47\pm0.09\,\rm{GeV}$ for the 2S $[uc]_S[\bar{d}\bar{c}]_A-[uc]_A[\bar{d}\bar{c}]_S$  tetraquark state exhibits  very good agreement with the experimental data $4475\pm7\,{_{-25}^{+15}}\,\rm{ MeV}$ from the   LHCb collaboration \cite{LHCb-Zc4430-PRL-2014}, which is in favor of assigning the $Z_c(4430)$ as the first radial  excitation of the   $[uc]_S[\bar{d}\bar{c}]_A-[uc]_A[\bar{d}\bar{c}]_S$ tetraquark state with the  $J^{PC}=1^{+-}$ \cite{WangZG-Zc4430-tetra-CTP-2015,WangZG-Zc4600-tetra-CPC-2020}.

The predicted mass $M_{Z}=4.60\pm0.09\,\rm{GeV}$ for the 2S     $[uc]_{\tilde{A}}[\bar{d}\bar{c}]_A-[uc]_A[\bar{d}\bar{c}]_{\tilde{A}}$ tetraquark state
 and  $M_{Z}=4.58\pm0.09\,\rm{GeV}$ for the 2S  $[uc]_A[\bar{d}\bar{c}]_A$ tetraquark state
 both  exhibit  very good agreement with the experimental data   $4600\,\rm{MeV}$ from the LHCb   collaboration \cite{LHCb-Zc4200-Zc4600-PRL-2019}, and the predicted mass $M_{Z}=4.66\pm0.10\,\rm{GeV}$ for the 1S   $[uc]_{\tilde{V}}[\bar{d}\bar{c}]_V+[uc]_V[\bar{d}\bar{c}]_{\tilde{V}}$ tetraquark state
is also compatible with the experimental data. In summary, there are three tetraquark state candidates with the $J^{PC}=1^{+-}$ for the $Z_c(4600)$.

\begin{table}
\begin{center}
\begin{tabular}{|c|c|c|c|c|c|c|c|c|}\hline\hline
 $Z_c$                                                & $T^2 (\rm{GeV}^2)$ & $s_0 $                      & $\mu(\rm{GeV})$   & pole                 \\ \hline
$[uc]_S[\bar{d}\bar{c}]_A-[uc]_A[\bar{d}\bar{c}]_S$   & $2.7-3.1$          & $(4.4\pm0.1\,\rm{GeV})^2$   & $1.4$             & $(40-63)\%$         \\ \hline
$[uc]_{\tilde{A}}[\bar{d}\bar{c}]_A-[uc]_A[\bar{d}\bar{c}]_{\tilde{A}}$   & $3.2-3.6$          & $21.0\pm1.0\,\rm{GeV}^2$    & $1.7$             & $(40-60)\%$         \\ \hline
$[uc]_{\tilde{V}}[\bar{d}\bar{c}]_V+[uc]_V[\bar{d}\bar{c}]_{\tilde{V}}$   & $3.7-4.1$          & $(5.25\pm0.10\,\rm{GeV})^2$ & $2.9$             & $(41-60)\%$         \\ \hline
$[uc]_A[\bar{d}\bar{c}]_A $                           & $3.2-3.6$          & $21.0\pm1.0\,\rm{GeV}^2$    & $1.7$             & $(41-61)\%$         \\ \hline
 \hline
\end{tabular}
\end{center}
\caption{ The Borel parameters, continuum threshold parameters, energy scales of the QCD spectral densities and pole contributions  for the QCDSR I \cite{WangZG-Zc4600-tetra-CPC-2020}. }\label{Borel-QCDSR-I-netative}
\end{table}

\begin{table}
\begin{center}
\begin{tabular}{|c|c|c|c|c|c|c|c|c|}\hline\hline
 $Z_c+Z^\prime_c$                                     & $T^2 (\rm{GeV}^2)$ & $s_0 $                      & $\mu(\rm{GeV})$   & pole ($Z_c$)                 \\ \hline
$[uc]_S[\bar{d}\bar{c}]_A-[uc]_A[\bar{d}\bar{c}]_S$   & $2.7-3.1$          & $(4.85\pm0.10\,\rm{GeV})^2$ & $2.6$             & $(72-88)\%$ ($(35-52)\%$) \\ \hline
$[uc]_{\tilde{A}}[\bar{d}\bar{c}]_A-[uc]_A[\bar{d}\bar{c}]_{\tilde{A}}$   & $3.2-3.6$          & $(4.95\pm0.10\,\rm{GeV})^2$ & $2.8$             & $(64-80)\%$ ($(30-44)\%$)\\ \hline
$[uc]_A[\bar{d}\bar{c}]_A $                           & $3.2-3.6$          & $(4.95\pm0.10\,\rm{GeV})^2$ & $2.8$             & $(64-81)\%$ ($(29-43)\%$)   \\ \hline
 \hline
\end{tabular}
\end{center}
\caption{ The Borel parameters, continuum threshold parameters, energy scales of the QCD spectral densities  and pole contributions  for the QCDSR II \cite{WangZG-Zc4600-tetra-CPC-2020}. }\label{Borel-QCDSR-II-netative}
\end{table}

\begin{table}
\begin{center}
\begin{tabular}{|c|c|c|c|c|c|c|c|c|}\hline\hline
 $Z_c$                                                    &$M_Z (\rm{GeV})$   &$\lambda_Z (\rm{GeV}^5)$    \\ \hline

$[uc]_S[\bar{d}\bar{c}]_A-[uc]_A[\bar{d}\bar{c}]_S$       &$3.90\pm0.08$      &$(2.09\pm0.33)\times 10^{-2}$    \\ \hline

$[uc]_{\tilde{A}}[\bar{d}\bar{c}]_A-[uc]_A[\bar{d}\bar{c}]_{\tilde{A}}$        &$4.01\pm0.09$      &$(5.96\pm0.94)\times 10^{-2}$  \\ \hline

$[uc]_{\tilde{V}}[\bar{d}\bar{c}]_V+[uc]_V[\bar{d}\bar{c}]_{\tilde{V}}$       &$4.66\pm0.10$      &$(1.18\pm0.22)\times 10^{-1}$   \\ \hline

$[uc]_A[\bar{d}\bar{c}]_A $          &$4.00\pm0.09$                     &$(2.91\pm0.46)\times 10^{-2}$  \\ \hline \hline
\end{tabular}
\end{center}
\caption{ The masses and pole residues of the 1S states  $Z_c$  from the QCDSR I \cite{WangZG-Zc4600-tetra-CPC-2020}. }\label{Ground-mass-Negatvie}
\end{table}

\begin{table}
\begin{center}
\begin{tabular}{|c|c|c|c|c|c|c|c|c|}\hline\hline
 $Z_c+Z_c^\prime$                                    &$M_Z (\rm{GeV})$  &$\lambda_Z (\rm{GeV}^5)$  &$M_{Z^\prime}(\rm{GeV})$ &$\lambda_{Z^\prime}(\rm{GeV}^5)$                \\ \hline
$[uc]_S[\bar{d}\bar{c}]_A-[uc]_A[\bar{d}\bar{c}]_S$  &$3.81 $           &$1.77\times10^{-2}$       &$4.47\pm0.09$            &$(6.02\pm0.80)\times10^{-2}$ \\ \hline
$[uc]_{\tilde{A}}[\bar{d}\bar{c}]_A-[uc]_A[\bar{d}\bar{c}]_{\tilde{A}}$  &$3.78$            &$3.94\times10^{-2}$       &$4.60\pm0.09$            &$(1.35\pm0.18)\times10^{-1}$  \\ \hline
$[uc]_A[\bar{d}\bar{c}]_A $                          &$3.73$            &$1.76\times10^{-2}$       &$4.58\pm0.09$            &$(6.55\pm0.85)\times10^{-2}$ \\ \hline
 \hline
\end{tabular}
\end{center}
\caption{ The masses and pole residues of the 1S  states $Z_c$ and 2S  states $Z_c^\prime$ from the QCDSR II \cite{WangZG-Zc4600-tetra-CPC-2020}. }\label{Excitation-mass-Negatvie}
\end{table}

The scheme of the QCDSR II and its modification have been applied extensively  to study the 2S tetraquark states,
such as the $Z_c(4430)$ \cite{WangZG-Zc4430-tetra-CTP-2015,WangZG-Zc4600-tetra-CPC-2020,
ChenHX-Z4600-PRD-2019,Azizi-Zc3900-Zc4430-PRD-2017,2S-Zc4430-Azizi-PLB-2020}, $Z_c(4600)$ \cite{WangZG-Zc4600-tetra-CPC-2020,ChenHX-Z4600-PRD-2019}, $X(4500)$ (as the 2S state of the $X(3915)$) \cite{2S-X4500-WZG-EPJC-2017,2S-X4500-WZG-EPJA-2017}, $X(4685)$ (as the 2S state of the $X(4140)$) \cite{2S-WZG-X4630-X4685-AHEP-2021}, and  $\Lambda_c(\rm 2S)$ and $\Xi_c(\rm 2S)$ \cite{2S-LambdaQ-CPC-2021}, etc.

\subsubsection{Tetraquark states with negative parity}\label{Tetraqurk-Negative}
In Sect.{\bf\ref{Tetra-Positive}}, we study the hidden-heavy tetraquark states with the positive parity, in this subsection, we would like to study the hidden-heavy tetraquark states with the negative  parity. Again we choose the diquark operators without explicit P-waves as the elementary building blocks, for detailed discussions about the diquark operators, see the beginning of  Sect.{\bf\ref{Tetra-Positive}}. Compared to the vector tetraquark states, it is easy to analyze the  pseudoscalar tetraquark states, and we would like to present the results directly.

Again, let us  adopt the correlation functions  $\Pi_{\mu\nu}(p)$ and $\Pi_{\mu\nu\alpha\beta}(p)$ defined in Eq.\eqref{CF-Pi}, and write down
 the currents
\begin{eqnarray}
J_\mu(x)&=&J^{PA}_{-,\mu}(x)\, ,\,\, J^{PA}_{+,\mu}(x)\, , \,\,J^{SV}_{-,\mu}(x)\, , \,\, J^{SV}_{+,\mu}(x)\, ,\,\,
J_{-,\mu}^{\widetilde{V}A}(x)\, ,\,\, J_{+,\mu}^{\widetilde{V}A}(x)\, , \,\, J_{-,\mu}^{\widetilde{A}V}(x)\, , \,\, J_{+,\mu}^{\widetilde{A}V}(x)\, , \nonumber\\
J_{\mu\nu}(x)&=&J^{S\widetilde{V}}_{-,\mu\nu}(x)\, , \,\, J^{S\widetilde{V}}_{+,\mu\nu}(x)\, , \,\, J^{P\widetilde{A}}_{-,\mu\nu}(x)\, , \,\,
J^{P\widetilde{A}}_{+,\mu\nu}(x)\, , \, \, J^{AA}_{-,\mu\nu}(x)\, ,
\end{eqnarray}
\begin{eqnarray}
J^{PA}_{-,\mu}(x)&=&\frac{\varepsilon^{ijk}\varepsilon^{imn}}{\sqrt{2}}\Big[u^{T}_j(x)Cc_k(x) \bar{d}_m(x)\gamma_\mu C \bar{c}^{T}_n(x)-u^{T}_j(x)C\gamma_\mu c_k(x)\bar{d}_m(x)C \bar{c}^{T}_n(x) \Big] \, ,\nonumber\\
J^{PA}_{+,\mu}(x)&=&\frac{\varepsilon^{ijk}\varepsilon^{imn}}{\sqrt{2}}\Big[u^{T}_j(x)Cc_k(x) \bar{d}_m(x)\gamma_\mu C \bar{c}^{T}_n(x)+u^{T}_j(x)C\gamma_\mu c_k(x)\bar{d}_m(x)C \bar{c}^{T}_n(x) \Big] \, ,\nonumber\\
J^{SV}_{-,\mu}(x)&=&\frac{\varepsilon^{ijk}\varepsilon^{imn}}{\sqrt{2}}\Big[u^{T}_j(x)C\gamma_5c_k(x) \bar{d}_m(x)\gamma_5\gamma_\mu C \bar{c}^{T}_n(x)+u^{T}_j(x)C\gamma_\mu\gamma_5 c_k(x)\bar{d}_m(x)\gamma_5C \bar{c}^{T}_n(x) \Big] \, ,\nonumber\\
J^{SV}_{+,\mu}(x)&=&\frac{\varepsilon^{ijk}\varepsilon^{imn}}{\sqrt{2}}\Big[u^{T}_j(x)C\gamma_5c_k(x) \bar{d}_m(x)\gamma_5\gamma_\mu C \bar{c}^{T}_n(x)-u^{T}_j(x)C\gamma_\mu\gamma_5 c_k(x)\bar{d}_m(x)\gamma_5C \bar{c}^{T}_n(x) \Big] \, ,\nonumber\\
\end{eqnarray}

\begin{eqnarray}
J_{-,\mu}^{\widetilde{V}A}(x)&=&\frac{\varepsilon^{ijk}\varepsilon^{imn}}{\sqrt{2}}\Big[u^{T}_j(x)C\sigma_{\mu\nu} c_k(x)\bar{d}_m(x)\gamma^\nu C \bar{c}^{T}_n(x)-u^{T}_j(x)C\gamma^\nu c_k(x)\bar{d}_m(x)\sigma_{\mu\nu} C \bar{c}^{T}_n(x) \Big] \, , \nonumber\\
J_{+,\mu}^{\widetilde{V}A}(x)&=&\frac{\varepsilon^{ijk}\varepsilon^{imn}}{\sqrt{2}}\Big[u^{T}_j(x)C\sigma_{\mu\nu} c_k(x)\bar{d}_m(x)\gamma^\nu C \bar{c}^{T}_n(x)+u^{T}_j(x)C\gamma^\nu c_k(x)\bar{d}_m(x)\sigma_{\mu\nu} C \bar{c}^{T}_n(x) \Big] \, , \nonumber\\
J_{-,\mu}^{\widetilde{A}V}(x)&=&\frac{\varepsilon^{ijk}\varepsilon^{imn}}{\sqrt{2}}\Big[u^{T}_j(x)C\sigma_{\mu\nu}\gamma_5 c_k(x)\bar{d}_m(x)\gamma_5\gamma^\nu C \bar{c}^{T}_n(x)+u^{T}_j(x)C\gamma^\nu\gamma_5 c_k(x)\bar{d}_m(x)\gamma_5\sigma_{\mu\nu} C \bar{c}^{T}_n(x) \Big] \, , \nonumber\\
J_{+,\mu}^{\widetilde{A}V}(x)&=&\frac{\varepsilon^{ijk}\varepsilon^{imn}}{\sqrt{2}}\Big[u^{T}_j(x)C\sigma_{\mu\nu}\gamma_5 c_k(x)\bar{d}_m(x)\gamma_5\gamma^\nu C \bar{c}^{T}_n(x)-u^{T}_j(x)C\gamma^\nu\gamma_5 c_k(x)\bar{d}_m(x)\gamma_5\sigma_{\mu\nu} C \bar{c}^{T}_n(x) \Big] \, , \nonumber\\
\end{eqnarray}

\begin{eqnarray}
J^{S\widetilde{V}}_{-,\mu\nu}(x)&=&\frac{\varepsilon^{ijk}\varepsilon^{imn}}{\sqrt{2}}\Big[u^{T}_j(x)C\gamma_5 c_k(x)  \bar{d}_m(x)\sigma_{\mu\nu} C \bar{c}^{T}_n(x)- u^{T}_j(x)C\sigma_{\mu\nu} c_k(x)  \bar{d}_m(x)\gamma_5 C \bar{c}^{T}_n(x) \Big] \, , \nonumber\\
J^{S\widetilde{V}}_{+,\mu\nu}(x)&=&\frac{\varepsilon^{ijk}\varepsilon^{imn}}{\sqrt{2}}\Big[u^{T}_j(x)C\gamma_5 c_k(x)  \bar{d}_m(x)\sigma_{\mu\nu} C \bar{c}^{T}_n(x)+ u^{T}_j(x)C\sigma_{\mu\nu} c_k(x)  \bar{d}_m(x)\gamma_5 C \bar{c}^{T}_n(x) \Big] \, , \nonumber\\
J^{P\widetilde{A}}_{-,\mu\nu}(x)&=&\frac{\varepsilon^{ijk}\varepsilon^{imn}}{\sqrt{2}}\Big[u^{T}_j(x)C c_k(x)  \bar{d}_m(x)\gamma_5\sigma_{\mu\nu} C \bar{c}^{T}_n(x)- u^{T}_j(x)C\sigma_{\mu\nu}\gamma_5 c_k(x)  \bar{d}_m(x)  C \bar{c}^{T}_n(x) \Big] \, , \nonumber\\
J^{P\widetilde{A}}_{+,\mu\nu}(x)&=&\frac{\varepsilon^{ijk}\varepsilon^{imn}}{\sqrt{2}}\Big[u^{T}_j(x)C c_k(x)  \bar{d}_m(x)\gamma_5\sigma_{\mu\nu} C \bar{c}^{T}_n(x)+ u^{T}_j(x)C\sigma_{\mu\nu}\gamma_5 c_k(x)  \bar{d}_m(x)  C \bar{c}^{T}_n(x) \Big] \, , \nonumber\\
\end{eqnarray}

\begin{eqnarray}
J^{AA}_{-,\mu\nu}(x)&=&\frac{\varepsilon^{ijk}\varepsilon^{imn}}{\sqrt{2}}\Big[u^{T}_j(x) C\gamma_\mu c_k(x) \bar{d}_m(x) \gamma_\nu C \bar{c}^{T}_n(x)  -u^{T}_j(x) C\gamma_\nu c_k(x) \bar{d}_m(x) \gamma_\mu C \bar{c}^{T}_n(x) \Big] \, ,  \nonumber\\
\end{eqnarray}
the subscripts $\pm$ denote the positive  and negative charge conjugation, respectively, the superscripts $P$, $S$, $A$($\widetilde{A}$) and $V$($\widetilde{V}$) denote the pseudoscalar, scalar, axialvector and vector diquark and antidiquark operators, respectively \cite{WangZG-formula-Vect-tetra-EPJC-2014,WangZG-tensor-diquark-CTP-2019,
WangZG-Y-tetra-EPJC-1601,WZG-HC-Vect-NPB-2021,WangZG-Y-tetra-EPJC-1803}.
 With the simple replacement,
 \begin{eqnarray}
 u &\to& s\, , \nonumber \\
 \bar{d} &\to& \bar{s}\, ,
 \end{eqnarray}
 we reach the corresponding currents for the  $c\bar{c}s\bar{s}$ states
 \cite{WangZG-formula-Vect-tetra-EPJC-2014,WangZG-Y-tetra-EPJC-1803,
 WZG-HC-ss-Vect-NPB-2024}.

 Under parity transformation  $\widehat{P}$, the current  operators $J_\mu(x)$ and $J_{\mu\nu}(x)$ have the  properties,
\begin{eqnarray}\label{J-parity-Vect}
\widehat{P} J_\mu(x)\widehat{P}^{-1}&=&+J^\mu(\tilde{x}) \, , \nonumber\\
\widehat{P} \tilde{J}_{\mu\nu}(x)\widehat{P}^{-1}&=&-\tilde{J}^{\mu\nu}(\tilde{x}) \, , \nonumber\\
\widehat{P} J^{AA}_{-,\mu\nu}(x)\widehat{P}^{-1}&=&+J_{AA}^{-,\mu\nu}(\tilde{x}) \, ,
\end{eqnarray}
according to the properties of the diquark operators,
\begin{eqnarray}
\widehat{P} \varepsilon^{ijk}q^T_j(x) C\Gamma c_k(x)\widehat{P}^{-1}&=&-\varepsilon^{ijk}q^T_j(\tilde{x}) C\gamma^0\Gamma\gamma^0 c_k(\tilde{x}) \, , \nonumber\\
\widehat{P} \varepsilon^{ijk}\bar{q}_j(x)\Gamma C \bar{c}^T_k(x)\widehat{P}^{-1}&=&-\varepsilon^{ijk}\bar{q}_j(\tilde{x}) \gamma^0\Gamma\gamma^0 C\bar{c}^T_k(\tilde{x}) \, ,
\end{eqnarray}
where $\tilde{J}_{\mu\nu}(x)=J^{S\widetilde{V}}_{-,\mu\nu}(x)$,
$J^{S\widetilde{V}}_{+,\mu\nu}(x)$,
$J^{P\widetilde{A}}_{-,\mu\nu}(x)$,
$J^{P\widetilde{A}}_{+,\mu\nu}(x)$, $x^\mu=(t,\vec{x})$ and $\tilde{x}^\mu=(t,-\vec{x})$. For $\Gamma=1$, $\gamma_5$, $\gamma_\mu$,  $\gamma_\mu\gamma_5$, $\sigma_{\mu\nu}$, $\sigma_{\mu\nu}\gamma_5$, we obtain $\gamma^0\Gamma\gamma^0=1$, $-\gamma_5$, $\gamma^\mu$,  $-\gamma^\mu\gamma_5$, $\sigma^{\mu\nu}$, $-\sigma^{\mu\nu}\gamma_5$.
 And we rewrite Eq.\eqref{J-parity-Vect} in  more explicit form,
\begin{eqnarray}
\widehat{P} J_i(x)\widehat{P}^{-1}&=&-J_i(\tilde{x}) \, , \nonumber\\
\widehat{P} \tilde{J}_{ij}(x)\widehat{P}^{-1}&=&-\tilde{J}_{ij}(\tilde{x}) \, , \nonumber\\
\widehat{P} J^{AA}_{-,0i}(x)\widehat{P}^{-1}&=&-J_{AA,0i}^{-}(\tilde{x}) \, ,
\end{eqnarray}
\begin{eqnarray}
\widehat{P} J_0(x)\widehat{P}^{-1}&=&+J_0(\tilde{x}) \, , \nonumber\\
\widehat{P} \tilde{J}_{0i}(x)\widehat{P}^{-1}&=&+\tilde{J}_{0i}(\tilde{x}) \, , \nonumber\\
\widehat{P} J^{AA}_{-,ij}(x)\widehat{P}^{-1}&=&+J_{AA,ij}^{-}(\tilde{x}) \, ,
\end{eqnarray}
where $i$, $j=1$, $2$, $3$.

Under charge conjugation transformation  $\widehat{C}$, the currents  $J_\mu(x)$ and $J_{\mu\nu}(x)$ have the properties,
\begin{eqnarray}
\widehat{C}J_{\pm,\mu}(x)\widehat{C}^{-1}&=&\pm J_{\pm,\mu}(x)\mid_{u\leftrightarrow d}  \, , \nonumber\\
\widehat{C}J_{\pm,\mu\nu}(x)\widehat{C}^{-1}&=&\pm J_{\pm,\mu\nu}(x)\mid_{u\leftrightarrow d}  \, .
\end{eqnarray}

 The currents  $J_\mu(x)$ and $J_{\mu\nu}(x)$ have the  symbolic quark structure  $\bar{c}c\bar{d}u$ with the isospin $I=1$ and $I_3=1$.  In the isospin limit, the currents with the  symbolic quark structures
 \begin{eqnarray}
 \bar{c}c\bar{d}u, \, \, \bar{c}c\bar{u}d, \, \, \bar{c}c\frac{\bar{u}u-\bar{d}d}{\sqrt{2}}, \, \, \bar{c}c\frac{\bar{u}u+\bar{d}d}{\sqrt{2}}
 \end{eqnarray}
 couple potentially  to the hidden-charm
tetraquark states with degenerated  masses, and the currents with the isospin $I=1$ and $0$ lead to the same QCD sum rules.
Only the currents with the symbolic quark structures $\bar{c}c\frac{\bar{u}u-\bar{d}d}{\sqrt{2}}$ and $\bar{c}c\frac{\bar{u}u+\bar{d}d}{\sqrt{2}}$ have definite
charge conjugation, again  we assume  that the  tetraquark states $\bar{c}c\bar{d}u$ have the same charge conjugation as their neutral  partners.

According to the current-hadron duality, we obtain the hadronic representation,
\begin{eqnarray}
\Pi_{\mu\nu}(p)&=&\frac{\lambda_{Y_{-}}^2}{M_{Y_{-}}^2-p^2}\left( -g_{\mu\nu}+\frac{p_{\mu}p_{\nu}}{p^2}\right) +\cdots \nonumber\\
&=&\Pi_{-}(p^2)\left( -g_{\mu\nu}+\frac{p_{\mu}p_{\nu}}{p^2}\right)+\cdots \, ,\nonumber\\
\widetilde{\Pi}_{\mu\nu\alpha\beta}(p)&=&\frac{\tilde{\lambda}_{Y_{-}}^2}{
M_{Y_{-}}^2-p^2}\left(p^2g_{\mu\alpha}g_{\nu\beta} -p^2g_{\mu\beta}g_{\nu\alpha} -g_{\mu\alpha}p_{\nu}p_{\beta}-g_{\nu\beta}p_{\mu}p_{\alpha}+g_{\mu\beta}p_{\nu}p_{\alpha}+g_{\nu\alpha}p_{\mu}p_{\beta}\right) \nonumber\\
&&+\frac{\tilde{\lambda}_{ Z_{+}}^2}{M_{Z_{+}}^2-p^2}\left( -g_{\mu\alpha}p_{\nu}p_{\beta}-g_{\nu\beta}p_{\mu}p_{\alpha}+g_{\mu\beta}p_{\nu}p_{\alpha}+g_{\nu\alpha}p_{\mu}p_{\beta}\right) +\cdots  \nonumber\\
&=&\widetilde{\Pi}_{-}(p^2)\left(p^2g_{\mu\alpha}g_{\nu\beta} -p^2g_{\mu\beta}g_{\nu\alpha} -g_{\mu\alpha}p_{\nu}p_{\beta}-g_{\nu\beta}p_{\mu}p_{\alpha}+g_{\mu\beta}p_{\nu}p_{\alpha}+g_{\nu\alpha}p_{\mu}p_{\beta}\right) \nonumber\\
&&+\widetilde{\Pi}_{+}(p^2)\left( -g_{\mu\alpha}p_{\nu}p_{\beta}-g_{\nu\beta}p_{\mu}p_{\alpha}+g_{\mu\beta}p_{\nu}p_{\alpha}+g_{\nu\alpha}p_{\mu}p_{\beta}\right) \, ,\nonumber\\
\Pi^{AA}_{\mu\nu\alpha\beta}(p)&=&\frac{\tilde{\lambda}_{ Z_{+}}^2}{M_{Z_{+}}^2-p^2}\left(p^2g_{\mu\alpha}g_{\nu\beta} -p^2g_{\mu\beta}g_{\nu\alpha} -g_{\mu\alpha}p_{\nu}p_{\beta}-g_{\nu\beta}p_{\mu}p_{\alpha}+g_{\mu\beta}p_{\nu}p_{\alpha}+g_{\nu\alpha}p_{\mu}p_{\beta}\right) \nonumber\\
&&+\frac{\tilde{\lambda}_{Y_{ -}}^2}{M_{Y_{-}}^2-p^2}\left( -g_{\mu\alpha}p_{\nu}p_{\beta}-g_{\nu\beta}p_{\mu}p_{\alpha}+g_{\mu\beta}p_{\nu}p_{\alpha}+g_{\nu\alpha}p_{\mu}p_{\beta}\right) +\cdots  \nonumber\\
&=&\widetilde{\Pi}_{+}(p^2)\left(p^2g_{\mu\alpha}g_{\nu\beta} -p^2g_{\mu\beta}g_{\nu\alpha} -g_{\mu\alpha}p_{\nu}p_{\beta}-g_{\nu\beta}p_{\mu}p_{\alpha}+g_{\mu\beta}p_{\nu}p_{\alpha}+g_{\nu\alpha}p_{\mu}p_{\beta}\right) \nonumber\\
&&+\widetilde{\Pi}_{-}(p^2)\left( -g_{\mu\alpha}p_{\nu}p_{\beta}-g_{\nu\beta}p_{\mu}p_{\alpha}+g_{\mu\beta}p_{\nu}p_{\alpha}+g_{\nu\alpha}p_{\mu}p_{\beta}\right) \, ,
\end{eqnarray}
where the pole residues are defined by
\begin{eqnarray}
  \langle 0|J_\mu(0)|Y_c^-(p)\rangle &=&\lambda_{Y_{-}}\varepsilon_\mu\, , \nonumber\\
  \langle 0|\tilde{J}_{\mu\nu}(0)|Y_c^-(p)\rangle &=& \tilde{\lambda}_{Y_{-}} \, \varepsilon_{\mu\nu\alpha\beta} \, \varepsilon^{\alpha}p^{\beta}\, , \nonumber\\
 \langle 0|\tilde{J}_{\mu\nu}(0)|Z_c^+(p)\rangle &=&\tilde{\lambda}_{Z_{+}} \left(\varepsilon_{\mu}p_{\nu}-\varepsilon_{\nu}p_{\mu} \right)\, , \nonumber\\
  \langle 0|J_{-,\mu\nu}^{AA}(0)|Z_c^+(p)\rangle &=& \tilde{\lambda}_{Z_{+}} \, \varepsilon_{\mu\nu\alpha\beta} \, \varepsilon^{\alpha}p^{\beta}\, , \nonumber\\
 \langle 0|J_{-,\mu\nu}^{AA}(0)|Y_c^-(p)\rangle &=&\tilde{\lambda}_{Y_{-}} \left(\varepsilon_{\mu}p_{\nu}-\varepsilon_{\nu}p_{\mu} \right)\, ,
\end{eqnarray}
$\lambda_{Z_{+}}=\tilde{\lambda}_{Z_{+}}M_{Z_{+}}$, $\lambda_{Y_{-}}=\tilde{\lambda}_{Y_{-}}M_{Y_{-}}$,
the  $\varepsilon_{\mu/\alpha}$  are the polarization vectors, we add the superscripts/subscripts $\pm$  to denote the positive and negative parity, respectively, and add the wide-tilde and superscript in the $\Pi_{\mu\nu\alpha\beta}(p)$ to denote
the currents $\tilde{J}_{\mu\nu}(x)$ and $J_{-,\mu\nu}^{AA}(x)$, respectively. We choose the components $\Pi_{-}(p^2)$ and $p^2\widetilde{\Pi}_{-}(p^2)$ to explore the negative parity hidden-charm tetraquark states \cite{WZG-HC-Vect-NPB-2021,WZG-HC-ss-Vect-NPB-2024}.

At the QCD side,  we calculate the vacuum condensates
up to dimension 10  and take into account the vacuum condensates $\langle\bar{q}q\rangle$, $\langle\frac{\alpha_{s}GG}{\pi}\rangle$, $\langle\bar{q}g_{s}\sigma Gq\rangle$, $\langle\bar{q}q\rangle^2$, $g_s^2\langle\bar{q}q\rangle^2$,
$\langle\bar{q}q\rangle \langle\frac{\alpha_{s}GG}{\pi}\rangle$,  $\langle\bar{q}q\rangle  \langle\bar{q}g_{s}\sigma Gq\rangle$,
$\langle\bar{q}g_{s}\sigma Gq\rangle^2$ and $\langle\bar{q}q\rangle^2 \langle\frac{\alpha_{s}GG}{\pi}\rangle$, which are vacuum expectations of the quark-gluon operators of the order $\mathcal{O}(\alpha_s^k)$ with $k\leq 1$ \cite{WZG-HC-Vect-NPB-2021,WZG-HC-ss-Vect-NPB-2024}.

We match  the hadron side with the QCD  side of the components $\Pi_{-}(p^2)$ and $p^2\widetilde{\Pi}_{-}(p^2)$ below the continuum thresholds   $s_0$ with the help of the spectral representation, and perform Borel transformation   with respect to
 $P^2=-p^2$ to obtain  the  QCD sum rules:
\begin{eqnarray}\label{QCDSR-Y}
\lambda^2_{Y}\, \exp\left(-\frac{M^2_{Y}}{T^2}\right)= \int_{4m_c^2}^{s_0} ds\, \rho_{QCD}(s) \, \exp\left(-\frac{s}{T^2}\right) \, ,
\end{eqnarray}
the  explicit expressions of the QCD spectral densities $\rho_{QCD}(s)$ are neglected for simplicity.

We derive Eq.\eqref{QCDSR-Y} with respect to  $\tau=\frac{1}{T^2}$,  and obtain the QCD sum rules for
 the masses of the  vector hidden-charm tetraquark states $Y_c$,
 \begin{eqnarray}\label{mass-QCDSR}
 M^2_{Y}&=& -\frac{\int_{4m_c^2}^{s_0} ds\frac{d}{d \tau}\rho_{QCD}(s)\exp\left(-\tau s \right)}{\int_{4m_c^2}^{s_0} ds \rho_{QCD}(s)\exp\left(-\tau s\right)}\, .
\end{eqnarray}
With a simple replacement $c \to b$, we obtain the corresponding QCD sum rules for the hidden-bottom tetraquark states directly.

We perform the same procedure as in the previous subsections,  and obtain the Borel windows, continuum threshold parameters, suitable energy scales of the QCD spectral densities and  pole contributions, which are shown explicitly in Tables \ref{BorelP-Y-cqcq}-\ref{BorelP-Y-cscs}.
From the Tables,  we can see clearly that the pole contributions are about $(40-60)\%$,  the central values are larger than $50\%$, the pole dominance is well satisfied.
In calculations, we observe  that the main contributions come from the perturbative terms, the higher dimensional condensates play a minor important role and the contributions $|D(10)|\ll 1\%$, the operator product  expansion converges  very good.

\begin{table}
\begin{center}
\begin{tabular}{|c|c|c|c|c|c|c|c|c|}\hline\hline
 $Y_c$                                                                     &$J^{PC}$  &$T^2(\rm{GeV}^2)$ &$\sqrt{s_0}(\rm GeV) $ &$\mu(\rm{GeV})$  &pole           \\ \hline

$[uc]_{P}[\overline{dc}]_{A}-[uc]_{A}[\overline{dc}]_{P}$                  &$1^{--}$  &$3.7-4.1$          &$5.15\pm0.10$          &$2.9$            &$(43-61)\%$  \\

$[uc]_{P}[\overline{dc}]_{A}+[uc]_{A}[\overline{dc}]_{P}$                  &$1^{-+}$  &$3.7-4.1$          &$5.10\pm0.10$          &$2.8$            &$(42-60)\%$ \\

$[uc]_{S}[\overline{dc}]_{V}+[uc]_{V}[\overline{dc}]_{S}$                  &$1^{--}$  &$3.2-3.6$          &$4.85\pm0.10$          &$2.4$            &$(42-62)\%$  \\

$[uc]_{S}[\overline{dc}]_{V}-[uc]_{V}[\overline{dc}]_{S}$                  &$1^{-+}$  &$3.7-4.1$          &$5.15\pm0.10$          &$2.9$            &$(41-60)\%$ \\

$[uc]_{\tilde{V}}[\overline{dc}]_{A}-[uc]_{A}[\overline{dc}]_{\tilde{V}}$  &$1^{--}$  &$3.6-4.0$          &$5.05\pm0.10$          &$2.7$            &$(42-60)\%$  \\

$[uc]_{\tilde{V}}[\overline{dc}]_{A}+[uc]_{A}[\overline{dc}]_{\tilde{V}}$  &$1^{-+}$  &$3.7-4.1$          &$5.15\pm0.10$          &$2.9$            &$(41-60)\%$ \\

$[uc]_{\tilde{A}}[\overline{dc}]_{V}+[uc]_{V}[\overline{dc}]_{\tilde{A}}$  &$1^{--}$  &$3.5-3.9$          &$5.00\pm0.10$          &$2.6$            &$(42-61)\%$    \\

$[uc]_{\tilde{A}}[\overline{dc}]_{V}-[uc]_{V}[\overline{dc}]_{\tilde{A}}$  &$1^{-+}$  &$3.6-4.0$          &$5.05\pm0.10$          &$2.7$            &$(42-61)\%$ \\

$[uc]_{S}[\overline{dc}]_{\tilde{V}}-[uc]_{\tilde{V}}[\overline{dc}]_{S}$  &$1^{--}$  &$3.4-3.8$          &$5.00\pm0.10$          &$2.6$            &$(42-61)\%$    \\

$[uc]_{S}[\overline{dc}]_{\tilde{V}}+[uc]_{\tilde{V}}[\overline{dc}]_{S}$  &$1^{-+}$  &$3.4-3.8$          &$5.00\pm0.10$          &$2.6$            &$(42-61)\%$ \\

$[uc]_{P}[\overline{dc}]_{\tilde{A}}-[uc]_{\tilde{A}}[\overline{dc}]_{P}$  &$1^{--}$  &$3.7-4.1$          &$5.10\pm0.10$          &$2.8$            &$(43-61)\%$    \\

$[uc]_{P}[\overline{dc}]_{\tilde{A}}+[uc]_{\tilde{A}}[\overline{dc}]_{P}$  &$1^{-+}$  &$3.7-4.1$          &$5.10\pm0.10$          &$2.8$            &$(43-61)\%$    \\

$[uc]_{A}[\overline{dc}]_{A}$                                              &$1^{--}$  &$3.8-4.2$          &$5.20\pm0.10$          &$3.0$            &$(42-60)\%$   \\

\hline\hline
\end{tabular}
\end{center}
\caption{ The Borel parameters, continuum threshold parameters, energy scales of the QCD spectral densities and  pole contributions  for the ground state vector hidden-charm tetraquark states \cite{WZG-HC-Vect-NPB-2021}. }\label{BorelP-Y-cqcq}
\end{table}

\begin{table}
\begin{center}
\begin{tabular}{|c|c|c|c|c|c|c|c|c|}\hline\hline
 $Y_c$                                                                     &$J^{PC}$  &$T^2(\rm{GeV}^2)$ &$\sqrt{s_0}(\rm GeV) $ &$\mu(\rm{GeV})$  &pole           \\ \hline

$[sc]_{P}[\overline{sc}]_{A}-[sc]_{A}[\overline{sc}]_{P}$                  &$1^{--}$  &$4.1-4.7$          &$5.35\pm0.10$          &$2.9$            &$(40-61)\%$  \\

$[sc]_{P}[\overline{sc}]_{A}+[sc]_{A}[\overline{sc}]_{P}$                  &$1^{-+}$  &$4.0-4.6$          &$5.30\pm0.10$          &$2.8$            &$(41-61)\%$ \\

$[sc]_{S}[\overline{sc}]_{V}+[sc]_{V}[\overline{sc}]_{S}$                  &$1^{--}$  &$3.5-4.0$          &$5.05\pm0.10$          &$2.5$            &$(41-62)\%$  \\

$[sc]_{S}[\overline{sc}]_{V}-[sc]_{V}[\overline{sc}]_{S}$                  &$1^{-+}$  &$4.0-4.6$          &$5.35\pm0.10$          &$2.9$            &$(40-60)\%$ \\

$[sc]_{\tilde{V}}[\overline{sc}]_{A}-[sc]_{A}[\overline{sc}]_{\tilde{V}}$  &$1^{--}$  &$3.9-4.5$          &$5.25\pm0.10$          &$2.7$            &$(40-61)\%$  \\

$[sc]_{\tilde{V}}[\overline{sc}]_{A}+[sc]_{A}[\overline{sc}]_{\tilde{V}}$  &$1^{-+}$  &$4.0-4.6$          &$5.35\pm0.10$          &$2.9$            &$(40-61)\%$ \\

$[sc]_{\tilde{A}}[\overline{sc}]_{V}+[sc]_{V}[\overline{sc}]_{\tilde{A}}$  &$1^{--}$  &$3.8-4.4$          &$5.20\pm0.10$          &$2.7$            &$(40-61)\%$    \\

$[sc]_{\tilde{A}}[\overline{sc}]_{V}-[sc]_{V}[\overline{sc}]_{\tilde{A}}$  &$1^{-+}$  &$3.9-4.5$          &$5.25\pm0.10$          &$2.7$            &$(40-61)\%$ \\

$[sc]_{S}[\overline{sc}]_{\tilde{V}}-[sc]_{\tilde{V}}[\overline{sc}]_{S}$  &$1^{--}$  &$3.7-4.2$          &$5.20\pm0.10$          &$2.7$            &$(41-62)\%$    \\

$[sc]_{S}[\overline{sc}]_{\tilde{V}}+[sc]_{\tilde{V}}[\overline{sc}]_{S}$  &$1^{-+}$  &$3.7-4.3$          &$5.20\pm0.10$          &$2.7$            &$(40-62)\%$ \\

$[sc]_{P}[\overline{sc}]_{\tilde{A}}-[sc]_{\tilde{A}}[\overline{sc}]_{P}$  &$1^{--}$  &$4.1-4.7$          &$5.30\pm0.10$          &$2.8$            &$(40-60)\%$    \\

$[sc]_{P}[\overline{sc}]_{\tilde{A}}+[sc]_{\tilde{A}}[\overline{sc}]_{P}$  &$1^{-+}$  &$4.1-4.7$          &$5.30\pm0.10$          &$2.8$            &$(40-60)\%$    \\

$[sc]_{A}[\overline{sc}]_{A}$                                              &$1^{--}$  &$4.2-4.9$          &$5.40\pm0.10$          &$3.0$            &$(40-60)\%$   \\

\hline\hline
\end{tabular}
\end{center}
\caption{ The Borel parameters, continuum threshold parameters, energy scales of the QCD spectral densities and  pole contributions  for the ground state hidden-charm-hidden-strange tetraquark states \cite{WZG-HC-ss-Vect-NPB-2024}. }\label{BorelP-Y-cscs}
\end{table}

We take  account of all the uncertainties of the relevant  parameters and obtain the masses and pole residues of the hidden-charm (and hidden-strange)   tetraquark states with the $J^{PC}=1^{--}$ and $1^{-+}$, which are  shown explicitly in Tables \ref{mass-residue-Y-cqcq}-\ref{mass-residue-Y-cscs}.
\begin{table}
\begin{center}
\begin{tabular}{|c|c|c|c|c|c|c|c|c|}\hline\hline
 $Y_c$                                                                     &$J^{PC}$   &$M_Y (\rm{GeV})$   &$\lambda_Y (\rm{GeV}^5) $            \\ \hline

$[uc]_{P}[\overline{dc}]_{A}-[uc]_{A}[\overline{dc}]_{P}$                  &$1^{--}$   &$4.66\pm0.07$      &$(7.19\pm0.84)\times 10^{-2}$   \\

$[uc]_{P}[\overline{dc}]_{A}+[uc]_{A}[\overline{dc}]_{P}$                  &$1^{-+}$   &$4.61\pm0.07$      &$(6.69\pm0.80)\times 10^{-2}$   \\

$[uc]_{S}[\overline{dc}]_{V}+[uc]_{V}[\overline{dc}]_{S}$                  &$1^{--}$   &$4.35\pm0.08$      &$(4.32\pm0.61)\times 10^{-2}$           \\

$[uc]_{S}[\overline{dc}]_{V}-[uc]_{V}[\overline{dc}]_{S}$                  &$1^{-+}$   &$4.66\pm0.09$      &$(6.67\pm0.82)\times 10^{-2}$           \\

$[uc]_{\tilde{V}}[\overline{dc}]_{A}-[uc]_{A}[\overline{dc}]_{\tilde{V}}$  &$1^{--}$   &$4.53\pm0.07$      &$(1.03\pm0.14)\times 10^{-1}$         \\

$[uc]_{\tilde{V}}[\overline{dc}]_{A}+[uc]_{A}[\overline{dc}]_{\tilde{V}}$  &$1^{-+}$   &$4.65\pm0.08$      &$(1.13\pm0.15)\times 10^{-1}$         \\

$[uc]_{\tilde{A}}[\overline{dc}]_{V}+[uc]_{V}[\overline{dc}]_{\tilde{A}}$  &$1^{--}$   &$4.48\pm0.08$      &$(9.47\pm1.27)\times 10^{-2}$     \\

$[uc]_{\tilde{A}}[\overline{dc}]_{V}-[uc]_{V}[\overline{dc}]_{\tilde{A}}$  &$1^{-+}$   &$4.55\pm0.07$      &$(1.06\pm0.14)\times 10^{-1}$     \\

$[uc]_{S}[\overline{dc}]_{\tilde{V}}-[uc]_{\tilde{V}}[\overline{dc}]_{S}$  &$1^{--}$   &$4.50\pm0.09$      &$(4.78\pm0.66)\times 10^{-2}$     \\

$[uc]_{S}[\overline{dc}]_{\tilde{V}}+[uc]_{\tilde{V}}[\overline{dc}]_{S}$  &$1^{-+}$   &$4.50\pm0.09$      &$(4.79\pm0.66)\times 10^{-2}$     \\

$[uc]_{P}[\overline{dc}]_{\tilde{A}}-[uc]_{\tilde{A}}[\overline{dc}]_{P}$  &$1^{--}$   &$4.60\pm0.07$      &$(6.32\pm0.74)\times 10^{-2}$     \\

$[uc]_{P}[\overline{dc}]_{\tilde{A}}+[uc]_{\tilde{A}}[\overline{dc}]_{P}$  &$1^{-+}$   &$4.61\pm0.08$      &$(6.36\pm0.74)\times 10^{-2}$     \\

$[uc]_{A}[\overline{dc}]_{A}$                                              &$1^{--}$   &$4.69\pm0.08$      &$(6.65\pm0.81)\times 10^{-2}$     \\

\hline\hline
\end{tabular}
\end{center}
\caption{ The masses and pole residues of the ground state vector  hidden-charm tetraquark states \cite{WZG-HC-Vect-NPB-2021}. }\label{mass-residue-Y-cqcq}
\end{table}

\begin{table}
\begin{center}
\begin{tabular}{|c|c|c|c|c|c|c|c|c|}\hline\hline
 $Y_c$                                                                     &$J^{PC}$   &$M_Y (\rm{GeV})$   &$\lambda_Y (\rm{GeV}^5) $            \\ \hline

$[sc]_{P}[\overline{sc}]_{A}-[sc]_{A}[\overline{sc}]_{P}$                  &$1^{--}$   &$4.80\pm0.08$      &$(8.97\pm1.09)\times 10^{-2}$   \\

$[sc]_{P}[\overline{sc}]_{A}+[sc]_{A}[\overline{sc}]_{P}$                  &$1^{-+}$   &$4.75\pm0.08$      &$(8.34\pm1.04)\times 10^{-2}$   \\

$[sc]_{S}[\overline{sc}]_{V}+[sc]_{V}[\overline{sc}]_{S}$                  &$1^{--}$   &$4.53\pm0.08$      &$(5.70\pm0.79)\times 10^{-2}$           \\

$[sc]_{S}[\overline{sc}]_{V}-[sc]_{V}[\overline{sc}]_{S}$                  &$1^{-+}$   &$4.83\pm0.09$      &$(8.39\pm1.05)\times 10^{-2}$           \\

$[sc]_{\tilde{V}}[\overline{sc}]_{A}-[sc]_{A}[\overline{sc}]_{\tilde{V}}$  &$1^{--}$   &$4.70\pm0.08$      &$(1.32\pm0.18)\times 10^{-1}$         \\

$[sc]_{\tilde{V}}[\overline{sc}]_{A}+[sc]_{A}[\overline{sc}]_{\tilde{V}}$  &$1^{-+}$   &$4.81\pm0.09$      &$(1.43\pm0.19)\times 10^{-1}$         \\

$[sc]_{\tilde{A}}[\overline{sc}]_{V}+[sc]_{V}[\overline{sc}]_{\tilde{A}}$  &$1^{--}$   &$4.65\pm0.08$      &$(1.23\pm0.17)\times 10^{-1}$     \\

$[sc]_{\tilde{A}}[\overline{sc}]_{V}-[sc]_{V}[\overline{sc}]_{\tilde{A}}$  &$1^{-+}$   &$4.71\pm0.08$      &$(1.34\pm0.17)\times 10^{-1}$     \\

$[sc]_{S}[\overline{sc}]_{\tilde{V}}-[sc]_{\tilde{V}}[\overline{sc}]_{S}$  &$1^{--}$   &$4.68\pm0.09$      &$(6.22\pm0.82)\times 10^{-2}$     \\

$[sc]_{S}[\overline{sc}]_{\tilde{V}}+[sc]_{\tilde{V}}[\overline{sc}]_{S}$  &$1^{-+}$   &$4.68\pm0.09$      &$(6.23\pm0.84)\times 10^{-2}$     \\

$[sc]_{P}[\overline{sc}]_{\tilde{A}}-[sc]_{\tilde{A}}[\overline{sc}]_{P}$  &$1^{--}$   &$4.75\pm0.08$      &$(7.93\pm0.98)\times 10^{-2}$     \\

$[sc]_{P}[\overline{sc}]_{\tilde{A}}+[sc]_{\tilde{A}}[\overline{sc}]_{P}$  &$1^{-+}$   &$4.75\pm0.08$      &$(7.94\pm0.97)\times 10^{-2}$     \\

$[sc]_{A}[\overline{sc}]_{A}$                                              &$1^{--}$   &$4.85\pm0.09$      &$(8.38\pm1.05)\times 10^{-2}$     \\

\hline\hline
\end{tabular}
\end{center}
\caption{ The masses and pole residues of the ground state  hidden-charm-hidden-strange tetraquark states \cite{WZG-HC-ss-Vect-NPB-2024}. }\label{mass-residue-Y-cscs}
\end{table}

In Tables \ref{Assignments-Table-Y-cqcq}-\ref{Assignments-Table-Y-cscs}, we present the possible assignments of the  hidden-charm tetraquark states with the $J^{PC}=1^{--}$ and $1^{-+}$ obtained in Refs.\cite{WZG-HC-Vect-NPB-2021,WZG-HC-ss-Vect-NPB-2024}. Considering the large uncertainties, it is possible to assign the
$X(4630)$ as the $[sc]_{S}[\overline{sc}]_{\tilde{V}}+[sc]_{\tilde{V}}[\overline{sc}]_{S}$ state with the   $J^{PC}=1^{-+}$, which has a mass $4.68\pm0.09\, \rm{GeV}$, see Table \ref{Assignments-Table-Y-cscs}.
In Ref.\cite{WZG-Landau-PRD-2020}, we prove that it is feasible and reliable to study  the multiquark states in the framework of  the QCD sum rules, and  obtain the prediction for  the mass of the $D_s^*\bar{D}_{s1}-D_{s1}\bar{D}_s^*$ molecular state with the exotic quantum numbers  $J^{PC}=1^{-+}$,  $M_{X}=4.67\pm0.08\,\rm{GeV}$, which was obtained before the LHCb data and is compatible with the LHCb data. The $X(4630)$ maybe have two important Fock components.

\begin{table}
\begin{center}
\begin{tabular}{|c|c|c|c|c|c|c|c|c|}\hline\hline
  $Y_c$                                                                    & $J^{PC}$  & $M_Y (\rm{GeV})$  & Assignments          \\ \hline

$[uc]_{P}[\overline{dc}]_{A}-[uc]_{A}[\overline{dc}]_{P}$                  &$1^{--}$   &$4.66\pm0.07$      & ?\,\,$Y(4660)$          \\

$[uc]_{P}[\overline{dc}]_{A}+[uc]_{A}[\overline{dc}]_{P}$                  &$1^{-+}$   &$4.61\pm0.07$      &                     \\

$[uc]_{S}[\overline{dc}]_{V}+[uc]_{V}[\overline{dc}]_{S}$                  &$1^{--}$   &$4.35\pm0.08$      & ?\,\,$Y(4360/4390)$     \\

$[uc]_{S}[\overline{dc}]_{V}-[uc]_{V}[\overline{dc}]_{S}$                  &$1^{-+}$   &$4.66\pm0.09$      &                     \\

$[uc]_{\tilde{V}}[\overline{dc}]_{A}-[uc]_{A}[\overline{dc}]_{\tilde{V}}$  &$1^{--}$   &$4.53\pm0.07$      &                     ? $Y(4500)$\\

$[uc]_{\tilde{V}}[\overline{dc}]_{A}+[uc]_{A}[\overline{dc}]_{\tilde{V}}$  &$1^{-+}$   &$4.65\pm0.08$      &                     \\

$[uc]_{\tilde{A}}[\overline{dc}]_{V}+[uc]_{V}[\overline{dc}]_{\tilde{A}}$  &$1^{--}$   &$4.48\pm0.08$      &                     ? $Y(4500)$\\

$[uc]_{\tilde{A}}[\overline{dc}]_{V}-[uc]_{V}[\overline{dc}]_{\tilde{A}}$  &$1^{-+}$   &$4.55\pm0.07$      &                     \\

$[uc]_{S}[\overline{dc}]_{\tilde{V}}-[uc]_{\tilde{V}}[\overline{dc}]_{S}$  &$1^{--}$   &$4.50\pm0.09$      &                     ? $Y(4500)$\\

$[uc]_{S}[\overline{dc}]_{\tilde{V}}+[uc]_{\tilde{V}}[\overline{dc}]_{S}$  &$1^{-+}$   &$4.50\pm0.09$      &                     \\

$[uc]_{P}[\overline{dc}]_{\tilde{A}}-[uc]_{\tilde{A}}[\overline{dc}]_{P}$  &$1^{--}$   &$4.60\pm0.07$      &                     \\

$[uc]_{P}[\overline{dc}]_{\tilde{A}}+[uc]_{\tilde{A}}[\overline{dc}]_{P}$  &$1^{-+}$   &$4.61\pm0.08$      &                     \\

$[uc]_{A}[\overline{dc}]_{A}$                                              &$1^{--}$   &$4.69\pm0.08$      & ?\,\,$Y(4660)$      \\
\hline\hline
\end{tabular}
\end{center}
\caption{ The possible assignments of the  hidden-charm tetraquark states, the isospin limit is implied \cite{WZG-HC-Vect-NPB-2021}. }\label{Assignments-Table-Y-cqcq}
\end{table}

\begin{table}
\begin{center}
\begin{tabular}{|c|c|c|c|c|c|c|c|c|}\hline\hline
 $Y_c$                                                                     &$J^{PC}$   &$M_Y (\rm{GeV})$   &Assignments            \\ \hline

$[sc]_{P}[\overline{sc}]_{A}-[sc]_{A}[\overline{sc}]_{P}$                  &$1^{--}$   &$4.80\pm0.08$    & ? $Y(4790)$ \\

$[sc]_{P}[\overline{sc}]_{A}+[sc]_{A}[\overline{sc}]_{P}$                  &$1^{-+}$   &$4.75\pm0.08$    &   \\

$[sc]_{S}[\overline{sc}]_{V}+[sc]_{V}[\overline{sc}]_{S}$                  &$1^{--}$   &$4.53\pm0.08$    & \\

$[sc]_{S}[\overline{sc}]_{V}-[sc]_{V}[\overline{sc}]_{S}$                  &$1^{-+}$   &$4.83\pm0.09$    &  \\

$[sc]_{\tilde{V}}[\overline{sc}]_{A}-[sc]_{A}[\overline{sc}]_{\tilde{V}}$  &$1^{--}$   &$4.70\pm0.08$    & ? $Y(4710)$ \\

$[sc]_{\tilde{V}}[\overline{sc}]_{A}+[sc]_{A}[\overline{sc}]_{\tilde{V}}$  &$1^{-+}$   &$4.81\pm0.09$    & \\

$[sc]_{\tilde{A}}[\overline{sc}]_{V}+[sc]_{V}[\overline{sc}]_{\tilde{A}}$  &$1^{--}$   &$4.65\pm0.08$    & ? $Y(4660)$     \\

$[sc]_{\tilde{A}}[\overline{sc}]_{V}-[sc]_{V}[\overline{sc}]_{\tilde{A}}$  &$1^{-+}$   &$4.71\pm0.08$    &   \\

$[sc]_{S}[\overline{sc}]_{\tilde{V}}-[sc]_{\tilde{V}}[\overline{sc}]_{S}$  &$1^{--}$   &$4.68\pm0.09$    & ? $Y(4660)$     \\

$[sc]_{S}[\overline{sc}]_{\tilde{V}}+[sc]_{\tilde{V}}[\overline{sc}]_{S}$  &$1^{-+}$   &$4.68\pm0.09$    &  ?? $X(4630)$ \\

$[sc]_{P}[\overline{sc}]_{\tilde{A}}-[sc]_{\tilde{A}}[\overline{sc}]_{P}$  &$1^{--}$   &$4.75\pm0.08$    &    \\

$[sc]_{P}[\overline{sc}]_{\tilde{A}}+[sc]_{\tilde{A}}[\overline{sc}]_{P}$  &$1^{-+}$   &$4.75\pm0.08$    &    \\

$[sc]_{A}[\overline{sc}]_{A}$                                              &$1^{--}$   &$4.85\pm0.09$    &     \\

\hline\hline
\end{tabular}
\end{center}
\caption{ The possible assignments of the hidden-charm-hidden-strange tetraquark states \cite{WZG-HC-ss-Vect-NPB-2024}. }\label{Assignments-Table-Y-cscs}
\end{table}

After Ref.\cite{WZG-HC-Vect-NPB-2021} was published, the $Y(4500)$ was observed by the BESIII collaboration \cite{BESIII-Y4500-KK-CPC-2022,BESIII-Y4500-DvDvpi-PRL-2023,BESIII-Y4544-omegachi-2024}.  At the energy about $4.5\,\rm{GeV}$, we obtain three hidden-charm tetraquark states with the $J^{PC}=1^{--}$, the $[uc]_{\tilde{V}}[\overline{uc}]_{A}+[dc]_{\tilde{V}}[\overline{dc}]_{A}
-[uc]_{A}[\overline{uc}]_{\tilde{V}}-[dc]_{A}[\overline{dc}]_{\tilde{V}}$,
 $[uc]_{\tilde{A}}[\overline{uc}]_{V}+[dc]_{\tilde{A}}[\overline{dc}]_{V}
 +[uc]_{V}[\overline{uc}]_{\tilde{A}}+[dc]_{V}[\overline{dc}]_{\tilde{A}}$ and
 $[uc]_{S}[\overline{uc}]_{\tilde{V}}+[dc]_{S}[\overline{dc}]_{\tilde{V}}
 -[uc]_{\tilde{V}}[\overline{uc}]_{S}-[dc]_{\tilde{V}}[\overline{dc}]_{S}$ tetraquark states have
 the masses $4.53\pm0.07\, \rm{GeV}$, $4.48\pm0.08\,\rm{GeV}$ and $4.50\pm0.09\,\rm{GeV}$, respectively \cite{WZG-HC-Vect-NPB-2021}. In Ref.\cite{WZG-Decay-Y4500-NPB-2024}, we study  the two-body strong decays systematically, i.e. we obtain thirty QCD sum rules for the hadronic coupling constants based on rigorous quark-hadron duality, then obtain the partial decay widths, therefore the total widths approximately, which are compatible with the experimental data of the
 $Y(4500)$  from the BESIII collaboration, see Sect.{\bf\ref{Subsect-Y4500}} for details.
  In Ref.\cite{LightCone-Y4500-NPB-2023}, we take the $Y(4500)$ as the  $[uc]_{\tilde{A}}[\overline{uc}]_{V}+[dc]_{\tilde{A}}[\overline{dc}]_{V}
  +[uc]_{V}[\overline{uc}]_{\tilde{A}}+[dc]_{V}[\overline{dc}]_{\tilde{A}}$ tetraquark state, and  study the three-body strong decay $Y(4500)\to D^{*-}D^{*0}\pi^+$ with the light-cone QCD sum rules, see Sect.{\bf\ref{LC-QCDSR-Y4500}} for details.

If only the mass is concerned, the $Y(4660)$ can be assigned as the $[sc]_{\tilde{A}}[\overline{sc}]_{V}+[sc]_{V}[\overline{sc}]_{\tilde{A}}$, $[sc]_{S}[\overline{sc}]_{\tilde{V}}-[sc]_{\tilde{V}}[\overline{sc}]_{S}$, $[uc]_{P}[\overline{uc}]_{A}+[dc]_{P}[\overline{dc}]_{A}-[uc]_{A}[\overline{uc}]_{P}-[dc]_{A}[\overline{dc}]_{P}$  or  $[uc]_{A}[\overline{uc}]_{A}+[dc]_{A}[\overline{dc}]_{A}$ tetraquark state, see Tables \ref{Assignments-Table-Y-cqcq}-\ref{Assignments-Table-Y-cscs}. In other words, the $Y(4660)$ maybe have several important Fock components, we have to study the strong decays in details to diagnose its nature. For example, if we assign the $Y(4660)$ as the $[sc]_{\tilde{A}}[\overline{sc}]_{V}+[sc]_{V}[\overline{sc}]_{\tilde{A}}$ or $[sc]_{S}[\overline{sc}]_{\tilde{V}}-[sc]_{\tilde{V}}[\overline{sc}]_{S}$ state, then the
strong decays $Y(4660)\to J/\psi f_0(980)$, $ \eta_c \phi$,    $ \chi_{c0}\phi$, $ D_s \bar{D}_s$, $ D_s^* \bar{D}^*_s$, $ D_s \bar{D}^*_s$,  $ D_s^* \bar{D}_s$, $  J/\psi \pi^+\pi^-$ and $  \psi^\prime \pi^+\pi^-$ are Okubo-Zweig-Iizuka super-allowed, considering the intermediate process  $f_0(980)\to \pi^+\pi^-$. Up to now, only the decays $Y(4660)\to J/\psi \pi^+\pi^-$, $\psi_2(3823)\pi^+\pi^-$, $\Lambda_c^+\Lambda_c^-$ and $D^+_sD^{-}_{s1}$ have been observed \cite{PDG-2024}, which cannot exclude the assignments $Y(4660)=[uc]_{P}[\overline{uc}]_{A}+[dc]_{P}[\overline{dc}]_{A}-[uc]_{A}[\overline{uc}]_{P}-[dc]_{A}[\overline{dc}]_{P}$  or  $[uc]_{A}[\overline{uc}]_{A}+[dc]_{A}[\overline{dc}]_{A}$, as the decay $Y(4660)\to D^+_sD^{-}_{s1}$ can take place through the re-scattering mechanism.
 We can investigate or search for the  neutral $Y_c$   tetraquark   states with the $J^{PC}=1^{--}$ and $1^{-+}$ through the two-body or three-body strong decays,
\begin{eqnarray}
  Y_c(1^{--}) &\to&  \chi_{c0}\rho/\omega \, ,\, J/\psi \pi^+\pi^- \, ,\,   J/\psi K\bar{K}\, ,\,  \eta_c\rho/\omega\, ,\, \chi_{c1}\rho/\omega\, , \nonumber\\
  Y_c(1^{-+}) &\to& J/\psi\rho/\omega \, ,\, h_c\rho/\omega \, .
\end{eqnarray}

From Tables \ref{Assignments-Table-Y-cqcq}-\ref{Assignments-Table-Y-cscs}, we observe that there is no room for the $Y(4260/4220)$. In Ref.\cite{WangZG-Y-tetra-EPJC-1803}, we choose  the currents,
\begin{eqnarray}
J^1_\mu(x) &=&J^{PA}_{-,\mu}(x)\mid_{u\bar{d}\to s\bar{s}}\, , \nonumber\\
J^4_\mu(x) &=&J^{SV}_{-,\mu}(x)\mid_{u\bar{d}\to \frac{u\bar{u}+d\bar{d}}{\sqrt{2}}}\, ,
\end{eqnarray}
to interpolate the $Y$ states, and fit the correlation functions,
\begin{eqnarray}
\Sigma_{Y=Y(4220), \,Y(4360), \,Y(4390), \, Y(4660)}\,\lambda^2_{Y}\, \exp\left(-\frac{M^2_{Y}}{T^2}\right)= \int_{4m_c^2}^{s_0} ds\, \rho_{QCD}^1(s) \, \exp\left(-\frac{s}{T^2}\right) \, , \nonumber
\end{eqnarray}
\begin{eqnarray}
\Sigma_{Y=Y(4220), \,Y(4360), \,Y(4390), \, Y(4660)}\,\lambda^2_{Y}\, \exp\left(-\frac{M^2_{Y}}{T^2}\right)= \int_{4m_c^2}^{s_0} ds\, \rho_{QCD}^4(s) \, \exp\left(-\frac{s}{T^2}\right) \, ,
\end{eqnarray}
by taking the $\lambda_Y$ as free parameters. We obtain the best values, which are shown in Table \ref{Fit-Y-Residue}, at the pertinent  energy scales  $\mu=2.4\,\rm{GeV}$ for the current $J^4_\mu(x)$ and $\mu=2.9\,\rm{GeV}$ for the current $J^1_\mu(x)$, the values of the pole residue $\lambda_{Y(4220)}$ are very small. Without introducing explicit P-waves, we cannot produce the experimental mass of the $Y(4260/4220)$ in the scenario of tetraquark state.

\begin{table}
\begin{center}
\begin{tabular}{|c|c|c|c|c|c|c|c|}\hline\hline
                 &$J_\mu^1(x)\, ,\mu=2.9\,\rm{GeV}$ &$J_\mu^1(x)\, ,\mu=2.4\,\rm{GeV}$ &$J_\mu^4(x)\, ,\mu=2.4\,\rm{GeV}$  &$J_\mu^4(x)\, ,\mu=2.9\,\rm{GeV}$ \\ \hline

$\lambda_{Y(4220)}$  &$0.38729$          &$0.29100$          &$0.02632$            &$1.56680$          \\ \hline

$\lambda_{Y(4320)}$  &$0.69720$          &$0.20867$          &$3.90360$            &$3.94290$          \\ \hline

$\lambda_{Y(4390)}$  &$0.41733$          &$0.41695$          &$0.00000$            &$0.00190$          \\ \hline

$\lambda_{Y(4660)}$  &$6.47460$          &$5.93670$          &$0.00000$            &$0.00000$          \\ \hline
 \hline
\end{tabular}
\end{center}
\caption{ The central values of the fitted   pole residues, where the unit is $10^{-2}\,\rm{GeV}^5$ \cite{WangZG-Y-tetra-EPJC-1803}. }\label{Fit-Y-Residue}
\end{table}

The $Y(4660)$ has been studied extensively via the QCD sum rules \cite{WangZG-formula-Vect-tetra-EPJC-2014,Y4660-Nielsen-NPA-2009,
Nielsen-Y-mole-PRD-2011,
WChen-Vect-axial-tetra-PRD-2011,
WZG-Y4660-CTP-2010,WangZG-Y-tetra-EPJC-1601,WZG-HC-Vect-NPB-2021,
WangZG-Y-tetra-EPJC-1803,WZG-HC-ss-Vect-NPB-2024,
Y4660-JRZhang-PRD-2011}, however, no definite conclusion can be obtained, more works are still needed to decipher its structure.

Now let us turn to the pseudoscalar tetraquark states and write down the local currents,
\begin{eqnarray}
J(x)&=&J^{+}_{AV}(x)\, ,\,\, J^{-}_{AV}(x)\, , \,\,J_{PS}^{+}(x)\, , \,\, J_{PS}^{-}(x)\, ,\,\,
J_{TT}^{+}(x)\, ,\,\,J_{TT}^{-}(x)\, ,
\end{eqnarray}
\begin{eqnarray}
J_{AV}^{+}(x)&=&\frac{\varepsilon^{ijk}\varepsilon^{imn}}{\sqrt{2}}\Big[q^{T}_j(x)C\gamma_{\mu}c_k(x) \bar{q}^{\prime}_m(x)\gamma_5\gamma^\mu C \bar{c}^{T}_n(x)-q^{T}_j(x)C\gamma_\mu\gamma_5 c_k(x)\bar{q}^{\prime }_m(x)\gamma^{\mu}C \bar{c}^{T}_n(x) \Big] \, ,\nonumber\\
J_{AV}^{-}(x)&=&\frac{\varepsilon^{ijk}\varepsilon^{imn}}{\sqrt{2}}\Big[q^{T}_j(x)C\gamma_{\mu}c_k(x) \bar{q}^{\prime}_m(x)\gamma_5\gamma^\mu C \bar{c}^{T}_n(x)+q^{T}_j(x)C\gamma_\mu\gamma_5 c_k(x)\bar{q}^{\prime }_m(x)\gamma^{\mu}C \bar{c}^{T}_n(x) \Big] \, ,\nonumber\\
J_{PS}^{+}(x)&=&\frac{\varepsilon^{ijk}\varepsilon^{imn}}{\sqrt{2}}\Big[q^{T}_j(x)C c_k(x) \bar{q}^{\prime }_m(x)\gamma_5 C \bar{c}^{T}_n(x)+q^{T}_j(x)C\gamma_5 c_k(x)\bar{q}^{\prime }_m(x)C \bar{c}^{T}_n(x) \Big] \, ,\nonumber\\
J_{PS}^{-}(x)&=&\frac{\varepsilon^{ijk}\varepsilon^{imn}}{\sqrt{2}}\Big[q^{T}_j(x)C c_k(x) \bar{q}^{\prime }_m(x)\gamma_5 C \bar{c}^{T}_n(x)-q^{T}_j(x)C\gamma_5 c_k(x)\bar{q}^{\prime }_m(x)C \bar{c}^{T}_n(x) \Big] \, ,\nonumber\\
J_{TT}^{+}(x)&=&\frac{\varepsilon^{ijk}\varepsilon^{imn}}{\sqrt{2}}\Big[q^{T}_j(x)C\sigma_{\mu\nu}c_k(x) \bar{q}^{\prime }_m(x)\gamma_5\sigma^{\mu\nu} C \bar{c}^{T}_n(x)+q^{T}_j(x)C\sigma_{\mu\nu}\gamma_5 c_k(x)\bar{q}^{\prime }_m(x)\sigma^{\mu\nu}C \bar{c}^{T}_n(x) \Big] \, ,\nonumber\\
J_{TT}^{-}(x)&=&\frac{\varepsilon^{ijk}\varepsilon^{imn}}{\sqrt{2}}\Big[q^{T}_j(x)C\sigma_{\mu\nu}c_k(x) \bar{q}^{\prime }_m(x)\gamma_5\sigma^{\mu\nu} C \bar{c}^{T}_n(x)-q^{T}_j(x)C\sigma_{\mu\nu}\gamma_5 c_k(x)\bar{q}^{\prime }_m(x)\sigma^{\mu\nu}C \bar{c}^{T}_n(x) \Big] \, ,\nonumber\\
\end{eqnarray}
 with $q$, $q^\prime=u$, $d$, $s$,  the superscripts $\pm$ symbolize the positive  and negative charge-conjugation, respectively, the subscripts $P$, $S$, $V$,  $A$ and $T$ stand for  the pseudoscalar, scalar, vector, axialvector and tensor diquark operators, respectively \cite{WZG-HC-Pseudo-NPB-2022}.

Under  parity transformation  $\widehat{P}$, the  $J(x)$ have the  property,
\begin{eqnarray}
\widehat{P} J(x)\widehat{P}^{-1}&=&-J(\tilde{x}) \, .
\end{eqnarray}
Under  charge-conjugation transformation  $\widehat{C}$, the  $J(x)$  have the property,
\begin{eqnarray}
\widehat{C}J^{\pm}(x)\widehat{C}^{-1}&=&\pm J^{\pm}(x)\mid_{q\leftrightarrow q^\prime}  \,   ,
\end{eqnarray}
and we can prove that the current $J_{TT}^{-}(x)=0$ through performing the Fierz-transformation.
Again, we take the isospin limit $m_u=m_d$, the four-quark currents with the  symbolic quark  structures,
 \begin{eqnarray}
 \bar{c}c\bar{d}u, \, \, \bar{c}c\bar{u}d, \, \, \bar{c}c\frac{\bar{u}u-\bar{d}d}{\sqrt{2}}, \, \, \bar{c}c\frac{\bar{u}u+\bar{d}d}{\sqrt{2}}\, ,
 \end{eqnarray}
 couple potentially  to the pseudoscalar  tetraquark states with  degenerated  masses. And the four-quark currents with the  symbolic quark structures,
 \begin{eqnarray}
 \bar{c}c\bar{u}s, \, \, \bar{c}c\bar{d}s, \, \, \bar{c}c\bar{s}u, \, \, \bar{c}c\bar{s}d\, ,
 \end{eqnarray}
also couple  potentially  to the pseudoscalar tetraquark states with degenerated  masses according to the isospin symmetry. And we obtain the QCD sum rules routinely.

\begin{table}
\begin{center}
\begin{tabular}{|c|c|c|c|c|c|c|c|c|}\hline\hline
 $Z_c$                                                        &$J^{PC}$  &$T^2(\rm{GeV}^2)$ &$\sqrt{s_0}(\rm GeV) $ &$\mu(\rm{GeV})$  &pole     \\ \hline

$[uc]_{A}[\bar{d}\bar{c}]_{V}-[uc]_{V}[\bar{d}\bar{c}]_{A}$   &$0^{-+}$  &$3.7-4.1$          &$5.10\pm0.10$          &$2.7$           &$(42-60)\%$ \\ \hline

$[uc]_{A}[\bar{d}\bar{c}]_{V}+[uc]_{V}[\bar{d}\bar{c}]_{A}$   &$0^{--}$  &$3.7-4.1$          &$5.10\pm0.10$          &$2.8$           &$(42-60)\%$ \\ \hline

$[uc]_{A}[\bar{s}\bar{c}]_{V}-[uc]_{V}[\bar{s}\bar{c}]_{A}$   &$0^{-+}$  &$3.7-4.1$          &$5.15\pm0.10$          &$2.7$           &$(43-61)\%$ \\ \hline

$[uc]_{A}[\bar{s}\bar{c}]_{V}+[uc]_{V}[\bar{s}\bar{c}]_{A}$   &$0^{--}$  &$3.7-4.1$          &$5.15\pm0.10$          &$2.8$           &$(43-61)\%$ \\ \hline

$[sc]_{A}[\bar{s}\bar{c}]_{V}-[sc]_{V}[\bar{s}\bar{c}]_{A}$   &$0^{-+}$  &$3.8-4.2$          &$5.20\pm0.10$          &$2.7$           &$(42-60)\%$ \\ \hline

$[sc]_{A}[\bar{s}\bar{c}]_{V}+[sc]_{V}[\bar{s}\bar{c}]_{A}$   &$0^{--}$  &$3.8-4.2$          &$5.20\pm0.10$          &$2.8$           &$(43-60)\%$ \\ \hline

$[uc]_{P}[\bar{d}\bar{c}]_{S}+[uc]_{S}[\bar{d}\bar{c}]_{P}$   &$0^{-+}$  &$3.7-4.1$          &$5.10\pm0.10$          &$2.8$           &$(42-60)\%$ \\ \hline

$[uc]_{P}[\bar{d}\bar{c}]_{S}-[uc]_{S}[\bar{d}\bar{c}]_{P}$   &$0^{--}$  &$3.7-4.1$          &$5.10\pm0.10$          &$2.8$           &$(42-60)\%$ \\ \hline

$[uc]_{P}[\bar{s}\bar{c}]_{S}+[uc]_{S}[\bar{s}\bar{c}]_{P}$   &$0^{-+}$  &$3.7-4.1$          &$5.15\pm0.10$          &$2.8$           &$(43-61)\%$ \\ \hline

$[uc]_{P}[\bar{s}\bar{c}]_{S}-[uc]_{S}[\bar{s}\bar{c}]_{P}$   &$0^{--}$  &$3.7-4.1$          &$5.15\pm0.10$          &$2.8$           &$(43-61)\%$ \\ \hline

$[sc]_{P}[\bar{s}\bar{c}]_{S}+[sc]_{S}[\bar{s}\bar{c}]_{P}$   &$0^{-+}$  &$3.8-4.2$          &$5.20\pm0.10$          &$2.8$           &$(43-61)\%$ \\ \hline

$[sc]_{P}[\bar{s}\bar{c}]_{S}-[sc]_{S}[\bar{s}\bar{c}]_{P}$   &$0^{--}$  &$3.8-4.2$          &$5.20\pm0.10$          &$2.8$           &$(43-61)\%$ \\ \hline

$[uc]_{T}[\bar{d}\bar{c}]_{T}+[uc]_{T}[\bar{d}\bar{c}]_{T}$   &$0^{-+}$  &$3.7-4.1$          &$5.10\pm0.10$          &$2.7$           &$(41-60)\%$ \\ \hline

$[uc]_{T}[\bar{s}\bar{c}]_{T}+[uc]_{T}[\bar{s}\bar{c}]_{T}$   &$0^{-+}$  &$3.7-4.1$          &$5.15\pm0.10$          &$2.7$           &$(43-61)\%$ \\ \hline

$[sc]_{T}[\bar{s}\bar{c}]_{T}+[sc]_{T}[\bar{s}\bar{c}]_{T}$   &$0^{-+}$  &$3.8-4.2$          &$5.20\pm0.10$          &$2.7$           &$(42-60)\%$ \\ \hline

\hline\hline
\end{tabular}
\end{center}
\caption{ The Borel parameters, continuum threshold parameters, energy scales of the QCD spectral densities and  pole contributions for the pseudoscalar hidden-charm tetraquark states \cite{WZG-HC-Pseudo-NPB-2022}. }\label{BorelP-Pseudu-HC}
\end{table}

\begin{table}
\begin{center}
\begin{tabular}{|c|c|c|c|c|c|c|c|c|}\hline\hline
 $Z_c$                                                        &$J^{PC}$  &$M_Z(\rm{GeV})$   &$\lambda_Z(\rm GeV^5) $     \\ \hline

$[uc]_{A}[\bar{d}\bar{c}]_{V}-[uc]_{V}[\bar{d}\bar{c}]_{A}$   &$0^{-+}$  &$4.56\pm0.08$     &$(1.33\pm0.18)\times 10^{-1}$      \\   \hline

$[uc]_{A}[\bar{d}\bar{c}]_{V}+[uc]_{V}[\bar{d}\bar{c}]_{A}$   &$0^{--}$  &$4.58\pm0.07$     &$(1.37\pm0.17)\times 10^{-1}$      \\   \hline

$[uc]_{A}[\bar{s}\bar{c}]_{V}-[uc]_{V}[\bar{s}\bar{c}]_{A}$   &$0^{-+}$  &$4.61\pm0.08$     &$(1.41\pm0.19)\times 10^{-1}$      \\   \hline

$[uc]_{A}[\bar{s}\bar{c}]_{V}+[uc]_{V}[\bar{s}\bar{c}]_{A}$   &$0^{--}$  &$4.63\pm0.08$     &$(1.45\pm0.19)\times 10^{-1}$      \\   \hline

$[sc]_{A}[\bar{s}\bar{c}]_{V}-[sc]_{V}[\bar{s}\bar{c}]_{A}$   &$0^{-+}$  &$4.66\pm0.08$     &$(1.50\pm0.20)\times 10^{-1}$      \\   \hline

$[sc]_{A}[\bar{s}\bar{c}]_{V}+[sc]_{V}[\bar{s}\bar{c}]_{A}$   &$0^{--}$  &$4.67\pm0.08$     &$(1.53\pm0.20)\times 10^{-1}$      \\   \hline

$[uc]_{P}[\bar{d}\bar{c}]_{S}+[uc]_{S}[\bar{d}\bar{c}]_{P}$   &$0^{-+}$  &$4.58\pm0.07$     &$(6.92\pm0.86)\times 10^{-2}$      \\   \hline

$[uc]_{P}[\bar{d}\bar{c}]_{S}-[uc]_{S}[\bar{d}\bar{c}]_{P}$   &$0^{--}$  &$4.58\pm0.07$     &$(6.91\pm0.86)\times 10^{-2}$      \\   \hline

$[uc]_{P}[\bar{s}\bar{c}]_{S}+[uc]_{S}[\bar{s}\bar{c}]_{P}$   &$0^{-+}$  &$4.63\pm0.07$     &$(7.30\pm0.90)\times 10^{-2}$      \\   \hline

$[uc]_{P}[\bar{s}\bar{c}]_{S}-[uc]_{S}[\bar{s}\bar{c}]_{P}$   &$0^{--}$  &$4.63\pm0.07$     &$(7.30\pm0.90)\times 10^{-2}$      \\   \hline

$[sc]_{P}[\bar{s}\bar{c}]_{S}+[sc]_{S}[\bar{s}\bar{c}]_{P}$   &$0^{-+}$  &$4.67\pm0.08$     &$(7.73\pm0.97)\times 10^{-2}$      \\   \hline

$[sc]_{P}[\bar{s}\bar{c}]_{S}-[sc]_{S}[\bar{s}\bar{c}]_{P}$   &$0^{--}$  &$4.67\pm0.08$     &$(7.73\pm0.96)\times 10^{-2}$      \\   \hline

$[uc]_{T}[\bar{d}\bar{c}]_{T}+[uc]_{T}[\bar{d}\bar{c}]_{T}$   &$0^{-+}$  &$4.57\pm0.08$     &$(4.62\pm0.61)\times 10^{-1}$      \\   \hline

$[uc]_{T}[\bar{s}\bar{c}]_{T}+[uc]_{T}[\bar{s}\bar{c}]_{T}$   &$0^{-+}$  &$4.62\pm0.08$     &$(4.89\pm0.63)\times 10^{-1}$      \\   \hline

$[sc]_{T}[\bar{s}\bar{c}]_{T}+[sc]_{T}[\bar{s}\bar{c}]_{T}$   &$0^{-+}$  &$4.67\pm0.08$     &$(5.19\pm0.67)\times 10^{-1}$      \\   \hline

\hline\hline
\end{tabular}
\end{center}
\caption{ The masses and pole residues  for the ground state pseudoscalar  hidden-charm tetraquark states \cite{WZG-HC-Pseudo-NPB-2022}. }\label{mass-residue-Pseudu-HC}
\end{table}

In Table \ref{BorelP-Pseudu-HC}, we present the Borel windows, continuum threshold parameters, energy scales of the QCD spectral densities and  pole contributions.
From the Table,  we can see distinctly that the pole contributions are about $(40-60)\%$ at the hadron side, while the central values are larger than $50\%$, the pole dominance criterion  is  satisfied  very good.
On the other hand,  the higher  vacuum condensates play a minor  important role, the operator product expansion converges  very well.

We take  all the uncertainties of the parameters into account  and acquire  the masses and pole residues,   see Table \ref{mass-residue-Pseudu-HC}. From  Tables   \ref{BorelP-Pseudu-HC}-\ref{mass-residue-Pseudu-HC}, we can see distinctly that the modified energy scale formula $\mu=\sqrt{M^2_{X/Y/Z}-(2{\mathbb{M}}_c)^2}-k\, m_s(\mu)$ with $k=0$, $1$ or $2$ is satisfied, where we subtract the small $s$-quark mass approximately to account for the small light-flavor $SU(3)$ mass-breaking effects, which is slightly different from Eq.\eqref{modify-formula}.

As can be seen distinctly from Table \ref{mass-residue-Pseudu-HC} that the lowest mass of the pseudoscalar hidden-charm tetraquark state with the symbolic quark constituents  $c\bar{c}u\bar{d}$ is about $4.56\pm0.08\,\rm{GeV}$, which is much larger than the value   $4239\pm18{}^{+45}_{-10}\,\rm{MeV}$ from the LHCb collaboration \cite{LHCb-Zc4430-PRL-2014}. In 2014, the LHCb collaboration  provided the first independent confirmation of the existence of the $Z_c(4430)$  in the $\psi^\prime \pi^-$ mass spectrum and established its spin-parity to be $J^P=1^+$ \cite{LHCb-Zc4430-PRL-2014}. Furthermore, the LHCb collaboration observed a weak evidence  for an additional resonance, the $Z_{c}(4240)$,  in the $\psi^\prime \pi^-$ mass spectrum with the preferred  spin-parity $J^P=0^-$ and the Breit-Wigner mass $4239\pm18{}^{+45}_{-10}\,\rm{MeV}$ and width $220\pm47\,{}^{+108}_{-\phantom{0}74}\,\rm{MeV}$, respectively with large uncertainties \cite{LHCb-Zc4430-PRL-2014}. If the $Z_{c}(4240)$ is confirmed by further experiments   in the future, it is an excellent candidate for the hidden-charm tetraquark state with the $J^{PC}=0^{--}$, and we should revisit the QCD sum rules for the discrepancy.

In Ref.\cite{WChen-JPC-0-tetra-PRD-2010},  Chen and Zhu study the hidden-charm tetraquark states with the symbolic quark constituents  $c\bar{c}u\bar{d}$ with   the QCD sum rules, and  obtain the ground state  masses $4.55\pm0.11\,\rm{GeV}$ for the tetraquark states with the $J^{PC}=0^{--}$, the masses $4.55\pm0.11\,\rm{GeV}$, $4.67\pm0.10\,\rm{GeV}$, $4.72\pm0.10\,\rm{GeV}$
 for the tetraquark states with the $J^{PC}=0^{-+}$. The present predictions are consistent with their calculations, again, we should bear in mind that their interpolating currents and  schemes  in treating the operator product expansion and input parameters at the QCD side   differ from the present work remarkably.
 Any current  with the same quantum numbers and same quark structure as a Fock state in a hadron couples potentially to this  hadron, so we can construct several currents to interpolate a hadron, or construct a current  to interpolate several hadrons.

From Table \ref{mass-residue-Pseudu-HC}, we can see explicitly  that the central values of the masses of the $J^{PC}=0^{-+}$ tetraquark states   with the symbolic quark constituents $uc\bar{d}\bar{c}$, $uc\bar{s}\bar{c}$, $sc\bar{s}\bar{c}$  are about $4.56\sim4.58\,\rm{GeV}$,  $4.61\sim4.62\,\rm{GeV}$ and $4.66\sim4.67\,\rm{GeV}$, respectively,
the central values of the masses of the $J^{PC}=0^{--}$ tetraquark states with the symbolic quark constituents $uc\bar{d}\bar{c}$, $uc\bar{s}\bar{c}$ and $sc\bar{s}\bar{c}$  are about $4.58\,\rm{GeV}$,  $4.63\,\rm{GeV}$  and $4.67\,\rm{GeV}$, respectively.
We obtain the conclusion tentatively that the currents $J^{+}_{AV}(x)$, $J_{PS}^{+}(x)$ and $J_{TT}^{+}(x)$ ($J^{-}_{AV}(x)$ and $J_{PS}^{-}(x)$) couple potentially to three (two) different pseudoscalar tetraquark states with almost degenerated masses,  or to one pseudoscalar tetraquark state with three (two) different Fock components.
As the currents with the same quantum numbers couple potentially to the pseudoscalar tetraquark  states with almost degenerated masses, the mixing effects cannot improve the predictions remarkably if only the tetraquark masses are concerned. All in all, we obtain reasonable predictions for the masses of the pseudoscalar tetraquark states without strange, with strange and with hidden-strange, the central values  are about $4.56\sim4.58\,\rm{GeV}$,  $4.61\sim4.63\,\rm{GeV}$ and $4.66\sim4.67\,\rm{GeV}$, respectively.

 The following two-body strong decays of the pseudoscalar hidden-charm tetraquark states,
 \begin{eqnarray}\label{Two-Body-Decay}
  Z_c(0^{--}) &\to&  \chi_{c1}\rho \, ,\, \eta_c\rho \, ,\,J/\psi a_1(1260)\, ,\, J/\psi \pi \, ,\,   D\bar{D}_0+h.c.\, ,\,  D^*\bar{D}_1+h.c.\, ,\,  D^*\bar{D}+h.c.\, , \nonumber\\
  Z_c(0^{-+}) &\to&  \chi_{c0}\pi \, ,\, \eta_c f_0(500)\, ,\,J/\psi \rho \, ,\,   D\bar{D}_0+h.c.\, ,\,  D^*\bar{D}_1+h.c.\, ,\,  D^*\bar{D}+h.c.\, , \nonumber\\
  Z_{cs}(0^{--}) &\to&  \chi_{c1}K^* \, ,\, \eta_cK^* \, ,\,J/\psi K_1\, ,\, J/\psi K \, ,\,   D_s\bar{D}_0+h.c.\, ,\, D\bar{D}_{s0}+h.c.\, ,\, D_s^*\bar{D}_1+h.c.\, ,\,\nonumber\\
  && D^*\bar{D}_{s1}+h.c.\, ,\, D_s^*\bar{D}+h.c.\, ,\, D^*\bar{D}_s+h.c.\, , \nonumber\\
  Z_{cs}(0^{-+}) &\to&  \chi_{c0}K \, ,\, \eta_c K^*_0(700) \, ,\,J/\psi K^*\, ,\,   D_s\bar{D}_0+h.c.\, ,\, D\bar{D}_{s0}+h.c.\, ,\, D_s^*\bar{D}_1+h.c.\, ,\,\nonumber\\
  && D^*\bar{D}_{s1}+h.c.\, ,\, D_s^*\bar{D}+h.c.\, ,\, D^*\bar{D}_s+h.c.\, ,\nonumber\\
  Z_{css}(0^{--}) &\to&  \chi_{c1}\phi \, ,\, \eta_c\phi \, ,\,J/\psi f_1\, ,\, J/\psi \eta \, ,\,   D_s\bar{D}_{s0}+h.c.\, ,\,  D_s^*\bar{D}_{s1}+h.c.\, ,\,  D_s^*\bar{D}_s+h.c.\, , \nonumber\\
  Z_{css}(0^{-+}) &\to&  \chi_{c0}\eta \, ,\, \eta_c f_0(980)\, ,\,J/\psi \phi \, ,\,    D_s\bar{D}_{s0}+h.c.\, ,\,  D_s^*\bar{D}_{s1}+h.c.\, ,\,  D_s^*\bar{D}_s+h.c.\, ,
\end{eqnarray}
can take place through the Okubo-Zweig-Iizuka super-allowed fall-apart mechanism, we suggest to search for the pseudoscalar hidden-charm tetraquark states in those channels.

The QCD sum rules obtained in this sub-section can be extended directly to study the tetraquark states in the bottom sector with the simple replacements $c \to b$ and $\bar{c} \to \bar{b}$.

\subsubsection{Tetraquark states with an explicit  P-wave}\label{Tetraquark-P-wave-explicit}

In the type-II diquark model \cite{Maiani-1405-Tetra-model-2},  Maiani et al assign  the $Y(4008)$, $Y(4260)$, $Y(4290/4220)$  and $Y(4630)$ as  four tetraquark  states with the $L=1$ based on the effective   spin-spin and spin-orbit  interactions, see Eq.\eqref{Maiani-model-2-L1-H}. In Ref.\cite{Ali-Maiani-Y-EPJC-2018}, A. Ali et al incorporate the dominant spin-spin, spin-orbit and tensor interactions, see Eq.\eqref{Ali-model-tensor-H}, and observe that the preferred  assignments of the tetraquark states with the  $L=1$ are the $Y(4220)$, $Y(4330)$, $Y(4390)$, $Y(4660)$.
In the diquark model, the quantum numbers of the $Y$ states are shown explicitly in Table \ref{Maiani-Ali-Y-assign}, where the $L$ is the angular momentum between the diquark and antidiquark, $\vec{S}=\vec{S}_{qc}+ \vec{S}_{\bar{q}\bar{c}}$, $\vec{J}=\vec{S}+ \vec{L}$, and $L=1$ denotes the explicit P-wave.

We take the isospin limit, and construct the interpolating currents according to the quantum numbers shown in Table \ref{Maiani-Ali-Y-assign},
\begin{eqnarray}\label{Currents-P-wave}
J^1_\mu(x)&=&\frac{\varepsilon^{ijk}\varepsilon^{imn}}{\sqrt{2}}u^{T}_j(x)C\gamma_5 c_k(x)\stackrel{\leftrightarrow}{\partial}_\mu \bar{d}_m(x)\gamma_5 C \bar{c}^{T}_n(x) \, , \nonumber \\
J^2_\mu(x)&=&\frac{\varepsilon^{ijk}\varepsilon^{imn}}{\sqrt{2}}u^{T}_j(x)C\gamma_\alpha c_k(x)\stackrel{\leftrightarrow}{\partial}_\mu \bar{d}_m(x)\gamma^\alpha C \bar{c}^{T}_n(x) \, , \nonumber \\
J^3_\mu(x)&=&\frac{\varepsilon^{ijk}\varepsilon^{imn}}{2}\left[u^{T}_j(x)C\gamma_\mu c_k(x)\stackrel{\leftrightarrow}{\partial}_\alpha \bar{d}_m(x)\gamma^\alpha C \bar{c}^{T}_n(x) \right.\nonumber\\
&&\left.+u^{T}_j(x)C\gamma^\alpha c_k(x)\stackrel{\leftrightarrow}{\partial}_\alpha \bar{d}_m(x)\gamma_\mu C \bar{c}^{T}_n(x)\right]\, , \nonumber \\
J_{\mu\nu}(x)&=&\frac{\varepsilon^{ijk}\varepsilon^{imn}}{2\sqrt{2}}\left[u^{T}_j(x)C\gamma_5 c_k(x)\stackrel{\leftrightarrow}{\partial}_\mu \bar{d}_m(x)\gamma_\nu C \bar{c}^{T}_n(x) \right.\nonumber\\
&&+u^{T}_j(x)C\gamma_\nu c_k(x)\stackrel{\leftrightarrow}{\partial}_\mu \bar{d}_m(x)\gamma_5 C \bar{c}^{T}_n(x) \nonumber  \\
&&-u^{T}_j(x)C\gamma_5 c_k(x)\stackrel{\leftrightarrow}{\partial}_\nu \bar{d}_m(x)\gamma_\mu C \bar{c}^{T}_n(x) \nonumber \\
&&\left.-u^{T}_j(x)C\gamma_\mu c_k(x)\stackrel{\leftrightarrow}{\partial}_\nu \bar{d}_m(x)\gamma_5 C \bar{c}^{T}_n(x)\right]\, ,
\end{eqnarray}
where $\stackrel{\leftrightarrow}{\partial}_\mu=
\stackrel{\rightarrow}{\partial}_\mu-\stackrel{\leftarrow}{\partial}_\mu$ embodies the explicit P-wave.

\begin{table}
\begin{center}
\begin{tabular}{|c|c|c|c|c|c|c|c|}\hline\hline
$|S_{qc}, S_{\bar{q}\bar{c}}; S, L; J\rangle$                                   &\cite{Maiani-1405-Tetra-model-2} &\cite{Ali-Maiani-Y-EPJC-2018} &Currents \\ \hline

$|0, 0; 0, 1; 1\rangle$                                                         &$Y(4008)$             &$Y(4220)$           &$J_\mu^1(x)$   \\ \hline

$\frac{1}{\sqrt{2}}\left(|1, 0; 1, 1; 1\rangle+|0, 1; 1, 1; 1\rangle\right)$    &$Y(4260)$             &$Y(4330)$           &$J_{\mu\nu}(x)$   \\ \hline

$|1, 1; 0, 1; 1\rangle$                                                         &$Y(4290/4220)$        &$Y(4390)$           &$J_\mu^2(x)$  \\ \hline

$|1, 1; 2, 1; 1\rangle$                                                         &$Y(4630)$             &$Y(4660)$           &$J_\mu^3(x)$  \\ \hline

$|1, 1; 2, 3; 1\rangle$                                                         &                      &                    &\\ \hline  \hline
\end{tabular}
\end{center}
\caption{ The vector tetraquark states, possible assignments  and  corresponding vector tetraquark currents, where the mixing effects are neglected \cite{WangZG-HC-Vect-Derive-EPJC-2019}.} \label{Maiani-Ali-Y-assign}
\end{table}
Under charge conjugation transformation  $\widehat{C}$, the currents $J_\mu(x)$ and $J_{\mu\nu}(x)$ have the property,
\begin{eqnarray}
\widehat{C}J_{\mu}(x)\widehat{C}^{-1}&=&- J_{\mu}(x) \, , \nonumber\\
\widehat{C}J_{\mu\nu}(x)\widehat{C}^{-1}&=&- J_{\mu\nu}(x) \, ,
\end{eqnarray}
the currents have definite charge conjugation.

We choose  the currents $J_\mu(x)=J^1_\mu(x)$, $J^2_\mu(x)$, $J^3_\mu(x)$ and $J_{\mu\nu}(x)$ and resort to the correlation functions in Eq.\eqref{CF-Pi} to study the vector tetraquark states using the modified  energy scale formula,
\begin{eqnarray}\label{Modify-Energy-formula-P}
 \mu&=&\sqrt{M^2_{X/Y/Z}-(2{\mathbb{M}}_c+0.5\,\rm{GeV})^2}\, ,\nonumber\\
 & =&\sqrt{M^2_{X/Y/Z}-(4.1\,\rm{GeV})^2}\, ,
\end{eqnarray}
   to determine the ideal  energy scales of the QCD spectral densities, and reexamine the possible assignments of the $Y$ states  \cite{WangZG-Lowest-Y-tetra-EPJC-1803,WangZG-HC-Vect-Derive-EPJC-2019}. The numerical results are shown explicitly in Tables \ref{Borel-P-wave-Y}-\ref{masses-P-wave-Y}.

The predicted mass $M_{Y}=4.24\pm0.10\,\rm{GeV}$ of the $|0, 0; 0, 1; 1\rangle$ tetraquark  state is in excellent agreement with the experimental data  $M_{Y(4220)}=4222.0\pm3.1\pm 1.4\,  \rm{MeV}$ from the BESIII    collaboration \cite{BES-Y4220-Y4320-PRL-2017}, or $M_{Y(4260)}=4230.0\pm 8.0\,  \rm{MeV}$ from the Particle Data Group \cite{PDG-2018}, which supports  assigning the $Y(4260/4220)$  as the  $C\gamma_5\otimes\stackrel{\leftrightarrow}{\partial}_\mu\otimes \gamma_5C$   type vector tetraquark state.
We obtain the {\bf  vector hidden-charm tetraquark state with the lowest mass} up to now.

There have been other possible assignments for the $Y(4260)$ states, such as the hybrid states \cite{Born-Braaten-PRD-2014,Y4260-ZhuSL-PLB-2005,Y4260-Close-PLB-2005,Y4260-Kou-PLB-2005,Y4260-LiuL-JHEP-2012},  molecular state \cite{ZhaoQ-Y4260-Zc3900-Tri-pole-PRL-2013,GuoFK-Y4260-mole-PRD-2014,OBE-Y4260-mole-DingGJ-PRD-2009,
FKGuo-mole-Review-Progr-2021,Y4260-Latt-Chiu-PRD-2006}, baryonium states \cite{Y4260-QiaoCF-PLB-2006,Y4260-QiaoCF-JPG-2008}, hadro-charmonium state \cite{Y4260-Voloshin-MPLA-2014}, interference effect \cite{Y4260-ChenDY-PRD-2011-NON,Y4260-ChenDY-PRD-2016-NON}, etc.

 \begin{table}
\begin{center}
\begin{tabular}{|c|c|c|c|c|c|c|c|}\hline\hline
$|S_{qc}, S_{\bar{q}\bar{c}}; S, L; J\rangle$                                &$\mu(\rm{GeV})$ &$T^2(\rm{GeV}^2)$ &$\sqrt{s_0}(\rm{GeV})$ &pole    &$D(10)$ \\ \hline

$|0, 0; 0, 1; 1\rangle$                                                      &$1.1$         &$2.2-2.8$       &$4.80\pm0.10$        &$(49-81)\%$   &$\leq 1\%$ \\ \hline

$|1, 1; 0, 1; 1\rangle$                                                      &$1.2$         &$2.2-2.8$       &$4.85\pm0.10$        &$(45-79)\%$   &$(1-5)\%$ \\ \hline

$\frac{1}{\sqrt{2}}\left(|1, 0; 1, 1; 1\rangle+|0, 1; 1, 1; 1\rangle\right)$ &$1.3$         &$2.6-3.2$       &$4.90\pm0.10$        &$(46-75)\%$   &$\ll 1\%$ \\ \hline

$|1, 1; 2, 1; 1\rangle$                                                      &$1.4$         &$2.6-3.2$       &$4.90\pm0.10$        &$(40-71)\%$   &$\leq 1\%$ \\ \hline \hline
\end{tabular}
\end{center}
\caption{ The Borel windows $T^2$, continuum threshold parameters $s_0$, ideal energy scales, pole contributions of the ground states and contributions of the vacuum condensates of dimension  $10$ \cite{WangZG-HC-Vect-Derive-EPJC-2019}.   }\label{Borel-P-wave-Y}
\end{table}

\begin{table}
\begin{center}
\begin{tabular}{|c|c|c|c|c|c|c|c|}\hline\hline
$|S_{qc}, S_{\bar{q}\bar{c}}; S, L; J\rangle$                                &$M_Y(\rm{GeV})$  &$\lambda_Y(10^{-2}\rm{GeV}^6)$ &Assignments  \\ \hline

$|0, 0; 0, 1; 1\rangle$                                                      &$4.24\pm0.10$    &$2.31 \pm0.45$  &$Y(4220)$           \\ \hline

$|1, 1; 0, 1; 1\rangle$                                                      &$4.28\pm0.10$    &$4.93 \pm1.00$  &$Y(4220/4320)$            \\ \hline

$\frac{1}{\sqrt{2}}\left(|1, 0; 1, 1; 1\rangle+|0, 1; 1, 1; 1\rangle\right)$ &$4.31\pm0.10$    &$2.99 \pm0.54$   &$Y(4320/4390)$             \\ \hline

$|1, 1; 2, 1; 1\rangle$                                                      &$4.33\pm0.10$    &$7.35 \pm1.39$   &$Y(4320/4390)$             \\ \hline \hline
\end{tabular}
\end{center}
\caption{ The masses, pole residues and possible assignments of the vector tetraquark states \cite{WangZG-HC-Vect-Derive-EPJC-2019}. }\label{masses-P-wave-Y}
\end{table}

From Tables \ref{Assignments-Table-Y-cqcq} and \ref{masses-P-wave-Y}, we can see explicitly that there are no rooms for the $Y(4008)$ and $Y(4750)$ in the  hidden-charm  tetraquark scenario. If we assign the $Y(4220/4230/4260)$  as the ground state, then we could assign the $Y(4750)$  as its first radial excitation  according to the mass gap $M_{Y(4750)}-M_{Y(4260)}=0.51\,\rm{GeV}$ \cite{WangZG-HC-Vect-Derive-CPC-2024}.

We carry out the calculations routinely to obtain two QCD sum rules,
\begin{eqnarray}\label{QCDST-1P}
\lambda^2_{Y}\, \exp\left(-\frac{M^2_{Y}}{T^2}\right)&=& \int_{4m_c^2}^{s_0} ds  \,\rho_{QCD}(s)  \exp\left(-\frac{s}{T^2}\right) \, ,
\end{eqnarray}
\begin{eqnarray}\label{QCDST-2P}
\lambda^2_{Y}\, \exp\left(-\frac{M^2_{Y}}{T^2}\right)+\lambda^2_{Y^\prime}\, \exp\left(-\frac{M^2_{Y^\prime}}{T^2}\right)&=& \int_{4m_c^2}^{s^\prime_0} ds  \,\rho_{QCD}(s) \exp\left(-\frac{s}{T^2}\right) \, ,
\end{eqnarray}
where the $s_0$  and $s_0^\prime$ correspond to the ground states $Y$ and first radial excitations $Y^\prime$, respectively \cite{WangZG-HC-Vect-Derive-CPC-2024}.

We  adopt the QCDSR II,  see  Eq.\eqref{QCDSR-II} and Eqs.\eqref{QCDSR-II-M1-N}-\eqref{QCDSR-II-M2-N}, to study the radially excited states,  and obtain the Borel windows, continuum threshold parameters, suitable energy scales and  pole contributions, which are shown explicitly in Tables \ref{Borel-1P}-\ref{Borel-1P-2P}.
From the tables,  we can see explicitly that the pole contributions of the 1P states (the 1P plus 2P states) are about $(40-60)\%$ ($(67-85)\%$),    the pole dominance is satisfied very well. On the other hand, the contributions from the highest dimensional condensates play a minor important role,   $|D(10)|< 3\%$ or $\ll 1\%$ ($< 1\%$ or $\ll 1\%$) for the 1P states (the 1P plus 2P states), the operator product  expansion converges  very good and better than that in our previous work \cite{WangZG-HC-Vect-Derive-EPJC-2019}, see Table \ref{Borel-P-wave-Y}.

The predicted masses and pole residues are presented in Table \ref{mass-1P-2P}. From Tables \ref{Borel-1P-2P}-\ref{mass-1P-2P}, we can see explicitly that the modified energy scale formula, see Eq.\eqref{Modify-Energy-formula-P},   can be well satisfied, and the relations  $\sqrt{s_0}=M_Y+0.50\sim0.55\pm 0.10\,\rm{GeV}$ and $\sqrt{s^\prime_0}=M_{Y^\prime}+0.40\pm 0.10\,\rm{GeV}$ are hold, see Eq.\eqref{1S-2S-3S-gaps}.

\begin{figure}
 \centering
 \includegraphics[totalheight=5cm,width=7cm]{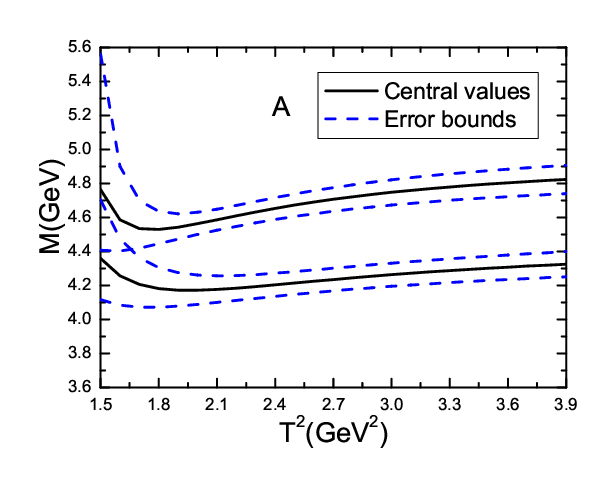}
 \includegraphics[totalheight=5cm,width=7cm]{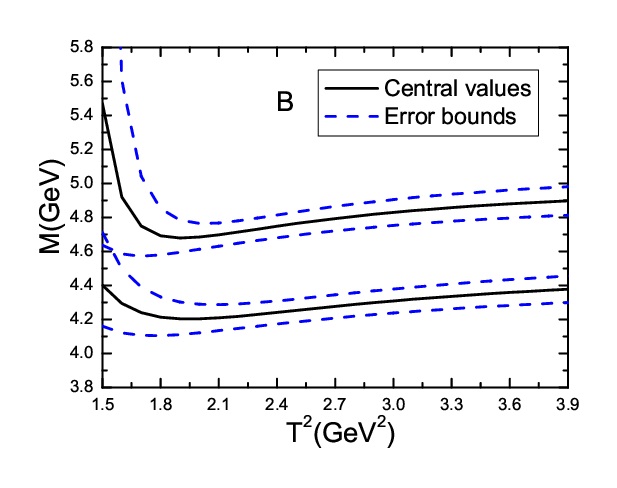}
 \includegraphics[totalheight=5cm,width=7cm]{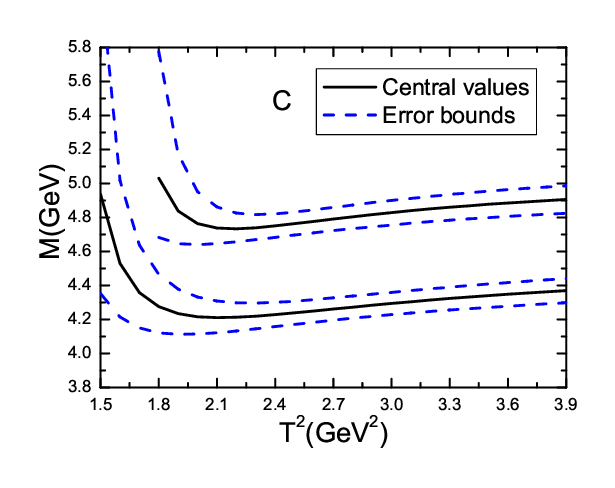}
 \includegraphics[totalheight=5cm,width=7cm]{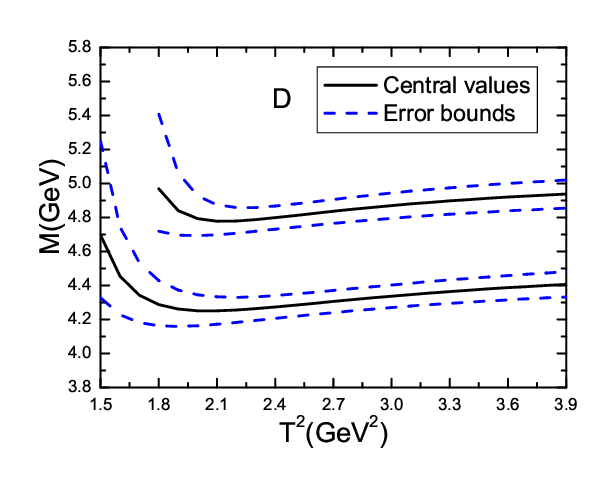}
  \caption{ The masses of the vector tetraquark states with variations of the Borel parameters $T^2$, where  the $A$, $B$, $C$ and $D$ stand for  the
  $|0, 0; 0, 1; 1\rangle$, $|1, 1; 0, 1; 1\rangle$, $\frac{1}{\sqrt{2}}\left(|1, 0; 1, 1; 1\rangle+|0, 1; 1, 1; 1\rangle\right)$ and $|1, 1; 2, 1; 1\rangle$  states,  respectively; the lower lines and upper lines stand for  the ground states and first radial excitations, respectively.   }\label{fig-mass-1P-2P}
\end{figure}

In Fig.\ref{fig-mass-1P-2P}, we plot the masses  of the 1P and 2P hidden-charm tetraquark states with the  $J^{PC}=1^{--}$. From the figure, we can see explicitly  that  there appear flat platforms in the Borel windows, the uncertainties come from the Borel parameters are rather small.

In Table \ref{Assignment-Y}, we present the possible assignments of the vector tetraquark states \cite{WangZG-HC-Vect-Derive-CPC-2024}. From the table, we can see explicitly that there is a room to accommodate the $Y(4750)$, i.e. the $Y(4220/4260)$ and $Y(4750)$ can be assigned as the ground state and first radial excited state of the $C\gamma_5\stackrel{\leftrightarrow}{\partial}_\mu \gamma_5C $ type tetraquark states with the $J^{PC}=1^{--}$, respectively \cite{WangZG-HC-Vect-Derive-CPC-2024}.

We can study the corresponding hidden-bottom tetraquark states with the simple replacement $c \to b$ in Eq.\eqref{Currents-P-wave}. In Ref.\cite{WangZG-Y10750-tetra-CPC-2019}, we observe that the $Y(10750)$ observed by the Belle collaboration \cite{Belle-Y10750-JHEP-2019} can be assigned as the $C\gamma_5\otimes\stackrel{\leftrightarrow}{\partial}_\mu\otimes \gamma_5C$ type hidden-bottom tetraquark state with the $J^{PC}=1^{--}$.

\begin{table}
\begin{center}
\begin{tabular}{|c|c|c|c|c|c|c|c|}\hline\hline
$|S_{qc}, S_{\bar{q}\bar{c}}; S, L; J\rangle$                                &$\mu(\rm{GeV})$ &$T^2(\rm{GeV}^2)$ &$\sqrt{s_0}(\rm{GeV})$ &pole    &$D(10)$ \\ \hline

$|0, 0; 0, 1; 1\rangle$                                                      &$1.1$         &$2.6-3.0$       &$4.75\pm0.10$        &$(40-65)\%$   &$< 1\%$ \\ \hline

$|1, 1; 0, 1; 1\rangle$                                                      &$1.2$         &$2.5-2.9$       &$4.80\pm0.10$        &$(39-64)\%$   &$< 3\%$ \\ \hline

$\frac{1}{\sqrt{2}}\left(|1, 0; 1, 1; 1\rangle+|0, 1; 1, 1; 1\rangle\right)$ &$1.3$         &$3.0-3.4$       &$4.85\pm0.10$        &$(38-60)\%$   &$\ll 1\%$ \\ \hline

$|1, 1; 2, 1; 1\rangle$                                                      &$1.3$         &$2.7-3.1$       &$4.85\pm0.10$        &$(39-63)\%$   &$< 1\%$ \\ \hline \hline
\end{tabular}
\end{center}
\caption{ The Borel windows $T^2$, continuum threshold parameters $s_0$, energy scales of the QCD spectral densities, contributions of the ground states, and values of the  $D(10)$ \cite{WangZG-HC-Vect-Derive-CPC-2024}.   }\label{Borel-1P}
\end{table}

\begin{table}
\begin{center}
\begin{tabular}{|c|c|c|c|c|c|c|c|}\hline\hline
$|S_{qc}, S_{\bar{q}\bar{c}}; S, L; J\rangle$                                &$\mu(\rm{GeV})$ &$T^2(\rm{GeV}^2)$ &$\sqrt{s^\prime_0}(\rm{GeV})$ &pole    &$D(10)$ \\ \hline

$|0, 0; 0, 1; 1\rangle$                                                      &$2.4$         &$2.8-3.2$       &$5.15\pm0.10$        &$(67-85)\%$   &$\ll 1\%$ \\ \hline

$|1, 1; 0, 1; 1\rangle$                                                      &$2.5$         &$2.6-3.0$       &$5.20\pm0.10$        &$(67-86)\%$   &$< 1\%$ \\ \hline

$\frac{1}{\sqrt{2}}\left(|1, 0; 1, 1; 1\rangle+|0, 1; 1, 1; 1\rangle\right)$ &$2.6$         &$3.0-3.4$       &$5.25\pm0.10$        &$(67-84)\%$   &$\ll 1\%$ \\ \hline

$|1, 1; 2, 1; 1\rangle$                                                      &$2.6$         &$2.7-3.1$       &$5.25\pm0.10$        &$(68-87)\%$   &$\ll 1\%$ \\ \hline \hline
\end{tabular}
\end{center}
\caption{ The Borel windows $T^2$, continuum threshold parameters $s_0^\prime$, energy scales of the QCD spectral densities, contributions of the ground states plus first radial excitations, and values of the  $D(10)$ \cite{WangZG-HC-Vect-Derive-CPC-2024}.   }\label{Borel-1P-2P}
\end{table}

\begin{table}
\begin{center}
\begin{tabular}{|c|c|c|c|c|c|c|c|}\hline\hline
$|S_{qc}, S_{\bar{q}\bar{c}}; S, L; J\rangle$                                &$M_Y(\rm{GeV})$  &$\lambda_Y(10^{-2}\rm{GeV}^6)$ &$M_Y(\rm{GeV})$  &$\lambda_Y(10^{-2}\rm{GeV}^6)$  \\ \hline

$|0, 0; 0, 1; 1\rangle$                                                      &$4.24\pm0.09$    &$2.28 \pm0.42$  &$4.75\pm0.10$    &$8.19 \pm1.23$           \\ \hline

$|1, 1; 0, 1; 1\rangle$                                                      &$4.28\pm0.09$    &$4.80 \pm0.95$   &$4.81\pm0.10$    &$18.3 \pm3.0$            \\ \hline

$\frac{1}{\sqrt{2}}\left(|1, 0; 1, 1; 1\rangle+|0, 1; 1, 1; 1\rangle\right)$ &$4.31\pm0.09$    &$2.94 \pm0.50$   &$4.85\pm0.09$    &$8.63 \pm1.22$              \\ \hline

$|1, 1; 2, 1; 1\rangle$                                                      &$4.33\pm0.09$    &$6.55 \pm1.19$   &$4.86\pm0.10$    &$21.7 \pm3.4$              \\ \hline \hline
\end{tabular}
\end{center}
\caption{ The masses and pole residues of the ground states and first radial excitations \cite{WangZG-HC-Vect-Derive-CPC-2024}.   }\label{mass-1P-2P}
\end{table}

\begin{table}
\begin{center}
\begin{tabular}{|c|c|c|c|c|c|c|c|}\hline\hline
$|S_{qc}, S_{\bar{q}\bar{c}}; S, L; J\rangle$                                &$M_Y(\rm{GeV})$     &Assignments       \\ \hline
$|0, 0; 0, 1; 1\rangle$\,(1P)                                                      &$4.24\pm0.09$       &$Y(4220/4260)$                  \\ \hline
$|0, 0; 0, 1; 1\rangle$\,(2P)
&$4.75\pm0.10$       &$Y(4750)$            \\ \hline

$|1, 1; 0, 1; 1\rangle$\,(1P)                                                      &$4.28\pm0.09$       &$Y(4220/4320)$             \\ \hline
$|1, 1; 0, 1; 1\rangle$\,(2P)                                                      &$4.81\pm0.10$       &               \\ \hline

$\frac{1}{\sqrt{2}}\left(|1, 0; 1, 1; 1\rangle+|0, 1; 1, 1; 1\rangle\right)$\,(1P) &$4.31\pm0.09$       &$Y(4320/4390)$               \\ \hline
$\frac{1}{\sqrt{2}}\left(|1, 0; 1, 1; 1\rangle+|0, 1; 1, 1; 1\rangle\right)$\,(2P) &$4.85\pm0.09$       &                \\ \hline

$|1, 1; 2, 1; 1\rangle$ \, (1P)                                                     &$4.33\pm0.09$       &$Y(4320/4390)$                 \\ \hline
$|1, 1; 2, 1; 1\rangle$ \, (2P)                                                     &$4.86\pm0.10$       &        \\ \hline
\end{tabular}
\end{center}
\caption{ The masses  of the vector tetraquark states and possible assignments, where the 1P and 2P denote the ground states and first radial excitations, respectively \cite{WangZG-HC-Vect-Derive-CPC-2024}.   }\label{Assignment-Y}
\end{table}

\subsection{Doubly heavy tetraquark states}\label{Doubly H tetraquark}
In 2016, the LHCb collaboration observed the doubly-charmed baryon state  $\Xi_{cc}^{++}$ in the $\Lambda_c^+ K^- \pi^+\pi^+$ mass spectrum  and measured the mass, but did not determine the spin \cite{LHCb-Xicc-PRL-2017}. The $\Xi_{cc}^{++}$ maybe have the spin $\frac{1}{2}$ or $\frac{3}{2}$, we can take the diquark $\varepsilon^{ijk}  c^T_iC\gamma_\mu c_j$ as basic constituent to construct the current
\begin{eqnarray}
J(x)&=& \varepsilon^{ijk}  c^T_i(x)C\gamma_\mu c_j(x)
\gamma_5\gamma^\mu u_k(x)  \, ,
\end{eqnarray}
or
\begin{eqnarray}
J_\mu(x)&=& \varepsilon^{ijk}  c^T_i(x)C\gamma_\mu c_j(x) u_k(x)  \, ,
\end{eqnarray}
to study it with the QCD sum rules \cite{WangZG-Xicc-penta-EPJC-2018}.
The observation of the doubly-charmed baryon state  $\Xi_{cc}^{++}$ has led to a renaissance on the doubly-heavy tetraquark spectroscopy \cite{Karliner-cc-tetra-2017-PRL,Eichten-cc-tetra-2017-PRL,QinQin-cc-tetra-2021-CPC}.
For a $QQ\bar{q}\bar{q}^\prime$ system,  if  the two $Q$-quarks are in  long
separation, the gluon exchange induced  force between them would be screened by the two $\bar{q}$-quarks, then a loosely $Q\bar{q}-Q\bar{q}^\prime$ type
bound state is formed. If the two $Q$-quarks are in  short separation, the
$QQ$ pair forms a compact point-like color source in heavy quark limit, and attracts a   $\bar{q}\bar{q}^\prime$ pair, which serves as another compact point-like color source,  then an exotic $QQ-\bar{q}\bar{q}^\prime$ type tetraquark state is formed.
The existence and stability of the $QQ\bar{q}\bar{q}^\prime$ tetraquark states have been extensively discussed in early literatures based on the potential models \cite{JMRichard-QQqq-PRD-1982,Heller-QQqq-PRD-1987,Carlson-QQqq-PRD-1988,
JMRichard-QQqq-ZPC-1986,Stancu-QQqq-PRD-1998,
Czarnecki-QQqq-PLB-2018} and heavy quark symmetry \cite{Manohar-QQqq-NPB-1993}.

In Ref.\cite{WZG-QQ-tetra-APPB-2018}, we  choose the currents $J_\mu(x)$ and $\eta_\mu(x)$ to study the doubly heavy tetraquark states with the $J^{P}=1^+$,
where
\begin{eqnarray}\label{Current-Tcc-APPB}
J_\mu(x)&=&\varepsilon^{ijk}\varepsilon^{imn} \, Q^{T}_j(x)C\gamma_\mu Q_k(x) \,\bar{u}_m(x)\gamma_5C \bar{s}^T_n(x) \, ,\nonumber  \\
\eta_\mu(x)&=&\varepsilon^{ijk}\varepsilon^{imn} \, Q^{T}_j(x)C\gamma_\mu Q_k(x) \,\bar{u}_m(x)\gamma_5C \bar{d}^T_n(x) \, ,
\end{eqnarray}
$Q=c,b$, again, we adopt  the correlation functions $\Pi_{\mu\nu}(p)$ in  Eq.\eqref{CF-Pi}.
The tetraquark states are spatial extended objects, not point-like objects, however,  we choose the local currents to interpolate them and take all the quarks and antiquarks as the color sources, and neglect the finite size effects.

We rewrite the current $J_\mu(x)$ as
\begin{eqnarray}
J_\mu(x)&=& Q^{T}_j(x)C\gamma_\mu Q_k(x) \left[\bar{u}_j(x)\gamma_5C \bar{s}^T_k(x)-\bar{u}_k(x)\gamma_5C \bar{s}^T_j(x)\right]   \nonumber\\
&=& \frac{1}{2}\left[Q^{T}_j(x)C\gamma_\mu Q_k(x)-Q^{T}_k(x)C\gamma_\mu Q_j(x) \right]\left[\bar{u}_j(x)\gamma_5C \bar{s}^T_k(x)-\bar{u}_k(x)\gamma_5C \bar{s}^T_j(x)\right] \, ,
\end{eqnarray}
according to the identity  $\varepsilon_{ijk}\,\varepsilon_{imn}=\delta_{jm}\delta_{kn}-\delta_{jn}\delta_{km}$ in the color space. The current $J_\mu(x)$ is of $\bar{\mathbf 3}\otimes \mathbf 3$ type  in  the color space, we can also construct the current $\widetilde{J}_\mu(x)$ satisfying the Fermi-Dirac statistics,
\begin{eqnarray}
\widetilde{J}_\mu(x)&=&\frac{1}{2} \left[Q^{T}_j(x)C\gamma_5 Q_k(x)+Q^{T}_k(x)C\gamma_5 Q_j(x)\right] \left[\bar{u}_j(x)\gamma_\mu C \bar{s}^T_k(x)+\bar{u}_k(x)\gamma_\mu C \bar{s}^T_j(x)\right] \, ,
\end{eqnarray}
 which is of  ${\mathbf 6}\otimes \bar{\mathbf 6}$ type in the color space, and differs from the corresponding current constructed in Ref.\cite{DuML-33-66-QQ-PRD-2013} slightly.

The color factor defined in Eq.\eqref{lambda-lambda} has the values,
  $\langle\widehat{C}_i \cdot \widehat{C}_j \rangle=-\frac{2}{3}$ and  $\frac{1}{3}$ for the $\bar{\mathbf{3}}$ and $\mathbf{6}$ diquark $[qq]$, respectively.
If we define $\widehat{C}_{12}\cdot \widehat{C}_{34}=(\widehat{C}_1+\widehat{C}_2 )\cdot (\widehat{C}_3+\widehat{C}_4 )$, then $\langle \widehat{C}_{12}\cdot \widehat{C}_{34}\rangle=-\frac{4}{3}$ and $-\frac{10}{3}$ for the $\bar{\mathbf{3}}\mathbf{3}$ and $\mathbf{6}\bar{\mathbf{6}}$ type tetraquark states.
The one-gluon exchange induced attractive (repulsive)  interaction  favors (disfavors) formation of the $\bar{\mathbf 3}$ ($\mathbf 6$) diquark state $Q^{T}_jC\gamma_\mu Q_k-Q^{T}_kC\gamma_\mu Q_j$ ($Q^{T}_jC\gamma_5 Q_k+Q^{T}_kC\gamma_5 Q_j$), while the ${\mathbf 6} \bar{\mathbf 6}$-type  tetraquark states are expected to have much  smaller  masses than that of the $\bar{\mathbf 3} \mathbf 3$-type tetraquark states according to the $\widehat{C}_{12}\cdot \widehat{C}_{34}$.
Furthermore, the color magnetic interaction, see Eq.\eqref{color-spin},
 leads to mixing between the  ${\mathbf 6} \bar{\mathbf 6}$ and
 $\bar{\mathbf 3} \mathbf 3$-type  tetraquark states.
In Ref.\cite{DuML-33-66-QQ-PRD-2013}, M. L. Du et al obtain degenerate masses for the ${\mathbf 6} \bar{\mathbf 6}$ and $\bar{\mathbf 3} \mathbf 3$-type tetraquark states based on the QCD sum rules. We should study this subject further.

After tedious but straightforward calculations, we obtain the QCD sum rules for the doubly-heavy
tetraquark states \cite{WZG-QQ-tetra-APPB-2018}, which are named as the $T$ states in {\it The Review of Particle Physics} \cite{PDG-2024}. According to the energy scale formula in Eq.\eqref{formula},
we suggest an energy scale formula,
\begin{eqnarray}\label{formula-T}
 \mu &=&\sqrt{M^2_{T}-(2{\mathbb{M}}_Q)^2}-\kappa\, m_s(\mu)\, ,
\end{eqnarray}
  to determine  the optimal   energy scales of the QCD spectral densities.

There was  no experimental candidate for the  doubly heavy tetraquark state when performing the calculations \cite{WZG-QQ-tetra-APPB-2018}. After careful examinations,  we choose the effective heavy quark masses ${\mathbb{M}}_c=1.84\,\rm{GeV}$ and ${\mathbb{M}}_b=5.12\,\rm{GeV}$, and take  account of the $SU(3)$ breaking effect by subtracting the $\kappa\,m_s(\mu)$. In Table \ref{Borel-QQ-tetra}, we present the Borel windows $T^2$, continuum threshold parameters $s_0$, optimal energy scales $\mu$, pole contributions of the ground states,  where  the same parameters as the ones in the QCD sum rules for  the $Z_c(3900)$ are chosen, see the last line.

\begin{table}
\begin{center}
\begin{tabular}{|c|c|c|c|c|c|c|c|}\hline\hline
                     &$T^2(\rm{GeV}^2)$   &$\sqrt{s_0}(\rm{GeV})$   &$\mu(\rm{GeV})$  &pole          &$M(\rm{GeV})$  &$\lambda(\rm{GeV}^5)$ \\ \hline

$cc\bar{u}\bar{d}$   &$2.6-3.0$           &$4.45\pm0.10$            &1.3              &$(39-63)\%$   &$3.90\pm0.09$  &$(2.64\pm0.42)\times10^{-2}$   \\ \hline

$cc\bar{u}\bar{s}$   &$2.6-3.0$           &$4.50\pm0.10$            &1.3              &$(41-64)\%$   &$3.95\pm0.08$  &$(2.88\pm0.46)\times10^{-2}$   \\ \hline

$bb\bar{u}\bar{d}$   &$6.9-7.7$           &$11.14\pm0.10$           &2.4              &$(41-60)\%$   &$10.52\pm0.08$ &$(1.30\pm0.20)\times10^{-1}$   \\ \hline

$bb\bar{u}\bar{s}$   &$6.8-7.6$           &$11.15\pm0.10$           &2.4              &$(41-61)\%$   &$10.55\pm0.08$ &$(1.33\pm0.20)\times10^{-1}$   \\ \hline

$cc\bar{u}\bar{d}$   &$2.6-3.0$           &$4.40\pm0.10$            &1.4              &$(39-62)\%$   &$3.85\pm0.09$  &$(2.60\pm0.42)\times10^{-2}$   \\ \hline
 \hline
\end{tabular}
\end{center}
\caption{ The Borel windows, continuum threshold parameters, ideal energy scales, pole contributions,   masses and pole residues for the doubly heavy  tetraquark states \cite{WZG-QQ-tetra-APPB-2018}. }\label{Borel-QQ-tetra}
\end{table}

\begin{figure}
 \centering
 \includegraphics[totalheight=5cm,width=7cm]{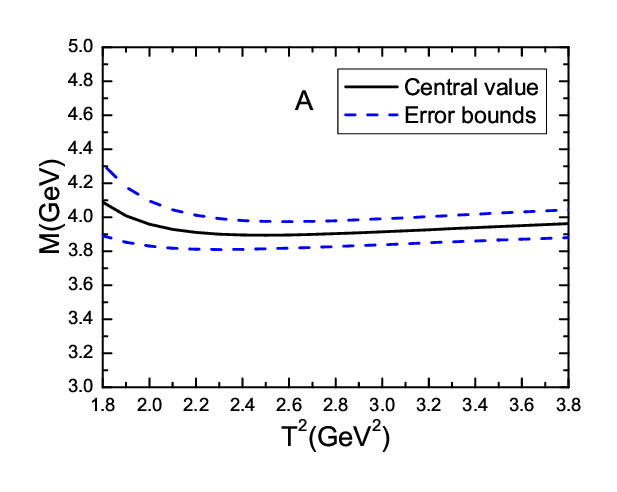}
  \includegraphics[totalheight=5cm,width=7cm]{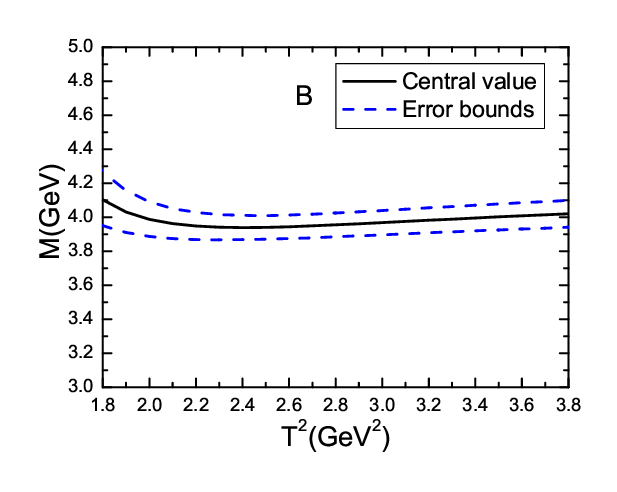}
  \caption{ The masses    with variations  of the Borel parameter $T^2$, where the  $A$ and $B$ denote the  $cc\bar{u}\bar{d}$ and $cc\bar{u}\bar{s}$ tetraquark states, respectively.  }\label{Mass-Borel-QQ}
\end{figure}

We take  into account all the uncertainties of the relevant  parameters,
and obtain the values of the masses and pole residues of
 the    $Z_{QQ}$, which are  shown explicitly in Table \ref{Borel-QQ-tetra} and Fig.\ref{Mass-Borel-QQ}.
From Fig.\ref{Mass-Borel-QQ}, we can see explicitly that there appear platforms in  the Borel windows shown in Table \ref{Borel-QQ-tetra} indeed.   And we suggested to search for the $Z_{QQ}$ states in the   Okubo-Zweig-Iizuka  super-allowed two-body strong decays
\begin{eqnarray}
Z_{cc\bar{u}\bar{d}} &\to& D^0D^{*+}\, , \,\, D^+D^{*0}\, ,\nonumber\\
Z_{cc\bar{u}\bar{s}} &\to& D^0D_s^{*+}\, , \,\, D^+_sD^{*0}\, ,
\end{eqnarray}
and weak decays  through $b\to c\bar{c}s$ at the quark level,
 \begin{eqnarray}
Z_{bb\bar{u}\bar{d}} &\to& \bar{B}^0B^{*-}\, , \,\, B^-\bar{B}^{*0}\to\, \gamma\, J/\psi K^-\,J/\psi \bar{K}^0 \,  , \nonumber\\
Z_{bb\bar{u}\bar{s}} &\to& \bar{B}_s^0B^{*-}\, , \,\, B^-\bar{B}_s^{*0}\to\, \gamma\, J/\psi \phi\, J/\psi K^-\,   .
\end{eqnarray}

In 2021, the LHCb collaboration  observed  the exotic state $T_{cc}^+(3875)$   just below  the $D^0D^{*+}$ threshold \cite{LHCb-Tcc-NatureP-2022,LHCb-Tcc-NatureC-2022}.  The Breit-Wigner mass and width are $\delta M_{BW} = -273\pm 61\pm 5^{+11}_{-14}~\text{KeV}$ below the $D^0D^{*+}$ threshold and $\Gamma_{BW} = 410\pm 165\pm 43^{+18}_{-38}~\text{KeV}$  \cite{LHCb-Tcc-NatureP-2022,LHCb-Tcc-NatureC-2022}. While the Particle Data Group fit the Breit-Wigner mass and width to be $3874.83\pm 0.11\,\rm{MeV}$ and $0.410\pm 0.165{}^{+0.047}_{-0.057}\,\rm{MeV} $, respectively \cite{PDG-2024}. The prediction $M_{Z_{cc\bar{u}\bar{d}}}=3.90\pm0.09\,\rm{GeV}$ is in excellent agreement with the LHCb data.

Before the LHCb data, several theoretical groups had made predictions for the $T_{cc}$ masses \cite{LSE-momentum-GengLS-PRD-2019,OBE-tetra-mole-ZhuSL-PRD-2013,
Tetra-33-66-RQM-LuQF-PRD-2020,Tetra-33-Ebert-PRD-2007,
CQM-Tetra-Vijande-EPJA-2004,CQM-Tetra-SHLee-PRD-2021,
FluxTube-Tetra-ChenH-EPJA-2020,Latt-QQ-tetra-Cheung-JHEP-2017,
Latt-QQ-tetra-Junnarkar-PRD-2019,
Karliner-cc-tetra-2017-PRL,Eichten-cc-tetra-2017-PRL,WZG-QQ-tetra-APPB-2018,
DuML-33-66-QQ-PRD-2013,
WZG-cc-tetra-EPJC-2018,
Before-Tcc-Moinester-ZPA-1996,Before-Tcc-Stancu-PLB-1997,Before-Tcc-Nussinov-PLB-2003,
Before-Tcc-Rosina-FewBody-2004,Before-Tcc-Nielsen-PLB-2007,WangZG-QQ-CTP-2011,Before-Tcc-Vijande-PRD-2007,
Before-Tcc-LeeSH-EPJC-2009,Before-Tcc-YangY-PRD-2009,Before-Tcc-LiuYR-EPJC-2017,
Before-Tcc-LeeSH-NPA-2019,Before-Tcc-Maiani-PRD-2019,Before-Tcc-YangG-PRD-2020,Before-Tcc-Braaten-PRD-2021,
Before-Tcc-JiaDJ-MPLA-2022,Before-Tcc-LiSY-CPC-2021,
Before-Tcc-Faustov-Universe-2021,
Before-Tcc-Ikeda-PLB-2014}, the predicted masses differ from each other in one way or the other.

In Ref.\cite{WZG-cc-tetra-EPJC-2018}, we resort to  the  correlation functions $\Pi(p)$ and  $\Pi_{\mu\nu\alpha\beta}(p)$ in Eq.\eqref{CF-Pi} and currents
\begin{eqnarray}
 J(x)&=&J_{\bar{u}\bar{d};0}(x),\,J_{\bar{u}\bar{s};0}(x),\,J_{\bar{s}\bar{s};0}(x)\, ,\nonumber\\
 J_{\mu\nu}(x)&=&J_{\mu\nu;1}(x)\, , \, J_{\mu\nu;2}(x) \, , \nonumber\\
 J_{\mu\nu;1/2}(x)&=&J_{\mu\nu;\bar{u}\bar{d};1/2}(x),
 \,J_{\mu\nu;\bar{u}\bar{s};1/2}(x),\,J_{\mu\nu;\bar{s}\bar{s};1/2}(x)\, ,
 \end{eqnarray}
 where
\begin{eqnarray}
J_{\bar{u}\bar{d};0}(x)&=&\varepsilon^{ijk}\varepsilon^{imn} \, c^{T}_j(x)C\gamma_\mu c_k(x) \,\bar{u}_m(x)\gamma^\mu C \bar{d}^T_n(x) \, , \nonumber\\
J_{\bar{u}\bar{s};0}(x)&=&\varepsilon^{ijk}\varepsilon^{imn} \, c^{T}_j(x)C\gamma_\mu c_k(x) \,\bar{u}_m(x)\gamma^\mu C \bar{s}^T_n(x) \, , \nonumber\\
J_{\bar{s}\bar{s};0}(x)&=&\varepsilon^{ijk}\varepsilon^{imn} \, c^{T}_j(x)C\gamma_\mu c_k(x) \,\bar{s}_m(x)\gamma^\mu C \bar{s}^T_n(x) \, , \nonumber\\
J_{\mu\nu;\bar{u}\bar{d};1/2}(x)&=&\varepsilon^{ijk}\varepsilon^{imn} \,\left[ c^{T}_j(x)C\gamma_\mu c_k(x) \,\bar{u}_m(x)\gamma_\nu C \bar{d}^T_n(x)\mp c^{T}_j(x)C\gamma_\nu c_k(x) \,\bar{u}_m(x)\gamma_\mu C \bar{d}^T_n(x) \right] \, , \nonumber\\
J_{\mu\nu;\bar{u}\bar{s};1/2}(x)&=&\varepsilon^{ijk}\varepsilon^{imn} \,\left[ c^{T}_j(x)C\gamma_\mu c_k(x) \,\bar{u}_m(x)\gamma_\nu C \bar{s}^T_n(x)\mp c^{T}_j(x)C\gamma_\nu c_k(x) \,\bar{u}_m(x)\gamma_\mu C \bar{s}^T_n(x) \right] \, , \nonumber\\
J_{\mu\nu;\bar{s}\bar{s};1/2}(x)&=&\varepsilon^{ijk}\varepsilon^{imn} \,\left[ c^{T}_j(x)C\gamma_\mu c_k(x) \,\bar{s}_m(x)\gamma_\nu C \bar{s}^T_n(x)\mp c^{T}_j(x)C\gamma_\nu c_k(x) \,\bar{s}_m(x)\gamma_\mu C \bar{s}^T_n(x) \right] \, , \nonumber\\
\end{eqnarray}
to study the mass spectrum of the $J^P=0^+$, $1^\pm$ and $2^+$ doubly charmed tetraquark states systematically, where the subscripts $0$, $1$ and $2$ denote the spins.

At the phenomenological side,  we obtain the hadronic representation and isolate the ground state contributions,
\begin{eqnarray}\label{CF-Hadron-cc-J0}
\Pi(p)&=&\frac{\lambda_{ Z}^2}{M_{Z}^2-p^2} +\cdots =\Pi_{0}(p^2)\,\, ,
\end{eqnarray}
\begin{eqnarray}\label{CF-Hadron-cc-J1}
\Pi_{\mu\nu\alpha\beta;1}(p)&=&\frac{\tilde{\lambda}_{ Z}^2}{M_{Z}^2-p^2}\left(p^2g_{\mu\alpha}g_{\nu\beta} -p^2g_{\mu\beta}g_{\nu\alpha} -g_{\mu\alpha}p_{\nu}p_{\beta}-g_{\nu\beta}p_{\mu}p_{\alpha}+g_{\mu\beta}p_{\nu}p_{\alpha}+g_{\nu\alpha}p_{\mu}p_{\beta}\right) \nonumber\\
&&+\frac{\tilde{\lambda}_{ Y}^2}{M_{Y}^2-p^2}\left( -g_{\mu\alpha}p_{\nu}p_{\beta}-g_{\nu\beta}p_{\mu}p_{\alpha}+g_{\mu\beta}p_{\nu}p_{\alpha}+g_{\nu\alpha}p_{\mu}p_{\beta}\right) +\cdots \, \, ,\nonumber\\
&=&\Pi_Z(p^2)\left(p^2g_{\mu\alpha}g_{\nu\beta} -p^2g_{\mu\beta}g_{\nu\alpha} -g_{\mu\alpha}p_{\nu}p_{\beta}-g_{\nu\beta}p_{\mu}p_{\alpha}+g_{\mu\beta}p_{\nu}p_{\alpha}+g_{\nu\alpha}p_{\mu}p_{\beta}\right) \nonumber\\
&&+\Pi_Y(p^2)\left( -g_{\mu\alpha}p_{\nu}p_{\beta}-g_{\nu\beta}p_{\mu}p_{\alpha}+g_{\mu\beta}p_{\nu}p_{\alpha}+g_{\nu\alpha}p_{\mu}p_{\beta}\right) \, ,
\end{eqnarray}
\begin{eqnarray}\label{CF-Hadron-cc-J2}
\Pi_{\mu\nu\alpha\beta;2} (p) &=&\frac{\lambda_{ Z}^2}{M_{Z}^2-p^2}\left( \frac{\widetilde{g}_{\mu\alpha}\widetilde{g}_{\nu\beta}+\widetilde{g}_{\mu\beta}\widetilde{g}_{\nu\alpha}}{2}-\frac{\widetilde{g}_{\mu\nu}\widetilde{g}_{\alpha\beta}}{3}\right) +\cdots \, \, , \nonumber\\
&=&\Pi_2(p^2)\left( \frac{\widetilde{g}_{\mu\alpha}\widetilde{g}_{\nu\beta}+\widetilde{g}_{\mu\beta}\widetilde{g}_{\nu\alpha}}{2}-\frac{\widetilde{g}_{\mu\nu}\widetilde{g}_{\alpha\beta}}{3}\right)\, , \end{eqnarray}
where $\widetilde{g}_{\mu\nu}=g_{\mu\nu}-\frac{p_{\mu}p_{\nu}}{p^2}$,  the  pole residues  $\lambda_{Z}$ and $\lambda_{Y}$ are defined  analogous to  Eq.\eqref{define-pole-residue} with the simple replacements $Z^- \to Y$ and $Z^+ \to Z$, and $\lambda_{Y}=\tilde{\lambda}_{Y}M_Y$, $\lambda_{Z}=\tilde{\lambda}_{Z}M_Z$.

We perform the calculations routinely, and obtain the QCD sum rules for the masses and pole residues from the components  $\Pi_0(p^2)$, $\Pi_{1;A}(p^2)$, $\Pi_{1;V}(p^2)$ and $\Pi_2(p^2)$, respectively,  where
$\Pi_{1;A}(p^2)=p^2\Pi_Z(p^2)$ and $\Pi_{1;V}(p^2)=p^2\Pi_Y(p^2)$. Again, we adopt  the modified  energy scale formula in Eq.\eqref{formula-T} to determine the best energy scales  of the QCD spectral densities.

After trial and error, we obtain the Borel windows $T^2$, continuum threshold parameters $s_0$, optimal energy scales $\mu$, pole contributions, see  Table \ref{Borel-mass-T-012}. From the Table, we can see clearly that the pole dominance can be well satisfied.
In calculations, we observe that for the $J^P=1^-$ tetraquark states, the operator product expansion is well convergent, while in the case of the $J^P=0^+$, $1^+$ and $2^+$ tetraquark states,  the contributions of the vacuum condensates of dimensions $6, \,8,\,10$ have the hierarchy $|D(6)|\gg |D(8)|\gg |D(10)|$, the operator product expansion is also convergent.
At last, we  take account of all uncertainties of the relevant  parameters,
and obtain the values of the masses and pole residues, which are  shown explicitly in Table \ref{Borel-mass-T-012}.

The centroids  of the masses of the $C\gamma_\mu\otimes \gamma_\nu C$ type tetraquark states are
\begin{eqnarray}\label{centroids-Cmu-Cmu}
M_{C\gamma_\mu\otimes \gamma_\nu C}(cc\bar{u}\bar{d})&=&\frac{M_{cc\bar{u}\bar{d};0^+}+3M_{cc\bar{u}\bar{d};1^+}+5M_{cc\bar{u}\bar{d};2^+}}{9}=3.92\,\rm{GeV}\, , \nonumber\\
M_{C\gamma_\mu\otimes \gamma_\nu C}(cc\bar{u}\bar{s})&=&\frac{M_{cc\bar{u}\bar{s};0^+}+3M_{cc\bar{u}\bar{s};1^+}+5M_{cc\bar{u}\bar{s};2^+}}{9}=3.99\,\rm{GeV}\, , \nonumber\\
M_{C\gamma_\mu\otimes \gamma_\nu C}(cc\bar{s}\bar{s})&=&\frac{M_{cc\bar{s}\bar{s};0^+}+3M_{cc\bar{s}\bar{s};1^+}+5M_{cc\bar{s}\bar{s};2^+}}{9}=4.04\,\rm{GeV}\, ,
\end{eqnarray}
which are slightly larger than the centroids of the masses of the corresponding $C\gamma_\mu\otimes \gamma_5 C$ type tetraquark states,
\begin{eqnarray}\label{centroids-C5-C5}
M_{C\gamma_\mu\otimes \gamma_5 C}(cc\bar{u}\bar{d})&=&3.90\,\rm{GeV}\, , \nonumber\\
M_{C\gamma_\mu\otimes \gamma_5 C}(cc\bar{u}\bar{s})&=&3.95\,\rm{GeV}\, ,
\end{eqnarray}
the lowest  states are the $C\gamma_\mu\otimes \gamma_5 C$ type tetraquark states, which is consistent with our naive expectation that the axialvector (anti)diquarks have larger masses than the corresponding scalar (anti)diquarks. The lowest centroids $M_{cc\bar{u}\bar{d};0^+}=3.87\,\rm{GeV}$ and $M_{cc\bar{u}\bar{s};0^+}=3.94\,\rm{GeV}$ originate from the spin splitting,  in other words, the spin-spin interaction between the doubly heavy diquark and  light antidiquark. In fact, the predicted masses have uncertainties, the centroids  of the masses are not the super values, all values within uncertainties make sense.

The QCD sum rules indicate that  the masses of the
 light  axialvector  diquark states lie about  $(150-200)\,\rm{MeV}$ above that of the  light scalar diquark states \cite{WangZG-Light-diquark-CTP-2013,Light-Diquark-Dosch-ZPC-1989,
Light-Diquark-Jamin-PLB-1990,Light-Diquark-ZhangAL-PRD-2007}, if they have the same valence quarks. Therefore,
the centroids  of the masses of the $C\gamma_\mu\otimes \gamma_\nu C$ type tetraquark states should be  larger than $4.0\,\rm{GeV}$, the present calculations maybe under-estimate the doubly-heavy tetraquark masses. If we take  the simple replacement $m_s(\mu) \to \mathbb{M}_s$ in the modified energy scale formula in Eq.\eqref{formula-T}, the predictions should be improved, about $+100\,\rm{MeV}$.

After observation of the $T_{cc}(3875)$, several new works on the $T_{cc}(3875)$ in the $\bar{\mathbf{3}}\mathbf{3}$ type tetraquark scenario appear \cite{Chromomagnetic-GuoT-PRD-2022,Azizi-Tcc-tetra-NPB-2022,
After-Tcc-tetra-SLZhu-CPC-2022,
After-Tcc-tetra-Azizi-PRD-2021,
After-Tcc-tetra-Oka-PRD-2022,After-Tcc-tetra-MaYL-PRD-2023,
After-Tcc-tetra-Karliner-PRD-2022}. Roughly speaking, the centroid of the $C\gamma_\alpha \otimes \gamma_\beta C$ type tetraquark states maybe lie about $100\,\rm{MeV}$ above the corresponding $C\gamma_\alpha \otimes \gamma_5 C$ type tetraquark states, and more works are still needed.

The doubly-charmed tetraquark states with the $J^P=0^+$, $1^+$ and $2^+$ lie near the corresponding  charmed meson pair thresholds, the decays  are  Okubo-Zweig-Iizuka  super-allowed,
\begin{eqnarray}
Z_{cc\bar{u}\bar{d};0^+} &\to& D^0D^{+}\, , \nonumber\\
Z_{cc\bar{u}\bar{s};0^+} &\to& D^0D_s^{+}\, , \nonumber\\
Z_{cc\bar{s}\bar{s};0^+} &\to& D^+_s D_s^{+}\, , \nonumber\\
Z_{cc\bar{u}\bar{d};1^+} &\to& D^0D^{*+}\, , \,\, D^+D^{*0}\, , \nonumber\\
Z_{cc\bar{u}\bar{s};1^+} &\to& D^0D_s^{*+}\, , \,\, D^+_sD^{*0}\, , \nonumber\\
Z_{cc\bar{s}\bar{s};1^+} &\to& D^+_sD_s^{*+}\, , \nonumber\\
Z_{cc\bar{u}\bar{d};2^+} &\to& D^0D^{+}\, , \,\,D^{*0}D^{*+}\, , \nonumber\\
Z_{cc\bar{u}\bar{s};2^+} &\to& D^0D_s^{+}\, , \nonumber\\
Z_{cc\bar{s}\bar{s};2^+} &\to& D^+_s D_s^{+}\, ,
\end{eqnarray}
but the available  phase spaces are very small, thus the  decays are kinematically depressed, the  doubly charmed tetraquark states with the $J^P=0^+$, $1^+$ and $2^+$
maybe have small widths. On the other hand,  the doubly charmed tetraquark states   with the $J^P=1^-$ lie above  the corresponding  charmed meson pair thresholds, the decays  are  Okubo-Zweig-Iizuka  super-allowed,
\begin{eqnarray}
Y_{cc\bar{u}\bar{d};1^-} &\to& D^0D^{+}\, , \,\,D^0D^{*+}\, , \,\, D^+D^{*0}\, , \nonumber\\
Y_{cc\bar{u}\bar{s};1^-} &\to& D^0D_s^{+}\, , \,\,D^0D_s^{*+}\, , \,\, D^+_sD^{*0}\, , \nonumber\\
Y_{cc\bar{s}\bar{s};1^-} &\to& D^+_sD_s^{+}\, , \,\, D^+_sD_s^{*+}\, ,
\end{eqnarray}
and  the available  phase spaces are large, thus the  decays are kinematically facilitated, the  doubly charmed tetraquark states with the $J^P=1^-$ should
 have large widths.

\begin{table}
\begin{center}
\begin{tabular}{|c|c|c|c|c|c|c|c|}\hline\hline
                           &$T^2(\rm{GeV}^2)$   &$\sqrt{s_0}(\rm{GeV})$   &$\mu(\rm{GeV})$  &pole          &$M(\rm{GeV})$  &$\lambda(\rm{GeV}^5)$ \\ \hline

$cc\bar{u}\bar{d}(0^+)$   &$2.4-2.8$           &$4.40\pm0.10$             &1.2              &$(38-63)\%$   &$3.87\pm0.09$  &$(3.90\pm0.63)\times10^{-2}$   \\ \hline

$cc\bar{u}\bar{s}(0^+)$   &$2.6-3.0$           &$4.50\pm0.10$             &1.3              &$(38-62)\%$   &$3.94\pm0.10$  &$(4.92\pm0.89)\times10^{-2}$   \\ \hline

$cc\bar{s}\bar{s}(0^+)$   &$2.6-3.0$           &$4.55\pm0.10$             &1.3              &$(39-63)\%$   &$3.99\pm0.10$  &$(5.31\pm0.99)\times10^{-2}$   \\ \hline

$cc\bar{u}\bar{d}(1^+)$   &$2.6-3.0$           &$4.45\pm0.10$             &1.3              &$(39-62)\%$   &$3.90\pm0.09$  &$(3.44\pm0.54)\times10^{-2}$   \\ \hline

$cc\bar{u}\bar{s}(1^+)$   &$2.6-3.0$           &$4.50\pm0.10$             &1.3              &$(40-64)\%$   &$3.96\pm0.08$  &$(3.78\pm0.59)\times10^{-2}$   \\ \hline

$cc\bar{s}\bar{s}(1^+)$   &$2.7-3.1$           &$4.55\pm0.10$             &1.3              &$(39-62)\%$   &$4.02\pm0.09$  &$(4.11\pm0.68)\times10^{-2}$   \\ \hline

$cc\bar{u}\bar{d}(2^+)$   &$2.7-3.1$           &$4.50\pm0.10$             &1.4              &$(39-62)\%$   &$3.95\pm0.09$  &$(5.67\pm0.90)\times10^{-2}$   \\ \hline

$cc\bar{u}\bar{s}(2^+)$   &$2.8-3.2$           &$4.55\pm0.10$             &1.4              &$(38-60)\%$   &$4.01\pm0.09$  &$(6.27\pm1.02)\times10^{-2}$   \\ \hline

$cc\bar{s}\bar{s}(2^+)$   &$2.8-3.2$           &$4.60\pm0.10$             &1.4              &$(39-61)\%$   &$4.06\pm0.09$  &$(6.78\pm1.12)\times10^{-2}$   \\ \hline

$cc\bar{u}\bar{d}(1^-)$   &$3.3-3.9$           &$5.20\pm0.10$             &2.9              &$(50-73)\%$   &$4.66\pm0.10$  &$(1.31\pm0.17)\times10^{-1}$   \\ \hline

$cc\bar{u}\bar{s}(1^-)$   &$3.4-4.0$           &$5.25\pm0.10$             &2.9              &$(49-71)\%$   &$4.73\pm0.11$  &$(1.40\pm0.19)\times10^{-1}$   \\ \hline

$cc\bar{s}\bar{s}(1^-)$   &$3.7-4.3$           &$5.30\pm0.10$             &2.9              &$(49-72)\%$   &$4.78\pm0.11$  &$(1.48\pm0.19)\times10^{-1}$   \\ \hline
 \hline
\end{tabular}
\end{center}
\caption{ The Borel windows, continuum threshold parameters, optimal  energy scales, pole contributions,   masses and pole residues for the doubly charmed  tetraquark states \cite{WZG-cc-tetra-EPJC-2018}. }\label{Borel-mass-T-012}
\end{table}

\subsection{Fully heavy tetraquark states} \label{QQQQ-tetraquark}

The exotic states $Z_c(3900)$,  $Z_c(4020)$,  $Z_c(4430)$, $T_{cc}(3875)$, $P_c(4312)$, $P_c(4380)$, $P_c(4440)$, $P_c(4457)$,
$Z_b(10610)$,  $Z_b(10650)$, $\cdots$  are excellent candidates   for  the multiquark states, which
  consist two heavy quarks and two or three light quarks, we have to deal with both the heavy and light degrees of freedom of the dynamics.
If there exist multiquark  configurations consist of fully heavy quarks, the dynamics is much simple  at first glance, and   the $QQ\bar{Q}\bar{Q}$ tetraquark states have been studied extensively before the LHCb data \cite{Chromomagnetic-WuJ-PRD-2018,Tetra-33-RQM-Ferretti-PRD-2018,
CQM-Tetra-Vijande-PRD-2006,CQM-Tetra-ZhuSL-PRD-2019,cccc-ChenW-PLB-2017,
JMRichard-QQqq-PRD-1982,cccc-Iwasaki-PRL-1976,cccc-KTChao-ZPC-1981,
cccc-Ioffe-NPB-1987,cccc-Berezhnoy-PRD-2012,cccc-Rosner-PRL-2017,cccc-Navarra-CPC-2019,
cccc-LUQF-PRD-2019,cccc-Lloyd-PRD-2004,cccc-WangZG-EPJC-2017,
cccc-WangZG-Di-APPB-2019,cccc-GuoFK-EPJC-2018,Eichten-bbbb-PRD-2018}.

The quarks have color $SU(3)$ symmetry,  we can construct the tetraquark states according to the routine  ${\rm quark}\to {\rm diquark}\to {\rm tetraquark}$,
\begin{eqnarray}
({\mathbf 3}\otimes {\mathbf 3})\otimes(\overline{{\mathbf 3}}\otimes \overline{{\mathbf 3}}) &\to&(\overline{{\mathbf 3}}\oplus {\mathbf 6})\otimes({\mathbf 3}\oplus \overline{{\mathbf 6}})\to(\overline{{\mathbf 3}}\otimes {\mathbf 3})\oplus ({\mathbf 6}\otimes\overline{{\mathbf 6}}) \to({\mathbf 1}\oplus{\mathbf 8}) \oplus \cdots \, .
\end{eqnarray}
For the $\bar{\mathbf{3}}$ diquarks, only the operators $\varepsilon^{ijk} Q^{T}_j C\gamma_\mu Q_k$ and $\varepsilon^{ijk} Q^{T}_j C\sigma_{\mu\nu} Q_k$ could  exist due to Fermi-Dirac statistics, and we usually take the operators $\varepsilon^{ijk} Q^{T}_j C\gamma_\mu Q_k$  to construct the four-quark currents $J(x)$ and $J_{\mu\nu}(x)$ \cite{cccc-WangZG-EPJC-2017,
cccc-WangZG-Di-APPB-2019}, where  $J_{\mu\nu}(x)=J^1_{\mu\nu}(x)$, $J^2_{\mu\nu}(x)$, and
\begin{eqnarray}
J(x)&=&\varepsilon^{ijk}\varepsilon^{imn}Q^T_j(x)C\gamma_\mu Q_k(x) \bar{Q}_m(x)\gamma^\mu C \bar{Q}^T_n(x) \, ,\nonumber \\
J^1_{\mu\nu}(x)&=&\varepsilon^{ijk}\varepsilon^{imn}\left\{Q^T_j(x)C\gamma_\mu Q_k(x) \bar{Q}_m(x) \gamma_\nu C \bar{Q}^T_n(x)- Q^T_j(x)C\gamma_\nu  Q_k(x) \bar{Q}_m(x)\gamma_\mu C \bar{Q}^T_n(x) \right\} \, , \nonumber \\
J^2_{\mu\nu}(x)&=&\frac{\varepsilon^{ijk}\varepsilon^{imn}}{\sqrt{2}}\left\{Q^T_j(x)C\gamma_\mu Q_k(x) \bar{Q}_m(x) \gamma_\nu C \bar{Q}^T_n(x)+ Q^T_j(x)C\gamma_\nu  Q_k(x) \bar{Q}_m(x)\gamma_\mu C \bar{Q}^T_n(x) \right\} \, , \nonumber\\
\end{eqnarray}
then resort to the correlation functions $\Pi(p)$ and $\Pi_{\mu\nu\alpha\beta}(p)$ shown in Eq.\eqref{CF-Pi} to obtain the QCD sum rules.

In Ref.\cite{cccc-ChenW-PLB-2017}, Chen et al construct  the currents $\eta^i(x)$, $\eta_\mu^k(x)$ and $\eta^j_{\mu\nu}(x)$ with $i=1,2,3,4,5$, $k=1,2$ and $j=1,2$,  to interpolate the $QQ\bar{Q}\bar{Q}$ tetraquark states with the $J^{PC}=0^{++}$, $1^{+-}$ and $2^{++}$, respectively,
\begin{eqnarray}
\eta^i(x)&=&Q^T_a(x)C\Gamma^i Q_b(x) \bar{Q}_a(x)\Gamma^i C \bar{Q}^T_b(x)\, , \nonumber\\
\eta^1_{\mu}(x)&=&Q^T_a(x)C\gamma_\mu\gamma_5 Q_b(x) \bar{Q}_a(x) C \bar{Q}^T_b(x)-Q^T_a(x)C Q_b(x) \bar{Q}_a(x)\gamma_\mu\gamma_5 C \bar{Q}^T_b(x)\, ,\nonumber\\
\eta^2_{\mu}(x)&=&Q^T_a(x)C\sigma_{\mu\nu}\gamma_5 Q_b(x) \bar{Q}_a(x) \gamma^{\nu}C \bar{Q}^T_b(x)-Q^T_a(x)C\gamma^{\nu} Q_b(x) \bar{Q}_a(x)\sigma_{\mu\nu}\gamma_5 C \bar{Q}^T_b(x)\, ,\nonumber\\
\eta^j_{\mu\nu}(x)&=&Q^T_a(x)C\Gamma^j_\mu Q_b(x) \bar{Q}_a(x)\Gamma^j_\nu C \bar{Q}^T_b(x)+Q^T_a(x)C\Gamma^j_\nu Q_b(x) \bar{Q}_a(x)\Gamma^j_\mu C \bar{Q}^T_b(x)\, ,
\end{eqnarray}
where
$\Gamma^1=\gamma_5$, $\Gamma^2=\gamma_\mu\gamma_5$, $\Gamma^3=\sigma_{\mu\nu}$, $\Gamma^4=\gamma_\mu$, $\Gamma^5=1$, $\Gamma^1_\mu=\gamma_\mu$, $\Gamma^2_\mu=\gamma_\mu\gamma_5$, the $a$ and $b$ are color indexes. The $C\gamma_5$, $C$, $C\gamma_\mu\gamma_5$ are antisymmetric, while the $C\gamma_\mu$, $C\sigma_{\mu\nu}$, $C\sigma_{\mu\nu}\gamma_5$ are symmetric. The currents $\eta^{1/2/5}(x)$, $\eta^{1}_\mu(x)$ and $\eta^2_{\mu\nu}(x)$ are in the color ${\mathbf 6}  \bar{\mathbf{6}}$ representation, while the currents $\eta^{3/4}(x)$, $\eta^{2}_\mu(x)$ and $\eta^1_{\mu\nu}(x)$ are in the color  $\bar{\mathbf{3}}{\mathbf 3}  $ representation. For more currents and predictions in {\bf Scheme II}, we can consult Ref.\cite{cccc-ChenW-PLB-2017}.

At the hadron  side, we  isolate  the ground state contributions and obtain the results \cite{cccc-WangZG-EPJC-2017,cccc-WangZG-Di-APPB-2019},
\begin{eqnarray}
\Pi (p) &=&\frac{\lambda_X^2}{M^2_X-p^2} +\cdots =\Pi_S(p^2)\, ,
\end{eqnarray}
\begin{eqnarray}
\Pi^1_{\mu\nu\alpha\beta}(p)&=&\frac{\tilde{\lambda}_{ Y^+}^2}{M_{Y^+}^2-p^2}\left(p^2g_{\mu\alpha}g_{\nu\beta} -p^2g_{\mu\beta}g_{\nu\alpha} -g_{\mu\alpha}p_{\nu}p_{\beta}-g_{\nu\beta}p_{\mu}p_{\alpha}+g_{\mu\beta}p_{\nu}p_{\alpha}+g_{\nu\alpha}p_{\mu}p_{\beta}\right) \nonumber\\
&&+\frac{\tilde{\lambda}_{ Y^-}^2}{M_{Y^-}^2-p^2}\left( -g_{\mu\alpha}p_{\nu}p_{\beta}-g_{\nu\beta}p_{\mu}p_{\alpha}+g_{\mu\beta}p_{\nu}p_{\alpha}+g_{\nu\alpha}p_{\mu}p_{\beta}\right) +\cdots \, \, ,\nonumber\\
&=&\Pi_{A}(p^2)\left(p^2g_{\mu\alpha}g_{\nu\beta} -p^2g_{\mu\beta}g_{\nu\alpha} -g_{\mu\alpha}p_{\nu}p_{\beta}-g_{\nu\beta}p_{\mu}p_{\alpha}+g_{\mu\beta}p_{\nu}p_{\alpha}+g_{\nu\alpha}p_{\mu}p_{\beta}\right) \nonumber\\
&&+\Pi_{V}(p^2)\left( -g_{\mu\alpha}p_{\nu}p_{\beta}-g_{\nu\beta}p_{\mu}p_{\alpha}+g_{\mu\beta}p_{\nu}p_{\alpha}+g_{\nu\alpha}p_{\mu}p_{\beta}\right) \, .
\end{eqnarray}
\begin{eqnarray}
\Pi^2_{\mu\nu\alpha\beta} (p) &=&\frac{\lambda_X^2}{M_X^2-p^2}\left( \frac{\widetilde{g}_{\mu\alpha}\widetilde{g}_{\nu\beta}+\widetilde{g}_{\mu\beta}\widetilde{g}_{\nu\alpha}}{2}-\frac{\widetilde{g}_{\mu\nu}\widetilde{g}_{\alpha\beta}}{3}\right) +\cdots \, \, ,  \nonumber\\
&=&\Pi_{T}(p^2)\left( \frac{\widetilde{g}_{\mu\alpha}\widetilde{g}_{\nu\beta}+\widetilde{g}_{\mu\beta}\widetilde{g}_{\nu\alpha}}{2}-\frac{\widetilde{g}_{\mu\nu}\widetilde{g}_{\alpha\beta}}{3}\right) +\cdots \, \, ,
\end{eqnarray}
where $\widetilde{g}_{\mu\nu}=g_{\mu\nu}-\frac{p_{\mu}p_{\nu}}{p^2}$, we add the superscripts $1$ and $2$ to denote the spins,  and define the pole residues  $\lambda_{X}$ and $\lambda_{Y}$ with $\lambda_{Y^\pm}=\tilde{\lambda}_{Y^\pm}M_{Y^\pm}$ analogous to Eq.\eqref{define-pole-residue} with the simple replacements $Z^+ \to X$, $Y^+$ and $Z^- \to Y^-$.

\begin{figure}
 \centering
 \includegraphics[totalheight=5cm,width=7cm]{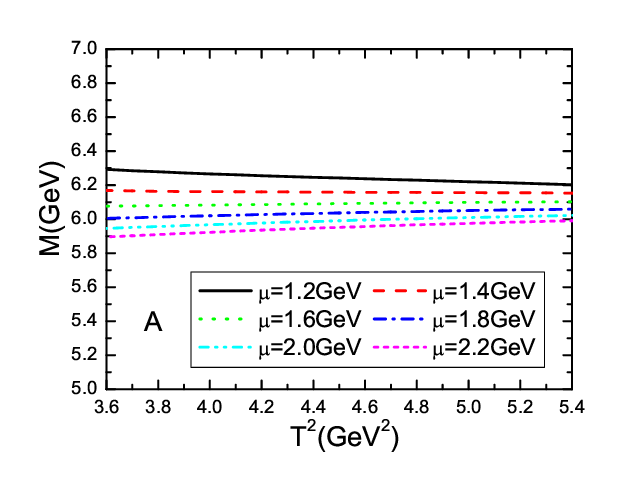}
 \includegraphics[totalheight=5cm,width=7cm]{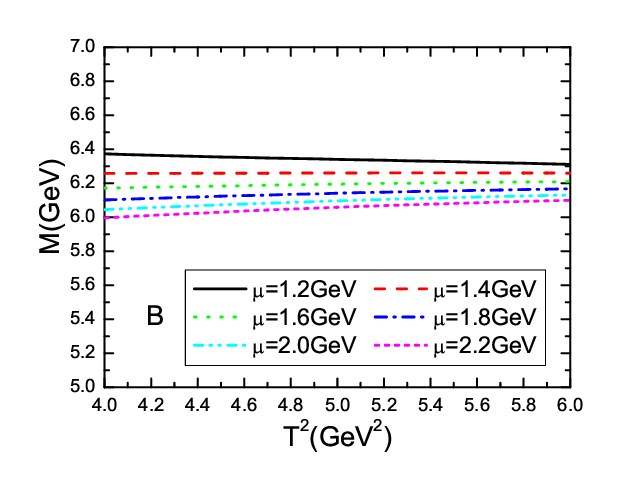}
   \caption{ The masses  of the $cc\bar{c}\bar{c}$ tetraquark states  with variations of the energy scales and Borel parameters,  where the $A$ and $B$ denote the $cc\bar{c}\bar{c}(0^{++})$ and $cc\bar{c}\bar{c}(2^{++})$, respectively \cite{cccc-WangZG-EPJC-2017}.  }\label{cccc-mu}
\end{figure}

If we take into account the first radial excited states, we obtain
\begin{eqnarray}
\Pi_{S/T}(p^2)&=&\frac{\lambda_X^2}{M^2_X-p^2} +\frac{\lambda_{X^\prime}^2}{M^2_{X^\prime}-p^2}+\cdots \, , \nonumber\\
\Pi_{A/V}(p^2)&=&\frac{\tilde{\lambda}_{ Y^\pm}^2}{M_{Y^\pm}^2-p^2}+\frac{\tilde{\lambda}_{ Y^{\prime\pm}}^2}{M_{Y^{\prime\pm}}^2-p^2}+\cdots\, .
\end{eqnarray}

Again we  take the quark-hadron duality below the continuum thresholds  $s_0$ and $s_0^\prime$, respectively,  and perform the Borel transformation   with respect to
the variable $P^2=-p^2$ to obtain  the QCD sum rules:
\begin{eqnarray}\label{QCDST-1S}
\lambda^2_{X/Y}\, \exp\left(-\frac{M^2_{X/Y}}{T^2}\right)&=& \int_{16m_c^2}^{s_0} ds \int_{z_i}^{z_f}dz \int_{t_i}^{t_f}dt \int_{r_i}^{r_f}dr\, \rho(s,z,t,r)  \exp\left(-\frac{s}{T^2}\right) \, ,
\end{eqnarray}
\begin{eqnarray}\label{QCDST-2S}
\lambda^2_{X/Y}\, \exp\left(-\frac{M^2_{X/Y}}{T^2}\right)+\lambda^2_{X^\prime/Y^\prime}\, \exp\left(-\frac{M^2_{X^\prime/Y^\prime}}{T^2}\right)&=& \int_{16m_c^2}^{s^\prime_0} ds \int_{z_i}^{z_f}dz \int_{t_i}^{t_f}dt \int_{r_i}^{r_f}dr\, \rho(s,z,t,r)  \nonumber\\
&& \exp\left(-\frac{s}{T^2}\right) \, ,
\end{eqnarray}
where the QCD spectral densities  $\rho(s,z,t,r) =\rho_S(s,z,t,r) $, $\rho_A(s,z,t,r) $, $\rho_V(s,z,t,r)$ and $\rho_T(s,z,t,r) $ \cite{cccc-WangZG-EPJC-2017,cccc-WangZG-Di-APPB-2019,WZG-cccc-CPC-2020}.

In Fig.\ref{cccc-mu}, we plot the masses of the  $cc\bar{c}\bar{c}$ tetraquark states with variations of the energy scales and Borel parameters for the threshold parameters $s_S^0=42\,\rm{GeV}^2$ and $s_T^0=44\,\rm{GeV}^2$. The predicted masses decrease  monotonously and slowly with increase of the energy scales, the QCD sum rules  are stable with variations of the Borel parameters at the energy scales $1.2\,{\rm{GeV}}<\mu<2.2\,{\rm GeV}$. We take  the largest  energy scale $\mu=2.0\,\rm{GeV}$ in Ref.\cite{cccc-WangZG-EPJC-2017}. In Refs.\cite{cccc-WangZG-EPJC-2017,cccc-WangZG-Di-APPB-2019}, we obtain the predictions for the masses of the  ground states, see Table \ref{QQQQ-Borel-mass}, where $M_{cc\bar{c}\bar{c}(0^{++})}< 2M_{J/\psi}$.

\begin{table}
\begin{center}
\begin{tabular}{|c|c|c|c|c|c|c|c|}\hline\hline
                            &$T^2(\rm{GeV}^2)$  &$s_0(\rm{GeV}^2)$ &$\mu(\rm{GeV})$ &pole         &$M_{X}(\rm{GeV})$    &$\lambda_{X}(\rm{GeV}^5)$ \\ \hline
$cc\bar{c}\bar{c}(0^{++})$  &$4.2-4.6$          &$42\pm1$          &$2.0$           &$(46-62)\%$  &$5.99\pm0.08$        &$(3.72\pm0.54)\times10^{-1}$  \\
$cc\bar{c}\bar{c}(1^{+-})$  &$4.5-4.9$          &$43\pm1$          &$2.0$           &$(46-61)\%$  &$6.05\pm0.08$        &$(2.97\pm0.44)\times10^{-1}$  \\
$cc\bar{c}\bar{c}(2^{++})$  &$4.6-5.2$          &$44\pm1$          &$2.0$           &$(46-63)\%$  &$6.09\pm0.08$        &$(3.36\pm0.45)\times10^{-1}$  \\
$cc\bar{c}\bar{c}(1^{--})$  &$4.2-4.6$          &$44\pm1$          &$2.0$           &$(46-62)\%$  &$6.11\pm0.08$        &$(1.82\pm0.33)\times10^{-1}$  \\
$bb\bar{b}\bar{b}(0^{++})$  &$13.0-13.6$        &$374\pm3$         &$3.1$           &$(49-61)\%$  &$18.84\pm0.09$       &$6.79\pm1.27$  \\
$bb\bar{b}\bar{b}(1^{+-})$  &$13.3-13.9$        &$374\pm3$         &$3.1$           &$(48-60)\%$  &$18.84\pm0.09$       &$5.45\pm1.01$  \\
$bb\bar{b}\bar{b}(2^{++})$  &$13.0-13.6$        &$375\pm3$         &$3.1$           &$(51-63)\%$  &$18.85\pm0.09$       &$5.55\pm1.00$  \\
$bb\bar{b}\bar{b}(1^{--})$  &$11.7-12.3$        &$376\pm3$         &$3.1$           &$(47-60)\%$  &$18.89\pm0.09$       &$1.64\pm0.36$  \\  \hline\hline
\end{tabular}
\end{center}
\caption{ The Borel parameters, continuum threshold parameters, energy scales,  pole contributions, masses and pole residues for  the 1S and 1P fully heavy tetraquark states \cite{cccc-WangZG-EPJC-2017,cccc-WangZG-Di-APPB-2019}. } \label{QQQQ-Borel-mass}
\end{table}

In 2020, the LHCb collaboration observed  a broad structure  above the $J/\psi J/\psi$ threshold ranging from 6.2 to 6.8 GeV and a narrow structure at about 6.9 GeV   in the $J/\psi J/\psi$  mass spectrum,  and they also observed  some vague structures around 7.2 GeV \cite{LHCb-cccc-SB-2020}. Accordingly, we obtain the masses of the 2S and 2P states from the QCDSR II in Eq.\eqref{QCDST-2S}, and obtain the 3/4S and 3/4P masses by fitting  the  Regge trajectories,
 \begin{eqnarray}
 M_n^2&=&\alpha (n-1)+\alpha_0\, ,
 \end{eqnarray}
 where the $\alpha$ and $\alpha_0$ are constants, see Tables \ref{cccc-2S-2P-Borel-mass}-\ref{mass-cccc-123}, which  support assigning the broad structure from 6.2 to 6.8 GeV in the di-$J/\psi$ mass spectrum as the 2S or 2P  tetraquark state, and assigning the narrow structure at about 6.9 GeV in the di-$J/\psi$ mass spectrum as the  3S  tetraquark state \cite{WZG-cccc-CPC-2020}.

\begin{table}
\begin{center}
\begin{tabular}{|c|c|c|c|c|c|c|c|}\hline\hline
$J^{PC}$        &$T^2(\rm{GeV}^2)$    &$\sqrt{s_0^\prime}(\rm{GeV})$ &$\mu(\rm{GeV})$ &pole       &$M_{X/Y}(\rm{GeV})$ &$\lambda_{X/Y}(10^{-1}\rm{GeV}^5)$ \\ \hline

$0^{++}(\rm 2S)$  &$4.4-4.8$          &$6.80\pm0.10$          &$2.5$           &$(65-79)\%$       &$6.48\pm0.08$        &$7.41\pm1.12$  \\ \hline

$1^{+-}(\rm 2S)$  &$4.4-4.8$          &$6.85\pm0.10$          &$2.5$           &$(69-82)\%$       &$6.52\pm0.08$        &$5.56\pm0.80$  \\ \hline

$2^{++}(\rm 2S)$  &$4.9-5.3$          &$6.90\pm0.10$          &$2.5$           &$(63-76)\%$       &$6.56\pm0.08$        &$5.92\pm0.83$  \\ \hline

$1^{--}(\rm 2P)$  &$4.5-4.9$          &$6.90\pm0.10$          &$2.2$           &$(57-73)\%$       &$6.58\pm0.09$        &$3.46\pm0.58$  \\ \hline
\hline

\end{tabular}
\end{center}
\caption{ The Borel parameters, continuum threshold parameters, energy scales,  pole contributions, masses and pole residues for  the 2S and 2P $cc\bar{c}\bar{c}$  tetraquark states \cite{WZG-cccc-CPC-2020}. }\label{cccc-2S-2P-Borel-mass}
\end{table}

\begin{table}
\begin{center}
\begin{tabular}{|c|c|c|c|c|c|c|c|}\hline\hline
$J^{PC}$                 &$M_{1}(\rm{GeV})$    &$M_{2}(\rm{GeV})$     &$M_{3}(\rm{GeV})$   &$M_{ 4}(\rm{GeV})$      \\ \hline

$0^{++} $                &$5.99\pm0.08$                       &$6.48\pm0.08$         &$6.94\pm0.08$       &$7.36\pm0.08$             \\ \hline

$1^{+-} $                &$6.05\pm0.08$                       &$6.52\pm0.08$         &$6.96\pm0.08$       &$7.37\pm0.08$              \\ \hline

$2^{++} $                &$6.09\pm0.08$                       &$6.56\pm0.08$         &$7.00\pm0.08$       &$7.41\pm0.08$             \\ \hline

$1^{--}$                 &$6.11\pm0.08$                       &$6.58\pm0.09$         &$7.02\pm0.09$       &$7.43\pm0.09$              \\ \hline
\hline

\end{tabular}
\end{center}
\caption{ The  masses of the $cc\bar{c}\bar{c}$ tetraquark states with the radial quantum numbers $n=1$, $2$, $3$ and $4$ \cite{WZG-cccc-CPC-2020}. }\label{mass-cccc-123}
\end{table}

In 2023, the ATLAS collaboration  observed a narrow resonance at about $6.9\,\rm{GeV}$ and a broader structure at much lower mass in the $J/\psi J/\psi$ channel,  moreover, they observed   a statistically significant excess at about $7.0\,\rm{GeV}$ in the $J/\psi \psi^\prime$ channel \cite{ATLAS-cccc-PRL-2023}.
In 2024, the CMS collaboration observed three resonant structures in the $J/\psi J/\psi$ mass spectrum,
\begin{flalign}
 & X(6600) : M = 6552\pm10\pm12 \mbox{ MeV}\, , \, \Gamma = 124^{+32}_{-26} \pm 33 \mbox{ MeV} \, ,\nonumber \\
 & X(6900) : M = 6927\pm9\pm4 \mbox{ MeV} \, ,\, \Gamma = 122^{+24}_{-21} \pm 18 \mbox{ MeV} \, , \nonumber \\
  & X(7300) : M = 7287^{+20}_{-18}\pm 5 \mbox{ MeV} \, ,\, \Gamma =95^{+59}_{-40} \pm 19 \mbox{ MeV} \, ,
\end{flalign}
in the no-interference model \cite{CMS-cccc-PRL-2024}.

In Ref.\cite{WangZG-X6600-NPB-2022}, we update the calculations by taking  the energy scale $\mu=1.4\,\rm{GeV}$, the lower bound in Fig.\ref{cccc-mu},  and adopt   the relation,
 \begin{eqnarray}
 M_1<\sqrt{s_0}<M_2<\sqrt{s_0^\prime}<M_3\, ,
  \end{eqnarray}
and obtain the predictions, see Tables \ref{cccc-mass-residue-1S-update}-\ref{cccc-mass-residue-2S-update}, and make the possible assignments, see Table \ref{cccc-mass-Regge-update}. From Table \ref{cccc-mass-Regge-update}, we can see explicitly that {\bf the lowest state lies at about $\mathbf {6.2}$ GeV}, which is consistent with the recent  coupled-channel analysis \cite{X6220-GuoFK-PRD-2025}.

\begin{table}
\begin{center}
\begin{tabular}{|c|c|c|c|c|c|c|c|}\hline\hline
$J^{PC}$        &$T^2(\rm{GeV}^2)$    &$\sqrt{s_0}(\rm{GeV})$ &$\mu(\rm{GeV})$ &pole       &$M_{X/Y}(\rm{GeV})$ &$\lambda_{X/Y}(10^{-1}\rm{GeV}^5)$ \\ \hline

$0^{++}(\rm 1S)$  &$3.6-4.0$          &$6.55\pm0.10$          &$1.4$           &$(39-61)\%$       &$6.20\pm0.10$        &$2.68\pm0.57$  \\ \hline

$1^{+-}(\rm 1S)$  &$3.8-4.2$          &$6.60\pm0.10$          &$1.4$           &$(40-61)\%$       &$6.24\pm0.10$        &$2.18\pm0.44$  \\ \hline

$2^{++}(\rm 1S)$  &$4.0-4.4$          &$6.65\pm0.10$          &$1.4$           &$(39-60)\%$       &$6.27\pm0.09$        &$2.35\pm0.46$  \\ \hline

$1^{--}(\rm 1P)$  &$3.2-3.6$          &$6.70\pm0.10$          &$1.2$           &$(39-63)\%$       &$6.33\pm0.10$        &$0.86\pm0.24$  \\ \hline
\hline

\end{tabular}
\end{center}
\caption{ The Borel parameters, continuum threshold parameters, energy scales,  pole contributions, masses and pole residues for  the 1S and 1P  $cc\bar{c}\bar{c}$  tetraquark states \cite{WangZG-X6600-NPB-2022}. }\label{cccc-mass-residue-1S-update}
\end{table}

\begin{table}
\begin{center}
\begin{tabular}{|c|c|c|c|c|c|c|c|}\hline\hline
$J^{PC}$        &$T^2(\rm{GeV}^2)$    &$\sqrt{s_0^\prime}(\rm{GeV})$ &$\mu(\rm{GeV})$ &pole       &$M_{X/Y}(\rm{GeV})$ &$\lambda_{X/Y}(10^{-1}\rm{GeV}^5)$ \\ \hline

$0^{++}(\rm 2S)$  &$4.8-5.2$          &$6.90\pm0.10$          &$2.4$           &$(61-75)\%$       &$6.57\pm0.09$        &$8.12\pm1.21$  \\ \hline

$1^{+-}(\rm 2S)$  &$5.2-5.6$          &$7.00\pm0.10$          &$2.4$           &$(61-75)\%$       &$6.64\pm0.09$        &$6.43\pm0.88$  \\ \hline

$2^{++}(\rm 2S)$  &$5.3-5.7$          &$7.05\pm0.10$          &$2.4$           &$(62-75)\%$       &$6.69\pm0.09$        &$6.84\pm0.92$  \\ \hline

$1^{--}(\rm 2P)$  &$5.0-5.4$          &$7.10\pm0.10$          &$2.4$           &$(60-74)\%$       &$6.74\pm0.09$        &$4.74\pm0.71$  \\ \hline
\hline

\end{tabular}
\end{center}
\caption{ The Borel parameters, continuum threshold parameters, energy scales,  pole contributions, masses and pole residues for the 2S and 2P  $cc\bar{c}\bar{c}$  tetraquark states \cite{WangZG-X6600-NPB-2022}. }\label{cccc-mass-residue-2S-update}
\end{table}

\begin{table}
\begin{center}
\begin{tabular}{|c|c|c|c|c|c|c|c|}\hline\hline
$J^{PC}$         &$M_{1}(\rm{GeV})$      &$M_{2}(\rm{GeV})$     &$M_{3}(\rm{GeV})$   &$M_{ 4}(\rm{GeV})$   \\ \hline

$0^{++} $        &$6.20\pm0.10$          &$6.57\pm0.09$         &$6.92\pm0.09$       &$7.25\pm0.09$       \\

                 &? $X(6220)$            &? $X(6600/6620)$      &? $X(6900)$         &? $X(7220/7300)$      \\ \hline

$1^{+-} $        &$6.24\pm0.10$          &$6.64\pm0.09$         &$7.03\pm0.09$       &$7.40\pm0.09$      \\

                 &? $X(6220)$            &? $X(6600/6620)$      &                    &         \\ \hline

$2^{++} $        &$6.27\pm0.09$          &$6.69\pm0.09$         &$7.09\pm0.09$       &$7.46\pm0.09$          \\ \hline

$1^{--}$         &$6.33\pm0.10$          &$6.74\pm0.09$         &$7.13\pm0.09$       &$7.50\pm0.09$           \\ \hline
\hline

\end{tabular}
\end{center}
\caption{ The  masses of the fully-charm  tetraquark states with the radial quantum numbers $n=1$, $2$, $3$ and $4$. In the lower lines, we present the  possible assignments \cite{WangZG-X6600-NPB-2022}.  }\label{cccc-mass-Regge-update}
\end{table}

The thresholds of the $J/\psi J/\psi$, $J/\psi \psi^\prime$ and $J/\psi \psi(3770)$ are $6194\,\rm{MeV}$, $6783\,\rm{MeV}$ and $6875\,\rm{MeV}$, respectively \cite{PDG-2024}, we cannot obtain a simple molecule scenario to interpret those $X$ states without introducing complex coupled-channel effects
\cite{After-cccc-mole-GuoFK-PRL-2021,After-cccc-mole-GuoZH-PRD-2021,
After-cccc-mole-ZRLiang-PRD-2021,After-cccc-mole-HQZheng-CPC-2021},
or just assign them as the $\bar{\mathbf{3}}\mathbf{3}$ and
$\mathbf{6}{\bar{\mathbf{6}}}$  type tetraquark states \cite{Tetra-33-RQM-Roberts-EPJC-2020,After-cccc-33-LuQF-EPJC-2020,
Tetra-33-QQQQ-Faustov-PRD-2020,CQM-Tetra-ZhaoZ-PRD-2021,
JRZhang-cccc-PRD-2021,Before-Tcc-Faustov-Universe-2021,
After-cccc-33-Lebed-PRD-2020,
After-cccc-33-Karliner-PRD-2020,After-cccc-33-ChenH-PRD-2021,
After-cccc-33-ZhuSL-PRD-2021,After-cccc-33-ZhuRL-NPB-2021,
After-cccc-33-Gordillo-PRD-2020,
After-cccc-33-LiuXH-EPJC-2021,After-cccc-33-Mutuk-EPJC-2021,
After-cccc-33-KTChao-SB-2020,After-cccc-33-GLWang-PRD-2021,
After-cccc-33-ZhongXH-PRD-2021}, or with gluonic constituent \cite{After-cccc-3G3-QiaoCF-PLB-2021}.

We perform Fierz transformation for the currents $J(x)$ and $J_{\mu\nu}(x)$, and obtain particular superpositions of a series of color $\mathbf{1}\mathbf{1}$ type currents,
\begin{eqnarray}
J &=& 2\bar{Q}Q\,\bar{Q}Q+2\bar{Q}i\gamma_5Q\,\bar{Q}i\gamma_5Q+\bar{Q}\gamma_{\alpha} Q\,\bar{Q}\gamma^{\alpha}Q-\bar{Q}\gamma_{\alpha}\gamma_5 Q\,\bar{Q}\gamma^{\alpha}\gamma_5Q \, ,\nonumber\\
J^1_{\mu\nu} &=& 2i\bar{Q}\sigma_{\mu\nu}Q \,\bar{Q}Q+ 2i\varepsilon_{\mu\nu\alpha\beta}\bar{Q}\gamma_{\alpha}\gamma_{5}Q\,
\bar{Q}\gamma_{\beta}Q -2\bar{Q}\sigma_{\mu\nu}\gamma_{5}Q\, \bar{Q}i\gamma_{5}Q \, ,\nonumber\\
\tilde{J}^2_{\mu\nu} &=&  2\bar{Q}\gamma_\mu\gamma_5Q\, \bar{Q}\gamma_\nu\gamma_5Q  -2\bar{Q}\gamma_\mu Q\, \bar{Q}\gamma_\nu Q   +g^{\alpha\beta}\left(\bar{Q}\sigma_{\mu\alpha}Q\, \bar{Q}\sigma_{\nu\beta}Q+\bar{Q}\sigma_{\nu\alpha}Q\, \bar{Q}\sigma_{\mu\beta}Q\right)+\nonumber\\
  && g_{\mu\nu}\Big( \bar{Q}Q\,\bar{Q}Q+\bar{Q}i\gamma_5Q\,\bar{Q}i\gamma_5Q +\bar{Q}\gamma_{\alpha} Q\,\bar{Q}\gamma^{\alpha}Q-\bar{Q}\gamma_{\alpha}\gamma_5 Q\,\bar{Q}\gamma^{\alpha}\gamma_5Q-\frac{1}{2}\bar{Q}\sigma_{\alpha\beta} Q\,\bar{Q}\sigma^{\alpha\beta}Q\Big)  \, ,\nonumber\\
\end{eqnarray}
with $\tilde{J}^2_{\mu\nu}=\sqrt{2}J^2_{\mu\nu}$, the components $\bar{Q}\Gamma Q \bar{Q}\Gamma^\prime Q$ couple potentially to the
molecular states, where the $\Gamma$ and $\Gamma^\prime$ stand for some Dirac $\gamma$-matrixes.
Therefore the $\bar{\mathbf 3}\mathbf{3}$ type tetraquark states have some important $\mathbf{1}\mathbf{1}$ type Fock components, which would decay to their constituents via the Okubo-Zweig-Iizuka super-allowed fall apart mechanism if they are kinematically permitted. There exists a term $\bar{Q}\gamma_\mu Q \bar{Q}\gamma_\nu Q$, the decay to the di-$J/\psi$ is super-allowed, which is consistent with the observations of the ATLAS, CMS and LHCb experiments.

If we insist on that the di-$J/\psi$ system should have positive charge conjugation, we would like to construct a cousin of currents, $J_{-,\mu}^{\widetilde{A}A}(x)$ and $J_{+,\mu}^{\widetilde{A}A}(x)$,
\begin{eqnarray}
J_{-,\mu}^{\widetilde{A}A}(x)&=&\frac{\varepsilon^{ijk}\varepsilon^{imn}}{\sqrt{2}}
\Big[c^{T}_j(x)C\sigma_{\mu\nu}\gamma_5 c_k(x)\bar{c}_m(x)\gamma^\nu C \bar{c}^{T}_n(x)-c^{T}_j(x)C\gamma^\nu c_k(x)\bar{c}_m(x)\gamma_5\sigma_{\mu\nu} C \bar{c}^{T}_n(x) \Big] \, , \nonumber\\
J_{+,\mu}^{\widetilde{A}A}(x)&=&\frac{\varepsilon^{ijk}\varepsilon^{imn}}{\sqrt{2}}
\Big[c^{T}_j(x)C\sigma_{\mu\nu}\gamma_5 c_k(x)\bar{c}_m(x)\gamma^\nu C \bar{c}^{T}_n(x)+c^{T}_j(x)C\gamma^\nu c_k(x)\bar{c}_m(x)\gamma_5\sigma_{\mu\nu} C \bar{c}^{T}_n(x) \Big] \, , \nonumber\\
\end{eqnarray}
which couple potentially to the fully-charm tetraquark states with the $J^{PC}=1^{+-}$ and $1^{++}$, respectively.

In Ref.\cite{WangZG-cccc-P-wave-IJMPA-2021}, we introduce a relative P-wave to construct the doubly-charmed vector diquark operator $\hat{V}$, then  construct the scalar and tensor four-quark currents,
\begin{eqnarray}
J(x)&=&\varepsilon^{ijk}\varepsilon^{imn}
c^{T}_j(x)C\gamma_5\stackrel{\leftrightarrow}{\partial}_\mu c_k(x)\, \bar{c}_m(x)\stackrel{\leftrightarrow}{\partial}_\nu \gamma_5C \bar{c}^{T}_n(x) \, g^{\mu\nu}\, , \nonumber\\
J^1_{\mu\nu}(x)&=&\varepsilon^{ijk}\varepsilon^{imn}\Big\{c^{T}_j(x)C
\gamma_5\stackrel{\leftrightarrow}{\partial}_\mu c_k(x)\, \bar{c}_m(x) \stackrel{\leftrightarrow}{\partial}_\nu \gamma_5C \bar{c}^{T}_n(x)\nonumber\\
&&-c^{T}_j(x)C\gamma_5\stackrel{\leftrightarrow}{\partial}_\nu  c_k(x)\, \bar{c}_m(x)\stackrel{\leftrightarrow}{\partial}_\mu \gamma_5C \bar{c}^{T}_n(x) \Big\} \, , \nonumber \\
J^2_{\mu\nu}(x)&=&\varepsilon^{ijk}\varepsilon^{imn}\Big\{c^{T}_j(x)C\gamma_5\stackrel{\leftrightarrow}{\partial}_\mu c_k(x) \,\bar{c}_m(x) \stackrel{\leftrightarrow}{\partial}_\nu \gamma_5C \bar{c}^{T}_n(x)\nonumber\\
&&+c^{T}_j(x)C\gamma_5\stackrel{\leftrightarrow}{\partial}_\nu  c_k(x)\, \bar{c}_m(x)\stackrel{\leftrightarrow}{\partial}_\mu \gamma_5C \bar{c}^{T}_n(x) \Big\} \, ,
\end{eqnarray}
to study the scalar, axialvector and tensor fully-charm tetraquark states with the QCD sum rules. And we observe that the  ground state  $\hat{V}\hat{V}$ type tetraquark states  and the first radial excited states of the $AA$ type tetraquark states have almost degenerated masses.

We can extend  this subsection directly to study the ${\bar{\mathbf 3}}{\bar{\mathbf 3}}{\bar{\mathbf 3}}$ type fully heavy pentaquark states and
${\mathbf 3}{\mathbf 3}{\mathbf 3}$ type fully heavy hexaquark states \cite{WangZG-5Q-NPB-2021,WangZG-6Q-IJMPA-2022}.

\section{$\mathbf 1\mathbf 1 $ type tetraquark states}\label{11-4-quark}
The $X$, $Y$, $Z$, $T$ and $P$ states always lie near the two-particle thresholds,
such as
\begin{eqnarray}
D\bar{D}^*/\bar{D}D^* &:& X(3872)\, ,\, Z_c(3885/3900)\, , \nonumber \\
DD^* &:& T_{cc}(3875)\, , \nonumber \\
D^*\bar{D}^*&:& Z_c(4020/4025) \, , \nonumber\\
 D\bar{D}_s^*/D^*\bar{D}_s &:& Z_{cs}(3985/4000) \, , \nonumber\\
 D_s^*\bar{D}_s^*&:& X(4140) \, , \nonumber\\
 D\bar{D}_1/\bar{D}D_1 &:& Y(4260/4220)\, , \, Z_c(4250)\, , \nonumber \\
D^*\bar{D}_0/\bar{D}^*D_0 &:& Y(4360/4320)\, , \nonumber \\
  \bar{D}\Sigma_c&:& P_c(4312) \, ,\nonumber \\
  \bar{D}\Xi_c&:& P_{cs}(4338) \, ,\nonumber \\
 \bar{D}\Xi^\prime_c/\bar{D}^*\Xi_c &:& P_{cs}(4459) \, ,\nonumber \\
  \bar{D}\Sigma^*_c&:& P_c(4380) \, ,\nonumber \\
  \bar{D}^*\Sigma_c&:& P_c(4440/4457) \, ,\nonumber \\
  \Lambda_c^+\Lambda_c^{-}/f_0(980)\psi^\prime &:& Y(4660)\, ,\nonumber\\
B\bar{B}^*/\bar{B}B^* &:& Z_b(10610)\, , \nonumber \\
B^*\bar{B}^*&:& Z_b(10650) \, ,
\end{eqnarray}
naively, we expect that they consist of two color-neutral clusters, and they are molecular states, more precisely, they are the $\mathbf 1\mathbf 1 $ type hidden-charm or doubly-charmed tetraquark or pentaquark states \cite{GuoFK-mole-CTP-2021}.
The establishment of the $J^{PC}=1^{++}$ of the $Y(4140)$ by the LHCb collaboration \cite{X4140-X4500-LHCb-PRL-16061,X4140-X4500-LHCb-PRD-16062} excludes  its  assignment as the $D_s^*\bar{D}^*_s$ molecular state with the $J^{PC}=0^{++}$
 \cite{Nielsen-Y4140-mole-PLB-2009,JRZhang-Y4140-mole-JPG-2010,WZG-Y4140-EPJC-2009,
 WZG-Y4140-EPJC-2009-2,
 X4140-mole-LiuX-PRD-2009,
 X4140-mole-DingGJ-EPJC-2009}, however, which does not mean non-existence of the $D_s^*\bar{D}^*_s$ molecular state with the $J^{PC}=0^{++}$.

The $\bar{\mathbf 3}{\mathbf 3}$ type four-quark currents could be reformed in a series of $\mathbf 1\mathbf 1 $ type four-quark currents through Fierz transformation \cite{WangZG-formula-Vect-tetra-EPJC-2014,WangZG-Zc4020-Vector-CTP-2015}, some useful  examples are given explicitly  in the appendix, see Eqs.\eqref{Fierz-Zc3900}-\eqref{Fierz-Scalar}.

According to the quark-hadron duality, the $\bar{\mathbf 3}{\mathbf 3}$ and ${\mathbf 1}{\mathbf 1}$ type local currents couple potentially to the $\bar{\mathbf 3}{\mathbf 3}$ and ${\mathbf 1}{\mathbf 1}$ type tetraquark states, respectively. The $\bar{\mathbf 3}{\mathbf 3}$ type tetraquark states could be taken as a particular superposition of a series of the ${\mathbf 1}{\mathbf 1}$ type tetraquark states, while the ${\mathbf 1}{\mathbf 1}$ type tetraquark states could decay through the Okubo-Zweig-Iizuka super-allowed fall-apart mechanism.
We usually use the identities in Eqs.\eqref{Fierz-Zc3900}-\eqref{Fierz-Scalar} to analyze the strong decays \cite{WangZG-formula-Vect-tetra-EPJC-2014,WangZG-Zc4020-Vector-CTP-2015}. For example, the current in Eq.\eqref{Fierz-Zc3900} couples potentially to the $Z_c(3900)$, in the nonrelativistic and heavy quark limit, the component $\bar{c}\sigma^{\mu\nu}\gamma_5u\,\bar{d}\gamma_\nu c$ can be  reduced to the form,
\begin{eqnarray}
\bar{c}\sigma^{0j}\gamma_5 u \, \bar{d}\gamma_j c&\propto&  \xi^{\dagger}_c\sigma^j\zeta_u\, \chi^{\dagger}_d\vec{\sigma}\cdot \vec{k}_{d}\sigma^j\xi_c\,\,\,\propto\,\,\, \xi^{\dagger}_c\frac{\sigma^j}{2}\zeta_u\, \chi^{\dagger}_d\frac{\sigma^j}{2}\xi_c=\vec{S}_{\bar{D}^*} \cdot \vec{S}_{D^*}  \, , \nonumber\\
\bar{c}\sigma^{ij}\gamma_5 u \, \bar{d}\gamma_j c&\propto& \epsilon^{ijk} \xi^{\dagger}_c\sigma^k\vec{\sigma}\cdot \vec{k}_{u}\zeta_u\, \chi^{\dagger}_d\vec{\sigma}\cdot \vec{k}_{d}\sigma^j\xi_c\,\,\, \propto\,\,\, \epsilon^{ijk} \xi^{\dagger}_c\frac{\sigma^k}{2}\zeta_u\, \chi^{\dagger}_d\frac{ \sigma^j}{2}\xi_c=\vec{S}_{D^*}\times \vec{S}_{\bar{D}^*} \, ,
\end{eqnarray}
where the $\xi$, $\zeta$ and  $\chi$ are two-component spinors of the  quark fields, the $\vec{k}$  are the three-vectors of the  quark fields,   the $\sigma^i$ are the pauli matrixes, and the $\vec{S}$ are the spin operators. It is obvious that the currents $\bar{c}\sigma^{\mu\nu}\gamma_5u\,\bar{d}\gamma_\nu c$ and $\bar{c}\gamma_\nu u\,\bar{d}\sigma^{\mu\nu}\gamma_5 c$
couple potentially   to the $J^P=0^+$ and $1^+$ $(D^*\bar{D}^*)^+$ states. However, the strong decays $Z^{\pm}_c(3900) \to(D^*\bar{D}^*)^{\pm}$ are kinematically forbidden.

\subsection{Hidden heavy tetraquark states}\label{11-tetra-states}
Again, let us  adopt  the  correlation functions $\Pi(p)$, $\Pi_{\mu\nu}(p)$ and $\Pi_{\mu\nu\alpha\beta}(p)$ defined in Eq.\eqref{CF-Pi} and write down the currents
\begin{eqnarray}\label{current-HC-mole}
J(x)&=&J_{D\bar{D}}(x)\, , \,\, J_{D\bar{D}_s}(x)\, , \, \, J_{D_s\bar{D}_s}(x)\, , \,\, J_{D^*\bar{D}^*}(x)\, ,\, \, J_{D^*\bar{D}_s^*}(x)\, , \, \, J_{D_s^*\bar{D}_s^*}(x)\, ,\nonumber\\
 J_\mu(x)&=&J_{D\bar{D}^*,\pm,\mu}(x)\, , \,\, J_{D\bar{D}_s^*,\pm,\mu}(x)\, , \,\, J_{D_s\bar{D}_s^*,\pm,\mu}(x)\, , \nonumber \\
 J_{\mu\nu}(x)&=&J_{\pm,\mu\nu}(x)\, , \nonumber\\
 J_{\pm,\mu\nu}(x)&=&J_{D^*\bar{D}^*,\pm,\mu\nu}(x)\, ,
\,\, J_{D^*\bar{D}_s^*,\pm,\mu\nu}(x)\, , \,\,J_{D_s^*\bar{D}_s^*,\pm,\mu\nu}(x)\, ,
\end{eqnarray}
and
\begin{eqnarray}
J_{D\bar{D}}(x)&=& \bar{q}(x)i\gamma_5  c(x)\,   \bar{c}(x)i\gamma_5  q(x) \, ,\nonumber \\
J_{D\bar{D}_s}(x)&=& \bar{q}(x)i\gamma_5  c(x)\,   \bar{c}(x)i\gamma_5  s(x) \, ,\nonumber \\
J_{D_s\bar{D}_s}(x)&=& \bar{s}(x)i\gamma_5  c(x)\,   \bar{c}(x)i\gamma_5  s(x) \, ,\nonumber \\
J_{D^*\bar{D}^*}(x)&=& \bar{q}(x)\gamma_\mu  c(x)\,   \bar{c}(x)\gamma^\mu  q(x) \, ,\nonumber \\
J_{D^*\bar{D}_s^*}(x)&=& \bar{q}(x)\gamma_\mu  c(x)\,   \bar{c}(x)\gamma^\mu  s(x) \, ,\nonumber \\
J_{D^*_s\bar{D}_s^*}(x)&=& \bar{s}(x)\gamma_\mu  c(x)\,   \bar{c}(x)\gamma^\mu  s(x) \, ,
\end{eqnarray}
\begin{eqnarray}
J_{D\bar{D}^*,\pm,\mu}(x)&=&\frac{1}{\sqrt{2}}\Big[\bar{u}(x)i\gamma_5c(x)\, \bar{c}(x)\gamma_\mu d(x)\mp\bar{u}(x)\gamma_\mu c(x)\, \bar{c}(x)i\gamma_5 d(x) \Big] \, ,\nonumber\\
J_{D\bar{D}_s^*,\pm,\mu}(x)&=&\frac{1}{\sqrt{2}}\Big[\bar{q}(x)i\gamma_5c(x)\, \bar{c}(x)\gamma_\mu s(x)\mp\bar{q}(x)\gamma_\mu c(x)\, \bar{c}(x)i\gamma_5 s(x) \Big] \, ,\nonumber\\
J_{D_s\bar{D}_s^*,\pm,\mu}(x)&=&\frac{1}{\sqrt{2}}\Big[\bar{s}(x)i\gamma_5c(x)\, \bar{c}(x)\gamma_\mu s(x)\mp\bar{s}(x)\gamma_\mu c(x)\, \bar{c}(x)i\gamma_5 s(x) \Big] \, ,
\end{eqnarray}
\begin{eqnarray}
J_{D^*\bar{D}^*,\pm,\mu\nu}(x)&=&\frac{1}{\sqrt{2}}\Big[\bar{u}(x)\gamma_\mu c(x)\, \bar{c}(x)\gamma_\nu d(x)\pm\bar{u}(x)\gamma_\nu c(x)\, \bar{c}(x) \gamma_\mu d(x) \Big] \, ,\nonumber\\
J_{D^*\bar{D}_s^*,\pm,\mu\nu}(x)&=&\frac{1}{\sqrt{2}}\Big[\bar{q}(x)\gamma_\mu c(x)\, \bar{c}(x)\gamma_\nu s(x)\pm\bar{q}(x)\gamma_\nu c(x)\, \bar{c}(x) \gamma_\mu s(x) \Big] \, ,\nonumber\\
J_{D_s^*\bar{D}_s^*,\pm,\mu\nu}(x)&=&\frac{1}{\sqrt{2}}\Big[\bar{s}(x)\gamma_\mu c(x)\, \bar{c}(x)\gamma_\nu s(x)\pm\bar{s}(x)\gamma_\nu c(x)\, \bar{c}(x) \gamma_\mu s(x) \Big] \, ,
\end{eqnarray}
and $q=u$, $d$. The subscripts $D\bar{D}$, $D\bar{D}_s$, $\cdots$ and $D_s^*\bar{D}_s^*$ stand for  the two color-neutral clusters; especially, the subscripts $D\bar{D}^*,\pm$, $D\bar{D}_s^*,\pm$  and $D_s\bar{D}_s^*,\pm$ correspond to  the two color-neutral clusters $D\bar{D}^*\mp D^*\bar{D}$, $D\bar{D}_s^*\mp D^*\bar{D}_s$ and $D_s\bar{D}_s^*\mp D_s^*\bar{D}_s$, respectively, etc.

Again, we take the isospin limit, the currents with the  symbolic quark structures $\bar{c}c\bar{d}u$, $\bar{c}c\bar{u}d$, $\bar{c}c\frac{\bar{u}u-\bar{d}d}{\sqrt{2}}$, $\bar{c}c\frac{\bar{u}u+\bar{d}d}{\sqrt{2}}$ couple potentially  to the hidden-charm
 molecular  states with degenerated  masses, the currents with the isospin $I=1$ and $0$ lead to the same QCD sum rules \cite{X3872-mole-WangZG-HT-EPJC-2014,WZG-tetra-mole-IJMPA-2021,
 WZG-tetra-mole-AAPPS-2022,WangZG-X4140-Zb10650-mole-EPJC-2014}.

 Under parity transformation  $\widehat{P}$, the currents $J(x)$, $J_\mu(x)$ and $J_{\mu\nu}(x)$  have the  properties,
\begin{eqnarray}
\widehat{P} J(x)\widehat{P}^{-1}&=&+J(\tilde{x}) \, , \nonumber\\
\widehat{P} J_\mu(x)\widehat{P}^{-1}&=&-J^\mu(\tilde{x}) \, , \nonumber\\
\widehat{P} J_{\pm,\mu\nu}(x)\widehat{P}^{-1}&=&+J_{\pm,}{}^{\mu\nu}(\tilde{x}) \, ,
\end{eqnarray}
where  $x^\mu=(t,\vec{x})$ and $\tilde{x}^\mu=(t,-\vec{x})$.
 Under charge conjugation transformation  $\widehat{C}$, the currents $J(x)$, $J_\mu(x)$ and $J_{\mu\nu}(x)$ have the properties,
\begin{eqnarray}
\widehat{C}J(x)\widehat{C}^{-1}&=&+ J(x) \, , \nonumber\\
\widehat{C}J_{\pm,\mu}(x)\widehat{C}^{-1}&=&\pm J_{\pm,\mu}(x)  \, , \nonumber\\
\widehat{C}J_{\pm,\mu\nu}(x)\widehat{C}^{-1}&=&\pm J_{\pm,\mu\nu}(x)  \, .
\end{eqnarray}

At the hadron side, we  obtain the hadronic representation and isolate the ground state hidden-charm molecule contributions \cite{WZG-tetra-mole-IJMPA-2021,WZG-tetra-mole-AAPPS-2022},
\begin{eqnarray}
\Pi(p)&=&\frac{\lambda_{Z_+}^2}{M_{Z_+}^2-p^2} +\cdots =\Pi_{+}(p^2) \, ,\nonumber \\
\Pi_{\mu\nu}(p)&=&\frac{\lambda_{Z_+}^2}{M_{Z_+}^2-p^2}\left( -g_{\mu\nu}+\frac{p_{\mu}p_{\nu}}{p^2}\right) +\cdots \nonumber\\
&=&\Pi_{+}(p^2)\left( -g_{\mu\nu}+\frac{p_{\mu}p_{\nu}}{p^2}\right)+\cdots \, ,\nonumber
\end{eqnarray}
\begin{eqnarray}
\Pi_{-,\mu\nu\alpha\beta}(p)&=&\frac{\tilde{\lambda}_{ Z_+}^2}{M_{Z_+}^2-p^2}\left(p^2g_{\mu\alpha}g_{\nu\beta} -p^2g_{\mu\beta}g_{\nu\alpha} -g_{\mu\alpha}p_{\nu}p_{\beta}-g_{\nu\beta}p_{\mu}p_{\alpha}+g_{\mu\beta}p_{\nu}p_{\alpha}+g_{\nu\alpha}p_{\mu}p_{\beta}\right) \nonumber\\
&&+\frac{\tilde{\lambda}_{ Z_-}^2}{M_{Z_-}^2-p^2}\left( -g_{\mu\alpha}p_{\nu}p_{\beta}-g_{\nu\beta}p_{\mu}p_{\alpha}+g_{\mu\beta}p_{\nu}p_{\alpha}+g_{\nu\alpha}p_{\mu}p_{\beta}\right) +\cdots  \nonumber\\
&=&\widetilde{\Pi}_{+}(p^2)\left(p^2g_{\mu\alpha}g_{\nu\beta} -p^2g_{\mu\beta}g_{\nu\alpha} -g_{\mu\alpha}p_{\nu}p_{\beta}-g_{\nu\beta}p_{\mu}p_{\alpha}+g_{\mu\beta}p_{\nu}p_{\alpha}+g_{\nu\alpha}p_{\mu}p_{\beta}\right) \nonumber\\
&&+\widetilde{\Pi}_{-}(p^2)\left( -g_{\mu\alpha}p_{\nu}p_{\beta}-g_{\nu\beta}p_{\mu}p_{\alpha}+g_{\mu\beta}p_{\nu}p_{\alpha}+g_{\nu\alpha}p_{\mu}p_{\beta}\right) \, ,\nonumber
\end{eqnarray}
\begin{eqnarray}
\Pi_{+,\mu\nu\alpha\beta}(p)&=&\frac{\lambda_{ Z_+}^2}{M_{Z_+}^2-p^2}\left( \frac{\widetilde{g}_{\mu\alpha}\widetilde{g}_{\nu\beta}+\widetilde{g}_{\mu\beta}\widetilde{g}_{\nu\alpha}}{2}-\frac{\widetilde{g}_{\mu\nu}\widetilde{g}_{\alpha\beta}}{3}\right) +\cdots \, \, , \nonumber \\
&=&\Pi_{+}(p^2)\left( \frac{\widetilde{g}_{\mu\alpha}\widetilde{g}_{\nu\beta}+\widetilde{g}_{\mu\beta}\widetilde{g}_{\nu\alpha}}{2}-\frac{\widetilde{g}_{\mu\nu}\widetilde{g}_{\alpha\beta}}{3}\right) +\cdots\, ,
\end{eqnarray}
where the $Z$ represents the molecular states $Z_c$, $X_c$, $Z_{cs}$, etc.  We add the subscripts $\pm$  in the hidden-charm  molecular states $Z_{\pm}$ and the components $\Pi_{\pm}(p^2)$ and $\widetilde{\Pi}_{\pm}(p^2)$ to represent  the positive  and negative parity contributions, respectively, and define
the   pole residues $\lambda_{Z_\pm}$ analogous to Eq.\eqref{define-pole-residue} with  $\lambda_{Z_\pm}=M_{Z_\pm}\tilde{\lambda}_{Z_\pm}$.
We choose the components $\Pi_{+}(p^2)$ and $p^2\widetilde{\Pi}_{+}(p^2)$ to study the scalar, axialvector and tensor hidden-charm  molecular states.  According to the discussions in Sect.{\bf \ref{reliable?}},
we have no reason to  worry about the contaminations from the two-meson scattering states.

At the QCD side, we carry out the operator product expansion up to the vacuum condensates of dimension 10, and take  account of the vacuum condensates $\langle\bar{q}q\rangle$, $\langle\frac{\alpha_{s}GG}{\pi}\rangle$, $\langle\bar{q}g_{s}\sigma Gq\rangle$, $\langle\bar{q}q\rangle^2$, $g_s^2\langle\bar{q}q\rangle^2$,
$\langle\bar{q}q\rangle \langle\frac{\alpha_{s}GG}{\pi}\rangle$,  $\langle\bar{q}q\rangle  \langle\bar{q}g_{s}\sigma Gq\rangle$,
$\langle\bar{q}g_{s}\sigma Gq\rangle^2$ and $\langle\bar{q}q\rangle^2 \langle\frac{\alpha_{s}GG}{\pi}\rangle$,  where $q=u$, $d$ or $s$ quark, just like in Sect.{\bf\ref{Tetra-Positive}}.

According to analogous routine of Sect.{\bf\ref{Tetra-Positive}}, we obtain  the  QCD sum rules:
\begin{eqnarray}\label{QCDSR-mole-tetra}
\lambda^2_{Z_+}\, \exp\left(-\frac{M^2_{Z_+}}{T^2}\right)= \int_{4m_c^2}^{s_0} ds\, \rho_{QCD}(s) \, \exp\left(-\frac{s}{T^2}\right) \, ,
\end{eqnarray}
 and
 \begin{eqnarray}\label{QCDSR-mass-mole-tetra}
 M^2_{Z_+}&=& -\frac{\int_{4m_c^2}^{s_0} ds\frac{d}{d \tau}\rho_{QCD}(s)\exp\left(-\tau s \right)}{\int_{4m_c^2}^{s_0} ds \rho_{QCD}(s)\exp\left(-\tau s\right)}\mid_{\tau=\frac{1}{T^2}}\, .
\end{eqnarray}

In the heavy quark limit, we describe the $Q\bar{Q} q\bar{q}$ systems by  a double-well potential model,  the heavy  quark $Q$ ($\bar{Q}$) serves as a static well potential and attracts  the light antiquark $\bar{q}$ ($q$) to form a color-neutral cluster. We introduce the effective heavy quark mass $\mathbb{M}_Q$ and divide the molecular states into both the heavy and light degrees of freedom, i.e.
$2{\mathbb{M}}_Q$ and $\mu=\sqrt{M^2_{X/Y/Z}-(2{\mathbb{M}}_Q)^2}$, respectively.

 Analysis of the $J/\psi$ and $\Upsilon$ mass spectrum
with the famous   Cornell potential or Coulomb-potential-plus-linear-potential leads to the constituent quark masses $m_c=1.84\,\rm{GeV}$ and $m_b=5.17\,\rm{GeV}$ \cite{Cornell}, we can set the effective $c$-quark mass  as  the constituent quark mass ${\mathbb{M}}_c=m_c=1.84\,\rm{GeV}$.   The old value ${\mathbb{M}}_c=1.84\,\rm{GeV}$ and updated value  ${\mathbb{M}}_c=1.85\,\rm{GeV}$, which are fitted  for the hidden-charm  molecular states,  are all consistent with the  constituent quark mass $m_c=1.84\,\rm{GeV}$ \cite{X3872-mole-WangZG-HT-EPJC-2014,WangZG-X4140-Zb10650-mole-EPJC-2014,WZG-Y4220-mole-CPC-2017}. We  choose the value ${\mathbb{M}}_c=1.84\,\rm{GeV}$ to determine the ideal energy scales of the QCD spectral densities, and add an uncertainty $\delta\mu=\pm0.1\,\rm{GeV}$ to account for the difference between the values ${\mathbb{M}}_c=1.84\,\rm{GeV}$ and $1.85\,\rm{GeV}$. Furthermore, we take the modified energy scale formula
\begin{eqnarray}\label{formula-mole-modify}
\mu&=&\sqrt{M^2_{X/Y/Z}-(2{\mathbb{M}}_c)^2}-k\,\mathbb{M}_s\, ,
 \end{eqnarray}
 with $k=0$, $1$, $2$ and $\mathbb{M}_s=0.2\,\rm{GeV}$  to account for the light flavor $SU(3)$ breaking effects.

After trial and error,  we obtain the Borel windows, continuum threshold parameters, energy scales of the QCD spectral densities,  pole contributions, and contributions of the vacuum condensates of dimension $10$,  which are shown explicitly in Table \ref{Borel-HC-mole-tetra}.
At the hadron side, the pole contributions are about $(40-60)\%$, while the central values are larger than $50\%$, the pole dominance condition  is well satisfied.
At the QCD side, the contributions of the vacuum condensates of dimension $10$ are $|D(10)|\leq 1 \%$ or $\ll 1\%$, the convergent behaviors  of the operator product  expansion are  very good.

\begin{table}
\begin{center}
\begin{tabular}{|c|c|c|c|c|c|c|c|c|}\hline\hline
 $Z_c$($X_c$)                            & $J^{PC}$ & $T^2 (\rm{GeV}^2)$ & $\sqrt{s_0}(\rm GeV) $      &$\mu(\rm{GeV})$   &pole         &$|D(10)|$ \\ \hline

$D\bar {D}$                             & $0^{++}$  & $2.7-3.1$         & $4.30\pm0.10$                &$1.3$  &$(40-63)\%$   &$\ll1\%$ \\

$D\bar {D}_s$                           & $0^{++}$  & $2.8-3.2$         & $4.40\pm0.10$                &$1.3$  &$(40-63)\%$   &$\ll1\%$\\

$D_s \bar {D}_s$                        & $0^{++}$  & $2.9-3.3$         & $4.50\pm0.10$                &$1.3$  &$(41-62)\%$   &$\ll1\%$\\ \hline

 $D^*\bar{D}^*$                          & $0^{++}$ & $2.8-3.2$          & $4.55\pm0.10$               &$1.6$             &$(40-62)\%$  &$\leq1\%$   \\

 $D^*\bar{D}_s^*$                        & $0^{++}$ & $2.9-3.3$          & $4.65\pm0.10$               &$1.6$             &$(41-63)\%$  &$<1\%$   \\

 $D_s^*\bar{D}_s^*$                      & $0^{++}$ & $3.1-3.5$          & $4.75\pm0.10$               &$1.6$             &$(40-61)\%$  &$\ll1\%$   \\ \hline

 $D\bar{D}^*-D^*\bar{D}$                  & $1^{++}$ & $2.7-3.1$          & $4.40\pm0.10$               &$1.3$             &$(40-63)\%$  &$\ll1\%$   \\

 $D\bar{D}_s^*-D^*\bar{D}_s$              & $1^{++}$ & $2.9-3.3$          & $4.55\pm0.10$               &$1.3$             &$(41-63)\%$  &$\ll1\%$   \\

 $D_s\bar{D}_s^*-D_s^*\bar{D}_s$          & $1^{++}$ & $3.0-3.4$          & $4.65\pm0.10$               &$1.3$             &$(42-63)\%$  &$\ll1\%$   \\ \hline

$D\bar{D}^*+D^*\bar{D}$                  & $1^{+-}$ & $2.7-3.1$          & $4.40\pm0.10$               &$1.3$             &$(40-63)\%$  &$\ll1\%$   \\

$D\bar{D}_s^*+D^*\bar{D}_s$              & $1^{+-}$ & $2.9-3.3$          & $4.55\pm0.10$               &$1.3$             &$(41-63)\%$  &$\ll1\%$   \\

$D_s\bar{D}_s^*+D_s^*\bar{D}_s$          & $1^{+-}$ & $3.0-3.4$          & $4.65\pm0.10$               &$1.3$             &$(42-63)\%$  &$\ll1\%$   \\  \hline

$D^*\bar{D}^*$                           & $1^{+-}$ & $3.0-3.4$          & $4.55\pm0.10$               &$1.6$             &$(42-63)\%$  &$<1\%$   \\

 $D^*\bar{D}_s^*$                        & $1^{+-}$ & $3.2-3.6$          & $4.65\pm0.10$               &$1.6$             &$(41-61)\%$  &$\ll1\%$   \\

 $D^*_s\bar{D}_s^*$                      & $1^{+-}$ & $3.3-3.7$          & $4.75\pm0.10$               &$1.6$             &$(42-61)\%$  &$\ll1\%$   \\ \hline

 $D^*\bar{D}^*$                          & $2^{++}$ & $3.0-3.4$          & $4.55\pm0.10$               &$1.6$             &$(41-62)\%$  &$<1\%$   \\

 $D^*\bar{D}_s^*$                        & $2^{++}$ & $3.2-3.6$          & $4.65\pm0.10$               &$1.6$             &$(40-60)\%$  &$\ll1\%$   \\

 $D_s^*\bar{D}_s^*$                      & $2^{++}$ & $3.3-3.7$          & $4.75\pm0.10$               &$1.6$             &$(41-61)\%$  &$\ll1\%$   \\

\hline\hline
\end{tabular}
\end{center}
\caption{ The Borel parameters, continuum threshold parameters, energy scales of the QCD spectral densities,  pole contributions and contributions of the vacuum condensates of dimension $10$  for the ground states \cite{WZG-tetra-mole-IJMPA-2021,WZG-tetra-mole-AAPPS-2022}. }\label{Borel-HC-mole-tetra}
\end{table}

\begin{table}
\begin{center}
\begin{tabular}{|c|c|c|c|c|c|c|c|c|}\hline\hline
$Z_c$($X_c$)                                                            & $J^{PC}$  & $M_Z (\rm{GeV})$   & $\lambda_Z (\rm{GeV}^5) $             \\ \hline

$D\bar {D}$                                                             & $0^{++}$  & $3.74\pm0.09$       &$(1.61\pm0.23)\times 10^{-2}$     \\

$D\bar {D}_s$                                                           & $0^{++}$  & $3.88\pm0.10$       &$(1.98\pm0.30)\times 10^{-2}$    \\

$D_s \bar {D}_s$                                                        & $0^{++}$  & $3.98\pm0.10$       &$(2.36\pm0.45)\times 10^{-2}$    \\ \hline

$D^*\bar{D}^*$                                                          & $0^{++}$  & $4.02\pm0.09$      & $(4.30\pm0.72)\times 10^{-2}$           \\

$D^*\bar{D}_s^*$                                                        & $0^{++}$  & $4.10\pm0.09$      & $(5.00\pm0.83)\times 10^{-2}$           \\

$D_s^*\bar{D}_s^*$                                                      & $0^{++}$  & $4.20\pm0.09$      & $(5.86\pm0.98)\times 10^{-2}$           \\ \hline

$D\bar{D}^*-D^*\bar{D}$                                                 & $1^{++}$  & $3.89\pm0.09$      & $(1.72\pm0.30)\times 10^{-2}$           \\

$D\bar{D}_s^*-D^*\bar{D}_s$                                             & $1^{++}$  & $3.99\pm0.09$      & $(1.96\pm0.35)\times 10^{-2}$           \\

$D_s\bar{D}_s^*-D_s^*\bar{D}_s$                                         & $1^{++}$  & $4.07\pm0.09$      & $(2.07\pm0.37)\times 10^{-2}$           \\ \hline

$D\bar{D}^*+D^*\bar{D}$                                                 & $1^{+-}$  & $3.89\pm0.09$      & $(1.72\pm0.30)\times 10^{-2}$           \\

$D\bar{D}_s^*+D^*\bar{D}_s$                                             & $1^{+-}$  & $3.99\pm0.09$      & $(1.96\pm0.35)\times 10^{-2}$           \\

$D_s\bar{D}_s^*+D_s^*\bar{D}_s$                                         & $1^{+-}$  & $4.07\pm0.09$      & $(2.07\pm0.37)\times 10^{-2}$           \\  \hline

$D^*\bar{D}^*$                                                          & $1^{+-}$  & $4.02\pm0.09$      & $(2.33\pm0.35)\times 10^{-2}$           \\

$D^*\bar{D}_s^*$                                                        & $1^{+-}$  & $4.11\pm0.09$      & $(2.71\pm0.41)\times 10^{-2}$           \\

$D_s^*\bar{D}_s^*$                                                      & $1^{+-}$  & $4.19\pm0.09$      & $(3.12\pm0.47)\times 10^{-2}$           \\ \hline

$D^*\bar{D}^*$                                                          & $2^{++}$  & $4.02\pm0.09$      & $(3.29\pm0.51)\times 10^{-2}$           \\

$D^*\bar{D}_s^*$                                                        & $2^{++}$  & $4.11\pm0.09$      & $(3.84\pm0.59)\times 10^{-2}$           \\

$D_s^*\bar{D}_s^*$                                                      & $2^{++}$  & $4.19\pm0.09$      & $(4.42\pm0.67)\times 10^{-2}$           \\
\hline\hline
\end{tabular}
\end{center}
\caption{ The masses and pole residues of the ground state hidden-charm  molecular states \cite{WZG-tetra-mole-IJMPA-2021,WZG-tetra-mole-AAPPS-2022}. }\label{mass-residue-HC-mole-tetra}
\end{table}

We take  account of  all the uncertainties of the relevant  parameters,  and obtain the masses and pole residues of the  molecular  states without strange, with strange and with hidden-strange, which are shown explicitly in Table \ref{mass-residue-HC-mole-tetra}. From  Tables \ref{Borel-HC-mole-tetra}--\ref{mass-residue-HC-mole-tetra}, it is  obvious that the modified energy scale formula in Eq.\eqref{formula-mole-modify} is well satisfied \cite{WZG-tetra-mole-IJMPA-2021,WZG-tetra-mole-AAPPS-2022}.

\begin{figure}
\centering
\includegraphics[totalheight=6cm,width=7cm]{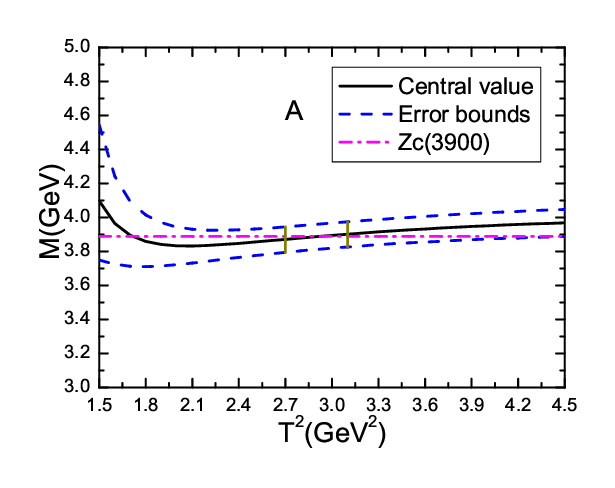}
\includegraphics[totalheight=6cm,width=7cm]{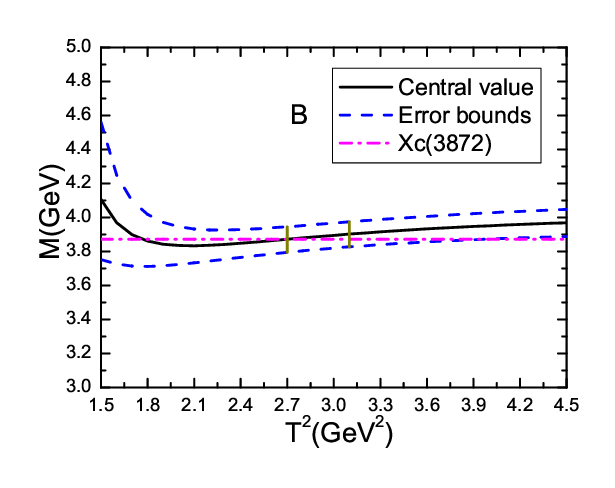}
\includegraphics[totalheight=6cm,width=7cm]{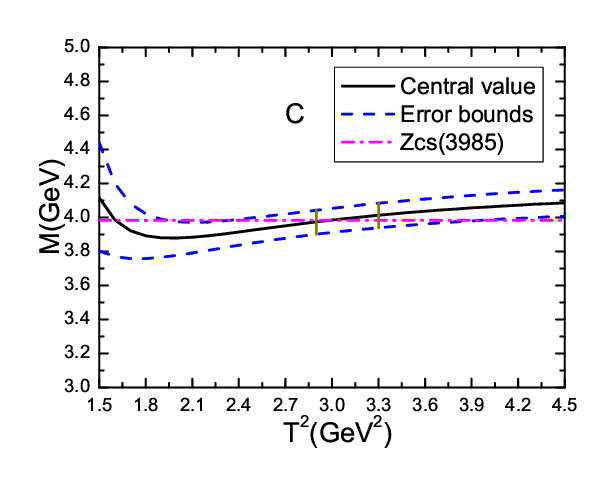}
\includegraphics[totalheight=6cm,width=7cm]{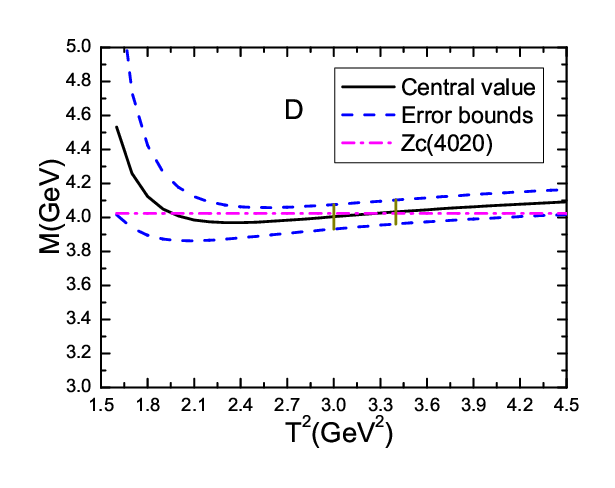}
  \caption{ The masses of the  molecular states with variations of the Borel parameters $T^2$, where the $A$, $B$, $C$  and $D$ correspond  to the axialvector  $D\bar{D}^*+D^*\bar{D}$, $D\bar{D}^*-D^*\bar{D}$, $D\bar{D}_s^*+D^*\bar{D}_s$   and $D^*\bar{D}^*$  molecular states, respectively. }\label{mass-Zc-Zcs-mole}
\end{figure}

 In  Fig.\ref{mass-Zc-Zcs-mole}, we plot the masses of the  axialvector  molecular states $D\bar{D}^*+D^*\bar{D}$, $D\bar{D}^*-D^*\bar{D}$, $D\bar{D}_s^*+D^*\bar{D}_s$   and $D^*\bar{D}^*$ with variations of the Borel parameters at much larger ranges than the Borel widows as an example. From the figure, we can see plainly that there appear very flat platforms in the Borel windows indeed, where the regions between the two short vertical lines are the Borel windows.

In Fig.\ref{mass-Zc-Zcs-mole}, we also present the experimental values of the masses of the $Z_c(3900)$, $X_c(3872)$, $Z_{cs}(3985)$ and $Z_c(4020)$ \cite{BES-Zcs3985-PRL-2021,PDG-2020},  the predicted masses are in excellent agreement   with the experimental data. The calculations support assigning the $Z_c(3900)$, $X_c(3872)$, $Z_{cs}(3985)$ and $Z_c(4020)$ to be the $D\bar{D}^*+D^*\bar{D}$, $D\bar{D}^*-D^*\bar{D}$, $D\bar{D}_s^*+D^*\bar{D}_s$   and $D^*\bar{D}^*$ tetraquark molecular states with the quantum numbers $J^{PC}=1^{+-}$, $1^{++}$, $1^{+-}$ and $1^{+-}$, respectively.
In Table \ref{Assignments-mole-tetra}, we present the possible assignments of the ground state hidden-charm  molecular states. However, the lattice QCD calculations do not favor the existence of the $Z_c(3900)$
  \cite{Latt-Zc3900-mole-No-Prelovsek-PRD-2015,Latt-Zc3900-mole-ChenY-PRD-2014,Latt-Zc3900-mole-Prelovsek-PRC-2013,Latt-Zc3900-mole-Ikeda-PRD-2016}.

We could reproduce the experimental masses of the  $X_c(3872)$, $Z_c(3900)$, $Z_{cs}(3985)$, $Z_{cs}(4123)$ and $Z_c(4020)$ both in the  scenarios  of tetraquark  states and molecular states, see Tables \ref{Identifications-Table-cqcq-positive}-\ref{Assignments-Zcs-mass} and \ref{Assignments-mole-tetra}, the tetraquark scenario  can  accommodate much more exotic $X$ and $Z$ states than the molecule scenario.
Even in the tetraquark scenario, there are no rooms to accommodate the $X(3940)$, $X(4160)$, $Z_c(4100)$ and $Z_c(4200)$ without resorting to fine tuning. The $X(3940)$ and $X(4160)$ might  be  the conventional $\eta_c(\rm3S)$ and $\eta_c(\rm 4S)$ states  with the   $J^{PC}=0^{-+}$, respectively. The $Z_c(4100)$ might be a mixing scalar  tetraquark state with the  $J^{PC}=0^{++}$, and the $Z_c(4200)$ might  be  an axialvector color $\mathbf{8}\mathbf{8}$  type tetraquark state with the  $J^{PC}=1^{+-}$. For detailed discussions about this subject, one can consult Ref.\cite{WZG-HC-PRD-2020}.

The $Z_c(3900)$ and $Z_c(3885)$ have almost degenerated masses but quite different decay widths \cite{BES-Z3900-PRL-2013,Belle-Z3900-PRL-2013,BES-Z3885-PRL-2014}, they are taken as the same particle by the Particle Data Group \cite{PDG-2024} (also Ref.\cite{X3872-tetra-WangZG-HuangT-PRD-2014}), however, it is difficult to  explain  the large  ratio,
\begin{eqnarray}
R_{exp} &=&\frac{\Gamma(Z_c(3885)\to D\bar{D}^*)}{\Gamma(Z_c(3900)\to J/\psi \pi)}=6.2 \pm 1.1 \pm 2.7 \, ,
\end{eqnarray}
from  the BESIII collaboration \cite{BES-Z3885-PRL-2014}. If we assign the $Z_c(3900)$ as the  $\bar{\mathbf{3}}\mathbf{3}$ type  tetraquark state, and assign the $Z_c(3885)$
as  the $D\bar{D}+D^{*}\bar{D}$  molecular state, it is easy to explain the large ratio $R_{exp}$.

The $Z_{cs}(3985)$ observed by the BESIII collaboration near the $D_s^- D^{*0}$ and $D^{*-}_s D^0$ thresholds in the $K^{+}$ recoil-mass spectrum  in the processes  $e^+e^-\to K^+ (D_s^- D^{*0}+ D^{*-}_s D^0)$, its Breit-Wigner  mass and width are  $3985.2^{+2.1}_{-2.0}\pm1.7\,\rm{MeV}$ and $13.8^{+8.1}_{-5.2}\pm4.9\,\rm{MeV}$ respectively with an assignment of the spin-parity $J^P=1^+$  \cite{BES-Zcs3985-PRL-2021}.  The $Z_{cs}(4000)$
observed in the $J/\psi K^+$ mass spectrum by the LHCb collaboration has a mass of $4003 \pm 6 {}^{+4}_{-14}\,\rm{MeV}$, a width of $131 \pm 15 \pm 26\,\rm{MeV}$, and the spin-parity
$J^P =1^+$ \cite{LHCb-Zcs4000-PRL-2021}. The $Z_{cs}(3885)$ and $Z_{cs}(4000)$ are two quite different particles, although they have almost degenerated masses.

In  the $J/\psi K^+$ mass spectrum from the LHCb collaboration, there is a hint of a dip at the energy about $4.1\,\rm{GeV}$ \cite{LHCb-Zcs4000-PRL-2021}, which maybe due to the $Z^-_{cs}(4123)$ observed by the BESIII collaboration  with  a mass about $(4123.5\pm0.7\pm4.7) \, \rm{MeV}$  \cite{BES-Zcs4123-CPC-2023}.
More experimental data are still needed to obtain a precise resolution.

After Ref.\cite{WZG-tetra-mole-IJMPA-2021} was published, the Belle  collaboration observed
weak evidences for two structures  in the $\gamma\psi(2S)$ invariant mass spectrum  in the two-photon process $\gamma\gamma \to \gamma\psi(2S)$, one at $3922.4\pm 6.5\pm 2.0 \,\rm{MeV}$ with a width $22\pm 17\pm 4 \,\rm{MeV}$, and another at $4014.3\pm 4.0\pm 1.5 \,\rm{MeV}$ with a width $4\pm 11\pm 6\,\rm{ MeV}$ \cite{X2-4014-Belle-PRD-2022}. The first structure is consistent with the $X(3915)$ or $\chi_{c2}(3930)$, the second one might  be an  exotic  charmonium-like state. We present its possible assignment in Table \ref{Assignments-mole-tetra}, see $\mathbf {X_2(4014)}$, which cannot be accommodated in the scenario of tetraquark state.

\begin{table}
\begin{center}
\begin{tabular}{|c|c|c|c|c|c|c|c|c|}\hline\hline
$Z_c$($X_c$)                                                            & $J^{PC}$   & $M_Z (\rm{GeV})$   & Assignments              \\ \hline

$D\bar {D}$                                                             & $0^{++}$  & $3.74\pm0.09$       &    \\

$D\bar {D}_s$                                                           & $0^{++}$  & $3.88\pm0.10$       &   \\

$D_s \bar {D}_s$                                                        & $0^{++}$  & $3.98\pm0.10$       & ? $X(3960)$\\

$D^*\bar{D}^*$                                                          & $0^{++}$   & $4.02\pm0.09$      &             \\

$D^*\bar{D}_s^*$                                                        & $0^{++}$   & $4.10\pm0.09$      &              \\

$D_s^*\bar{D}_s^*$                                                      & $0^{++}$   & $4.20\pm0.09$      &             \\ \hline

$D\bar{D}^*-D^*\bar{D}$                                                 & $1^{++}$   & $3.89\pm0.09$       & ? $X_c(3872)$           \\

$D\bar{D}_s^*-D^*\bar{D}_s$                                             & $1^{++}$   & $3.99\pm0.09$       &              \\

$D_s\bar{D}_s^*-D_s^*\bar{D}_s$                                         & $1^{++}$   & $4.07\pm0.09$       &             \\ \hline

$D\bar{D}^*+D^*\bar{D}$                                                 & $1^{+-}$   & $3.89\pm0.09$       & ? $Z_c(3900)$             \\

$D\bar{D}_s^*+D^*\bar{D}_s$                                             & $1^{+-}$   & $3.99\pm0.09$       & ? $Z_{cs}(3985/4000)$             \\

$D_s\bar{D}_s^*+D_s^*\bar{D}_s$                                         & $1^{+-}$   & $4.07\pm0.09$       &               \\  \hline

$D^*\bar{D}^*$                                                          & $1^{+-}$   & $4.02\pm0.09$       & ? $Z_c(4020)$             \\

$D^*\bar{D}_s^*$                                                        & $1^{+-}$   & $4.11\pm0.09$       & ? $Z_{cs}(4100/4123)$      \\

$D_s^*\bar{D}_s^*$                                                      & $1^{+-}$   & $4.19\pm0.09$       &             \\ \hline

$D^*\bar{D}^*$                                                          & $2^{++}$   & $4.02\pm0.09$       &                ? $\bf X_2(4014)$\\

$D^*\bar{D}_s^*$                                                        & $2^{++}$   & $4.11\pm0.
09$       &              \\

$D_s^*\bar{D}_s^*$                                                      & $2^{++}$   & $4.19\pm0.09$       &               \\
\hline\hline
\end{tabular}
\end{center}
\caption{ The possible assignments of the ground state hidden-charm  molecular states, the isospin limit is implied \cite{WZG-tetra-mole-IJMPA-2021,WZG-tetra-mole-AAPPS-2022}.  }\label{Assignments-mole-tetra}
\end{table}

In Table \ref{mass-residue-HC-mole-tetra}, the central values of the predicted molecule masses lie at the thresholds of the corresponding two-meson scattering states, where we have taken the currents having two color-neutral clusters, in each cluster, the constituents $q$ and $\bar{Q}$ (or $Q$ and $\bar{q}$) are in relative S-wave, see Eq.\eqref{current-HC-mole}.

Now let us see the possible outcomes, if one of the color-neutral clusters has a relative P-wave between  the constituents $q$ and $\bar{Q}$ (or $Q$ and $\bar{q}$), as one of the possible assignments of the $Y(4260)$ is the  $D\bar{D}_1$ molecular state with the $J^{PC}=1^{--}$ \cite{ZhaoQ-Y4260-Zc3900-Tri-pole-PRL-2013,GuoFK-Y4260-mole-PRD-2014,OBE-Y4260-mole-DingGJ-PRD-2009,
FKGuo-mole-Review-Progr-2021,Y4260-Latt-Chiu-PRD-2006}.

For example, in the isospin limit, we write the valence quarks of the  $D\bar{D}_1(2420)$ and $D^*\bar{D}_0^*(2400)$ molecular states  symbolically  as
\begin{eqnarray}
\bar{u}d \bar{c}c\, ,\,\,\, \frac{\bar{u}u-\bar{d}d }{\sqrt{2}}\bar{c}c\, ,\,\,\,\bar{d}u \bar{c}c\, ,\,\,\,\frac{\bar{u}u+\bar{d}d }{\sqrt{2}}\bar{c}c\, ,
\end{eqnarray}
the isospin triplet  and  singlet  have degenerate masses. We  take the isospin limit to study those molecular states \cite{WZG-Y4220-mole-CPC-2017}.

Again we resort to the  correlation functions $\Pi_{\mu\nu}(p)$  in Eq.\eqref{CF-Pi} with the currents
 $J_\mu(x)=J_\mu^1(x)$, $J_\mu^2(x)$, $J_\mu^3(x)$ and $J_\mu^4(x)$ to study the $D\bar{D}_1(2420)$ and $D^*\bar{D}_0^*(2400)$ molecular states,
where
\begin{eqnarray}
J^{1/2}_\mu(x)&=&\frac{1}{\sqrt{2}}\left\{ \bar{u}(x)i\gamma_5c(x)\bar{c}(x)\gamma_\mu \gamma_5 d(x)\mp \bar{u}(x)\gamma_\mu \gamma_5c(x)\bar{c}(x)i\gamma_5 d(x)\right\} \, , \nonumber \\
J^{3/4}_\mu(x)&=&\frac{1}{\sqrt{2}}\left\{ \bar{u}(x)c(x)\bar{c}(x)\gamma_\mu  d(x)\pm\bar{u}(x)\gamma_\mu c(x)\bar{c}(x) d(x)\right\}  \, .
\end{eqnarray}
 Under charge conjugation transformation  $\widehat{C}$, the currents $J_\mu(x)$ have the properties,
\begin{eqnarray}
\widehat{C}J^{1/3}_{\mu}(x)\widehat{C}^{-1}&=& - J^{1/3}_\mu(x) |_{u\leftrightarrow d} \, , \nonumber\\
\widehat{C}J^{2/4}_{\mu}(x)\widehat{C}^{-1}&=&+ J^{2/4}_\mu(x)|_{u\leftrightarrow d} \, .
\end{eqnarray}

According to the quark-hadron duality, we  isolate the ground state
contributions of the vector molecular  states,
\begin{eqnarray}
\Pi_{\mu\nu}(p)&=&\frac{\lambda_{Y}^2}{M_{Y}^2-p^2}\left(-g_{\mu\nu} +\frac{p_\mu p_\nu}{p^2}\right) +\cdots \, \, ,
\end{eqnarray}
where the $\lambda_{Y}$ are the pole residues.

We carry out the operator product expansion  up to the vacuum condensates of dimension 10 in a consistent way routinely \cite{WZG-tetra-mole-IJMPA-2021,WZG-tetra-mole-AAPPS-2022,WZG-Y4220-mole-CPC-2017}, and obtain the QCD sum rules for the masses and pole residues.

We adopt the energy scale formula in Eq.\eqref{formula-mole-modify} to choose the best energy scales of the QCD spectral densities, and take  the updated value ${\mathbb{M}}_c=1.85\,\rm{GeV}$ \cite{WZG-Y4220-mole-CPC-2017}.

After trial and error, we obtain the Borel parameters, continuum threshold parameters, pole contributions and energy scales, see Table \ref{Borel-mass-Vector-mole}, where the central values of the pole contributions are larger than $50\%$, the pole dominance is well  satisfied.
In the Borel windows, the  contributions  $D_{10}\ll 1\% $, the operator product expansion is well convergent.

We take  account of  all the  uncertainties of the relevant parameters,
and obtain the values of the masses and pole residues, which are also shown  Table \ref{Borel-mass-Vector-mole} \cite{WZG-Y4220-mole-CPC-2017}.

The prediction  $M_{D\bar{D}_1(1^{--})}=4.36\pm0.08\,\rm{GeV}$ is consistent with the experimental data  $M_{Y(4390)}=4391.6\pm6.3\pm1.0\,\rm{MeV}$ from the BESIII collaboration within uncertainties \cite{BES-Y4390-PRL-2017}, while the predictions  $M_{D\bar{D}_1(1^{-+})}=4.60\pm 0.08\,\rm{GeV}$, $M_{D^*\bar{D}_0^*(1^{--})}=4.78\pm0.07\,\rm{GeV}$ and $M_{D^*\bar{D}_0^*(1^{-+})}=4.73\pm0.07\,\rm{GeV}$   are much larger than upper bound of the experimental data $M_{Y(4220)}=4218.4\pm4.0\pm0.9\,\rm{MeV}$ and $M_{Y(4390)}=4391.6\pm6.3\pm1.0\,\rm{MeV}$ \cite{BES-Y4390-PRL-2017}, moreover, they are much larger than the near thresholds $M_{D^+D_1(2420)^{-}}=4293\,\rm{MeV}$, $M_{D^0D_1(2420)^{0}}=4285\,\rm{MeV}$, $M_{D^{*+}D_0^{*}(2400)^-}=4361\,\rm{MeV}$,   $M_{D^{*0}D_0^{*}(2400)^0}=4325\,\rm{MeV}$ \cite{PDG-2016}.
The present predictions only support  assigning the $Y(4390)$   to be the $D\bar{D}_1(1^{--})$ molecular state. Or the $Y(4260)$ has sizable non $D\bar{D}_1$ component \cite{Y4260-non-DD1-GuoFK-PRD-2019}.

\begin{table}
\begin{center}
\begin{tabular}{|c|c|c|c|c|c|c|c|}\hline\hline
                            &$T^2(\rm{GeV}^2)$ &$\sqrt{s_0}(\rm{GeV})$ &pole        &$\mu(\rm{GeV})$  &$M_{Y}(\rm{GeV})$ &$\lambda_{Y}(10^{-2}\rm{GeV}^5)$\\ \hline
$D\bar{D}_1$ ($1^{--}$)     &$3.2-3.6$         &$4.9\pm0.1$            &$(45-65)\%$ &$2.3$            &$4.36\pm0.08$     &$3.97\pm 0.54 $    \\ \hline

$D\bar{D}_1$ ($1^{-+}$)     &$3.5-3.9$         &$5.1\pm0.1$            &$(44-63)\%$ &$2.7$
&$4.60\pm 0.08$    &$5.26 \pm0.65$     \\ \hline

$D^*\bar{D}_0^*$ ($1^{--}$) &$4.0-4.4$         &$5.3\pm0.1$            &$(44-61)\%$ &$3.0$            &$4.78\pm0.07$     &$7.56\pm 0.84$    \\ \hline

$D^*\bar{D}_0^*$ ($1^{-+}$) &$3.8-4.2$         &$5.2\pm0.1$            &$(44-61)\%$ &$2.9$            &$4.73\pm0.07$     &$6.83\pm 0.84$    \\ \hline \hline
\end{tabular}
\end{center}
\caption{ The Borel parameters, continuum threshold parameters, pole contributions, energy scales, masses and pole residues of the vector molecular  states \cite{WZG-Y4220-mole-CPC-2017}. }\label{Borel-mass-Vector-mole}
\end{table}

In Ref.\cite{JRZhang-mole-PRD-2009}, Zhang and Huang study the $Q\bar{q}\,\bar{Q}^{\prime}q$ type scalar, vector and axialvector molecular states with the QCD sum rules systematically by calculating the operator product expansion up to the vacuum condensates of dimension 6. The predicted molecule masses $M_{D^*\bar{D}^*_0}=4.26 \pm 0.07\,\rm{GeV}$ and $M_{D\bar{D}_1}=4.34 \pm 0.07\,\rm{GeV}$ are consistent with the $Y(4220)$ and $Y(4390)$, respectively.  However, they do not distinguish the charge conjugation of the molecular states and neglect the higher dimensional vacuum condensates.

In Ref.\cite{Nielsen-Z4050-mole-NPA-2009},  Lee, Morita and Nielsen distinguish the  charge conjugation, calculate the operator product expansion up to the vacuum condensates of dimension 6 including dimension 8 partly. They obtain the mass of the $D\bar{D}_1(2420)$ molecular state with the $J^{PC}=1^{-+}$,  $M_{D\bar{D}_1}=4.19 \pm 0.22\,\rm{GeV}$, which differs from the prediction $M_{D\bar{D}_1}=4.34 \pm 0.07\,\rm{GeV}$ significantly \cite{JRZhang-mole-PRD-2009}.

\begin{figure}
 \centering
 \includegraphics[totalheight=5cm,width=14cm]{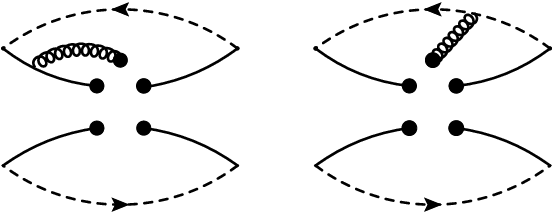}\\
 \vspace{1cm}
  \includegraphics[totalheight=5cm,width=14cm]{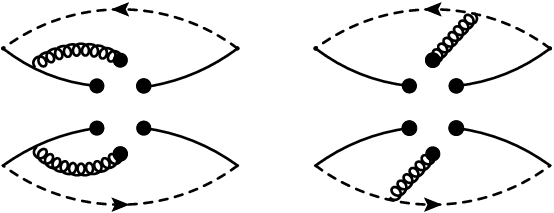}
    \caption{ The  Feynman diagrams  contribute  to  the    $\langle\bar{q}q\rangle\langle\bar{q}g_s\sigma Gq\rangle$
    and $\langle\bar{q}g_s\sigma Gq\rangle^2$. Other diagrams obtained by interchanging of the light  and  heavy quark lines are implied.
       }\label{OPE-qqg-qqg}
\end{figure}

In Refs.\cite{Nielsen-Z4050-mole-NPA-2009,JRZhang-mole-PRD-2009}, {\bf Scheme II} is chosen,   some higher dimensional  vacuum condensates are neglected, which are associated  with $\frac{1}{T^2}$, $\frac{1}{T^4}$, $\frac{1}{T^6}$ in the QCD spectral densities and manifest themselves at small values of the $T^2$, thus  we have to choose large values of  $T^2$ to warrant convergence of the operator product expansion.
The higher dimensional vacuum condensates, see Fig.\ref{OPE-qqg-qqg}, play an important role in determining the Borel windows therefore the ground state  masses and pole residues, we should take them into account consistently. All in all, we observe that the QCD sum rules favor much larger masses than the two-meson thresholds if there exists a P-wave in one constituent, and we should bear in mind that the continuum threshold parameters should not be large enough to include contaminations from the higher resonances \cite{WZG-comment-Narison}, i.e. if a bound state really exists, the continuum threshold $\sqrt{s_0}$   should be less than $M_{ground}+0.7\,\rm{GeV}$ in the case of the hidden-charm four-quark systems.

Besides the $Y(4260)$, it is also difficult to reproduce the mass of the $Y(4660)$ with a $c\bar{q}-q\bar{c}$ type current, however, we can reproduce its mass with a $c\bar{c}-q\bar{q}$ type current \cite{WZG-pseud-Y4660-EPJC-2010,WZG-Y4660-CTP-2010}, i.e. it might be a $\psi^\prime f_0(980)$ bound state \cite{WZG-pseud-Y4660-EPJC-2010,WZG-Y4660-CTP-2010,Y4660-psif0-GuoFK-PLB-2008,Y4660-psif0-GuoFK-PRD-2010}.

With the simple replacement $c\to b$, we obtain the corresponding QCD sum rules for the hidden-bottom tetraquark molecular states from Eq.\eqref{QCDSR-mass-mole-tetra}.
We can reproduce the experimental masses of the $Z_b(10610)$ and $Z_b(10650)$ as the $B\bar{B}^*-B^*\bar{B}$ and $B^*\bar{B}^*$ molecular states respectively with the $J^{PC}=1^{+-}$ via the QCD sum rules
\cite{X3872-mole-WangZG-HT-EPJC-2014,WangZG-X4140-Zb10650-mole-EPJC-2014,
JRZhang-Zb10610-mole-PLB-2011,YLLiu-Zb10610-PRD-2012,WChen-Zc-Zb-mole-PRD-2015}, although those QCD sum rules suffer from shortcomings in one way or the other.
It is more easy to reproduce a weak bound state if there exist weak attractive interactions  between the two constituents \cite{X3872-mole-Nieves-PRD-2012,One-pion-Nieves-PRD-2013,
BSE-tetra-mole-LiXQ-JHEP-2012,
LSE-momentum-Nieves-PRD-2011,
OBE-Zb-mole-ZhuSL-PRD-2011,Zb-Mehen-PRD-2011,Zb-WangQ-PRD-2018,
Zb-Ohkoda-PRD-2012,Zb-Baru-PRD-2019,Simple-mole-Karliner-JHEP-2013}.

In Ref.\cite{Y4260-QiaoCF-PLB-2006}, Qiao assigns the $Y(4260)$ as the $\bar{\Lambda}_c\Lambda_c$ baryonium state with the $J^{PC}=1^{--}$. In Ref.\cite{Y4260-QiaoCF-JPG-2008}, Qiao includes  the $\Sigma_c^0$ constituent, and   suggests    a triplet and a singlet baryonium states,
 \begin{eqnarray}
B^+_1\equiv |\Lambda_c^+ \; \bar{\Sigma}_c^0>\, ,~~~~~~~~~\nonumber\\
{\rm Triplet:}\;\;\;\;\; B^0_1\equiv \frac{1}{\sqrt{2}}(|\Lambda_c^+
\;\bar{\Lambda}^-_c>\; -\; |{\Sigma}_c^0 \bar{\Sigma}_c^0>)\, ,\\
B^-_1\equiv |\Lambda^-_c\; {\Sigma}_c^0>\, ,~~~~~~~~~\nonumber
 \end{eqnarray}
and
\begin{eqnarray}
{\rm Singlet:}\;\;\;\;\; B^0_0\equiv \frac{1}{\sqrt{2}}(|\Lambda_c^+
\;\bar{\Lambda}^-_c>\; + \; |{\Sigma}_c^0 \bar{\Sigma}_c^0>)\, ,
\end{eqnarray}
then he assigns the $Z_c^\pm(4430)$ as the 2S $B_1^\pm$ state with the $J^{PC}=1^{+-}$, and the $Y(4360)$ ($Y(4660)$) as the 2S $\bar{\Lambda}_c\Lambda_c$ ($\bar{\Sigma}_c\Sigma_c$) baryonium state with the $J^{PC}=1^{--}$.

We study the $\bar{\Lambda}_c\Lambda_c$ and $\bar{\Sigma}_c\Sigma_c$ type hidden-charm  baryonium states via the QCD sum rules consistently by carrying out the operator product expansion up to the vacuum condensates of dimension 16 according to the counting roles in Sects.{\bf\ref{Tetra-QCDSR}} and
{\bf\ref{Tetra-Positive}}, and observe that the $\bar{\Lambda}_c\Lambda_c$ state with the $J^P=1^-$  and $\bar{\Sigma}_c\Sigma_c$ states with the $J^P=0^-$, $1^-$ lie at the corresponding baryon-antibaryon thresholds, respectively, while the $\bar{\Lambda}_c\Lambda_c$ states with the $J^P=0^-$, $1^+$ and $\bar{\Sigma}_c\Sigma_c$ states with the $J^P=0^+$, $1^+$ lie above the corresponding baryon-antibaryon thresholds, respectively, whose masses are all much larger than the $Y(4260)$ \cite{WangXW-Sigm-Sigm-AHEP-2022}. In Ref.\cite{QiaoCF-LamLam-EPJC-2020}, Wan, Tang and Qiao study the $\bar{\Lambda}_Q\Lambda_Q$ states with the $J^{PC}=0^{++}$, $0^{-+}$, $1^{++}$ and $1^{--}$ via the QCD sum rules by taking account of the vacuum condensates up to dimension 12, and observe that only the baryonium states with the  $J^{PC}=0^{++}$ and $1^{--}$ could exist, they also obtain masses much larger than the corresponding  $\bar{\Lambda}_Q\Lambda_Q$ thresholds, respectively.

In Ref.\cite{WangZG-XiccXcc-PLB-2021}, we construct the color singlet-singlet type six-quark
current,
\begin{eqnarray}
J(x)&=&\bar{J}_{cc}(x) i\gamma_5 J_{cc}(x)\, , \nonumber\\
J_{cc}(x)&=&\varepsilon^{ijk} c^T_{i}(x) C\gamma_\alpha c_j(x)\gamma^\alpha\gamma^5 q_k(x)\, ,
\end{eqnarray}
with $q=u$ or $d$,
to study
 the $\overline{\Xi}_{cc}\Xi_{cc}$ hexaquark molecular state by calculating the vacuum condensates up to dimension 14,  the predicted mass $M_X \sim 7.2\,\rm{GeV}$ supports assigning the $X(7200)$ to be the $\overline{\Xi}_{cc}\Xi_{cc}$ hexaquark molecular state
 with the $J^{PC}=0^{-+}$.  However, the pole contribution is not larger than $40\%$, which would weaken the predictive ability.

In Ref.\cite{DiZY-DDK-AHEP-2019},  we construct the color singlet-singlet-singlet type six-quark current with the $I(J^P)=\frac{3}{2}(1^-)$
to study the $D\bar{D}^*K$ system via the QCD sum rules by considering the contributions of the vacuum condensates up to dimension-16, and observe that there indeed exists  a resonance state which lies
above the $D\bar{D}^*K$ threshold, and suggest to search for it in the
 $J/\psi \pi K$ mass spectrum.

Such subjects need further studies to obtain definite conclusion, at the present time, they are open problems.

\subsection{Doubly heavy tetraquark states}\label{11-doubly-tetra-states}
Before and after the observation of the doubly-charmed tetraquark candidate $T_{cc}^+(3875)$, especially after the observation,  there have been many  works on the doubly-charmed tetraquark (molecular) states with different theoretical approaches \cite{Latt-Tc-mole-Padmanath-PRL-2022,Latt-Tc-mole-ChenY-PLB-2022,
Latt-Tc-mole-DuML-PRL-2023,
HALQCD-Aoki-PRL-2023,Azizi-Tcc-mole-JHEP-2022,
Azizi-Tcc-tetra-NPB-2022,WZG-cc-mole-EPJA-2022,
After-Tcc-mole-DuML-PRD-2022,After-Tcc-mole-Oset-PRD-2021,
After-Tcc-mole-Albaladejo-PLB-2022,After-Tcc-mole-LiuMZ-PLB-2022,
After-Tcc-mole-MengL-PRD-2021,After-Tcc-mole-DengCR-PRD-2022,
After-Tcc-mole-Fleming-PRD-2021,After-Tcc-mole-MJYan-PRD-2022,
After-Tcc-mole-RChen-PRD-2021,After-Tcc-mole-RLZhu-AHEP-2022}.

If we perform Fierz rearrangements for the  four-quark axialvector  current $J_\mu(x)$ \cite{WZG-QQ-tetra-APPB-2018}, see Eq.\eqref{Current-Tcc-APPB}, we  obtain a special superposition of the color singlet-singlet type currents,
\begin{eqnarray}\label{Fierz-Tcc}
J_\mu(x)&=&\varepsilon^{ijk}\varepsilon^{imn} \, Q^{T}_j(x)C\gamma_\mu Q_k(x) \,\bar{u}_m(x)\gamma_5C \bar{d}^T_n(x) \, , \nonumber\\
&=&\frac{i}{2}\left[\bar{u}i\gamma_5Q \bar{d}\gamma_\mu Q -\bar{d}i\gamma_5 Q\bar{u}\gamma_\mu Q \right]+\frac{1}{2}\left[\bar{u}Q \bar{d}\gamma_\mu\gamma_5 Q -\bar{d} Q\bar{u}\gamma_\mu \gamma_5Q \right]\nonumber\\
&&-\frac{i}{2}\left[\bar{u}\sigma_{\mu\nu}\gamma_5Q \bar{d}\gamma^\nu Q -\bar{d}\sigma_{\mu\nu}\gamma_5 Q\bar{u}\gamma^\nu Q \right]+\frac{i}{2}\left[\bar{u}\sigma_{\mu\nu}Q \bar{d}\gamma^\nu \gamma_5Q -\bar{d}\sigma_{\mu\nu} Q\bar{u}\gamma^\nu\gamma_5 Q \right]\, , \nonumber\\
&=&\frac{i}{2}J_\mu^1(x)+\frac{1}{2}J_\mu^2(x)-\frac{i}{2}J_\mu^3(x)+\frac{i}{2}J_\mu^4(x)\, .
\end{eqnarray}
The currents $J_\mu^1(x)$, $J_\mu^2(x)$, $J_\mu^3(x)$ and $J_\mu^4(x)$ couple potentially to the color singlet-singlet type tetraquark states or two-meson scattering states.
In fact, there exist spatial separations between the diquark and antidiquark pair, the currents $J_\mu(x)$ should be modified to $J_\mu(x,\epsilon)$,
\begin{eqnarray}
J_\mu(x,\epsilon)&=&\varepsilon^{ijk}\varepsilon^{imn} \, Q^{T}_j(x)C\gamma_\mu Q_k(x) \,\bar{u}_m(x+\epsilon)\gamma_5C \bar{d}^T_n(x+\epsilon) \, ,
\end{eqnarray}
where the four-vector $\epsilon^\alpha=(0,\vec{\epsilon})$. The spatial distance between the diquark and antidiquark pair maybe frustrate the Fierz rearrangements or recombination, although  we usually take the local limit $\epsilon \to 0$, we should not take it for granted that the Fierz rearrangements are feasible \cite{Two-particle-Zc3900-WangZG-IJMPA-2020}, we cannot obtain the conclusion that the $T_{cc}^+$ has the $D^{*}D -DD^{*}$ Fock component according to the component  $\bar{q}i\gamma_5c\otimes\bar{q}\gamma_\mu c$ in Eq.\eqref{Fierz-Tcc}. According to the predictions in Table \ref{Borel-QQ-tetra}, we can only obtain the conclusion tentatively that the $T_{cc}^+$ has a diquark-antidiquark type tetraquark Fock component with the spin-parity $J^P=1^+$ and isospin $I=0$ \cite{WZG-QQ-tetra-APPB-2018}. It is interesting to explore whether or not there exists a color $\mathbf 1\mathbf 1 $ type  Fock component indeed.

Again, we resort to the correlation functions $\Pi(p)$, $\Pi_{\mu\nu}(p)$ and $\Pi_{\mu\nu\alpha\beta}(p)$ defined  in Eq.\eqref{CF-Pi}, and construct the currents $J(x)$, $J_\mu(x)$ and $J_{\mu\nu}(x)$,
\begin{eqnarray}
J(x)&=&J_{{D}{D}}(x)\, , \,J_{D{D}_s}(x)\, , \,J_{{D}_s{D}_s}(x)\, , \,J_{{D}^*{D}^*}(x)\, , \,J_{D^*{D}_s^*}(x)\, , \,J_{{D}_s^*{D}_s^*}(x)\, ,
\end{eqnarray}
\begin{eqnarray}
J_{\mu}(x)&=&J_{D{D}^*,L,\mu}(x)\, , \, J_{D{D}^*,H,\mu}(x)\, ,
 J_{D{D}_s^*,L,\mu}(x)\, , \, J_{D{D}_s^*,H,\mu}(x)\, , \,  J_{D_s{D}_s^*,\mu}(x)\, , \nonumber\\
 && J_{D_1D_0^*,L,\mu}(x)\, , \,  J_{D_1D_0^*,H,\mu}(x)\, , \,  J_{D_{s1}D_{0}^*,L,\mu}(x)\, , \, J_{D_{s1}D_{0}^*,H,\mu}(x)\, , \,  J_{D_{s1}D_{s0}^*,\mu}(x)\, ,
 \end{eqnarray}
\begin{eqnarray}
 J_{\mu\nu}(x)&=&J_{{D}^*{D}^*,L,\mu\nu}(x)\, , \, J_{{D}^*{D}^*,H,\mu\nu}(x)\, , \,   J_{{D}^*{D}_s^*,L,\mu\nu}(x)\, , \, J_{{D}^*{D}_s^*,H,\mu\nu}(x)\, , \,J_{{D}_s^*{D}_s^*,L,\mu\nu}(x)\, ,  \nonumber\\
 &&  J_{{D}_s^*{D}_s^*,H,\mu\nu}(x)\, ,
\end{eqnarray}
\begin{eqnarray}
J_{{D}{D}}(x)&=&\bar{u}(x)i\gamma_5 c(x)\, \bar{d}(x)i\gamma_5 c(x) \, ,\nonumber\\
J_{{D}{D}_s}(x)&=&\bar{q}(x)i\gamma_5 c(x)\, \bar{s}(x)i\gamma_5 c(x) \, ,\nonumber\\
J_{{D}_s{D}_s}(x)&=&\bar{s}(x)i\gamma_5 c(x)\, \bar{s}(x)i\gamma_5 c(x) \, , \nonumber\\
J_{{D}^*{D}^*}(x)&=&\bar{u}(x)\gamma_\mu c(x)\, \bar{d}(x)\gamma^\mu c(x) \, ,\nonumber\\
J_{{D}^*{D}_s^*}(x)&=&\bar{q}(x)\gamma_\mu c(x)\, \bar{s}(x)\gamma^\mu c(x) \, ,\nonumber\\
J_{{D}^*_s{D}_s^*}(x)&=&\bar{s}(x)\gamma_\mu c(x)\, \bar{s}(x)\gamma^\mu c(x) \, ,
\end{eqnarray}
\begin{eqnarray}
J_{D{D}^*,L/H,\mu}(x)&=&\frac{1}{\sqrt{2}}\Big[\bar{u}(x)i\gamma_5c(x)\, \bar{d}(x)\gamma_\mu c(x) \mp\bar{u}(x)\gamma_\mu c(x)\,\bar{d}(x)i\gamma_5 c(x)\,  \Big] \, ,\nonumber\\
J_{D{D}_s^*,L/H,\mu}(x)&=&\frac{1}{\sqrt{2}}\Big[\bar{q}(x)i\gamma_5c(x)\, \bar{s}(x)\gamma_\mu c(x)\mp\bar{q}(x)\gamma_\mu c(x)\, \bar{s}(x)i\gamma_5 c(x) \Big] \, ,\nonumber\\
J_{D_s{D}_s^*,\mu}(x)&=&\bar{s}(x)i\gamma_5c(x)\, \bar{s}(x)\gamma_\mu c(x) \, ,
\end{eqnarray}
\begin{eqnarray}
J_{D_1D_0^*,L/H,\mu}(x)&=&\frac{1}{\sqrt{2}}\Big[\bar{u}(x)c(x)\, \bar{d}(x)\gamma_\mu \gamma_5c(x) \pm\bar{u}(x)\gamma_\mu \gamma_5  c(x)\, \bar{d}(x)c(x) \Big] \, ,\nonumber\\
J_{D_{s1}D_{0}^*,L/H,\mu}(x)&=&\frac{1}{\sqrt{2}}\Big[\bar{q}(x)c(x)\, \bar{s}(x)\gamma_\mu \gamma_5 c(x)\pm\bar{q}(x)\gamma_\mu \gamma_5 c(x)\, \bar{s}(x) c(x) \Big] \, ,\nonumber\\
J_{D_{s1}D_{s0}^*,\mu}(x)&=&\bar{s}(x)c(x)\, \bar{s}(x)\gamma_\mu \gamma_5 c(x) \, ,
\end{eqnarray}
\begin{eqnarray}
J_{{D}^*{D}^*,L/H,\mu\nu}(x)&=&\frac{1}{\sqrt{2}}\Big[\bar{u}(x)\gamma_\mu c(x)\, \bar{d}(x)\gamma_\nu c(x) \mp\bar{u}(x)\gamma_\nu c(x)\,\bar{d}(x)\gamma_\mu c(x)  \Big] \, ,\nonumber\\
J_{{D}^*{D}_s^*,L/H,\mu\nu}(x)&=&\frac{1}{\sqrt{2}}\Big[\bar{q}(x)\gamma_\mu c(x)\, \bar{s}(x)\gamma_\nu c(x)\mp\bar{q}(x)\gamma_\nu c(x)\, \bar{s}(x)\gamma_\mu c(x) \Big] \, ,\nonumber\\
J_{{D}^*_s{D}_s^*,L/H,\mu\nu}(x)&=&\frac{1}{\sqrt{2}}\Big[\bar{s}(x)\gamma_\mu c(x)\, \bar{s}(x)\gamma_\nu c(x)\mp\bar{s}(x)\gamma_\nu c(x)\, \bar{s}(x)\gamma_\mu c(x) \Big] \, ,
\end{eqnarray}
and $q=u$, $d$, the subscripts $DD$, $DD^*$, $\cdots$ and $D_s^*D_s^*$ stand for the two color-neutral clusters,  we  add the subscripts $L$ and $H$ to distinguish the lighter and heavier  states in the same doublet due to the mixing effects, as direct calculations indicate that there exists such a tendency.

 Under parity transformation  $\widehat{P}$, the currents have the  properties,
\begin{eqnarray}\label{J-parity-Tcc-mole}
\widehat{P} J(x)\widehat{P}^{-1}&=&+J(\tilde{x}) \, , \nonumber\\
\widehat{P} J_\mu(x)\widehat{P}^{-1}&=&-J^\mu(\tilde{x}) \, , \nonumber\\
\widehat{P} J_{\mu\nu}(x)\widehat{P}^{-1}&=&+J^{\mu\nu}(\tilde{x}) \, ,
\end{eqnarray}
where $x^\mu=(t,\vec{x})$ and $\tilde{x}^\mu=(t,-\vec{x})$.
We rewrite Eq.\eqref{J-parity-Tcc-mole} in  more explicit form,
\begin{eqnarray}
\widehat{P} J_i(x)\widehat{P}^{-1}&=&+J_i(\tilde{x}) \, ,\nonumber\\
\widehat{P} J_{ij}(x)\widehat{P}^{-1}&=&+J_{ij}(\tilde{x}) \, ,
\end{eqnarray}
\begin{eqnarray}
\widehat{P} J_0(x)\widehat{P}^{-1}&=&-J_0(\tilde{x}) \, , \nonumber\\
\widehat{P} J_{0i}(x)\widehat{P}^{-1}&=&-J_{0i}(\tilde{x}) \, ,
\end{eqnarray}
where the space indexes $i$, $j=1$, $2$, $3$. There are both positive and negative components, and they couple potentially to the axialvector/tensor and pseudoscalar/vector  molecular states, respectively. We will introduce an superscript $-$ to denote the negative parity.

According to the quark-hadron duality, we obtain the hadronic representation
and isolate the ground state  contributions  of the scalar, axialvector and tensor molecular states \cite{WZG-tetra-mole-AAPPS-2022,WZG-cc-mole-EPJA-2022},
\begin{eqnarray}
\Pi(p)&=&\frac{\lambda_{T}^2}{M_{T}^2-p^2} +\cdots =\Pi^{0}_{T}(p^2)+\cdots \, ,\nonumber
\end{eqnarray}
\begin{eqnarray}
\Pi_{\mu\nu}(p)&=&\frac{\lambda_{T}^2}{M_{T}^2-p^2}\left( -g_{\mu\nu}+\frac{p_{\mu}p_{\nu}}{p^2}\right) +\cdots \nonumber\\
&=&\Pi_{T}^{1}(p^2)\left( -g_{\mu\nu}+\frac{p_{\mu}p_{\nu}}{p^2}\right)+\cdots \, ,\nonumber
\end{eqnarray}
\begin{eqnarray}
\Pi_{L,\mu\nu\alpha\beta}(p)&=&\frac{\tilde{\lambda}_{ T}^2}{M_{T}^2-p^2}\left(p^2g_{\mu\alpha}g_{\nu\beta} -p^2g_{\mu\beta}g_{\nu\alpha} -g_{\mu\alpha}p_{\nu}p_{\beta}-g_{\nu\beta}p_{\mu}p_{\alpha}+g_{\mu\beta}p_{\nu}p_{\alpha}+g_{\nu\alpha}p_{\mu}p_{\beta}\right) \nonumber\\
&&+\frac{\tilde{\lambda}_{ T^-}^2}{M_{T^-}^2-p^2}\left( -g_{\mu\alpha}p_{\nu}p_{\beta}-g_{\nu\beta}p_{\mu}p_{\alpha}+g_{\mu\beta}p_{\nu}p_{\alpha}+g_{\nu\alpha}p_{\mu}p_{\beta}\right) +\cdots  \nonumber\\
&=&\widetilde{\Pi}^{1}_{T}(p^2)\left(p^2g_{\mu\alpha}g_{\nu\beta} -p^2g_{\mu\beta}g_{\nu\alpha} -g_{\mu\alpha}p_{\nu}p_{\beta}-g_{\nu\beta}p_{\mu}p_{\alpha}+g_{\mu\beta}p_{\nu}p_{\alpha}+g_{\nu\alpha}p_{\mu}p_{\beta}\right) \nonumber\\
&&+\widetilde{\Pi}_{T}^{1,-}(p^2)\left( -g_{\mu\alpha}p_{\nu}p_{\beta}-g_{\nu\beta}p_{\mu}p_{\alpha}+g_{\mu\beta}p_{\nu}p_{\alpha}+g_{\nu\alpha}p_{\mu}p_{\beta}\right) \, ,\nonumber
\end{eqnarray}
\begin{eqnarray}
\Pi_{H,\mu\nu\alpha\beta}(p)&=&\frac{\lambda_{ T}^2}{M_{T}^2-p^2}\left( \frac{\widetilde{g}_{\mu\alpha}\widetilde{g}_{\nu\beta}+\widetilde{g}_{\mu\beta}\widetilde{g}_{\nu\alpha}}{2}-\frac{\widetilde{g}_{\mu\nu}\widetilde{g}_{\alpha\beta}}{3}\right) +\cdots \, \, , \nonumber \\
&=&\Pi_{T}^{2}(p^2)\left( \frac{\widetilde{g}_{\mu\alpha}\widetilde{g}_{\nu\beta}+\widetilde{g}_{\mu\beta}\widetilde{g}_{\nu\alpha}}{2}-\frac{\widetilde{g}_{\mu\nu}\widetilde{g}_{\alpha\beta}}{3}\right) +\cdots\, ,
\end{eqnarray}
where the  pole residues   $\lambda_{T}$ and $\lambda_{T^{-}}$ are defined analogous to Eq.\eqref{define-pole-residue} with the simple replacements  $Z^+ \to T$ and $Z^-\to T^-$, and $\lambda_{T}=\tilde{\lambda}_{T}M_{T}$ and $\lambda_{T^-}=\tilde{\lambda}_{T^-}M_{T^-}$, we introduce the superscripts $0$, $1$ and $2$ denote the spins of the molecular states.
Thereafter, we choose the components $\Pi_{T}^{0/2}(p^2)$ and $p^2\widetilde{\Pi}_{T}^1(p^2)$.

We accomplish  the operator product expansion  up to the vacuum condensates of dimension $10$  and take  account of the vacuum condensates $\langle\bar{q}q\rangle$, $\langle\frac{\alpha_{s}GG}{\pi}\rangle$, $\langle\bar{q}g_{s}\sigma Gq\rangle$, $\langle\bar{q}q\rangle^2$,
$\langle\bar{q}q\rangle \langle\frac{\alpha_{s}GG}{\pi}\rangle$,  $\langle\bar{q}q\rangle  \langle\bar{q}g_{s}\sigma Gq\rangle$,
$\langle\bar{q}g_{s}\sigma Gq\rangle^2$ and $\langle\bar{q}q\rangle^2 \langle\frac{\alpha_{s}GG}{\pi}\rangle$ with the vacuum saturation in a consistent way \cite{WZG-Saturation},  where  $q=u$, $d$ or $s$ \cite{X3872-tetra-WangZG-HuangT-PRD-2014,WZG-HC-PRD-2020,
WZG-tetra-mole-IJMPA-2021,WZG-tetra-mole-AAPPS-2022,
WangZG-Zcs4123-tetra-CPC-2022,
WangZG-Zcs3985-mass-tetra-CPC-2021,
WangZG-Regge-Ms-CPC-2021,WZG-HC-ss-NPB-2024,WZG-cc-mole-EPJA-2022,
WangZG-Pcs4459-penta-IJMPA-2021,WZG-HC-Pseudo-NPB-2022}.
 Again, we neglect the $u$ and $d$ quark masses, and take account of  the terms $\propto m_s$ according to the light-flavor $SU(3)$ breaking effects.

Then we match  the hadron side with the QCD  side below the continuum thresholds  $s_0$, perform the Borel transformation, and obtain the  QCD sum rules for the masses and pole residues.
With a simple replacement $c \to b$, we obtain the corresponding QCD sum rules for the doubly-bottom tetraquark states, the calculations are straightforward.

We take the modified energy scale formula shown in Eq.\eqref{formula-mole-modify} and  the effective $c/s$-quark masses $\mathbb{M}_c=1.82\,\rm{GeV}$ and $\mathbb{M}_s=0.2\,\rm{GeV}$ to determine the optimal energy scales of the QCD spectral densities.
After trial and error, we obtain the Borel parameters, continuum threshold parameters, energy scales of the QCD spectral densities,  pole contributions and contributions of the  vacuum condensates of dimension 10, which are shown  in Table \ref{Borel-Tcc-mole} \cite{WZG-tetra-mole-AAPPS-2022,WZG-cc-mole-EPJA-2022}. The pole contributions
 are  about $(40-60)\%$ at the hadron side, the central values are larger than $50\%$, the pole dominance is  satisfied very good.  Moreover, the contributions of the vacuum condensates of dimension $10$ are $|D(10)|< 1 \%$ or $\ll 1\%$ at the QCD side, the operator product expansion converges  very good \cite{WZG-tetra-mole-AAPPS-2022,WZG-cc-mole-EPJA-2022}.

We take  account of  all the uncertainties of the relevant  parameters,  and obtain the masses and pole residues of the  doubly-charmed molecular  states without strange, with strange and with doubly-strange, which are presented  explicitly in Table \ref{mass-residue-Tcc-mole-Table} \cite{WZG-tetra-mole-AAPPS-2022,WZG-cc-mole-EPJA-2022}.

\begin{figure}
\centering
\includegraphics[totalheight=6cm,width=7cm]{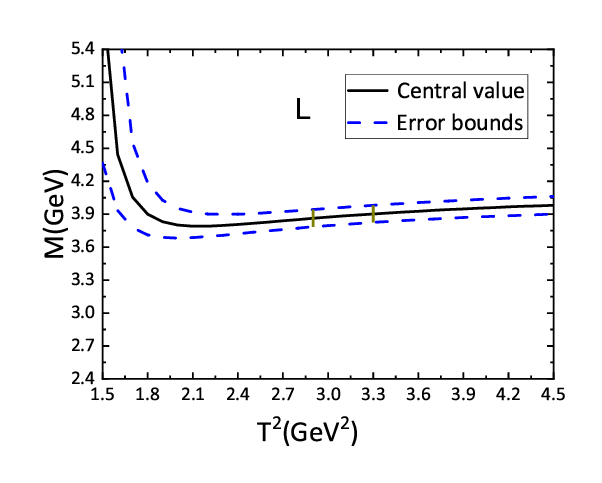}
\includegraphics[totalheight=6cm,width=7cm]{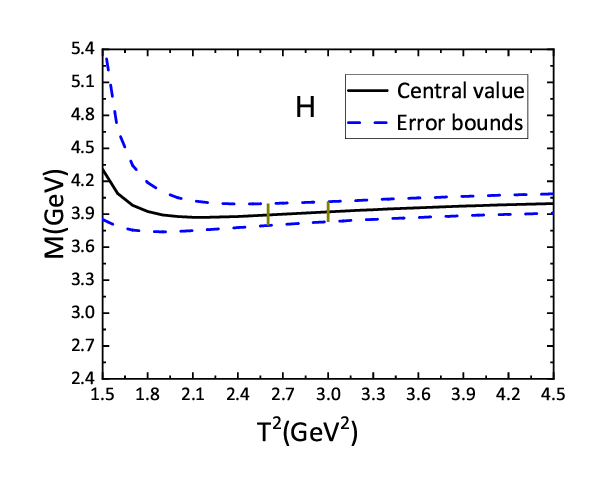}
  \caption{ The masses  with variations of the  Borel parameters for the axialvector tetraquark  molecular states, where the $L$ and $H$ denote the lighter and heavier $DD^{*}$ states,  respectively.   } \label{mass-Borel-Tcc-mole}
\end{figure}

 In  Fig.\ref{mass-Borel-Tcc-mole}, as an example, we plot the masses of the  axialvector $(D^{*}D -DD^{*})_L$ and
$(D^{*}D +DD^{*})_H$  molecular states   with variations of the Borel parameters at much larger ranges than the Borel widows. There appear very flat platforms in the Borel windows,  the regions between the two short perpendicular lines.

There exist both a lighter and a heavier state for the $cc\bar{u}\bar{d}$ and $cc\bar{q}\bar{s}$  molecular states, the lighter state
$(D^{*}D -DD^{*})_L$  with the isospin $(I,I_3)=(0,0)$  has a  mass $3.88\pm0.11\,\rm{GeV}$, which is in excellent agreement with the mass of the doubly-charmed tetraquark candidate $T_{cc}^+$ from the LHCb collaboration \cite{LHCb-Tcc-NatureP-2022,LHCb-Tcc-NatureC-2022}, and supports assigning the $T_{cc}^+$ to be the $(D^{*}D -DD^{*})_L$ molecular state, as the $T_{cc}^+$ has the isospin $I=0$. In other words, the exotic state $T_{cc}^+$ maybe have  a $(D^{*}D -DD^{*})_L$ Fock component.  The heavier state $(D^{*}D +DD^{*})_H$ with the isospin $(I,I_3)=(1,0)$ has a  mass  $3.90\pm0.11\,\rm{GeV}$, the central value lies slightly above the $DD^*$ threshold, the strong decays to the final states $DD\pi$ are  kinematically allowed but with small phase-space.
If we choose the same input parameters, the $DD^*$ molecular state with the isospin $I=1$ has slightly larger mass than the corresponding molecule with the isospin $I=0$, it is  indeed that the isoscalar $DD^*$ molecular state is lighter.

For the $(D_0^*D_1 +D_1D_0^*)_L$, $(D_0^*D_1 -D_1D_0^*)_H$,
$(D_{0}^*D_{s1} +D_{s0}^*D_1)_L$ and
$(D_{0}^*D_{s1} -D_{s0}^*D_1)_H$ molecular states, there exists  a  P-wave in the color-singlet constituents, the P-wave is embodied implicitly in the underlined $\underline{\gamma_5}$ in the scalar currents $\bar{q}i\gamma_5 \underline{\gamma_5}c$, $\bar{s}i\gamma_5 \underline{\gamma_5}c$ and axialvector currents  $\bar{q}\gamma_\mu \underline{\gamma_5} c$, $\bar{s}\gamma_\mu \underline{\gamma_5} c$, as multiplying $\gamma_5$ to the pseudoscalar  currents $\bar{q}i\gamma_5 c$, $\bar{s}i\gamma_5 c$ and vector currents  $\bar{q}\gamma_\mu  c$, $\bar{s}\gamma_\mu  c$ changes their parity. We should introduce the spin-orbit interactions to account for the large mass gaps between the lighter and heavier states $(L,H)$, i.e. $((D_0^*D_1 +D_1D_0^*)_L, \,(D_0^*D_1 -D_1D_0^*)_H)$,
$((D_{0}^*D_{s1} +D_{s0}^*D_1)_L, \,(D_{0}^*D_{s1} -D_{s0}^*D_1)_H)$ \cite{WZG-tetra-mole-AAPPS-2022,WZG-cc-mole-EPJA-2022}.

From Table \ref{mass-residue-Tcc-mole-Table}, we can see explicitly that the $D_0^*D_1 -D_1D_0^*$, $D_0^*D_1 +D_1D_0^*$, $D_{0}^*D_{s1} -D_{s0}^*D_1$, $D_{0}^*D_{s1} +D_{s0}^*D_1$ and $D_{s1}D_{s0}^*$
 molecular states have much larger masses than the corresponding two-meson thresholds, just like in the case of the hidden-charm molecular states studied in the previous sub-section, where there exist a relative P-wave in one of the color-neutral clusters.

\begin{table}
\begin{center}
\begin{tabular}{|c|c|c|c|c|c|c|c|c|}\hline\hline
$T_{cc}$                           &Isospin        &$T^2(\rm{GeV}^2)$   & $\sqrt{s_0}(\rm GeV) $   &$\mu(\rm{GeV})$  &pole          &$|D(10)|$        \\ \hline

$D{D}$                             &1              &$2.6-3.0$          &$4.30\pm0.10$             &$1.4$  &$(40-64)\%$   &$\ll1\%$      \\

$D{D}_s$                           &$\frac{1}{2}$  &$2.7-3.1$          &$4.40\pm0.10$             &$1.4$ &$(41-64)\%$   &$\ll1\%$       \\

$D_s {D}_s$                        &0              &$2.8-3.2$          &$4.50\pm0.10$             &$1.4$ &$(41-63)\%$   &$\ll1\%$    \\

$D^{*}D^{*}$                       &1              & $2.8-3.2$          & $4.55\pm0.10$            &$1.7$            &$(41-61)\%$   &$<1\%$            \\

$D_{s}^*D^{*}$                     &$\frac{1}{2}$  & $2.9-3.3$          & $4.65\pm0.10$            &$1.7$            &$(42-62)\%$   &$\ll1\%$        \\

$D_{s}^*D_{s}^*$                   & 0             & $3.2-3.5$          & $4.80\pm0.10$            &$1.8$           &$(42-61)\%$    &$\ll1\%$       \\

$D^{*}D -DD^{*}$                   &0              & $2.9-3.3$          & $4.45\pm0.10$            &$1.4$            &$(42-62)\%$   &$\ll1\%$     \\

$D^{*}D +DD^{*}$                   &1              & $2.6-3.0$          & $4.40\pm0.10$            &$1.4$            &$(42-63)\%$   &$\ll1\%$   \\

$D_s^*D -D_sD^*$                   &$\frac{1}{2}$  & $3.0-3.4$          & $4.50\pm0.10$            &$1.5$            &$(40-62)\%$   &$<1\%$   \\

$D_s^*D +D_sD^*$                   &$\frac{1}{2}$  & $2.9-3.3$          & $4.50\pm0.10$            &$1.5$            &$(40-60)\%$   &$\ll1\%$ \\

$D_s^*D_s$                         & 0             & $3.0-3.4$          & $4.60\pm0.10$            &$1.5$            &$(41-63)\%$   &$\ll1\%$  \\

$D_0^*D_1 -D_1D_0^*$               & 0             & $5.6-7.0$          & $6.35\pm0.10$            &$4.6$            &$(41-60)\%$   &$\ll1\%$   \\

$D_0^*D_1 +D_1D_0^*$               & 1             & $4.7-6.1$          & $5.90\pm0.10$            &$4.0$            &$(42-61)\%$   &$\ll1\%$   \\

$D_{0}^*D_{s1} -D_{s0}^*D_1$       &$\frac{1}{2}$  & $5.8-7.2$          & $6.50\pm0.10$            &$4.6$             &$(43-60)\%$  &$<1\%$   \\

$D_{0}^*D_{s1} +D_{s0}^*D_1$       &$\frac{1}{2}$  & $4.7-6.1$          & $6.05\pm0.10$            &$4.0$             &$(42-62)\%$  &$<1\%$   \\

$D_{s1}D_{s0}^*$                   & 0             & $4.9-6.3$          & $6.20\pm0.10$            &$4.0$             &$(43-61)\%$  &$<1\%$   \\

$D^{*}D^{*}-D^{*}D^{*}$            &0              & $3.2-3.6$          & $4.55\pm0.10$            &$1.7$             &$(42-61)\%$  &$<1\%$ \\

$D^{*}D^{*}+D^{*}D^{*}$            &1              & $3.0-3.4$          & $4.55\pm0.10$            &$1.7$             &$(41-60)\%$  &$<1\%$  \\

$D_{s}^*D^{*}-D_{s}^{*}D^*$        &$\frac{1}{2}$  & $3.3-3.7$          & $4.65\pm0.10$            &$1.7$             &$(40-59)\%$  &$\ll1\%$ \\

$D_{s}^*D^{*}+D_{s}^{*}D^*$        &$\frac{1}{2}$  & $3.1-3.5$          & $4.65\pm0.10$            &$1.7$             &$(42-61)\%$  &$<1\%$   \\

$D_{s}^*D_{s}^*-D_{s}^*D_{s}^*$    & 0             & $3.6-4.0$          & $4.80\pm0.10$            &$1.8$             &$(40-60)\%$  &$\ll1\%$ \\

$D_{s}^*D_{s}^*+D_{s}^*D_{s}^*$    & 0             & $3.4-3.9$          & $4.80\pm0.10$            &$1.8$             &$(41-61)\%$  &$<1\%$    \\
\hline\hline
\end{tabular}
\end{center}
\caption{ The Borel parameters, continuum threshold parameters, energy scales of the QCD spectral densities,  pole contributions  and  contributions of the vacuum condensates of dimension $10$  for the doubly-charmed  molecular states \cite{WZG-tetra-mole-AAPPS-2022,WZG-cc-mole-EPJA-2022}. }\label{Borel-Tcc-mole}
\end{table}

\begin{table}
\begin{center}
\begin{tabular}{|c|c|c|c|c|c|c|c|c|}\hline\hline
$T_{cc}$                           &Isospin          & $M_T (\rm{GeV})$   & $\lambda_T (\rm{GeV}^5) $        \\ \hline

$D{D}$                             &1                &$3.75\pm0.09$       &$(1.48 \pm0.23)\times 10^{-2}$     \\

$D{D}_s$                           &$\frac{1}{2}$    &$3.85\pm0.09$       &$(1.69\pm0.26)\times 10^{-2}$      \\

$D_s {D}_s$                        &0                &$3.95\pm0.09$       &$(2.00\pm0.32)\times 10^{-2}$   \\

$D^{*}D^{*}$                       & 1               & $4.04\pm0.11$      &$(4.99\pm0.66)\times 10^{-2}$     \\

$D_{s}^*D^{*}$                     &$\frac{1}{2}$    & $4.12\pm0.10$      & $(5.74\pm0.78)\times 10^{-2}$      \\

$D_{s}^*D_{s}^*$                   & 0               & $4.22\pm0.10$      & $(7.46\pm0.89)\times 10^{-2}$    \\

$D^{*}D -DD^{*}$                   & 0               & $3.88\pm0.11$      & $(1.92\pm0.29)\times 10^{-2}$    \\

$D^{*}D +DD^{*}$                   & 1               & $3.90\pm0.11$      & $(1.50\pm0.22)\times 10^{-2}$     \\

$D_s^*D -D_sD^*$                   &$\frac{1}{2}$    & $3.97\pm0.10$      & $(2.40\pm0.41)\times 10^{-2}$     \\

$D_s^*D +D_sD^*$                   &$\frac{1}{2}$    & $3.98\pm0.11$      & $(2.06\pm0.30)\times 10^{-2}$      \\

$D_s^*D_s$                         & 0               & $4.10\pm0.12$      & $(2.31\pm0.45)\times10^{-2}$      \\

$D_0^*D_1 -D_1D_0^*$               & 0               & $5.79\pm0.15$      & $(2.13\pm0.19)\times 10^{-1}$     \\

$D_0^*D_1 +D_1D_0^*$               & 1               & $5.37\pm0.13$      & $(1.30\pm0.11)\times 10^{-1}$    \\

$D_{0}^*D_{s1} -D_{s0}^*D_1$       &$\frac{1}{2}$    & $5.93\pm0.27$      & $(2.80\pm0.33)\times 10^{-1}$    \\

$D_{0}^*D_{s1} +D_{s0}^*D_1$       &$\frac{1}{2}$    & $5.54\pm0.20$      & $(1.51\pm0.16)\times 10^{-1}$     \\

$D_{s1}D_{s0}^*$                   & 0               & $5.67\pm0.27$      & $(1.77\pm0.27)\times 10^{-1}$     \\

$D^{*}D^{*}-D^{*}D^{*}$            & 0               & $4.00\pm0.11$      & $(2.47\pm0.32)\times 10^{-2}$    \\

$D^{*}D^{*}+D^{*}D^{*}$            & 1               & $4.02\pm0.11$      & $(2.83\pm0.30)\times 10^{-2}$     \\

$D_{s}^*D^{*}-D_{s}^{*}D^*$        &$\frac{1}{2}$    & $4.08\pm0.10$      & $(2.81\pm0.40)\times 10^{-2}$     \\

$D_{s}^*D^{*}+D_{s}^{*}D^*$        &$\frac{1}{2}$    & $4.10\pm0.11$      & $(3.19\pm0.44)\times 10^{-2}$      \\

$D_{s}^*D_{s}^*-D_{s}^*D_{s}^*$    & 0               & $4.19\pm0.09$      & $(3.49\pm0.49)\times 10^{-2}$      \\

$D_{s}^*D_{s}^*+D_{s}^*D_{s}^*$    & 0               & $4.20\pm0.10$      & $(4.00\pm0.53)\times 10^{-2}$    \\
\hline\hline
\end{tabular}
\end{center}
\caption{ The masses and pole residues of the ground state doubly-charmed  molecular states \cite{WZG-tetra-mole-AAPPS-2022,WZG-cc-mole-EPJA-2022}. }\label{mass-residue-Tcc-mole-Table}
\end{table}

Beyond those color $\mathbf{1}\mathbf{1}$ tetraquark states, there maybe also exist   some corresponding hexaquark states. The QCD sum rules indicate that there exist the $\Lambda_c\Lambda_c$, $\Sigma_c\Sigma_c$  and $\Xi_{cc}\Sigma_c$  dibaryon states \cite{WangZG-dibaryon-PRD-2020,WangXW-Sigm-Sigm-AHEP-2022,QiaoCF-LamLam-EPJC-2020},
 the $\bar{\mathbf 3}\bar{\mathbf 3}\bar{\mathbf 3}$ type triply-charmed hexaquark states \cite{WangZG-hexaquark-ccc-IJMPA-2020},
and the color $\mathbf{1}\mathbf{1}\mathbf{1}$ type triply-charmed hexaquark molecular states
\cite{WangZG-Regge-CTP-2021}. However, for the light dibaryon/baryonium  states, we can only obtain very small pole contributions at the hadron side or bad convergent behaviors  of the operator product expansion at the QCD side    \cite{d2380-HXChen-PRC-2015,H-dibaryon-QCDSR-1994,X1835-WangZG-JPG-2007}, which weakens  the predictive ability. If we adopt the truncation rule in Sects.{\bf\ref{Tetra-QCDSR}} and {\bf\ref{Tetra-Positive}}, i.e. each heavy quark line emits a gluon and each light quark contributes a quark-antiquark pair, which leads to a quark-gluon operator  to reach  the highest dimensional vacuum condensates, the two basic criteria of the QCD sum rules are difficult to satisfy. Such subjects need further studies.

On the other hand, Lattice calculations indicate that there maybe exist the $N\Omega$, $\Omega\Omega$, $\Omega(ccc)\Omega(ccc)$, $\Omega(bbb)\Omega(bbb)$, $\cdots$  dibaryon states
 \cite{N-Omega-Iritani-PLB-2019,Omega-Omega-Gongyo-PRL-2018,Deuteronlike-Junnarkar-PRL-2019,
 Omega-Omega-MengJ-PRL-2021,Omega-Omega-Mathur-PRL-2023,Omega-Omega-Mathur-PRD-2025}.
While the heavy-antiquark-diquark symmetry implies that there exists
 a model-independent relation between
the spin-splitting in the masses of the hidden-charm pentaquark states and  corresponding splitting
for the  triply-charmed dibaryon states \cite{dibaryon-ccc-GengLS-PRD-2020}.

\section{Hidden heavy pentaquark states}\label{333-11-5-quark}
If a baryon current $J(x)$ has the spin-parity $J^P={1 \over 2}^+$, then the
current $i\gamma_5J(x)$ would have the spin-parity $J^P={1 \over 2}^-$, as multiplying $i\gamma_5$
changes the parity of the current $J(x)$ \cite{Chung82}.
In 1993, Bagan et al took the infinite heavy quark limit to  separate the contributions of
the positive and negative parity heavy baryon states unambiguously \cite{Bagan93}.
In 1996, Jido et al introduced a novel approach  to separate the contributions of   the
negative-parity light-flavor  baryons from the positive-parity ones
 \cite{Oka96}.

At first, we write down the correlation functions $\Pi_{\pm}(p)$,
\begin{eqnarray}
\Pi_{\pm}(p)&=&i\int d^4x e^{ip \cdot x} \langle
0|T\left\{J_{\pm}(x)\bar{J}_{\pm}(0)\right\}|0\rangle \, ,
\end{eqnarray}
where we add the subscripts $\pm$ to denote the positive and negative parity, respectively,
 $J_{-} =i\gamma_{5} J_{+}$.
 We  decompose the correlation functions $\Pi_{\pm}(p)$,
\begin{equation}
    \Pi_{\pm}(p) = \!\not\!{p} \Pi_{1}(p^{2}) \pm \Pi_{0}(p^{2})\, ,
\end{equation}
according  to Lorentz covariance,   because
\begin{equation}
    \Pi_{-}(p) = -\gamma_{5} \Pi_{+}(p)\gamma_{5}   \, .
\end{equation}
The currents $J_{+}$ couple potentially  to both the positive-  and
negative-parity baryons \cite{Chung82},
\begin{eqnarray}\label{JP-JN}
    \langle{0}|J_{+}| B^{\pm}\rangle \langle B^{\pm}|\bar{J}_{+}|0\rangle =
    - \gamma_{5}\langle 0|J_{-}| B^{\pm}\rangle \langle B^{\pm}| \bar{J}_{-}|0\rangle \gamma_{5} \, ,
\end{eqnarray}
where the $B^{\pm}$ denote the positive and negative parity baryons, respectively.

According to the relation in Eq.\eqref{JP-JN}, we obtain the hadronic representation
 \cite{Oka96},
\begin{eqnarray}
    \Pi_{+}(p)     & = &   \lambda_+^2 {\!\not\!{p} +
    M_{+} \over M^{2}_+ -p^{2} } + \lambda_{-}^2
    {\!\not\!{p} - M_{-} \over M_{-}^{2}-p^{2}  } +\cdots \, ,
    \end{eqnarray}
where the $M_{\pm}$ are the baryon masses, the pole residues are defined by $\langle 0|J_\pm (0)|B^\pm(p)\rangle=\lambda_{\pm}U_{\pm}$, the $U_{\pm}$ are the Dirac spinors.

 If we take $\vec{p} = 0$, then
\begin{eqnarray}
  \rm{limit}_{\epsilon\rightarrow0}\frac{{\rm Im}  \Pi_+(p_{0}+i\epsilon)}{\pi} & = &
    \lambda_+^2 {\gamma_{0} + 1\over 2} \delta(p_{0} - M_+) +
    \lambda_{-}^{2} {\gamma_{0} - 1\over 2} \delta(p_{0} - M_{-})+\cdots \nonumber \\
  & = & \gamma_{0} A(p_{0}) + B(p_{0})+\cdots \, ,
\end{eqnarray}
where
\begin{eqnarray}
  A(p_{0}) & = & {1 \over 2} \left[ \lambda_+^{2}
  \delta(p_{0} - M_+)  + \lambda_-^{2} \delta(p_{0} -
  M_{-})\right] \, , \nonumber \\
   B(p_{0}) & = & {1 \over 2} \left[ \lambda_+^{2}
  \delta(p_{0} - M_+)  - \lambda_-^{2} \delta(p_{0} -
  M_{-})\right] \, ,
\end{eqnarray}
the contribution $A(p_{0}) + B(p_{0})$ ($A(p_{0}) - B(p_{0})$)
contains contributions  from the positive parity (negative parity)
states only.

We carry out the operator product expansion at large $P^2=-p_0^2$
region, then use the dispersion relation to obtain the spectral densities $\rho^A(p_0)$ and $\rho^B(p_0)$
at the  quark-gluon level. At last, we introduce the weight functions $\exp\left[-\frac{p_0^2}{T^2}\right]$,
$p_0^2\exp\left[-\frac{p_0^2}{T^2}\right]$,   and obtain the QCD
 sum rules,
\begin{eqnarray}
  \int_{\Delta}^{\sqrt{s_0}}dp_0 \left[
A(p_0)+B(p_0)\right]\exp\left[-\frac{p_0^2}{T^2}\right]&=&\int_{\Delta}^{\sqrt{s_0}}dp_0
\left[\rho^A(p_0)
+\rho^B(p_0)\right]\exp\left[-\frac{p_0^2}{T^2}\right] \,
,\nonumber \\
\end{eqnarray}
\begin{eqnarray}
  \int_{\Delta}^{\sqrt{s_0}}dp_0 \left[
A(p_0)+B(p_0)\right]
p_0^2\exp\left[-\frac{p_0^2}{T^2}\right]&=&\int_{\Delta}^{\sqrt{s_0}}dp_0
\left[\rho^A(p_0)
+\rho^B(p_0)\right]p_0^2\exp\left[-\frac{p_0^2}{T^2}\right] \,
,\nonumber \\
\end{eqnarray}
where the $\Delta$ and $\sqrt{s_0}$ are the threshold and continuum threshold  respectively,  the $T^2$ is the Borel parameter \cite{Oka96}.
Thereafter, such semi-analytical method was applied to study the heavy, doubly-heavy and triply-heavy baryons
with the $J^P={\frac{1}{2}}^\pm$ and ${\frac{3}{2}}^\pm$ \cite{WangHbaryon-PLB-2010,WangHbaryon-EPJA-2010,WangHbaryon-EPJC-2010,WangHbaryon-EPJC-2010-2,WangHbaryon-EPJA-2011,
WangHbaryon-CTP-2012}.

As the procedure introduced in Ref.\cite{Oka96} is semi-analytical, in 2016, we suggested an analytical procedure to study the pentaquark states \cite{WangZG-Penta-EPJC-2016-70}.
For a correlation function $\Pi(p^2)=\Pi_{-}(p)$, at the hadron side,  we obtain the  spectral densities through the dispersion relation,
\begin{eqnarray}
\frac{{\rm Im}\Pi(s)}{\pi}&=&\!\not\!{p} \left[\lambda^{2}_{-} \delta\left(s-M_{-}^2\right)+\lambda^{2}_{+} \delta\left(s-M_{+}^2\right)\right] +\left[M_{-}\lambda^{2}_{-} \delta\left(s-M_{-}^2\right)-M_{+}\lambda^{2}_{+} \delta\left(s-M_{+}^2\right)\right]\, , \nonumber\\
&=&\!\not\!{p}\, \rho^1_{H}(s)+\rho^0_{H}(s) \, ,
\end{eqnarray}
where the subscript $H$ denotes  the hadron side,
then we introduce the weight function $\exp\left(-\frac{s}{T^2}\right)$ to obtain the QCD sum rules at the hadron side,
\begin{eqnarray}
\int_{\Delta^2}^{s_0}ds \left[\sqrt{s}\rho^1_{H}(s)\pm\rho^0_{H}(s)\right]\exp\left( -\frac{s}{T^2}\right)
&=&2M_{\mp}\lambda^{2}_{\mp}\exp\left( -\frac{M_{\mp}^2}{T^2}\right) \, ,
\end{eqnarray}
where the $s_0$ are the continuum threshold parameters.
We separate the  contributions  of the negative parity pentaquark states from that of the positive parity ones unambiguously \cite{WangZG-Penta-EPJC-2016-70}. The calculations at the QCD side are analytical as we do not set $\vec{p}=0$.

For the early works on the pentaquark states, one can consult the QCD sum rules on the $\Theta(1540)$, where the currents $J_Z(x)$ \cite{Cita1540-SLZhu-PRL-2003}, $J_O(x)$  \cite{Cita1540-MOka-PLB-2004} and $J_N(x)$ \cite{Cita1540-Nielsen-PLB-2004},
\begin{eqnarray}
J_{Z}(x)&=&{1\over \sqrt{2}} \varepsilon^{ijk} u^T_i(x) C\gamma_5
d_j (x) \left\{ u_m (x){\bar s}_m (x) i\gamma_5 d_k(x) - \left( u\leftrightarrow
d\right) \right\} \, , \nonumber\\
J_{O}(x)&=&\varepsilon^{ijk}\varepsilon^{lmn}\varepsilon^{knb}
   u_i^T(x)Cd_j(x) u_l^T(x)C\gamma_5 d_m(x)C\bar{s}_b^T(x) \, , \nonumber\\
J_N(x)&=&{\cos\theta\over\sqrt{2}}\varepsilon^{ijk}u_i^T(x) C \gamma_5 d_j(x)
u_k^T(x) C
\gamma_5 d_m(x)C\bar{s}^T_m(x)- (u\leftrightarrow d) \nonumber\\
 &&+ {\sin\theta\over\sqrt{2}}\varepsilon^{ijk}u_i^T(x) C  d_j(x)
u_k^T(x) C  d_m(x)C\bar{s}^T_m(x) - (u\leftrightarrow d)\, ,
\end{eqnarray}
were constructed to interpolate it.
The current $J_O(x)$ was studied in the semi-analytical method \cite{Cita1540-MOka-PLB-2004}.

\subsection{$\bar{\mathbf 3}\bar{\mathbf 3} \bar{\mathbf 3}$ type  pentaquark states}\label{333-penta-Sect}
Now, let us turn to the pentaquark states completely  and  write down  the correlation functions $\Pi(p)$, $\Pi_{\mu\nu}(p)$ and $\Pi_{\mu\nu\alpha\beta}(p)$,
\begin{eqnarray}\label{CF-Pi-Pi-Pi}
\Pi(p)&=&i\int d^4x e^{ip \cdot x} \langle0|T\left\{J(x)\bar{J}(0)\right\}|0\rangle \, , \nonumber \\
\Pi_{\mu\nu}(p)&=&i\int d^4x e^{ip \cdot x} \langle0|T\left\{J_{\mu}(x)\bar{J}_{\nu}(0)\right\}|0\rangle \, , \nonumber \\
\Pi_{\mu\nu\alpha\beta}(p)&=&i\int d^4x e^{ip \cdot x} \langle0|T\left\{J_{\mu\nu}(x)\bar{J}_{\alpha\beta}(0)\right\}|0\rangle \, ,
\end{eqnarray}
where
\begin{eqnarray}\label{current-3-3-3}
 J(x)&=&J^1(x)\, , \, J^2(x)\, , \, J^3(x)\, , \, J^4(x)\, , \nonumber\\
 J_\mu(x)&=&J^1_\mu(x)\, , \, J^2_\mu(x)\, , \, J^3_\mu(x)\, , \, J^4_\mu(x)\, , \nonumber\\
 J_{\mu\nu}(x)&=&J^1_{\mu\nu}(x)\, , \, J^2_{\mu\nu}(x)\, ,
 \end{eqnarray}
\begin{eqnarray}
 J^1(x)&=&\varepsilon^{ila} \varepsilon^{ijk}\varepsilon^{lmn}  u^T_j(x) C\gamma_5 d_k(x)\,u^T_m(x) C\gamma_5 c_n(x)\,  C\bar{c}^{T}_{a}(x) \, , \nonumber\\
J^2(x)&=&\varepsilon^{ila} \varepsilon^{ijk}\varepsilon^{lmn}  u^T_j(x) C\gamma_5 d_k(x)\,u^T_m(x) C\gamma_\mu c_n(x)\,\gamma_5 \gamma^\mu C\bar{c}^{T}_{a}(x) \, , \nonumber\\
J^{3}(x)&=&\frac{\varepsilon^{ila} \varepsilon^{ijk}\varepsilon^{lmn}}{\sqrt{3}} \left[ u^T_j(x) C\gamma_\mu u_k(x) d^T_m(x) C\gamma_5 c_n(x)+2u^T_j(x) C\gamma_\mu d_k(x) u^T_m(x) C\gamma_5 c_n(x)\right] \gamma_5 \gamma^\mu  C\bar{c}^{T}_{a}(x) \, ,  \nonumber\\
J^{4}(x)&=&\frac{\varepsilon^{ila} \varepsilon^{ijk}\varepsilon^{lmn}}{\sqrt{3}} \left[ u^T_j(x) C\gamma_\mu u_k(x)d^T_m(x) C\gamma^\mu c_n(x)+2u^T_j(x) C\gamma_\mu d_k(x)u^T_m(x) C\gamma^\mu c_n(x) \right] C\bar{c}^{T}_{a}(x) \, , \nonumber
\end{eqnarray}
\begin{eqnarray}
J^1_{\mu}(x)&=&\varepsilon^{ila} \varepsilon^{ijk}\varepsilon^{lmn}  u^T_j(x) C\gamma_5 d_k(x)\,u^T_m(x) C\gamma_\mu c_n(x)\, C\bar{c}^{T}_{a}(x) \, , \nonumber \\
J^{2}_{\mu}(x)&=&\frac{\varepsilon^{ila} \varepsilon^{ijk}\varepsilon^{lmn}}{\sqrt{3}} \left[ u^T_j(x) C\gamma_\mu u_k(x) d^T_m(x) C\gamma_5 c_n(x)+2u^T_j(x) C\gamma_\mu d_k(x) u^T_m(x) C\gamma_5 c_n(x)\right]    C\bar{c}^{T}_{a}(x) \, ,  \nonumber\\
J^{3}_{\mu}(x)&=&\frac{\varepsilon^{ila} \varepsilon^{ijk}\varepsilon^{lmn}}{\sqrt{3}} \left[ u^T_j(x) C\gamma_\mu u_k(x)d^T_m(x) C\gamma_\alpha c_n(x)+2u^T_j(x) C\gamma_\mu d_k(x)u^T_m(x) C\gamma_\alpha c_n(x) \right] \gamma_5\gamma^\alpha C\bar{c}^{T}_{a}(x) \, , \nonumber\\
 J^{4}_{\mu}(x)&=&\frac{\varepsilon^{ila} \varepsilon^{ijk}\varepsilon^{lmn}}{\sqrt{3}} \left[ u^T_j(x) C\gamma_\alpha u_k(x)d^T_m(x) C\gamma_\mu c_n(x)+2u^T_j(x) C\gamma_\alpha d_k(x)u^T_m(x) C\gamma_\mu c_n(x) \right] \gamma_5\gamma^\alpha C\bar{c}^{T}_{a}(x) \, , \nonumber
 \end{eqnarray}
 \begin{eqnarray}
J^1_{\mu\nu}(x)&=&\frac{\varepsilon^{ila} \varepsilon^{ijk}\varepsilon^{lmn}}{\sqrt{6}} \left[ u^T_j(x) C\gamma_\mu u_k(x)d^T_m(x) C\gamma_\nu c_n(x)+2u^T_j(x) C\gamma_\mu d_k(x)u^T_m(x) C\gamma_\nu c_n(x) \right]    \nonumber\\
 &&C\bar{c}^{T}_{a}(x)+\left( \mu\leftrightarrow\nu\right)\, ,  \nonumber\\
J^2_{\mu\nu}(x)&=&\frac{1}{\sqrt{2}}\varepsilon^{ila} \varepsilon^{ijk}\varepsilon^{lmn}  u^T_j(x) C\gamma_5 d_k(x)\left[u^T_m(x) C\gamma_\mu c_n(x)\, \gamma_5\gamma_{\nu}C\bar{c}^{T}_{a}(x)\right.\nonumber\\
&&\left.+u^T_m(x) C\gamma_\nu c_n(x)\,\gamma_5 \gamma_{\mu}C\bar{c}^{T}_{a}(x)\right] \, ,
\end{eqnarray}
we choose the $C\gamma_5$ and $C\gamma_\mu$ diquarks in the color $\bar{\mathbf 3}$, the most stable diquark configurations,  as the basic constituents to construct the  diquark-diquark-antiquark type currents $J(x)$, $J_\mu(x)$ and $J_{\mu\nu}(x)$ with the spin-parity $J^{P}={\frac{1}{2}}^-$, ${\frac{3}{2}}^-$ and ${\frac{5}{2}}^-$, respectively,  which are expected to couple potentially to the lowest pentaquark states \cite{WZG-HC-penta-IJMPA-2020,WangZG-Penta-EPJC-2016-70,WangZG-Pcs4459-penta-IJMPA-2021}.

In the currents $J(x)$, $J_\mu(x)$ and $J_{\mu\nu}(x)$, there are diquark constituents $\varepsilon^{ijk}u^T_jC\gamma_5d_k$, $\varepsilon^{ijk}u^T_jC\gamma_{\mu}d_k$, $\varepsilon^{ijk}u^T_jC\gamma_{\mu}u_k$, $\varepsilon^{ijk}q^T_jC\gamma_5c_k$, $\varepsilon^{ijk}q^T_jC\gamma_{\mu}c_k$ with $q=u$, $d$. If we use the $S_L$ and $S_H$
to represent  the spins of the light  and heavy diquarks, respectively, the light diquarks  $\varepsilon^{ijk}u^T_jC\gamma_5d_k$, $\varepsilon^{ijk}u^T_jC\gamma_{\mu}d_k$ and $\varepsilon^{ijk}u^T_jC\gamma_{\mu}u_k$ have the spins $S_L=0$, $1$ and $1$, respectively,  the heavy
diquarks $\varepsilon^{ijk}q^T_jC\gamma_5c_k$ and $\varepsilon^{ijk}q^T_jC\gamma_{\mu}c_k$ have the spins $S_H=0$ and $1$, respectively. A light diquark and a heavy diquark form a charmed tetraquark in the color $\mathbf 3$ with the  angular momentum $\vec{J}_{LH}=\vec{S}_L+\vec{S}_H$, which has the values $J_{LH}=0$, $1$ or $2$.
The $\bar{c}$-quark operator $C\bar{c}_a^T$ has the spin-parity $J^P={\frac{1}{2}}^-$,
 the $\bar{c}$-quark operator $\gamma_5\gamma_{\mu}C\bar{c}_a^T$ has the spin-parity $J^P={\frac{3}{2}}^-$ due to  the  factor $\gamma_5\gamma_{\mu}$. The total angular momentums  are $\vec{J}=\vec{J}_{LH}+\vec{J}_{\bar{c}}$ with the values $J=\frac{1}{2}$, $\frac{3}{2}$ or $\frac{5}{2}$, which are shown explicitly in Table \ref{current-pentaQ}. In Table \ref{current-pentaQ}, we present the quark structures of the  currents   explicitly \cite{WZG-HC-penta-IJMPA-2020}. For example, in the current $J^2_{\mu\nu}(x)$, there are a scalar diquark
$\varepsilon^{ijk}  u^T_j(x) C\gamma_5 d_k(x)$ with the spin-parity $J^P=0^+$, an axialvector diquark  $\varepsilon^{lmn} u^T_m(x) C\gamma_\mu c_n(x)$ with the spin-parity $J^P=1^+$, and an antiquark $\gamma_5\gamma_{\nu}C\bar{c}^T_a(x)$ with the spin-parity $J^{P}={\frac{3}{2}}^-$,  the total angular momentum is $J={\frac{5}{2}}$.  For more intuitive and simple diquark models for the pentaquark states, one can consult Refs.\cite{Penta-Maiani-PLB-2015,Penta-Ali-PLB-2019,
di-tri-penta-Lebed-PLB-2015,di-tri-penta-QiaoCF-PLB-2016,
Penta-Ali-PRD-2016,Penta-dibaryon-Maiani-PLB-2015}.

\begin{table}
\begin{center}
\begin{tabular}{|c|c|c|c|c|c|c|c|c|}\hline\hline
$[qq^\prime][q^{\prime\prime}c]\bar{c}$ ($S_L$, $S_H$, $J_{LH}$, $J$)& $J^{P}$              & Currents              \\ \hline

$[ud][uc]\bar{c}$ ($0$, $0$, $0$, $\frac{1}{2}$)                     & ${\frac{1}{2}}^{-}$  & $J^1(x)$              \\

$[ud][uc]\bar{c}$ ($0$, $1$, $1$, $\frac{1}{2}$)                     & ${\frac{1}{2}}^{-}$  & $J^2(x)$              \\

$[uu][dc]\bar{c}+2[ud][uc]\bar{c}$ ($1$, $0$, $1$, $\frac{1}{2}$)    & ${\frac{1}{2}}^{-}$  & $J^3(x)$              \\

$[uu][dc]\bar{c}+2[ud][uc]\bar{c}$ ($1$, $1$, $0$, $\frac{1}{2}$)    & ${\frac{1}{2}}^{-}$  & $J^4(x)$              \\

$[ud][uc]\bar{c}$ ($0$, $1$, $1$, $\frac{3}{2}$)                     & ${\frac{3}{2}}^{-}$  & $J^1_\mu(x)$           \\

$[uu][dc]\bar{c}+2[ud][uc]\bar{c}$ ($1$, $0$, $1$, $\frac{3}{2}$)    & ${\frac{3}{2}}^{-}$  & $J^2_\mu(x)$          \\

$[uu][dc]\bar{c}+2[ud][uc]\bar{c}$ ($1$, $1$, $2$, $\frac{3}{2}$)    & ${\frac{3}{2}}^{-}$  & $J^3_\mu(x)$           \\

$[uu][dc]\bar{c}+2[ud][uc]\bar{c}$ ($1$, $1$, $2$, $\frac{3}{2}$)    & ${\frac{3}{2}}^{-}$  & $J^4_\mu(x)$           \\

$[uu][dc]\bar{c}+2[ud][uc]\bar{c}$ ($1$, $1$, $2$, $\frac{5}{2}$)    & ${\frac{5}{2}}^{-}$  & $J^1_{\mu\nu}(x)$           \\

$[ud][uc]\bar{c}$ ($0$, $1$, $1$, $\frac{5}{2}$)                     & ${\frac{5}{2}}^{-}$  & $J^2_{\mu\nu}(x)$     \\ \hline\hline
\end{tabular}
\end{center}
\caption{ The quark structures of the  currents, where the $S_L$ and $S_H$ denote the spins of the light  and heavy diquarks respectively, $\vec{J}_{LH}=\vec{S}_L+\vec{S}_H$, $\vec{J}=\vec{J}_{LH}+\vec{J}_{\bar{c}}$, the $\vec{J}_{\bar{c}}$ is the angular momentum of the $\bar{c}$-quark \cite{WZG-HC-penta-IJMPA-2020}.  }\label{current-pentaQ}
\end{table}

Although the currents $J(x)$, $J_\mu(x)$ and $J_{\mu\nu}(x)$ have negative parity, but they couple potentially to  the   pentaquark states with positive parity, as  multiplying $i \gamma_{5}$ to the currents  $J(x)$, $J_\mu(x)$ and $J_{\mu\nu}(x)$ changes their parity \cite{Chung82,Bagan93,Oka96,WangHbaryon-PLB-2010,WangHbaryon-EPJA-2010,WangHbaryon-EPJC-2010,WangHbaryon-EPJC-2010-2,WangHbaryon-EPJA-2011,
WangHbaryon-CTP-2012,WangZG-Baryon-D-wave-NPB-2018,WangZG-Baryon-P-wave-CPC-2023,
YuGL-Baryon-2022-CPC}.

Now we write down the current-pentaquark couplings explicitly,
\begin{eqnarray}\label{Coupling12}
\langle 0| J (0)|P_{\frac{1}{2}}^{-}(p)\rangle &=&\lambda^{-}_{\frac{1}{2}} U^{-}(p,s) \, , \nonumber \\
\langle 0| J (0)|P_{\frac{1}{2}}^{+}(p)\rangle &=&\lambda^{+}_{\frac{1}{2}} i\gamma_5 U^{+}(p,s) \, ,
\end{eqnarray}
\begin{eqnarray}
\langle 0| J_{\mu} (0)|P_{\frac{3}{2}}^{-}(p)\rangle &=&\lambda^{-}_{\frac{3}{2}} U^{-}_\mu(p,s) \, ,  \nonumber \\
\langle 0| J_{\mu} (0)|P_{\frac{3}{2}}^{+}(p)\rangle &=&\lambda^{+}_{\frac{3}{2}}i\gamma_5 U^{+}_\mu(p,s) \, ,  \nonumber \\
\langle 0| J_{\mu} (0)|P_{\frac{1}{2}}^{+}(p)\rangle &=&f^{+}_{\frac{1}{2}}p_\mu U^{+}(p,s) \, , \nonumber \\
\langle 0| J_{\mu} (0)|P_{\frac{1}{2}}^{-}(p)\rangle &=&f^{-}_{\frac{1}{2}}p_\mu i\gamma_5 U^{-}(p,s) \, ,
\end{eqnarray}
\begin{eqnarray}\label{Coupling52}
\langle 0| J_{\mu\nu} (0)|P_{\frac{5}{2}}^{-}(p)\rangle &=&\sqrt{2}\lambda^{-}_{\frac{5}{2}} U^{-}_{\mu\nu}(p,s) \, ,\nonumber\\
\langle 0| J_{\mu\nu} (0)|P_{\frac{5}{2}}^{+}(p)\rangle &=&\sqrt{2}\lambda^{+}_{\frac{5}{2}}i\gamma_5 U^{+}_{\mu\nu}(p,s) \, ,\nonumber\\
\langle 0| J_{\mu\nu} (0)|P_{\frac{3}{2}}^{+}(p)\rangle &=&f^{+}_{\frac{3}{2}} \left[p_\mu U^{+}_{\nu}(p,s)+p_\nu U^{+}_{\mu}(p,s)\right] \, , \nonumber\\
\langle 0| J_{\mu\nu} (0)|P_{\frac{3}{2}}^{-}(p)\rangle &=&f^{-}_{\frac{3}{2}}i\gamma_5 \left[p_\mu U^{-}_{\nu}(p,s)+p_\nu U^{-}_{\mu}(p,s)\right] \, , \nonumber\\
\langle 0| J_{\mu\nu} (0)|P_{\frac{1}{2}}^{-}(p)\rangle &=&g^{-}_{\frac{1}{2}}p_\mu p_\nu U^{-}(p,s) \, , \nonumber\\
\langle 0| J_{\mu\nu} (0)|P_{\frac{1}{2}}^{+}(p)\rangle &=&g^{+}_{\frac{1}{2}}p_\mu p_\nu i\gamma_5 U^{+}(p,s) \, ,
\end{eqnarray}
where the superscripts $\pm$ denote the positive parity and negative parity, respectively, the subscripts $\frac{1}{2}$, $\frac{3}{2}$ and $\frac{5}{2}$ denote the spins of the pentaquark states,     the $\lambda$, $f$ and $g$ are the pole residues.

The spinors $U^\pm(p,s)$ satisfy the Dirac equations  $(\not\!\!p-M_{\pm})U^{\pm}(p)=0$, while the spinors $U^{\pm}_\mu(p,s)$ and $U^{\pm}_{\mu\nu}(p,s)$ satisfy the Rarita-Schwinger equations $(\not\!\!p-M_{\pm})U^{\pm}_\mu(p)=0$ and $(\not\!\!p-M_{\pm})U^{\pm}_{\mu\nu}(p)=0$,  and relations $\gamma^\mu U^{\pm}_\mu(p,s)=0$,
$p^\mu U^{\pm}_\mu(p,s)=0$, $\gamma^\mu U^{\pm}_{\mu\nu}(p,s)=0$,
$p^\mu U^{\pm}_{\mu\nu}(p,s)=0$, $ U^{\pm}_{\mu\nu}(p,s)= U^{\pm}_{\nu\mu}(p,s)$, respectively \cite{WangZG-Penta-EPJC-2016-70,WangZG-Xicc-penta-EPJC-2018}.

 At the hadron  side, we insert a complete set of intermediate pentaquark states with the same quantum numbers as the currents  $J(x)$, $i\gamma_5 J(x)$, $J_\mu(x)$, $i\gamma_5 J_\mu(x)$, $J_{\mu\nu}(x)$ and $i\gamma_5 J_{\mu\nu}(x)$ into the correlation functions
$\Pi(p)$, $\Pi_{\mu\nu}(p)$ and $\Pi_{\mu\nu\alpha\beta}(p)$ to obtain the hadronic representation, and  isolate the lowest
states of the hidden-charm  pentaquark states with negative and positive parity \cite{WZG-HC-penta-IJMPA-2020,WangZG-Penta-EPJC-2016-70,
WangZG-Pcs4459-penta-IJMPA-2021}:
\begin{eqnarray}\label{CF-Hadron-12}
\Pi(p) & = & {\lambda^{-}_{\frac{1}{2}}}^2  {\!\not\!{p}+ M_{-} \over M_{-}^{2}-p^{2}  }+  {\lambda^{+}_{\frac{1}{2}}}^2  {\!\not\!{p}- M_{+} \over M_{+}^{2}-p^{2}  } +\cdots  \, ,\nonumber\\
&=&\Pi_{\frac{1}{2}}^1(p^2)\!\not\!{p}+\Pi_{\frac{1}{2}}^0(p^2)\, ,
 \end{eqnarray}
\begin{eqnarray}\label{CF-Hadron-32}
 \Pi_{\mu\nu}(p) & = & {\lambda^{-}_{\frac{3}{2}}}^2  {\!\not\!{p}+ M_{-} \over M_{-}^{2}-p^{2}  } \left(- g_{\mu\nu}+\frac{\gamma_\mu\gamma_\nu}{3}+\frac{2p_\mu p_\nu}{3p^2}-\frac{p_\mu\gamma_\nu-p_\nu \gamma_\mu}{3\sqrt{p^2}}
\right)\nonumber\\
&&+  {\lambda^{+}_{\frac{3}{2}}}^2  {\!\not\!{p}- M_{+} \over M_{+}^{2}-p^{2}  } \left(- g_{\mu\nu}+\frac{\gamma_\mu\gamma_\nu}{3}+\frac{2p_\mu p_\nu}{3p^2}-\frac{p_\mu\gamma_\nu-p_\nu \gamma_\mu}{3\sqrt{p^2}}
\right)   \nonumber \\
& &+ {f^{+}_{\frac{1}{2}}}^2  {\!\not\!{p}+ M_{+} \over M_{+}^{2}-p^{2}  } p_\mu p_\nu+  {f^{-}_{\frac{1}{2}}}^2  {\!\not\!{p}- M_{-} \over M_{-}^{2}-p^{2}  } p_\mu p_\nu  +\cdots  \, , \nonumber\\
&=&\left[\Pi_{\frac{3}{2}}^1(p^2)\!\not\!{p}+\Pi_{\frac{3}{2}}^0(p^2)\right]\left(- g_{\mu\nu}\right)+\cdots\, ,
\end{eqnarray}
\begin{eqnarray}\label{CF-Hadron-52}
\Pi_{\mu\nu\alpha\beta}(p) & = &2{\lambda^{-}_{\frac{5}{2}}}^2  {\!\not\!{p}+ M_{-} \over M_{-}^{2}-p^{2}  } \left[\frac{ \widetilde{g}_{\mu\alpha}\widetilde{g}_{\nu\beta}+\widetilde{g}_{\mu\beta}\widetilde{g}_{\nu\alpha}}{2}-\frac{\widetilde{g}_{\mu\nu}\widetilde{g}_{\alpha\beta}}{5}-\frac{1}{10}\left( \gamma_{\mu}\gamma_{\alpha}+\frac{\gamma_{\mu}p_{\alpha}-\gamma_{\alpha}p_{\mu}}{\sqrt{p^2}}-\frac{p_{\mu}p_{\alpha}}{p^2}\right)\widetilde{g}_{\nu\beta}\right.\nonumber\\
&&\left.-\frac{1}{10}\left( \gamma_{\nu}\gamma_{\alpha}+\frac{\gamma_{\nu}p_{\alpha}-\gamma_{\alpha}p_{\nu}}{\sqrt{p^2}}-\frac{p_{\nu}p_{\alpha}}{p^2}\right)\widetilde{g}_{\mu\beta}
+\cdots\right]\nonumber\\
&&+  2 {\lambda^{+}_{\frac{5}{2}}}^2  {\!\not\!{p}- M_{+} \over M_{+}^{2}-p^{2}  } \left[\frac{ \widetilde{g}_{\mu\alpha}\widetilde{g}_{\nu\beta}+\widetilde{g}_{\mu\beta}\widetilde{g}_{\nu\alpha}}{2}
-\frac{\widetilde{g}_{\mu\nu}\widetilde{g}_{\alpha\beta}}{5}-\frac{1}{10}\left( \gamma_{\mu}\gamma_{\alpha}+\frac{\gamma_{\mu}p_{\alpha}-\gamma_{\alpha}p_{\mu}}{\sqrt{p^2}}-\frac{p_{\mu}p_{\alpha}}{p^2}\right)\widetilde{g}_{\nu\beta}\right.\nonumber\\
&&\left.
-\frac{1}{10}\left( \gamma_{\nu}\gamma_{\alpha}+\frac{\gamma_{\nu}p_{\alpha}-\gamma_{\alpha}p_{\nu}}{\sqrt{p^2}}-\frac{p_{\nu}p_{\alpha}}{p^2}\right)\widetilde{g}_{\mu\beta}
 +\cdots\right]   \nonumber\\
 && +{f^{+}_{\frac{3}{2}}}^2  {\!\not\!{p}+ M_{+} \over M_{+}^{2}-p^{2}  } \left[ p_\mu p_\alpha \left(- g_{\nu\beta}+\frac{\gamma_\nu\gamma_\beta}{3}+\frac{2p_\nu p_\beta}{3p^2}-\frac{p_\nu\gamma_\beta-p_\beta \gamma_\nu}{3\sqrt{p^2}}
\right)+\cdots \right]\nonumber\\
&&+  {f^{-}_{\frac{3}{2}}}^2  {\!\not\!{p}- M_{-} \over M_{-}^{2}-p^{2}  } \left[ p_\mu p_\alpha \left(- g_{\nu\beta}+\frac{\gamma_\nu\gamma_\beta}{3}+\frac{2p_\nu p_\beta}{3p^2}-\frac{p_\nu\gamma_\beta-p_\beta \gamma_\nu}{3\sqrt{p^2}}
\right)+\cdots \right]   \nonumber \\
& &+ {g^{-}_{\frac{1}{2}}}^2  {\!\not\!{p}+ M_{-} \over M_{-}^{2}-p^{2}  } p_\mu p_\nu p_\alpha p_\beta+  {g^{+}_{\frac{1}{2}}}^2  {\!\not\!{p}- M_{+} \over M_{+}^{2}-p^{2}  } p_\mu p_\nu p_\alpha p_\beta  +\cdots \, , \nonumber\\
& = & \left[\Pi_{\frac{5}{2}}^1(p^2)\!\not\!{p}+\Pi_{\frac{5}{2}}^0(p^2)\right]\left( g_{\mu\alpha}g_{\nu\beta}+g_{\mu\beta}g_{\nu\alpha}\right)  +\cdots \, ,
 \end{eqnarray}
where we have used the summations of the spinors \cite{HuangShiZhong},
\begin{eqnarray}\label{Sum-UU}
\sum_s U \overline{U}&=&\left(\!\not\!{p}+M_{\pm}\right) \,  ,  \nonumber\\
\sum_s U_\mu \overline{U}_\nu&=&\left(\!\not\!{p}+M_{\pm}\right)\left( -g_{\mu\nu}+\frac{\gamma_\mu\gamma_\nu}{3}+\frac{2p_\mu p_\nu}{3p^2}-\frac{p_\mu
\gamma_\nu-p_\nu \gamma_\mu}{3\sqrt{p^2}} \right) \,  ,  \nonumber\\
\sum_s U_{\mu\nu}\overline {U}_{\alpha\beta}&=&\left(\!\not\!{p}+M_{\pm}\right)\left\{\frac{\widetilde{g}_{\mu\alpha}\widetilde{g}_{\nu\beta}+\widetilde{g}_{\mu\beta}\widetilde{g}_{\nu\alpha}}{2} -\frac{\widetilde{g}_{\mu\nu}\widetilde{g}_{\alpha\beta}}{5}-\frac{1}{10}\left( \gamma_{\mu}\gamma_{\alpha}+\frac{\gamma_{\mu}p_{\alpha}-\gamma_{\alpha}p_{\mu}}{\sqrt{p^2}}-\frac{p_{\mu}p_{\alpha}}{p^2}\right)\widetilde{g}_{\nu\beta}\right. \nonumber\\
&&-\frac{1}{10}\left( \gamma_{\nu}\gamma_{\alpha}+\frac{\gamma_{\nu}p_{\alpha}-\gamma_{\alpha}p_{\nu}}{\sqrt{p^2}}-\frac{p_{\nu}p_{\alpha}}{p^2}\right)\widetilde{g}_{\mu\beta}
-\frac{1}{10}\left( \gamma_{\mu}\gamma_{\beta}+\frac{\gamma_{\mu}p_{\beta}-\gamma_{\beta}p_{\mu}}{\sqrt{p^2}}-\frac{p_{\mu}p_{\beta}}{p^2}\right)\widetilde{g}_{\nu\alpha}\nonumber\\
&&\left.-\frac{1}{10}\left( \gamma_{\nu}\gamma_{\beta}+\frac{\gamma_{\nu}p_{\beta}-\gamma_{\beta}p_{\nu}}{\sqrt{p^2}}-\frac{p_{\nu}p_{\beta}}{p^2}\right)\widetilde{g}_{\mu\alpha} \right\} \, ,
\end{eqnarray}
and $p^2=M^2_{\pm}$ on mass-shell.
We study the components $\Pi_{\frac{1}{2}}^1(p^2)$, $\Pi_{\frac{1}{2}}^0(p^2)$, $\Pi_{\frac{3}{2}}^1(p^2)$, $\Pi_{\frac{3}{2}}^0(p^2)$, $\Pi_{\frac{5}{2}}^1(p^2)$ and  $\Pi_{\frac{5}{2}}^0(p^2)$ to avoid possible contaminations from other pentaquark states with different spins.

We obtain the spectral densities at the hadron  side through  dispersion relation,
\begin{eqnarray}
\frac{{\rm Im}\Pi_{j}^1(s)}{\pi}&=& {\lambda^{-}_{j}}^2 \delta\left(s-M_{-}^2\right)+{\lambda^{+}_{j}}^2 \delta\left(s-M_{+}^2\right) =\, \rho^1_{j,H}(s) \, , \\
\frac{{\rm Im}\Pi^0_{j}(s)}{\pi}&=&M_{-}{\lambda^{-}_{j}}^2 \delta\left(s-M_{-}^2\right)-M_{+}{\lambda^{+}_{j}}^2 \delta\left(s-M_{+}^2\right)
=\rho^0_{j,H}(s) \, ,
\end{eqnarray}
where $j=\frac{1}{2}$, $\frac{3}{2}$, $\frac{5}{2}$, the subscript $H$ denotes  the hadron side,
then we introduce the  weight functions $\sqrt{s}\exp\left(-\frac{s}{T^2}\right)$ and $\exp\left(-\frac{s}{T^2}\right)$ to obtain the QCD sum rules
at the hadron side,
\begin{eqnarray}
\int_{4m_c^2}^{s_0}ds \left[\sqrt{s}\,\rho^1_{j,H}(s)\pm\rho^0_{j,H}(s)\right]\exp\left( -\frac{s}{T^2}\right)
&=&2M_{\mp}{\lambda^{\mp}_{j}}^2\exp\left( -\frac{M_{\mp}^2}{T^2}\right) \, ,
\end{eqnarray}
where the $s_0$ are the continuum threshold parameters, and the $T^2$ are the Borel parameters.
We distinguish  the  contributions  of the negative and positive parity pentaquark  states unambiguously, and there is no contamination.

\begin{figure}
 \centering
 \includegraphics[totalheight=3.0cm,width=15cm]{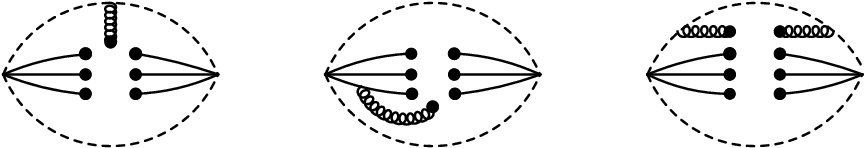}
  \vglue+3mm
 \includegraphics[totalheight=3.0cm,width=15cm]{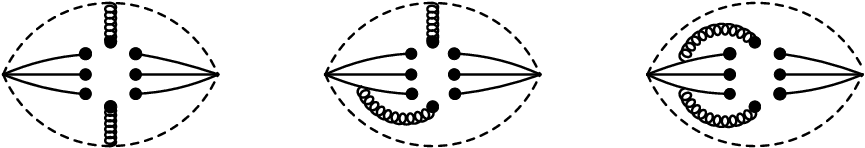}
    \caption{The diagrams contribute  to the  condensates $\langle\bar{q} q\rangle^2\langle\bar{q}g_s\sigma Gq\rangle $, $\langle\bar{q} q\rangle \langle\bar{q}g_s\sigma Gq\rangle^2 $, $\langle \bar{q}q\rangle^3\langle \frac{\alpha_s}{\pi}GG\rangle$. Other
diagrams obtained by interchanging of the $c$ quark lines (dashed lines) or light quark lines (solid lines) are implied \cite{WZG-HC-penta-IJMPA-2020}. }\label{Feynman-penta-13}
\end{figure}

Now we briefly outline  the operator product expansion. Firstly,  we contract the $u$, $d$ and $c$ quark fields in the correlation functions $\Pi(p)$, $\Pi_{\mu\nu}(p)$ and $\Pi_{\mu\nu\alpha\beta}(p)$
 with Wick theorem, for example,
\begin{eqnarray}\label{CF-Pi-penta-12}
\Pi(p)&=&i\,\varepsilon^{ila}\varepsilon^{ijk}\varepsilon^{lmn}\varepsilon^{i^{\prime}l^{\prime}a^{\prime}}\varepsilon^{i^{\prime}j^{\prime}k^{\prime}}
\varepsilon^{l^{\prime}m^{\prime}n^{\prime}}\int d^4x e^{ip\cdot x} \nonumber\\
&&\Big\{  - Tr\Big[\gamma_5 D_{kk^\prime}(x) \gamma_5 C U^{T}_{jj^\prime}(x)C\Big] \,Tr\Big[\gamma_5 C_{nn^\prime}(x) \gamma_5 C U^{T}_{mm^\prime}(x)C\Big] C C_{a^{\prime}a}^T(-x)C   \nonumber\\
&&+  Tr \Big[\gamma_5 D_{kk^\prime}(x) \gamma_5 C U^{T}_{mj^\prime}(x)C \gamma_5 C_{nn^\prime}(x) \gamma_5 C U^{T}_{jm^\prime}(x)C\Big]   C C_{a^{\prime}a}^T(-x)C   \Big\} \, ,
\end{eqnarray}
for the current $J(x)=J^1(x)$, where
the $U_{ij}(x)$, $D_{ij}(x)$ and $C_{ij}(x)$ are the full $u$, $d$ and $c$ quark propagators, respectively, see
Eqs.\eqref{S-progator}-\eqref{Q-progator}.
  Then we compute all the integrals  in the coordinate and momentum spaces  sequentially  to obtain the $\Pi(p)$, $\Pi_{\mu\nu}(p)$ and $\Pi_{\mu\nu\alpha\beta}(p)$   at the quark-gluon  level, and finally we obtain the QCD spectral densities through   dispersion relation,
\begin{eqnarray}\label{QCD-rho1-2}
 \rho^1_{j,QCD}(s) &=&\frac{{\rm Im}\Pi_{j}^1(s)}{\pi}\, , \nonumber\\
\rho^0_{j,QCD}(s) &=&\frac{{\rm Im}\Pi_{j}^0(s)}{\pi}\, ,
\end{eqnarray}
where $j=\frac{1}{2}$, $\frac{3}{2}$, $\frac{5}{2}$.
  In computing the integrals, we draw up all the Feynman diagrams from Eq.\eqref{CF-Pi-penta-12} and calculate them one by one.  In Eq.\eqref{CF-Pi-penta-12},  there are two $c$-quark propagators and three light quark propagators, if each $c$-quark line emits a gluon and each light quark line contributes  a quark-antiquark  pair, we obtain an operator $G_{\mu\nu}G_{\alpha\beta}\bar{u}u\bar{u}u\bar{d}d$ according to the counting role in Eq.\eqref{quark-gluon-operator}, which is of dimension 13, see Fig.\ref{Feynman-penta-13}.  We should take account of the vacuum condensates at least up to dimension $13$ in stead of dimension $10$, which is adopted in most literatures.  The vacuum condensates  $\langle\bar{q} q\rangle^2\langle\bar{q}g_s\sigma Gq\rangle $, $\langle\bar{q} q\rangle \langle\bar{q}g_s\sigma Gq\rangle^2 $, $\langle \bar{q}q\rangle^3\langle \frac{\alpha_s}{\pi}GG\rangle$ are of dimension $11$ and $13$ respectively, and come from the Feynman diagrams shown in Fig.\ref{Feynman-penta-13} \cite{WZG-HC-penta-IJMPA-2020}.  Those vacuum condensates are  associated with the $\frac{1}{T^2}$, $\frac{1}{T^4}$ and $\frac{1}{T^6}$, which manifest themselves at the small $T^2$ and play an important role in choosing  the Borel windows.
   We take the truncations $n\leq 13$ and $k\leq 1$ in a consistent way,
the quark-gluon operators of the orders $\mathcal{O}( \alpha_s^{k})$ with $k> 1$ and dimension $n>13$ are  discarded.

In Refs.\cite{WangZG-Penta-EPJC-2016-70,WangZG-Penta-EPJC-2016-43,
WangZG-Penta-EPJC-2016-142,WangZG-Penta-NPB-2016-163,WangZG-Penta-APPB-2017-2013},  we take  the truncations $n\leq 10$ and $k\leq 1$ and discard the quark-gluon operators of the orders $\mathcal{O}( \alpha_s^{k})$ with $k> 1$ and dimension $n>10$.  Sometimes  we also neglected the vacuum condensates   $\langle \frac{\alpha_sGG}{\pi}\rangle$,
 $\langle \bar{q}q\rangle\langle \frac{\alpha_sGG}{\pi}\rangle$, $\langle \bar{s}s\rangle\langle \frac{\alpha_sGG}{\pi}\rangle$, $\langle \bar{q}q\rangle^2\langle \frac{\alpha_sGG}{\pi}\rangle$, $\langle \bar{q}q\rangle \langle \bar{s}s\rangle\langle \frac{\alpha_sGG}{\pi}\rangle$,
 $ \langle \bar{s}s\rangle^2\langle \frac{\alpha_sGG}{\pi}\rangle$, which are not associated with the $\frac{1}{T^2}$, $\frac{1}{T^4}$ and $\frac{1}{T^6}$ to manifest themselves at the small $T^2$. Such an approximation would impair the predictive ability.

 Then let us  match the hadron side with the QCD side of the correlation functions, take the quark-hadron duality below the continuum thresholds  $s_0$, and  obtain  the  QCD sum rules:
\begin{eqnarray}\label{QCDSR-penta}
2M_{\mp}\lambda^{\mp}_j{}^2\exp\left( -\frac{M_{\mp}^2}{T^2}\right)&=& \int_{4m_c^2}^{s_0}ds \,\rho_{QCD,j}(s)\,\exp\left( -\frac{s}{T^2}\right)\,  ,
\end{eqnarray}
where $\rho_{QCD,j}(s)=\sqrt{s}\rho_{QCD,j}^1(s)\pm\rho_{QCD,j}^{0}(s)$.

We derive   Eq.\eqref{QCDSR-penta} with respect to  $\frac{1}{T^2}$, then eliminate the
 pole residues $\lambda^{\mp}_{j}$ and obtain the QCD sum rules for
 the masses of the hidden-charm  pentaquark states,
 \begin{eqnarray}\label{QCDSR-penta-masses}
 M^2_{\mp} &=& \frac{-\int_{4m_c^2}^{s_0}ds \frac{d}{d(1/T^2)}\, \rho_{QCD,j}(s)\,\exp\left( -\frac{s}{T^2}\right)}{\int_{4m_c^2}^{s_0}ds \, \rho_{QCD,j}(s)\,\exp\left( -\frac{s}{T^2}\right)}\,  .
\end{eqnarray}

With a simple replacement $c \to b$, we obtain the corresponding  QCD sum rules for the hidden-bottom pentaquark states.

According to the discussions in Sect.{\bf\ref{Energy-scale-dependence}}, we take the energy scale formula,
\begin{eqnarray}\label{formula-penta}
\mu&=&\sqrt{M^2_{X/Y/Z/P}-(2{\mathbb{M}}_Q)^2}\, ,
\end{eqnarray}
to determine the best energy scales of the QCD spectral densities \cite{WZG-HC-penta-IJMPA-2020,  WangZG-Penta-EPJC-2016-70,WangZG-Penta-EPJC-2016-43,
WangZG-Penta-EPJC-2016-142,WangZG-Penta-NPB-2016-163,WangZG-Penta-APPB-2017-2013}, and choose the updated value of the effective $c$-quark mass  ${\mathbb{M}}_c=1.82\,\rm{GeV}$ \cite{WangZG-Y-tetra-EPJC-1601}.

After trial and error, we obtain the Borel windows $T^2$, continuum threshold parameters $s_0$, ideal energy scales of the QCD spectral densities, pole contributions of the ground states and contributions of the vacuum condensates of dimension $13$, which   are shown   explicitly in Table \ref{Borel-penta-Pc4312} \cite{WZG-HC-penta-IJMPA-2020}.

\begin{figure}
\centering
\includegraphics[totalheight=7cm,width=10cm]{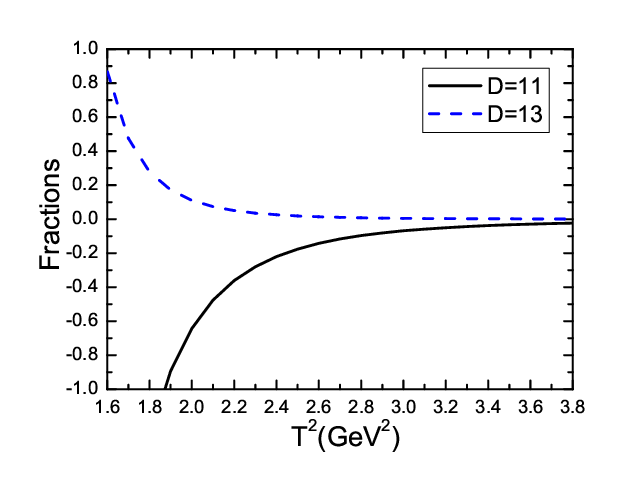}
  \caption{ The contributions of the vacuum condensates of dimension $11$ and $13$  with variation of the  Borel parameter $T^2$ for  the hidden-charm pentaquark state $[ud][uc]\bar{c}$ ($0$, $0$, $0$, $\frac{1}{2}$) \cite{WZG-HC-penta-IJMPA-2020}. }\label{fr-D13-fig}
\end{figure}

In Fig.\ref{fr-D13-fig}, we plot the contributions of the  vacuum condensates of dimension $11$ and $13$ with variation of the  Borel parameter $T^2$ for  the hidden-charm pentaquark state $[ud][uc]\bar{c}$ ($0$, $0$, $0$, $\frac{1}{2}$) with the central values of the parameters shown in Table \ref{Borel-penta-Pc4312} as an example. The vacuum condensates of dimension $13$ manifest themselves at the region $T^2< 2\,\rm{GeV}^2$, we should choose the value $T^2> 2\,\rm{GeV}^2$. While the vacuum condensates of dimension $11$ manifest themselves at the region $T^2< 2.6\,\rm{GeV}^2$, which requires  a larger Borel parameter $T^2> 2.6\,\rm{GeV}^2$ to warrant the  convergence of the  operator product expansion.
The higher dimensional vacuum condensates play an important role in choosing the Borel windows,  where they play an minor important role as the operator product expansion should be convergent, we should take them into account in a consistent way, for example,  the $|D(13)|$ is less than $1\%$ \cite{WZG-HC-penta-IJMPA-2020}.

\begin{figure}
\centering
\includegraphics[totalheight=7cm,width=10cm]{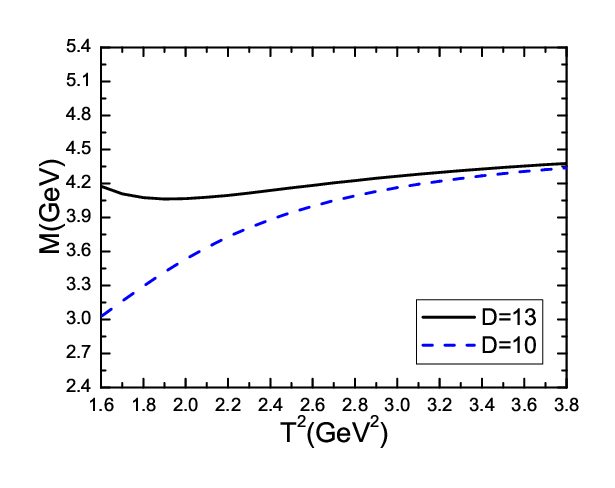}
  \caption{ The mass  with variation of the  Borel parameter $T^2$ for  the hidden-charm pentaquark state $[ud][uc]\bar{c}$ ($0$, $0$, $0$, $\frac{1}{2}$), the $D=10$, $13$ denote truncations of the operator product expansion \cite{WZG-HC-penta-IJMPA-2020}. }\label{mass-D10-fig}
\end{figure}

In Fig.\ref{mass-D10-fig}, we plot the mass of the hidden-charm pentaquark state $[ud][uc]\bar{c}$ ($0$, $0$, $0$, $\frac{1}{2}$) with   variation of the  Borel parameter $T^2$ for truncations   of the operator product expansion up to the  vacuum condensates of dimension $10$ and $13$, respectively \cite{WZG-HC-penta-IJMPA-2020}.   The vacuum condensates of dimension $11$ and $13$ play an important role to obtain stable QCD sum rules, we should take them into account.

In Table \ref{Borel-penta-Pc4312},  the pole contributions are about $(40-60)\%$ and the contributions of the vacuum condensates of dimension $13$ are $\leq 1\%$ or $\ll 1\%$,  the pole dominance and convergence of the operator product expansion are all satisfied,  {\bf the two basic criteria of the QCD sum rules  are satisfied  in the case of the hidden-charm pentaquark states for the first time}.

We take  account of all uncertainties  of the relevant  parameters,
and obtain  the masses and pole residues of
 the    hidden-charm pentaquark states with negative parity, which are shown explicitly in Table  \ref{mass-penta-Pc4312}.

The predicted masses $M_{P}=4.31\pm0.11\,\rm{GeV}$ for the ground state $[ud][uc]\bar{c}$ ($0$, $0$, $0$, $\frac{1}{2}$) pentaquark state and
$M_{P}=4.34\pm0.14\,\rm{GeV}$ for the ground state $[uu][dc]\bar{c}+2[ud][uc]\bar{c}$ ($1$, $1$, $0$, $\frac{1}{2}$) pentaquark state
are both in excellent agreement  with the experimental data   $M_{P(4312)}=4311.9\pm0.7^{+6.8}_{-0.6} \,\rm{MeV}$ from the LHCb   collaboration \cite{LHCb-Pc4312}, and support assigning the $P_c(4312)$ to be the hidden-charm pentaquark state with the  $J^{P}={\frac{1}{2}}^-$.

After the work was published \cite{WZG-HC-penta-IJMPA-2020},  the LHCb collaboration observed an evidence for a structure $P_c(4337)$ in the $J/\psi p$ and $J/\psi \bar{p}$ systems  with  a significance about $3.1-3.7\sigma$ depending on the $J^P$ hypothesis \cite{LHCb-Pc4337}, the Breit-Wigner mass and width are $4337^{+7}_{-4} {}^{+2}_{-2}\, \mbox{ MeV}$ and $29^{+26}_{-12} {}^{+14}_{-14}\, \mbox{ MeV} $ respectively.
The $P_c(4337)$ can be assigned as the ground state $[uu][dc]\bar{c}+2[ud][uc]\bar{c}$ ($1$, $1$, $0$, $\frac{1}{2}$) pentaquark state with the mass $4.34\pm0.14\,\rm{GeV}$.

The predicted masses
$M_{P}=4.45\pm0.11\,\rm{GeV}$ for the ground state $[ud][uc]\bar{c}$ ($0$, $1$, $1$, $\frac{1}{2}$)  pentaquark state,
$M_{P}=4.46\pm0.11\,\rm{GeV}$ for the ground state $[uu][dc]\bar{c}+2[ud][uc]\bar{c}$ ($1$, $0$, $1$, $\frac{1}{2}$)  pentaquark state and
$M_{P}=4.39\pm0.11$ for the ground state $[ud][uc]\bar{c}$ ($0$, $1$, $1$, $\frac{3}{2}$), $[uu][dc]\bar{c}+2[ud][uc]\bar{c}$ ($1$, $1$, $2$, $\frac{5}{2}$), $[ud][uc]\bar{c}$ ($0$, $1$, $1$, $\frac{5}{2}$)    pentaquark states
are in excellent agreement (or  compatible with) the experimental data   $M_{P(4440)}=4440.3\pm1.3^{+4.1}_{-4.7} \,\rm{MeV}$ from the LHCb   collaboration \cite{LHCb-Pc4312}, and support assigning the $P_c(4440)$ to be the hidden-charm pentaquark state with $J^{P}={\frac{1}{2}}^-$, ${\frac{3}{2}}^-$ or ${\frac{5}{2}}^-$.

  The predicted masses
$M_{P}=4.45\pm0.11\,\rm{GeV}$ for the ground state $[ud][uc]\bar{c}$ ($0$, $1$, $1$, $\frac{1}{2}$)  pentaquark state,
$M_{P}=4.46\pm0.11\,\rm{GeV}$ for the ground state $[uu][dc]\bar{c}+2[ud][uc]\bar{c}$ ($1$, $0$, $1$, $\frac{1}{2}$)  pentaquark state   and $M_{P}=4.47\pm0.11\,\rm{GeV}$ for the ground state
 $[uu][dc]\bar{c}+2[ud][uc]\bar{c}$ ($1$, $0$, $1$, $\frac{3}{2}$)    pentaquark states
are in excellent agreement  the experimental data   $M_{P(4457)}=4457.3\pm0.6^{+4.1}_{-1.7} \,\rm{MeV}$ from the LHCb   collaboration \cite{LHCb-Pc4312}, and support assigning the $P_c(4457)$ to be the hidden-charm pentaquark state with $J^{P}={\frac{1}{2}}^-$ or ${\frac{3}{2}}^-$.

In Table \ref{mass-penta-Pc4312}, we present the possible assignments of the $P_c(4312)$, $P_c(4440)$ and $P_c(4457)$ explicitly as a summary. In Table \ref{mass-1508-et al}, we compare the present predictions with our previous calculations \cite{WangZG-Penta-EPJC-2016-70,WangZG-Penta-EPJC-2016-43,
WangZG-Penta-EPJC-2016-142,WangZG-Penta-NPB-2016-163}, where the vacuum condensates of dimension $11$ and $13$ were  neglected, sometimes the vacuum condensates $\langle\frac{\alpha_sGG}{\pi}\rangle$,
$\langle \bar{q}q\rangle\langle\frac{\alpha_sGG}{\pi}\rangle$ and $\langle \bar{q}q\rangle^2\langle\frac{\alpha_sGG}{\pi}\rangle$ were also neglected. From the Table \ref{mass-1508-et al}, we can see that in some cases the predicted masses change  remarkably, while in other cases the predicted masses change  slightly. In Ref.\cite{WangZG-Penta-EPJC-2016-70}, we construct the current $\gamma_5 J_{\mu\nu}^2(x)$ to interpolate the
hidden-charm tetraquark state with the $J^P={5 \over 2}^+$, which should be updated.

\begin{table}
\begin{center}
\begin{tabular}{|c|c|c|c|c|c|c|c|}\hline\hline
                  &$T^2 \rm{GeV}^2)$     &$\sqrt{s_0}(\rm{GeV})$    &$\mu(\rm{GeV})$  &pole          &$D_{13}$         \\ \hline

$J^1(x)$          &$3.1-3.5$             &$4.96\pm0.10$             &$2.3$            &$(41-62)\%$   &$<1\%$      \\ \hline

$J^2(x)$          &$3.2-3.6$             &$5.10\pm0.10$             &$2.6$            &$(42-63)\%$   &$<1\%$      \\ \hline

$J^3(x)$          &$3.2-3.6$             &$5.11\pm0.10$             &$2.6$            &$(42-63)\%$   &$\ll1\%$      \\ \hline

$J^4(x)$          &$2.9-3.3$             &$5.00\pm0.10$             &$2.4$            &$(40-64)\%$   &$\leq1\%$     \\ \hline

$J^1_\mu(x)$      &$3.1-3.5$             &$5.03\pm0.10$             &$2.4$            &$(42-63)\%$   &$\leq1\%$     \\ \hline

$J^2_\mu(x)$      &$3.3-3.7$             &$5.11\pm0.10$             &$2.6$            &$(40-61)\%$   &$\ll1\%$     \\ \hline

$J^3_\mu(x)$      &$3.4-3.8$             &$5.26\pm0.10$             &$2.8$            &$(42-62)\%$   &$\ll1\%$     \\ \hline

$J^4_\mu(x)$      &$3.3-3.7$             &$5.17\pm0.10$             &$2.7$            &$(41-61)\%$   &$<1\%$     \\ \hline

$J^1_{\mu\nu}(x)$ &$3.2-3.6$             &$5.03\pm0.10$             &$2.4$            &$(40-61)\%$   &$\leq1\%$     \\ \hline

$J^2_{\mu\nu}(x)$ &$3.1-3.5$             &$5.03\pm0.10$             &$2.4$            &$(42-63)\%$   &$\leq1\%$     \\ \hline\hline
\end{tabular}
\end{center}
\caption{ The Borel  windows, continuum threshold parameters, ideal energy scales, pole contributions and   contributions of the vacuum condensates of dimension 13 for the hidden-charm pentaquark states \cite{WZG-HC-penta-IJMPA-2020}. }\label{Borel-penta-Pc4312}
\end{table}

\begin{table}
\begin{center}
\begin{tabular}{|c|c|c|c|c|c|c|c|c|}\hline\hline
$[qq^\prime][q^{\prime\prime}c]\bar{c}$ ($S_L$, $S_H$, $J_{LH}$, $J$) &$M(\rm{GeV})$   &$\lambda(10^{-3}\rm{GeV}^6)$ &Assignments        &Currents \\ \hline

$[ud][uc]\bar{c}$ ($0$, $0$, $0$, $\frac{1}{2}$)                      &$4.31\pm0.11$   &$1.40\pm0.23$                &$?\,P_c(4312)$      &$J^1(x)$    \\

$[ud][uc]\bar{c}$ ($0$, $1$, $1$, $\frac{1}{2}$)                      &$4.45\pm0.11$   &$3.02\pm0.48$                &$?\,P_c(4440/4457)$ &$J^2(x)$    \\

$[uu][dc]\bar{c}+2[ud][uc]\bar{c}$ ($1$, $0$, $1$, $\frac{1}{2}$)     &$4.46\pm0.11$   &$4.32\pm0.71$                &$?\,P_c(4440/4457)$ &$J^3(x)$    \\

$[uu][dc]\bar{c}+2[ud][uc]\bar{c}$ ($1$, $1$, $0$, $\frac{1}{2}$)     &$4.34\pm0.14$   &$3.23\pm0.61$                &$?\,P_c(4312/{\mathbf{4337}})$      &$J^4(x)$   \\

$[ud][uc]\bar{c}$ ($0$, $1$, $1$, $\frac{3}{2}$)                      &$4.39\pm0.11$   &$1.44\pm0.23$                &$?\,P_c(4440)$      &$J^1_\mu(x)$ \\

$[uu][dc]\bar{c}+2[ud][uc]\bar{c}$ ($1$, $0$, $1$, $\frac{3}{2}$)     &$4.47\pm0.11$   &$2.41\pm0.38$                &$?\,P_c(4440/4457)$ &$J^2_\mu(x)$  \\

$[uu][dc]\bar{c}+2[ud][uc]\bar{c}$ ($1$, $1$, $2$, $\frac{3}{2}$)     &$4.61\pm0.11$   &$5.13\pm0.79$                &                    &$J^3_\mu(x)$  \\

$[uu][dc]\bar{c}+2[ud][uc]\bar{c}$ ($1$, $1$, $2$, $\frac{3}{2}$)     &$4.52\pm0.11$   &$4.49\pm0.72$                &                    &$J^4_\mu(x)$ \\

$[uu][dc]\bar{c}+2[ud][uc]\bar{c}$ ($1$, $1$, $2$, $\frac{5}{2}$)     &$4.39\pm0.11$   &$1.94\pm0.31$                &$?\,P_c(4440)$      &$J^1_{\mu\nu}(x)$ \\

$[ud][uc]\bar{c}$ ($0$, $1$, $1$, $\frac{5}{2}$)                      &$4.39\pm0.11$   &$1.44\pm0.23$                &$?\,P_c(4440)$      &$J^2_{\mu\nu}(x)$ \\ \hline\hline
\end{tabular}
\end{center}
\caption{ The masses  and pole residues of the hidden-charm pentaquark states \cite{WZG-HC-penta-IJMPA-2020}. }\label{mass-penta-Pc4312}
\end{table}

\begin{table}
\begin{center}
\begin{tabular}{|c|c|c|c|c|c|c|c|c|}\hline\hline
$[qq^\prime][q^{\prime\prime}c]\bar{c}$ ($S_L$, $S_H$, $J_{LH}$, $J$) &Ref.\cite{WZG-HC-penta-IJMPA-2020}       &Old calculations                 &Currents \\ \hline

$[ud][uc]\bar{c}$ ($0$, $0$, $0$, $\frac{1}{2}$)                      &$4.31\pm0.11$   &$4.29\pm 0.13$                &$J^1(x)$    \\

$[ud][uc]\bar{c}$ ($0$, $1$, $1$, $\frac{1}{2}$)                      &$4.45\pm0.11$   &$4.30 \pm 0.13$               &$J^2(x)$    \\

$[uu][dc]\bar{c}+2[ud][uc]\bar{c}$ ($1$, $0$, $1$, $\frac{1}{2}$)     &$4.46\pm0.11$   &$4.42 \pm 0.12$               &$J^3(x)$    \\

$[uu][dc]\bar{c}+2[ud][uc]\bar{c}$ ($1$, $1$, $0$, $\frac{1}{2}$)     &$4.34\pm0.14$   &$4.35\pm 0.15$                &$J^4(x)$   \\

$[ud][uc]\bar{c}$ ($0$, $1$, $1$, $\frac{3}{2}$)                      &$4.39\pm0.11$   &$4.38 \pm 0.13$               &$J^1_\mu(x)$ \\

$[uu][dc]\bar{c}+2[ud][uc]\bar{c}$ ($1$, $0$, $1$, $\frac{3}{2}$)     &$4.47\pm0.11$   &$4.39\pm 0.13$                &$J^2_\mu(x)$  \\

$[uu][dc]\bar{c}+2[ud][uc]\bar{c}$ ($1$, $1$, $2$, $\frac{3}{2}$)     &$4.61\pm0.11$   &$4.39 \pm 0.14$               &$J^3_\mu(x)$  \\

$[uu][dc]\bar{c}+2[ud][uc]\bar{c}$ ($1$, $1$, $2$, $\frac{3}{2}$)     &$4.52\pm0.11$   &$4.39 \pm 0.14$               &$J^4_\mu(x)$ \\

$[uu][dc]\bar{c}+2[ud][uc]\bar{c}$ ($1$, $1$, $2$, $\frac{5}{2}$)     &$4.39\pm0.11$   &                              &$J^1_{\mu\nu}(x)$ \\

$[ud][uc]\bar{c}$ ($0$, $1$, $1$, $\frac{5}{2}$)                      &$4.39\pm0.11$   &                              &$J^2_{\mu\nu}(x)$ \\ \hline\hline
\end{tabular}
\end{center}
\caption{ The masses (in unit of GeV) are compared with the old calculations  \cite{WangZG-Penta-EPJC-2016-70,WangZG-Penta-EPJC-2016-43,
WangZG-Penta-EPJC-2016-142,WangZG-Penta-NPB-2016-163}.   }\label{mass-1508-et al}
\end{table}

In Ref.\cite{WangZG-Penta-EPJC-2016-142}, we construct the $J_{q_1q_2q_3}^{j_Lj_H}(x)$ with the spin-parity $J^P={1\over 2}^-$ to study the hidden-charm pentaquark states with the $J^P={1\over 2}^\pm$  according to the
rules,
\begin{eqnarray}
1^+_{q_1q_2}\otimes 0^+_{q_3c} \otimes {\frac{1}{2}}^-_{\bar{c}} &=&  \underline{{\frac{1}{2}}^-_{q_1q_2q_3c\bar{c}}} \oplus {\frac{3}{2}}^-_{q_1q_2q_3c\bar{c}} \, , \nonumber
\end{eqnarray}
\begin{eqnarray}
1^+_{q_1q_2}\otimes 1^+_{q_3c} \otimes {\frac{1}{2}}^-_{\bar{c}} &=& \left[0^+_{q_1q_2q_3c}\oplus 1^+_{q_1q_2q_3c}\oplus  2^+_{q_1q_2q_3c}\right]\otimes {\frac{1}{2}}^-_{\bar{c}} \nonumber \\
&=&\underline{{\frac{1}{2}}^-_{q_1q_2q_3c\bar{c}}} \oplus \left[{\frac{1}{2}}^-_{q_1q_2q_3c\bar{c}} \oplus {\frac{3}{2}}^-_{q_1q_2q_3c\bar{c}}\right]\oplus \left[ {\frac{3}{2}}^-_{q_1q_2q_3c\bar{c}}\oplus {\frac{5}{2}}^-_{q_1q_2q_3c\bar{c}}\right] \, , \nonumber
\end{eqnarray}
\begin{eqnarray}
1^+_{q_1q_2}\otimes 0^+_{q_3c} \otimes \left[1^-\otimes {\frac{1}{2}}^-_{\bar{c}}\right] &=& 1^+_{q_1q_2}\otimes 0^+_{q_3c} \otimes  \left[{\frac{1}{2}}^+_{\bar{c}}\oplus {\frac{3}{2}}^+_{\bar{c}}\right]  \nonumber \\
&=& \left[\underline{{\frac{1}{2}}^+_{q_1q_2q_3c\bar{c}}}\oplus {\frac{3}{2}}^+_{q_1q_2q_3c\bar{c}}\right]\oplus \left[ {\frac{1}{2}}^+_{q_1q_2q_3c\bar{c}}\oplus {\frac{3}{2}}^+_{q_1q_2q_3c\bar{c}} \oplus {\frac{5}{2}}^+_{q_1q_2q_3c\bar{c}} \right] \, ,  \nonumber
\end{eqnarray}
\begin{eqnarray}
1^+_{q_1q_2}\otimes 1^+_{q_3c} \otimes \left[1^-\otimes {\frac{1}{2}}^-_{\bar{c}} \right]&=& \left[0^+_{q_1q_2q_3c}\oplus 1^+_{q_1q_2q_3c}\oplus  2^+_{q_1q_2q_3c}\right]\otimes \left[{\frac{1}{2}}^+_{\bar{c}}\oplus {\frac{3}{2}}^+_{\bar{c}}\right]\nonumber \\
&=&\underline{{\frac{1}{2}}^+_{q_1q_2q_3c\bar{c}}} \oplus \left[{\frac{1}{2}}^+_{q_1q_2q_3c\bar{c}} \oplus {\frac{3}{2}}^+_{q_1q_2q_3c\bar{c}}\right]\oplus \left[ {\frac{3}{2}}^+_{q_1q_2q_3c\bar{c}}\oplus {\frac{5}{2}}^+_{q_1q_2q_3c\bar{c}}\right]  \nonumber\\
&&  \oplus {\frac{3}{2}}^+_{q_1q_2q_3c\bar{c}} \oplus\left[{\frac{1}{2}}^+_{q_1q_2q_3c\bar{c}} \oplus {\frac{3}{2}}^+_{q_1q_2q_3c\bar{c}}\oplus {\frac{5}{2}}^+_{q_1q_2q_3c\bar{c}}\right] \nonumber \\
&&\oplus\left[{\frac{1}{2}}^+_{q_1q_2q_3c\bar{c}} \oplus{\frac{3}{2}}^+_{q_1q_2q_3c\bar{c}}\oplus {\frac{5}{2}}^+_{q_1q_2q_3c\bar{c}}\oplus {\frac{7}{2}}^+_{q_1q_2q_3c\bar{c}}\right]\, ,
\end{eqnarray}
where the $1^-$ denotes the additional P-wave and is embodied  by a $\gamma_5$ in constructing the currents,  the subscripts $q_1q_2$, $q_3c$, $\cdots$ denote the quark   constituents. The quark and antiquark have opposite parity,   we usually take it for granted that the quarks (antiquarks) have positive (negative) parity, and the $\bar{c}$-quark has the $J^P={\frac{1}{2}}^-$.

We write down the currents $J_{q_1q_2q_3}^{j_Lj_H}(x)$ explicitly,
\begin{eqnarray}
J^{11}_{uuu}(x)&=&\varepsilon^{ila} \varepsilon^{ijk}\varepsilon^{lmn}  u^T_j(x) C\gamma_\mu u_k(x)u^T_m(x) C\gamma^\mu c_n(x)  C\bar{c}^{T}_{a}(x) \, , \nonumber\\
J^{11}_{uud}(x)&=&\frac{\varepsilon^{ila} \varepsilon^{ijk}\varepsilon^{lmn}}{\sqrt{3}} \left[ u^T_j(x) C\gamma_\mu u_k(x)d^T_m(x) C\gamma^\mu c_n(x)+2u^T_j(x) C\gamma_\mu d_k(x)u^T_m(x) C\gamma^\mu c_n(x) \right] C\bar{c}^{T}_{a}(x) \, , \nonumber\\
J^{11}_{udd}(x)&=&\frac{\varepsilon^{ila} \varepsilon^{ijk}\varepsilon^{lmn}}{\sqrt{3}} \left[ d^T_j(x) C\gamma_\mu d_k(x)u^T_m(x) C\gamma^\mu c_n(x)+2d^T_j(x) C\gamma_\mu u_k(x)d^T_m(x) C\gamma^\mu c_n(x) \right] C\bar{c}^{T}_{a}(x) \, , \nonumber\\
J^{11}_{ddd}(x)&=&\varepsilon^{ila} \varepsilon^{ijk}\varepsilon^{lmn}  d^T_j(x) C\gamma_\mu d_k(x)d^T_m(x) C\gamma^\mu c_n(x)  C\bar{c}^{T}_{a}(x) \, ,
\end{eqnarray}
\begin{eqnarray}
 J^{11}_{uus}(x)&=&\frac{\varepsilon^{ila} \varepsilon^{ijk}\varepsilon^{lmn}}{\sqrt{3}} \left[ u^T_j(x) C\gamma_\mu u_k(x)s^T_m(x) C\gamma^\mu c_n(x)+2u^T_j(x) C\gamma_\mu s_k(x)u^T_m(x) C\gamma^\mu c_n(x) \right] C\bar{c}^{T}_{a}(x) \, , \nonumber\\
  J^{11}_{uds}(x)&=&\frac{\varepsilon^{ila} \varepsilon^{ijk}\varepsilon^{lmn}}{\sqrt{3}} \left[ u^T_j(x) C\gamma_\mu d_k(x)s^T_m(x) C\gamma^\mu c_n(x)+u^T_j(x) C\gamma_\mu s_k(x)d^T_m(x) C\gamma^\mu c_n(x) \right. \nonumber\\
 &&\left.+d^T_j(x) C\gamma_\mu s_k(x)u^T_m(x) C\gamma^\mu c_n(x) \right] C\bar{c}^{T}_{a}(x) \, , \nonumber\\
  J^{11}_{dds}(x)&=&\frac{\varepsilon^{ila} \varepsilon^{ijk}\varepsilon^{lmn}}{\sqrt{3}} \left[ d^T_j(x) C\gamma_\mu d_k(x)s^T_m(x) C\gamma^\mu c_n(x)+2d^T_j(x) C\gamma_\mu s_k(x)d^T_m(x) C\gamma^\mu c_n(x) \right] C\bar{c}^{T}_{a}(x) \, , \nonumber\\
 \end{eqnarray}
\begin{eqnarray}
  J^{11}_{uss}(x)&=&\frac{\varepsilon^{ila} \varepsilon^{ijk}\varepsilon^{lmn}}{\sqrt{3}} \left[ s^T_j(x) C\gamma_\mu s_k(x)u^T_m(x) C\gamma^\mu c_n(x)+2s^T_j(x) C\gamma_\mu u_k(x)s^T_m(x) C\gamma^\mu c_n(x) \right] C\bar{c}^{T}_{a}(x) \, , \nonumber\\
   J^{11}_{dss}(x)&=&\frac{\varepsilon^{ila} \varepsilon^{ijk}\varepsilon^{lmn}}{\sqrt{3}} \left[ s^T_j(x) C\gamma_\mu s_k(x)d^T_m(x) C\gamma^\mu c_n(x)+2s^T_j(x) C\gamma_\mu d_k(x)s^T_m(x) C\gamma^\mu c_n(x) \right] C\bar{c}^{T}_{a}(x) \, , \nonumber\\
   \end{eqnarray}
\begin{eqnarray}
  J^{11}_{sss}(x)&=&\varepsilon^{ila} \varepsilon^{ijk}\varepsilon^{lmn}  s^T_j(x) C\gamma_\mu s_k(x)s^T_m(x) C\gamma^\mu c_n(x)  C\bar{c}^{T}_{a}(x) \, ,
\end{eqnarray}
\begin{eqnarray}
J_{uuu}^{10}(x)&=&\varepsilon^{ila} \varepsilon^{ijk}\varepsilon^{lmn}   u^T_j(x) C\gamma_\mu u_k(x) u^T_m(x) C\gamma_5 c_n(x) \gamma_5  \gamma^\mu C\bar{c}^{T}_{a}(x) \, , \nonumber\\
J^{10}_{uud}(x)&=&\frac{\varepsilon^{ila} \varepsilon^{ijk}\varepsilon^{lmn}}{\sqrt{3}} \left[ u^T_j(x) C\gamma_\mu u_k(x) d^T_m(x) C\gamma_5 c_n(x)+2u^T_j(x) C\gamma_\mu d_k(x) u^T_m(x) C\gamma_5 c_n(x)\right] \gamma_5 \gamma^\mu  C\bar{c}^{T}_{a}(x) \, ,  \nonumber\\
J^{10}_{udd}(x)&=&\frac{\varepsilon^{ila} \varepsilon^{ijk}\varepsilon^{lmn}}{\sqrt{3}} \left[ d^T_j(x) C\gamma_\mu d_k(x) u^T_m(x) C\gamma_5 c_n(x)+2d^T_j(x) C\gamma_\mu u_k(x) d^T_m(x) C\gamma_5 c_n(x)\right]  \gamma_5 \gamma^\mu C\bar{c}^{T}_{a}(x) \, , \nonumber \\
J_{ddd}^{10}(x)&=&\varepsilon^{ila} \varepsilon^{ijk}\varepsilon^{lmn}   d^T_j(x) C\gamma_\mu d_k(x) d^T_m(x) C\gamma_5 c_n(x) \gamma_5  \gamma^\mu C\bar{c}^{T}_{a}(x) \, ,
\end{eqnarray}
\begin{eqnarray}
J^{10}_{uus}(x)&=&\frac{\varepsilon^{ila} \varepsilon^{ijk}\varepsilon^{lmn}}{\sqrt{3}} \left[ u^T_j(x) C\gamma_\mu u_k(x) s^T_m(x) C\gamma_5 c_n(x)+2u^T_j(x) C\gamma_\mu s_k(x) u^T_m(x) C\gamma_5 c_n(x)\right] \gamma_5 \gamma^\mu  C\bar{c}^{T}_{a}(x) \, ,  \nonumber\\
J^{10}_{uds}(x)&=&\frac{\varepsilon^{ila} \varepsilon^{ijk}\varepsilon^{lmn}}{\sqrt{3}} \left[ u^T_j(x) C\gamma_\mu d_k(x) s^T_m(x) C\gamma_5 c_n(x)+u^T_j(x) C\gamma_\mu s_k(x) d^T_m(x) C\gamma_5 c_n(x)\right.  \nonumber\\
&&\left.+ d^T_j(x) C\gamma_\mu s_k(x) u^T_m(x) C\gamma_5 c_n(x)\right] \gamma_5 \gamma^\mu  C\bar{c}^{T}_{a}(x) \, ,\nonumber\\
J^{10}_{dds}(x)&=&\frac{\varepsilon^{ila} \varepsilon^{ijk}\varepsilon^{lmn}}{\sqrt{3}} \left[ d^T_j(x) C\gamma_\mu d_k(x) s^T_m(x) C\gamma_5 c_n(x)+2d^T_j(x) C\gamma_\mu s_k(x) d^T_m(x) C\gamma_5 c_n(x)\right] \gamma_5 \gamma^\mu  C\bar{c}^{T}_{a}(x) \, ,  \nonumber\\
\end{eqnarray}
\begin{eqnarray}
J^{10}_{uss}(x)&=&\frac{\varepsilon^{ila} \varepsilon^{ijk}\varepsilon^{lmn}}{\sqrt{3}} \left[ s^T_j(x) C\gamma_\mu s_k(x) u^T_m(x) C\gamma_5 c_n(x)+2s^T_j(x) C\gamma_\mu u_k(x) s^T_m(x) C\gamma_5 c_n(x)\right]  \gamma_5 \gamma^\mu C\bar{c}^{T}_{a}(x) \, , \nonumber \\
J^{10}_{dss}(x)&=&\frac{\varepsilon^{ila} \varepsilon^{ijk}\varepsilon^{lmn}}{\sqrt{3}} \left[ s^T_j(x) C\gamma_\mu s_k(x) d^T_m(x) C\gamma_5 c_n(x)+2s^T_j(x) C\gamma_\mu d_k(x) s^T_m(x) C\gamma_5 c_n(x)\right]  \gamma_5 \gamma^\mu C\bar{c}^{T}_{a}(x) \, , \nonumber \\
\end{eqnarray}
\begin{eqnarray}
J^{10}_{sss}(x)&=&\varepsilon^{ila} \varepsilon^{ijk}\varepsilon^{lmn}   s^T_j(x) C\gamma_\mu s_k(x) s^T_m(x) C\gamma_5 c_n(x) \gamma_5  \gamma^\mu C\bar{c}^{T}_{a}(x) \, ,
 \end{eqnarray}
where the superscripts $j_L$ and $j_H$  are  the spins of the light  and heavy diquarks,  respectively.

We take the isospin limit and classify the currents  which couple potentially to the pentaquark states with degenerate masses  into the following 8 types,
\begin{eqnarray}
&&J^{11}_{uuu}(x)\, , \, \, \, J^{11}_{uud}(x)\, , \, \, \, J^{11}_{udd}(x)\, , \, \, \,J^{11}_{ddd}(x) \, ; \nonumber\\
&&J^{11}_{uus}(x)\, , \, \, \,  J^{11}_{uds}(x)\, , \, \, \,  J^{11}_{dds}(x) \, ; \nonumber\\
&&  J^{11}_{uss}(x)\, , \, \, \,   J^{11}_{dss}(x)\, ; \nonumber\\
&&  J^{11}_{sss}(x) \, ; \nonumber\\
&& J_{uuu}^{10}(x)\, , \, \, \, J^{10}_{uud}(x)\, , \, \, \, J^{10}_{udd}(x)\, , \, \, \,J_{ddd}^{10}(x)\, ; \nonumber\\
&&J^{10}_{uus}(x)\, , \, \, \, J^{10}_{uds}(x)\, , \, \, \, J^{10}_{dds}(x) \, ; \nonumber\\
&&J^{10}_{uss}(x)\, , \, \, \, J^{10}_{dss}(x) \, ; \nonumber \\
&&J^{10}_{sss}(x) \, ,
 \end{eqnarray}
and perform the operator product expansion up to the vacuum condensates of dimension 10 to obtain the QCD sum rules, the  predictions for the hidden-charm pentaquark states $P_{q_1q_2q_3}^{j_Lj_H \frac{1}{2}}$ with the spin-parity $J^P={\frac{1}{2}}^\pm$ are presented in Table \ref{mass-EPJC-2016-142} in the Appendix.

From Table \ref{mass-EPJC-2016-142}, we can see that the two-body strong decays to the  $J/\psi B_{10}$,
\begin{eqnarray}
P_{q_1q_2q_3}^{11\frac{1}{2}}\left({\frac{1}{2}^-}\right),\,P_{q_1q_2q_3}^{10\frac{1}{2}}\left({\frac{1}{2}^-}\right) &\to& J/\psi B_{10}\, ,
\end{eqnarray}
for example,
\begin{eqnarray}
P_{uuu}^{11\frac{1}{2}}\left({\frac{1}{2}^-}\right),\,P_{uuu}^{10\frac{1}{2}}\left({\frac{1}{2}^-}\right) &\to& J/\psi \Delta^{++} \, , \nonumber\\
P_{uus}^{11\frac{1}{2}}\left({\frac{1}{2}^-}\right),\,P_{uus}^{10\frac{1}{2}}\left({\frac{1}{2}^-}\right) &\to& J/\psi \Sigma^{*+} \, , \nonumber\\
P_{uss}^{11\frac{1}{2}}\left({\frac{1}{2}^-}\right),\,P_{uss}^{10\frac{1}{2}}\left({\frac{1}{2}^-}\right) &\to& J/\psi \Xi^{*0} \, , \nonumber\\
P_{sss}^{11\frac{1}{2}}\left({\frac{1}{2}^-}\right),\,P_{sss}^{10\frac{1}{2}}\left({\frac{1}{2}^-}\right) &\to& J/\psi \Omega^{-} \, ,
\end{eqnarray}
could take place, but the decay  widths are rather small due to the small available phase-spaces; on the other hand, the two-body strong decays,
\begin{eqnarray}
P_{q_1q_2q_3}^{11\frac{1}{2}}\left({\frac{1}{2}^-}\right),\,P_{q_1q_2q_3}^{10\frac{1}{2}}\left({\frac{1}{2}^-}\right) &\to& J/\psi B_{8}\, ,\\
P_{q_1q_2q_3}^{11\frac{1}{2}}\left({\frac{1}{2}^+}\right) &\to& J/\psi B_{10}\, , J/\psi B_{8}\, ,
\end{eqnarray}
for example,
\begin{eqnarray}
P_{uud}^{11\frac{1}{2}}\left({\frac{1}{2}^\pm}\right),\,P_{uud}^{10\frac{1}{2}}\left({\frac{1}{2}^-}\right) &\to& J/\psi p \, , \nonumber\\
P_{uus}^{11\frac{1}{2}}\left({\frac{1}{2}^\pm}\right),\,P_{uus}^{10\frac{1}{2}}\left({\frac{1}{2}^-}\right) &\to& J/\psi \Sigma^{+} \, , \nonumber\\
P_{uss}^{11\frac{1}{2}}\left({\frac{1}{2}^\pm}\right),\,P_{uss}^{10\frac{1}{2}}\left({\frac{1}{2}^-}\right) &\to& J/\psi \Xi^{0} \,   , \nonumber\\
P_{uuu}^{11\frac{1}{2}}\left({\frac{1}{2}^+}\right) &\to& J/\psi \Delta^{++} \, , \nonumber\\
P_{uus}^{11\frac{1}{2}}\left({\frac{1}{2}^+}\right) &\to& J/\psi \Sigma^{*+} \, , \nonumber\\
P_{uss}^{11\frac{1}{2}}\left({\frac{1}{2}^+}\right) &\to& J/\psi \Xi^{*0} \, , \nonumber\\
P_{sss}^{11\frac{1}{2}}\left({\frac{1}{2}^+}\right) &\to& J/\psi \Omega^{-} \, ,
\end{eqnarray}
can take place more easily,   the decay widths are larger due to the larger available phase-spaces; furthermore, the two-body strong decays
\begin{eqnarray}
P_{q_1q_2q_3}^{10\frac{1}{2}}\left({\frac{1}{2}^+}\right) &\to&J/\psi B_{8}\, , J/\psi B_{10}\, ,
\end{eqnarray}
for example,
\begin{eqnarray}
P_{uud}^{10\frac{1}{2}}\left({\frac{1}{2}^+}\right) &\to& J/\psi p \, , \nonumber\\
P_{uus}^{10\frac{1}{2}}\left({\frac{1}{2}^+}\right) &\to& J/\psi \Sigma^{+} \, , \nonumber\\
P_{uss}^{10\frac{1}{2}}\left({\frac{1}{2}^+}\right) &\to& J/\psi \Xi^{0} \,   , \nonumber\\
P_{uuu}^{10\frac{1}{2}}\left({\frac{1}{2}^+}\right) &\to& J/\psi \Delta^{++} \, , \nonumber\\
P_{uus}^{10\frac{1}{2}}\left({\frac{1}{2}^+}\right) &\to& J/\psi \Sigma^{*+} \, , \nonumber\\
P_{uss}^{10\frac{1}{2}}\left({\frac{1}{2}^+}\right) &\to& J/\psi \Xi^{*0} \, , \nonumber\\
P_{sss}^{10\frac{1}{2}}\left({\frac{1}{2}^+}\right) &\to& J/\psi \Omega^{-} \, ,
\end{eqnarray}
can take place fluently,   the decay  widths are rather large due to the large available phase-spaces.
We can search for those pentaquark states in the $J/\psi B_8$ and $J/\psi B_{10}$ mass spectra  in the decays of the bottom baryons to the final states $J/\psi B_8$ and $J/\psi B_{10}$ associated with the light vector mesons or pseudoscalar mesons \cite{Penta-Maiani-PLB-2015,HeXG-penta-JHEP-2015,ChengHY-penta-PRD-2015},
for example,
\begin{eqnarray}
\Omega_b^- &\to& P_{uss}^{11\frac{1}{2}}\left({\frac{1}{2}}^{\pm}\right)K^- \to J/\psi \Xi^{*0} K^-\, , \nonumber\\
\Omega_b^- &\to& P_{sss}^{11\frac{1}{2}}\left({\frac{1}{2}}^{\pm}\right)\phi \to J/\psi \Omega^- \phi\, .
\end{eqnarray}

In Ref.\cite{WangZG-Penta-NPB-2016-163}, we construct the currents  $J_{q_1q_2q_3,\mu}^{j_Lj_Hj}(x)$ to interpolate the $ J^P={\frac{3}{2}}^{\pm}$ hidden-charm tetraquark states, where the superscripts $j_L$ and $j_H$  are  the spins of the light and heavy diquarks,  respectively, $\vec{j}=\vec{j}_H +\vec{j}_{\bar{c}}$, the $j_{\bar{c}}$ is the spin of the heavy antiquark, the subscripts $q_1$, $q_2$, $q_3$ are the light quark constituents $u$, $d$ or $s$. We write down the currents $J_{q_1q_2q_3,\mu}^{j_Lj_Hj}(x)$ explicitly,
\begin{eqnarray}
J_{uuu,\mu}^{10\frac{1}{2}}(x)&=&\varepsilon^{ila} \varepsilon^{ijk}\varepsilon^{lmn}   u^T_j(x) C\gamma_\mu u_k(x) u^T_m(x) C\gamma_5 c_n(x) C\bar{c}^{T}_{a}(x) \, , \nonumber \\
J^{10\frac{1}{2}}_{uud,\mu}(x)&=&\frac{\varepsilon^{ila} \varepsilon^{ijk}\varepsilon^{lmn}}{\sqrt{3}} \left[ u^T_j(x) C\gamma_\mu u_k(x) d^T_m(x) C\gamma_5 c_n(x)+2u^T_j(x) C\gamma_\mu d_k(x) u^T_m(x) C\gamma_5 c_n(x)\right]    C\bar{c}^{T}_{a}(x) \, ,  \nonumber\\
J^{10\frac{1}{2}}_{udd,\mu}(x)&=&\frac{\varepsilon^{ila} \varepsilon^{ijk}\varepsilon^{lmn}}{\sqrt{3}} \left[ d^T_j(x) C\gamma_\mu d_k(x) u^T_m(x) C\gamma_5 c_n(x)+2d^T_j(x) C\gamma_\mu u_k(x) d^T_m(x) C\gamma_5 c_n(x)\right]    C\bar{c}^{T}_{a}(x) \, , \nonumber \\
J_{ddd,\mu}^{10\frac{1}{2}}(x)&=&\varepsilon^{ila} \varepsilon^{ijk}\varepsilon^{lmn}   d^T_j(x) C\gamma_\mu d_k(x) d^T_m(x) C\gamma_5 c_n(x) C\bar{c}^{T}_{a}(x) \, ,
\end{eqnarray}

\begin{eqnarray}
J^{10\frac{1}{2}}_{uus,\mu}(x)&=&\frac{\varepsilon^{ila} \varepsilon^{ijk}\varepsilon^{lmn}}{\sqrt{3}} \left[ u^T_j(x) C\gamma_\mu u_k(x) s^T_m(x) C\gamma_5 c_n(x)+2u^T_j(x) C\gamma_\mu s_k(x) u^T_m(x) C\gamma_5 c_n(x)\right]    C\bar{c}^{T}_{a}(x) \, ,  \nonumber\\
J^{10\frac{1}{2}}_{uds,\mu}(x)&=&\frac{\varepsilon^{ila} \varepsilon^{ijk}\varepsilon^{lmn}}{\sqrt{3}} \left[ u^T_j(x) C\gamma_\mu d_k(x) s^T_m(x) C\gamma_5 c_n(x)+u^T_j(x) C\gamma_\mu s_k(x) d^T_m(x) C\gamma_5 c_n(x)\right.  \nonumber\\
&&\left.+d^T_j(x) C\gamma_\mu s_k(x) u^T_m(x) C\gamma_5 c_n(x)\right]    C\bar{c}^{T}_{a}(x) \, ,  \nonumber\\
J^{10\frac{1}{2}}_{dds,\mu}(x)&=&\frac{\varepsilon^{ila} \varepsilon^{ijk}\varepsilon^{lmn}}{\sqrt{3}} \left[ d^T_j(x) C\gamma_\mu d_k(x) s^T_m(x) C\gamma_5 c_n(x)+2d^T_j(x) C\gamma_\mu s_k(x) d^T_m(x) C\gamma_5 c_n(x)\right]    C\bar{c}^{T}_{a}(x) \, ,  \nonumber\\
\end{eqnarray}

\begin{eqnarray}
J^{10\frac{1}{2}}_{uss,\mu}(x)&=&\frac{\varepsilon^{ila} \varepsilon^{ijk}\varepsilon^{lmn}}{\sqrt{3}} \left[ s^T_j(x) C\gamma_\mu s_k(x) u^T_m(x) C\gamma_5 c_n(x)+2s^T_j(x) C\gamma_\mu u_k(x) s^T_m(x) C\gamma_5 c_n(x)\right]    C\bar{c}^{T}_{a}(x) \, , \nonumber \\
J^{10\frac{1}{2}}_{dss,\mu}(x)&=&\frac{\varepsilon^{ila} \varepsilon^{ijk}\varepsilon^{lmn}}{\sqrt{3}} \left[ s^T_j(x) C\gamma_\mu s_k(x) d^T_m(x) C\gamma_5 c_n(x)+2s^T_j(x) C\gamma_\mu d_k(x) s^T_m(x) C\gamma_5 c_n(x)\right]    C\bar{c}^{T}_{a}(x) \, , \nonumber \\
\end{eqnarray}

\begin{eqnarray}
J^{10\frac{1}{2}}_{sss,\mu}(x)&=&\varepsilon^{ila} \varepsilon^{ijk}\varepsilon^{lmn}   s^T_j(x) C\gamma_\mu s_k(x) s^T_m(x) C\gamma_5 c_n(x) C\bar{c}^{T}_{a}(x) \, ,
\end{eqnarray}

\begin{eqnarray}
J^{11\frac{1}{2}}_{uuu,\mu}(x)&=&\varepsilon^{ila} \varepsilon^{ijk}\varepsilon^{lmn}  u^T_j(x) C\gamma_\mu u_k(x)u^T_m(x) C\gamma_\alpha c_n(x)  \gamma_5\gamma^\alpha C\bar{c}^{T}_{a}(x) \, , \nonumber\\
J^{11\frac{1}{2}}_{uud,\mu}(x)&=&\frac{\varepsilon^{ila} \varepsilon^{ijk}\varepsilon^{lmn}}{\sqrt{3}} \left[ u^T_j(x) C\gamma_\mu u_k(x)d^T_m(x) C\gamma_\alpha c_n(x)+2u^T_j(x) C\gamma_\mu d_k(x)u^T_m(x) C\gamma_\alpha c_n(x) \right] \gamma_5\gamma^\alpha C\bar{c}^{T}_{a}(x) \, , \nonumber\\
J^{11\frac{1}{2}}_{udd,\mu}(x)&=&\frac{\varepsilon^{ila} \varepsilon^{ijk}\varepsilon^{lmn}}{\sqrt{3}} \left[ d^T_j(x) C\gamma_\mu d_k(x)u^T_m(x) C\gamma_\alpha c_n(x)+2d^T_j(x) C\gamma_\mu u_k(x)d^T_m(x) C\gamma_\alpha c_n(x) \right] \gamma_5\gamma^\alpha C\bar{c}^{T}_{a}(x) \, , \nonumber\\
J^{11\frac{1}{2}}_{ddd,\mu}(x)&=&\varepsilon^{ila} \varepsilon^{ijk}\varepsilon^{lmn}  d^T_j(x) C\gamma_\mu d_k(x)d^T_m(x) C\gamma_\alpha c_n(x)  \gamma_5\gamma^\alpha C\bar{c}^{T}_{a}(x) \, ,
\end{eqnarray}

\begin{eqnarray}
 J^{11\frac{1}{2}}_{uus,\mu}(x)&=&\frac{\varepsilon^{ila} \varepsilon^{ijk}\varepsilon^{lmn}}{\sqrt{3}} \left[ u^T_j(x) C\gamma_\mu u_k(x)s^T_m(x) C\gamma_\alpha c_n(x)+2u^T_j(x) C\gamma_\mu s_k(x)u^T_m(x) C\gamma_\alpha c_n(x) \right] \gamma_5\gamma^\alpha C\bar{c}^{T}_{a}(x) \, , \nonumber\\
 J^{11\frac{1}{2}}_{uds,\mu}(x)&=&\frac{\varepsilon^{ila} \varepsilon^{ijk}\varepsilon^{lmn}}{\sqrt{3}} \left[ u^T_j(x) C\gamma_\mu d_k(x)s^T_m(x) C\gamma_\alpha c_n(x)+u^T_j(x) C\gamma_\mu s_k(x)d^T_m(x) C\gamma_\alpha c_n(x) \right.\nonumber\\
 &&\left.+d^T_j(x) C\gamma_\mu s_k(x)u^T_m(x) C\gamma_\alpha c_n(x) \right] \gamma_5\gamma^\alpha C\bar{c}^{T}_{a}(x) \, ,\nonumber\\
 J^{11\frac{1}{2}}_{dds,\mu}(x)&=&\frac{\varepsilon^{ila} \varepsilon^{ijk}\varepsilon^{lmn}}{\sqrt{3}} \left[ d^T_j(x) C\gamma_\mu d_k(x)s^T_m(x) C\gamma_\alpha c_n(x)+2d^T_j(x) C\gamma_\mu s_k(x)d^T_m(x) C\gamma_\alpha c_n(x) \right] \gamma_5\gamma^\alpha C\bar{c}^{T}_{a}(x) \, , \nonumber\\
\end{eqnarray}

\begin{eqnarray}
  J^{11\frac{1}{2}}_{uss,\mu}(x)&=&\frac{\varepsilon^{ila} \varepsilon^{ijk}\varepsilon^{lmn}}{\sqrt{3}} \left[ s^T_j(x) C\gamma_\mu s_k(x)u^T_m(x) C\gamma_\alpha c_n(x)+2s^T_j(x) C\gamma_\mu u_k(x)s^T_m(x) C\gamma_\alpha c_n(x) \right] \gamma_5\gamma^\alpha C\bar{c}^{T}_{a}(x) \, , \nonumber\\
    J^{11\frac{1}{2}}_{dss,\mu}(x)&=&\frac{\varepsilon^{ila} \varepsilon^{ijk}\varepsilon^{lmn}}{\sqrt{3}} \left[ s^T_j(x) C\gamma_\mu s_k(x)d^T_m(x) C\gamma_\alpha c_n(x)+2s^T_j(x) C\gamma_\mu d_k(x)s^T_m(x) C\gamma_\alpha c_n(x) \right] \gamma_5\gamma^\alpha C\bar{c}^{T}_{a}(x) \, , \nonumber\\
 \end{eqnarray}

\begin{eqnarray}
  J^{11\frac{1}{2}}_{sss,\mu}(x)&=&\varepsilon^{ila} \varepsilon^{ijk}\varepsilon^{lmn}  s^T_j(x) C\gamma_\mu s_k(x)s^T_m(x) C\gamma_\alpha c_n(x)  \gamma_5\gamma^\alpha C\bar{c}^{T}_{a}(x) \, ,
 \end{eqnarray}

\begin{eqnarray}
  J^{11\frac{3}{2}}_{uuu,\mu}(x)&=&\varepsilon^{ila} \varepsilon^{ijk}\varepsilon^{lmn}  u^T_j(x) C\gamma_\alpha u_k(x)u^T_m(x) C\gamma_\mu c_n(x)  \gamma_5\gamma^\alpha C\bar{c}^{T}_{a}(x) \, , \nonumber\\
  J^{11\frac{3}{2}}_{uud,\mu}(x)&=&\frac{\varepsilon^{ila} \varepsilon^{ijk}\varepsilon^{lmn}}{\sqrt{3}} \left[ u^T_j(x) C\gamma_\alpha u_k(x)d^T_m(x) C\gamma_\mu c_n(x)+2u^T_j(x) C\gamma_\alpha d_k(x)u^T_m(x) C\gamma_\mu c_n(x) \right] \gamma_5\gamma^\alpha C\bar{c}^{T}_{a}(x) \, , \nonumber\\
  J^{11\frac{3}{2}}_{udd,\mu}(x)&=&\frac{\varepsilon^{ila} \varepsilon^{ijk}\varepsilon^{lmn}}{\sqrt{3}} \left[ d^T_j(x) C\gamma_\alpha d_k(x)u^T_m(x) C\gamma_\mu c_n(x)+2d^T_j(x) C\gamma_\alpha u_k(x)d^T_m(x) C\gamma_\mu c_n(x) \right] \gamma_5\gamma^\alpha C\bar{c}^{T}_{a}(x) \, , \nonumber\\
  J^{11\frac{3}{2}}_{ddd,\mu}(x)&=&\varepsilon^{ila} \varepsilon^{ijk}\varepsilon^{lmn}  d^T_j(x) C\gamma_\alpha d_k(x)d^T_m(x) C\gamma_\mu c_n(x)  \gamma_5\gamma^\alpha C\bar{c}^{T}_{a}(x) \, ,
\end{eqnarray}

\begin{eqnarray}
 J^{11\frac{3}{2}}_{uus,\mu}(x)&=&\frac{\varepsilon^{ila} \varepsilon^{ijk}\varepsilon^{lmn}}{\sqrt{3}} \left[ u^T_j(x) C\gamma_\alpha u_k(x)s^T_m(x) C\gamma_\mu c_n(x)+2u^T_j(x) C\gamma_\alpha s_k(x)u^T_m(x) C\gamma_\mu c_n(x) \right] \gamma_5\gamma^\alpha C\bar{c}^{T}_{a}(x) \, , \nonumber\\
  J^{11\frac{3}{2}}_{uds,\mu}(x)&=&\frac{\varepsilon^{ila} \varepsilon^{ijk}\varepsilon^{lmn}}{\sqrt{3}} \left[ u^T_j(x) C\gamma_\alpha d_k(x)s^T_m(x) C\gamma_\mu c_n(x)+u^T_j(x) C\gamma_\alpha s_k(x)d^T_m(x) C\gamma_\mu c_n(x) \right. \nonumber\\
  &&\left.+d^T_j(x) C\gamma_\alpha s_k(x)u^T_m(x) C\gamma_\mu c_n(x) \right] \gamma_5\gamma^\alpha C\bar{c}^{T}_{a}(x) \, , \nonumber\\
   J^{11\frac{3}{2}}_{dds,\mu}(x)&=&\frac{\varepsilon^{ila} \varepsilon^{ijk}\varepsilon^{lmn}}{\sqrt{3}} \left[ d^T_j(x) C\gamma_\alpha d_k(x)s^T_m(x) C\gamma_\mu c_n(x)+2d^T_j(x) C\gamma_\alpha s_k(x)d^T_m(x) C\gamma_\mu c_n(x) \right] \gamma_5\gamma^\alpha C\bar{c}^{T}_{a}(x) \, , \nonumber\\
 \end{eqnarray}

\begin{eqnarray}
  J^{11\frac{3}{2}}_{uss,\mu}(x)&=&\frac{\varepsilon^{ila} \varepsilon^{ijk}\varepsilon^{lmn}}{\sqrt{3}} \left[ s^T_j(x) C\gamma_\alpha s_k(x)u^T_m(x) C\gamma_\mu c_n(x)+2s^T_j(x) C\gamma_\alpha u_k(x)s^T_m(x) C\gamma_\mu c_n(x) \right] \gamma_5\gamma^\alpha C\bar{c}^{T}_{a}(x) \, , \nonumber\\
   J^{11\frac{3}{2}}_{dss,\mu}(x)&=&\frac{\varepsilon^{ila} \varepsilon^{ijk}\varepsilon^{lmn}}{\sqrt{3}} \left[ s^T_j(x) C\gamma_\alpha s_k(x)d^T_m(x) C\gamma_\mu c_n(x)+2s^T_j(x) C\gamma_\alpha d_k(x)s^T_m(x) C\gamma_\mu c_n(x) \right] \gamma_5\gamma^\alpha C\bar{c}^{T}_{a}(x) \, , \nonumber\\
 \end{eqnarray}

\begin{eqnarray}
  J^{11\frac{3}{2}}_{sss,\mu}(x)&=&\varepsilon^{ila} \varepsilon^{ijk}\varepsilon^{lmn}  s^T_j(x) C\gamma_\alpha s_k(x)s^T_m(x) C\gamma_\mu c_n(x)  \gamma_5\gamma^\alpha C\bar{c}^{T}_{a}(x) \, .
 \end{eqnarray}

We take the isospin limit and classify the  thirty currents  couple potentially  to the hidden-charm pentaquark states with degenerate masses  into the following  twelve types,
\begin{eqnarray}
 && J_{uuu,\mu}^{10\frac{1}{2}}(x)\, , \, \, \, J^{10\frac{1}{2}}_{uud,\mu}(x)\, , \, \, \, J^{10\frac{1}{2}}_{udd,\mu}(x)\, , \, \, \,
J_{ddd,\mu}^{10\frac{1}{2}}(x)\, ; \nonumber\\
&&J^{10\frac{1}{2}}_{uus,\mu}(x)\, , \, \, \, J^{10\frac{1}{2}}_{uds,\mu}(x)\, , \, \, \, J^{10\frac{1}{2}}_{dds,\mu}(x) \, ; \nonumber\\
&&J^{10\frac{1}{2}}_{uss,\mu}(x)\, , \, \, \, J^{10\frac{1}{2}}_{dss,\mu}(x) \, ; \nonumber \\
&&J^{10\frac{1}{2}}_{sss,\mu}(x) \, ; \nonumber\\
&&J^{11\frac{1}{2}}_{uuu,\mu}(x)\, , \, \, \, J^{11\frac{1}{2}}_{uud,\mu}(x)\, , \, \, \, J^{11\frac{1}{2}}_{udd,\mu}(x)\, , \, \, \,
J^{11\frac{1}{2}}_{ddd,\mu}(x) \, ; \nonumber\\
&&J^{11\frac{1}{2}}_{uus,\mu}(x)\, , \, \, \,  J^{11\frac{1}{2}}_{uds,\mu}(x)\, , \, \, \,  J^{11\frac{1}{2}}_{dds,\mu}(x) \, ; \nonumber\\
&&  J^{11\frac{1}{2}}_{uss,\mu}(x)\, , \, \, \,   J^{11\frac{1}{2}}_{dss,\mu}(x)\, ; \nonumber\\
&&  J^{11\frac{1}{2}}_{sss,\mu}(x) \, ; \nonumber\\
&&  J^{11\frac{3}{2}}_{uuu,\mu}(x)\, , \, \, \, J^{11\frac{3}{2}}_{uud,\mu}(x)\, , \, \, \,  J^{11\frac{3}{2}}_{udd,\mu}(x)\, , \, \, \, J^{11\frac{3}{2}}_{ddd,\mu}(x) \, ; \nonumber\\
&& J^{11\frac{3}{2}}_{uus,\mu}(x)\, , \, \, \,  J^{11\frac{3}{2}}_{uds,\mu}(x)\, , \, \, \,  J^{11\frac{3}{2}}_{dds,\mu}(x) \, ; \nonumber\\
&&  J^{11\frac{3}{2}}_{uss,\mu}(x)\, , \, \, \,  J^{11\frac{3}{2}}_{dss,\mu}(x) \, ; \nonumber\\
&&  J^{11\frac{3}{2}}_{sss,\mu}(x) \, ,
 \end{eqnarray}
and perform the operator product expansion up to the vacuum condensates of dimension 10 to obtain the QCD sum rules, the  predictions are presented in Table \ref{mass-NPB-2016-163} in the Appendix.

According to Figs.\ref{fr-D13-fig}-\ref{mass-D10-fig}, the vacuum condensates of the dimensions 11 and 13 play an important role, we should update the old calculations, just like in Ref.\cite{WZG-HC-penta-IJMPA-2020}.

In Ref.\cite{WangZG-Pcs4459-penta-IJMPA-2021}, we extend our previous works \cite{WZG-HC-penta-IJMPA-2020,
WangZG-Penta-EPJC-2016-43} to  explore the possible assignment of the $P_{cs}(4459)$ as the $[ud][sc]\bar{c}$ ($0$, $0$, $0$, $\frac{1}{2}$) pentaquark state with the spin-parity $J^P={\frac{1}{2}}^-$. We choose the current $J(x)$ and carry out the operator product expansion up to the vacuum condensates of dimension 13, where
\begin{eqnarray}
 J(x)&=&\varepsilon^{ila} \varepsilon^{ijk}\varepsilon^{lmn}  u^T_j(x) C\gamma_5 d_k(x)\,s^T_m(x) C\gamma_5 c_n(x)\,  C\bar{c}^{T}_{a}(x) \, .
\end{eqnarray}
In calculations, we take the modified energy scale formula   $\mu=\sqrt{M_{P}-(2{\mathbb{M}}_c)^2}-m_s(\mu)=2.4\,\rm{GeV}$ to constraint the QCD spectral densities.

\begin{figure}
\centering
\includegraphics[totalheight=7cm,width=10cm]{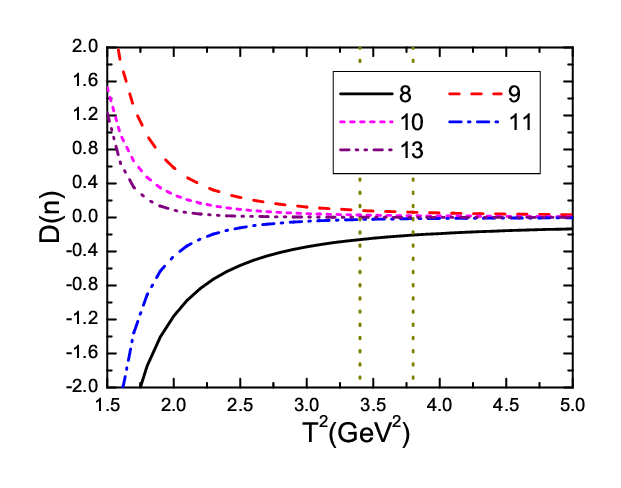}
  \caption{ The  contributions  of the higher dimensional  vacuum condensates $D(n)$ with variation of the  Borel parameter $T^2$, where the $n=8$, $9$, $10$, $11$ and $13$,   the region between the two vertical lines is the Borel window. }\label{OPE-Borel-Pcs4459}
\end{figure}

 We obtain the  Borel  window $T^2=3.4-3.8\,\rm{GeV}^2$ via trial  and error, the pole contribution is about $(40-61)\%$, which is large enough to extract the pentaquark mass reliably.
In Fig.\ref{OPE-Borel-Pcs4459}, we plot the contributions of the higher dimensional  vacuum condensates with variation of the  Borel parameter $T^2$. The higher dimensional  vacuum condensates  manifest themselves at the region $T^2< 2.5\,\rm{GeV}^2$, we should choose the value $T^2> 2.5\,\rm{GeV}^2$. Their values decrease monotonously and quickly with the increase of the Borel parameter, in the Borel window $T^2=3.4-3.8\,\rm{GeV}^2$, the contributions of the higher dimensional vacuum condensates are  $D(8)=-(21-26)\%$, $D(9)=(6-8)\%$, $D(10)=(2-3)\%$, $D(11)=-(1-2)\%$, $D(13)\ll1\%$, the convergent behavior is very good \cite{WangZG-Pcs4459-penta-IJMPA-2021}.

\begin{figure}
\centering
\includegraphics[totalheight=7cm,width=10cm]{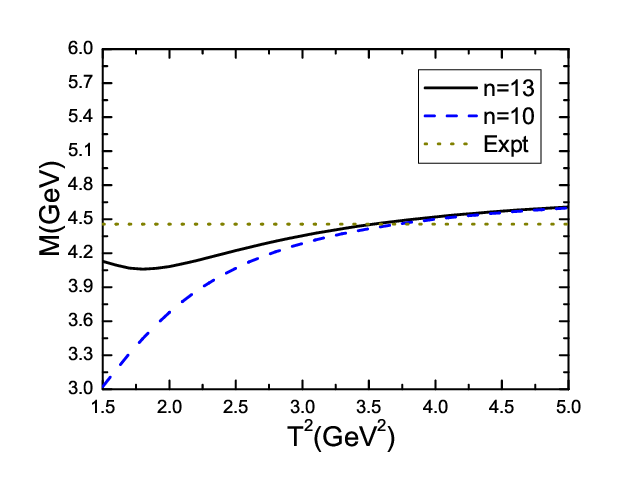}
  \caption{ The predicted  mass  with variation of the  Borel parameter $T^2$, where the $n=10$ and $13$ denote truncations of the operator product expansion, the expt denotes the experimental value of the mass of the $P_{cs}(4459)$. }\label{mass-D10-fig-Pcs4459}
\end{figure}

 In Fig.\ref{mass-D10-fig-Pcs4459}, we plot the predicted pentaquark mass  with variation of the  Borel parameter $T^2$ with the  truncations of the operator product expansion up to the vacuum condensates of dimensions $n=10$ and $13$, respectively.  From the figure, we can see explicitly that the vacuum condensates of dimensions $11$ and $13$ play an important role to obtain the flat platform, we should take them into account in a consistent way, just like in the case of  Ref.\cite{WZG-HC-penta-IJMPA-2020}, where no $s$-quark is present. All in all, the higher dimensional vacuum condensates play an important role in obtaining the flat platform, especially those associated with the inverse Borel parameters  $\frac{1}{T^2}$, $\frac{1}{T^4}$, $\frac{1}{T^6}$ and $\frac{1}{T^8}$.

At last, we obtain   the mass and pole residue \cite{WangZG-Pcs4459-penta-IJMPA-2021},
\begin{eqnarray}
M_P&=&4.47\pm0.11\,\rm{GeV}\, ,\nonumber\\
\lambda_P&=&\left(1.86\pm0.31\right)\times 10^{-3}\,\rm{GeV^6}\, .
\end{eqnarray}
The predicted mass  $M_P=4.47\pm0.11\,\rm{GeV}$ is in excellent agreement with the experimental data $ 4458.8 \pm 2.9 {}^{+4.7}_{-1.1} \mbox{ MeV}$ from the  LHCb collaboration  \cite{LHCb-Pcs4459}, and supports assigning the $P_{cs}(4459)$   as the $[ud][sc]\bar{c}$ ($0$, $0$, $0$, $\frac{1}{2}$) pentaquark state
with the spin-parity $J^P={\frac{1}{2}}^-$. In Ref.\cite{WZG-HC-penta-IJMPA-2020}, we observe that the $P_c(4312)$ can be assigned to be the $[ud][uc]\bar{c}$ ($0$, $0$, $0$, $\frac{1}{2}$) pentaquark state with the spin-parity $J^P={\frac{1}{2}}^-$. The light-flavor  $SU(3)$ mass-breaking effect is about $\Delta=m_s=147\,\rm{MeV}$, the estimations presented in Table \ref{mass-1508-et al-Pcs} are reasonable and reliable, where we take the experimental value of the mass of the $P_{cs}(4459)$. In Refs.\cite{WangZG-Penta-EPJC-2016-142,WangZG-Penta-NPB-2016-163}, we study the $J^P={\frac{1}{2}}^\pm$ and ${\frac{3}{2}}^\pm$ hidden-charm pentaquark states with the strangeness $S=0$, $-1$, $-2$ and $-3$ in a systematic way by carrying out the operator product expansion up to the vacuum condensates of dimension $10$ and choosing the old value ${\mathbb{M}}_c=1.80\,\rm{GeV}$, and observe that the light-flavor  $SU(3)$ mass-breaking effects  are $\Delta=(90-130)\,\rm{MeV}$ for the negative-parity pentaquark states. The new analysis supports a larger light-flavor  $SU(3)$ mass-breaking effect.

\begin{table}
\begin{center}
\begin{tabular}{|c|c|c|c|c|c|c|c|c|}\hline\hline
$[qq^\prime][q^{\prime\prime}c]\bar{c}$ ($S_L$, $S_H$, $J_{LH}$, $J$) &New analysis \cite{WZG-HC-penta-IJMPA-2020}    &$u\,{\rm or}\,d \to s$  &Assignments        \\ \hline

$[ud][uc]\bar{c}$ ($0$, $0$, $0$, $\frac{1}{2}$)                      &$4.31\pm0.11$   &$4.46\pm0.11$           &$?\,P_{cs}(4459)$         \\

$[ud][uc]\bar{c}$ ($0$, $1$, $1$, $\frac{1}{2}$)                      &$4.45\pm0.11$   &$4.60\pm0.11$           &     \\

$[uu][dc]\bar{c}+2[ud][uc]\bar{c}$ ($1$, $0$, $1$, $\frac{1}{2}$)     &$4.46\pm0.11$   &$4.61\pm0.11$           &      \\

$[uu][dc]\bar{c}+2[ud][uc]\bar{c}$ ($1$, $1$, $0$, $\frac{1}{2}$)     &$4.34\pm0.14$   &$4.49\pm0.14$           &      \\

$[ud][uc]\bar{c}$ ($0$, $1$, $1$, $\frac{3}{2}$)                      &$4.39\pm0.11$   &$4.54\pm0.11$           &       \\

$[uu][dc]\bar{c}+2[ud][uc]\bar{c}$ ($1$, $0$, $1$, $\frac{3}{2}$)     &$4.47\pm0.11$   &$4.62\pm0.11$           &    \\

$[uu][dc]\bar{c}+2[ud][uc]\bar{c}$ ($1$, $1$, $2$, $\frac{3}{2}$)     &$4.61\pm0.11$   &$4.76\pm0.11$           &                      \\

$[uu][dc]\bar{c}+2[ud][uc]\bar{c}$ ($1$, $1$, $2$, $\frac{3}{2}$)     &$4.52\pm0.11$   &$4.67\pm0.11$           &                     \\

$[uu][dc]\bar{c}+2[ud][uc]\bar{c}$ ($1$, $1$, $2$, $\frac{5}{2}$)     &$4.39\pm0.11$   &$4.54\pm0.11$           &      \\

$[ud][uc]\bar{c}$ ($0$, $1$, $1$, $\frac{5}{2}$)                      &$4.39\pm0.11$   &$4.54\pm0.11$           &     \\ \hline\hline
\end{tabular}
\end{center}
\caption{ The masses (in unit of GeV) of the pentaquark states with the strangeness $S=-1$, where the $S_L$ and $S_H$ denote the spins of the light and heavy diquarks respectively, $\vec{J}_{LH}=\vec{S}_L+\vec{S}_H$, $\vec{J}=\vec{J}_{LH}+\vec{J}_{\bar{c}}$, the $\vec{J}_{\bar{c}}$ is the angular momentum of the $\bar{c}$-quark \cite{WangZG-Pcs4459-penta-IJMPA-2021}.  }\label{mass-1508-et al-Pcs}
\end{table}

We can extend this section directly to study the ${\bar{\mathbf 3}}{\bar{\mathbf 3}}{\bar{\mathbf 3}}$ type triply-heavy pentaquark states \cite{WangZG-Penta-ccc-EPJC-2018}.

\subsection{$\mathbf 1\mathbf 1 $ type  pentaquark states}\label{11-penta-Sect}
In Ref.\cite{ZhuSL-mole-penta-PRL-2015}, Chen et al construct the color singlet-singlet type currents,
\begin{eqnarray}\label{Chen-32}
J^{\bar D^*\Sigma_c}_{\mu}(x) &=& \bar{c}(x) \gamma_\mu d(x)\,  \varepsilon^{ijk} u^T_i(x) C \gamma_\nu u_j(x) \gamma^\nu \gamma_5 c_k(x)  \, ,\nonumber\\
J^{\bar D\Sigma_c^*}_{\mu} (x)&=& \bar{c}(x) \gamma_5 d(x) \, \varepsilon^{ijk} u^T_i(x) C \gamma_\mu u_j(x) c_k(x)  \, ,
\end{eqnarray}
\begin{eqnarray}\label{Chen-52}
J^{\bar D^*\Sigma_c^*}_{\mu\nu}(x) &=& \bar{c}(x) \gamma_\mu d(x) \, \varepsilon^{ijk} u^T_i(x) C \gamma_\nu u_j(x) \gamma_5 c_k(x) + ( \mu \leftrightarrow \nu ) \, , \nonumber\\
J^{\bar D\Sigma_c^*}_{\mu\nu}(x) &=& \bar{c}(x) \gamma_\mu \gamma_5 d(x)\, \varepsilon^{ijk} u^T_i(x) C \gamma_\nu u_j(x) c_k(x) + ( \mu \leftrightarrow \nu ) \, ,\nonumber\\
 J^{\bar D^*\Lambda_c}_{\mu\nu}(x) &=& \bar{c}(x) \gamma_\mu u(x)\,\varepsilon^{ijk} u^T_i(x) C \gamma_\nu \gamma_5 d_j(x) c(x) + ( \mu \leftrightarrow \nu ) \, ,
\end{eqnarray}
for the first time, and take the currents $J^{\bar D^*\Sigma_c}_{\mu}(x)$ and $J^{\bar D^*\Sigma_c^*}_{\mu\nu}(x)$ to study the $P_c(4380)$ and $P_c(4450)$ by carrying out the operator product expansion up to the vacuum condensates of dimension $8$. Thereafter, the $\mathbf 1\mathbf 1 $ type currents were used to interpolate the pentaquark molecular states \cite{WangZG-Regge-Ms-CPC-2021,WangZG-Scheme-1-2-3-IJMPA-2019,HXChen-mole-Pc-EPJC-2016,Azizi-mole-Pc4380-PRD-2017,Azizi-mole-Pc4380-EPJC-2018,
HXChen-mole-Pc4312-PRD-2019,JRZhang-Pc4312-mole-EPJC-2019,
HXChen-mole-Pcs-EPJC-2021,Azizi-mole-Pc412-CPC-2021,Penta-mole-No-Iso-Azizi-PRD-2017,
Penta-mole-No-Iso-ChenHX-CPC-2019}. However, in those works, the isospins of the currents are not specified and should be improved, as the two-body strong decays to the final states $J/\psi p$ and $J/\psi \Lambda$ conserve isospins.

The $u$ and $d$ quarks have the isospin $I=\frac{1}{2}$, i.e. $\widehat{I}u=\frac{1}{2}u$ and $\widehat{I}d=-\frac{1}{2}d$, where the $\widehat{I}$ is the isospin operator. Then the $\bar{D}^0$, $\bar{D}^{*0}$, $\bar{D}^-$, $\bar{D}^{*-}$, $\bar{D}_s^-$, $\bar{D}_s^{*-}$, $\Sigma_c^+$, $\Sigma_c^{*+}$, $\Sigma_c^{++}$, $\Sigma_c^{*++}$,
$\Xi_c^{\prime 0}$, $\Xi_c^{*0}$, $\Xi_c^{\prime +}$, $\Xi_c^{*+}$, $\Xi_c^{0}$, $\Xi_c^{+}$ and $\Lambda_c^{ +}$
 correspond to the eigenstates $|\frac{1}{2},\frac{1}{2}\rangle$, $|\frac{1}{2},\frac{1}{2}\rangle$, $|\frac{1}{2},-\frac{1}{2}\rangle$, $|\frac{1}{2},-\frac{1}{2}\rangle$, $|0,0\rangle$, $|0,0\rangle$,  $|1,0\rangle$, $|1,0\rangle$, $|1,1\rangle$, $|1,1\rangle$, $|\frac{1}{2},-\frac{1}{2}\rangle$, $|\frac{1}{2},-\frac{1}{2}\rangle$, $|\frac{1}{2},\frac{1}{2}\rangle$, $|\frac{1}{2},\frac{1}{2}\rangle$,
 $ |\frac{1}{2},-\frac{1}{2}\rangle$, $|\frac{1}{2},\frac{1}{2}\rangle$ and $|0,0\rangle$ in the isospin space $|I,I_3\rangle$, respectively.
We construct the color-singlet currents to interpolate them,
\begin{eqnarray}
J^{\bar{D}^{0/-}}(x)&=&\bar{c}(x)i\gamma_5u/d(x)\, ,\nonumber \\
J^{\bar{D}_s^-}(x)&=&\bar{c}(x)i\gamma_5s(x)\, ,\nonumber \\
J^{\bar{D}^{*0/-}}_{\mu}(x)&=&\bar{c}(x)\gamma_{\mu} u/d(x)\, , \nonumber\\
J^{\bar{D}_s^{*-}}_{\mu}(x)&=&\bar{c}(x)\gamma_\mu s(x)\, ,
\end{eqnarray}

\begin{eqnarray}
J^{\Sigma_c^+}(x)&=&\varepsilon^{ijk}u^{T}_i(x)C\gamma_{\mu}d_j(x)\gamma^{\mu}\gamma_5c_k(x)\, ,\nonumber\\
J^{\Sigma_c^{++}}(x)&=&\varepsilon^{ijk}u^{T}_i(x)C\gamma_{\mu}u_j(x)\gamma^{\mu}\gamma_5c_k(x)\, , \nonumber\\
J^{\Sigma_c^{*+}}_{\mu}(x)&=&\varepsilon^{ijk}u^{T}_i(x)C\gamma_{\mu}d_j(x)c_k(x)\, , \nonumber\\
J^{\Sigma_c^{*++}}_{\mu}(x)&=&\varepsilon^{ijk}u^{T}_i(x)C\gamma_{\mu}u_j(x)c_k(x)\, ,
\end{eqnarray}

\begin{eqnarray}
J^{\Xi_c^{\prime 0}}(x)&=&\varepsilon^{ijk}d^{T}_i(x)C\gamma_{\mu}s_j(x)\gamma^{\mu}\gamma_5c_k(x)\, ,\nonumber\\
J^{\Xi_c^{*0}}_{\mu}(x)&=&\varepsilon^{ijk}d^{T}_i(x)C\gamma_{\mu}s_j(x)c_k(x)\, , \nonumber\\
J^{\Xi_c^{\prime +}}(x)&=&\varepsilon^{ijk}u^{T}_i(x)C\gamma_{\mu}s_j(x)\gamma^{\mu}\gamma_5c_k(x)\, , \nonumber\\
J^{\Xi_c^{*+}}_{\mu}(x)&=&\varepsilon^{ijk}u^{T}_i(x)C\gamma_{\mu}s_j(x)c_k(x)\, ,
\end{eqnarray}

\begin{eqnarray}
J^{\Xi_c^{ 0}}(x)&=&\varepsilon^{ijk}d^{T}_i(x)C\gamma_{5}s_j(x) c_k(x)\, ,\nonumber\\
J^{\Xi_c^{ +}}(x)&=&\varepsilon^{ijk}u^{T}_i(x)C\gamma_{5}s_j(x) c_k(x)\, ,\nonumber\\
J^{\Lambda_c^{ +}}(x)&=&\varepsilon^{ijk}u^{T}_i(x)C\gamma_{5}d_j(x) c_k(x)\, .
\end{eqnarray}

Accordingly, we construct the $\mathbf 1\mathbf 1 $  type five-quark currents to interpolate the $\bar{D}^{(*)}\Sigma_c^{(*)}$, $\bar{D}^{(*)}\Xi_c^{(*)}$,  $\bar{D}^{(*)}\Xi_c^{\prime}$  and $\bar{D}^{(*)}\Lambda_c$ type pentaquark  sates,
where the $\bar{D}^{(*)}$, $\Sigma_c^{(*)}$, $\Xi_c^{(*)}$, $\Xi_c^{\prime}$  and $\Lambda_c$ represent the color-singlet clusters  having the same quantum numbers as the physical states  $\bar{D}^{(*)}$, $\Sigma_c^{(*)}$, $\Xi_c^{(*)}$,  $\Xi_c^{\prime}$  and $\Lambda_c$ respectively except for the masses,
\begin{eqnarray}
J(x)&=&J_{\frac{1}{2}}^{\bar{D}\Sigma_c}(x)\, , \,\,\,J_{\frac{3}{2}}^{\bar{D}\Sigma_c}(x)\, , \,\,\,J_{0}^{\bar{D}\Xi_c^{\prime}}(x)\, , \,\,\,J_{1}^{\bar{D}\Xi_c^{\prime}}(x)\, , \,\,\,\nonumber\\
&&J_{0}^{\bar{D}\Xi_c}(x)\, , \,\,\, J_{1}^{\bar{D}\Xi_c}(x)\, , \,\,\, J_{\frac{1}{2}}^{\bar{D}\Lambda_c}(x)\, , \,\,\, J_{\frac{1}{2}}^{\bar{D}_s\Xi_c}(x)\, , \, \,\,J_{0}^{\bar{D}_s\Lambda_c}(x)\, ,\nonumber \\
J_{\mu} (x)&=&J_{\frac{1}{2};\mu}^{\bar{D}\Sigma_c^*}(x)\, , \,\,\, J_{\frac{3}{2};\mu}^{\bar{D}\Sigma_c^*}(x)\, ,\,\,\, J_{\frac{1}{2};\mu}^{\bar{D}^{*}\Sigma_c}(x)\,, \,\,\,J_{\frac{3}{2};\mu}^{\bar{D}^{*}\Sigma_c}(x)\, , \,\,\,\nonumber\\
&&J_{0;\mu}^{\bar{D}\Xi_c^*}(x)\, , \,\,\, J_{1;\mu}^{\bar{D}\Xi_c^*}(x)\, ,\,\,\, J_{0;\mu}^{\bar{D}^{*}\Xi_c^{\prime}}(x)\,, \,\,\,J_{1;\mu}^{\bar{D}^{*}\Xi_c^{\prime}}(x)\, , \,\,\,\nonumber\\
&&J_{0;\mu}^{\bar{D}^*\Xi_c}(x)\, , \,\,\,J_{1;\mu}^{\bar{D}^*\Xi_c}(x)\, , \,\,\,J_{\frac{1}{2};\mu}^{\bar{D}^*\Lambda_c}(x)\, , \,\,\,J_{\frac{1}{2};\mu}^{\bar{D}_s^*\Xi_c}(x)\, , \,\,\, J_{0;\mu}^{\bar{D}_s^*\Lambda_c}(x)\, ,\nonumber\\
 J_{\mu\nu} (x)&=&J_{\frac{1}{2};\mu\nu}^{\bar{D}^{*}\Sigma_c^*}(x)\, , \, \,\, J_{\frac{3}{2};\mu\nu}^{\bar{D}^{*}\Sigma_c^*}(x)\, ,\,\,\,J_{0;\mu\nu}^{\bar{D}^{*}\Xi_c^*}(x)\, , \, \,\, J_{1;\mu\nu}^{\bar{D}^{*}\Xi_c^*}(x)\, ,
  \end{eqnarray}

\begin{eqnarray}\label{J-12-32-DSigma}
J_{\frac{1}{2}}^{\bar{D}\Sigma_c}(x)&=&\frac{1}{\sqrt{3}}J^{\bar{D}^0}(x)J^{\Sigma_c^+}(x)-\sqrt{\frac{2}{3}}J^{\bar{D}^-}(x)J^{\Sigma_c^{++}}(x) \, , \nonumber\\
J_{\frac{3}{2}}^{\bar{D}\Sigma_c}(x)&=&\sqrt{\frac{2}{3}}J^{\bar{D}^0}(x)J^{\Sigma_c^+}(x)+\frac{1}{\sqrt{3}}J^{\bar{D}^-}(x)J^{\Sigma_c^{++}}(x)\, ,\nonumber\\
J_{\frac{1}{2};\mu}^{\bar{D}\Sigma_c^*}(x)&=&\frac{1}{\sqrt{3}}J^{\bar{D}^0}(x)J^{\Sigma_c^{*+}}_{\mu}(x)-\sqrt{\frac{2}{3}}J^{\bar{D}^-}(x)J^{\Sigma_c^{*++}}_{\mu}(x)\, ,\nonumber\\
J_{\frac{3}{2};\mu}^{\bar{D}\Sigma_c^*}(x)&=&\sqrt{\frac{2}{3}}J^{\bar{D}^0}(x)J^{\Sigma_c^{*+}}_{\mu}(x)+\frac{1}{\sqrt{3}}J^{\bar{D}^-}(x)J^{\Sigma_c^{*++}}_{\mu}(x)\, ,
\end{eqnarray}

\begin{eqnarray}
J_{\frac{1}{2};\mu}^{\bar{D}^{*}\Sigma_c}(x)&=&\frac{1}{\sqrt{3}}J^{\bar{D}^{*0}}_{\mu}(x)J^{\Sigma_c^+}(x)-\sqrt{\frac{2}{3}}J^{\bar{D}^{*-}}_{\mu}(x)J^{\Sigma_c^{++}}(x)\, ,\nonumber\\
J_{\frac{3}{2};\mu}^{\bar{D}^{*}\Sigma_c}(x)&=&=\sqrt{\frac{2}{3}}J^{\bar{D}^{*0}}_{\mu}(x)J^{\Sigma_c^+}(x)+\frac{1}{\sqrt{3}}J^{\bar{D}^{*-}}_{\mu}(x)J^{\Sigma_c^{++}}(x)\, ,\nonumber\\
J_{\frac{1}{2};\mu\nu}^{\bar{D}^{*}\Sigma_c^*}(x)&=&\frac{1}{\sqrt{3}}J^{\bar{D}^{*0}}_{\mu}(x)J^{\Sigma_c^{*+}}_{\nu}(x)-\sqrt{\frac{2}{3}}J^{\bar{D}^{*-}}_{\mu}(x)J^{\Sigma_c^{*++}}_{\nu}(x)+(\mu\leftrightarrow\nu)\, ,\nonumber\\
J_{\frac{3}{2};\mu\nu}^{\bar{D}^{*}\Sigma_c^*}(x)&=&\sqrt{\frac{2}{3}}J^{\bar{D}^{*0}}_{\mu}(x)J^{\Sigma_c^{*+}}_{\nu}(x)+\frac{1}{\sqrt{3}}J^{\bar{D}^{*-}}_{\mu}(x)J^{\Sigma_c^{*++}}_{\nu}(x)+(\mu\leftrightarrow\nu)\, ,
\end{eqnarray}

\begin{eqnarray}
J_{0}^{\bar{D}\Xi_c^{\prime}}(x)&=&\frac{1}{\sqrt{2}}J^{\bar{D}^0}(x)J^{\Xi_c^{\prime 0}}(x)-\frac{1}{\sqrt{2}}J^{\bar{D}^-}(x)J^{\Xi_c^{\prime +}}(x) \, , \nonumber\\
J_{1}^{\bar{D}\Xi_c^{\prime}}(x)&=&\frac{1}{\sqrt{2}}J^{\bar{D}^0}(x)J^{\Xi_c^{\prime0}}(x)+\frac{1}{\sqrt{2}}J^{\bar{D}^-}(x)J^{\Xi_c^{\prime+}}(x)\, ,\nonumber\\
J_{0;\mu}^{\bar{D}\Xi_c^*}(x)&=&\frac{1}{\sqrt{2}}J^{\bar{D}^0}(x)J^{\Xi_c^{*0}}_{\mu}(x)-\frac{1}{\sqrt{2}}J^{\bar{D}^-}(x)J^{\Xi_c^{*+}}_{\mu}(x)\, ,\nonumber\\
J_{1;\mu}^{\bar{D}\Xi_c^*}(x)&=&\frac{1}{\sqrt{2}}J^{\bar{D}^0}(x)J^{\Xi_c^{*0}}_{\mu}(x)+\frac{1}{\sqrt{2}}J^{\bar{D}^-}(x)J^{\Xi_c^{*+}}_{\mu}(x)\, ,
\end{eqnarray}

\begin{eqnarray}
J_{0;\mu}^{\bar{D}^{*}\Xi_c^{\prime}}(x)&=&\frac{1}{\sqrt{2}}J^{\bar{D}^{*0}}_{\mu}(x)J^{\Xi_c^{\prime0}}(x)-\frac{1}{\sqrt{2}}J^{\bar{D}^{*-}}_{\mu}(x)J^{\Xi_c^{\prime+}}(x)\, ,\nonumber\\
J_{1;\mu}^{\bar{D}^{*}\Xi_c^{\prime}}(x)&=&\frac{1}{\sqrt{2}}J^{\bar{D}^{*0}}_{\mu}(x)J^{\Xi_c^{\prime0}}(x)+\frac{1}{\sqrt{2}}J^{\bar{D}^{*-}}_{\mu}(x)J^{\Xi_c^{\prime+}}(x)\, ,\nonumber\\
J_{0;\mu\nu}^{\bar{D}^{*}\Xi_c^*}(x)&=&\frac{1}{\sqrt{2}}J^{\bar{D}^{*0}}_{\mu}(x)J^{\Xi_c^{*0}}_{\nu}(x)-\frac{1}{\sqrt{2}}J^{\bar{D}^{*-}}_{\mu}(x)J^{\Xi_c^{*+}}_{\nu}(x)+(\mu\leftrightarrow\nu)\, ,\nonumber\\
J_{1;\mu\nu}^{\bar{D}^{*}\Xi_c^*}(x)&=&\frac{1}{\sqrt{2}}J^{\bar{D}^{*0}}_{\mu}(x)J^{\Xi_c^{*0}}_{\nu}(x)+\frac{1}{\sqrt{2}}J^{\bar{D}^{*-}}_{\mu}(x)J^{\Xi_c^{*+}}_{\nu}(x)+(\mu\leftrightarrow\nu)\, ,
\end{eqnarray}

\begin{eqnarray}
J_{0}^{\bar{D}\Xi_c}(x)&=&\frac{1}{\sqrt{2}}J^{\bar{D}^0}(x)J^{\Xi_c^{0}}(x)-\frac{1}{\sqrt{2}}J^{\bar{D}^-}(x)J^{\Xi_c^{+}}(x) \, , \nonumber\\
J_{1}^{\bar{D}\Xi_c}(x)&=&\frac{1}{\sqrt{2}}J^{\bar{D}^0}(x)J^{\Xi_c^{0}}(x)+\frac{1}{\sqrt{2}}J^{\bar{D}^-}(x)J^{\Xi_c^{+}}(x) \, , \nonumber\\
J_{\frac{1}{2}}^{\bar{D}\Lambda_c}(x)&=&J^{\bar{D}^0}(x)J^{\Lambda_c^{+}}(x) \, , \nonumber\\
J_{\frac{1}{2}}^{\bar{D}_s\Xi_c}(x)&=&J^{\bar{D}^-_s}(x)J^{\Xi_c^{+}}(x) \, , \nonumber\\
J_{0}^{\bar{D}_s\Lambda_c}(x)&=&J^{\bar{D}^{-}_s}(x)J^{\Lambda_c^{+}}(x) \, ,
\end{eqnarray}

\begin{eqnarray}
J_{0;\mu}^{\bar{D}^*\Xi_c}(x)&=&\frac{1}{\sqrt{2}}J_\mu^{\bar{D}^{*0}}(x)J^{\Xi_c^{0}}(x)
-\frac{1}{\sqrt{2}}J_\mu^{\bar{D}^{*-}}(x)J^{\Xi_c^{+}}(x) \, , \nonumber\\
J_{1;\mu}^{\bar{D}^*\Xi_c}(x)&=&\frac{1}{\sqrt{2}}J_\mu^{\bar{D}^{*0}}(x)J^{\Xi_c^{0}}(x)
+\frac{1}{\sqrt{2}}J_\mu^{\bar{D}^{*-}}(x)J^{\Xi_c^{+}}(x) \, , \nonumber\\
J_{\frac{1}{2};\mu}^{\bar{D}^*\Lambda_c}(x)&=&J_\mu^{\bar{D}^{*0}}(x)J^{\Lambda_c^{+}}(x) \, , \nonumber\\
J_{\frac{1}{2};\mu}^{\bar{D}^*_s\Xi_c}(x)&=&J_\mu^{\bar{D}^{*-}_s}(x)J^{\Xi_c^{+}}(x) \, , \nonumber\\
J_{0;\mu}^{\bar{D}^*_s\Lambda_c}(x)&=&J_\mu^{\bar{D}^{*-}_s}(x)J^{\Lambda_c^{+}}(x) \, ,
\end{eqnarray}
 the subscripts $\frac{1}{2}$, $\frac{3}{2}$, $0$ and $1$ represent the isospins $I$ \cite{WangZG-Regge-Ms-CPC-2021,WangXW-penta-mole-SCPMA-2022,
 WangXW-penta-mole-IJMPA-2022,WangXW-penta-mole-CPC-2023}.  They are the isospin eigenstates $|I,I_3\rangle=$
$|\frac{1}{2},\frac{1}{2}\rangle$, $|\frac{3}{2},\frac{1}{2}\rangle$, $|1,0\rangle$ or $|0,0\rangle$.

For example, from Eq.\eqref{J-12-32-DSigma}, we obtain the relations,
\begin{eqnarray}
J^{\bar{D}^0}(x)J^{\Sigma_c^+}(x)&=&\frac{1}{\sqrt{3}}J_{\frac{1}{2}}^{\bar{D}\Sigma_c}(x)
+\sqrt{\frac{2}{3}}J_{\frac{3}{2}}^{\bar{D}\Sigma_c}(x)\, ,\nonumber\\
J^{\bar{D}^-}(x)J^{\Sigma_c^{++}}(x)&=&\frac{1}{\sqrt{3}}J_{\frac{3}{2}}^{\bar{D}\Sigma_c}(x)
-\sqrt{\frac{2}{3}}J_{\frac{1}{2}}^{\bar{D}\Sigma_c}(x)\, ,
\end{eqnarray}
the currents $J^{\bar{D}^0}(x)J^{\Sigma_c^+}(x)$ and $ J^{\bar{D}^-}(x)J^{\Sigma_c^{++}}(x)$ have both the isospin  $(I,I_3)=(\frac{1}{2},\frac{1}{2})$ and $(\frac{3}{2},\frac{1}{2})$ components, and couple potentially to the
pentaquark molecular states with the isospins $(\frac{1}{2},\frac{1}{2})$ and $(\frac{3}{2},\frac{1}{2})$, which decay to the final states $J/\psi p$ and $J/\psi \Delta^+$, respectively. As the $P_c(4312)$, $P_c(4380)$, $P_c(4440)$, $P_c(4457)$ and $P_c(4337)$ are observed in the $J/\psi p$ mass spectrum, it is better to choose the current $J_{\frac{1}{2}}^{\bar{D}\Sigma_c}(x)$  with the definite isospin, we prefer the color singlet-singlet type currents with the definite isospins \cite{WangZG-Regge-Ms-CPC-2021,WangXW-penta-mole-SCPMA-2022,
WangXW-penta-mole-IJMPA-2022,WangXW-penta-mole-CPC-2023}, and thereafter those currents are adopted in  Refs.\cite{Penta-mole-Iso-Ozdem-PLB-2023,Penta-mole-Iso-Ozdem-PLB-2023-2,
Penta-mole-Iso-Ozdem-EPJC-2024}.
Although the mass splitting  between the isospin  cousins is of several 10-MeV in most cases, in some cases, the mass splitting can be as large as 150 MeV, see Table \ref{mass-residue-penta-mole}. Phenomenologically, the molecule-type $P$ states have been studied extensively with the help of heavy quark symmetry \cite{One-pion-Hosaka-PRD-2017,BSE-Oset-penta-mole-PRD-2015,
Quasi-BSE-HeJ-Pc-PLB-2016,Quasi-BSE-HeJ-Pc-EPJC-2019,LSE-momentum-GengLS-PRL-2019,
OBE-penta-mole-WangBo-PRD-2019,OBE-penta-mole-WangBo-JHEP-2019,
OBE-penta-mole-GengLS-PRD-2021,
Penta-mole-SLZhu-PRL-2015,Penta-mole-FKGuo-PRD-2015,
Penta-mole-ChenR-PRD-2019,Penta-mole-GengLS-PRD-2019,
Penta-mole-Burns-EPJA-2015,Penta-mole-FKGuo-PRD-2019,
Penta-mole-HXChen-PRC-2016,Penta-mole-Burns-PRD-2019,
Penta-mole-Hosaka-PRD-2020,Penta-mole-Shimizu-PRD-2016,
Penta-mole-Yamaguchi-PRD-2017,Penta-mole-MLDu-PRL-2020,Penta-mole-MLDu-JHEP-2021}, it is more easy to apply isospin eigenstates in the effective field theory than in QCD.

We resort to the correlation functions $\Pi(p)$, $\Pi_{\mu\nu}(p)$ and $\Pi_{\mu\alpha\beta}(p)$ in Eq.\eqref{CF-Pi-Pi-Pi}, and perform analogous analysis to obtain the QCD sum rules for the masses and pole residues like Eqs.\eqref{QCDSR-penta}-\eqref{QCDSR-penta-masses},
\begin{eqnarray}\label{QCDSR-penta-mole}
2M_{-}\lambda^{-}_j{}^2\exp\left( -\frac{M_{-}^2}{T^2}\right)&=& \int_{4m_c^2}^{s_0}ds \,\rho_{QCD,j}(s)\,\exp\left( -\frac{s}{T^2}\right)\,  ,
\end{eqnarray}
\begin{eqnarray}\label{QCDSR-penta-mole-masses}
 M^2_{-} &=& \frac{-\int_{4m_c^2}^{s_0}ds \frac{d}{d(1/T^2)}\, \rho_{QCD,j}(s)\,\exp\left( -\frac{s}{T^2}\right)}{\int_{4m_c^2}^{s_0}ds \, \rho_{QCD,j}(s)\,\exp\left( -\frac{s}{T^2}\right)}\,  ,
\end{eqnarray}
where $j=\frac{1}{2}$, $\frac{3}{2}$ and $\frac{5}{2}$, the pole residues are defined analogous to Eqs.\eqref{Coupling12}-\eqref{Coupling52}.

As we study the $\mathbf 1\mathbf 1 $  type  pentaquark  states,  it is better to  choose the updated values  $\mathbb{M}_c=1.85\pm0.01 \,\rm GeV$  and $\mathbb{M}_s=0.2\,\rm{GeV}$ to determine the optimal energy scales of the QCD spectral densities with
the formula \cite{WZG-Y4220-mole-CPC-2017},
\begin{eqnarray}
\mu=\sqrt{M_{X/Y/Z/P}^2-(2\mathbb{M}_c)^2}-k\,\mathbb{M}_s\, .
\end{eqnarray}

After trial and error,  we obtain  the Borel windows, continuum threshold parameters, energy scales and pole contributions,  see Table \ref{BorelP-pole-penta-mole} \cite{WangXW-penta-mole-SCPMA-2022,
 WangXW-penta-mole-IJMPA-2022,WangXW-penta-mole-CPC-2023},  the pole contributions  are about  or slightly larger than $(40-60)\%$,  we obtain the {\bf largest  pole contributions} up to now.
In the Borel windows,  the highest dimensional condensate contributions $|D(12)|$ and $|D(13)|$ are approximately zero,  the most important contributions are mainly from the lowest order contributions $\langle \bar{q}q\rangle$, $\langle \bar{q}q\rangle^2$ and $\langle\bar qg_s\sigma Gq\rangle\langle\bar qq\rangle$. The  operator product expansion converges very well.

Then we calculate the uncertainties of the  masses and pole residues, which are shown in  Table \ref{mass-residue-penta-mole} \cite{WangXW-penta-mole-SCPMA-2022,
 WangXW-penta-mole-IJMPA-2022,WangXW-penta-mole-CPC-2023}.
The central value of the mass of the $\bar{D}\Sigma_c$ molecular    state with the $IJ^P=\frac{1}{2}\frac{1}{2}^-$  is $4.31$ $\rm{GeV}$, it is only about $10$ $\rm{MeV}$ below the $\bar{D}^0\Sigma_c^+$ threshold,  we assign it as the $P_c(4312)$ tentatively. For the $\bar{D}\Sigma_c$ molecular state with the $IJ^P=\frac{3}{2}\frac{1}{2}^-$, the central value of the mass is $4.33$ $\rm{GeV}$, which is about $10$ $\rm{MeV}$ above the  $\bar{D}^-\Sigma_c^{++}$ threshold,   we  tentatively assign it as the $\bar{D}\Sigma_c$ resonance, the isospin cousin of the $P_c(4312)$.

In a similar way,  we  assign the  $P_c(4380)$, $P_c(4440)$ and $P_c(4457)$ as the  $\bar{D}\Sigma_c^*$, $\bar{D}^*\Sigma_c$ and $\bar{D}^*\Sigma_c^*$ molecular  states with the  $IJ^P=\frac{1}{2}\frac{3}{2}^-$,  $\frac{1}{2}\frac{3}{2}^-$ and   $\frac{1}{2}\frac{5}{2}^-$, respectively. For the molecular  states (resonances) $\bar{D}\Sigma_c^*$ with the $IJ^P=\frac{3}{2}\frac{3}{2}^-$, $\bar{D}^*\Sigma_c$ with the $IJ^P=\frac{3}{2}\frac{3}{2}^-$ and $\bar{D}^*\Sigma_c^*$ with the $IJ^P=\frac{3}{2}\frac{5}{2}^-$, the central values of the  masses are about $20$ $\rm{MeV}$, $10$ $\rm{MeV}$ and $90$ $\rm{MeV}$ above the corresponding meson-baryon thresholds, respectively.

If we choose the same input parameters, the $\mathbf 1\mathbf 1 $  type pentaquark states with the isospin $I=\frac{3}{2}$ ($1$) have slightly larger masses than the corresponding ones with the isospin $I=\frac{1}{2}$ ($0$).

The $P_c(4312)$, $P_c(4380)$, $P_c(4440)$ and $P_c(4457)$ can be assigned as  the  $\bar{D}\Sigma_c$, $\bar{D}\Sigma_c^*$, $\bar{D}^*\Sigma_c$ and $\bar{D}^*\Sigma_c^*$ molecular   states with the isospin $I=\frac{1}{2}$ respectively,  since the two-body strong decays $P_c \to J/\psi p$ conserve isospin. If  the assignments are robust, there exist four slightly higher molecular  states $\bar{D}\Sigma_c$, $\bar{D}\Sigma_c^*$, $\bar{D}^*\Sigma_c$ and $\bar{D}^*\Sigma_c^*$ with the isospin $I=\frac{3}{2}$, we can search for the four resonances in the $J/\psi \Delta^+$  mass spectrum, as the two-body strong decays $P_c \to J/\psi \Delta^+$ also conserve isospin, the $J/\psi$, $p$ and $\Delta$ have the isospins $I=0$, $\frac{1}{2}$ and $\frac{3}{2}$, respectively.

The $\bar{D}^*\Xi_c^{\prime}$ and $\bar{D}^*\Xi_c^{*}$ molecular states lie  about $0.1\,\rm{GeV}$  and  $0.2\,\rm{GeV}$  above the $P_{cs}(4459)$ respectively, the $P_{cs}(4459)$ is unlikely to be the $\bar{D}^*\Xi_c^{\prime}$ or $\bar{D}^*\Xi_c^{*}$ molecular state. The mass of the $\bar{D}\Xi^{\prime}$ molecular state with the isospin $I=1$ is $4.45_{-0.08}^{+0.07}\,\rm{GeV}$, which is near the value $4459\,\rm{MeV}$, but  lies slightly above the corresponding  meson-baryon threshold, it is  a resonance,  from the decay channel $P_{cs}(4459)\to J/\psi \Lambda$ \cite{LHCb-Pcs4459}, the isospin of the  $P_{cs}(4459)$ is zero, which  excludes assigning the $P_{cs}(4459)$ as the $\bar{D}\Xi_c^{*}$ molecular state with the isospin $I=1$. The mass of the $\bar{D}\Xi_c^{*}$ molecular state with the isospin $I=0$ is $4.46^{+0.07}_{-0.07}\,\rm{GeV}$, which is in very good agreement with the  $P_{cs}(4459)$,  it is very good to assign the $P_{cs}(4459)$ as the $\bar{D}\Xi_c^{*}$ molecular state with the isospin $I=0$ and the spin-parity  $J^P={\frac{3}{2}}^-$.  The predications  also favor assigning  the $P_{cs}(4459)$ as the $\bar{D}^*\Xi_c$  molecular state with the spin-parity  $J^P={\frac{3}{2}}^-$ and isospin $I=0$.

\begin{table}
\begin{center}
\begin{tabular}{|c|c|c|c|c|c|c|c|c|c|c|c|c|}\hline\hline
                         &$IJ^P$                        &$T^2({\rm GeV}^2)$   &$\sqrt{s_0}({\rm GeV})$    &$\mu ({\rm GeV})$   &$\rm PC$  \\ \hline

$\bar{D}\Sigma_c$        &$\frac{1}{2}\frac{1}{2}^-$    &$3.2-3.8$  &$5.00\pm0.10$  &$2.2$   &$(42-60)\% $ \\ \hline
$\bar{D}\Sigma_c$        &$\frac{3}{2}\frac{1}{2}^-$    &$2.8-3.4$  &$4.98\pm0.10$  &$2.2$   &$(44-65)\% $ \\ \hline

$\bar{D}\Sigma_c^*$      &$\frac{1}{2}\frac{3}{2}^-$    &$3.3-3.9$  &$5.06\pm0.10$  &$2.3$   &$(42-60)\% $ \\ \hline
$\bar{D}\Sigma_c^*$      &$\frac{3}{2}\frac{3}{2}^-$    &$2.9-3.5$  &$5.03\pm0.10$  &$2.4$   &$(44-64)\% $ \\ \hline

$\bar{D}^*\Sigma_c$      &$\frac{1}{2}\frac{3}{2}^-$    &$3.3-3.9$  &$5.12\pm0.10$  &$2.5$   &$(42-60)\% $ \\ \hline
$\bar{D}^*\Sigma_c$      &$\frac{3}{2}\frac{3}{2}^-$    &$3.0-3.6$  &$5.10\pm0.10$  &$2.5$   &$(41-61)\% $ \\ \hline

$\bar{D}^*\Sigma_c^*$    &$\frac{1}{2}\frac{5}{2}^-$    &$3.2-3.8$  &$5.08\pm0.10$  &$2.5$   &$(43-60)\% $ \\ \hline
$\bar{D}^*\Sigma_c^*$    &$\frac{3}{2}\frac{5}{2}^-$    &$3.0-3.6$  &$5.24\pm0.10$  &$2.8$   &$(42-61)\% $ \\
\hline

$\bar{D}\Xi_c^{\prime}$  &$0{\frac{1}{2}}^-$            &$3.4-4.0$  &$5.12\pm0.10$  &$2.2$   &$(41-58)\% $
\\ \hline
$\bar{D}\Xi_c^{\prime}$  &$1{\frac{1}{2}}^-$            &$3.2-3.8$  &$5.14\pm0.10$  &$2.3$   &$(43-61)\% $
 \\ \hline

$\bar{D}\Xi_c^*$         &$0{\frac{3}{2}}^-$            &$3.4-4.0$  &$5.15\pm0.10$  &$2.3$   &$(43-60)\% $
 \\ \hline
$\bar{D}\Xi_c^*$         &$1{\frac{3}{2}}^-$            &$3.3-3.9$  &$5.22\pm0.10$  &$2.4$   &$(44-62)\% $
\\ \hline

$\bar{D}^*\Xi_c^{\prime}$&$0{\frac{3}{2}}^-$            &$3.5-4.1$  &$5.26\pm0.10$  &$2.5$   &$(42-59)\% $
\\ \hline
$\bar{D}^*\Xi_c^{\prime}$&$1{\frac{3}{2}}^-$            &$3.4-4.0$  &$5.31\pm0.10$  &$2.6$   &$(43-60)\% $
\\ \hline

$\bar{D}^*\Xi_c^*$       &$0{\frac{5}{2}}^-$            &$3.6-4.2$  &$5.31\pm0.10$  &$2.6$   &$(42-58)\% $
\\ \hline
$\bar{D}^*\Xi_c^*$       &$1{\frac{5}{2}}^-$            &$3.4-4.0$  &$5.35\pm0.10$  &$2.6$   &$(44-61)\% $
\\ \hline
$\bar{D}\,\Xi_c$         &$0{\frac{1}{2}}^-$            &$3.2-3.8$  &$5.00\pm0.10$  &2.1     &$(41-60)\%$  \\   \hline

$\bar{D}\,\Xi_c$         &$1{\frac{1}{2}}^-$            &$3.1-3.7$  &$5.09\pm0.10$  &2.3     &$(42-61)\%$  \\   \hline

$\bar{D}\,\Lambda_c$     &$\frac{1}{2}{\frac{1}{2}}^-$  &$3.2-3.8$  &$5.11\pm0.10$  &2.5     &$(42-60)\%$  \\   \hline

$\bar{D}_s\,\Xi_c$       &$\frac{1}{2}{\frac{1}{2}}^-$  &$3.2-3.8$  &$5.15\pm0.10$  &2.2     &$(41-59)\%$  \\   \hline

$\bar{D}_s\,\Lambda_c$   &$0{\frac{1}{2}}^-$            &$3.2-3.8$  &$5.13\pm0.10$  &2.3     &$(43-61)\%$  \\   \hline

$\bar{D}^*\,\Xi_c$       &$0{\frac{3}{2}}^-$            &$3.2-3.8$  &$5.10\pm0.10$  &2.3     &$(43-61)\%$  \\   \hline

$\bar{D}^*\,\Xi_c$       &$1{\frac{3}{2}}^-$            &$3.3-3.9$  &$5.27\pm0.10$  &2.6     &$(43-61)\%$  \\   \hline

$\bar{D}^*\,\Lambda_c$   &$\frac{1}{2}{\frac{3}{2}}^-$  &$3.3-3.9$  &$5.23\pm0.10$  &2.7     &$(41-61)\%$  \\   \hline

$\bar{D}^*_s\,\Xi_c$     &$\frac{1}{2}{\frac{3}{2}}^-$  &$3.3-3.9$  &$5.28\pm0.10$  &2.4     &$(42-59)\%$  \\   \hline

$\bar{D}^*_s\,\Lambda_c$ &$0{\frac{3}{2}}^-$            &$3.2-3.8$  &$5.14\pm0.10$  &2.4     &$(42-60)\%$  \\   \hline
\hline
\end{tabular}
\end{center}
\caption{ The Borel parameters, continuum threshold parameters, energy scales and pole contributions for the $\mathbf 1\mathbf 1 $  type  pentaquark states \cite{WangXW-penta-mole-SCPMA-2022,
 WangXW-penta-mole-IJMPA-2022,WangXW-penta-mole-CPC-2023}. }\label{BorelP-pole-penta-mole}
\end{table}

\begin{table}
\begin{center}
\begin{tabular}{|c|c|c|c|c|c|c|c|c|c|c|c|c|}\hline\hline
                         &$IJ^P$                        &$M({\rm GeV})$  &$\lambda(10^{-3}{\rm GeV}^6)$     & Assignments  &Thresholds (MeV)\\ \hline

$\bar{D}\Sigma_c$        &$\frac{1}{2}\frac{1}{2}^-$    &$4.31^{+0.07}_{-0.07}$  &$3.25^{+0.43}_{-0.41}$   &$P_c(4312)$  &$4321$ \\ \hline
$\bar{D}\Sigma_c$        &$\frac{3}{2}\frac{1}{2}^-$    &$4.33^{+0.09}_{-0.08}$  &$1.97^{+0.28}_{-0.26}$
&             &$4321$ \\ \hline

$\bar{D}\Sigma_c^*$      &$\frac{1}{2}\frac{3}{2}^-$    &$4.38^{+0.07}_{-0.07}$  &$1.97^{+0.26}_{-0.24}$   &$P_c(4380)$  &$4385$ \\ \hline
$\bar{D}\Sigma_c^*$      &$\frac{3}{2}\frac{3}{2}^-$    &$4.41^{+0.08}_{-0.08}$  &$1.24^{+0.17}_{-0.16}$
&             &$4385$\\ \hline

$\bar{D}^*\Sigma_c$      &$\frac{1}{2}\frac{3}{2}^-$    &$4.44^{+0.07}_{-0.08}$  &$3.60^{+0.47}_{-0.44}$   &$P_c(4440)$ &$4462$\\ \hline
$\bar{D}^*\Sigma_c$      &$\frac{3}{2}\frac{3}{2}^-$    &$4.47^{+0.09}_{-0.09}$  &$2.31^{+0.33}_{-0.31}$
&            &$4462$\\ \hline

$\bar{D}^*\Sigma_c^*$    &$\frac{1}{2}\frac{5}{2}^-$    &$4.46^{+0.08}_{-0.08}$  &$4.05^{+0.54}_{-0.50}$   &$P_c(4457)$  &$4527$\\ \hline
$\bar{D}^*\Sigma_c^*$    &$\frac{3}{2}\frac{5}{2}^-$    &$4.62^{+0.09}_{-0.09}$  &$2.40^{+0.37}_{-0.35}$
&             &$4527$\\ \hline

$\bar{D}\Xi_c^{\prime}$  &$0{\frac{1}{2}}^-$            &$4.43^{+0.07}_{-0.07}$  &$3.02^{+0.39}_{-0.37}$
&             &$4446$ \\ \hline
$\bar{D}\Xi_c^{\prime}$  &$1{\frac{1}{2}}^-$            &$4.45^{+0.07}_{-0.08}$  &$2.50^{+0.33}_{-0.31}$
&             &$4446$ \\ \hline

$\bar{D}\Xi_c^*$         &$0{\frac{3}{2}}^-$            &$4.46^{+0.07}_{-0.07}$  &$1.71^{+0.22}_{-0.21}$   &$P_{cs}(4459)$   &$4513$ \\ \hline
$\bar{D}\Xi_c^*$         &$1{\frac{3}{2}}^-$            &$4.53^{+0.07}_{-0.07}$  &$1.56^{+0.20}_{-0.19}$
&                 &$4513$ \\ \hline

$\bar{D}^*\Xi_c^{\prime}$&$0{\frac{3}{2}}^-$            &$4.57^{+0.07}_{-0.07}$  &$3.41^{+0.43}_{-0.41}$
&                 &$4588$\\ \hline
$\bar{D}^*\Xi_c^{\prime}$&$1{\frac{3}{2}}^-$            &$4.62^{+0.08}_{-0.08}$  &$3.05^{+0.39}_{-0.37}$
&                 &$4588$\\ \hline

$\bar{D}^*\Xi_c^*$       &$0{\frac{5}{2}}^-$            &$4.64^{+0.07}_{-0.07}$  &$4.36^{+0.54}_{-0.51}$
&                 &$4655$\\ \hline
$\bar{D}^*\Xi_c^*$       &$1{\frac{5}{2}}^-$            &$4.67^{+0.08}_{-0.08}$  &$3.25^{+0.41}_{-0.39}$
&                 &$4655$ \\ \hline

$\bar{D}\,\Xi_c$         &$0{\frac{1}{2}}^-$            &$4.34_{-0.07}^{+0.07}$  &$1.43^{+0.19}_{-0.18}$
&? $P_{cs}(4338)$ &4337                   \\      \hline

$\bar{D}\,\Xi_c$         &$1{\frac{1}{2}}^-$            &$4.46_{-0.07}^{+0.07}$  &$1.37^{+0.19}_{-0.18}$
&                 &4337                  \\      \hline

$\bar{D}\,\Lambda_c$     &$\frac{1}{2}{\frac{1}{2}}^-$  &$4.46_{-0.08}^{+0.07}$  &$1.47^{+0.20}_{-0.18}$
&                 &4151                   \\  \hline

$\bar{D}_s\,\Xi_c$       &$\frac{1}{2}{\frac{1}{2}}^-$  &$4.54_{-0.07}^{+0.07}$  &$1.58^{+0.21}_{-0.20}$
&                 &4437                     \\  \hline

$\bar{D}_s\,\Lambda_c$   &$0{\frac{1}{2}}^-$            &$4.48_{-0.07}^{+0.07}$  &$1.57^{+0.21}_{-0.20}$
&                 &4255                    \\ \hline
$\bar{D}^*\,\Xi_c$       &$0{\frac{3}{2}}^-$            &$4.46_{-0.07}^{+0.07}$  &$1.55^{+0.20}_{-0.19}$
&? $P_{cs}(4459)$ &4479                   \\      \hline

$\bar{D}^*\,\Xi_c$       &$1{\frac{3}{2}}^-$            &$4.63_{-0.08}^{+0.08}$  &$1.69^{+0.22}_{-0.21}$
&                 &4479                  \\      \hline

$\bar{D}^*\,\Lambda_c$   &$\frac{1}{2}{\frac{3}{2}}^-$  &$4.59_{-0.08}^{+0.08}$  &$1.67^{+0.22}_{-0.21}$
&                 &4293                   \\  \hline

$\bar{D}^*_s\,\Xi_c$     &$\frac{1}{2}{\frac{3}{2}}^-$  &$4.65_{-0.08}^{+0.08}$  &$1.66^{+0.22}_{-0.21}$
&                 &4580                     \\  \hline

$\bar{D}^*_s\,\Lambda_c$ &$0{\frac{3}{2}}^-$            &$4.50_{-0.07}^{+0.07}$  &$1.52^{+0.21}_{-0.19}$
&                 &4398                    \\ \hline
\hline
\end{tabular}
\end{center}
\caption{ The  masses, pole residues and possible assignments for the $\mathbf 1\mathbf 1 $  type  pentaquark states, where the thresholds denote the corresponding thresholds of the  meson-baryon scattering states \cite{WangXW-penta-mole-SCPMA-2022,
 WangXW-penta-mole-IJMPA-2022,WangXW-penta-mole-CPC-2023}. }\label{mass-residue-penta-mole}
\end{table}

 The  predictions
   support  assigning  the $P_{cs}(4338)$ as the $\bar{D}\Xi_c$  molecular state with the spin-parity  $J^P={\frac{1}{2}}^-$ and isospin $I=0$, the observation of its cousin with the isospin $I=1$ in the $J/\psi\Sigma^0/\eta_c\Sigma^0$   mass spectrum   would
decipher the inner structure of the $P_{cs}(4338)$. However, there exists no candidate for the $P_c(4337)$ \cite{LHCb-Pc4337}.

Beyond the $\bar{{\mathbf 3}}\bar{{\mathbf 3}}\bar{{\mathbf 3}}$ type currents,  in Ref.\cite{Penta-88-Pimikov-PRD-2020}, Pimikov, Lee and  Zhang construct the color ${\mathbf 8}{\mathbf 8}$ type currents $J^{A}(\Gamma_2,\Gamma_3)$ and $J^S_\mu(\Gamma_2,\Gamma_3)$ to interpolate the hidden-charm pentaquark states,
\begin{eqnarray}
J^{A}(\Gamma_2,\Gamma_3)&=&\varepsilon_{f_1f_2f_3}
\varepsilon_{c_1c_3c} t^m_{c c_2} \left[q^T_1C q_2 \Gamma_2q_3-q^T_1C\gamma_5 q_2 \gamma_5\Gamma_2q_3 \right] \left[\bar q_5t^m\Gamma_3 q_4 \right]
\, , \nonumber\\
J_{\mu}^{S}(\Gamma_2,\Gamma_3)&=&\varepsilon_{f_1f_2f_3}\varepsilon_{c_1c_3c}t^m_{c c_2}
q^T_1C\gamma_\mu q_2 \Gamma_2q_3 \left[\bar q_5t^m\Gamma_3 q_4 \right]\, ,
\end{eqnarray}
where the quark fields have the flavors $f_i$ and color $c_i$, the
 $\Gamma_2$ and $\Gamma_3$ are some  Dirac matrixes.  Then, they calculate the mass spectrum by taking account of the vacuum condensates $\langle \bar{q}q\rangle$, $ \langle\frac{\alpha_sGG}{\pi}\rangle$,
 $\langle \bar{q}g_s \sigma Gq\rangle $, $\langle \bar{q}q\rangle^2$,
 $\langle \bar{q}q\rangle\langle  \bar{q}g_s \sigma Gq\rangle$,
 $\langle \bar{q}q\rangle^3$ and $\langle \bar{q}g_s\sigma Gq\rangle^2 $.

\section{Singly-heavy exotic states}

\subsection{Singly-heavy tetraquark states}\label{Singly-Q-tetraquark}
The $X_0(2900)$ and $X_1(2900)$  observed in the $D^- K^+$  mass spectrum  are the first exotic structures
  with fully open flavor \cite{LHCb-X3930-PRL-2020,LHCb-X3930-PRD-2020}, they have the valence quarks $ud\bar{c}\bar{s}$.
    The $T_{c\bar{s}}^{0}(2900)$ and $T_{c\bar{s}}^{++}(2900)$ are observed in the $D_s^+ \pi^-$ and $ D_s^+ \pi^+$ mass spectra, respectively \cite{Tcs2900-LHCb-PRL-2023,Tcs2900-LHCb-PRD-2023}, they have  the valence quarks  $c\bar{s}u\bar{d}$ and $c\bar{s}\bar{u}d$, respectively.

Based on the predicted masses of the $AA$-type  scalar tetraquark states from the QCD sum rules \cite{WZG-HC-PRD-2020,WangZG-ssss-AHEP-2020},
\begin{eqnarray}
M_{qq\bar{q}\bar{q}}&=&1.86\pm0.11\, \rm{GeV}\, ,\nonumber\\
M_{ss\bar{s}\bar{s}}&=&2.08\pm0.13\, \rm{GeV}\, ,\nonumber\\
M_{cq\bar{c}\bar{q}}&=&3.95\pm 0.09\, \rm{GeV}\, ,
\end{eqnarray}
 we estimate the mass of the $AA$-type $cs\bar{u}\bar{d}$ tetraquark state crudely,
\begin{eqnarray}
M_{cs\bar{u}\bar{d}}&=&\frac{M_{qq\bar{q}\bar{q}}+M_{ss\bar{s}\bar{s}}+2M_{cq\bar{c}\bar{q}}}{4}=2.96\pm0.11\,\rm{GeV}\, ,
\end{eqnarray}
which is consistent with the mass of the $X_0(2900)$ within uncertainties \cite{X2900-tetra-WangZG-IJMPA-2020}.

In Ref.\cite{X2900-tetra-WangZG-IJMPA-2020}, we construct the $AA$  and $SS$-type scalar four-quark currents,
\begin{eqnarray}
 J_{AA}(x)&=&\varepsilon^{ijk} \varepsilon^{imn} s^T_j(x) C\gamma_{\alpha} c_k(x)\,\bar{u}_m(x) \gamma^{\alpha} C \bar{d}^T_n(x)\,  , \nonumber\\
  J_{SS}(x)&=&\varepsilon^{ijk} \varepsilon^{imn} s^T_j(x) C\gamma_5 c_k(x)\,\bar{u}_m(x) \gamma_5 C \bar{d}^T_n(x)\,  ,
\end{eqnarray}
to study the $cs\bar{u}\bar{d}$ tetraquark states with the correlation function $\Pi(p)$, see Eq.\eqref{CF-Pi}.
We carry out the
operator product expansion up to the vacuum condensates  of dimension-11 and
assume vacuum saturation for the  higher dimensional vacuum condensates according to the routine in Sect.{\bf\ref{Tetra-QCDSR}} and Sect.{\bf\ref{Tetra-Positive}}. As there are three $q$-quark lines and one $Q$-quark line, if each $Q$-quark line
emits a gluon and each $q$-quark line contributes a quark-antiquark pair, we obtain a quark-gluon  operator $g_sG_{\mu\nu} \bar{q}q\bar{q}q\bar{s}s$, which is of dimension 11, and leads  to the vacuum condensates
 $\langle\bar{q}q\rangle^2\langle\bar{s}g_s\sigma Gs\rangle$ and $\langle\bar{q}q\rangle \langle\bar{s}s\rangle\langle\bar{q}g_s\sigma Gq\rangle$.

We obtain the QCD sum rules routinely, at the QCD side, we choose the flavor numbers $n_f=4$ and the typical energy scale $\mu=1\,\rm{GeV}$.
After trial and error, we obtain the  Borel windows and continuum threshold parameters, therefore  the pole contributions of the ground states and  convergent behaviors of the operator product expansion, see Table \ref{Borel-X2900}. In the Borel windows, the pole contributions are about $(38-67)\%$,  the central values exceed $52\%$. The contributions of the vacuum condensates $|D(11)|$ are about $(2-4)\%$ and $(0-1)\%$ for the  $AA$ and $SS$-type tetraquark states, respectively.

\begin{figure}
 \centering
 \includegraphics[totalheight=5cm,width=7cm]{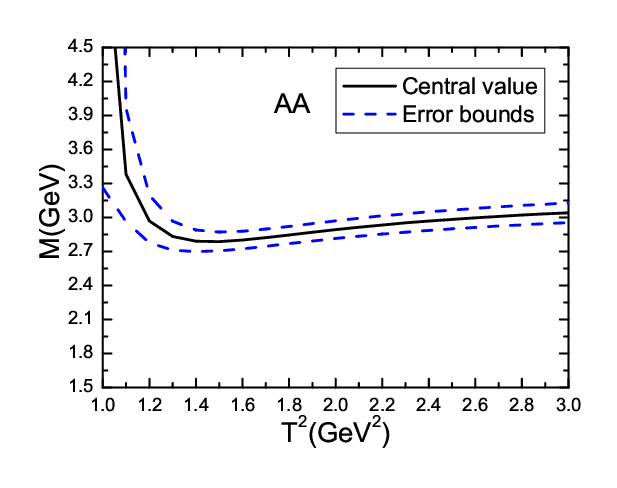}
  \includegraphics[totalheight=5cm,width=7cm]{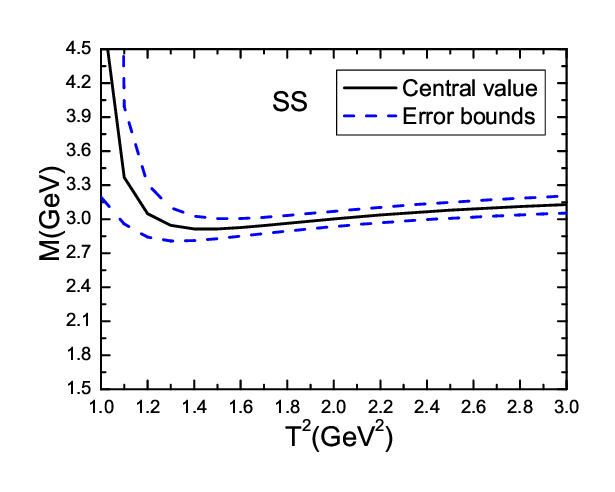}
 \caption{ The masses of the $AA$ and $SS$-type tetraquark states with variations  of the Borel parameters $T^2$.   }\label{mass-fig-X2900}
\end{figure}

At last, we obtain the values of the masses and pole residues, which are also shown  in Table \ref{Borel-X2900}.
 In Fig.\ref{mass-fig-X2900}, we plot the masses  of the  $AA$ and $SS$-type scalar $cs\bar{u}\bar{d}$ tetraquark states with variations  of the Borel parameters $T^2$ in  much larger  ranges than the Borel windows,  there appear  platforms in the Borel windows indeed.

The predicted mass $M_{AA}=2.91\pm0.12\,\rm{GeV}$ is consistent with the experimental value $2866\pm7~{\rm MeV}$ from the LHCb collaboration \cite{LHCb-X3930-PRL-2020,LHCb-X3930-PRD-2020}, and supports assigning the $X_0(2900)$ to be the $AA$-type  $cs\bar{u}\bar{d}$ tetraquark state with the spin-parity $J^P=0^+$. While the predicted mass $M_{SS}=3.05\pm0.10\,\rm{GeV}$ lies above the experimental value $2866\pm7~{\rm MeV}$  \cite{LHCb-X3930-PRL-2020,LHCb-X3930-PRD-2020}.

\begin{table}
\begin{center}
\begin{tabular}{|c|c|c|c|c|c|c|c|}\hline\hline
                              &$T^2(\rm{GeV}^2)$   &$\sqrt{s_0}(\rm{GeV})$  &pole         &$|D(11)|$  &$M(\rm{GeV})$  &$\lambda(\rm{GeV}^5)$ \\ \hline

$[cs]_A[\bar{u}\bar{d}]_{A}$  &$1.9-2.3$           &$3.5\pm0.1$             &$(38-67)\%$  &$(2-4)\%$  &$2.91\pm0.12$  &$(1.60\pm0.33)\times10^{-2}$   \\ \hline

$[cs]_S[\bar{u}\bar{d]}_{S}$  &$2.1-2.5$           &$3.6\pm0.1$             &$(39-66)\%$  &$(0-1)\%$  &$3.05\pm0.10$  &$(1.20\pm0.21)\times10^{-2}$   \\ \hline\hline
\end{tabular}
\end{center}
\caption{ The Borel windows, continuum threshold parameters, pole contributions, contributions of the vacuum condensates of dimension $11$,  masses and pole residues for the scalar  tetraquark states \cite{X2900-tetra-WangZG-IJMPA-2020}. } \label{Borel-X2900}
\end{table}

In Ref.\cite{WangZG-Tcs2900-IJMPA-2023}, we  construct the $A\bar{A}$-type  currents to study the ground state  mass spectrum of the tetraquark states with strange and doubly-strange via the QCD sum rules to verify the inner structures of the $T_{c\bar{s}}(2900)$,
where
\begin{eqnarray}
 J(x)&=&J^0_s(x)\, , \,\, J^0_{ss}(x)\, , \nonumber\\
  J_{\mu\nu}(x)&=&J_{s,\mu\nu}^{1}(x)\, , \, \, J_{ss,\mu\nu}^{1}(x)\, , \,\, J_{s,\mu\nu}^{2}(x)\, , \, \, J_{ss,\mu\nu}^{2}(x)\, ,
\end{eqnarray}

\begin{eqnarray}
J^0_s(x)&=&\varepsilon^{ijk}\varepsilon^{imn}u^T_j (x)C\gamma_\mu c_k(x) \bar{d}_m(x)\gamma^\mu C \bar{s}^T_n(x) \, ,\nonumber \\
J^0_{ss}(x)&=&\varepsilon^{ijk}\varepsilon^{imn}u^T_j (x)C\gamma_\mu c_k(x) \bar{s}_m(x)\gamma^\mu C \bar{s}^T_n(x) \, ,\nonumber \\
J^1_{s,\mu\nu}(x)&=&\varepsilon^{ijk}\varepsilon^{imn}\left[u^T_j (x)C\gamma_\mu c_k(x) \bar{d}_m(x)\gamma_\nu C \bar{s}^T_n(x)-u^T_j (x)C\gamma_\nu c_k(x) \bar{d}_m(x)\gamma_\mu C \bar{s}^T_n(x)\right] \, , \nonumber \\
J^1_{ss,\mu\nu}(x)&=&\varepsilon^{ijk}\varepsilon^{imn}\left[u^T_j (x)C\gamma_\mu c_k(x) \bar{s}_m(x)\gamma_\nu C \bar{s}^T_n(x)-u^T_j (x)C\gamma_\nu c_k(x) \bar{s}_m(x)\gamma_\mu C \bar{s}^T_n(x)\right] \, , \nonumber \\
J^2_{s,\mu\nu}(x)&=&\varepsilon^{ijk}\varepsilon^{imn}\left[u^T_j (x)C\gamma_\mu c_k(x) \bar{d}_m(x)\gamma_\nu C \bar{s}^T_n(x)+u^T_j (x)C\gamma_\nu c_k(x) \bar{d}_m(x)\gamma_\mu C \bar{s}^T_n(x)\right] \, , \nonumber \\
J^2_{ss,\mu\nu}(x)&=&\varepsilon^{ijk}\varepsilon^{imn}\left[u^T_j (x)C\gamma_\mu c_k(x) \bar{s}_m(x)\gamma_\nu C \bar{s}^T_n(x)+u^T_j (x)C\gamma_\nu c_k(x) \bar{s}_m(x)\gamma_\mu C \bar{s}^T_n(x)\right] \, , \nonumber\\
\end{eqnarray}
 the superscripts $0$, $1$ and $2$ denote the spins. With a simple replacement $u\leftrightarrow d$, we obtain the corresponding currents in the same isospin multiplets. In the isospin limit, the tetraquark states in the same multiplets have the same masses.

Again, we resort to the correlation functions $\Pi(p)$ and $\Pi_{\mu\nu\alpha\beta}(p)$, see Eq.\eqref{CF-Pi}, and obtain the ground state contributions according to the hadron representations in Eqs.\eqref{CF-Hadron-cc-J0}-\eqref{CF-Hadron-cc-J2}, and obtain the QCD sum rules routinely, again we take the flavor numbers $n_f=4$  and the typical energy scale $\mu=1.0\,\rm{GeV}$.

After trial and error, we obtain the Borel windows, continuum threshold parameters and pole contributions, see Table \ref{Borel-mass-Tcs2900}, where the pole contributions are $38\% - 66\%$ and the central values for  the six states are larger than $50\%$,  furthermore,  the contributions of the vacuum condensates   show a descending trend  $|D(6)|>|D(8)|>|D(9)|>|D(10)|\sim |D(11)|\sim 0$.  At last, we obtain  the masses and pole residues,  which are also shown in Table \ref{Borel-mass-Tcs2900} \cite{WangZG-Tcs2900-IJMPA-2023}.

In Table \ref{Borel-mass-Tcs2900}, the predicted mass of the $J^P=0^+$ state $cu\bar{d}\bar{s}$, $M = 2.92\pm0.12\,\rm{GeV}$, is in very good agreement with the experimental values  $M=2.892\pm0.014\pm0.015\,\rm{GeV}$ and $2.921\pm0.017\pm0.020\,\rm{GeV}$ from the LHCb collaboration \cite{Tcs2900-LHCb-PRL-2023,Tcs2900-LHCb-PRD-2023}, and supports assigning the $T_{c\bar{s}}(2900)$ to be the $A\bar{A}$-type $c\bar{s}q\bar{q}$ tetraquark states with the spin-parity  $J^P=0^+$.

\begin{table}
\begin{center}
\begin{tabular}{|c|c|c|c|c|c|c|c|}\hline\hline
                   &$J^P$    &$T^2 (\rm{GeV}^2)$  &$\sqrt{s_0}(\rm GeV) $  &pole        &$M_T(\rm{GeV})$  &$\lambda_T(10^{-2}\rm{GeV}^5)$ \\ \hline

$cu\bar{d}\bar{s}$ &$0^+$    &$1.9-2.3$           &$3.50\pm0.10$           &$(40-65)\%$  &$2.92\pm0.12$   &$1.62\pm0.33 $  \\ \hline

$cu\bar{d}\bar{s}$ &$1^+$    &$2.2-2.6$           &$3.65\pm0.10$           &$(40-62)\%$  &$3.10\pm0.10$   &$1.59\pm0.27$  \\ \hline

$cu\bar{d}\bar{s}$ &$2^+$    &$2.5-2.9$           &$3.95\pm0.10$           &$(42-61)\%$  &$3.40\pm0.10$   &$3.51\pm0.55$  \\ \hline

$cu\bar{s}\bar{s}$ &$0^+$    &$2.0-2.4$           &$3.60\pm0.10$           &$(38-66)\%$  &$3.02\pm0.12$   &$2.72\pm0.43$  \\ \hline

$cu\bar{s}\bar{s}$ &$1^+$    &$2.3-2.7$           &$3.75\pm0.10$           &$(40-65)\%$  &$3.20\pm0.10$   &$2.64\pm0.33$  \\ \hline

$cu\bar{s}\bar{s}$ &$2^+$    &$2.6-3.0$           &$4.05\pm0.10$           &$(41-64)\%$  &$3.49\pm0.10$   &$5.70\pm0.65$  \\ \hline
\end{tabular}
\end{center}
\caption{ The spin-parity, Borel parameters, continuum threshold parameters, pole contributions, masses and pole residues for the singly-charmed tetraquark states \cite{WangZG-Tcs2900-IJMPA-2023}. }\label{Borel-mass-Tcs2900}
\end{table}

Those typical singly-charmed tetraquark candidates, which lie at $2.9\,\rm{GeV}$,  have attracted many theoretical works, and are assigned as the $\bar{\mathbf{3}}\mathbf{3}$ type tetraquark states  \cite{X2900-tetra-Karliner-PRD-2020,X2900-tetra-WangZG-IJMPA-2020,
X2900-tetra-ZhangJR-PRD-2021,X2900-tetra-mole-ChenHX-CPL-2020,
X2900-tetra-GJWang-EPJC-2021}, or their radial/orbital excitations \cite{X2900-tetra-2S-HeXG-EPJC-2020}, or non-tetraquark states \cite{X2900-No-tetra-LuQF-PRD-2020},
  $D^*\bar{K}^*$ molecular states \cite{X2900-tetra-mole-ChenHX-CPL-2020,X2900-DvKv-LiuMZ-PRD-2020,X2900-mole-WangQ-CPC-2021,
 X2900-mole-Oset-PLB-2020,X2900-mole-Azizi-JPG-2021,X2900-mole-Narison-NPA-2021,
 X2900-mole-JJXie-EPJC-2020,X2900-mole-DYChen-PRD-2021,X2900-mole-BWang-EPJC-2022},  triangle singularities  \cite{X2900-Triangle-LiuXH-PRD-2020,X2900-Triangle-Swanson-PLB-2021,
 X2900-Triangle-Swanson-PRD-2021}, etc.
 We could  only obtain  a mass about $2.3\,\rm{GeV}$ for the singly-charmed tetraquark states at the cost of sacrificing the pole dominance
 \cite{Ds2317-trtra-Nielsen-PLB-2005,Ds2317-trtra-HKim-PRD-2005,
Ds2317-trtra-WangZG-NPA-2006}.

 At the bottom sector, the
 singly-bottom tetraquark candidate  $X(5568)$
is not confirmed \cite{X5568-No-Burns-PLB-2016,X5568-No-GuoFK-CTP-2016}. The lowest mass  of the ground state $b\bar{s}\bar{u}d$ might have a mass $M_{T/X}+m_b(m_b)-m_c(m_c)\approx 5.81\,\rm{GeV}$, which lies above the $X(5568)$.

\subsection{Singly-heavy pentaquark states}
The experimental candidates for the singly-charmed pentaquark states are not as robust as that of the singly-charmed  tetraquark states, the assignments in the scenario of pentaquark states are only conjectures.

 In 2017,  the LHCb collaboration observed  five narrow structures $\Omega_c(3000)$, $\Omega_c(3050)$, $\Omega_c(3066)$, $\Omega_c(3090)$ and $\Omega_c(3119)$ \cite{LHCb2017-omegac}. Also in 2017, the Belle collaboration confirmed the $\Omega_c(3000)$, $\Omega_c(3050)$, $\Omega_c(3066)$ and  $\Omega_c(3090)$ in the $\Xi_c^+K^-$ decay mode \cite{Belle2017-omegac}.
In 2023, the LHCb collaboration observed the $\Omega_c(3185)$ and $\Omega_c(3327)$  in the  $\Xi_c^+K^-$ mass spectrum \cite{LHCb2023-omegac}, which lie  near the  $D\Xi$ and $D^*\Xi$ thresholds, respectively,  the measured Breit-Wigner masses and decay widths are
\begin{flalign}
 &\Omega_c(3185) : M = 3185.1\pm1.7^{+7.4}_{-0.9}\pm0.2 \mbox{ MeV}\, , \, \Gamma =50\pm7 ^{+10}_{-20}\mbox{ MeV}\, , \nonumber \\
 & \Omega_c(3327) : M = 3327.1\pm1.2^{+0.1}_{-1.3} \pm0.2\mbox{ MeV}\, , \, \Gamma = 20\pm5 ^{+13}_{-1}\mbox{ MeV} \, .
\end{flalign}

As early as 2005, the Belle collaboration tentatively assigned the $\Sigma_c^0(2800)$, $\Sigma_c^+(2800)$ and $\Sigma_c^{++}(2800)$ in the $\Lambda_c^+\pi^{-/0/+}$ mass spectra as the isospin triplet states with the spin-parity $J^P=\frac{3}{2}^-$  \cite{Belle-2005-sig2800}, the measured masses  and decay widths are
\begin{flalign}
 &\Sigma_c^{0}(2800) : M = M_{\Lambda^+}+515.4^{+3.2}_{-3.1}{}^{+2.1}_{-6.0}\, \mbox{MeV}\, , \, \Gamma =61^{+18}_{-13}{}^{+22}_{-13} \mbox{ MeV}\, , \nonumber \\
 &\Sigma_c^{+}(2800) : M = M_{\Lambda^+}+505.4^{+5.8}_{-4.6}{}^{+12.4}_{-2.0} \,\mbox{MeV}\, , \, \Gamma =62^{+37}_{-23}{}^{+52}_{-38} \mbox{ MeV}\, , \nonumber \\
 & \Sigma_c^{++}(2800) : M = M_{\Lambda^+}+514.5^{+3.4}_{-3.1}{}^{+2.8}_{-4.9} \,\mbox{MeV}\, , \, \Gamma = 75^{+18}_{-13}{}^{+12}_{-11}\mbox{ MeV}\, .
\end{flalign}

In 2008, the BaBar collaboration observed the $\Sigma_c^{0}(2800)$ in the $\Lambda_c^+\pi^-$ mass spectrum  with the possible spin-parity  $J^P=\frac{1}{2}^- $ \cite{BaBar-2008-sig2800}, the mass and decay width are,
\begin{flalign}
&\Sigma_c^0(2800): M = 2846\pm8\pm10 \mbox{ MeV}\,,\,\Gamma = 86{}^{+33}_{-22}\pm7\mbox{ MeV} \, .
\end{flalign}

In 2007, the BaBar collaboration observed  the $\Lambda_c(2940)$  in the $D^0p$ invariant mass spectrum  \cite{BaBar-2940-2007}. Subsequently, the Belle collaboration  verified the $\Lambda_c(2940)$ in the decay mode $\Lambda_c(2940) \!\!\to\!\Sigma_c(2455) \pi$ \cite{Belle-2940-2007}. In 2017, the LHCb collaboration determined the spin-parity  of the $\Lambda(2940)^+$ to be $J^P=\frac{3}{2}^-$ by  analyzing the process $\Lambda_b^0 \!\!\to\!D^0p \pi^-$ \cite{LHCb-2017-2940}. The measured masses and decay widths  are,
\begin{flalign}
&\Lambda_c(2940): M = 2939.8 \pm1.3 \pm1.0 \mbox{ MeV}\, ,\,\Gamma = 17.5\pm5.0\pm5.9 \mbox{ MeV} \,\,\,\,\,\,{\rm(BaBar)}\, , \nonumber \\
&\Lambda_c(2940): M = 2938.0 \pm1.3 ^{+2.0}_{-4.0} \mbox{ MeV}\, ,\,\Gamma = 13{}^{+8}_{-5}{}^{+27}_{-7} \mbox{ MeV} \,\,\,\,\,\,{\rm(Belle)}\, , \nonumber \\
&\Lambda_c(2940): M = 2944.8 ^{+3.5}_{-2.5} \pm0.4 {}^{+0.1}_{-4.6} \mbox{ MeV}\, ,\,\Gamma = 27.7^{+8.2}_{-6.0}\pm0.9{}^{+5.2}_{-10.4} \mbox{ MeV} \,\,\,\,\,\,{\rm(LHCb)}\,.
\end{flalign}

In 2023,  the Belle collaboration studied  the decays $\bar{B}^0 \!\!\to\!\Sigma_c(2455)^{0,++} \pi^{\pm}\bar{p}$ and found a new structure $\Lambda^+_c(2910)$ in the $\Sigma_c(2455)^{0,++}\pi^{\pm}$ mass spectrum  \cite{Belle2022-2910}, the mass and decay width are,
\begin{flalign}
&\Lambda_c(2910): M = 2913.8 \pm5.6 \pm3.8 \mbox{ Mev}\,,\,\Gamma = 51.8\pm20.0\pm18.8 \mbox{ MeV} \,\, .
\end{flalign}
The $\Sigma_c(2800)$, $\Lambda_c(2940)$ and $\Lambda_c(2910)$ lie near the  $D^{(*)}N$ thresholds, and they might be the $D^{(*)}N$ molecular states. For example, we can assign  the $\Sigma_c(2800)$ as the $DN$ molecular (bound) state \cite{2800-GFK,PJL-2800-2940-1,GXH-2800,Valderrama-2800-2940-3185-3327}, and  assign the $\Lambda_c(2940)$ as the $D^*N$  molecular state \cite{PJL-2800-2940-1,Valderrama-2800-2940-3185-3327,PJL-2800-2940-2,ZJR-siglam,
ZLZ-2910-2940,ZSL-2940,HY-2940}.
We would like to study the $\mathbf{1}\mathbf{1}$ type charmed pentaquark states and explore the possible assignments in the scenario of molecular states.

Again, we resort to  the correlation functions $\Pi(p)$, $\Pi_{\mu\nu}(p)$
and $\Pi_{\mu\nu\alpha\beta}(p)$ defined in Eq.\eqref{CF-Pi-Pi-Pi}, and write down the currents,
\begin{eqnarray}
J(x)&=&J^{DN}_{\mid 1,0 \rangle}(x),\, J^{DN}_{\mid 0,0 \rangle}(x),\, J^{D\Xi}_{\mid 1,0 \rangle}(x),\, J^{D\Xi}_{\mid 0,0 \rangle}(x), \,J^{D_s\Xi}_{\mid \frac{1}{2},\pm\frac{1}{2} \rangle}(x) \,,\nonumber \\
J_\mu(x)&=&J^{D^{*}N}_{\mid 1,0 \rangle}(x),\, J^{D^{*}N}_{\mid 0,0 \rangle}(x), \, J^{D^{*}\Xi}_{\mid 1,0 \rangle}(x),\, J^{D^{*}\Xi}_{\mid 0,0 \rangle}(x), \,J^{D_s^*\Xi}_{\mid \frac{1}{2},\pm\frac{1}{2} \rangle}(x),\, J^{D\Xi^{*}}_{\mid 1,0 \rangle}(x),\, J^{D\Xi^{*}}_{\mid 0,0 \rangle}(x),\, J^{D_s\Xi^*}_{\mid \frac{1}{2},\pm\frac{1}{2} \rangle}(x)\,,\nonumber \\
J_{\mu\nu}(x)&=&J^{D^{*}\Xi^{*}}_{\mid 1,0 \rangle}(x), \,J^{D^{*}\Xi^{*}}_{\mid 0,0 \rangle}(x),\,  J^{D_s^*\Xi^*}_{\mid \frac{1}{2},\pm\frac{1}{2} \rangle}(x) \,  ,
\end{eqnarray}
\begin{eqnarray}\label{current-DN}
J^{DN}_{\mid 1,0 \rangle}(x) &=& \frac{1}{\sqrt{2}}  J^{D^0}_{\mid \frac{1}{2},-\frac{1}{2} \rangle}(x) J^{N^+}_{\mid \frac{1}{2},\frac{1}{2} \rangle}(x) +\frac{1}{\sqrt{2}}J^{D^+}_{\mid \frac{1}{2},\frac{1}{2} \rangle}(x)J^{N^0}_{\mid \frac{1}{2},-\frac{1}{2} \rangle}  (x)  \,,\nonumber \\
J^{DN}_{\mid 0,0 \rangle}(x) &=& \frac{1}{\sqrt{2}}  J^{D^0}_{\mid \frac{1}{2},-\frac{1}{2} \rangle}(x) J^{N^+}_{\mid \frac{1}{2},\frac{1}{2} \rangle}(x)        -\frac{1}{\sqrt{2}}J^{D^+}_{\mid \frac{1}{2},\frac{1}{2} \rangle}(x)J^{N^0}_{\mid \frac{1}{2},-\frac{1}{2} \rangle}  (x)       \,,\nonumber \\
J^{D\Xi}_{\mid 1,0 \rangle}(x) &=& \frac{1}{\sqrt{2}}J^{D^0}_{\mid \frac{1}{2},-\frac{1}{2} \rangle}(x) J^{\Xi^0}_{\mid \frac{1}{2},\frac{1}{2} \rangle}(x)+\frac{1}{\sqrt{2}} J^{D^+}_{\mid \frac{1}{2},\frac{1}{2} \rangle}(x)J^{\Xi^-}_{\mid \frac{1}{2},-\frac{1}{2} \rangle}(x)    \,,\nonumber \\
J^{D\Xi}_{\mid 0,0 \rangle}(x) &=& \frac{1}{\sqrt{2}}J^{D^0}_{\mid \frac{1}{2},-\frac{1}{2} \rangle}(x) J^{\Xi^0}_{\mid \frac{1}{2},\frac{1}{2} \rangle}(x)-\frac{1}{\sqrt{2}} J^{D^+}_{\mid \frac{1}{2},\frac{1}{2} \rangle}(x)J^{\Xi^-}_{\mid \frac{1}{2},-\frac{1}{2} \rangle} (x)          \,,\nonumber \\
J^{D_s\Xi}_{\mid \frac{1}{2},\pm\frac{1}{2} \rangle}(x) &=& J^{D_s^+}_{\mid 0,0 \rangle} (x) J^{\Xi^{0/-}} _{\mid \frac{1}{2},\pm\frac{1}{2} \rangle} (x)       \, ,
\end{eqnarray}

\begin{eqnarray}\label{current-DN-A}
J^{D^{*}N}_{\mid 1,0 \rangle}(x) &=& \frac{1}{\sqrt{2}}  J^{D^{*0}}_{\mid \frac{1}{2},-\frac{1}{2} \rangle}(x) J^{N^+}_{\mid \frac{1}{2},\frac{1}{2} \rangle}(x)         +\frac{1}{\sqrt{2}}J^{D^{*+}}_{\mid \frac{1}{2},\frac{1}{2} \rangle}(x)J^{N^0}_{\mid \frac{1}{2},-\frac{1}{2} \rangle} (x)\,,\nonumber \\
J^{D^{*}N}_{\mid 0,0 \rangle}(x) &=& \frac{1}{\sqrt{2}}  J^{D^{*0}}_{\mid \frac{1}{2},-\frac{1}{2} \rangle}(x) J^{N^+}_{\mid \frac{1}{2},\frac{1}{2} \rangle}(x)         -\frac{1}{\sqrt{2}}J^{D^{*+}}_{\mid \frac{1}{2},\frac{1}{2} \rangle}(x)J^{N^0}_{\mid \frac{1}{2},-\frac{1}{2} \rangle} (x)\,,\nonumber \\
J^{D^{*}\Xi}_{\mid 1,0 \rangle}(x) &=& \frac{1}{\sqrt{2}}J^{D^{*0}}_{\mid \frac{1}{2},-\frac{1}{2} \rangle}(x) J^{\Xi^0}_{\mid \frac{1}{2},\frac{1}{2} \rangle}(x)+\frac{1}{\sqrt{2}} J^{D^{*+}}_{\mid \frac{1}{2},\frac{1}{2} \rangle}(x)J^{\Xi^-}_{\mid \frac{1}{2},-\frac{1}{2} \rangle} (x)          \,,\nonumber \\
J^{D^{*}\Xi}_{\mid 0,0 \rangle}(x) &=& \frac{1}{\sqrt{2}}J^{D^{*0}}_{\mid \frac{1}{2},-\frac{1}{2} \rangle}(x) J^{\Xi^0}_{\mid \frac{1}{2},\frac{1}{2} \rangle}(x)-\frac{1}{\sqrt{2}} J^{D^{*+}}_{\mid \frac{1}{2},\frac{1}{2} \rangle}(x)J^{\Xi^-}_{\mid \frac{1}{2},-\frac{1}{2} \rangle} (x)              \,, \nonumber \\
J^{D_s^*\Xi}_{\mid \frac{1}{2},\pm\frac{1}{2} \rangle}(x) &=& J^{D_s^{*+}}_{\mid 0,0\rangle} (x)  J^{\Xi^{0/-}}_{\mid \frac{1}{2},\pm\frac{1}{2} \rangle} (x)       \, ,      \nonumber \\
J^{D\Xi^{*}}_{\mid 1,0 \rangle}(x) &=& \frac{1}{\sqrt{2}}J^{D^{0}}_{\mid \frac{1}{2},-\frac{1}{2} \rangle}(x) J^{\Xi^{*0}}_{\mid \frac{1}{2},\frac{1}{2} \rangle}(x)+\frac{1}{\sqrt{2}} J^{D^{+}}_{\mid \frac{1}{2},\frac{1}{2} \rangle}(x)J^{\Xi^{*-}}_{\mid \frac{1}{2},-\frac{1}{2} \rangle} (x)          \,,\nonumber \\
J^{D\Xi^{*}}_{\mid 0,0 \rangle}(x) &=& \frac{1}{\sqrt{2}}J^{D^{0}}_{\mid \frac{1}{2},-\frac{1}{2} \rangle}(x) J^{\Xi^{*0}}_{\mid \frac{1}{2},\frac{1}{2} \rangle}(x)-\frac{1}{\sqrt{2}} J^{D^{+}}_{\mid \frac{1}{2},\frac{1}{2} \rangle}(x)J^{\Xi^{*-}}_{\mid \frac{1}{2},-\frac{1}{2} \rangle} (x)            \,, \nonumber \\
J^{D_s\Xi^*}_{\mid \frac{1}{2},\pm\frac{1}{2} \rangle}(x) &=& J^{D_s^{+}}_{\mid 0,0\rangle} (x)  J^{\Xi^{*0/-}}_{\mid \frac{1}{2},\pm\frac{1}{2} \rangle} (x)       \, ,
\end{eqnarray}

\begin{eqnarray}\label{current-DN-T}
J^{D^{*}\Xi^{*}}_{\mid 1,0 \rangle}(x) &=& \frac{1}{\sqrt{2}}J^{D^{*0}}_{\mid \frac{1}{2},-\frac{1}{2} \rangle}(x) J^{\Xi^{*0}}_{\mid \frac{1}{2},\frac{1}{2} \rangle}(x)+\frac{1}{\sqrt{2}} J^{D^{*+}}_{\mid \frac{1}{2},\frac{1}{2} \rangle}(x)J^{\Xi^{*-}}_{\mid \frac{1}{2},-\frac{1}{2} \rangle} (x)          \,,\nonumber \\
J^{D^{*}\Xi^{*}}_{\mid 0,0 \rangle}(x) &=& \frac{1}{\sqrt{2}}J^{D^{*0}}_{\mid \frac{1}{2},-\frac{1}{2} \rangle}(x) J^{\Xi^{*0}}_{\mid \frac{1}{2},\frac{1}{2} \rangle}(x)-\frac{1}{\sqrt{2}} J^{D^{*+}}_{\mid \frac{1}{2},\frac{1}{2} \rangle}(x)J^{\Xi^{*-}}_{\mid \frac{1}{2},-\frac{1}{2} \rangle} (x)            \,, \nonumber \\
J^{D_s^*\Xi^*}_{\mid \frac{1}{2},\pm\frac{1}{2} \rangle}(x) &=& J^{D_s^{*+}}_{\mid 0,0\rangle} (x)  J^{\Xi^{*0/-}}_{\mid \frac{1}{2},\pm\frac{1}{2} \rangle} (x)       \, ,
\end{eqnarray}
and
\begin{eqnarray}
J^{D^0}_{\mid \frac{1}{2},-\frac{1}{2} \rangle}(x) &=& \bar{u}(x)i\gamma_5 c(x)\, ,      \nonumber \\
J^{D^+}_{\mid \frac{1}{2},\frac{1}{2} \rangle}(x) &=& -\bar{d}(x)i\gamma_5 c(x)\, ,      \nonumber \\
J^{D_s^+}_{\mid 0,0 \rangle}(x) &=& \bar{s}(x)i\gamma_5 c(x)\, ,      \nonumber \\
J^{D^{*0}}_{\mid \frac{1}{2},-\frac{1}{2} \rangle}(x) &=& \bar{u}(x)\gamma_\mu c(x)\, ,      \nonumber \\
J^{D^{*+}}_{\mid \frac{1}{2},\frac{1}{2} \rangle}(x) &=& -\bar{d}(x)\gamma_\mu c(x)\, ,      \nonumber \\
J^{D_s^{*+}}_{\mid 0,0 \rangle}(x) &=& \bar{s}(x)\gamma_\mu c(x)\, ,
\end{eqnarray}

\begin{eqnarray}
J^{N^+}_{\mid \frac{1}{2},\frac{1}{2} \rangle}(x) &=&\varepsilon^{ijk}u_i^T(x)C\gamma_\mu u_j(x)\gamma^\mu\gamma_5 d_k(x)\, ,      \nonumber \\
J^{N^0}_{\mid \frac{1}{2},-\frac{1}{2} \rangle}(x) &=&\varepsilon^{ijk}d_i^T(x)C\gamma_\mu d_j(x)\gamma^\mu\gamma_5 u_k(x)\, ,      \nonumber \\
J^{\Xi^0}_{\mid \frac{1}{2},\frac{1}{2} \rangle}(x) &=&\varepsilon^{ijk}s_i^T(x)C\gamma_\mu s_j(x)\gamma^\mu\gamma_5 u_k(x)\, ,      \nonumber \\
J^{\Xi^-}_{\mid \frac{1}{2},-\frac{1}{2} \rangle}(x) &=&\varepsilon^{ijk}s_i^T(x)C\gamma_\mu s_j(x)\gamma^\mu\gamma_5 d_k(x)\, ,      \nonumber \\
J^{\Xi^{*0}}_{\mid \frac{1}{2},\frac{1}{2} \rangle}(x) &=&\varepsilon^{ijk}s_i^T(x)C\gamma_\mu s_j(x) u_k(x)\, ,      \nonumber \\
J^{\Xi^{*-}}_{\mid \frac{1}{2},-\frac{1}{2} \rangle}(x) &=&\varepsilon^{ijk}s_i^T(x)C\gamma_\mu s_j(x) d_k(x)\, ,
\end{eqnarray}
 the subscripts  $| 1,0 \rangle$, $|0,0 \rangle$,
$|\frac{1}{2},\frac{1}{2} \rangle$ and $|\frac{1}{2},-\frac{1}{2} \rangle$ are isospin indexes ${| I,I_3 \rangle}$, we choose the convention $|\frac{1}{2},\frac{1}{2} \rangle=-\bar{d}$ \cite{XinQ-Single-Heavy-IJMPA-2023}.

The currents $J(0)$, $J_\mu(0)$ and $J_{\mu\nu}(0)$ couple potentially to the $J^P={\frac{1}{2}}^\mp$, ${\frac{3}{2}}^\mp$ and ${\frac{5}{2}}^\mp$  singly-charmed  molecular  states $P_{\frac{1}{2}}^\mp$, $P_{\frac{3}{2}}^\mp$, and $P_{\frac{5}{2}}^\mp$, respectively,
for more details, see Sect.{\bf \ref{333-penta-Sect}}.
At the hadron side, we isolate the ground state contributions,
\begin{eqnarray}
\Pi(p) & = & {\lambda^{-}_{\frac{1}{2}}}^2  {\!\not\!{p}+ M_{-} \over M_{-}^{2}-p^{2}  } +  {\lambda^{+}_{\frac{1}{2}}}^2  {\!\not\!{p}- M_{+} \over M_{+}^{2}-p^{2}  } +\cdots  \, ,\nonumber\\
  &=&\Pi_{\frac{1}{2}}^1(p^2)\!\not\!{p}+\Pi_{\frac{1}{2}}^0(p^2)\, ,
\end{eqnarray}

\begin{eqnarray}
\Pi_{\mu\nu}(p) & = & {\lambda^{-}_{\frac{3}{2}}}^2  {\!\not\!{p}+ M_{-} \over M_{-}^{2}-p^{2}  } \left(- g_{\mu\nu}+\frac{\gamma_\mu\gamma_\nu}{3}+\frac{2p_\mu p_\nu}{3p^2}-\frac{p_\mu\gamma_\nu-p_\nu \gamma_\mu}{3\sqrt{p^2}}\right)\nonumber\\
&&+ {\lambda^{+}_{\frac{3}{2}}}^2  {\!\not\!{p}- M_{+} \over M_{+}^{2}-p^{2}  } \left(- g_{\mu\nu}+\frac{\gamma_\mu\gamma_\nu}{3}+\frac{2p_\mu p_\nu}{3p^2}-\frac{p_\mu\gamma_\nu-p_\nu \gamma_\mu}{3\sqrt{p^2}}\right)   +\cdots  \, ,\nonumber\\
   &=&-\Pi_{\frac{3}{2}}^1(p^2)\!\not\!{p}\,g_{\mu\nu}-\Pi_{\frac{3}{2}}^0(p^2)\,g_{\mu\nu}+\cdots\, ,
\end{eqnarray}

\begin{eqnarray}
\Pi_{\mu\nu\alpha\beta}(p) & = & {\lambda^{-}_{\frac{5}{2}}}^2  {\!\not\!{p}+ M_{-} \over M_{-}^{2}-p^{2}  } \left[\frac{ \widetilde{g}_{\mu\alpha}\widetilde{g}_{\nu\beta}+\widetilde{g}_{\mu\beta}\widetilde{g}_{\nu\alpha}}{2}-\frac{\widetilde{g}_{\mu\nu}\widetilde{g}_{\alpha\beta}}{5}
-\frac{1}{10}\left( \gamma_{\alpha}\gamma_{\mu}+\cdots\right)\widetilde{g}_{\nu\beta}
+\cdots\right]\nonumber\\
&&+   {\lambda^{+}_{\frac{5}{2}}}^2  {\!\not\!{p}- M_{+} \over M_{+}^{2}-p^{2}  }  \left[\frac{ \widetilde{g}_{\mu\alpha}\widetilde{g}_{\nu\beta}+\widetilde{g}_{\mu\beta}\widetilde{g}_{\nu\alpha}}{2}-\frac{\widetilde{g}_{\mu\nu}\widetilde{g}_{\alpha\beta}}{5}
+\cdots\right]   +\cdots \, , \nonumber\\
&=&\Pi_{\frac{5}{2}}^1(p^2)\!\not\!{p}\frac{g_{\mu\alpha}g_{\nu\beta}+g_{\mu\beta}g_{\nu\alpha}}{2} +\Pi_{\frac{5}{2}}^0(p^2)\,\frac{g_{\mu\alpha}g_{\nu\beta}+g_{\mu\beta}g_{\nu\alpha}}{2} + \cdots \, .
\end{eqnarray}
 We choose the components $\Pi_{\frac{1}{2}}^1(p^2)$, $\Pi_{\frac{1}{2}}^0(p^2)$, $\Pi_{\frac{3}{2}}^1(p^2)$, $\Pi_{\frac{3}{2}}^0(p^2)$, $\Pi_{\frac{5}{2}}^1(p^2)$ and $\Pi_{\frac{5}{2}}^0(p^2)$ to explore the spin-parity $J^{P}={\frac{1}{2}}^-$, ${\frac{3}{2}}^-$ and ${\frac{5}{2}}^-$ molecular states, respectively. Again see Sect.{\bf \ref{333-penta-Sect}}  for technical details in tensor analysis.

Again  we obtain the spectral densities through dispersion relation,
\begin{eqnarray}
\frac{{\rm Im}\Pi_{j}^1(s)}{\pi}&=& {\lambda^{-}_{j}}^2 \delta\left(s-M_{-}^2\right)+{\lambda^{+}_{j}}^2 \delta\left(s-M_{+}^2\right) =\, \rho^1_{j,H}(s) \, ,\nonumber \\
\frac{{\rm Im}\Pi^0_{j}(s)}{\pi}&=&M_{-}{\lambda^{-}_{j}}^2 \delta\left(s-M_{-}^2\right)-M_{+}{\lambda^{+}_{j}}^2 \delta\left(s-M_{+}^2\right)
=\rho^0_{j,H}(s) \, ,
\end{eqnarray}
where $j=\frac{1}{2}$, $\frac{3}{2}$, $\frac{5}{2}$, the subscript $H$ denotes  the hadron side,
then we introduce the  weight functions $\sqrt{s}\exp\left(-\frac{s}{T^2}\right)$ and $\exp\left(-\frac{s}{T^2}\right)$ to obtain the QCD sum rules at the hadron side,
\begin{eqnarray}\label{QCDSR-M}
\int_{m_c^2}^{s_0}ds \left[\sqrt{s}\rho^1_{j,H}(s)+\rho^0_{j,H}(s)\right]\exp\left( -\frac{s}{T^2}\right)
&=&2M_{-}{\lambda^{-}_{j}}^2\exp\left( -\frac{M_{-}^2}{T^2}\right) \, ,
\end{eqnarray}
which are free from  contaminations of the positive-parity molecular states.

At the QCD side, we accomplish the operator product expansion with the full light/heavy-quark propagators  up to the vacuum condensates of dimension 13 in a consistent way,  and  obtain   the QCD spectral densities  through  dispersion relation,
\begin{eqnarray}
\frac{{\rm Im}\Pi^{1/0}_{j}(s)}{\pi}&=& \rho^{1/0}_{j,QCD}(s) \, ,
\end{eqnarray}
where $j=\frac{1}{2}$, $\frac{3}{2}$, $\frac{5}{2}$. See Sect.{\bf \ref{Tetra-Positive}} for technical details.

At last, we obtain the QCD sum rules,
\begin{eqnarray}\label{QCDSR-Single-penta}
2M_{-}{\lambda^{-}_{j}}^2\exp\left( -\frac{M_{-}^2}{T^2}\right)
&=& \int_{m_c^2}^{s_0}ds \left[\sqrt{s}\rho^1_{j,QCD}(s)+\rho^0_{j,QCD}(s)\right]\exp\left( -\frac{s}{T^2}\right)\, .
\end{eqnarray}

We differentiate   Eq.\eqref{QCDSR-Single-penta} with respect to  $\tau=\frac{1}{T^2}$, then eliminate the pole residues  to obtain the QCD sum rules for the masses,
\begin{eqnarray}\label{QCDSR-Single-penta-mass}
M^2_{-} &=& \frac{-\frac{d}{d \tau}\int_{m_c^2}^{s_0}ds \,\left[\sqrt{s}\,\rho^1_{QCD}(s)+\,\rho^0_{QCD}(s)\right]\exp\left(- \tau s\right)}{\int_{m_c^2}^{s_0}ds \left[\sqrt{s}\,\rho_{QCD}^1(s)+\,\rho^0_{QCD}(s)\right]\exp\left( -\tau s\right)}\, ,
\end{eqnarray}
where the spectral densities $\rho_{QCD}^1(s)=\rho_{j,QCD}^1(s)$ and $\rho^0_{QCD}(s)=\rho^0_{j,QCD}(s)$.

We take the quark flavor number $n_f = 4$ and resort to  the modified  energy scale formula
\begin{eqnarray}
 \mu &=&\sqrt{M_{P}^2-{\mathbb{M}}_c^2}-k\,{\mathbb{M}}_s \, ,
 \end{eqnarray}
to choose the ideal energy scales of the QCD spectral densities, where $M_P=M_{-}$, the effective $c$-quark mass  ${\mathbb{M}}_c=1.82\,\rm{GeV}$ and effective $s$-quark mass  ${\mathbb{M}}_s=0.2\,\rm{GeV}$, the $k$ is the number of the $s$-quark in the currents/states, see Sect.{\bf \ref{11-tetra-states}} for details.

Routinely, we obtain the Borel windows, continuum threshold parameters, energy scales of the QCD spectral densities and contributions of the $D(13)$, see Table \ref{Borel-Single-heavy}, where the pole dominance is well satisfied.  On the other hand, the highest dimensional vacuum  condensate contributions have the relation $|D(11)|\gg|D(12)|>|D(13)|$,  the operator product expansion is convergent very well.

\begin{table}
\begin{center}
\begin{tabular}{|c|c|c|c|c|c|c|c|c|}\hline\hline
$$      &$J^P$  &\rm{Isospin}      &$\mu$                &$T^2 (\rm{GeV}^2)$   &$\sqrt{s_0}(\rm GeV) $    &pole          &$|D(13)|$

\\\hline

$DN$                 &$\frac{1}{2}^-$   &$0$            &$2.1$        & $1.9-2.3$            &$3.45\pm0.10$       &$(41-70)\%$       &$4.9\%$           \\ \hline

$DN$                 &$\frac{1}{2}^-$   &$1$            &$2.1$        & $1.9-2.3$            &$3.45\pm0.10$       &$(41-70)\%$       &$4.9\%$     \\ \hline

$D\Xi$               &$\frac{1}{2}^-$   &$0$            &$2.2$        &$2.3-2.7$           &$3.85\pm0.10$       &$(41-67)\%$       &$1.2\%$   \\ \hline

$D\Xi$               &$\frac{1}{2}^-$   &$1$            &$2.2$        &$2.3-2.7$          &$3.85\pm0.10$       &$(41-67)\%$       &$1.2\%$    \\ \hline

$D_s\Xi$             &$\frac{1}{2}^-$   &$\frac{1}{2}$  &$2.2$       &$2.4-2.8$             &$3.95\pm0.10$       &$(41-65)\%$       &$0.9\%$       \\ \hline

$D^*N$               &$\frac{3}{2}^-$   &$0$             &$2.3$      & $2.1-2.5$           &$3.60\pm0.10$       &$(37-64)\%$       &$1.2\%$         \\ \hline

$D^*N$               &$\frac{3}{2}^-$   &$1$             &$2.3$      & $2.1-2.5$           &$3.60\pm0.10$       &$(37-64)\%$       &$1.2\%$       \\ \hline

$D^*\Xi$             &$\frac{3}{2}^-$   &$0$             &$2.4$      & $2.5-2.9$            &$4.00\pm0.10$       &$(38-62)\%$       &$0.3\%$         \\ \hline

$D^*\Xi$             &$\frac{3}{2}^-$   &$1$              &$2.4$      & $2.5-2.9$            &$4.00\pm0.10$       &$(38-62)\%$       &$0.3\%$ \\ \hline

$D_s^*\Xi$           &$\frac{3}{2}^-$   &$\frac{1}{2}$    &$2.4$     & $2.5-2.9$            &$4.05\pm0.10$       &$(38-62)\%$       &$0.3\%$\\ \hline

$D\Xi^*$             &$\frac{3}{2}^-$   &$0$              &$2.5$     & $2.5-2.9$           &$4.10\pm0.10$       &$(41-66)\%$       &$0.5\%$     \\ \hline

$D\Xi^*$             &$\frac{3}{2}^-$   &$1$              &$2.5$     & $2.5-2.9$           &$4.10\pm0.10$       &$(41-66)\%$       &$0.5\%$  \\ \hline

$D_s\Xi^*$           &$\frac{3}{2}^-$   &$\frac{1}{2}$    &$2.5$     & $2.6-3.0$            &$4.20\pm0.10$       &$(40-64)\%$       &$0.5\%$ \\  \hline

$D^*\Xi^*$           &$\frac{5}{2}^-$   &$0$              &$2.6$     & $2.6-3.0$            &$4.20\pm0.10$       &$(41-64)\%$       &$0.4\%$ \\  \hline

$D^*\Xi^*$           &$\frac{5}{2}^-$   &$1$              &$2.6$     & $2.6-3.0$            &$4.20\pm0.10$       &$(41-64)\%$       &$0.4\%$ \\ \hline

$D_s^*\Xi^*$        &$\frac{5}{2}^-$    &$\frac{1}{2}$    &$2.6$     & $2.8-3.2$         &$4.30\pm0.10$      &$(40-63)\%$        &$0.2\%$                   \\
\hline\hline
\end{tabular}
\end{center}
\caption{ The spin-parity, isospin, energy scales $\mu$, Borel parameters $T^2$, continuum threshold parameters $s_0$, pole contributions and  vacuum condensate contributions  $|D(13)|$ for the singly-charmed  pentaquark molecular states \cite{XinQ-Single-Heavy-IJMPA-2023}. }\label{Borel-Single-heavy}
\end{table}

\begin{table}
\begin{center}
\begin{tabular}{|c|c|c|c|c|c|c|c|c|}\hline\hline
$$       &$J^P$  &\rm{Isospin}      &$M (\rm{GeV})$    &$\lambda (10^{-4}\rm{GeV}^6) $
&Thresholds\,(\rm{MeV}) &Assignments

\\ \hline

$DN$              &$\frac{1}{2}^-$           &$0$        &$2.81^{+0.13}_{-0.16}$   &$7.98^{+1.98}_{-1.56}$           &$2806$       &                \\ \hline

$DN$              &$\frac{1}{2}^-$           &$1$        &$2.81^{+0.13}_{-0.16}$   &$7.98^{+1.98}_{-1.56}$           &$2806$       & ? $\Sigma_c(2800)$           \\ \hline

$D\Xi$            &$\frac{1}{2}^-$           &$0$        & $3.19^{+0.11}_{-0.13}$           &$14.22^{+2.74}_{-2.40}$          &$3186$       & ? $\Omega_c(3185)$        \\ \hline

$D\Xi$            &$\frac{1}{2}^-$           &$1$        & $3.19^{+0.11}_{-0.13}$           &$14.22^{+2.74}_{-2.40}$          &$3186$       &                \\ \hline

$D_s\Xi$          &$\frac{1}{2}^-$      &$\frac{1}{2}$   &$3.28^{+0.12}_{-0.12}$           &$16.04^{+3.04}_{-2.66}$         &$3287$    &                          \\ \hline

$D^*N$            &$\frac{3}{2}^-$           &$0$        &$2.96^{+0.13}_{-0.14}$           &$8.87^{+1.84}_{-1.59}$          &$2947$     & ?? $\Lambda_c(2940)/\Lambda_c(2910)$
\\ \hline

$D^*N$            &$\frac{3}{2}^-$           &$1$         &$2.96^{+0.13}_{-0.14}$           &$8.87^{+1.84}_{-1.59}$          &$2947$     &                       \\ \hline

$D^*\Xi$          &$\frac{3}{2}^-$           &$0$         &$3.35^{+0.11}_{-0.13}$          &$15.89^{+2.93}_{-2.56}$         &$3327$     & ? $\Omega_c(3327)$    \\ \hline

$D^*\Xi$          &$\frac{3}{2}^-$           &$1$         & $3.35^{+0.11}_{-0.13}$          &$15.89^{+2.93}_{-2.56}$         &$3327$     &                                 \\ \hline

$D_s^*\Xi$        &$\frac{3}{2}^-$     &$\frac{1}{2}$     &$3.44^{+0.11}_{-0.11}$             &$17.24^{+3.06}_{-2.66}$
&$3430$    &          \\  \hline

$D\Xi^*$          &$\frac{3}{2}^-$          &$0$          & $3.41^{+0.11}_{-0.12}$
&$10.67^{+1.92}_{-1.68}$         &$3399$    &   \\ \hline

$D\Xi^*$          &$\frac{3}{2}^-$          &$1$          & $3.41^{+0.11}_{-0.12}$
&$10.67^{+1.92}_{-1.68}$         &$3399$       &   \\ \hline

$D_s\Xi^*$        &$\frac{3}{2}^-$     &$\frac{1}{2}$      & $3.51^{+0.11}_{-0.13}$
&$11.93^{+2.11}_{-1.85}$         &$3500$    &       \\  \hline

$D^*\Xi^*$        &$\frac{5}{2}^-$         &$0$             &$3.54^{+0.11}_{-0.12}$                &$10.93^{+1.94}_{-1.71}$         &$3540$    &                       \\  \hline

$D^*\Xi^*$        &$\frac{5}{2}^-$         &$1$             &$3.54^{+0.11}_{-0.12}$                &$10.93^{+1.94}_{-1.71}$         &$3540$             &              \\ \hline

$D_s^*\Xi^*$      &$\frac{5}{2}^-$      &$\frac{1}{2}$      & $3.65^{+0.11}_{-0.11}$        &$12.91^{+2.19}_{-1.93}$         &$3644$              &                       \\
\hline\hline
\end{tabular}
\end{center}
\caption{ The masses, pole residues, (corresponding meson-baryon) thresholds and possible assignments for the singly-charmed pentaquark  molecular states \cite{XinQ-Single-Heavy-IJMPA-2023}. }\label{mass-residue-single-heavy}
\end{table}

At last, we obtain the masses and pole residues of the pentaquark molecular states, see Table \ref{mass-residue-single-heavy} \cite{XinQ-Single-Heavy-IJMPA-2023}. From Tables \ref{Borel-Single-heavy}-\ref{mass-residue-single-heavy}, it is obvious that the  modified  energy scale formula $\mu= \sqrt{M_{P}^2-{\mathbb{M}}_c^2}-k\,{\mathbb{M}}_s$ is satisfied in most cases.   The energy scales $ \sqrt{M_{P}^2-{\mathbb{M}}_c^2}-3{\mathbb{M}}_s$ for the $D_s\Xi$, $D_s^*\Xi$ and $D_s\Xi^*$ states are smaller than the  energy scales $ \sqrt{M_{P}^2-{\mathbb{M}}_c^2}-2{\mathbb{M}}_s$ for the corresponding  $D\Xi$, $D^*\Xi$ and $D\Xi^*$ states, respectively,  we choose the same energy scales $ \sqrt{M_{P}^2-{\mathbb{M}}_c^2}-2{\mathbb{M}}_s$ for those cousins, as larger masses correspond to larger energy scales naively.

The predicted masses $2.81^{+0.13}_{-0.16}~\rm{GeV}$ and $3.19^{+0.11}_{-0.13}~\rm{GeV}$ for the $DN$ and $D\Xi$ molecular states (respectively) with the $J^P =\frac{1}{2}^-$ support assigning  the $\Sigma_c(2800)$ and $\Omega_c(3185)$ as the $DN$ and $D\Xi$  molecular states, respectively. The predicted mass  $3.35^{+0.11}_{-0.13}~\rm{GeV}$  for  the $D^*\Xi$ molecular  state with the $J^P =\frac{3}{2}^-$  supports  assigning  the $\Omega_c(3327)$  as the $D^*\Xi$ molecular  state. The predicted mass  $2.96^{+0.14}_{-0.13}~\rm{GeV}$ for the $D^*N$ molecular state is consistent with the $\Lambda_c(2940)/\Lambda_c(2910)$ within uncertainties, which does not exclude the possibility that they are molecular states.

In Ref.\cite{ZJR-siglam}, J. R. Zhang studies the $\Sigma_{c}(2800)$ and $\Lambda_{c}(2940)$ as the $S$-wave $DN$ and $D^{*}N$ molecule candidates respectively with the QCD sum rules,  obtains the masses $3.64\pm0.33~\mbox{GeV}$ and $3.73\pm0.35~\mbox{GeV}$
for the $S$-wave $DN$ state with the $J^{P}=\frac{1}{2}^{-}$
 and  $S$-wave $D^{*}N$ state with the  $J^{P}=\frac{3}{2}^{-}$, respectively, which  are somewhat bigger than the experimental data for the $\Sigma_{c}(2800)$ and $\Lambda_{c}(2940)$,
respectively, and differ from the present calculations significantly.  We should bear in mind that  J. R. Zhang chooses {\bf Scheme II}  while we choose {\bf Scheme I}.

Taking the $D\Xi$ ($D^*\Xi$)  pentaquark molecular states with the $J^P={\frac{1}{2}}^-$ (${\frac{3}{2}}^-$)   as an example,  we obtain a symmetric
isotriplet  $| 1,1 \,\rangle$, $|1,0 \rangle$, $|1,-1\,\rangle$ and an antisymmetric isosinglet $| 0,0 \,\rangle$. The $\Omega_c(3185)$ ($\Omega_c(3327)$) is a good candidate for the  $D\Xi$ ($D^*\Xi$) molecular state with the isospin $| 0,0\, \rangle$. We expect to search for  the  molecular states with the isospin $| 1,1 \,\rangle$ and $| 1,-1 \,\rangle$  in  the $\Xi_c^+\bar{K}^0$  and $\Xi_c^0\bar{K}^-$ mass spectra respectively to shed light on the nature of the $\Omega_c(3185)$ ($\Omega_c(3327)$).

Other interpretations also exist, the $\Sigma_c(2800)$ could be assigned as overlap of the $\Sigma_c (2813)$ and $\Sigma_c(2840)$ \cite{ZXH-2800-1} or P-wave charmed baryon state with the   $J^P=\frac{1}{2}^- /\frac{3}{2}^-/\frac{5}{2}^-$ \cite{ZXH-2800-2,CHX-2800,JDJ-2800,YGL-2800-2940}. Cheng et al  suggest  that the $\Sigma_c(2800)$  is not possible to be a $J^P=\frac{1}{2}^-$ charmed baryon state \cite{CHY-2800-2940}.
The $\Lambda_c^+(2940)$ is most likely to be the $J^P=\frac{1}{2}^-$ or $\frac{3}{2}^-$ (2P) state \cite{YGL-2800-2940,CHY-2800-2940,LX-2940,LQF-2940-1}.
And the $\Lambda_c^+(2910)$ is probably the $J^P=\frac{1}{2}^-$ (2P) \cite{Azizi-2910} or $\frac{5}{2}^-$ (1P) state \cite{ZXH-2910}.
The $\Omega_c(3185)$ lies in the region of the 2S state \cite{YGL-baryonQ,Karliner-3185-3327}, while the $\Omega_c(3327)$ is a good candidate for the 2S $\Omega_c$ state \cite{Karliner-3185-3327} or D-wave  $\Omega_c$ baryon state with the $J^P=\frac{1}{2}^+/\frac{3}{2}^+/\frac{5}{2}^+$ \cite{YGL-baryonQ,LX-3327,WZG-3327}.

With a simple replacement $c\to b$, we obtain the corresponding QCD sum rules for the singly-bottom pentaquark molecular states, see Eqs.\eqref{QCDSR-Single-penta}-\eqref{QCDSR-Single-penta-mass}.
In Ref.\cite{WangHJ-Xi6227-mole-IJTP-2020}, we choose  the current,
\begin{eqnarray}
   J(x)&=&\bar{u}(x)i\gamma_{5}s(x)\varepsilon^{i j k} u_{i}^T(x)C\gamma_{\alpha}d_{j}(x)\gamma^{\alpha}\gamma_{5}b_{k}(x) \, ,
\end{eqnarray}
to interpolate   the  molecular states with the $J^{P}=\frac{1}{2}^{\pm}$, the prediction supports assigning the $\Xi_b(6227)$ as the pentaquark molecular state with the $J^{P}=\frac{1}{2}^{-}$.

In Ref.\cite{WDW-omegac-penta-CTP-2021},  we construct the $\bar{\mathbf 3}\bar{\mathbf 3}\bar{\mathbf 3}$ type five-quark currents,
 \begin{eqnarray}
 J_{SS}(x)&=&\varepsilon^{i l a}\varepsilon^{i j k}\varepsilon^{l m n} s_{j}^{ T} (x)C\gamma_{5} u_{k}(x) s_{m}^{ T} (x)C\gamma_{5} c_{n}(x)C \bar{u}_{a}^{ T}(x) \, ,\nonumber\\
   J_{AA}(x)&=&\varepsilon^{i l a}\varepsilon^{i j k}\varepsilon^{l m n} s_{j}^{ T} (x)C\gamma_{\mu} u_{k}(x) s_{m}^{ T} (x)C\gamma^{\mu} c_{n}(x)C \bar{u}_{a}^{ T}(x)\, ,
\end{eqnarray}
to interpolate  the singly-charmed pentaquark states $susc\bar{u}$ with the $J^{P}=\frac{1}{2}^{\pm}$. With a simple replacement,
\begin{eqnarray}
susc\bar{u} &\to& \frac{1}{\sqrt{2}}\left(susc\bar{u}+sdsc\bar{d} \right)\, ,
\end{eqnarray}
we obtain the isospin singlet current, the expressions  of the QCD sum rules survive. The predictions  support  assigning the excited $\Omega_c$ states from the LHCb collaboration as the $SS$ or $AA$-type pentaquark states with the $J^P={1\over 2}^-$ and the mass about $3.08\,\rm{GeV}$.
We group the quark flavors as $[su][sc]\bar{u}$ to construct the  currents, if we group the quark flavors as $[ss][cu]\bar{u}$, the $SS$-type current will vanish  due to Fermi-Dirac statistics, but the $AA$-type current survives and leads to almost degenerated mass  according to the light-flavor $SU(3)$ symmetry.

In Ref.\cite{ZGWJXZ-omegac-penta-mole},  we  study  the $\bar{\mathbf 3}\bar{\mathbf 3}\bar{\mathbf 3}$ type singly-charmed pentaquark states with the $J^P={\frac{3}{2}}^\pm$
via the QCD sum rules. We distinguish the isospins in constructing the interpolating currents via two clusters, a diquark $q^{\prime T}_j C\gamma_\mu q^{\prime\prime}_k$ ($D_i$) plus a triquark $q^{\prime\prime\prime T}_m C\gamma_5 c_n C\bar{q}^{T}_{a}$ ($T_{mna}$), which have the properties,
\begin{eqnarray}
\hat{I}^2 \,\varepsilon_{ijk} q^T_j C\gamma_\mu q^\prime_j &=& 1(1+1)\,\varepsilon_{ijk} q^T_j C\gamma_\mu q^\prime_j \, , \nonumber\\
\hat{I}^2 \,\varepsilon_{ijk} q^T_j C\gamma_\mu s_j &=& \frac{1}{2}\left(\frac{1}{2}+1\right)\,\varepsilon_{ijk} q^T_j C\gamma_\mu s_j \, , \nonumber\\
\hat{I}^2 \,\varepsilon_{ijk} s^T_j C\gamma_\mu s_j &=& 0(0+1)\,\varepsilon_{ijk} s^T_j C\gamma_\mu s_j \, ,
\end{eqnarray}
\begin{eqnarray}
\hat{I}^2 \, u^T_m C\gamma_5 c_n C \bar{d}^T_a &=& 1\left(1+1\right)\,u^T_m C\gamma_5 c_n C \bar{d}^T_a \, ,  \nonumber\\
\hat{I}^2 \, d^T_m C\gamma_5 c_n C \bar{u}^T_a &=& 1\left(1+1\right)\,d^T_m C\gamma_5 c_n C \bar{u}^T_a \, , \nonumber\\
\hat{I}^2 \,\left[ u^T_m C\gamma_5 c_n C \bar{u}^T_a -d^T_m C\gamma_5 c_n C \bar{d}^T_a\right]&=& 1\left(1+1\right)\,\left[ u^T_m C\gamma_5 c_n C \bar{u}^T_a -d^T_m C\gamma_5 c_n C \bar{d}^T_a\right]\, , \nonumber\\
\hat{I}^2 \,\left[ u^T_m C\gamma_5 c_n C \bar{u}^T_a +d^T_m C\gamma_5 c_n C \bar{d}^T_a\right]&=& 0\left(0+1\right)\,\left[ u^T_m C\gamma_5 c_n C \bar{u}^T_a +d^T_m C\gamma_5 c_n C \bar{d}^T_a\right]\, , \nonumber\\
\hat{I}^2 \, q^T_m C\gamma_5 c_n C \bar{s}^T_a &=& \frac{1}{2}\left(\frac{1}{2}+1\right)\,q^T_m C\gamma_5 c_n C \bar{s}^T_a \, , \nonumber\\
\hat{I}^2 \, s^T_m C\gamma_5 c_n C \bar{q}^T_a &=& \frac{1}{2}\left(\frac{1}{2}+1\right)\,s^T_m C\gamma_5 c_n C \bar{q}^T_a \, , \nonumber\\
\hat{I}^2 \, s^T_m C\gamma_5 c_n C \bar{s}^T_a &=& 0\left(0+1\right)\,s^T_m C\gamma_5 c_n C \bar{s}^T_a \, ,
\end{eqnarray}
with $q$, $q^\prime=u$, $d$, the $\hat{I}^2$ is the isospin operator. The diquark clusters $D_i$ and triquark clusters $T_{mna}$ have the isospins $I=1$, $\frac{1}{2}$ or $0$.
We could  obtain some mass relations based on the $SU(3)$ breaking effects of the $u$, $d$, $s$ quarks, the mass relation among the diquark clusters $D_i$ is $m_{I=0}-m_{I=\frac{1}{2}}=m_{I=\frac{1}{2}}-m_{I=1}$,
while the mass relation among the triquark clusters $T_{mna}$ is $m_{I=0}-m_{I=\frac{1}{2}}=m_{I=\frac{1}{2}}-m_{I=1}$, if the hidden-flavor (for $u$ and $d$) isospin singlet
$u^T_m C\gamma_5 c_n C \bar{u}^T_a +d^T_m C\gamma_5 c_n C \bar{d}^T_a$ is excluded. Moreover,  the isospin triplet $u^T_m C\gamma_5 c_n C \bar{u}^T_a -d^T_m C\gamma_5 c_n C \bar{d}^T_a$
and isospin singlet $u^T_m C\gamma_5 c_n C \bar{u}^T_a +d^T_m C\gamma_5 c_n C \bar{d}^T_a$ are expected to have degenerate  masses,
which can be inferred from the tiny mass difference between the
vector mesons $\rho^0(770)$ and $\omega(780)$. In fact, if we choose the currents $J_\mu(x)=\bar{u}(x)\gamma_\mu u(x)-\bar{d}(x)\gamma_\mu d(x)$ and
$\bar{u}(x)\gamma_\mu u(x)+\bar{d}(x)\gamma_\mu d(x)$ to interpolate the $\rho^0(770)$ and $\omega(780)$, respectively, we obtain the same  QCD sum rules.

We write down the currents explicitly,
\begin{eqnarray}
J_{q^{\prime}q^{\prime\prime}q^{\prime\prime\prime}\bar{q},\mu}(x)&=&\varepsilon^{ila} \varepsilon^{ijk}\varepsilon^{lmn}
q^{\prime T}_j(x) C\gamma_\mu q^{\prime\prime}_k(x) q^{\prime\prime\prime T}_m(x) C\gamma_5 c_n(x) C\bar{q}^{T}_{a}(x) \, ,
\end{eqnarray}
and  study the  singly-charmed pentaquark states $uuuc\bar{u}$ and $sssc\bar{s}$
 with the QCD sum rules in details.  Then we estimate the masses of the singly-charmed pentaquark states $ssuc\bar{u}$, $susc\bar{u}$, $ssdc\bar{d}$ and $sdsc\bar{d}$\
  with $J^P={\frac{3}{2}}^-$ to be $3.15\pm0.13\,\rm{GeV}$ according to the
  light-flavor $SU(3)$ breaking effects, which is compatible with the experimental values of
  the masses of the $\Omega_c(3050)$, $\Omega_c(3066)$, $\Omega_c(3090)$,
  $\Omega_c(3119)$.

\section{Strong decays of exotic  states}
In Refs.\cite{WangZhang-Solid,WangZG-Y4660-tetra-decay-EPJC-2019}, we suggest rigorous quark-hadron duality
 to calculate the hadronic coupling constants in the two-body strong
 decays of the tetraquark states with the QCD sum rules. At first,
we write down the  three-point correlation functions $\Pi(p,q)$,
\begin{eqnarray}
\Pi(p,q)&=&i^2\int d^4xd^4y e^{ip\cdot x}e^{iq\cdot y}\langle 0|T\left\{J_{B}(x)J_{C}(y)J_{A}^{\dagger}(0)\right\}|0\rangle\, ,
\end{eqnarray}
where the currents $J_A(0)$ interpolate the tetraquark states $A$, the currents $J_B(x)$ and $J_C(y)$ interpolate the  conventional mesons $B$ and $C$,  respectively,
\begin{eqnarray}
\langle0|J_{A}(0)|A(p^\prime)\rangle&=&\lambda_{A} \,\, , \nonumber \\
\langle0|J_{B}(0)|B(p)\rangle&=&\lambda_{B} \,\, , \nonumber \\
\langle0|J_{C}(0)|C(q)\rangle&=&\lambda_{C} \,\, ,
\end{eqnarray}
the $\lambda_A$, $\lambda_B$  and $\lambda_{C}$ are the pole residues or   decay constants.

At the hadron side,  we insert  a complete set of intermediate hadronic states with the same quantum numbers as the currents $J_A(0)$, $J_B(x)$, $J_{C}(y)$ into the three-point correlation functions $\Pi(p,q)$ and  isolate the ground state
contributions to obtain the  result \cite{WangZG-Y4660-tetra-decay-EPJC-2019},
\begin{eqnarray}
\Pi(p,q)&=& \frac{\lambda_{A}\lambda_{B}\lambda_{C}G_{ABC} }{(m_{A}^2-p^{\prime2})(m_{B}^2-p^2)(m_{C}^2-q^2)}+ \frac{1}{(m_{A}^2-p^{\prime2})(m_{B}^2-p^2)} \int_{s^0_C}^\infty dt\frac{\rho_{AC^\prime}(p^{\prime 2},p^2,t)}{t-q^2}\nonumber\\
&& + \frac{1}{(m_{A}^2-p^{\prime2})(m_{C}^2-q^2)} \int_{s^0_{B}}^\infty dt\frac{\rho_{AB^\prime}(p^{\prime 2},t,q^2)}{t-p^2}  \nonumber\\
&& + \frac{1}{(m_{B}^2-p^{2})(m_{C}^2-q^2)} \int_{s^0_{A}}^\infty dt\frac{\rho_{A^{\prime}B}(t,p^2,q^2)+\rho_{A^{\prime}C}(t,p^2,q^2)}{t-p^{\prime2}}+\cdots \nonumber\\
&=&\Pi(p^{\prime2},p^2,q^2) \, ,
\end{eqnarray}
where $p^\prime=p+q$,  the $G_{ABC}$  are the hadronic coupling constants defined by
\begin{eqnarray}
\langle B(p)C(q)|A(p^{\prime})\rangle&=&i G_{ABC}  \, ,
\end{eqnarray}
the four   functions $\rho_{AC^\prime}(p^{\prime 2},p^2,t)$, $ \rho_{AB^\prime}(p^{\prime 2},t,q^2)$,
$ \rho_{A^{\prime}B}(t^\prime,p^2,q^2)$ and $\rho_{A^{\prime}C}(t^\prime,p^2,q^2)$
   have complex dependence on the transitions between the ground states and the higher resonances  or  continuum states.

We rewrite the correlation functions  $\Pi_H(p^{\prime 2},p^2,q^2)$ at the hadron  side as
\begin{eqnarray}
\Pi_{H}(p^{\prime 2},p^2,q^2)&=&\int_{\Delta^2}^{s_{A}^0}ds^\prime \int_{\Delta_s^2}^{s^0_{B}}ds \int_{\Delta_u^2}^{u^0_{C}}du  \frac{\rho_H(s^\prime,s,u)}{(s^\prime-p^{\prime2})(s-p^2)(u-q^2)}\nonumber\\
&&+\int_{s^0_A}^{\infty}ds^\prime \int_{\Delta_s^2}^{s^0_{B}}ds \int_{\Delta_u^2}^{u^0_{C}}du  \frac{\rho_H(s^\prime,s,u)}{(s^\prime-p^{\prime2})(s-p^2)(u-q^2)}+\cdots\, ,
\end{eqnarray}
 through triple-dispersion relation, where the $\rho_H(s^\prime,s,u)$   are the hadronic spectral densities,
\begin{eqnarray}
\rho_H(s^\prime,s,u)&=&{\lim_{\epsilon_3\to 0}}\,\,{\lim_{\epsilon_2\to 0}} \,\,{\lim_{\epsilon_1\to 0}}\,\,\frac{ {\rm Im}_{s^\prime}\, {\rm Im}_{s}\,{\rm Im}_{u}\,\Pi_H(s^\prime+i\epsilon_3,s+i\epsilon_2,u+i\epsilon_1) }{\pi^3} \, ,
\end{eqnarray}
where the $\Delta^2$, $\Delta_s^2$ and $\Delta_u^2$ are the thresholds, the  $s_{A}^0$, $s_{B}^0$, $u_{C}^0$ are the continuum thresholds.

Now we carry out the operator product expansion at the QCD side, and write the correlation functions  $\Pi_{QCD}(p^{\prime 2},p^2,q^2)$  as
\begin{eqnarray}
\Pi_{QCD}(p^{\prime 2},p^2,q^2)&=&  \int_{\Delta_s^2}^{s^0_{B}}ds \int_{\Delta_u^2}^{u^0_{C}}du  \frac{\rho_{QCD}(p^{\prime2},s,u)}{(s-p^2)(u-q^2)}+\cdots\, ,
\end{eqnarray}
through double-dispersion relation, where the $\rho_{QCD}(p^{\prime 2},s,u)$   are the QCD spectral densities,
\begin{eqnarray}
\rho_{QCD}(p^{\prime 2},s,u)&=& {\lim_{\epsilon_2\to 0}} \,\,{\lim_{\epsilon_1\to 0}}\,\,\frac{  {\rm Im}_{s}\,{\rm Im}_{u}\,\Pi_{QCD}(p^{\prime 2},s+i\epsilon_2,u+i\epsilon_1) }{\pi^2} \, .
\end{eqnarray}
As the QCD spectral densities $\rho_{QCD}(s^\prime,s,u)$ do  not exist,
\begin{eqnarray}
\rho_{QCD}(s^\prime,s,u)&=&{\lim_{\epsilon_3\to 0}}\,\,{\lim_{\epsilon_2\to 0}} \,\,{\lim_{\epsilon_1\to 0}}\,\,\frac{ {\rm Im}_{s^\prime}\, {\rm Im}_{s}\,{\rm Im}_{u}\,\Pi_{QCD}(s^\prime+i\epsilon_3,s+i\epsilon_2,u+i\epsilon_1) }{\pi^3} \nonumber\\
&=&0\, ,
\end{eqnarray}
because
\begin{eqnarray}
{\lim_{\epsilon_3\to 0}}\,\,\frac{ {\rm Im}_{s^\prime}\,\Pi_{QCD}(s^\prime+i\epsilon_3,p^2,q^2) }{\pi} &=&0\, .
\end{eqnarray}
Thereafter we will write the QCD spectral densities  $\rho_{QCD}(p^{\prime 2},s,u)$ as $\rho_{QCD}(s,u)$ for simplicity.

We match the hadron side  with the QCD side of the correlation functions,
and accomplish  the integral over $ds^\prime$  firstly to obtain the rigorous quark-hadron  duality \cite{WangZhang-Solid},
\begin{eqnarray}
\int_{\Delta_s^2}^{s_B^0} ds \int_{\Delta_u^2}^{u_C^0} du \frac{\rho_{QCD}(s,u)}{(s-p^2)(u-q^2)}&=&\int_{\Delta_s^2}^{s_B^0} ds \int_{\Delta_u^2}^{u_C^0} du \frac{1}{(s-p^2)(u-q^2)}\left[ \int_{\Delta^2}^{\infty} ds^\prime \frac{\rho_{H}(s^{\prime},s,u)}{s^\prime-p^{\prime2}}\right]\, , \nonumber\\
\end{eqnarray}
 the  $\Delta^2$ denotes the thresholds $(\Delta_{s}+\Delta_{u})^2$.
 Now we write down  the quark-hadron duality explicitly,
 \begin{eqnarray}\label{Solid-explicit}
  \int_{\Delta_s^2}^{s^0_{B}}ds \int_{\Delta_u^2}^{u^0_{C}}du  \frac{\rho_{QCD}(s,u)}{(s-p^2)(u-q^2)}&=& \int_{\Delta_s^2}^{s^0_{B}}ds \int_{\Delta_u^2}^{u^0_{C}}du   \int_{\Delta^2}^{\infty}ds^\prime \frac{\rho_H(s^\prime,s,u)}{(s^\prime-p^{\prime2})(s-p^2)(u-q^2)} \nonumber\\
  &=&\frac{\lambda_{A}\lambda_{B}\lambda_{C}G_{ABC} }{(m_{A}^2-p^{\prime2})(m_{B}^2-p^2)(m_{C}^2-q^2)} +\frac{C_{A^{\prime}B}+C_{A^{\prime}C}}{(m_{B}^2-p^{2})(m_{C}^2-q^2)} \, . \nonumber\\
\end{eqnarray}
 No approximation is needed, we do not need the continuum threshold parameter $s^0_{A}$ in the $s^\prime$ channel, as we  match the hadron side with the QCD side  below the continuum thresholds $s_0$ and $u_0$ to obtain rigorous quark-hadron  duality, and we take account of the continuum contributions in the $s^\prime$ channel.

 In Eq.\eqref{Solid-explicit}, we introduce the parameters $C_{AC^\prime}$, $C_{AB^\prime}$, $C_{A^\prime B}$ and $C_{A^\prime C}$   to parameterize the net effects,
\begin{eqnarray}
C_{AC^\prime}&=&\int_{s^0_C}^\infty dt\frac{ \rho_{AC\prime}(p^{\prime 2},p^2,t)}{t-q^2}\, ,\nonumber\\
C_{AB^\prime}&=&\int_{s^0_{B}}^\infty dt\frac{\rho_{AB^\prime}(p^{\prime 2},t,q^2)}{t-p^2}\, ,\nonumber\\
C_{A^\prime B}&=&\int_{s^0_{A}}^\infty dt\frac{ \rho_{A^\prime B}(t,p^2,q^2)}{t-p^{\prime2}}\, ,\nonumber\\
C_{A^\prime C}&=&\int_{s^0_{A}}^\infty dt\frac{ \rho_{A^\prime C}(t,p^2,q^2)}{t-p^{\prime2}}\, .
\end{eqnarray}
In numerical calculations,   we   take the relevant functions $C_{A^\prime B}$ and $C_{A^\prime C}$  as free parameters, and choose suitable values  to eliminate the contaminations from the higher resonances and continuum states to obtain the stable QCD sum rules with the variations of
the Borel parameters $T^2$.

 According to the discussions in Sect.{\bf\ref{reliable?}}, the quantum field theory  does not forbid the couplings between the four-quark currents $J_A(0)$ and two-meson scattering states  $BC$, if they have the same quantum numbers.
 The local currents $J_A(0)$ have direct non-vanishing   couplings to the two-meson scattering states $BC$, although the overlaps of the wave-functions are very small  \cite{WZG-local-current}, which leads to a finite  width to modify the dispersion relation, see Eqs.\eqref{Self-Energy}-\eqref{Modify-width}.

There exists another term $\Pi_{H}^{D}(p^{\prime 2},p^2,q^2)$ at the hadron side beyond that shown in Eq.\eqref{Solid-explicit},
\begin{eqnarray}\label{Direct-solid}
\Pi_{H}^{D}(p^{\prime 2},p^2,q^2)&=&\frac{\lambda_{B}\lambda_{C}\lambda_{BC}}{(m_{B}^2-p^2)(m_{C}^2-q^2)}+\cdots,
\end{eqnarray}
where
\begin{eqnarray}
\langle 0|J_A(0)|B(p)C(q)\rangle &=&\lambda_{BC}\, .
\end{eqnarray}
Such terms $\Pi_{H}^{D}(p^{\prime 2},p^2,q^2)$ shown  in Eq.\eqref{Direct-solid}
could be absorbed into the parameters $C_{A^{\prime}B}+C_{A^{\prime}C}$ with the
simple replacement,
\begin{eqnarray}
C_{A^{\prime}B}+C_{A^{\prime}C} &\to& C_{A^{\prime}B}+C_{A^{\prime}C}+\lambda_{B}\lambda_{C}\lambda_{BC}\, .
\end{eqnarray}

In Ref.\cite{Nielsen-Zc3900-decay-PRD-2013,Nielsen-X3873-decay-PLB-2006},
 Nielsen et al approximate  the hadron side of the correlation functions as
\begin{eqnarray}\label{Nielsen-decay-hadron-side}
\Pi_{H}(p^{\prime2},p^2,q^2)&=& \frac{\lambda_{A}\lambda_{B}\lambda_{C}G_{ABC} }{(m_{A}^2-p^{\prime2})(m_{B}^2-p^2)(m_{C}^2-q^2)} +\frac{B_{H}}{(s_0^\prime-p^{\prime2})(m_{C}^2-q^2)}\, ,
\end{eqnarray}
then match them with the QCD side below the continuum threshold $s_0$ by taking the chiral limit $m_C^2\to 0$ and $q^2 \to 0$ sequentially, where the $B_{H}$ stands  for the pole-continuum transitions, and we have rewritten their notations  into the present form for
convenience. Although Nielsen et al take account of the continuum contributions by introducing a parameter $s^\prime_0$ in the $s^\prime$ channel phenomenologically, they neglect the continuum contributions in the $u$ channel at the hadron side by hand. Such  an approximation is also adopted in Refs.\cite{Review-QCDSR-Nielsen-JPG-2019,WangZG-Zb10610-tetra-NPA-2014,Nielsen-X5568-decay-PLB-2016,WChen-Zc4200-decay-EPJC-2015}.

There is another scheme to study the strong hadronic coupling constants with the correlation functions $\Pi(p,p^\prime)$,
\begin{eqnarray}
\Pi(p,p^\prime)&=&i^2\int d^4xd^4y e^{-ip \cdot x}e^{ip^{\prime}\cdot y}\langle 0|T\left\{J_{B}(y)J_{C}(0)J_{A}^{\dagger}(x)\right\}|0\rangle\, .
\end{eqnarray}
At the QCD side, a double-dispersion relation,
\begin{eqnarray}
\Pi(p^2,p^{\prime2},q^2)&=&\int_{\Delta_s^2}^{s_0} ds \int_{\Delta_{s^\prime}^2}^{s^\prime_0} ds^\prime \frac{\rho(s,s^\prime,q^2)}{(s-p^2)(s^\prime-p^{\prime2})}
+\cdots \, ,
\end{eqnarray}
with $q=p^\prime-p$ is adopted \cite{Azizi-Z4100-tetra-EPJC-2019,Azizi-Tcc-tetra-NPB-2022}. However, we cannot obtain the QCD spectral densities $\rho(s,s^\prime,q^2)$. Such an scheme is also adopted in the case of pentaquark states \cite{Azizi-Pc4459-decay-PRD-2021,Azizi-Pcsss-decay-PRD-2023,
Azizi-Pcss-decay-EPJC-2022,Azizi-Pcs4338-decay-PRD-2023}.

Let us turn to Eq.\eqref{Solid-explicit} again. If the  $B$  are charmonium or bottomonium  states, we set  $p^{\prime2}=p^2$  and perform the double Borel transformation   with respect to the variables $P^2=-p^2$ and $Q^2=-q^2$, respectively  to obtain the  QCD sum rules,
\begin{eqnarray}
&& \frac{\lambda_{A}\lambda_{B}\lambda_{C}G_{ABC}}{m_{A}^2-m_{B}^2} \left[ \exp\left(-\frac{m_{B}^2}{T_1^2} \right)-\exp\left(-\frac{m_{A}^2}{T_1^2} \right)\right]\exp\left(-\frac{m_{C}^2}{T_2^2} \right) +\nonumber\\
&&\left(C_{A^{\prime}B}+C_{A^{\prime}C}\right) \exp\left(-\frac{m_{B}^2}{T_1^2} -\frac{m_{C}^2}{T_2^2} \right)=\int_{\Delta_s^2}^{s_B^0} ds \int_{\Delta_u^2}^{u_C^0} du\, \rho_{QCD}(s,u)\exp\left(-\frac{s}{T_1^2} -\frac{u}{T_2^2} \right)\, ,
\end{eqnarray}
where the $T_1^2$ and $T_2^2$ are the Borel parameters.
If the  $B$   are open-charm or open-bottom mesons, we set  $p^{\prime2}=4p^2$  and perform the double Borel transformation   with respect to the variables $P^2=-p^2$ and $Q^2=-q^2$, respectively  to obtain the  QCD sum rules,
\begin{eqnarray}
&& \frac{\lambda_{A}\lambda_{B}\lambda_{C}G_{ABC}}{4\left(\widetilde{m}_{A}^2-m_{B}^2\right)} \left[ \exp\left(-\frac{m_{B}^2}{T_1^2} \right)-\exp\left(-\frac{\widetilde{m}_{A}^2}{T_1^2} \right)\right]\exp\left(-\frac{m_{C}^2}{T_2^2} \right) +\nonumber\\
&&\left(C_{A^{\prime}B}+C_{A^{\prime}C}\right) \exp\left(-\frac{m_{B}^2}{T_1^2} -\frac{m_{C}^2}{T_2^2} \right)=\int_{\Delta_s^2}^{s_B^0} ds \int_{\Delta_u^2}^{u_C^0} du\, \rho_{QCD}(s,u)\exp\left(-\frac{s}{T_1^2} -\frac{u}{T_2^2} \right)\, ,
\end{eqnarray}
where $\widetilde{m}_A^2=\frac{m_A^2}{4}$. Or set $p^{\prime2}=p^2$, just like in the first case.
The scheme based on the rigorous quark-hadron duality is adopted in Refs.\cite{WZG-X3872-decay-PRD-2024,WangZG-Zcs4123-tetra-CPC-2022,
WangZG-X4140-decay-EPJC-2019,WangZG-X4274-decay-APPB-2020,
WangZG-Zc4600-tetra-decay-IJMPA-2019,
WangZG-Pc4312-decay-CPC-2020,
WangZG-Pc4312-decay-CPC-2021,YLLiu-Pc4312-decay-EPJC-2021,
WangZG-Zcs3985-tetra-decay-CPC-2022,WangZG-QQQQ-decay-AAPPS-2024,
WangXW-Pc4312-decay-CPC-2024,WangXW-Pcs4338-decay-PRD-2024,
LiuYL-Pc4312-decay-PRD-2020,WangZG-Zc3900-Momemt-EPJC-2018}.

\subsection{Strong decays of the $Y(4500)$ as an example}\label{Subsect-Y4500}
In this sub-section, we would like to use a typical example to illustrate the procedure in details.

After Ref.\cite{WZG-HC-Vect-NPB-2021} was published, the $Y(4500)$ was observed by the BESIII collaboration \cite{BESIII-Y4500-KK-CPC-2022,BESIII-Y4500-DvDvpi-PRL-2023,BESIII-Y4544-omegachi-2024}.  At the energy about $4.5\,\rm{GeV}$, we obtain three hidden-charm tetraquark states with the $J^{PC}=1^{--}$, the $[uc]_{\tilde{V}}[\overline{uc}]_{A}+[dc]_{\tilde{V}}[\overline{dc}]_{A}
-[uc]_{A}[\overline{uc}]_{\tilde{V}}-[dc]_{A}[\overline{dc}]_{\tilde{V}}$,
 $[uc]_{\tilde{A}}[\overline{uc}]_{V}+[dc]_{\tilde{A}}[\overline{dc}]_{V}
 +[uc]_{V}[\overline{uc}]_{\tilde{A}}+[dc]_{V}[\overline{dc}]_{\tilde{A}}$ and
 $[uc]_{S}[\overline{uc}]_{\tilde{V}}+[dc]_{S}[\overline{dc}]_{\tilde{V}}
 -[uc]_{\tilde{V}}[\overline{uc}]_{S}-[dc]_{\tilde{V}}[\overline{dc}]_{S}$ tetraquark states have
 the masses $4.53\pm0.07\, \rm{GeV}$, $4.48\pm0.08\,\rm{GeV}$ and $4.50\pm0.09\,\rm{GeV}$, respectively \cite{WZG-HC-Vect-NPB-2021}, see Table \ref{Assignments-Table-Y-cqcq}. In Ref.\cite{WZG-Decay-Y4500-NPB-2024}, we study  their two-body strong decays systematically with the three-point correlation functions,
\begin{eqnarray}
\Pi^{\bar{D}D\widetilde{A}V}_{\mu}(p,q)&=&i^2\int d^4xd^4y \, e^{ip\cdot x}e^{iq\cdot y}\, \langle 0|T\left\{J^{\bar{D}}(x)J^{D}(y)J_{-,\mu}^{\widetilde{A}V}{}^\dagger(0)\right\}|0\rangle\, ,
\end{eqnarray}

\begin{eqnarray}
\Pi^{\bar{D}^*D\widetilde{A}V}_{\alpha\mu}(p,q)&=&i^2\int d^4xd^4y \, e^{ip\cdot x}e^{iq\cdot y}\, \langle 0|T\left\{J_{\alpha}^{\bar{D}^*}(x)J^{D}(y)J_{-,\mu}^{\widetilde{A}V}{}^\dagger(0)\right\}|0\rangle\, ,
\end{eqnarray}

\begin{eqnarray}
\Pi^{\bar{D}^*D^*\widetilde{A}V}_{\alpha\beta\mu}(p,q)&=&i^2\int d^4xd^4y \, e^{ip\cdot x}e^{iq\cdot y}\, \langle 0|T\left\{J_{\alpha}^{\bar{D}^*}(x)J_\beta^{D^*}(y)J_{-,\mu}^{\widetilde{A}V}{}^\dagger(0)\right\}|0\rangle\, ,
\end{eqnarray}

\begin{eqnarray}
\Pi^{\bar{D}_0D^*\widetilde{A}V}_{\alpha\mu}(p,q)&=&i^2\int d^4xd^4y \, e^{ip\cdot x}e^{iq\cdot y}\, \langle 0|T\left\{J^{\bar{D}_0}(x)J_\alpha^{D^*}(y)J_{-,\mu}^{\widetilde{A}V}{}^\dagger(0)\right\}|0\rangle\, ,
\end{eqnarray}

\begin{eqnarray}
\Pi^{\bar{D}_1D\widetilde{A}V}_{\alpha\mu}(p,q)&=&i^2\int d^4xd^4y \, e^{ip\cdot x}e^{iq\cdot y}\, \langle 0|T\left\{J_{\alpha}^{\bar{D}_1}(x)J^{D}(y)J_{-,\mu}^{\widetilde{A}V}{}^\dagger(0)\right\}|0\rangle\, ,
\end{eqnarray}

\begin{eqnarray}
\Pi^{\eta_c\omega\widetilde{A}V}_{\alpha\mu}(p,q)&=&i^2\int d^4xd^4y \, e^{ip\cdot x}e^{iq\cdot y}\, \langle 0|T\left\{J^{\eta_c}(x)J_\alpha^{\omega}(y)J_{-,\mu}^{\widetilde{A}V}{}^\dagger(0)\right\}|0\rangle\, ,
\end{eqnarray}

\begin{eqnarray}
\Pi^{J/\psi\omega\widetilde{A}V}_{\alpha\beta\mu}(p,q)&=&i^2\int d^4xd^4y \, e^{ip\cdot x}e^{iq\cdot y}\, \langle 0|T\left\{J_{\alpha}^{J/\psi}(x)J_\beta^{\omega}(y)J_{-,\mu}^{\widetilde{A}V}{}^\dagger(0)\right\}|0\rangle\, ,
\end{eqnarray}

\begin{eqnarray}
\Pi^{\chi_{c0}\omega\widetilde{A}V}_{\alpha\mu}(p,q)&=&i^2\int d^4xd^4y \, e^{ip\cdot x}e^{iq\cdot y}\, \langle 0|T\left\{J^{\chi_{c0}}(x)J_\alpha^{\omega}(y)J_{-,\mu}^{\widetilde{A}V}{}^\dagger(0)\right\}|0\rangle\, ,
\end{eqnarray}

\begin{eqnarray}
\Pi^{\chi_{c1}\omega\widetilde{A}V}_{\alpha\beta\mu}(p,q)&=&i^2\int d^4xd^4y \, e^{ip\cdot x}e^{iq\cdot y}\, \langle 0|T\left\{J_\alpha^{\chi_{c1}}(x)J_\beta^{\omega}(y)J_{-,\mu}^{\widetilde{A}V}{}^\dagger(0)\right\}|0\rangle\, ,
\end{eqnarray}

\begin{eqnarray}
\Pi^{J/\psi f_0\widetilde{A}V}_{\alpha\mu}(p,q)&=&i^2\int d^4xd^4y \, e^{ip\cdot x}e^{iq\cdot y}\, \langle 0|T\left\{J_\alpha^{J/\psi}(x)J^{f_0}(y)J_{-,\mu}^{\widetilde{A}V}{}^\dagger(0)\right\}|0\rangle\, .
\end{eqnarray}
With the simple replacement $\widetilde{A}V \to \widetilde{V}A$, we obtain the
corresponding correlation functions for the current $J_{-,\mu}^{\widetilde{V}A}$.
And with the simple replacements $\widetilde{A}V \to S\widetilde{V}$ and ${}_\mu \to {}_{\mu\nu}$, we obtain the
corresponding correlation functions for the current $J_{-,\mu\nu}^{S\widetilde{V}}$,
where the currents
\begin{eqnarray}
J^{\bar{D}}(x)&=&\bar{c}(x)i\gamma_{5} u(x)  \, ,\nonumber \\
J^{D}(y)&=&\bar{u}(y)i\gamma_{5} c(y) \, ,\nonumber \\
J_{\alpha}^{\bar{D}^*}(x)&=&\bar{c}(x)\gamma_{\alpha} u(x)  \, ,\nonumber \\
J_{\beta}^{D^*}(y)&=&\bar{u}(y)\gamma_{\beta} c(y) \, ,\nonumber \\
J^{\bar{D}_0}(x)&=&\bar{c}(x) u(x) \, ,\nonumber \\
J_{\alpha}^{\bar{D}_1}(x)&=&\bar{c}(x)\gamma_\alpha \gamma_5 u(x) \, ,
\end{eqnarray}
\begin{eqnarray}
J^{\eta_c}(x)&=&\bar{c}(x)i \gamma_5 c(x) \, ,\nonumber \\
J_{\alpha}^{J/\psi}(x)&=&\bar{c}(x)\gamma_{\alpha} c(x)  \, ,\nonumber \\
J^{\chi_{c0}}(x)&=&\bar{c}(x) c(x) \, ,\nonumber \\
J_{\alpha}^{\chi_{c1}}(x)&=&\bar{c}(x)\gamma_{\alpha}\gamma_5 c(x)  \, , \end{eqnarray}
\begin{eqnarray}
J_{\alpha}^{\omega}(y)&=&\frac{\bar{u}(y) \gamma_\alpha u(y)+\bar{d}(y) \gamma_\alpha d(y)}{\sqrt{2}} \, ,\nonumber \\
J^{f_0}(y)&=&\frac{\bar{u}(y)  u(y)+\bar{d}(y)  d(y)}{\sqrt{2}} \, ,
\end{eqnarray}
\begin{eqnarray}\label{AV-Current}
J_{-,\mu}^{\widetilde{A}V}(x)&=&\frac{\varepsilon^{ijk}\varepsilon^{imn}}{2}
\Big[u^{T}_j(x)C\sigma_{\mu\nu}\gamma_5 c_k(x)\bar{u}_m(x)\gamma_5\gamma^\nu C \bar{c}^{T}_n(x)+d^{T}_j(x)C\sigma_{\mu\nu}\gamma_5 c_k(x)\bar{d}_m(x)\gamma_5\gamma^\nu C \bar{c}^{T}_n(x) \nonumber \\
&&+u^{T}_j(x)C\gamma^\nu\gamma_5 c_k(x)\bar{u}_m(x)\gamma_5\sigma_{\mu\nu} C \bar{c}^{T}_n(x)+d^{T}_j(x)C\gamma^\nu\gamma_5 c_k(x)\bar{d}_m(x)\gamma_5\sigma_{\mu\nu} C \bar{c}^{T}_n(x) \Big] \, ,
\end{eqnarray}
\begin{eqnarray}\label{VA-Current}
J_{-,\mu}^{\widetilde{V}A}(x)&=&\frac{\varepsilon^{ijk}\varepsilon^{imn}}{2}
\Big[u^{T}_j(x)C\sigma_{\mu\nu} c_k(x)\bar{u}_m(x)\gamma^\nu C \bar{c}^{T}_n(x)
+d^{T}_j(x)C\sigma_{\mu\nu} c_k(x)\bar{d}_m(x)\gamma^\nu C \bar{c}^{T}_n(x)\nonumber\\
&&
-u^{T}_j(x)C\gamma^\nu c_k(x)\bar{u}_m(x)\sigma_{\mu\nu} C \bar{c}^{T}_n(x)
-d^{T}_j(x)C\gamma^\nu c_k(x)\bar{d}_m(x)\sigma_{\mu\nu} C \bar{c}^{T}_n(x) \Big] \, ,
\end{eqnarray}
\begin{eqnarray}\label{SV-Current}
J^{S\widetilde{V}}_{-,\mu\nu}(x)&=&\frac{\varepsilon^{ijk}\varepsilon^{imn}}{2}
\Big[u^{T}_j(x)C\gamma_5 c_k(x)  \bar{u}_m(x)\sigma_{\mu\nu} C \bar{c}^{T}_n(x)+d^{T}_j(x)C\gamma_5 c_k(x)  \bar{d}_m(x)\sigma_{\mu\nu} C \bar{c}^{T}_n(x)\nonumber\\
&&- u^{T}_j(x)C\sigma_{\mu\nu} c_k(x)  \bar{u}_m(x)\gamma_5 C \bar{c}^{T}_n(x) -d^{T}_j(x)C\sigma_{\mu\nu} c_k(x)  \bar{d}_m(x)\gamma_5 C \bar{c}^{T}_n(x)\Big] \, .
\end{eqnarray}

According to quark-hadron duality, we obtain the hadron representation  and  isolate the ground state contributions explicitly,
\begin{eqnarray}\label{CF-structure-first}
\Pi^{\bar{D}D\widetilde{A}V}_{\mu}(p,q)&=&\Pi_{\bar{D}D\widetilde{A}V}(p^{\prime2},p^2,q^2)
\,i\left(p-q\right)_\mu+\cdots\, ,
\end{eqnarray}

\begin{eqnarray}
\Pi^{\bar{D}^*D\widetilde{A}V}_{\alpha\mu}(p,q)&=&
\Pi_{\bar{D}^*D\widetilde{A}V}(p^{\prime2},p^2,q^2)
\,\left(-i\varepsilon_{\alpha\mu\lambda\tau}p^\lambda q^\tau\right)+\cdots\, ,
\end{eqnarray}

\begin{eqnarray}
\Pi^{\bar{D}^*D^*\widetilde{A}V}_{\alpha\beta\mu}(p,q)&=&
\Pi_{\bar{D}^*D^*\widetilde{A}V}(p^{\prime2},p^2,q^2)
\,\left(-ig_{\alpha\beta}p_\mu\right)+\cdots\, ,
\end{eqnarray}

\begin{eqnarray}
\Pi^{\bar{D}_0D^*\widetilde{A}V}_{\alpha\mu}(p,q)&=&
\Pi_{\bar{D}_0D^*\widetilde{A}V}(p^{\prime2},p^2,q^2)
\,\left(-ig_{\alpha\mu}p \cdot q\right)+\cdots\, ,
\end{eqnarray}

\begin{eqnarray}
\Pi^{\bar{D}_1D\widetilde{A}V}_{\alpha\mu}(p,q)&=&
\Pi_{\bar{D}_1D\widetilde{A}V}(p^{\prime2},p^2,q^2)
\,g_{\alpha\mu}+\cdots\, ,
\end{eqnarray}

\begin{eqnarray}
\Pi^{\eta_c\omega\widetilde{A}V}_{\alpha\mu}(p,q)&=&
\Pi_{\eta_c\omega\widetilde{A}V}(p^{\prime2},p^2,q^2)
\,\left(i\varepsilon_{\alpha\mu\lambda\tau}p^\lambda q^\tau\right)+\cdots\, ,
\end{eqnarray}

\begin{eqnarray}
\Pi^{J/\psi\omega\widetilde{A}V}_{\alpha\beta\mu}(p,q)&=&
\Pi_{J/\psi\omega\widetilde{A}V}(p^{\prime2},p^2,q^2)\,
\left(ig_{\alpha\beta}p_\mu\right)+\cdots\, ,
\end{eqnarray}

\begin{eqnarray}
\Pi^{\chi_{c0}\omega\widetilde{A}V}_{\alpha\mu}(p,q)&=&
\Pi_{\chi_{c0}\omega\widetilde{A}V}(p^{\prime2},p^2,q^2)
\,ig_{\alpha\mu}+\cdots\, ,
\end{eqnarray}

\begin{eqnarray}
\Pi^{\chi_{c1}\omega\widetilde{A}V}_{\alpha\beta\mu}(p,q)&=&
\Pi_{\chi_{c1}\omega\widetilde{A}V}(p^{\prime2},p^2,q^2)
\,\left(\varepsilon_{\alpha\beta\mu\lambda}p^\lambda \,p \cdot q\right)+\cdots\, ,
\end{eqnarray}

\begin{eqnarray}
\Pi^{J/\psi f_0\widetilde{A}V}_{\alpha\mu}(p,q)&=&
\Pi_{J/\psi f_0\widetilde{A}V}(p^{\prime2},p^2,q^2)
\,\left(-ig_{\alpha\mu}\right)+\cdots\, ,
\end{eqnarray}
other ground state contributions are given in Ref.\cite{WZG-Decay-Y4500-NPB-2024},
where
\begin{eqnarray}
\Pi_{\bar{D}D\widetilde{A}V}(p^{\prime2},p^2,q^2)&=&
\frac{\lambda_{\bar{D}D\widetilde{A}V}}{(m_Y^2-p^{\prime2})(m_D^2-p^2)(m_D^2-q^2)}
+\frac{C_{\bar{D}D\widetilde{A}V}}{(m_D^2-p^2)(m_D^2-q^2)}\nonumber\\
&&+\cdots\, ,
\end{eqnarray}
\begin{eqnarray}
\Pi_{\bar{D}^*D\widetilde{A}V}(p^{\prime2},p^2,q^2)&=&
\frac{\lambda_{\bar{D}^*D\widetilde{A}V}}{(m_Y^2-p^{\prime2})(m_{D^*}^2-p^2)(m_D^2-q^2)}
+\frac{C_{\bar{D}^*D\widetilde{A}V}}{(m_{D^*}^2-p^2)(m_D^2-q^2)}\nonumber\\
&&+\cdots\, ,
\end{eqnarray}
\begin{eqnarray}
\Pi_{\bar{D}^*D^*\widetilde{A}V}(p^{\prime2},p^2,q^2)&=&
\frac{\lambda_{\bar{D}^*D^*\widetilde{A}V}}{(m_Y^2-p^{\prime2})(m_{D^*}^2-p^2)(m_{D^*}^2-q^2)}
+\frac{C_{\bar{D}^*D^*\widetilde{A}V}}{(m_{D^*}^2-p^2)(m_{D^*}^2-q^2)}\nonumber\\
&&+\cdots\, ,
\end{eqnarray}
\begin{eqnarray}
\Pi_{\bar{D}_0D^*\widetilde{A}V}(p^{\prime2},p^2,q^2)&=&
\frac{\lambda_{\bar{D}_0D^*\widetilde{A}V}}{(m_Y^2-p^{\prime2})(m_{D_0}^2-p^2)(m_{D^*}^2-q^2)}
+\frac{C_{\bar{D}_0D^*\widetilde{A}V}}{(m_{D_0}^2-p^2)(m_{D^*}^2-q^2)}\nonumber\\
&&+\cdots\, ,
\end{eqnarray}
\begin{eqnarray}
\Pi_{\bar{D}_1D\widetilde{A}V}(p^{\prime2},p^2,q^2)&=&
\frac{\lambda_{\bar{D}_1D\widetilde{A}V}}{(m_Y^2-p^{\prime2})(m_{D_1}^2-p^2)(m_{D}^2-q^2)}
+\frac{C_{\bar{D}_1D\widetilde{A}V}}{(m_{D_1}^2-p^2)(m_{D}^2-q^2)}\nonumber\\
&&+\cdots\, ,
\end{eqnarray}

\begin{eqnarray}
\Pi_{\eta_c\omega \widetilde{A}V}(p^{\prime2},p^2,q^2)&=&
\frac{\lambda_{\eta_c\omega\widetilde{A}V}}{(m_Y^2-p^{\prime2})
(m_{\eta_c}^2-p^2)(m_{\omega}^2-q^2)}
+\frac{C_{\eta_c\omega\widetilde{A}V}}{(m_{\eta_c}^2-p^2)(m_{\omega}^2-q^2)}\nonumber\\
&&+\cdots\, ,
\end{eqnarray}
with the simple replacements $\eta_c \to J/\psi$, $\chi_{c0}$ and $\chi_{c1}$, we obtain the hadronic  representation  for the
$J/\psi \omega\widetilde{A}V $, $\chi_{c0} \omega\widetilde{A}V $ and $\chi_{c1}\omega\widetilde{A}V $ channels, respectively,
\begin{eqnarray}
\Pi_{J/\psi f_0 \widetilde{A}V}(p^{\prime2},p^2,q^2)&=&
\frac{\lambda_{J/\psi f_0\widetilde{A}V}}{(m_Y^2-p^{\prime2})
(m_{J/\psi}^2-p^2)(m_{f_0}^2-q^2)}
+\frac{C_{J/\psi f_0\widetilde{A}V}}{(m_{J/\psi}^2-p^2)(m_{f_0}^2-q^2)}\nonumber\\
&&+\cdots\, .
\end{eqnarray}
With the simple replacements $\widetilde{A}V\to \widetilde{V}A$ and $S\widetilde{V}$, we obtain the corresponding components $\Pi(p^{\prime2},p^2,q^2)$ for the currents $J_{-,\mu}^{\widetilde{V}A}(0)$ and $J_{-,\mu\nu}^{S\widetilde{V}}(0)$, except for the component  $\Pi_{\eta_c \omega S \widetilde{V}}(p^{\prime2},p^2,q^2)$,
\begin{eqnarray}\label{etacomegaSV}
\Pi_{\eta_c\omega S\widetilde{V}}(p^{\prime2},p^2,q^2)&=&
\frac{\lambda_{\eta_c\omega S\widetilde{V}}}{(m_Y^2-p^{\prime2})
(m_{\eta_c}^2-p^2)(m_{\omega}^2-q^2)}
+\frac{C_{\eta_c\omega S\widetilde{V}}}{(m_{\eta_c}^2-p^2)(m_{\omega}^2-q^2)}\nonumber\\
&&+\frac{\bar{\lambda}_{\eta_c\omega S\widetilde{A}}}{(m_X^2-p^{\prime2})
(m_{\eta_c}^2-p^2)(m_{\omega}^2-q^2)}+\cdots\, .
\end{eqnarray}
We introduce the collective notations to simplify the expressions,
\begin{eqnarray}
\lambda_{\bar{D}D\widetilde{A}V}&=&\frac{f_D^2m_D^4}{m_c^2}
\lambda_{\widetilde{A}V}G_{\bar{D}D\widetilde{A}V} \, , \nonumber \\
\lambda_{\bar{D}^*D\widetilde{A}V}&=&\frac{f_{D^*}m_{D^*}f_D m_D^2}{m_c}
\lambda_{\widetilde{A}V}G_{\bar{D}^*D\widetilde{A}V} \, ,\nonumber \\
\lambda_{\bar{D}^*D^*\widetilde{A}V}&=&f_{D^*}^2m_{D^*}^2
\lambda_{\widetilde{A}V}G_{\bar{D}^*D^*\widetilde{A}V}\, ,\nonumber \\
\lambda_{\bar{D}_0D^*\widetilde{A}V}&=&f_{D_0}m_{D_0}f_{D^*}m_{D^*}
\lambda_{\widetilde{A}V}G_{\bar{D}_0D^*\widetilde{A}V} \, ,\nonumber \\
\lambda_{\bar{D}_1D\widetilde{A}V}&=&\frac{f_{D_1}m_{D_1}f_{D}m_{D}^2}{m_c}
\lambda_{\widetilde{A}V}G_{\bar{D}_1D\widetilde{A}V} \, ,
\end{eqnarray}

\begin{eqnarray}
\lambda_{\eta_c\omega\widetilde{A}V}&=&\frac{f_{\eta_c}m_{\eta_c}^2f_{\omega}
m_{\omega}}{2m_c}
\lambda_{\widetilde{A}V}G_{\eta_c\omega\widetilde{A}V} \, ,\nonumber \\
\lambda_{J/\psi\omega\widetilde{A}V}&=&f_{J/\psi}m_{J/\psi}f_{\omega}
m_{\omega}\lambda_{\widetilde{A}V}G_{J/\psi\omega\widetilde{A}V}
\left(1+\frac{m_\omega^2}{m_Y^2}-\frac{m_{J/\psi}^2}{m_Y^2} \right)\, ,\nonumber \\
\lambda_{\chi_{c0}\omega\widetilde{A}V}&=&f_{\chi_{c0}}m_{\chi_{c0}}f_{\omega}
m_{\omega}\lambda_{\widetilde{A}V}G_{\chi_{c0}\omega\widetilde{A}V}\, ,\nonumber \\
\lambda_{\chi_{c1}\omega\widetilde{A}V}&=&f_{\chi_{c1}}m_{\chi_{c1}}f_{\omega}
m_{\omega}\lambda_{\widetilde{A}V}G_{\chi_{c1}\omega\widetilde{A}V}\, ,\nonumber \\
\lambda_{J/\psi f_0\widetilde{A}V}&=&f_{J/\psi}m_{J/\psi}f_{f_0}
m_{f_0}\lambda_{\widetilde{A}V}G_{J/\psi f_0\widetilde{A}V}\, .
\end{eqnarray}
With the simple replacements $\widetilde{A}V\to \widetilde{V}A$ and $S\widetilde{V}$, we obtain the  collective notations  $\lambda$ for the currents $J_{-,\mu}^{\widetilde{V}A}(0)$ and $J_{-,\mu\nu}^{S\widetilde{V}}(0)$, except for the $\lambda_{\eta_{c} \omega S \widetilde{V}}$,  $\lambda_{J/\psi \omega S \widetilde{V}}$, $\lambda_{\chi_{c1} \omega S \widetilde{V}}$,
where
\begin{eqnarray}
\lambda_{\eta_c\omega S\widetilde{V}}&=&\frac{f_{\eta_c}m_{\eta_c}^2f_{\omega}
m_{\omega}^3}{2m_c}
\lambda_{S \widetilde{V}}G_{\eta_c\omega S \widetilde{V}} \, ,\nonumber \\
\lambda_{J/\psi\omega S\widetilde{V}}&=&f_{J/\psi}m_{J/\psi}f_{\omega}
m_{\omega}\lambda_{S\widetilde{V}}G_{J/\psi\omega S\widetilde{V}}\, , \nonumber \\
\lambda_{\chi_{c1}\omega S\widetilde{V}}&=&f_{\chi_{c1}}m_{\chi_{c1}}^3f_{\omega}
m_{\omega}\lambda_{S \widetilde{V}}G_{\chi_{c1}\omega S \widetilde{V}}\, ,
\end{eqnarray}
and
\begin{eqnarray}
\bar{\lambda}_{\eta_c\omega S\widetilde{A}}&=&\frac{f_{\eta_c}m_{\eta_c}^2f_{\omega}
m_{\omega}}{2m_c}\bar{\lambda}_{S\widetilde{A}}\bar{G}_{\eta_c\omega S\widetilde{A}}\, .
\end{eqnarray}
We adopt the standard definitions for the  decay constants and pole residues \cite{WZG-Decay-Y4500-NPB-2024},
and we define the hadronic coupling constants,
\begin{eqnarray}
\langle \bar{D}(p)D(q)|Y_{\widetilde{A}V}(p^\prime)\rangle&=&(p-q)\cdot \varepsilon \,G_{\bar{D}D\widetilde{A}V}\, , \nonumber\\
\langle \bar{D}(p)D(q)|Y_{\widetilde{V}A}(p^\prime)\rangle&=&(p-q)\cdot \varepsilon \,G_{\bar{D}D\widetilde{V}A}\, , \nonumber\\
\langle \bar{D}(p)D(q)|Y_{S\widetilde{V}}(p^\prime)\rangle&=&-i(p-q)\cdot \varepsilon \,G_{\bar{D}DS\widetilde{V}}\, ,
\end{eqnarray}

\begin{eqnarray}
\langle \bar{D}^*(p)D(q)|Y_{\widetilde{A}V}(p^\prime)\rangle&=&-\varepsilon^{\lambda\tau\rho\sigma}
p_\lambda \xi^*_\tau p^\prime_\rho \varepsilon_\sigma \,G_{\bar{D}^*D\widetilde{A}V}\, , \nonumber\\
\langle \bar{D}^*(p)D(q)|Y_{\widetilde{V}A}(p^\prime)\rangle&=&\varepsilon^{\lambda\tau\rho\sigma}
p_\lambda \xi^*_\tau p^\prime_\rho \varepsilon_\sigma \,G_{\bar{D}^*D\widetilde{V}A}\, , \nonumber\\
\langle \bar{D}^*(p)D(q)|Y_{S\widetilde{V}}(p^\prime)\rangle&=&-i
\varepsilon^{\lambda\tau\rho\sigma}
p_\lambda \xi^*_\tau p^\prime_\rho \varepsilon_\sigma \,G_{\bar{D}^*DS\widetilde{V}}\, ,
\end{eqnarray}
etc \cite{WZG-Decay-Y4500-NPB-2024}.

In Eq.\eqref{etacomegaSV}, there are  contributions come from  the $J^{PC}=1^{+-}$ and $1^{--}$ tetraquark states, and we cannot choose the pertinent structures to exclude the contaminations from the $J^{PC}=1^{+-}$ tetraquark state $X$, so we include it  at the hadron side.  The unknown parameters $C_{\bar{D}D\widetilde{A}V}$, $C_{\bar{D}^*D\widetilde{A}V}$, $C_{\bar{D}^*D^*\widetilde{A}V}$, etc, parameterize the complex interactions among the excitations in the $p^{\prime2}$ channels and the ground state charmed meson pairs or charmonium plus $\omega/f_0$. It is difficult to choose the pertinent tensor structures in
Eqs.\eqref{CF-structure-first}-\eqref{etacomegaSV} to obtain good QCD sum rules without contaminations, we have to reach the satisfactory results via trial and error.

\begin{figure}
 \centering
  \includegraphics[totalheight=11cm,width=12cm]{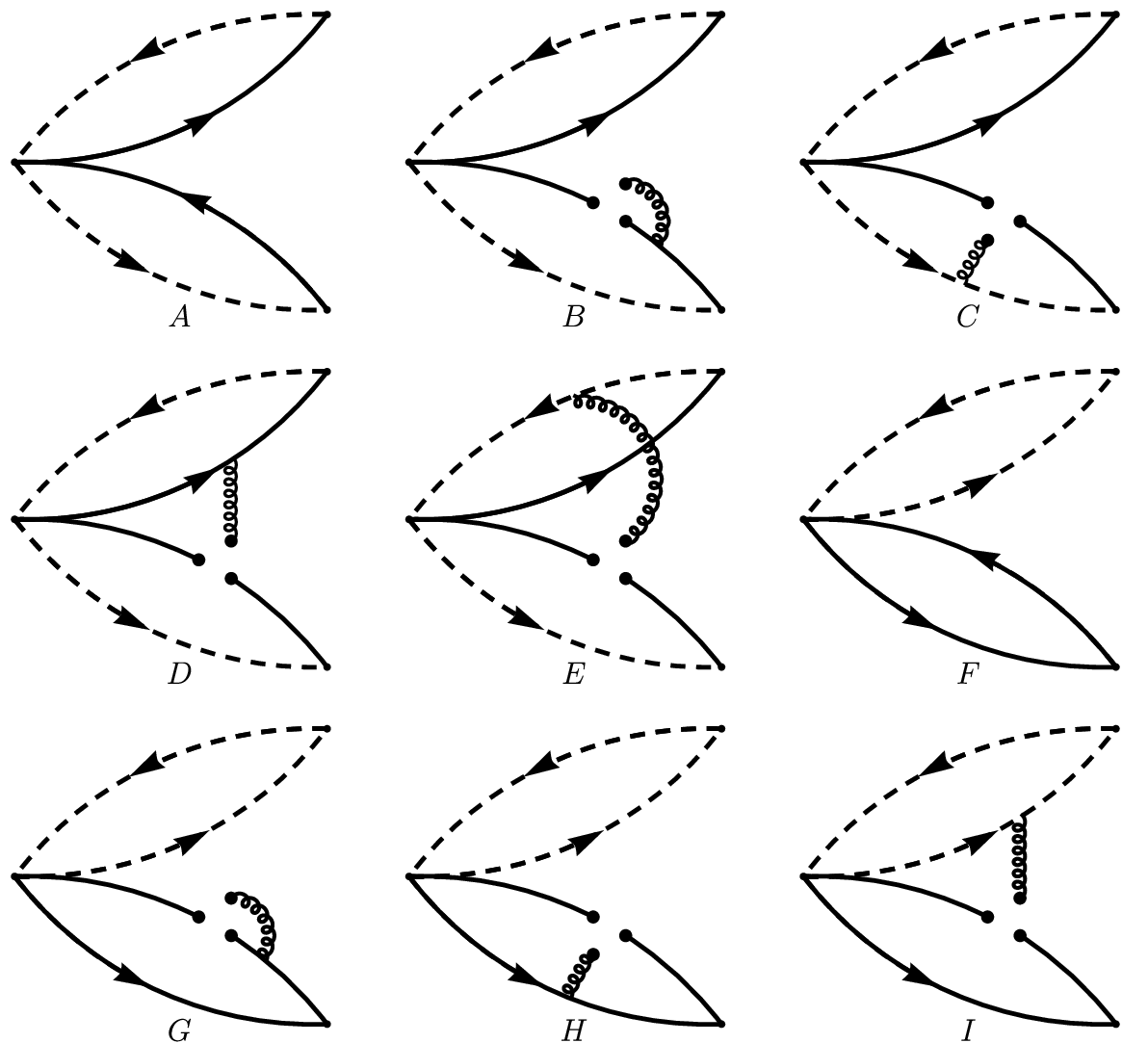}
 \caption{ The  connected and disconnected Feynman diagrams, where the dashed (solid) lines denote the heavy (light) quark lines, the interchanges of the heavy (light) quark lines are implied.}\label{Y4500-Two-decay-Feyn-fig}
\end{figure}

At the QCD side, we accomplish  the operator product expansion up to the vacuum condensates of dimension 5 and take account of both the connected  and disconnected  Feynman diagrams in the color space, see Fig.\ref{Y4500-Two-decay-Feyn-fig} (in Refs.\cite{Review-QCDSR-Nielsen-JPG-2019,Nielsen-Zc3900-decay-PRD-2013,
WangZG-Zb10610-tetra-NPA-2014,
Nielsen-X3873-decay-PLB-2006,
Nielsen-X5568-decay-PLB-2016}, only the connected Feynman  diagrams are taken into account, i.e. the $D$, $E$ and $I$ in Fig.\ref{Y4500-Two-decay-Feyn-fig}), and choose the components $\Pi_H(p^{\prime2},p^2,q^2)$ to study  the hadronic coupling constants $G_{H}$ based on the rigorous quark-hadron duality
\cite{WangZhang-Solid,WangZG-Y4660-tetra-decay-EPJC-2019}.

Then we set $p^{\prime2}=p^2$ in the components $\Pi(p^{\prime 2},p^2,q^2)$, and carry out  the double Borel transformation  with respect to the variables $P^2=-p^2$ and $Q^2=-q^2$ respectively, and set  $T_1^2=T_2^2=T^2$  to obtain thirty QCD sum rules,
\begin{eqnarray}
&&\frac{\lambda_{\bar{D}D\widetilde{A}V}}{m_Y^2-m_D^2}
\left[\exp\left(-\frac{m_D^2}{T^2} \right)-\exp\left(-\frac{m_Y^2}{T^2} \right) \right]\exp\left(-\frac{m_D^2}{T^2}\right)+C_{\bar{D}D\widetilde{A}V}\exp\left( -\frac{m_D^2}{T^2}-\frac{m_D^2}{T^2}\right)\nonumber\\
&&=\Pi^{QCD}_{\bar{D}D\widetilde{A}V}(T^2)\, ,
\end{eqnarray}

\begin{eqnarray}
&&\frac{\lambda_{\bar{D}^*D\widetilde{A}V}}{m_Y^2-m_{D^*}^2}
\left[\exp\left(-\frac{m_{D^*}^2}{T^2} \right)-\exp\left(-\frac{m_Y^2}{T^2} \right) \right]\exp\left(-\frac{m_D^2}{T^2}\right)+C_{\bar{D}^*D\widetilde{A}V}\exp\left( -\frac{m_{D^*}^2}{T^2}-\frac{m_D^2}{T^2}\right)\nonumber\\
&&=\Pi^{QCD}_{\bar{D}^*D\widetilde{A}V}(T^2)\, ,
\end{eqnarray}

\begin{eqnarray}
&&\frac{\lambda_{\bar{D}^*D^*\widetilde{A}V}}{m_Y^2-m_{D^*}^2}
\left[\exp\left(-\frac{m_{D^*}^2}{T^2} \right)-\exp\left(-\frac{m_Y^2}{T^2} \right) \right]\exp\left(-\frac{m_{D^*}^2}{T^2}\right)+C_{\bar{D}^*D^*\widetilde{A}V}\exp\left( -\frac{m_{D^*}^2}{T^2}-\frac{m_{D^*}^2}{T^2}\right)\nonumber\\
&&=\Pi^{QCD}_{\bar{D}^*D^*\widetilde{A}V}(T^2)\, ,
\end{eqnarray}

\begin{eqnarray}
&&\frac{\lambda_{\bar{D}_0D^*\widetilde{A}V}}{m_Y^2-m_{D_0}^2}
\left[\exp\left(-\frac{m_{D_0}^2}{T^2} \right)-\exp\left(-\frac{m_Y^2}{T^2} \right) \right]\exp\left(-\frac{m_{D^*}^2}{T^2}\right)+C_{\bar{D}_0D^*\widetilde{A}V}\exp\left( -\frac{m_{D_0}^2}{T^2}-\frac{m_{D^*}^2}{T^2}\right)\nonumber\\
&&=\Pi^{QCD}_{\bar{D}_0D^*\widetilde{A}V}(T^2)\, ,
\end{eqnarray}

\begin{eqnarray}
&&\frac{\lambda_{\bar{D}_1D\widetilde{A}V}}{m_Y^2-m_{D_1}^2}
\left[\exp\left(-\frac{m_{D_1}^2}{T^2} \right)-\exp\left(-\frac{m_Y^2}{T^2} \right) \right]\exp\left(-\frac{m_{D}^2}{T^2}\right)+C_{\bar{D}_1D\widetilde{A}V}\exp\left( -\frac{m_{D_1}^2}{T^2}-\frac{m_{D}^2}{T^2}\right)\nonumber\\
&&=\Pi^{QCD}_{\bar{D}_1D\widetilde{A}V}(T^2)\, ,
\end{eqnarray}

\begin{eqnarray}
&&\frac{\lambda_{\eta_c \omega\widetilde{A}V}}{m_Y^2-m_{\eta_c}^2}
\left[\exp\left(-\frac{m_{\eta_c}^2}{T^2} \right)-\exp\left(-\frac{m_Y^2}{T^2} \right) \right]\exp\left(-\frac{m_{\omega}^2}{T^2}\right)+C_{\eta_c \omega\widetilde{A}V}\exp\left( -\frac{m_{\eta_c}^2}{T^2}-\frac{m_{\omega}^2}{T^2}\right)\nonumber\\
&&=\Pi^{QCD}_{\eta_c \omega\widetilde{A}V}(T^2)\, ,
\end{eqnarray}
with the simple replacements $\eta_c \to J/\psi$, $\chi_{c0}$ and $\chi_{c1}$, we obtain the  QCD sum rules for the
$J/\psi \omega\widetilde{A}V $, $\chi_{c0} \omega\widetilde{A}V $ and $\chi_{c1}\omega\widetilde{A}V $ channels, respectively,
\begin{eqnarray}
&&\frac{\lambda_{J/\psi f_0\widetilde{A}V}}{m_Y^2-m_{J/\psi}^2}
\left[\exp\left(-\frac{m_{J/\psi}^2}{T^2} \right)-\exp\left(-\frac{m_Y^2}{T^2} \right) \right]\exp\left(-\frac{m_{f_0}^2}{T^2}\right)+C_{J/\psi f_0\widetilde{A}V}\exp\left( -\frac{m_{J/\psi}^2}{T^2}-\frac{m_{f_0}^2}{T^2}\right)\nonumber\\
&&=\Pi^{QCD}_{J/\psi f_0\widetilde{A}V}(T^2)\, .
\end{eqnarray}
With the simple replacements $\widetilde{A}V\to \widetilde{V}A$ and $S\widetilde{V}$, we obtain the QCD sum rules   for the  $J_{-,\mu}^{\widetilde{V}A}(0)$ and $J_{-,\mu\nu}^{S\widetilde{V}}(0)$, except for the $\eta_c \omega $ channel,
\begin{eqnarray}
&&\frac{\lambda_{\eta_c \omega S\widetilde{V}}}{m_Y^2-m_{\eta_c}^2}
\left[\exp\left(-\frac{m_{\eta_c}^2}{T^2} \right)-\exp\left(-\frac{m_Y^2}{T^2} \right) \right]\exp\left(-\frac{m_{\omega}^2}{T^2}\right)+C_{\eta_c \omega S\widetilde{V}}\exp\left( -\frac{m_{\eta_c}^2}{T^2}-\frac{m_{\omega}^2}{T^2}\right)\nonumber\\
&&+\frac{\bar{\lambda}_{\eta_c \omega S\widetilde{A}}}{m_X^2-m_{\eta_c}^2}
\left[\exp\left(-\frac{m_{\eta_c}^2}{T^2} \right)-\exp\left(-\frac{m_X^2}{T^2} \right) \right]\exp\left(-\frac{m_{\omega}^2}{T^2}\right)=\Pi^{QCD}_{\eta_c \omega S\widetilde{V}}(T^2)\, ,
\end{eqnarray}
the explicit expressions of the QCD side  are given in Ref.\cite{WZG-Decay-Y4500-NPB-2024}.

We  take the unknown parameters $C_{\bar{D}D\widetilde{A}V}$, $C_{\bar{D}^*D\widetilde{A}V}$, $C_{\bar{D}^*D^*\widetilde{A}V}$, $\cdots$ as free parameters, and adjust the suitable values to obtain flat Borel platforms for the hadronic coupling constants  \cite{WZG-Decay-Y4500-NPB-2024},
\begin{eqnarray}
C_{\bar{D}D\widetilde{A}V}&=&0.00045\,{\rm GeV^5}\times T^2\, , \nonumber\\
C_{\bar{D}^*D\widetilde{A}V}&=&-0.000003\,{\rm GeV^4}\times T^2\, , \nonumber\\
C_{\bar{D}^*D^*\widetilde{A}V}&=&0.0001\,{\rm GeV^5}\times T^2\, , \nonumber\\
C_{\bar{D}_0D^*\widetilde{A}V}&=&0.000055\,{\rm GeV^4}\times T^2\, , \nonumber\\
C_{\bar{D}_1D\widetilde{A}V}&=&0.0031\,{\rm GeV^6}\times T^2\, , \nonumber\\
C_{\eta_c\omega\widetilde{A}V}&=&-0.000082\,{\rm GeV^4}\times T^2\, , \nonumber\\
C_{J/\psi\omega\widetilde{A}V}&=&0.0\, , \nonumber\\
C_{\chi_{c0}\omega\widetilde{A}V}&=&0.00085\,{\rm GeV^6}\times T^2\, , \nonumber\\
C_{\chi_{c1}\omega\widetilde{A}V}&=&-0.000019\,{\rm GeV^3}\times T^2\, , \nonumber\\
C_{J/\psi f_0\widetilde{A}V}&=&0.00085\,{\rm GeV^6}\times T^2\, ,
\end{eqnarray}
\begin{eqnarray}
C_{\bar{D}D \widetilde{V}A}&=&0.0000015\,{\rm GeV^5}\times T^2\, , \nonumber\\
C_{\bar{D}^*D\widetilde{V}A}&=&0.000038\,{\rm GeV^4}\times T^2\, , \nonumber\\
C_{\bar{D}^*D^*\widetilde{V}A}&=&0.0000054\,{\rm GeV^5}\times T^2\, , \nonumber\\
C_{\bar{D}_0D^*\widetilde{V}A}&=&0.00264\,{\rm GeV^6}\times T^2\, , \nonumber\\
C_{\bar{D}_1D\widetilde{V}A}&=&0.000055\,{\rm GeV^4}\times T^2\, , \nonumber\\
C_{\eta_c\omega \widetilde{V}A}&=&-0.00006\,{\rm GeV^4}\times T^2\, , \nonumber\\
C_{J/\omega\omega \widetilde{V}A}&=&0.0\, , \nonumber\\
C_{\chi_{c0}\omega \widetilde{V}A}&=&0.00087\,{\rm GeV^6}\times T^2\, , \nonumber\\
C_{\chi_{c1}\omega \widetilde{V}A}&=&-0.000018\,{\rm GeV^3}\times T^2\, , \nonumber\\
C_{J/\psi f_0 \widetilde{V}A}&=&0.00075\,{\rm GeV^6}\times T^2\, ,
\end{eqnarray}
\begin{eqnarray}
C_{\bar{D}D S\widetilde{V}}&=&0.00014\, {\rm GeV^6}+0.00002\,{\rm GeV^4}\times T^2\, ,\nonumber\\
C_{\bar{D}^*D S\widetilde{V}}&=&0.0000006\,{\rm GeV^3}\times T^2\, , \nonumber\\
C_{\bar{D}^*D^* S\widetilde{V}}&=&-0.000002\,{\rm GeV^4}\times T^2\, , \nonumber\\
C_{\bar{D}_0D^* S\widetilde{V}}&=&0.0012\, {\rm GeV^7}+0.000164\,{\rm GeV^5}\times T^2\, , \nonumber\\
C_{\bar{D}_1D S\widetilde{V}}&=&0.0017\, {\rm GeV^7}+0.000205\,{\rm GeV^5}\times T^2\, , \nonumber\\
C_{\eta_c \omega S\widetilde{V}}&=&0.00014\, {\rm GeV^7}\, , \nonumber\\
\bar{\lambda}_{\eta_c\omega S\widetilde{A}}&=&0.0012\, {\rm GeV^7}\times T^2\, , \nonumber\\
C_{J/\psi\omega S\widetilde{V}}&=&-0.0005\, {\rm GeV^6}-0.0000216\,{\rm GeV^4}\times T^2\, , \nonumber\\
C_{\chi_{c0}\omega S\widetilde{V}}&=&-0.0018\, {\rm GeV^7}-0.000156\,{\rm GeV^5}\times T^2\, , \nonumber\\
C_{\chi_{c1}\omega S\widetilde{V}}&=&0.0012\, {\rm GeV^8}+0.00015\,{\rm GeV^6}\times T^2\, , \nonumber\\
C_{J/\psi f_0 S\widetilde{V}}&=&0.0005\, {\rm GeV^7}+0.00006\,{\rm GeV^5}\times T^2\, ,
\end{eqnarray}
 the Borel windows are shown explicitly in Table \ref{BorelP-Y4500-decay}.
We obtain uniform flat platforms  $T^2_{max}-T^2_{min}=1\,\rm{GeV}^2$, where the max and min denote the maximum and minimum, respectively. In calculations, we  choose quark flavor numbers  $n_f=4$, and evolve all the input parameters to the energy scale  $\mu=1\,\rm{GeV}$. For detailed information about the parameters, one can consult Ref.\cite{WZG-Decay-Y4500-NPB-2024}.

\begin{figure}
\centering
\includegraphics[totalheight=5cm,width=7cm]{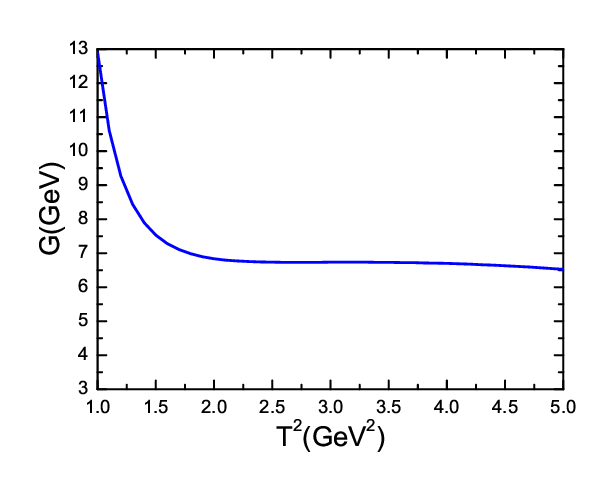}
\caption{The hadron-coupling constant $G_{\bar{D}_1 D \widetilde{A}V}$ with  the  Borel parameter.}\label{hadron-coupling-fig}
\end{figure}

In Fig.\ref{hadron-coupling-fig}, we plot the  $G_{\bar{D}_1 D \widetilde{A} V}$ with variation of the Borel parameter at a large interval as an example, in the Borel window, there appears very flat platform  indeed.

We estimate the uncertainties of the hadronic  coupling constants routinely. For an input parameter $\xi$, $\xi= \bar{\xi} +\delta \xi$,  the left side can be written as  $\lambda_{\widetilde{A}V}f_{\bar{D}}f_{D}G_{\bar{D}D\widetilde{A}V} = \bar{\lambda}_{\widetilde{A}V}\bar{f}_{\bar{D}}\bar{f}_{D}\bar{G}_{\bar{D}D\widetilde{A}V}
+\delta\,\lambda_{\widetilde{A}V}f_{\bar{D}}f_{D}G_{\bar{D}D\widetilde{A}V}$, $C_{\bar{D}D\widetilde{A}V} = \bar{C}_{\bar{D}D\widetilde{A}V}+\delta C_{\bar{D}D\widetilde{A}V}$, $\cdots$, where
\begin{eqnarray}\label{Uncertainty-4}
\delta\,\lambda_{\widetilde{A}V}f_{\bar{D}}f_{D}G_{\bar{D}D\widetilde{A}V} &=&\bar{\lambda}_{\widetilde{A}V}\bar{f}_{\bar{D}}\bar{f}_{D}\bar{G}_{\bar{D}D\widetilde{A}V}
\left( \frac{\delta f_{\bar{D}}}{\bar{f}_{\bar{D}}} +\frac{\delta f_{D}}{\bar{f}_{D}}+\frac{\delta \lambda_{\widetilde{A}V}}{\bar{\lambda}_{\widetilde{A}V}}
+\frac{\delta G_{\bar{D}D\widetilde{A}V}}{\bar{G}_{\bar{D}D\widetilde{A}V}}\right)\, ,
\end{eqnarray}
$\cdots$.
We set $\delta C_{\bar{D}D\widetilde{A}V}=0$,  $\frac{\delta f_{\bar{D}}}{\bar{f}_{\bar{D}}} =\frac{\delta f_{D}}{\bar{f}_{D}}=\frac{\delta \lambda_{\widetilde{A}V}}{\bar{\lambda}_{\widetilde{A}V}}
=\frac{\delta G_{\bar{D}D\widetilde{A}V}}{\bar{G}_{\bar{D}D\widetilde{A}V}}$, $\cdots$ approximately.

After taking into account the uncertainties, we obtain the values of the hadronic coupling constants, which are shown explicitly in Table \ref{BorelP-Y4500-decay}, then we obtain the partial decay widths directly, and show them explicitly in Table \ref{Width-Part}.

\begin{table}
\begin{center}
\begin{tabular}{|c|c|c|c|c|c|c|c|c|}\hline\hline
Channels                         &$T^2(\rm{GeV}^2)$ &$G $    \\ \hline

$\bar{D}D\widetilde{A}V$         &$3.8-4.8$  &$2.11\pm0.10$   \\

$\bar{D}^*D\widetilde{A}V$       &$4.7-5.7$  &$(4.49\pm0.15)\times\rm{10^{-2} \,GeV^{-1}}$ \\

$\bar{D}^*D^*\widetilde{A}V$     &$4.1-5.1$  &$0.95\pm0.04$      \\

$\bar{D}_0D^*\widetilde{A}V$     &$4.4-5.4$  &$0.30\pm0.01\,\rm{GeV^{-1}}$        \\

$\bar{D}_1D\widetilde{A}V$       &$2.5-3.5$  &$6.73\pm0.31\,\rm{GeV}$   \\

$\eta_c\omega\widetilde{A}V$     &$3.7-4.7$  &$-(0.35\pm0.03)\,\rm{GeV^{-1}}$     \\

$J/\psi\omega\widetilde{A}V$     &$---$      &$0.0$     \\

$\chi_{c0}\omega\widetilde{A}V$  &$3.8-4.8$  &$2.72\pm0.20\,\rm{GeV}$     \\

$\chi_{c1}\omega\widetilde{A}V$  &$5.2-6.2$  &$-(0.25\pm0.01)\,\rm{GeV}^{-2}$     \\

$J/\psi f_0\widetilde{A}V$       &$3.8-4.8$  &$2.51\pm0.15\,\rm{GeV}$     \\ \hline

$\bar{D}D \widetilde{V}A$        &$4.4-5.4$  &$-(4.02\pm0.16)\times\rm{10^{-2}}$ \\

$\bar{D}^*D\widetilde{V}A$       &$4.0-5.0$  &$0.30\pm0.01\,\rm{ GeV^{-1}}$ \\

$\bar{D}^*D^* \widetilde{V}A$    &$4.6-5.6$  &$-(6.00\pm0.20)\times\rm{10^{-2}}$ \\

$\bar{D}_0D^*\widetilde{V}A$     &$2.6-3.6$  &$8.00\pm0.37\,\rm{ GeV}$ \\

$\bar{D}_1D\widetilde{V}A$       &$2.8-3.8$  &$0.30\pm0.01\,\rm{ GeV}^{-1}$ \\

$\eta_c\omega \widetilde{V}A$    &$5.3-6.3$  &$-(0.40\pm0.03)\,\rm{GeV^{-1}}$     \\

$J/\psi\omega \widetilde{V}A$    &$---$      &$0.0$     \\

$\chi_{c0}\omega \widetilde{V}A$ &$3.0-4.0$  &$2.56\pm0.19\,\rm{GeV}$     \\

$\chi_{c1}\omega \widetilde{V}A$ &$5.4-6.4$  &$-(0.24\pm0.01)\,\rm{GeV}^{-2}$     \\

$J/\psi f_0 \widetilde{V}A$      &$4.6-5.6$  &$2.60\pm0.15\,\rm{GeV}$     \\ \hline

$\bar{D}D S\widetilde{V}$        &$2.5-3.5$  &$0.60\pm0.06 $     \\

$\bar{D}^*D S\widetilde{V}$      &$4.7-5.7$  &$-(7.00\pm0.24)\times 10^{-2}\,\rm{GeV}^{-1} $   \\

$\bar{D}^*D^* S\widetilde{V}$    &$4.0-5.0$  &$0.18\pm0.01$     \\

$\bar{D}_0D^* S\widetilde{V}$    &$3.3-4.3$  &$2.23\pm0.24\,\rm{GeV} $   \\

$\bar{D}_1D S\widetilde{V}$      &$3.5-4.5$  &$4.21\pm0.37\,\rm{GeV} $   \\

$\eta_c\omega S\widetilde{V}$    &$2.8-3.8$  &$1.71\pm0.31\,\rm{GeV}^{-1} $   \\

$J/\psi\omega S\widetilde{V}$    &$3.7-4.7$  &$-(1.08\pm0.19) $   \\

$\chi_{c0}\omega S\widetilde{V}$ &$4.4-5.4$  &$-(5.56\pm0.97)\,\rm{GeV} $   \\

$\chi_{c1}\omega S\widetilde{V}$ &$3.1-4.1$  &$0.29\pm0.04 $   \\

$J/\psi f_0 S\widetilde{V}$      &$3.5-4.5$  &$0.47\pm0.14\,\rm{GeV} $   \\

\hline\hline
\end{tabular}
\end{center}
\caption{ The Borel parameters and hadronic coupling constants \cite{WZG-Decay-Y4500-NPB-2024}. }\label{BorelP-Y4500-decay}
\end{table}

\begin{table}
\begin{center}
\begin{tabular}{|c|c|c|c|c|c|c|c|c|}\hline\hline
Channels                                               &$\Gamma(\rm{MeV})$ \\ \hline

$Y_{\widetilde{A}V}\to \bar{D}^0D^0$, $\bar{D}^-D^+$   &$22.5\pm2.1$      \\

$Y_{\widetilde{A}V}\to \frac{\bar{D}^{0*}D^0-\bar{D}^{0}D^{*0}}{\sqrt{2}}$, $\frac{\bar{D}^{-*}D^+-\bar{D}^{-}D^{*+}}{\sqrt{2}}$   &$0.08\pm0.01 $ \\

$Y_{\widetilde{A}V}\to \bar{D}^{*0}D^{*0}$, $\bar{D}^{*-}D^{*+}$  &$9.93\pm0.84$      \\

$Y_{\widetilde{A}V}\to\frac{\bar{D}^0_0D^{*0}-\bar{D}^{*0}D^{0}_0}{\sqrt{2}}$, $\frac{\bar{D}^-_0D^{*+}-\bar{D}^{*-}D^{+}_0}{\sqrt{2}}$     &$1.92\pm0.13$        \\

$Y_{\widetilde{A}V}\to\frac{\bar{D}^0_1D^{0}-\bar{D}^{0}D^{0}_1}{\sqrt{2}}$, $\frac{\bar{D}^-_1D^{+}-\bar{D}^{-}D^{+}_1}{\sqrt{2}}$  &$59.7\pm5.5$     \\

$Y_{\widetilde{A}V}\to\eta_c\omega$     &$3.83\pm0.66$     \\

$Y_{\widetilde{A}V}\to J/\psi\omega$    &$0.0$     \\

$Y_{\widetilde{A}V}\to\chi_{c0}\omega$  &$11.3\pm1.7$     \\

$Y_{\widetilde{A}V}\to\chi_{c1}\omega$  &$24.4\pm1.9$     \\

$Y_{\widetilde{A}V}\to J/\psi f_0(500)$ &$13.9\pm1.7$     \\ \hline

$Y_{\widetilde{V}A}\to \bar{D}^0D^0$, $\bar{D}^-D^+$   &$0.009\pm0.001$ \\

$Y_{\widetilde{V}A}\to \frac{\bar{D}^{0*}D^0-\bar{D}^{0}D^{*0}}{\sqrt{2}}$, $\frac{\bar{D}^{-*}D^+-\bar{D}^{-}D^{*+}}{\sqrt{2}}$   &$3.88\pm0.26$ \\

$Y_{\widetilde{V}A}\to \bar{D}^{*0}D^{*0}$, $\bar{D}^{*-}D^{*+}$  &$0.047\pm0.003$ \\

$Y_{\widetilde{V}A}\to\frac{\bar{D}^0_0D^{*0}-\bar{D}^{*0}D^{0}_0}{\sqrt{2}}$, $\frac{\bar{D}^-_0D^{*+}-\bar{D}^{*-}D^{+}_0}{\sqrt{2}}$  &$66.3\pm6.1$ \\

$Y_{\widetilde{V}A}\to\frac{\bar{D}^0_1D^{0}-\bar{D}^{0}D^{0}_1}{\sqrt{2}}$, $\frac{\bar{D}^-_1D^{+}-\bar{D}^{-}D^{+}_1}{\sqrt{2}}$   &$4.10\pm0.27$ \\

$Y_{\widetilde{V}A}\to\eta_c\omega $                     &$5.65\pm0.85$     \\

$Y_{\widetilde{V}A}\to J/\psi\omega $                    &$0.0$     \\

$Y_{\widetilde{V}A}\to\chi_{c0}\omega$                   &$11.1\pm1.6$     \\

$Y_{\widetilde{V}A}\to\chi_{c1}\omega $                  &$29.9\pm2.5$     \\

$Y_{\widetilde{V}A}\to J/\psi f_0(500)$                  &$15.2\pm1.8$     \\ \hline

$Y_{S\widetilde{V}}\to \bar{D}^0D^0$, $\bar{D}^-D^+$     &$1.88\pm0.38 $     \\

$Y_{S\widetilde{V}}\to \frac{\bar{D}^{0*}D^0-\bar{D}^{0}D^{*0}}{\sqrt{2}}$, $\frac{\bar{D}^{-*}D^+-\bar{D}^{-}D^{*+}}{\sqrt{2}}$  &$0.20\pm0.01 $   \\

$Y_{S\widetilde{V}}\to \frac{\bar{D}^{0*}D^0-\bar{D}^{0}D^{*0}}{\sqrt{2}}$, $\frac{\bar{D}^{-*}D^+-\bar{D}^{-}D^{*+}}{\sqrt{2}}$  &$0.38\pm0.04 $     \\

$Y_{S\widetilde{V}}\to\frac{\bar{D}^0_0D^{*0}-\bar{D}^{*0}D^{0}_0}{\sqrt{2}}$, $\frac{\bar{D}^-_0D^{*+}-\bar{D}^{*-}D^{+}_0}{\sqrt{2}}$   &$4.51\pm0.97 $   \\

$Y_{S\widetilde{V}}\to\frac{\bar{D}^0_1D^{0}-\bar{D}^{0}D^{0}_1}{\sqrt{2}}$, $\frac{\bar{D}^-_1D^{+}-\bar{D}^{-}D^{+}_1}{\sqrt{2}}$  &$24.4\pm4.3$   \\

$Y_{S\widetilde{V}}\to\eta_c\omega $      &$96.0\pm34.8 $   \\

$Y_{S\widetilde{V}}\to J/\psi\omega $     &$18.0\pm6.3 $   \\

$Y_{S\widetilde{V}}\to\chi_{c0}\omega $   &$49.4\pm17.2 $   \\

$Y_{S\widetilde{V}}\to \chi_{c1}\omega$   &$2.76\pm0.76 $   \\

$Y_{S\widetilde{V}}\to J/\psi f_0(500)$   &$0.49\pm0.29 $   \\

\hline\hline
\end{tabular}
\end{center}
\caption{ The partial decay widths \cite{WZG-Decay-Y4500-NPB-2024}. }\label{Width-Part}
\end{table}

At last, we saturate the total widths with the summary of partial decay widths,
\begin{eqnarray}\label{Widths-Y4500}
\Gamma\left(Y_{\widetilde{A}V}\right)&=&241.6\pm 9.0\, \rm{MeV}\, , \nonumber \\
\Gamma\left(Y_{\widetilde{V}A}\right)&=&210.6\pm 9.4\, \rm{MeV}\, , \nonumber \\
\Gamma\left(Y_{S\widetilde{V}}\right)&=&229.4\pm 39.9\, \rm{MeV}\, .
\end{eqnarray}
The widths of the $Y(4484)$, $Y(4469)$ and $Y(4544)$
are $111.1\pm30.1\pm15.2\,\rm{MeV}$,
 $246.3\pm36.7\pm9.4\,{\rm MeV}$ and $116.1\pm33.5\pm1.7\, \rm{MeV}$, respectively,  from the BESIII collaboration
 \cite{BESIII-Y4500-KK-CPC-2022,BESIII-Y4500-DvDvpi-PRL-2023,BESIII-Y4544-omegachi-2024}, which are compatible with the theoretical predictions in magnitude.

From Table \ref{Width-Part}, we obtain the typical decay modes. For the $Y_{\widetilde{A}V}$ state, the decays,
\begin{eqnarray}
Y_{\widetilde{A}V}&\to&\frac{\bar{D}^0_1D^{0}-\bar{D}^{0}D^{0}_1}{\sqrt{2}}\, , \,\frac{\bar{D}^-_1D^{+}-\bar{D}^{-}D^{+}_1}{\sqrt{2}}\, ,
\end{eqnarray}
have the largest partial decay width $59.7\pm5.5\,\rm{MeV}$; while the decay,
\begin{eqnarray}
Y_{\widetilde{A}V} &\to& J/\psi\omega\, ,
\end{eqnarray}
has zero partial decay width.
For the $Y_{\widetilde{V}A}$ state, the decays,
\begin{eqnarray}
Y_{\widetilde{V}A}&\to&\frac{\bar{D}^0_0D^{*0}-\bar{D}^{*0}D^{0}_0}{\sqrt{2}}\, , \, \frac{\bar{D}^-_0D^{*+}-\bar{D}^{*-}D^{+}_0}{\sqrt{2}}\, ,
\end{eqnarray}
have the largest partial decay width $66.3\pm6.1\,\rm{MeV}$; while the decay,
\begin{eqnarray}
Y_{\widetilde{V}A} &\to& J/\psi\omega\, ,
\end{eqnarray}
has zero partial decay width.
For the $Y_{S\widetilde{V}}$ state, the decay,
\begin{eqnarray}
Y_{S\widetilde{V}}&\to&\eta_c\omega\, ,
\end{eqnarray}
has the largest partial decay width $96.0\pm34.8 \,\rm{MeV}$; while the decay,
\begin{eqnarray}
Y_{S\widetilde{V}} &\to& J/\psi\omega\, ,
\end{eqnarray}
has the partial decay width $18.0\pm6.3\,\rm{MeV}$. We can search for the $Y(4500)$ in  those typical decays to diagnose its nature.

\subsection{Light-cone QCD sum rules for  the  $Y(4500)$ as an example}\label{LC-QCDSR-Y4500}
The light-cone QCD sum rules have been applied extensively to study the two-body strong decays of the tetraquark (molecular) states \cite{X2900-mole-Azizi-JPG-2021,Azizi-Zc3900-decay-PRD-2016,Azizi-Review-2020,LC-Azizi-PRD-2017,LC-Azizi-EPJC-2017,
LC-Azizi-PRD-2018,LC-HSun-NPB-2024,LC-Ozdem-EPJC-2018}, where only the ground state contributions are isolated, and an unknown parameter is (or not) introduced by hand to parameterize the higher resonance contributions, then the Ioffe-Smilga-type trick,
\begin{eqnarray}
1-T^2\frac{d}{dT^2}\, ,
\end{eqnarray}
is (or not) adopted to subtract this parameter \cite{Ioffe-NPB-1984,Belyaev-PRD-1995}. The Ioffe-Smilga-type trick was suggested to deal with the traditional hadrons, where there exists a triangle Feynman diagram. In the case of tetraquark (molecular) states, we deal with two disconnected loop diagrams approximately, see Fig.\ref{Y4500-Two-decay-Feyn-fig}, we would like not to resort to the Ioffe-Smilga trick, and write down the higher resonance contributions explicitly.

We would like to use an example to illustrate how to study the strong decays of the tetraquark states via the light-cone QCD sum rules, and  write down  the three-point correlation function  $\Pi_{\mu\alpha\beta}(p)$,
\begin{eqnarray}
\Pi_{\mu\alpha\beta}(p,q)&=&i^2\int d^4xd^4y \, e^{-ip\cdot x}e^{-iq\cdot y}\, \langle 0|T\left\{J_{\mu}^{Y}(0)J_\alpha^{D^{*+}}(x)J^{\bar{D}^{*0}}_{\beta}(y)\right\}|\pi(r)\rangle\, ,
\end{eqnarray}
where
\begin{eqnarray}\label{current-JY}
J_{\mu}^{Y}(0)&=&\frac{\varepsilon^{ijk}\varepsilon^{imn}}{2}\Big[u^{T}_j(0)C\sigma_{\mu\nu}\gamma_5 c_k(0)\bar{u}_m(0)\gamma_5\gamma^\nu C \bar{c}^{T}_n(0)+u^{T}_j(0)C\gamma^\nu\gamma_5 c_k(0)\bar{u}_m(0)\gamma_5\sigma_{\mu\nu} C \bar{c}^{T}_n(0) \nonumber\\
&&+d^{T}_j(x)C\sigma_{\mu\nu}\gamma_5 c_k(0)\bar{d}_m(0)\gamma_5\gamma^\nu C \bar{c}^{T}_n(0)+d^{T}_j(0)C\gamma^\nu\gamma_5 c_k(0)\bar{d}_m(0)\gamma_5\sigma_{\mu\nu} C \bar{c}^{T}_n(0) \Big] \, , \nonumber\\
J_{\alpha}^{D^{*+}}(y)&=&\bar{d}(x)\gamma_{\alpha} c(x) \, ,\nonumber \\
J_{\beta}^{\bar{D}^{*0}}(x)&=&\bar{c}(y)\gamma_{\beta} u(y) \, ,
\end{eqnarray}
interpolate the $Y(4500)$,  $\bar{D}^*$ and $D^*$, respectively \cite{LightCone-Y4500-NPB-2023}, the $|\pi(r)\rangle$ is the external $\pi$ state.

At the hadron side, we insert  a complete set of intermediate hadronic states with the same quantum numbers as the interpolating currents  into the three-point correlation function, and  isolate the ground state contributions,
\begin{eqnarray}\label{Hadron-CT-LC}
\Pi_{\mu\alpha\beta}(p,q)&=& \lambda_Y f_{D^*}^2M_{D^*}^2 \frac{-iG_{\pi}r_\tau+iG_{Y}p^\prime_\tau}{(M_{Y}^2-p^{\prime2})(M_{\bar{D}^*}^2-p^2)(M_{D^*}^2-q^2)}\varepsilon^{\rho\sigma\lambda\tau}  \left( -g_{\mu\rho}+\frac{p^\prime_\mu p^\prime_\rho}{p^{\prime2}}\right)\nonumber\\
&&\left( -g_{\alpha\sigma}+\frac{p_\alpha p_\sigma}{p^2}\right)\left( -g_{\lambda\beta}+\frac{q_\lambda q_\beta}{q^2}\right) + \cdots\, ,
\end{eqnarray}
where $p^\prime=p+q+r$, we adopt the standard definitions for the  decay constants $\lambda_Y$, $f_{D^*}$, $f_{\bar{D}^*}$, and define the hadronic coupling constants $G_\pi$ and $G_Y$,
\begin{eqnarray}\label{define-G-pi-X}
\langle Y_c(p^\prime)|\bar{D}^*(p)D^*(q)\pi(r)\rangle&=&G_{\pi}\varepsilon^{\rho\sigma\lambda\tau}\varepsilon^*_\rho\xi_\sigma\zeta_\lambda r_\tau-G_{Y}\varepsilon^{\rho\sigma\lambda\tau}\varepsilon^*_\rho\xi_\sigma\zeta_\lambda p^\prime_\tau \,\, ,
\end{eqnarray}
 the $\varepsilon_{\mu}$, $\xi_\alpha$ and $\zeta_\beta$  are polarization vectors of the $Y(4500)$, $\bar{D}^*$ and  $D^*$, respectively. In the isospin limit, $m_u=m_d$, $f_{D^*}=f_{\bar{D}^*}$ and $M_{D^*}=M_{\bar{D}^*}$.

We multiply Eq.\eqref{Hadron-CT-LC} with the tensor $\varepsilon_{\theta\omega}{}^{\alpha\beta}$ and obtain
\begin{eqnarray}\label{Hadron-CT-W}
\widetilde{\Pi}_{\mu\theta\omega}(p,q)&=& \varepsilon_{\theta\omega}{}^{\alpha\beta}\,\Pi_{\mu\alpha\beta}(p,q)\nonumber\\
&=&\lambda_Y f_{D^*}^2M_{D^*}^2 \frac{iG_{\pi}\left(g_{\mu\omega}r_\theta-g_{\mu\theta}r_\omega\right)-iG_{Y}
\left(g_{\mu\omega}p^\prime_\theta-g_{\mu\theta}p^\prime_\omega\right)}{(M_{Y}^2-p^{\prime2})(M_{\bar{D}^*}^2-p^2)
(M_{D^*}^2-q^2)}  + \cdots\, .
\end{eqnarray}
Again, we take the isospin limit,  then $\widetilde{\Pi}_{\mu\theta\omega}(p,q)=\widetilde{\Pi}_{\mu\theta\omega}(q,p)$,  and we write down the relevant components explicitly,
\begin{eqnarray}
\widetilde{\Pi}_{\mu\theta\omega}(p,q)&=&\left[i\Pi_{\pi}(p^{\prime2},p^2,q^2)-i\Pi_{Y}(p^{\prime2},p^2,q^2)\right]
\left(g_{\mu\omega}r_\theta-g_{\mu\theta}r_\omega\right) \nonumber\\
&&+i\Pi_{Y}(p^{\prime2},p^2,q^2)
\left(g_{\mu\omega}q_\theta-g_{\mu\theta}q_\omega\right)+\cdots \, ,
\end{eqnarray}
where
\begin{eqnarray}
\Pi_{\pi}(p^{\prime2},p^2,q^2)&=& \frac{\lambda_Y f_{D^*}^2M_{D^*}^2 G_{\pi}}{(M_{Y}^2-p^{\prime2})(M_{\bar{D}^*}^2-p^2)
(M_{D^*}^2-q^2)}  + \cdots\, ,\nonumber\\
\Pi_{Y}(p^{\prime2},p^2,q^2)&=& \frac{\lambda_Y f_{D^*}^2M_{D^*}^2 G_{Y}}{(M_{Y}^2-p^{\prime2})(M_{\bar{D}^*}^2-p^2)
(M_{D^*}^2-q^2)}  + \cdots\, .
\end{eqnarray}
Then we choose the tensor structures $g_{\mu\omega}r_\theta-g_{\mu\theta}r_\omega$ and $g_{\mu\omega}q_\theta-g_{\mu\theta}q_\omega$ to study the hadronic coupling constants  $G_\pi$ and $G_{Y}$, respectively. We obtain the hadronic  spectral densities $\rho_H(s^\prime,s,u)$ through triple  dispersion relation,
\begin{eqnarray}
\Pi_{H}(p^{\prime2},p^2,q^2)&=&\int_{\Delta_s^{\prime2}}^\infty ds^{\prime} \int_{\Delta_s^2}^\infty ds \int_{\Delta_u^2}^\infty du \frac{\rho_{H}(s^\prime,s,u)}{(s^\prime-p^{\prime2})(s-p^2)(u-q^2)}\, ,
\end{eqnarray}
where the $\Delta_{s}^{\prime2}$, $\Delta_{s}^{2}$ and
$\Delta_{u}^{2}$ are the thresholds, and we add the subscript $H$ to represent the hadron side.

\begin{figure}
 \centering
  \includegraphics[totalheight=5cm,width=12cm]{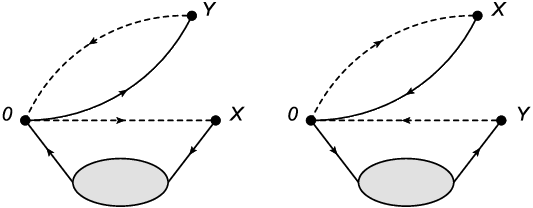}
 \caption{ The lowest order  Feynman diagrams, where the dashed (solid) lines denote the heavy (light) quark lines, the ovals denote the external $\pi^+$ meson.}\label{Y-DvDvpi-fig}
\end{figure}

We carry out   the operator product expansion up to the vacuum condensates of dimension 5 and neglect the tiny gluon condensate contributions \cite{WangZhang-Solid,WangZG-Y4660-tetra-decay-EPJC-2019},
\begin{eqnarray}\label{QCD-CT-1}
\Pi_{\pi}(p^2,q^{\prime2},q^2)&=& f_\pi m_c \int_0^1 du \varphi_\pi(u) \left[\int_0^1 dx x\bar{x}\frac{\Gamma(\epsilon-1)}{2\pi^2(p^2-\tilde{m}_c^2)^{\epsilon-1}}
-\frac{2m_c\langle\bar{q}q\rangle}{3(p^2-m_c^2)}\right.\nonumber\\
&&\left.+\frac{m_c^3\langle\bar{q}g_s\sigma Gq\rangle}{3(p^2-m_c^2)^3} \right]\frac{1}{(q+ur)^2-m_c^2}\nonumber\\
&&+ \frac{f_\pi m_\pi^2}{m_u+m_d} \int_0^1 du \varphi_5(u)\bar{u} \left[\int_0^1 dx x\bar{x}\frac{\Gamma(\epsilon-1)}{2\pi^2(p^2-\tilde{m}_c^2)^{\epsilon-1}}
-\frac{2m_c\langle\bar{q}q\rangle}{3(p^2-m_c^2)}\right. \nonumber\\
&&\left.+\frac{m_c^3\langle\bar{q}g_s\sigma Gq\rangle}{3(p^2-m_c^2)^3} \right]\frac{1}{(q+ur)^2-m_c^2}\nonumber\\
&&-\frac{f_\pi m_c^2\langle\bar{q}g_s\sigma Gq\rangle}{36} \int_0^1 du \varphi_\pi(u) \frac{1}{(p^2-m_c^2)((q+ur)^2-m_c^2)^2} \nonumber\\
&&+\frac{f_\pi m_\pi^2 m_c\langle\bar{q}g_s\sigma Gq\rangle}{36(m_u+m_d)} \int_0^1 du \varphi_5(u)\bar{u} \frac{1}{(p^2-m_c^2)((q+ur)^2-m_c^2)^2} \, ,
\end{eqnarray}

\begin{eqnarray}\label{QCD-CT-2}
\Pi_{Y}(p^2,q^{\prime2},q^2)&=& \frac{f_\pi m_\pi^2}{m_u+m_d} \int_0^1 du \varphi_5(u) \left[\int_0^1 dx x\bar{x}\frac{\Gamma(\epsilon-1)}{2\pi^2(p^2-\tilde{m}_c^2)^{\epsilon-1}}
-\frac{2m_c\langle\bar{q}q\rangle}{3(p^2-m_c^2)}\right. \nonumber\\
&&\left.+\frac{m_c^3\langle\bar{q}g_s\sigma Gq\rangle}{3(p^2-m_c^2)^3} \right]\frac{1}{(q+ur)^2-m_c^2}\nonumber\\
&&+\frac{f_\pi m_\pi^2 m_c\langle\bar{q}g_s\sigma Gq\rangle}{36(m_u+m_d)} \int_0^1 du \varphi_5(u) \frac{1}{(p^2-m_c^2)((q+ur)^2-m_c^2)^2} \nonumber\\
&&+f_{3\pi}m_\pi^2 \int_0^1 dx \bar{x} \left[\frac{3\Gamma(\epsilon)}{8\pi^2(p^2-\tilde{m}_c^2)^\epsilon}-\frac{p^2}{2\pi^2(p^2-\tilde{m}_c^2)}  \right]\frac{1}{q^2-m_c^2} \nonumber\\
&&-f_{3\pi}m_\pi^2 \int_0^1 dx x\bar{x} \left[\frac{\Gamma(\epsilon-1)}{2\pi^2(p^2-\tilde{m}_c^2)^{\epsilon-1}}+\frac{p^2\Gamma(\epsilon)}{2\pi^2(p^2-\tilde{m}_c^2)^{\epsilon}}  \right]\frac{1}{(q^2-m_c^2)^2} \nonumber\\
&&-f_{3\pi}m_\pi^2 \int_0^1 dx x \left[\frac{3\Gamma(\epsilon)}{8\pi^2(p^2-\tilde{m}_c^2)^\epsilon}+\frac{p^2}{4\pi^2(p^2-\tilde{m}_c^2)}  \right]\frac{1}{q^2-m_c^2} \, ,
\end{eqnarray}
where $q^\prime=q+r$, $\bar{u}=1-u$, $\bar{x}=1-x$, $\tilde{m}_c^2=\frac{m_c^2}{x}$, $(q-ur)^2-m_c^2=(1-u)q^2+u(q+r)^2-u\bar{u}m_\pi^2-m_c^2$.
And we have used the $\pi$ light-cone distribution functions \cite{PBall-LCDF},
\begin{eqnarray}
\langle 0|\bar{d}(0)\gamma_\mu\gamma_5 u(x)|\pi(r)\rangle &=&if_\pi r_\mu \int_0^1 du e^{-iur\cdot x} \varphi_{\pi}(u)+\cdots\, , \nonumber\\
\langle 0|\bar{d}(0)\sigma_{\mu\nu}\gamma_5 u(x)|\pi(r)\rangle &=&\frac{i}{6}\frac{f_\pi m_\pi^2}{m_u+m_d} \left(r_\mu x_\nu -r_\nu x_\mu \right) \int_0^1 du e^{-iur\cdot x} \varphi_{\sigma}(u) \, , \nonumber\\
\langle 0|\bar{d}(0)i\gamma_5 u(x)|\pi(r)\rangle &=& \frac{f_\pi m_\pi^2}{m_u+m_d}  \int_0^1 du e^{-iur\cdot x} \varphi_{5}(u) \, ,
\end{eqnarray}
and the approximation,
\begin{eqnarray} \label{qGq}
\langle 0|\bar{d}(x_1)\sigma_{\mu\nu}\gamma_5g_sG_{\alpha\beta}(x_2) u(x_3)|\pi(r)\rangle &=&if_{3\pi}\left( r_\mu r_\alpha g_{\nu\beta}+r_\nu r_\beta g_{\mu\alpha}-r_\nu r_\alpha g_{\mu\beta}-r_\mu r_\beta g_{\nu\alpha}\right) \, ,\nonumber\\
\end{eqnarray}
for the twist-3 quark-gluon light-cone distribution functions with the value $f_{3\pi}=0.0035\,\rm{GeV}^2$ at the energy scale $\mu=1\,\rm{GeV}$ \cite{PBall-LCDF,Braun-f3pi}.
Such terms proportional to $m_\pi^2$ and their contributions are greatly suppressed, and we also neglect the twist-4 light-cone distribution functions due to their small contributions. According to the Gell-Mann-Oakes-Renner relation $\frac{f_\pi m_\pi^2}{m_u+m_d}=-\frac{2\langle\bar{q}q\rangle}{f_\pi}$, we take account of the Chiral enhanced contributions fully in Eqs.\eqref{QCD-CT-1}-\eqref{QCD-CT-2}.

In Fig.\ref{Y-DvDvpi-fig}, we draw the lowest order Feynman diagrams as an example to illustrate the operator product expansion.

In the soft limit $r_\mu \to 0$, $(q+r)^2=q^2$, we can set $\Pi_{\pi/Y}(p^2,q^{\prime2},q^2)=\Pi_{\pi/Y}(p^2,q^2)$, then we obtain the QCD spectral densities $\rho_{QCD}(s,u)$  through double dispersion relation,
\begin{eqnarray}
\Pi^{QCD}_{\pi/Y}(p^2,q^2)&=& \int_{\Delta_s^2}^\infty ds \int_{\Delta_u^2}^\infty du \frac{\rho_{QCD}(s,u)}{(s-p^2)(u-q^2)}\, ,
\end{eqnarray}
 we add the superscript (subscript) $QCD$ to stand for  the QCD side.

We match the hadron side with the QCD side  below  the continuum thresholds $s_0$ and $u_0$ to acquire   rigorous quark-hadron  duality  \cite{WangZhang-Solid,WangZG-Y4660-tetra-decay-EPJC-2019},
 \begin{eqnarray}
  \int_{\Delta_s^2}^{s_{0}}ds \int_{\Delta_u^2}^{u_0}du  \frac{\rho_{QCD}(s,u)}{(s-p^2)(u-q^2)}&=& \int_{\Delta_s^2}^{s_0}ds \int_{\Delta_u^2}^{u_0}du  \left[ \int_{\Delta_{s}^{\prime2}}^{\infty}ds^\prime  \frac{\rho_H(s^\prime,s,u)}{(s^\prime-p^{\prime2})(s-p^2)(u-q^2)} \right]\, , \nonumber\\
\end{eqnarray}
and  we carry out  the integral over $ds^\prime$ firstly,
then
\begin{eqnarray}
\Pi_{H}(p^{\prime2},p^2,q^2)&=& \frac{\lambda_Y f_{D^*}^2M_{D^*}^2 G_{\pi/Y}}{(M_{Y}^2-p^{\prime2})(M_{\bar{D}^*}^2-p^2)
(M_{D^*}^2-q^2)}  +\int_{s^\prime_0}^{\infty}ds^\prime\frac{\tilde{\rho}_{H}(s^\prime,M_{\bar{D}^*}^2,M_{D^*}^2)}
{(s^\prime-p^{\prime2})(M_{\bar{D}^*}^2-p^2)
(M_{D^*}^2-q^2)}\nonumber\\
&& +\cdots\, ,\nonumber\\
&=& \frac{\lambda_Y f_{D^*}^2M_{D^*}^2 G_{\pi/Y}}{(M_{Y}^2-p^{\prime2})(M_{\bar{D}^*}^2-p^2)
(M_{D^*}^2-q^2)}  +\frac{C_{\pi/Y}}{(M_{\bar{D}^*}^2-p^2)(M_{D^*}^2-q^2)}+\cdots\, ,
\end{eqnarray}
where $\rho_{H}(s^\prime,s,u)=\tilde{\rho}_{H}(s^\prime,s,u)\delta(s-M_{\bar{D}^*}^2)\delta(u-M_{D^*}^2)$,
and we introduce the parameters $C_{\pi/Y}$ to parameterize the contributions concerning  the higher resonances and continuum states in the $s^\prime$ channel,
\begin{eqnarray}
C_{\pi/Y}&=&\int_{s^\prime_0}^{\infty}ds^\prime\frac{\tilde{\rho}_{H}(s^\prime,M_{\bar{D}^*}^2,M_{D^*}^2)}{
s^\prime-p^{\prime2}}\, .
\end{eqnarray}
As the strong interactions among the ground states $\pi$, $D^*$, $\bar{D}^*$ and excited $Y^\prime$ states are complex, and we have no knowledge about the corresponding four-hadron contact vertex. In practical calculations, we can take the unknown functions $C_{\pi/Y}$ as free parameters and adjust the values to acquire flat platforms  for the hadronic coupling constants $G_{\pi/Y}$ with variations of the Borel parameters. Such a method works well in the case of three-hadron  contact vertexes \cite{WangZhang-Solid,WangZG-Y4660-tetra-decay-EPJC-2019}, and we expect it also works here.

In Eq.\eqref{Hadron-CT-LC} and Eq.\eqref{Hadron-CT-W}, there exist three poles in the limit
$p^{\prime2} \to M_{Y}^2$, $p^2 \to M_{\bar{D}^*}^2$ and $q^2 \to M_{D^*}^2$. According to the relation $M_Y\approx M_{\bar{D}^*}+M_{D^*}$, we set $p^{\prime2}=4q^2$  and  perform  double Borel transformation  with respect  to the variables $P^2=-p^2$ and $Q^2=-q^2$ respectively, then we set  $T_1^2=T_2^2=T^2$  to obtain two QCD sum rules,
\begin{eqnarray} \label{pi-SR}
&&\frac{\lambda_{YD^*D^*}G_{\pi}}{4\left(\widetilde{M}_{Y}^2-M_{D^*}^2\right)} \left[ \exp\left(-\frac{M_{D^*}^2}{T^2} \right)-\exp\left(-\frac{\widetilde{M}_{Y}^2}{T^2} \right)\right]\exp\left(-\frac{M_{\bar{D}^*}^2}{T^2} \right)+C_{\pi} \exp\left(-\frac{M_{D^*}^2+M_{\bar{D}^*}^2}{T^2}  \right) \nonumber\\
&&= f_\pi m_c \int_{m_c^2}^{s_0}ds\int_0^1 du \varphi_\pi(u) \left[\frac{1}{2\pi^2}\int_{x_i}^1 dx x\bar{x}(s-\tilde{m}_c^2)
-\left(\frac{2m_c\langle\bar{q}q\rangle}{3}-\frac{m_c^3\langle\bar{q}g_s \sigma Gq\rangle}{6T^4}\right)\delta(s-m_c^2)\right]\nonumber\\
&&\exp\left(-\frac{s+m_c^2+u\bar{u}m_\pi^2}{T^2} \right)\nonumber\\
&&+ \frac{f_\pi m_\pi^2}{m_u+m_d}\int_{m_c^2}^{s_0}ds \int_0^1 du \varphi_5(u)\bar{u} \left[\frac{1}{2\pi^2}\int_{x_i}^1 dx x\bar{x}(s-\tilde{m}_c^2)
-\left(\frac{2m_c\langle\bar{q}q\rangle}{3}-\frac{m_c^3\langle\bar{q}g_s \sigma Gq\rangle}{6T^4}\right)\delta(s-m_c^2)\right]\nonumber\\
&&\exp\left(-\frac{s+m_c^2+u\bar{u}m_\pi^2}{T^2} \right)+\frac{f_\pi m_c^2\langle\bar{q}g_s\sigma Gq\rangle}{36T^2} \int_0^1 du \varphi_\pi(u) \exp\left(-\frac{2m_c^2+u\bar{u}m_\pi^2}{T^2} \right) \nonumber\\
&&-\frac{f_\pi m_\pi^2 m_c\langle\bar{q}g_s\sigma Gq\rangle}{36(m_u+m_d)T^2} \int_0^1 du \varphi_5(u)\bar{u} \exp\left(-\frac{2m_c^2+u\bar{u}m_\pi^2}{T^2} \right) \, ,
\end{eqnarray}

\begin{eqnarray} \label{X-SR}
&&\frac{\lambda_{YD^*D^*}G_{Y}}{4\left(\widetilde{M}_{Y}^2-M_{D^*}^2\right)} \left[ \exp\left(-\frac{M_{D^*}^2}{T^2} \right)-\exp\left(-\frac{\widetilde{M}_{Y}^2}{T^2} \right)\right]\exp\left(-\frac{M_{\bar{D}^*}^2}{T^2} \right)+C_{Y} \exp\left(-\frac{M_{D^*}^2+M_{\bar{D}^*}^2}{T^2}  \right) \nonumber\\
&&= \frac{f_\pi m_\pi^2}{m_u+m_d} \int_{m_c^2}^{s_0}ds\int_0^1 du \varphi_5(u) \left[\frac{1}{2\pi^2}\int_{x_i}^1 dx x\bar{x}(s-\tilde{m}_c^2)
-\left(\frac{2m_c\langle\bar{q}q\rangle}{3}-\frac{m_c^3\langle\bar{q}g_s \sigma Gq\rangle}{6T^4}\right)\delta(s-m_c^2)\right]\nonumber\\
&&\exp\left(-\frac{s+m_c^2+u\bar{u}m_\pi^2}{T^2} \right) -\frac{f_\pi m_\pi^2 m_c\langle\bar{q}g_s\sigma Gq\rangle}{36(m_u+m_d)T^2} \int_0^1 du \varphi_5(u) \exp\left(-\frac{2m_c^2+u\bar{u}m_\pi^2}{T^2} \right)  \nonumber\\
&&-\frac{f_{3\pi}m_\pi^2}{2\pi^2}\int_{m_c^2}^{s_0}ds \int_{x_i}^1 dx \bar{x} \left[\frac{3}{4}+s\,\delta(s-\tilde{m}_c^2) \right]\exp\left(-\frac{s+m_c^2}{T^2} \right)\nonumber\\
&&  -\frac{f_{3\pi}m_\pi^2}{2\pi^2 T^2}\int_{m_c^2}^{s_0}ds \int_{x_i}^1 dx x\bar{x} \,\tilde{m}_c^2\exp\left(-\frac{s+m_c^2}{T^2} \right) \nonumber\\
&&+\frac{f_{3\pi}m_\pi^2}{4\pi^2}\int_{m_c^2}^{s_0}ds \int_{x_i}^1 dx x \left[\frac{3}{2}-s\,\delta(s-\tilde{m}_c^2) \right]\exp\left(-\frac{s+m_c^2}{T^2} \right) \, ,
\end{eqnarray}
where $\lambda_{YD^*D^*}=\lambda_{Y}f_{D^*}^2M^2_{D^*}$, $\widetilde{M}_Y^2=\frac{M_Y^2}{4}$ and $x_i=\frac{m_c^2}{s}$.

\begin{figure}
\centering
\includegraphics[totalheight=9cm,width=15cm]{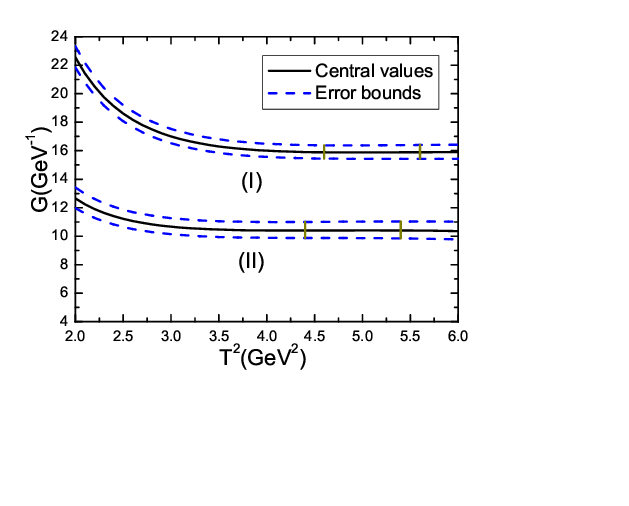}
  \caption{ The hadronic coupling constants with variations of the Borel parameters $T^2$, where the (I) and (II) denote the $G_{\pi}$ and $G_{Y}$, respectively, the regions between the two vertical lines are the Borel windows.  }\label{G-pi-X}
\end{figure}

In calculations, we fit the free parameters as $C_{\pi}=0.00101(T^2-3.6\,\rm{GeV}^2)\,\rm{GeV}^4$ and
$C_{Y}=0.00089(T^2-3.2\,\rm{GeV}^2)\,\rm{GeV}^4$
  to acquire uniform flat Borel platforms  $T^2_{max}-T^2_{min}=1\,\rm{GeV}^2$.
The Borel windows  are $T^2_{\pi}=(4.6-5.6)\,\rm{GeV}^2$ and $T^2_{Y}=(4.4-5.4)\,\rm{GeV}^2$, where  the subscripts $\pi$ and $Y$ represent the corresponding  channels. In Fig.\ref{G-pi-X}, we plot the hadronic coupling constants $G_{\pi}$ and $G_{Y}$ with variations of the Borel parameters. In the Borel windows, there appear very flat platforms indeed.

We obtain the hadronic coupling constants routinely,
\begin{eqnarray} \label{HCC-values}
G_{\pi} &=&15.9 \pm 0.5\,\rm{GeV}^{-1}\, , \nonumber\\
G_{Y} &=&10.4\pm 0.6\,\rm{GeV}^{-1}\, ,
\end{eqnarray}
by setting
\begin{eqnarray}\label{Uncertainty-5}
\delta\,\bar{\lambda}_{Y}\bar{f}_{D^*}\bar{f}_{\bar{D}^*}\bar{G}_{\pi/Y}  &=&\bar{\lambda}_{Y}\bar{f}_{D^*}\bar{f}_{\bar{D}^*}\bar{G}_{\pi/Y}\frac{4\delta G_{\pi/Y}}{\bar{G}_{\pi/Y}}\, .
\end{eqnarray}
It is direct to obtain the partial decay width,
\begin{eqnarray} \label{Partial-with}
\Gamma\left(Y(4500)\to D^*\bar{D}^* \pi^+\right)&=&\frac{1}{24\pi M_Y} \int dk^2 (2\pi)^4\delta^4(p^\prime-k-p)\frac{d^3 \vec{k}}{(2\pi)^3 2k_0}\frac{d^3 \vec{p}}{(2\pi)^3 2p_0} \nonumber\\
&& (2\pi)^4\delta^4(k-q-r)\frac{d^3 \vec{q}}{(2\pi)^3 2q_0}\frac{d^3 \vec{r}}{(2\pi)^3 2r_0}\Sigma |T|^2 \nonumber\\
&=&6.43^{+0.80}_{-0.76}\,\rm{MeV}\, ,
\end{eqnarray}
where $T=\langle Y_c(p^\prime)|\bar{D}^*(p)D^*(q)\pi(r)\rangle$.

The partial decay width  $\Gamma\left(Y(4500)\to D^*\bar{D}^* \pi^+\right) =6.43^{+0.80}_{-0.76}\,\rm{MeV}$ is much smaller than the total width  $\Gamma=246.3\pm 36.7\pm 9.4\,\rm{MeV}$ from the BESIII collaboration \cite{BESIII-Y4500-DvDvpi-PRL-2023}.

The three-body strong decays of the $Y(4230)$ are also studied in this scheme \cite{LightCone-Y4230-IJMPA-2023},
this scheme could be applied to the two-body strong decays of the tetraquark (molecular) states straightforwardly.

We write down the  two-point correlation functions
$\Pi(p,r)$,
\begin{eqnarray}
\Pi(p,r)&=&i\int d^4x \, e^{-ip\cdot x}\, \langle 0|T\left\{J_{A}(0)J_B(x)\right\}|P(r)\rangle\, ,
\end{eqnarray}
where the currents $J_A(0)$ and $J_B(x)$ interpolate the tetraquark (molecular) states and traditional mesons, respectively, the $P(r)$ are external states.

At the hadron side, we obtain
\begin{eqnarray}
\Pi(p,r)&=&\frac{\lambda_{A}\lambda_B G_{ABP}}{(M_A^2-(p+r)^2)(M_B^2-p^2)}+\cdots\, ,\nonumber\\
&=&\Pi_H(p^{\prime 2},p^2)\, ,
\end{eqnarray}
where $p^\prime=p+r$, and we rewrite the correlation functions  $\Pi_H(p^{\prime 2},p^2)$  as
\begin{eqnarray}
\Pi_{H}(p^{\prime 2},p^2)&=&\int_{\Delta^2}^{s_{A}^0}ds^\prime \int_{\Delta_s^2}^{s^0_{B}}ds   \frac{\rho_H(s^\prime,s)}{(s^\prime-p^{\prime2})(s-p^2)}+\int_{s^0_A}^{\infty}ds^\prime \int_{\Delta_s^2}^{s^0_{B}}ds \frac{\rho_H(s^\prime,s)}{(s^\prime-p^{\prime2})(s-p^2)}+\cdots\, ,\nonumber\\
\end{eqnarray}
 through double-dispersion relation, where the $\rho_H(s^\prime,s)$   are the hadronic spectral densities,
\begin{eqnarray}
\rho_H(s^\prime,s)&=&{\lim_{\epsilon_2\to 0}}\,\,{\lim_{\epsilon_1\to 0}} \frac{ {\rm Im}_{s^\prime}\, {\rm Im}_{s}\,\Pi_H(s^\prime+i\epsilon_2,s+i\epsilon_1) }{\pi^2} \, ,
\end{eqnarray}
where the $\Delta^2$ and $\Delta_s^2$  are the thresholds, the  $s_{A}^0$ and $s_{B}^0$ are the continuum thresholds.

Then we carry out the operator product expansion (not necessary on the light-cone) at the QCD side, and write the correlation functions  $\Pi_{QCD}(p^{\prime 2},p^2)$  as
\begin{eqnarray}
\Pi_{QCD}(p^{\prime 2},p^2)&=&  \int_{\Delta_s^2}^{s^0_{B}}ds \frac{\rho_{QCD}(p^{\prime2},s)}{s-p^2}+\cdots\, ,
\end{eqnarray}
through single-dispersion relation, where the $\rho_{QCD}(p^{\prime 2},s)$   are the QCD spectral densities,
\begin{eqnarray}
\rho_{QCD}(p^{\prime 2},s)&=& {\lim_{\epsilon_1\to 0}}\,\,\frac{  {\rm Im}_{s}\,\Pi_{QCD}(p^{\prime 2},s+i\epsilon_1) }{\pi} \, .
\end{eqnarray}
As the QCD spectral densities $\rho_{QCD}(s^\prime,s)$ do  not exist,
\begin{eqnarray}
\rho_{QCD}(s^\prime,s)&=&{\lim_{\epsilon_2\to 0}}\,\,{\lim_{\epsilon_1\to 0}} \,\,\frac{ {\rm Im}_{s^\prime}\, {\rm Im}_{s}\,\Pi_{QCD}(s^\prime+i\epsilon_2,s+i\epsilon_1) }{\pi^2} \nonumber\\
&=&0\, ,
\end{eqnarray}
because
\begin{eqnarray}
{\lim_{\epsilon_2\to 0}}\,\,\frac{ {\rm Im}_{s^\prime}\,\Pi_{QCD}(s^\prime+i\epsilon_2,p^2) }{\pi} &=&0\, .
\end{eqnarray}
And we will write the QCD spectral densities  $\rho_{QCD}(p^{\prime 2},s)$ as $\rho_{QCD}(s)$ for simplicity.

We match the hadron side  with the QCD side of the correlation functions,
and accomplish  the integral over $ds^\prime$  firstly to obtain the rigorous quark-hadron  duality \cite{WangZhang-Solid},
\begin{eqnarray}
\int_{\Delta_s^2}^{s_B^0} ds \frac{\rho_{QCD}(s)}{s-p^2}&=&\int_{\Delta_s^2}^{s_B^0} ds \frac{1}{s-p^2}\left[ \int_{\Delta^2}^{\infty} ds^\prime \frac{\rho_{H}(s^{\prime},s)}{s^\prime-p^{\prime2}}\right]\, .
\end{eqnarray}
 And we write down  the quark-hadron duality explicitly,
 \begin{eqnarray}\label{Solid-explicit}
  \int_{\Delta_s^2}^{s^0_{B}}ds \frac{\rho_{QCD}(s)}{s-p^2}
  &=&\frac{\lambda_{A}\lambda_{B}G_{ABP} }{(m_{A}^2-p^{\prime2})(m_{B}^2-p^2)} +\frac{C_{A^{\prime}B}}{m_{B}^2-p^{2}} \, .
\end{eqnarray}
 Again, we introduce the parameters  $C_{A^\prime B}$  to parameterize the net effects. In numerical calculations,   we   take the $C_{A^\prime B}$  as free parameters, and choose suitable values  to  obtain the stable QCD sum rules with the variations of
the Borel parameters $T^2$.

\section{Conclusion and Perspective}
At the present time, we can only say confidently that the tetraquark and pentaquark states are established in sense of that there are four and five valence quarks, respectively. The under-structures are still under hot debates, more experimental and theoretical works are still needed before reaching definite conclusion. The QCD sum rules method is a reliable and powerful theoretical tool in studying the multiquark states and has given many successful descriptions, however, the predictions have arbitrariness depending on the treating schemes, only comprehensive and systematic works would work.

\section*{Appendix}
Some typical and useful examples of the Borel transformations,
\begin{eqnarray}\label{Some-Borel}
\mathcal{B}[(P^2)^k]&=&0\,, \nonumber\\
\mathcal{B}\Big[\frac{\Gamma(k)}{(P^2)^k}\Big]&=&\frac{1}{(T^2)^{k-1}}\,, \nonumber \\
\mathcal{B}\Big[\frac{\Gamma(k)}{(s+P^2)^k}\Big]&=&\frac{1}{(T^2)^{k-1}}\exp\left(-\frac{s}{T^2} \right) \, .
\end{eqnarray}

Some useful examples of the Fierz transformations.
\begin{eqnarray}\label{Fierz-Zc3900}
2\sqrt{2}J_{1^{+-}}^{\mu}&=&2\varepsilon^{ijk}\varepsilon^{imn}\left\{u^T_jC\gamma_5 c_k \bar{d}_m\gamma^\mu C \bar{c}^T_n-u^T_jC\gamma^\mu c_k\bar{d}_m\gamma_5 C \bar{c}^T_n \right\} \, , \nonumber\\
 &=&i\bar{c}i\gamma_5 c\,\bar{d}\gamma^\mu u-i\bar{c} \gamma^\mu c\,\bar{d}i\gamma_5 u+\bar{c} u\,\bar{d}\gamma^\mu\gamma_5 c-\bar{c} \gamma^\mu \gamma_5u\,\bar{d}c- i\bar{c}\gamma_\nu\gamma_5c\, \bar{d}\sigma^{\mu\nu}u \nonumber\\
&& +i\bar{c}\sigma^{\mu\nu}c\, \bar{d}\gamma_\nu\gamma_5u
- i \bar{c}\sigma^{\mu\nu}\gamma_5u\,\bar{d}\gamma_\nu c+i\bar{c}\gamma_\nu u\, \bar{d}\sigma^{\mu\nu}\gamma_5c   \, ,
\end{eqnarray}
\begin{eqnarray}
2\sqrt{2}J_{1^{+-}}^{\mu\nu}&=&2\varepsilon^{ijk}\varepsilon^{imn}\left\{u^T_jC\gamma^\mu c_k \bar{d}_m\gamma^\nu C \bar{c}^T_n-u^T_jC\gamma^\nu c_k\bar{d}_m\gamma^\mu C \bar{c}^T_n \right\} \, , \nonumber\\
 &=&i\bar{d}u\, \bar{c}\sigma^{\mu\nu}c +i\bar{d}\sigma^{\mu\nu}u \,\bar{c}c+i\bar{d}c\, \bar{c}\sigma^{\mu\nu}u +i\bar{d}\sigma^{\mu\nu}c \,\bar{c}u  \nonumber\\
 &&-\bar{c}\sigma^{\mu\nu}\gamma_5c\,\bar{d}i\gamma_5u-\bar{c}i\gamma_5 c\,\bar{d}\sigma^{\mu\nu}\gamma_5u -\bar{d}\sigma^{\mu\nu}\gamma_5c\,\bar{c}i\gamma_5u-\bar{d}i\gamma_5 c\,\bar{c}\sigma^{\mu\nu}\gamma_5u\nonumber\\
 &&+i\varepsilon^{\mu\nu\alpha\beta}\bar{c}\gamma^\alpha\gamma_5c\, \bar{d}\gamma^\beta u-i\varepsilon^{\mu\nu\alpha\beta}\bar{c}\gamma^\alpha c\, \bar{d}\gamma^\beta \gamma_5u\nonumber\\
 &&+i\varepsilon^{\mu\nu\alpha\beta}\bar{c}\gamma^\alpha\gamma_5u\, \bar{d}\gamma^\beta c-i\varepsilon^{\mu\nu\alpha\beta}\bar{c}\gamma^\alpha u\, \bar{d}\gamma^\beta \gamma_5c  \, ,
\end{eqnarray}

\begin{eqnarray}
2\sqrt{2}J_{1^{--}}^{\mu}&=&2\varepsilon^{ijk}\varepsilon^{imn}\left\{u^T_jC c_k \bar{d}_m\gamma^\mu C \bar{c}^T_n-u^T_jC\gamma^\mu c_k\bar{d}_m C \bar{c}^T_n \right\} \, , \nonumber\\
 &=&\bar{c} \gamma^\mu c\,\bar{d} u-\bar{c} c\,\bar{d}\gamma^\mu u+i\bar{c}\gamma^\mu\gamma_5 u\,\bar{d}i\gamma_5 c-i\bar{c} i\gamma_5 u\,\bar{d}\gamma^\mu \gamma_5c \nonumber\\
&& - i\bar{c}\gamma_\nu\gamma_5c\, \bar{d}\sigma^{\mu\nu}\gamma_5u+i\bar{c}\sigma^{\mu\nu}\gamma_5c\, \bar{d}\gamma_\nu\gamma_5u
- i\bar{d}\gamma_\nu c\, \bar{c}\sigma^{\mu\nu}u+i \bar{d}\sigma^{\mu\nu}c \,\bar{c}\gamma_\nu u   \, ,
\end{eqnarray}

\begin{eqnarray}
2\sqrt{2}J_{1^{-+}}^{\mu}&=&2\varepsilon^{ijk}\varepsilon^{imn}\left\{u^T_jC c_k \bar{d}_m\gamma^\mu C \bar{c}^T_n+u^T_jC\gamma^\mu c_k\bar{d}_m C \bar{c}^T_n \right\} \, , \nonumber\\
 &=&i\bar{c}i\gamma_5 c\,\bar{d}\gamma^\mu\gamma_5 u-i\bar{c} \gamma^\mu\gamma_5 c\,\bar{d}i\gamma_5 u-\bar{c}\gamma^\mu u\,\bar{d} c+\bar{c} u\,\bar{d}\gamma^\mu c \nonumber\\
&&+i\bar{c}\sigma^{\mu\nu} c\, \bar{d}\gamma_\nu u - i\bar{c}\gamma_\nu c\, \bar{d}\sigma^{\mu\nu} u
- i\bar{d}\gamma_\nu\gamma_5 c\, \bar{c}\sigma^{\mu\nu}\gamma_5u+i \bar{d}\sigma^{\mu\nu}\gamma_5 c \,\bar{c}\gamma_\nu\gamma_5 u  \, ,
\end{eqnarray}

\begin{eqnarray}
2\sqrt{2}J_{1^{--}}^{\mu}&=&2\varepsilon^{ijk}\varepsilon^{imn}\left\{s^T_jC c_k \bar{s}_m\gamma^\mu C \bar{c}^T_n-s^T_jC\gamma^\mu c_k\bar{s}_m C \bar{c}^T_n \right\} \, , \nonumber\\
 &=&\bar{c} \gamma^\mu c\,\bar{s} s-\bar{c} c\,\bar{s}\gamma^\mu s+i\bar{c}\gamma^\mu\gamma_5 s\,\bar{s}i\gamma_5 c-i\bar{c} i\gamma_5 s\,\bar{s}\gamma^\mu \gamma_5c\nonumber\\
&& - i\bar{c}\gamma_\nu\gamma_5c\, \bar{s}\sigma^{\mu\nu}\gamma_5s+i\bar{c}\sigma^{\mu\nu}\gamma_5c\, \bar{s}\gamma_\nu\gamma_5s
- i\bar{s}\gamma_\nu c\, \bar{c}\sigma^{\mu\nu}s+i \bar{s}\sigma^{\mu\nu}c \,\bar{c}\gamma_\nu s   \, , \end{eqnarray}

\begin{eqnarray}
2\sqrt{2}J_{1^{-+}}^{\mu}&=&2\varepsilon^{ijk}\varepsilon^{imn} \left\{s^T_jC c_k \bar{s}_m\gamma^\mu C \bar{c}^T_n+s^T_jC\gamma^\mu c_k\bar{s}_m C \bar{c}^T_n \right\} \, , \nonumber\\
 &=&i\bar{c}i\gamma_5 c\,\bar{s}\gamma^\mu\gamma_5 s-i\bar{c} \gamma^\mu\gamma_5 c\,\bar{s}i\gamma_5 s-\bar{c}\gamma^\mu s\,\bar{s} c+\bar{c} s\,\bar{s}\gamma^\mu c \nonumber\\
&&+i\bar{c}\sigma^{\mu\nu} c\, \bar{s}\gamma_\nu s - i\bar{c}\gamma_\nu c\, \bar{s}\sigma^{\mu\nu} s
- i\bar{s}\gamma_\nu\gamma_5 c\, \bar{c}\sigma^{\mu\nu}\gamma_5s+i \bar{s}\sigma^{\mu\nu}\gamma_5 c \,\bar{c}\gamma_\nu\gamma_5 s   \, ,
\end{eqnarray}

\begin{eqnarray}
2\sqrt{2}J_{2^{++}}^{\mu\nu}&=&2 \varepsilon^{ijk}\varepsilon^{imn}\left\{u^T_jC\gamma^\mu c_k \bar{d}_m\gamma^\nu C \bar{c}^T_n+u^T_jC\gamma^\nu c_k\bar{d}_m\gamma^\mu C \bar{c}^T_n \right\} \, , \nonumber\\
 &=& \bar{c}\gamma^\mu\gamma_5c\, \bar{d}\gamma^\nu\gamma_5u+\bar{c}\gamma^\nu\gamma_5c\, \bar{d}\gamma^\mu\gamma_5u -\bar{c}\gamma^\mu c\, \bar{d}\gamma^\nu u-\bar{c}\gamma^\nu c\, \bar{d}\gamma^\mu u +\bar{c}\gamma^\mu\gamma_5u\, \bar{d}\gamma^\nu\gamma_5c\nonumber\\
 &&+\bar{c}\gamma^\nu\gamma_5u\, \bar{d}\gamma^\mu\gamma_5c -\bar{c}\gamma^\mu u\, \bar{d}\gamma^\nu c-\bar{c}\gamma^\nu u\, \bar{d}\gamma^\mu c +g_{\alpha\beta}\bar{c}\sigma^{\mu\alpha}c\, \bar{d}\sigma^{\nu\beta}u+g_{\alpha\beta}\bar{c}\sigma^{\nu\alpha}c\, \bar{d}\sigma^{\mu\beta}u \nonumber\\
 &&+g_{\alpha\beta}\bar{c}\sigma^{\mu\alpha}u\, \bar{d}\sigma^{\nu\beta}c+g_{\alpha\beta}\bar{c}\sigma^{\nu\alpha}u\, \bar{d}\sigma^{\mu\beta}c +g^{\mu\nu} \bar{c}c\,\bar{d}u+g^{\mu\nu}\bar{c}i\gamma_5c\,\bar{d}i\gamma_5u+g^{\mu\nu}\bar{c}\gamma_{\alpha} c\,\bar{d}\gamma^{\alpha}u\nonumber\\
 &&-g^{\mu\nu}\bar{c}\gamma_{\alpha}\gamma_5 c\,\bar{d}\gamma^{\alpha}\gamma_5u-\frac{1}{2}g^{\mu\nu}\bar{c}\sigma_{\alpha\beta} c\,\bar{d}\sigma^{\alpha\beta}u+g^{\mu\nu}\bar{c}u\,\bar{d}c+g^{\mu\nu}\bar{c}i\gamma_5u\,\bar{d}i\gamma_5c \nonumber\\
 &&+g^{\mu\nu}\bar{c}\gamma_{\alpha} u\,\bar{d}\gamma^{\alpha}c-g^{\mu\nu}\bar{c}\gamma_{\alpha}\gamma_5 u\,\bar{d}\gamma^{\alpha}\gamma_5c-\frac{1}{2}g^{\mu\nu}\bar{c}\sigma_{\alpha\beta} u\,\bar{d}\sigma^{\alpha\beta}c \, ,
\end{eqnarray}

\begin{eqnarray}\label{Fierz-Scalar}
J_{0^{++}}&=&\varepsilon^{ijk}\varepsilon^{imn}u^T_jC\gamma_\mu c_k \bar{d}_m\gamma^\mu C \bar{c}^T_n \, , \nonumber\\
 &=&  \bar{c}c\,\bar{d}u+\bar{c}i\gamma_5c\,\bar{d}i\gamma_5u+\frac{1}{2}\bar{c}\gamma_{\alpha} c\,\bar{d}\gamma^{\alpha}u-\frac{1}{2}\bar{c}\gamma_{\alpha}\gamma_5 c\,\bar{d}\gamma^{\alpha}\gamma_5u \nonumber\\
 &&+\bar{c}u\,\bar{d}c+\bar{c}i\gamma_5u\,\bar{d}i\gamma_5c+\frac{1}{2}\bar{c}\gamma_{\alpha} u\,\bar{d}\gamma^{\alpha}c-\frac{1}{2}\bar{c}\gamma_{\alpha}\gamma_5 u\,\bar{d}\gamma^{\alpha}\gamma_5c \, .
\end{eqnarray}

\begin{table}
\begin{center}
\begin{tabular}{|c|c|c|c|c|c|c|c|}\hline\hline
                                                      &$M_{P}(\rm{GeV})$ &$\lambda_{P}(\rm{GeV}^6)$    &$M_{B_{10}J/\psi(B_8J/\psi)}(\rm{GeV})$\\ \hline
$P_{uuu}^{11\frac{1}{2}}\left({\frac{1}{2}^-}\right)$            &$4.35\pm0.15$     &$(3.72\pm0.76)\times10^{-3}$  & 4.33 (4.04)\\ \hline
$P_{uus}^{11\frac{1}{2}}\left({\frac{1}{2}^-}\right)$            &$4.47\pm0.15$     &$(4.50\pm0.85)\times10^{-3}$  & 4.48 (4.29)\\ \hline
$P_{uss}^{11\frac{1}{2}}\left({\frac{1}{2}^-}\right)$            &$4.58\pm0.14$     &$(5.43\pm0.96)\times10^{-3}$  & 4.63 (4.41)\\ \hline
$P_{sss}^{11\frac{1}{2}}\left({\frac{1}{2}^-}\right)$            &$4.68\pm0.13$     &$(6.47\pm1.10)\times10^{-3}$  & 4.77\\ \hline

$P_{uuu}^{10\frac{1}{2}}\left({\frac{1}{2}^-}\right)$            &$4.42\pm0.12$     &$(4.14\pm0.70)\times10^{-3}$  & 4.33 (4.04)\\ \hline
$P_{uus}^{10\frac{1}{2}}\left({\frac{1}{2}^-}\right)$            &$4.51\pm0.11$     &$(4.97\pm0.79)\times10^{-3}$  & 4.48 (4.29)\\ \hline
$P_{uss}^{10\frac{1}{2}}\left({\frac{1}{2}^-}\right)$            &$4.60\pm0.11$     &$(5.87\pm0.89)\times10^{-3}$  & 4.63 (4.41) \\ \hline
$P_{sss}^{10\frac{1}{2}}\left({\frac{1}{2}^-}\right)$            &$4.71\pm0.11$     &$(6.84\pm1.00)\times10^{-3}$  & 4.77\\ \hline

$P_{uuu}^{11\frac{1}{2}}\left({\frac{1}{2}^+}\right)$            &$4.56\pm0.15$     &$(1.97\pm0.40)\times10^{-3}$  & 4.33 (4.04)\\ \hline
$P_{uus}^{11\frac{1}{2}}\left({\frac{1}{2}^+}\right)$            &$4.67\pm0.14$     &$(2.42\pm0.47)\times10^{-3}$  & 4.48 (4.29)\\ \hline
$P_{uss}^{11\frac{1}{2}}\left({\frac{1}{2}^+}\right)$            &$4.78\pm0.13$     &$(2.88\pm0.53)\times10^{-3}$  & 4.63 (4.41)\\ \hline
$P_{sss}^{11\frac{1}{2}}\left({\frac{1}{2}^+}\right)$            &$4.89\pm0.13$     &$(3.44\pm0.61)\times10^{-3}$  & 4.77\\ \hline

$P_{uuu}^{10\frac{1}{2}}\left({\frac{1}{2}^+}\right)$            &$5.12\pm0.08$     &$(8.01\pm0.92)\times10^{-3}$  & 4.33 (4.04)\\ \hline
$P_{uus}^{10\frac{1}{2}}\left({\frac{1}{2}^+}\right)$            &$5.19\pm0.08$     &$(8.92\pm1.05)\times10^{-3}$  & 4.48 (4.29)\\ \hline
$P_{uss}^{10\frac{1}{2}}\left({\frac{1}{2}^+}\right)$            &$5.26\pm0.08$     &$(9.93\pm1.21)\times10^{-3}$  & 4.63 (4.41)\\ \hline
$P_{sss}^{10\frac{1}{2}}\left({\frac{1}{2}^+}\right)$            &$5.40\pm0.08$     &$(12.17\pm1.28)\times10^{-3}$ & 4.77\\ \hline
   \hline
\end{tabular}
\end{center}
\caption{ The    masses and pole residues of the hidden-charm pentaquark states with the $J^P={\frac{1}{2}}^\pm$, where the $B_{10}$ and $B_8$ denote the decuplet  and octet baryons with the quark constituents $q_1q_2q_3$ respectively \cite{WangZG-Penta-EPJC-2016-142}.    }\label{mass-EPJC-2016-142}
\end{table}

\begin{table}
\begin{center}
\begin{tabular}{|c|c|c|c|c|c|c|c|}\hline\hline
                                                          &$M_P(\rm{GeV})$  &$\lambda_P(10^{-3}\rm{GeV}^6)$   \\ \hline

$P_{uuu}^{10\frac{1}{2}}\left({\frac{3}{2}^-}\right)$     &$4.39\pm0.13$    &$2.25\pm0.40$ \\ \hline
$P_{uus}^{10\frac{1}{2}}\left({\frac{3}{2}^-}\right)$     &$4.51\pm0.12$    &$2.75\pm0.45$ \\ \hline
$P_{uss}^{10\frac{1}{2}}\left({\frac{3}{2}^-}\right)$     &$4.60\pm0.11$    &$3.19\pm0.50$ \\ \hline
$P_{sss}^{10\frac{1}{2}}\left({\frac{3}{2}^-}\right)$     &$4.70\pm0.11$    &$3.78\pm0.57$ \\ \hline

$P_{uuu}^{11\frac{1}{2}}\left({\frac{3}{2}^-}\right)$     &$4.39\pm0.14$    &$3.75\pm0.68$ \\ \hline
$P_{uus}^{11\frac{1}{2}}\left({\frac{3}{2}^-}\right)$     &$4.51\pm0.12$    &$4.64\pm0.77$ \\ \hline
$P_{uss}^{11\frac{1}{2}}\left({\frac{3}{2}^-}\right)$     &$4.62\pm0.11$    &$5.60\pm0.88$ \\ \hline
$P_{sss}^{11\frac{1}{2}}\left({\frac{3}{2}^-}\right)$     &$4.71\pm0.11$    &$6.47\pm1.00$ \\ \hline

$P_{uuu}^{11\frac{3}{2}}\left({\frac{3}{2}^-}\right)$     &$4.39\pm0.14$    &$3.74\pm0.70$ \\ \hline
$P_{uus}^{11\frac{3}{2}}\left({\frac{3}{2}^-}\right)$     &$4.52\pm0.12$    &$4.64\pm0.79$ \\ \hline
$P_{uss}^{11\frac{3}{2}}\left({\frac{3}{2}^-}\right)$     &$4.61\pm0.12$    &$5.52\pm0.90$   \\ \hline
$P_{sss}^{11\frac{3}{2}}\left({\frac{3}{2}^-}\right)$     &$4.72\pm0.11$    &$6.52\pm1.02$  \\ \hline

$P_{uuu}^{10\frac{1}{2}}\left({\frac{3}{2}^+}\right)$     &$4.51\pm0.13$    &$0.99\pm0.19$  \\ \hline
$P_{uus}^{10\frac{1}{2}}\left({\frac{3}{2}^+}\right)$     &$4.60\pm0.12$    &$1.21\pm0.22$ \\ \hline
$P_{uss}^{10\frac{1}{2}}\left({\frac{3}{2}^+}\right)$     &$4.72\pm0.11$    &$1.45\pm0.25$ \\ \hline
$P_{sss}^{10\frac{1}{2}}\left({\frac{3}{2}^+}\right)$     &$4.81\pm0.12$    &$1.71\pm0.29$ \\ \hline

$P_{uuu}^{11\frac{1}{2}}\left({\frac{3}{2}^+}\right)$     &$4.91\pm0.09$    &$5.04\pm0.68$ \\ \hline
$P_{uus}^{11\frac{1}{2}}\left({\frac{3}{2}^+}\right)$     &$4.99\pm0.09$    &$5.83\pm0.82$ \\ \hline
$P_{uss}^{11\frac{1}{2}}\left({\frac{3}{2}^+}\right)$     &$5.08\pm0.09$    &$6.72\pm0.98$  \\ \hline
$P_{sss}^{11\frac{1}{2}}\left({\frac{3}{2}^+}\right)$     &$5.18\pm0.09$    &$7.73\pm1.10$ \\ \hline

$P_{uuu}^{11\frac{3}{2}}\left({\frac{3}{2}^+}\right)$     &$5.14\pm0.08$    &$7.86\pm0.91$   \\ \hline
$P_{uus}^{11\frac{3}{2}}\left({\frac{3}{2}^+}\right)$     &$5.21\pm0.08$    &$8.79\pm1.07$   \\ \hline
$P_{uss}^{11\frac{3}{2}}\left({\frac{3}{2}^+}\right)$     &$5.28\pm0.08$    &$9.91\pm1.25$   \\ \hline
$P_{sss}^{11\frac{3}{2}}\left({\frac{3}{2}^+}\right)$     &$5.36\pm0.08$    &$11.20\pm1.42$   \\ \hline
 \hline
\end{tabular}
\end{center}
\caption{ The masses and pole residues of the hidden-charm pentaquark states with the $J^P={3\over 2}^\pm$ \cite{WangZG-Penta-NPB-2016-163}. }\label{mass-NPB-2016-163}
\end{table}

\section*{Acknowledgements}
This  work is supported by National Natural Science Foundation, Grant Number  11775079.

\end{document}